\documentclass[11pt]{article}
\usepackage{eurosym}
\usepackage{makeidx}
\usepackage{amsfonts}
\usepackage{amssymb}
\usepackage{amsmath}
\usepackage{geometry}
\usepackage{appendix}

\newcommand{\func}[1]{\operatorname{#1}}

\DeclareMathSizes{10.95}{10}{7}{6}
\begin{document}

\title{Financial Interactions and Collective States\\
Part I. Investors and Firms}
\author{Pierre Gosselin\thanks{%
Pierre Gosselin : Institut Fourier, UMR 5582 CNRS-UGA, Universit\'{e}
Grenoble Alpes, BP 74, 38402 St Martin d'H\`{e}res, France.\ E-Mail:
Pierre.Gosselin@univ-grenoble-alpes.fr} \and A\"{\i}leen Lotz\thanks{%
A\"{\i}leen Lotz: Cerca Trova, BP 114, 38001 Grenoble Cedex 1, France.\
E-mail: a.lotz@cercatrova.eu}}
\date{September 2025}
\maketitle

\begin{abstract}
In a previous work, we applied a field formalism to analyze capital
allocation and accumulation within a network of investors and firms. In that
framework, financial agents could invest in firms or in other investors, and
banks--introduced as investors with a credit multiplier--
could play a stabilizing or destabilizing role. Collective states emerged
from these interactions, reflecting different configurations of capital
distribution and stability in the financial system. However, these results
relied on the assumption that financial connections were exogenous.

The present paper removes this assumption by modeling financial connections
as dynamic endogenous variables. Specifically, we extend the framework by
introducing a field representation of the network of financial connections.
The collective states previously identified are now embedded in a broader
class of states, characterized by the structure of investment stakes among
investors. We show that these collective states consist of inter-connected
groups of agents, along with their returns and disposable capital. The model
reveals the emergence of two investor classes: high- and
low-return (and capital) agents. High-return investors, in particular, act
as a source of instability within the system, enabling transitions between
configurations. In each collective state, some sectors may experience
defaults. When the collective state exhibits specific structural conditions,
defaults may spread across a significant share of the group.

Key words: Financial Markets, Real Economy, Capital Allocation, Statistical
Field Theory, Background fields, Collective states, Multi-Agent Model,
Interactions.

JEL Classification: B40, C02, C60, E00, E1, G10
\end{abstract}

\section{Introduction}

In a series of papers\footnote{%
Gosselin, Lotz and Wambst (2017, 2020, 2021)}, we developed a field
formalism that demonstrates how several collective states may emerge from a
landscape of microeconomic interactions. These collective states are neither
an ad hoc macroeconomic description, nor a representative agent aggregate;
they nonetheless constitute structurally static equilibria that shape
individual dynamics and may appear - or vanish - through transitions from
one collective state to another.

In Gosselin and Lotz (2024), we described the diffusion of capital across
sectors between three types of agents: firms, investors, and banks. The
resulting collective states, which gave the levels of capital and returns
for each sector, emerged from the dynamics of capital flows and, indirectly,
by the connections between agents.

However, the stakes between agents - that is their connections - remained
fixed: sectors did not reorganize in response to capital flows, no internal
mechanism of transition emerged between collective states. Whereas
individual dynamics should lead to transitions across collective states, the
collective states obtained exhibited no intrinsic dynamical character.

Since collective states emerge from the network of connections between the
agents that compose them, endogenizing these connections should naturally
give rise to transitions between collective states.

This paper refines Gosselin and Lotz (2024) in this direction. Stakes
between agents, here, the stakes of each agent, now derive from the
optimization by each agent of their objective function.\ In terms of fields,
this translates into the introduction of an additional field, whose
variables are the stakes, and whose action functional reflects the agents'
microeconomic objective functions\footnote{%
See Gosselin and Lotz (2024) for the translation method.}.\ The minimization
of this field action functional, along with the investors' return equations
of Gosselin and Lotz (2024), will yield the collective states of the
system.\ Each of these collective states will be described by the investors'
stakes in firms or investors between sectors, the investors' and firms'
returns and the disposable capital per sector.

Our results reveal the underlying dynamics behind static collective states.\
They also provide the parameters and conditions that are likely to trigger
transitions between them.

We show that agents tend to organize into several groups of investors. These
groups may be weakly interacting with each other, and their size and shape
depend on the uncertainty in investors' cross-sectoral investments. The
partition into groups is not unique, which reveals the multiplicity of
collective states.

Two distinct types of collective states emerge: no-default states, in which
all firms hold sufficient capital, and default states, characterized by the
undercapitalization of certain firms, potentially leading their sectors and
investors to default.

For these two types of collective states, and within any given group,
several distributions of stakes exist.\ Thus, there are an infinite number
of collective states.\ This reveals that the mechanism of capital and return
diffusion described in Gosselin and Lotz (2024) is inherently unstable. It
also calls for a more thorough analysis of the transitions, stability, and
fluctuations of these collective states, as well as the mechanisms driving
transitions from no-default to default states.

We conduct this analysis by considering variations in the stakes and returns
within the constitutive groups of the collective state. We show that the
stability of each group depends on the homogeneity of the distribution of
capital within this group. Discrepancies in investment induce sectoral
instability. In particular, strong and mutual cross-investments among
investors may drag groups towards instability or even defaults.

This work is organized as follows. Section 3 presents the basic notations
used in this study. Section 4 derives the field model of endogenous stakes.
Section 5 presents the resolution of the model and derives the saddle-point
equation for this field. Section 6 studies the stability of the solutions
found.\ Section 7 synthesizes the results, and section 8 discusses these
results. Section 9 concludes.

\section{Literature review}

Three major directions are related to our approach.

The first direction addresses heterogeneity among agents through
distributions of agents modeled by probability densities. In mean field
games (MFG) and mean field control, individual agents are negligible
relative to the population but interact through aggregate variables (see
Bensoussan, Frehse, and Yam (2018) and Lasry and Lions (2007, 2010a,
2010b)).\ This approach has been applied in the dynamic programming
framework developed by Gomes, Vilanova, and others (see Gomes et al., 2015;
Achdou et al., 2014). Heterogeneous Agent New Keynesian (HANK) models
incorporate similar heterogeneity into macroeconomic structures.\ An
equilibrium probability distribution is derived from a set of optimizing
heterogeneous agents\ in a new Keynesian context (see Kaplan and Violante
(2018), as well as Kaplan, Moll, and Violante (2018) for quantitative
implementations). Information-theoretic approaches build on Sims' (2003,
2006) rational inattention theory to model agents optimizing under
informational constraints (Yang (2018) and Matejka and McKay (2015)). This
information theoretic approach considers probabilistic states around the
equilibrium and replaces the Walrasian equilibrium with a statistical
equilibrium derived from an entropy maximisation program.\ In these three
types of models,\ probability distributions can be seen as particular types
of collective states postulated \textit{a priori} as equilibria of the
microeconomics set-up.

Field economics, on the contrary, builds on the interactions between agents
at the microeconomic level. We do not postulate an equilibrium probability
distribution for each type of agent.\ Rather we consider any possible
probability density for the entire system of $N$ agents and their
interactions, and translate these probability densities in terms of fields.
Since the fields encompass all the possible densities of agents as their
realizations\footnote{%
In our formalism, the notion of fields refers to some abstract complex
functions defined on the state space and is similar to the
second-quantized-wave functions of quantum theory.}, the state-space in
field economics is thus much larger than those considered in the above
approaches . This allows studying the agents' economic structural relations
and the emergence of the collective states induced by these specific
micro-relations, that will in turn impact each agent's stochastic dynamics
at the microeconomic level. These emerging collective states are in general
multiple with their own characteristic, average quantities, distribution of
agents...

Interacting agents with heterogeneous behavioral rules has been dealt by
multi-agent systems, particularly agent-based models (ABMs), with an
emphasis on non-equilibrium dynamics and bounded rationality (Gaffard and
Napoletano (2012), Delli Gatti et al. (2005)). Mandel, Landini, and
Gallegati (2010, 2012) further develop ABMs to capture innovation, sectoral
dynamics, and macroeconomic fluctuations. The field of economic networks,
notably Jackson (2010, 2014), focuses on the structural properties of agent
interactions within economic systems. Both approaches are highly numerical
and model-dependent.\ They also rely on microeconomic relations, such as 
\textit{ad hoc }reaction functions, that may be too simplistic. Field
economics, on the contrary, accounts for transitions between scales.
Macroeconomic patterns do not emerge from the sole dynamics of a large set
of agents: they are grounded in behaviours and interaction structures.
Describing these structures in terms of field theory allows for the
emergence of phases at the macro scale, and the study of their impact at the
individual level.

Econophysics applies statistical physics methods to socio-economic systems
(see Abergel et al. (2011a, 2011b) for reviews of the field's developments).
Lux (2008, 2016) explores stylized facts and agent-based dynamics, while
Kleinert (2009) uses path integral formulations to analyze financial
markets. Other relevant works include Bardoscia et al. (2017), Bouchaud and M%
\'{e}zard (2000), Chakraborti et al. (2011), Chakraborti et al. (2013).
However, Econophysics does not apply the full potential of field theory to
economic systems.\ It rather seeks to reveal empirical laws in economic
systems. Field economics, in contrast, keeps track of usual microeconomic
concepts, such as utility functions, expectations, and forward-looking
behaviors.\ It includes these behaviors into the analytical treatment of
multi-agent systems by translating the main characteristics of optimizing
agents in terms of statistical systems. Closer to our approach, Bardoscia et
al (2017) study a general equilibrium model for a large economy in the
context of statistical mechanics, and show that phase transitions may occur
in the system. Our problematic is similar, but our use of field theory deals
with a larger class of dynamic models.

The second direction addresses interactions between Finance and the Real
Economy. The interaction between financial frictions and capital
accumulation have typically been modeled within DSGE frameworks, enriched
with credit constraints, incomplete markets, or firm-level heterogeneity.
Cochrane (2006) provides a review of asset pricing in macro models.
Bernanke, Gertler, and Gilchrist (1999) model the financial accelerator,
linking firm balance sheets to investment dynamics. Holmstrom and Tirole
(1997) study liquidity provision under moral hazard. More recent work
includes Campello et al. (2010) and Quadrini (2012). Other contributions
extend this framework to heterogeneous agents and endogenous risk (Grassetti
et al. (2022), Grosshans and Zeisberger (2018), B\"{o}hm, Kuehn, and
Schmedders (2008), Khan and Thomas (2013), Monacelli et al. (2011) and Moll
(2014)).

Field economics differ from DSGE lodel in that their models are micro models
that stand for the entire set of agents.\ This does not allow to study the
diffusion and circulation of of capital among agents. Field economics, on
the contrary, studies the entire system of agents.\ When dealing with
capital, it describes the different states that result from this diffusion
and circulation of capital. Besides, the relative (un)stability of capital
allocation can be assessed globally: it is the relative distribution of
returns or external conditions that determine investors' allocation of
capital to firms. We can study how a local interest rate or a sector change
in returns would impact the equilibrium of one sector, and the whole system.

The third direction covers the literature on default and systemic risk, and
their contagion within financial networks. Early theoretical models by Allen
and Gale (2000), Cifuentes, Ferrucci, and Shin (2005), and Gai and Kapadia
(2010). The network-based approach has been formalized by Acemoglu,
Ozdaglar, and Tahbaz-Salehi (2015) and extended by Bardoscia et al. (2019)
and Glasserman and Young (2015) using graph-theoretic stress-testing
techniques and feedback loop dynamics (see also Battiston et al. (2012,
2020) for recursive losses in interconnected systems and Haldane and May
(2011) for an ecology and epidemiology approach to financial instability).
Empirically, contributions include Reinhart and Rogoff (2009), Gennaioli,
Martin, and Rossi (2012, 2018), Adrian and Brunnermeier (2016), Langfield,
Liu, and Ota (2020).

This paper adds to these studies by considering both firms and investors in
the analysis. Moreover, collective states are described both in terms of
global averages and sectoral quantities,\ allowing disparities in firms'
returns and borrowing conditions across agents to be explicitly accounted
for. Our analysis therefore moves back and forth between the micro and macro
levels to identify the conditions under which micro-level defaults propagate
to the macroeconomic level.

\section{Preliminary: basic notations}

We briefly recall the notations and the basic assumptions of our model.

The economic space comprises an infinite number of \emph{sectors}. Each
sector is labeled by a position, denoted by $X$, $X^{\prime }$, $X^{\prime
\prime }$ and so on. Each sector includes both investors and firms.

Since we are working within a field model, agents are, among other things,
characterized by the sector they occupy. Agents are not indexed
individually; rather, they are distinguished by their economic variables and
by their sector. We consider two types of agents: investors and firms. We
denote the investors located in sector $X$ as \emph{investor }$X$, and the
firms located in sector $X$ as firm $X$. Variables pertaining to investors
will be denoted with a hat.

In this model, only investors engage in investment. They may invest either
in other investors or in firms, whether within their own sector or in other
sectors. Their investments or \emph{stakes} can take two forms: equities
(i.e., the purchase of shares) or loans. An investor is thus characterized
by the stakes it owns. We denote the stakes by $S$, and use the index $\eta $%
, which takes the value $E$ for equities and $L$ for loans.

Investments made by one investor in another investor are denoted $\hat{S}$,
and is the sum of $\hat{S}_{E}$ and $\hat{S}_{L}$, and investments made in a
firm, denoted $S$, which is similarly the sum of $S_{E}$ and $S_{L}$.

In our field model of investment, two sectors must be distinguished: the
sector in which the investor is located (the origin sector of the
investment), and the sector in which the investment is made (the destination
sector).

To differentiate origin and destination sectors, we will use notations such
as $X$ and $X^{\prime }$ , or $X^{\prime }$ and $X^{\prime \prime }$,
respectively.\ When talking about a stake, we may write in parenthesis, on
the right, the sector of origin, on the left, the sector of destination, and
in between, when needed, the capital of the beneficiary. For instance, the
stake $\hat{S}_{\eta }\left( X^{\prime },\hat{K}^{\prime },X\right) $ is
read: the stake of type $\eta $ made by investor $X$ (located in sector $X$)
towards investor $X^{\prime }$, who has capital $K^{\prime }$. When capital
is not specified, as in $\hat{S}_{\eta }\left( X^{\prime },X\right) $, we
refer to the average stake from investor $X$ to investor $X^{\prime }$, the
average being taken over the capital of the beneficiaries of the sector $%
X^{\prime }$

An \emph{average stake} is defined as the inter- and intra-sectoral average
over all sectors $X^{\prime }$ and $X$, of the stakes taken by investors in
those sectors. It is denoted by $\left\langle \hat{S}_{\eta }\left(
X^{\prime },X\right) \right\rangle $ and $\left\langle S_{\eta }\left(
X^{\prime },X\right) \right\rangle $.

We will use the term \emph{aggregate} specifically to denote a sum over all
sectors connected to one specific sector.

The \emph{inward aggregate stake} is the total amount of stakes invested by
all investors in the economic space into a given investor or firm in sector $%
X^{\prime }$, relative to the capital available to the investors or firms of
that sector, respectively. It is denoted $\hat{S}_{\eta }\left( X^{\prime
}\right) $ for investors and $S_{\eta }\left( X^{\prime }\right) $ for firms.

The\textbf{\ }\emph{outward aggregate stakes}\textbf{\ }$\left\langle \hat{S}%
_{\eta }\left( X^{\prime },X\right) \right\rangle _{X^{\prime }}$\textbf{\ }%
compute the average stakes taken by an investors $X$\ in investors of all
sectors.

The \emph{average aggregate stake} is the average, across all sectors, of
the aggregate stakes taken in investors and firms in each sector, denoted $%
\left\langle \hat{S}_{\eta }\left( X^{\prime }\right) \right\rangle $, and $%
\left\langle S_{\eta }\left( X^{\prime }\right) \right\rangle $
respectively. Summing both equity and loan investments into investors is
denoted $\left\langle \hat{S}\left( X^{\prime }\right) \right\rangle $ and $%
\left\langle S\left( X^{\prime }\right) \right\rangle $ in firms.

Average stakes are not necessarily equal to average aggregate stakes,
because the latter are measured relative to the capital of the destination
sectors.

The connections between agents are non-normalized proportions of
investment.\ They are denoted $\hat{k}_{\eta }$ and $k_{\eta }$ according to
the same notation rules than the stakes\footnote{%
The indices $\eta =E$ or $L$ replace $\eta =1$ or $2$ in Gosselin and Lotz
(2024).}. The proportion of aggregated stake $\hat{S}_{\eta }\left(
X^{\prime }\right) $, $\hat{S}\left( X^{\prime }\right) $, $S_{\eta }\left(
X^{\prime }\right) $ and $S\left( X^{\prime }\right) $ will be denoted $%
\underline{\hat{k}}_{\eta }\left( X^{\prime }\right) $, $\underline{\hat{k}}%
\left( X^{\prime }\right) $, $\underline{k}_{\eta }\left( X^{\prime }\right) 
$ and $\underline{k}_{\eta }\left( X^{\prime }\right) $ respectively.

The \emph{agent's disposable capital}, the capital used to produce or
invest, denoted $K$, is the sum of agent's private capital along with the
capital he is allocated to, either through shares or loans.

The \emph{agents' average disposable capital in sector }$X$\emph{\ }is the
average of the agents' disposable capital in sector $X$.\ It is denoted $%
\hat{K}_{X}$ for investors, and $K_{X}$ for firms.\ 

The \emph{aggregate disposable capital in sector} $X$ will be denoted by $%
\hat{K}\left[ X\right] $ and $K\left[ X\right] $ for investors and firms,
respectively.

The \emph{average disposable capital }per investor and per firm across all
sectors are the averages of $\hat{K}_{X}$ and $K_{X}$ across all sectors.
They will be denoted by $\left\langle \hat{K}\right\rangle $ and $%
\left\langle K\right\rangle $.

We will denote the \emph{background fields}, the fields minimizing the
action functionals of investors and firms, $\hat{\Psi}\left( X\right) $ and $%
\Psi \left( X\right) $,\ respectively, and the \emph{densities }of investors
and of firms $X$, $\left\vert \hat{\Psi}\left( X\right) \right\vert ^{2}$and 
$\left\vert \Psi \left( X\right) \right\vert ^{2}$, respectively. The total
number, across all sectors, of investors and of firms will be denoted$%
\left\Vert \hat{\Psi}\right\Vert ^{2}$ and $\left\Vert \Psi \right\Vert ^{2}$%
, respectively. Note that, by construction, the disposable capital per
sector for investors and firms\footnote{%
Expressions of $\hat{K}\left[ \hat{X}^{\prime }\right] $\ and $K^{\prime }%
\left[ X^{\prime }\right] $\ as functions of returns and shares are given in
Appendix 1.} satisfy the identities: 
\begin{equation*}
\hat{K}\left[ X\right] =\hat{K}_{\hat{X}}\left\vert \hat{\Psi}\left(
X\right) \right\vert ^{2}
\end{equation*}%
\begin{equation*}
K\left[ X\right] =K_{X}\left\vert \Psi \left( X\right) \right\vert ^{2}
\end{equation*}%
The (inverse) \emph{uncertainty }of investor $X$\ about its stakes in
investors $X^{\prime }$\ and firms $X^{\prime }$\ will be measured by $\hat{w%
}_{\eta }\left( X^{\prime },X\right) $\ and $w_{\eta }\left( X^{\prime
},X\right) $\ respectively. When averaged across sectors $X^{\prime }$, they
measure the \emph{average uncertainty }among investors $X$ about their
stakes. They will be denoted $\hat{w}_{\eta }\left( X\right) $\ and $w_{\eta
}\left( X\right) $.

\section{The general set-up}

In Gosselin and Lotz (2024), we designed a field model for investors and
firms using two fields, one for each type of agent. However, the investor's
field return equation depended on exogenous connections\footnote{%
The full expression including default is given in Appendix 1 and studied
later.}. To endogeneize this inital model, we need to complete this basic
formulation by adding a third field, linking the investors investment
decisions and their returns. To do so, we will reexpress the original return
equation in terms of stakes (section 4.1), then endogeneize it by
introducing a third field in the model, the field of stakes (section 4.2).
Ultimately, the process of endogeneizing investment decisions calls for
modeling the structure of uncertainty (section 4.3).

\subsection{The investors' field return equations}

\subsubsection{Basic formulation}

The average return of an investor $X$, denoted $\hat{f}\left( X\right) $,
defines the excess return of this investor $X$ over interest rates interest
rate as $\hat{f}\left( X\right) -\bar{r}$. Since, it is computed over
investor's disposable capital, loans included, we must divide this excess
return by a factor $1+\underline{\hat{k}}_{L}\left( \hat{X}\right) $ to
account for the impact of loan repayments on investor yield.\footnote{%
See Gosselin and Lotz (2024) for details.}. The real average excess return
of an investor is therefore: 
\begin{equation*}
\frac{\hat{f}\left( X\right) -\bar{r}}{1+\underline{\hat{k}}_{L}\left(
X\right) }
\end{equation*}%
Similarly, the real excess return of a firm has the form: 
\begin{equation*}
\frac{\hat{f}\left( X^{\prime }\right) -\bar{r}}{1+\underline{\hat{k}}%
_{L}\left( X^{\prime }\right) }
\end{equation*}%
Here again, it is computed with respect to its disposable capital.

The investors' return equations in each sector $X$ relates the excess
returns of all investors and firms across sectors. Under a no-default
scenario, they can be written\footnote{%
See Gosselin and Lotz (2024) for details.} as:

\begin{eqnarray}
&&\frac{\hat{f}\left( X\right) -\bar{r}}{1+\underline{\hat{k}}_{L}\left(
X\right) }=\int \frac{\hat{k}_{E}\left( X^{\prime },X\right) \hat{K}\left[
X^{\prime }\right] }{1+\underline{\hat{k}}\left( X^{\prime }\right) }\frac{%
\hat{f}\left( X^{\prime }\right) -\bar{r}}{1+\underline{\hat{k}}_{L}\left(
X^{\prime }\right) }dX^{\prime }  \label{Rtq} \\
&&+\int \frac{k_{E}\left( X^{\prime },X\right) K\left[ X^{\prime }\right] }{%
1+\underline{k}\left( X^{\prime }\right) }\left( f\left( X^{\prime }\right) -%
\bar{r}\right) dX^{\prime }  \notag
\end{eqnarray}%
In the above equation, the excess return $\frac{\hat{f}\left( X\right) -\bar{%
r}}{1+\underline{\hat{k}}_{L}\left( X\right) }$ can be decomposed into two
components: first, the stakes taken by an investor $X$ in other investors%
\footnote{%
The factor $\frac{\hat{K}\left[ \hat{X}^{\prime }\right] }{1+\underline{\hat{%
k}}\left( \hat{X}^{\prime }\right) }$ represents the investor's private
capital, i.e. their share of disposable capital, which determines the
proportional size of participations.},\footnote{%
The coefficient $\underline{k}\left( X\right) $, defined by $\underline{\hat{%
k}}\left( \hat{X}^{\prime }\right) =\hat{k}\left( \hat{X}^{\prime }\right) 
\frac{\hat{K}\left[ \hat{X}^{\prime }\right] }{\left\langle \hat{K}%
\right\rangle \left\vert \hat{\Psi}_{0}\left( \hat{X}\right) \right\vert ^{2}%
}$, represents the ratio of investors' capital in a sector relative to this
sector's firms private capital (see Gosselin and Lotz (2024) for details).
\par
{}}:

\begin{equation*}
\frac{\hat{k}_{E}\left( X^{\prime },X\right) \hat{K}\left[ X^{\prime }\right]
}{1+\underline{\hat{k}}\left( X^{\prime }\right) }
\end{equation*}%
multiplied by their return $\frac{\hat{f}\left( X^{\prime }\right) -\bar{r}}{%
1+\underline{\hat{k}}_{L}\left( X^{\prime }\right) }$; and second, the
returns of investor $X$ from firms $X^{\prime }$:%
\begin{equation*}
f\left( X^{\prime }\right) -\bar{r}
\end{equation*}%
multiplied by the share invested by investor $X$ in those firms:%
\begin{equation*}
\frac{k_{E}\left( X^{\prime },X\right) K^{\prime }\left[ X^{\prime }\right] 
}{1+\underline{k}\left( X^{\prime }\right) }
\end{equation*}%
This share is proportional to $k_{E}\left( X^{\prime },X\right) $, the
proportion of connections\footnote{%
The coefficient $\underline{k}\left( X\right) $, defined by:%
\begin{equation*}
\underline{k}\left( X\right) =k\left( X\right) \frac{\hat{K}_{X}\left\vert 
\hat{\Psi}\left( \hat{X}\right) \right\vert ^{2}}{\left\langle
K\right\rangle \left\vert \Psi _{0}\left( X\right) \right\vert ^{2}}
\end{equation*}%
represents the ratio of investors' capital allocated to a sector relative to
the private capital of firms in that sector.} between an investor $X$ and a
firm $X^{\prime }$, multiplied by the firm's private capital $\frac{K\left[
X^{\prime }\right] }{1+\underline{k}\left( X^{\prime }\right) }$.

In this basic formulation, the intersectoral links $k_{E}$, $\hat{k}_{E}$, $%
k_{L}$ and $\hat{k}_{L}$ are modeled as exogenous and non-normalized.
However, investors' return equation can also be expressed in terms of stakes
owned by each agent.

\subsubsection{Formulation in terms of stakes}

To study the dynamics of interacting groups of investors, we must translate
the return equation (\ref{Rtq}) into a form suited to deal with the presence
of endogeneous stakes in the model. To implement it, we will need to
consider several measures of stakes between agents.\ Specifically, we will
consider the stakes according to the sector of origin of the investor that
owns them, but also according to the sector they are allocated in.

\paragraph{Stakes according to their investor's sector of origin}

The stakes invested by an investor $X$ in investor $X^{\prime }$ with
capital $\hat{K}^{\prime }$ are defined by: 
\begin{equation}
\hat{S}_{\eta }\left( X^{\prime },\hat{K}^{\prime },X\right) =\frac{\hat{k}%
_{\eta }\left( X^{\prime },X\right) \hat{K}^{\prime }\left\vert \hat{\Psi}%
\left( \hat{K}^{\prime },X^{\prime }\right) \right\vert ^{2}}{1+\underline{%
\hat{k}}\left( X^{\prime }\right) }  \label{Sh1}
\end{equation}%
and the stakes invested by an investor $X$ in a firm $X^{\prime }$ with
capital $K^{\prime }$ as:%
\begin{equation}
S_{\eta }\left( X^{\prime },K^{\prime },X\right) =\frac{k_{\eta }\left(
X^{\prime },X\right) \left\vert \Psi \left( K^{\prime },X^{\prime }\right)
\right\vert ^{2}K^{\prime }}{1+\underline{k}\left( X^{\prime }\right) }
\label{Sh2}
\end{equation}%
The total stakes invested by an investor $X$ in both investors\ $X^{\prime }$
and firms $X^{\prime }$, denoted $\hat{S}_{\eta }\left( X^{\prime },X\right) 
$ and $S_{\eta }\left( X^{\prime },X\right) $, respectively, are the
individual investments $\hat{S}_{\eta }\left( X^{\prime },\hat{K}^{\prime
},X\right) $ and $S_{\eta }\left( X^{\prime },K^{\prime },X\right) $
aggregated over the capital levels $\hat{K}^{\prime }$.\ They write:%
\begin{equation*}
\hat{S}_{\eta }\left( X^{\prime },X\right) \equiv \int \frac{\hat{k}_{\eta
}\left( X^{\prime },X\right) \hat{K}^{\prime }\left\vert \hat{\Psi}\left(
X^{\prime }\right) \right\vert ^{2}}{1+\underline{\hat{k}}\left( X^{\prime
}\right) }=\frac{\hat{k}_{\eta }\left( X^{\prime },X\right) \hat{K}_{\hat{X}%
^{\prime }}\left\vert \hat{\Psi}\left( X^{\prime }\right) \right\vert ^{2}}{%
1+\underline{\hat{k}}\left( X^{\prime }\right) }
\end{equation*}%
and:%
\begin{equation*}
S_{\eta }\left( X^{\prime },X\right) \equiv \frac{k_{\eta }\left( X^{\prime
},X\right) K_{X^{\prime }}\left\vert \Psi \left( X^{\prime }\right)
\right\vert ^{2}}{1+\underline{k}\left( X^{\prime }\right) }
\end{equation*}%
The total stakes allocated by investors $X$ in investors $X^{\prime }$,
denoted $\hat{S}\left( X^{\prime },X\right) $, are the amount of their
shares and loans: 
\begin{equation*}
\hat{S}\left( X^{\prime },X\right) =\hat{S}_{E}\left( X^{\prime },X\right) +%
\hat{S}_{L}\left( X^{\prime },X\right)
\end{equation*}%
and similarly, the total stakes allocated by investors $X$ in firms $%
X^{\prime }$, denoted $S\left( X^{\prime },X\right) $, will be: 
\begin{equation*}
S\left( X^{\prime },X\right) =S_{E}\left( X^{\prime },X\right) +S_{L}\left(
X^{\prime },X\right)
\end{equation*}%
Since all disposable capital is invested, these quantities satisfy the
constraint:%
\begin{equation}
\int \left( \hat{S}_{E}\left( X^{\prime },X\right) +\hat{S}_{L}\left(
X^{\prime },X\right) \right) dX^{\prime }+\int \left( S_{E}\left( X^{\prime
},X\right) +S_{L}\left( X^{\prime },X\right) \right) dX^{\prime }=1
\label{CSt}
\end{equation}%
We also define the outward aggregate stakes, the average stakes of an
investor $X$ in investors $X^{\prime }$ by:%
\begin{equation}
\left\langle \hat{S}_{\eta }\left( X^{\prime },X\right) \right\rangle
_{X^{\prime }}  \label{Twn}
\end{equation}%
and:%
\begin{equation}
\left\langle \hat{S}\left( X^{\prime },X\right) \right\rangle _{X^{\prime }}
\label{Tws}
\end{equation}%
where the bracket $\left\langle {}\right\rangle _{X^{\prime }}$ denotes the
average over investors $X^{\prime }$.

\paragraph{Stakes according to their sector of destination: inward aggregate
stakes}

We derive $\hat{S}_{\eta }\left( X^{\prime }\right) $, the inward aggregate
stakes, the aggregate stakes allocated in investors $X^{\prime }$ with
respect to the disposable capital in sector $X^{\prime }$. Similarly, we
compute $S_{\eta }\left( X^{\prime }\right) $, the inward aggregate stakes
allocated in firms of sector $X^{\prime }$ with respect to the disposable
capital in sector $X^{\prime }$. They write:

\begin{eqnarray}
\hat{S}_{\eta }\left( X^{\prime }\right) &=&\int \hat{S}_{\eta }\left(
X^{\prime },X\right) \frac{\hat{K}_{X}\left\vert \hat{\Psi}\left( X\right)
\right\vert ^{2}}{\hat{K}_{X^{\prime }}\left\vert \hat{\Psi}\left( X^{\prime
}\right) \right\vert ^{2}}dX  \label{Gsn} \\
S_{\eta }\left( X^{\prime }\right) &=&\int S_{\eta }\left( X^{\prime
},X\right) \frac{\hat{K}_{X}\left\vert \hat{\Psi}\left( X\right) \right\vert
^{2}}{K_{X^{\prime }}\left\vert \Psi \left( X^{\prime }\right) \right\vert
^{2}}dX  \label{Gst}
\end{eqnarray}%
so that $\hat{S}_{E}\left( X^{\prime }\right) $ and $\hat{S}_{L}\left(
X^{\prime }\right) $ are the proportions of aggregate cross-sectoral equity
or debt invested into investors $\hat{X}^{\prime }$, within the sector
disposable capital $\hat{K}_{\hat{X}^{\prime }}\left\vert \hat{\Psi}\left(
X^{\prime }\right) \right\vert ^{2}$, and $S_{E}\left( X^{\prime }\right) $
and $S_{L}\left( X^{\prime }\right) $ are the same proportions invested into
firms $X^{\prime }$ within their own disposable capital $K_{X^{\prime
}}\left\vert \Psi \left( X^{\prime }\right) \right\vert ^{2}$.

Ultimately, we define the global proportion of aggregate cross-sectoral
equity- or debt- investment in each sector: 
\begin{equation}
\hat{S}\left( X^{\prime }\right) =\hat{S}_{E}\left( X^{\prime }\right) +\hat{%
S}_{L}\left( X^{\prime }\right)  \label{Gsv}
\end{equation}%
and:%
\begin{equation}
S\left( X^{\prime }\right) =S_{E}\left( X^{\prime }\right) +S_{L}\left(
X^{\prime }\right)  \label{Gsw}
\end{equation}

\paragraph{The return equation expressed in terms of stakes}

Assuming\footnote{%
As in Gosselin and Lotz (2024).} investors invest in neighbouring firms, we
can write:%
\begin{eqnarray*}
S_{E}\left( X^{\prime },X\right) &=&S_{E}\left( X,X\right) \delta \left(
X^{\prime }-X\right) \\
S_{L}\left( X^{\prime },X\right) &=&S_{L}\left( X,X\right) \delta \left(
X^{\prime }-X\right)
\end{eqnarray*}%
where $\delta \left( X^{\prime }-X\right) $\ is the Dirac function. It is
equal to $0$\ for $X^{\prime }\neq X$. Under these assumptions, the
constraint (\ref{CSt}) simplifies to:%
\begin{equation*}
\int \left( \hat{S}_{E}\left( X^{\prime },X\right) +\hat{S}_{L}\left(
X^{\prime },X\right) \right) dX^{\prime }+S_{E}\left( X,X\right)
+S_{L}\left( X,X\right) =1
\end{equation*}%
and the return equation (\ref{Rtq}) thus writes\footnote{%
See Appendix 1 for details. The case of no default, or default are studied
in Appendices 1.2, 1.3.}:%
\begin{equation}
0=\int \left( \delta \left( X^{\prime }-X\right) -\hat{S}_{E}\left(
X^{\prime },X\right) \right) \widehat{DF}\left( X^{\prime }\right) \hat{R}%
_{exc}\left( X^{\prime }\right) dX^{\prime }-S_{E}\left( X,X\right)
R_{exc}\left( X\right)  \label{QDL}
\end{equation}%
where $\widehat{DF}\left( X\right) $ is the investor's discount factor for
debt repayment and writes\footnote{%
In the following, we will also encounter the firm's discount factor for debt
repayment: 
\begin{equation*}
DF\left( X\right) =\frac{1-S\left( X\right) }{1-S_{E}\left( X\right) }
\end{equation*}%
}:%
\begin{equation*}
\widehat{DF}\left( X\right) =\frac{1-\hat{S}\left( X\right) }{1-\hat{S}%
_{E}\left( X\right) }
\end{equation*}%
and:%
\begin{equation*}
\hat{R}_{exc}\left( X^{\prime }\right) =f\left( X^{\prime }\right) -\bar{r}
\end{equation*}%
\begin{equation*}
R_{exc}\left( X\right) =f\left( X\right) -\bar{r}
\end{equation*}%
are the investors and firms excess returns respectively.

\subsection{The field of stakes}

In the above, we have translated the initial investors' return equations in
terms of investment decisions, the stakes of investors. However, this is not
sufficient to consider these decisions as endogenous. To endogenize the
variables pertaining to these stakes, we must design a field theory for the
stakes themselves\footnote{%
This field theory will comprise the field of the stakes, and its action
functional, the function whose minimization will give the most probable
shape for the stakes.}.

To do so, we follow our general approach to translate a micro set-up into a
field formalism.\ We will first formulate a model in which investment
decisions, i.e. the investors' stakes, result from the optimization of a
benefit/uncertainty trade-off (section 4.2.1). We will then translate this
microeconomic framework into a field-theoretic formalism (section 4.2.2).
This formalism will be characterized by a field $\Gamma $ whose arguments
will be the investors' stakes along with its field action functional, that
encompasses the micro behaviors.

\subsubsection{Micro set-up for endogenous stakes}

Let us consider a set of investors, indexed by $i$\ or $j$,\ taking stakes
in other investors and firms, indexed by $k$. Since we will study these
agents within a micro setup, the economic variables will be indexed by the
agents.\ Their respective position within the sector space will be denoted $%
X_{i}$, $X_{j}$ and $X_{k}$, respectively.\ We will further denote $\hat{S}%
_{\eta ij}$\ the stakes taken by an investor $i$\ in an other investor $j$,\
and $S_{\eta ik}$\ the stakes taken by an investor $i$\ in a firm $k$. This
difference notwithstanding, in what follows, the notation will be similar to
that previously adopted.

In a classical framework, each investor optimizes their investments based on
their respective expected returns, and the uncertainty, or risk, associated
with each investment. Assuming agents adjust instantaneously to any change
in their environment, an investor indexed by $j$ maximizes the objective
function:%
\begin{equation*}
\sum_{j}\hat{S}_{Eij}\hat{f}_{j}+\sum_{j}\hat{S}_{Lij}\hat{r}_{j}-\frac{1}{2}%
\sum \frac{\left( \hat{S}_{\eta ij}\right) ^{2}}{\hat{w}_{\eta _{i}}\left( 
\hat{X}_{j}\right) }+\sum_{k}S_{Eik}f_{k}+\sum_{k}S_{Lik}\bar{r}_{k}-\frac{1%
}{2}\sum_{k}\frac{\left( S_{\eta ik}\right) ^{2}}{w_{\eta ik}\left(
X_{k}\right) }
\end{equation*}%
where the coefficients $\hat{w}_{\eta }$ and $w_{\eta }$ represent the
(inverse) uncertainties in returns.

This objective function is maximized under the constraint:%
\begin{equation*}
\sum_{j}\left( \hat{S}_{Eij}+\hat{S}_{Lij}\right) +\sum_{k}\left(
S_{Eik}+S_{Lik}\right) =1
\end{equation*}%
which is implemented by a Lagrange multiplier $\lambda _{i}$. The solutions
of the instantaneous optimization are given by:%
\begin{eqnarray*}
\hat{S}_{Eij} &=&\hat{w}_{Eij}\left( \hat{f}_{j}+\lambda _{i}\right) \\
\hat{S}_{Lij} &=&\hat{w}_{Lij}\left( \hat{r}_{j}+\lambda _{i}\right) \\
S_{Eik} &=&w_{Eik}\left( f_{k}+\lambda _{i}\right) \\
S_{Lik} &=&w_{Lik}\left( \bar{r}_{k}+\lambda _{i}\right)
\end{eqnarray*}%
or, taking into account some inertia in the allocation:%
\begin{eqnarray*}
\alpha \frac{d^{2}}{dt^{2}}\hat{S}_{Eij} &=&-\frac{\hat{S}_{Eij}}{\hat{w}%
_{Eij}}+\left( \hat{f}_{j}+\lambda _{i}\right) \\
\alpha \frac{d^{2}}{dt^{2}}\hat{S}_{Lij} &=&-\frac{\hat{S}_{Lij}}{\hat{w}%
_{Lij}}+\left( \hat{r}_{j}+\lambda _{i}\right) \\
\alpha \frac{d^{2}}{dt^{2}}S_{Eik} &=&-\frac{S_{Eik}}{w_{Eik}}+\left( \hat{r}%
_{j}+\lambda _{i}\right) \\
\alpha \frac{d^{2}}{dt^{2}}S_{Lik} &=&-\frac{S_{Lik}}{w_{Lik}}+\left( \hat{r}%
_{j}+\lambda _{i}\right)
\end{eqnarray*}%
with $\alpha <1$ in general\footnote{%
Note that in the optimization equations, the coefficients uncertainty
coefficients $\hat{w}_{\eta }$\ and $w_{\eta }$\ are perceived as exogenous
by any individual agent. We will see below that, in the context of the field
description, these coefficients are in fact endogenous to the system as a
whole.}.

\subsubsection{Field translation of the set-up}

To translate this micro set-up in terms of field, we must assume a field
whose variables are the stakes taken by investors, their sector, and the
sectors they invest in: 
\begin{equation*}
\Gamma \left( \hat{S}^{\left( T\right) },X^{\prime },X\right) \equiv \Gamma
\left( S_{E},\hat{S}_{E},S_{L},\hat{S}_{L},X^{\prime },X\right)
\end{equation*}%
where the arguments $S_{E}$ and $S_{L}$\ are the stakes of investor $X$ in
firms $\ $of the same sector $X$ through shares and loans respectively,
while $\hat{S}_{E}$ and $\hat{S}_{L}$ are those taken in an investor $%
X^{\prime }$.\ We gather the four possible types of stakes in a vector $\hat{%
S}^{\left( T\right) }$:%
\begin{equation*}
\hat{S}^{\left( T\right) }=\left( S_{E},\hat{S}_{E},S_{L},\hat{S}_{L}\right)
\end{equation*}%
The translation of the micro set-up in terms of fields yields the action
functional $S\left( \Gamma \right) $\footnote{%
See Appendix 2 for the details of the computation.},\footnote{%
An equivalent formulation of (\ref{GCt}) is given in Appendix 2.2.}:%
\begin{eqnarray}
S\left( \Gamma \right) &=&-\int \sigma _{\hat{K}}^{2}\Gamma ^{\dag }\left( 
\hat{S}^{\left( T\right) },X^{\prime },X\right) \nabla _{\hat{S}_{\eta
}^{\left( T\right) }}^{2}\Gamma \left( \hat{S}^{\left( T\right) },X^{\prime
},X\right) d\left( \hat{S}^{\left( T\right) },X^{\prime },X\right)
\label{GCt} \\
&&-\int \beta \left\vert \Gamma \left( \hat{S}^{\left( T\right) },X^{\prime
},X\right) \right\vert ^{2}d\left( \hat{S}^{\left( T\right) },X^{\prime
},X\right)  \notag \\
&&+\sum_{\eta }\int \left( \frac{\left( \hat{S}_{\eta }^{\left( T\right)
}\right) ^{2}}{2\hat{w}_{\eta }^{T}\left( X^{\prime },X\right) }-\hat{V}%
_{\eta }\hat{S}_{\eta }^{\left( T\right) }\right) \left\vert \Gamma \left( 
\hat{S}^{\left( T\right) },X^{\prime },X\right) \right\vert ^{2}d\left( \hat{%
S}^{\left( T\right) },X^{\prime },X\right)  \notag \\
&&+\int \lambda \left( X\right) \left( \sum_{\eta }\int \hat{S}_{\eta
}^{\left( T\right) }\left\vert \Gamma \left( \hat{S}^{\left( T\right)
},X^{\prime },X\right) \right\vert ^{2}dX^{\prime }d\hat{S}^{\left( T\right)
}-1\right) \left\vert \Gamma \left( \hat{S}^{\left( T\right) },X^{\prime
},X\right) \right\vert ^{2}d\left( \hat{S}^{\left( T\right) },X^{\prime
},X\right)  \notag
\end{eqnarray}%
where the functions $\hat{V}_{\eta }\left( \hat{S}_{\eta }\right) $ describe
the returns\footnote{%
See Appendix 2.1.}, and the components of the vector $\left( \hat{w}_{\eta
}^{T}\left( X^{\prime },X\right) \right) $ are the field-translation of the
inverse uncertainty coefficients $\hat{w}_{\eta ij}$ and $w_{\eta ik}$:%
\begin{equation*}
\left( \hat{w}_{\eta }^{T}\left( X^{\prime },X\right) \right) =\left( \hat{w}%
_{E}\left( X^{\prime },X\right) ,\hat{w}_{E}\left( X^{\prime },X\right)
,w_{E}\left( X,X\right) ,w_{L}\left( X,X\right) \right)
\end{equation*}%
which reflect the uncertainty perceived by investors $X$ about their stakes
(either equity or loans) in investors $X^{\prime }$\ and in firms $X$.

The Lagrange multipliers $\lambda \left( X\right) $ implement the constraint:%
\begin{equation*}
\sum_{\eta }\int \hat{S}_{\eta }^{\left( T\right) }\left\vert \Gamma \left( 
\hat{S}^{\left( T\right) },X^{\prime },X\right) \right\vert ^{2}dX^{\prime }d%
\hat{S}^{\left( T\right) }=1
\end{equation*}%
where the integral accounts for the sum of stakes an investor $X$ invests in
various sectors $X^{\prime }$, and the factor $\left\vert \Gamma \left( \hat{%
S}^{\left( T\right) },X^{\prime },X\right) \right\vert ^{2}$ weighs the
stakes $\hat{S}_{\eta }^{\left( T\right) }$ by the number of investors $%
X^{\prime }$ in which they are invested.

We define the partial averages:%
\begin{eqnarray*}
S_{\eta }\left( X,X\right) &=&\int S_{\eta }\left\vert \Gamma \left( \hat{S}%
^{\left( T\right) },X^{\prime },X\right) \right\vert ^{2}d\left( \hat{S}%
^{\left( T\right) },X^{\prime }\right) \\
\hat{S}_{\eta }\left( X^{\prime },X\right) &=&\int \hat{S}_{\eta }\left\vert
\Gamma \left( \hat{S}^{\left( T\right) },X^{\prime },X\right) \right\vert
^{2}d\left( \hat{S}^{\left( T\right) }\right)
\end{eqnarray*}%
so that the constraint on allocation writes:%
\begin{equation*}
\int \left( \hat{S}_{E}\left( X^{\prime },X\right) +\hat{S}_{L}\left(
X^{\prime },X\right) \right) dX^{\prime }+\int \left( S_{E}\left( X^{\prime
},X\right) +S_{L}\left( X^{\prime },X\right) \right) dX^{\prime }=1
\end{equation*}%
Ultimately, the minimization equations of (\ref{GCt}) will be written%
\footnote{%
Derived in Appendix 2.} in terms of the sectoral stakes, that are defined by:%
\begin{equation*}
\hat{S}_{\eta }^{\left( T\right) }\left( \hat{X}^{\prime },\hat{X}\right) =%
\frac{\int \hat{S}_{\eta }^{\left( T\right) }\left\vert \Gamma \left( \hat{S}%
^{\left( T\right) },\hat{X}^{\prime },\hat{X}\right) \right\vert ^{2}d\hat{S}%
^{\left( T\right) }}{\int \left\vert \Gamma \left( \hat{S}^{\left( T\right)
},\hat{X}^{\prime },\hat{X}\right) \right\vert ^{2}d\hat{S}^{\left( T\right)
}}
\end{equation*}

\subsection{Modeling the uncertainty}

In Gosselin and Lotz (2024), investments risk was supposed to be exogenous
and the nature of risk was unspecified. To fully endogeneize the model, we
must now fill this gap and close the system by deriving the investment risk
coefficients $\hat{w}\left( X^{\prime },X\right) $ and $w\left( X\right) $
from some additional assumptions about risk propagation\footnote{%
To be fully precise, the coefficients $\hat{w}\left( \hat{X}^{\prime },\hat{X%
}\right) $ and $w\left( \hat{X}\right) $ are a measure of the inverse of
risk.}. Because investors also invest in each other, their investment risk
is inherently non-local, and emerges endogenously from the structure of
intermediation: risk is composed through layers of exposure and depends on
the investment decisions of others.

\subsubsection{Nature of uncertainty}

In the present model, firms' returns are observed by all investors, albeit
with a delay. This lag generates anticipation errors, and thus uncertainty
in the model.

All investors have a minimal and uniform level of uncertainty regarding the
firms in their own sector. That is, investors will always be more confident
about firms in their own sector than firms in other sectors. Furthermore, no
investor is assumed more confident in their own judgment than any other
investor within the same sector. However, this assumption can be easily
relaxed.

Uncertainty is measured by the variance an investor assigns to his
investment. It depends on the model's parameters and on the investor's
allocation choices. Uncertainty may relate, on the one hand, to other
investors' potential returns: these are \emph{errors of appraisal} about
other investors' strategy, acumen, etc. On the other hand, they can relate
to firms' returns: these are investors' \emph{errors of anticipation }about
firms, first within their own sector --- which is set minimal and uniform by
assumption --- and second about firms in other sectors. Finally, uncertainty
may extend to the overall stability of the whole set of returns.

We assume that investors' uncertainty regarding firms within their own
sector is minimal and uniformly distributed across those firms; i.e. that
the average anticipation error regarding intra-sector firms is on average
zero. In contrast, both the evaluation errors investors from one sector make
about the returns of investors in other sectors, and their anticipation
errors regarding these returns, depend on the model's parameters and the
system's current state.

\subsubsection{Structure of investment risk}

To analyze the interconnected structure of investment risk, we consider the
return equations (\ref{RT}) in the absence of default:%
\begin{equation*}
\left( \delta \left( X-X^{\prime }\right) -\hat{S}_{E}\left( X^{\prime
},X\right) \right) \widehat{DF}\left( X^{\prime }\right) \hat{R}_{exc}\left(
X^{\prime }\right) =S_{E}\left( X,X\right) R_{exc}\left( X\right)
\end{equation*}%
and its series expansion:%
\begin{eqnarray*}
\hat{R}_{exc}\left( X\right) &=&\left( \widehat{DF}\left( X\right) \right)
^{-1}\left[ S_{E}\left( X,X\right) R_{exc}\left( X\right) \right] \\
&&+\left( \widehat{DF}\left( X\right) \right) ^{-1}\sum_{m}\hat{S}%
_{E}^{m}\left( X^{\prime },X\right) \left[ S_{E}\left( X^{\prime },X^{\prime
}\right) R_{exc}\left( X^{\prime }\right) \right]
\end{eqnarray*}%
This last equation reveals a diffusion effect: investing in other investors
initiates a chain of increasingly remote investments, in which the return $%
\hat{f}\left( X^{\prime }\right) $ of investor $X^{\prime }$ comes with an
ever increasing risk to investor $X$. Assuming investment risk increases
multiplicatively along each investment path, the risk associated to each
path, written $\rho _{p}$ can be represented as the product of local risks
along the path of length $m$:%
\begin{eqnarray}
&&\rho _{p}\left( \left( \widehat{DF}\left( X\right) \right) ^{-1}\hat{S}%
_{E}^{m}\left( \left( X^{\prime }\right) ^{\prime },X^{\prime }\right) \left[
\hat{S}_{E}\left( \left( X^{\prime }\right) ^{\prime }\right) R_{exc}\left(
\left( X^{\prime }\right) ^{\prime }\right) \right] \right)  \label{NC} \\
&\rightarrow &\zeta ^{2}\left( \hat{w}_{E}^{\left( 0\right) }\left( \left(
X^{\prime }\right) ^{\prime },X_{m-1}\right) ...\hat{w}_{E}^{\left( 0\right)
}\left( X_{1},X^{\prime }\right) \right) ^{-1}\hat{S}_{E}^{2m}\left(
X^{\prime },X\right)  \notag
\end{eqnarray}%
where the term $\zeta ^{2}$ denotes the variance of the term:%
\begin{equation*}
\hat{S}_{E}\left( \left( X^{\prime }\right) ^{\prime }\right) R_{exc}\left(
\left( X^{\prime }\right) ^{\prime }\right)
\end{equation*}%
and represents the uncertainty attached to a direct investment into a firm $%
X^{\prime }$.\ This parameter $\zeta ^{2}$ will be considered constant in
first-order approximation to focus on the propagation of risk rather than on
the intrinsic variance of firm-level returns.

The coefficients $\hat{w}_{E}^{\left( 0\right) }\left( X_{1},X^{\prime
}\right) $ are the local investment risk between neighboring agents. They
depend on the distance between sectors, but in first approximation, they too
will be considered constant.

We further assume that investment risk is additive for disconnected paths,
up to a normalization factor, so that first-order estimate of the total
investment risk $\rho \left( \hat{X},\hat{f}\left( X^{\prime }\right)
\right) $ can be obtained by summing the contributions of all distinct paths.%
\begin{eqnarray*}
\rho \left( \hat{X},\hat{f}\left( X^{\prime }\right) \right)
&=&\sum_{paths}\rho _{p} \\
&\rightarrow &\sum \zeta ^{2}\left( \hat{w}_{E}^{\left( 0\right) }\left(
X^{\prime },X_{m-1}\right) ...\hat{w}_{E}^{\left( 0\right) }\left( X_{m-1},%
\hat{X}\right) \right) ^{-1}\hat{S}_{E}^{2m}\left( X^{\prime },X\right)
\end{eqnarray*}%
Using these assumptions and under a minimal investment condition, agents tend%
\footnote{%
See Appendices 3.2 and 3.3.} to self-organize into relatively closed
investment groups.\ 

To the first approximation, loans and participations in any given investment
will be assumed to carry the same level of uncertainty\footnote{%
See Appendix 3.1 for details.}, so that the uncertainty-dependent
coefficients for participations and loans are equal:%
\begin{equation}
\hat{w}_{E}\left( X^{\prime },X\right) =\hat{w}_{L}\left( X^{\prime
},X\right) =\frac{1}{2}\hat{w}\left( X^{\prime },X\right)  \label{Sn}
\end{equation}%
and:%
\begin{equation}
w_{E}\left( X,X\right) =w_{L}\left( X,X\right) =\frac{1}{2}w\left( X,X\right)
\label{Sd}
\end{equation}%
In this setting, the risk coefficients for investment in investors are given
by:%
\begin{equation}
\hat{w}\left( X^{\prime },X\right) =\frac{2\left( 1-\left( \gamma
\left\langle \hat{S}_{E}\left( X\right) \right\rangle \right) ^{2}\right) 
\hat{w}_{E}^{\left( 0\right) }\left( X^{\prime },X\right) }{1+\hat{w}%
_{E}^{\left( 0\right) }\left( X^{\prime },X\right) \left( 1-\left( \gamma
\left\langle \hat{S}_{E}\left( X\right) \right\rangle \right) ^{2}\right)
+\Delta \left( \gamma \left\langle \hat{S}_{E}\left( X_{1},X^{\prime
}\right) \right\rangle _{\hat{X}_{1}}\right) ^{2}}  \label{hb}
\end{equation}%
where $\gamma $ represents the average uncertainty associated with
distance-dependent investment paths, and writes: 
\begin{equation*}
\gamma ^{2}\simeq \left( \hat{w}_{E}^{\left( 0\right) }\left( \left(
X^{\prime }\right) ^{\prime },X_{m-1}\right) ...\hat{w}_{E}^{\left( 0\right)
}\left( X_{1},X^{\prime }\right) \right) ^{-\frac{1}{m}}
\end{equation*}%
while the risk coefficients for investments in firms are given by:%
\begin{equation}
w\left( X,X\right) =1-\left\langle w\left( X^{\prime },X\right)
\right\rangle _{X^{\prime }}  \label{hc}
\end{equation}%
where the average $\gamma $ is taken over the set of sectors in which sector 
$X$ allocates capital.

In formula (\ref{hb}), the term:%
\begin{equation*}
\left( \gamma \left\langle \hat{S}_{E}\left( X\right) \right\rangle \right)
^{2}
\end{equation*}%
describes the risk associated to an average share of $\left\langle \hat{S}%
_{E}\left( X\right) \right\rangle $ between investors. The higher the
shares, the higher the risk and the lower the coefficient $\hat{w}\left(
X^{\prime },X\right) $. The contribution:\textbf{\ }%
\begin{equation*}
\Delta \left( \gamma \left\langle \hat{S}_{E}\left( X_{1},X^{\prime }\right)
\right\rangle _{\hat{X}_{1}}\right) ^{2}=\left( \gamma \left\langle \hat{S}%
_{E}\left( X_{1},X^{\prime }\right) \right\rangle _{\hat{X}_{1}}\right)
^{2}-\left( \gamma \left\langle \hat{S}_{E}\left( X\right) \right\rangle
\right) ^{2}
\end{equation*}%
in equation (\ref{hb}) is the gap between the risk perception of investor $%
X^{\prime }$\ and that of the market. It increases with the risk perception
of investor $X^{\prime }$, so that the higher this gap, the lower the
coefficient $\hat{w}\left( X^{\prime },X\right) $, and the lower the
investment of investor $X^{\prime }$ in investor $X$.

The coefficient $\hat{w}_{E}^{\left( 0\right) }\left( X^{\prime },X\right) $%
\ is a factor of local inverse uncertainty: it is a factor of confidence of
investors $X$\ in investors $X^{\prime }$. The higher this factor, the
higher the investments in sector $X^{\prime }$. As such, this coefficient
captures the characteristics of local uncertainty that induce deviations
from average behavior.

The functions $\hat{w}$ and $w$\ defined in equations (\ref{hb}) and (\ref%
{hc}) depend on exogeneous parameters but also on the stakes $\hat{S}%
_{E}\left( X^{\prime },X\right) $, $\hat{S}\left( X^{\prime },X\right) $, $%
S_{E}\left( X,X\right) $, $S\left( X,X\right) $.

\section{The equations of the field model}

Three types of equations characterize the full model with endogenous stakes.
First, the field return equation (\ref{QDL}), expressed in terms of excess
return and discount factors. Second, the stakes' field equation, which we
will first derive by minimizing the field of stakes' action functional, and
then express in terms of stakes. And third, the endogenous equations
governing uncertainties, as measured by the coefficients $\hat{w}_{\eta
}\left( X^{\prime },X\right) $ and $\hat{w}\left( X^{\prime },X\right) $.

\subsection{The field return equations for investors}

The minimization equations for the investor field are the return equations (%
\ref{QDL}).\ They link the distribution of stakes and the disposable capital
across the various sectors.\ Under a no-default scenario, they are\footnote{%
The default scenario will be examined later, in section 5.3.}:%
\begin{equation}
0=\int \left( \delta \left( X^{\prime }-X\right) -\hat{S}_{E}\left(
X^{\prime },X\right) \right) \widehat{DF}\left( X^{\prime }\right) \hat{R}%
_{exc}\left( X^{\prime }\right) dX^{\prime }-S_{E}\left( X,X\right)
R_{exc}\left( X\right)  \label{QDM}
\end{equation}%
where investors' and firms' excess returns are respectively defined by:%
\begin{eqnarray*}
\hat{R}_{exc}\left( X\right) &=&\hat{f}\left( X\right) -\hat{r}\left(
X\right) \\
R_{exc}\left( X\right) &=&f\left( X\right) -\bar{r}\left( X\right)
\end{eqnarray*}%
\ Note that the firm's excess returns $R_{exc}\left( X\right) $ include the
dividends and the share price variations, which themselves depend on
productivity and some exogenous factors\footnote{%
See Gosselin and Lotz (2024).}.

\subsection{The equation for the field of stakes}

We will first derive the minimization equation for the field of stakes, then
re-express this equation in terms of sectoral stakes in investors and firms.

\subsubsection{Stakes' field minimization equation}

The collective states satisfy the minimization equations associated with the
action functional $S\left( \Gamma \right) $. The various stakes taken by
investors will be proportional to the inverse uncertainty coefficients $\hat{%
w}_{E}\left( X\right) $, $\hat{w}_{L}\left( X\right) $, $w_{E}\left(
X\right) $ and $w_{L}\left( X\right) $. Assuming capital is fully invested,
we can normalize these coefficients and impose:%
\begin{equation*}
\hat{w}_{E}\left( X\right) +\hat{w}_{L}\left( X\right) +w_{E}\left( X\right)
+w_{L}\left( X\right) =1
\end{equation*}%
where:%
\begin{equation}
\hat{w}_{\eta }\left( X\right) =\int \hat{w}_{\eta }\left( X^{\prime
},X\right) dX^{\prime }  \label{Krf}
\end{equation}%
and:%
\begin{equation}
w_{\eta }\left( X\right) =\int w_{\eta }\left( X^{\prime },X\right)
dX^{\prime }  \label{Krh}
\end{equation}%
Nevertheless, the coefficients depend on the uncertainty in returns, which
themselves are functions of the shares $\underline{\hat{S}}_{\eta }$and $%
S_{\eta }$. The coefficients $\hat{w}_{\alpha }\left( X^{\prime },X\right) $
and $w_{\alpha }\left( X,X\right) $ are therefore endogeneous, and the
resolution of the system will thus require a detailed characterization of
these uncertainties.

Using the above simplifying assumptions, the minimization equations for the
field of stakes $\Gamma \left( \hat{S}^{\left( T\right) },X^{\prime
},X\right) $ satisfy\footnote{%
See Appendix 3.}:%
\begin{eqnarray}
0 &=&\left( \sum_{\eta }\left( -\sigma _{\hat{K}}^{2}\nabla _{\hat{S}_{\eta
}^{\left( T\right) }}^{2}+\frac{\left( \hat{S}_{\eta }^{\left( T\right)
}\right) ^{2}}{2\hat{w}_{\eta }}-\hat{V}_{\eta }\hat{S}_{\eta }^{\left(
T\right) }+\lambda \left( X\right) \left\Vert \Gamma \left( \hat{S}^{\left(
T\right) },X^{\prime },X\right) \right\Vert _{\hat{X}}^{2}\hat{S}_{\eta
}^{\left( T\right) }\right) -\beta \right)  \label{abc} \\
&&\times \Gamma \left( \hat{S}^{\left( T\right) },X^{\prime },X\right) 
\notag
\end{eqnarray}%
In first approximation, the solutions take the form\footnote{%
See Appendix 2.}:%
\begin{equation*}
\Gamma _{0,X^{\prime },X}\left( \hat{S}^{\left( T\right) }\right) =N\exp
\left( -\sum_{\eta }\frac{\left( \hat{S}_{\eta }^{\left( T\right) }-\hat{S}%
_{\eta }^{\left( T\right) }\left( X^{\prime },X\right) \right) ^{2}}{2\sigma
_{\hat{K}}^{2}}\right)
\end{equation*}%
where $N$\ is a normalization factor. Solving the minimization equations
thus reduces to finding the stakes invested by sector $X$\ into sector $%
X^{\prime }$, i.e. the sectoral averages $\hat{S}_{\eta }^{\left( T\right)
}\left( X^{\prime },X\right) $.\ They satisfy\footnote{%
See Appendix 2.}:%
\begin{equation*}
\hat{S}_{\eta }^{\left( T\right) }\left( X^{\prime },X\right) =\hat{w}_{\eta
}\left( X^{\prime },X\right) \left( \hat{V}_{\eta }\left( X^{\prime
},X\right) +\lambda \left( X\right) \right)
\end{equation*}%
and depend on the Lagrange multipliers\footnote{%
Given in Appendix 2.}.

\subsubsection{Equations in terms of sectoral stakes}

Solving for the Lagrange multiplier, equation (\ref{abc}) can be written in
an expanded form, as a function of the sectoral stakes.

\paragraph{Stakes taken in investors}

We first define the coefficients $\hat{w}\left( X\right) $\ and $w\left(
X\right) $\ as the average uncertainty of investor $X$\ over its stakes in
other investors and firms $X$\ respectively:%
\begin{equation}
\hat{w}\left( X\right) =\int \hat{w}\left( X^{\prime },X\right) dX^{\prime }
\label{Grf}
\end{equation}%
and:%
\begin{equation}
w\left( X\right) =1-\hat{w}\left( X\right)  \label{Grh}
\end{equation}%
We also define $R\left( X\right) $, $\hat{R}\left( X^{\prime }\right) $ and $%
\hat{R}$, the returns from all stakes (shares and loans) taken in firms $X$,
in investors $X^{\prime }$, and across all investors, respectively as:%
\begin{equation}
R\left( X\right) =\frac{1}{2}\left( f\left( X\right) +r\left( X\right)
\right)  \label{Frl}
\end{equation}%
\begin{equation*}
\hat{R}\left( X^{\prime }\right) =\frac{1}{2}\left( \hat{f}\left( X^{\prime
}\right) +\hat{r}\left( X^{\prime }\right) \right)
\end{equation*}%
and:%
\begin{equation}
\hat{R}_{X}=\frac{1}{2}\left( \left\langle \hat{f}\left( X^{\prime }\right)
\right\rangle _{\hat{w}\left( X\right) }+\left\langle \hat{r}\left(
X^{\prime }\right) \right\rangle _{\hat{w}\left( X\right) }\right)
\label{Frh}
\end{equation}%
The computation of $\hat{R}_{X}$ involves a risk-weighted average of
returns, whose definition, for any function $F\left( X^{\prime }\right) $,
is:%
\begin{equation}
\left\langle F\left( X^{\prime }\right) \right\rangle _{\hat{w}\left(
X\right) }=\frac{\int F\left( X^{\prime }\right) \hat{w}\left( X^{\prime
},X\right) }{\hat{w}\left( X\right) }  \label{Wv}
\end{equation}%
Given these definitions, the expression:\textbf{\ }%
\begin{equation*}
R^{w,\hat{w}}\left( X\right) =\hat{w}\left( X\right) \hat{R}_{X}+w\left(
X\right) R\left( X\right) \mathbf{\ }
\end{equation*}%
computes the return that investor $X$\ can expect from a diversified
investment between firms $X$\ and the market.

Using these notations, the equations for the stakes taken by investors in
other investors are:%
\begin{equation}
\hat{S}_{E}\left( X^{\prime },X\right) =\frac{1}{2}\underline{\hat{S}}\left(
X^{\prime },X\right) +\frac{1}{2}\hat{w}\left( X^{\prime },X\right) \hat{%
\Delta}^{E}\left( X^{\prime },X\right)  \label{Shn}
\end{equation}%
\begin{equation}
\hat{S}_{L}\left( X^{\prime },X\right) =\frac{1}{2}\underline{\hat{S}}\left(
X^{\prime },X\right) +\frac{1}{2}\hat{w}\left( X^{\prime },X\right) \hat{%
\Delta}^{L}\left( X^{\prime },X\right)  \label{Sht}
\end{equation}%
and:%
\begin{equation}
\hat{S}\left( X^{\prime },X\right) =\underline{\hat{S}}\left( X^{\prime
},X\right) +\hat{w}\left( X^{\prime },X\right) \hat{\Delta}\left( X^{\prime
},X\right)  \label{Shv}
\end{equation}%
where:%
\begin{eqnarray*}
\hat{\Delta}^{E}\left( X^{\prime },X\right) &=&\hat{f}\left( X^{\prime
}\right) -R^{w,\hat{w}}\left( X\right) \\
\hat{\Delta}^{L}\left( X^{\prime },X\right) &=&\hat{r}\left( X^{\prime
}\right) -R^{w,\hat{w}}\left( X\right) \\
\hat{\Delta}\left( X^{\prime },X\right) &=&\hat{R}\left( X^{\prime }\right)
-R^{w,\hat{w}}\left( X\right)
\end{eqnarray*}%
and:%
\begin{equation*}
\underline{\hat{S}}_{E}\left( X^{\prime },X\right) =\underline{\hat{S}}%
_{L}\left( X^{\prime },X\right) =\frac{1}{2}\underline{\hat{S}}\left(
X^{\prime },X\right) =\frac{1}{2}\hat{w}\left( X^{\prime },X\right)
\end{equation*}%
As expected, equations (\ref{Shn}) and (\ref{Sht})\ for $\hat{S}_{E}\left(
X^{\prime },X\right) $ and $\hat{S}_{L}\left( X^{\prime },X\right) $ show
that investments depend inversely on uncertainty through the coefficients $%
\hat{w}\left( X^{\prime },X\right) $, and vary linearly with the difference
between expected returns on shares, $\hat{f}\left( X^{\prime }\right) $, or
loans, $\hat{r}\left( X^{\prime }\right) $ and the weighted returns $R^{w,%
\hat{w}}\left( X\right) $.

\paragraph{Stakes taken in firms}

Similarly, we derive the equations for stakes taken by investors in firms
using the uncertainty coefficient is $w\left( X\right) $: 
\begin{equation}
S_{E}\left( X,X\right) =\frac{1}{2}\underline{S}\left( X,X\right) +\frac{1}{2%
}w\left( X\right) \Delta ^{E}\left( X\right)  \label{SNp}
\end{equation}%
\begin{equation*}
S_{L}\left( X,X\right) =\frac{1}{2}\underline{S}\left( X,X\right) +\frac{1}{2%
}w\left( X\right) \Delta ^{L}\left( X\right)
\end{equation*}%
and:%
\begin{equation}
S\left( X,X\right) =\underline{S}\left( X,X\right) +w\left( X\right) \left( 
\hat{w}\left( X\right) \Delta \left( X\right) \right)  \label{STp}
\end{equation}%
with:%
\begin{equation*}
\underline{S}_{E}\left( X,X\right) =\underline{S}_{L}\left( X,X\right) =%
\frac{1}{2}\underline{S}\left( X,X\right) =\frac{1}{2}w\left( X,X\right)
\end{equation*}%
and:%
\begin{equation*}
\Delta ^{E}\left( X\right) =f\left( X\right) -R^{w,\hat{w}}\left( X\right)
\end{equation*}%
\begin{equation*}
\Delta ^{L}\left( X\right) =\bar{r}\left( X\right) -R^{w,\hat{w}}\left(
X\right)
\end{equation*}%
\begin{equation*}
\Delta \left( X\right) =R\left( X\right) -\hat{R}_{X}
\end{equation*}%
Here too, the stakes are inversely proportional to the uncertainty of such
investments and depend on the difference between the expected returns of
these shares or loans and the weighted returns that can be expected from the
market and firms.

The equations (\ref{Shn}), (\ref{Sht}), (\ref{Shv}), (\ref{SNp}) and (\ref%
{STp}) relate the stakes, the returns, and each sector average disposable
capital, along with the coefficients $\hat{w}\left( X^{\prime },X\right) $, $%
w\left( X\right) $ and their averages. Note that the cofficients $\hat{w}%
\left( X^{\prime },X\right) $ and $w\left( X\right) $ are themselves
partially endogenous, reflecting the subjective uncertainties between
sectors, but also between agents: as investors invest in one another,
uncertainty propagates through the entire system. Besides, each sector
disposable capital depends on the structure of connections, derived in
Gosselin and Lotz (2024).\ We will recall their form below.

\paragraph{Inward aggregate stakes in investors}

The return equations (\ref{QDM}) involve the inward aggregate stakes (\ref%
{Gsn}), (\ref{Gst}), (\ref{Gsv}) and (\ref{Gsw}). Defining the coefficients $%
\hat{w}$\ \ and $w$\ as the averages over the sector space of the
coefficients\footnote{%
\ See equations (\ref{Grf}) and (\ref{Grh}).} $\hat{w}\left( X^{\prime
}\right) $\ and $w\left( X\right) $, these inward aggregate stakes can be
written:%
\begin{equation}
\hat{S}_{E}\left( X^{\prime }\right) \simeq \left( \frac{1}{2}\underline{%
\hat{S}}\left( X^{\prime }\right) +\frac{1}{2}\hat{w}\left( X^{\prime
}\right) \hat{\Delta}^{E}\left( X^{\prime }\right) \right) \frac{%
\left\langle \hat{K}\right\rangle \left\Vert \hat{\Psi}\right\Vert ^{2}}{%
\hat{K}_{X^{\prime }}\left\vert \hat{\Psi}\left( X^{\prime }\right)
\right\vert ^{2}}  \label{Grstn}
\end{equation}%
\begin{equation}
\hat{S}_{L}\left( X^{\prime }\right) \simeq \left( \frac{1}{2}\underline{%
\hat{S}}\left( X^{\prime }\right) +\frac{1}{2}\hat{w}\left( X^{\prime
}\right) \hat{\Delta}^{L}\left( X^{\prime }\right) \right) \frac{%
\left\langle \hat{K}\right\rangle \left\Vert \hat{\Psi}\right\Vert ^{2}}{%
\hat{K}_{X^{\prime }}\left\vert \hat{\Psi}\left( X^{\prime }\right)
\right\vert ^{2}}  \label{Grstw}
\end{equation}%
and:%
\begin{equation}
\hat{S}\left( X^{\prime }\right) \simeq \left( \underline{\hat{S}}\left(
X^{\prime }\right) +\hat{w}\left( X^{\prime }\right) \hat{\Delta}\left(
X^{\prime }\right) \right) \frac{\left\langle \hat{K}\right\rangle
\left\Vert \hat{\Psi}\right\Vert ^{2}}{\hat{K}_{X^{\prime }}\left\vert \hat{%
\Psi}\left( X^{\prime }\right) \right\vert ^{2}}  \label{Grstd}
\end{equation}%
where $\hat{\Delta}^{E}\left( X^{\prime }\right) $, $\hat{\Delta}^{L}\left(
X^{\prime }\right) $ and $\hat{\Delta}\left( X^{\prime }\right) $ are the
aggregate of $\hat{\Delta}^{E}\left( X^{\prime },X\right) $, $\hat{\Delta}%
^{L}\left( X^{\prime },X\right) $ and $\hat{\Delta}\left( X^{\prime
},X\right) $ over the variable $X$. They write:%
\begin{equation}
\hat{\Delta}^{E}\left( X^{\prime }\right) =\hat{f}\left( X^{\prime }\right)
-\left\langle R^{w,\hat{w}}\left( X\right) \right\rangle  \label{dtf}
\end{equation}%
\begin{equation}
\hat{\Delta}^{L}\left( X^{\prime }\right) =\hat{r}\left( X^{\prime }\right)
-\left\langle R^{w,\hat{w}}\left( X\right) \right\rangle  \label{dtr}
\end{equation}%
and:%
\begin{equation}
\hat{\Delta}\left( X^{\prime }\right) =\hat{R}\left( X^{\prime }\right)
-\left\langle R^{w,\hat{w}}\left( X\right) \right\rangle  \label{dtR}
\end{equation}%
with:%
\begin{eqnarray*}
\hat{R} &=&\left\langle \hat{R}_{X}\right\rangle \\
R &=&\left\langle R\left( X\right) \right\rangle
\end{eqnarray*}%
and:%
\begin{equation*}
\left\langle R^{w,\hat{w}}\left( X\right) \right\rangle =\left\langle \hat{w}%
\left( X^{\prime }\right) \right\rangle \hat{R}+\left\langle w\left(
X\right) \right\rangle R
\end{equation*}%
which computes the average of the weighed return $R^{w,\hat{w}}\left(
X\right) $.

\paragraph{Inward aggregate stakes in firms}

Similarly, the inward aggregate stakes\ (\ref{Gst}) invested in firms $X$
are of the form: 
\begin{equation}
S_{\eta }\left( X\right) =S_{\eta }\left( X,X\right) \frac{\hat{K}%
_{X}\left\vert \hat{\Psi}\left( X\right) \right\vert ^{2}}{K_{X}\left\vert
\Psi \left( X\right) \right\vert ^{2}}  \label{Grstf}
\end{equation}%
so that, using the expression for the stakes invested in firms (\ref{SNp})
and (\ref{STp}), we find:%
\begin{equation}
S_{E}\left( X\right) =\left( \frac{1}{2}\underline{S}\left( X,X\right) +%
\frac{1}{2}w\left( X\right) \Delta ^{E}\left( X\right) \right) \frac{\hat{K}%
_{X}\left\vert \hat{\Psi}\left( X\right) \right\vert ^{2}}{K_{X}\left\vert
\Psi \left( X\right) \right\vert ^{2}}
\end{equation}%
\begin{equation*}
S_{L}\left( X\right) =\left( \frac{1}{2}\underline{S}\left( X,X\right) +%
\frac{1}{2}w\left( X\right) \Delta ^{E}\left( X\right) \right) \frac{\hat{K}%
_{X}\left\vert \hat{\Psi}\left( X\right) \right\vert ^{2}}{K_{X}\left\vert
\Psi \left( X\right) \right\vert ^{2}}
\end{equation*}%
and:%
\begin{equation}
S\left( X,X\right) =\left( \underline{S}\left( X,X\right) +w\left( X\right)
\left( \hat{w}\left( X\right) \Delta \left( X\right) \right) \right) \frac{%
\hat{K}_{X}\left\vert \hat{\Psi}\left( X\right) \right\vert ^{2}}{%
K_{X}\left\vert \Psi \left( X\right) \right\vert ^{2}}
\end{equation}

\subsection{The equations for uncertainty}

The equation that defines the weight of inverse uncertainty for
cross-investment between investors, derived in section 4, is:%
\begin{equation}
\hat{w}\left( X^{\prime },X\right) =\frac{2\left( 1-\left( \gamma
\left\langle \hat{S}_{E}\left( X\right) \right\rangle \right) ^{2}\right) 
\hat{w}_{E}^{\left( 0\right) }\left( X^{\prime },X\right) }{1+\hat{w}%
_{E}^{\left( 0\right) }\left( X^{\prime },X\right) \left( 1-\left( \gamma
\left\langle \hat{S}_{E}\left( X\right) \right\rangle \right) ^{2}\right)
+\Delta \left( \gamma \left\langle \hat{S}_{E}\left( X_{1},X^{\prime
}\right) \right\rangle _{\hat{X}_{1}}\right) ^{2}}
\end{equation}%
Recall that $\gamma $ is the average uncertainty of the distance-dependent
investment paths and writes: 
\begin{equation*}
\gamma ^{2}\simeq \left( \hat{w}_{E}^{\left( 0\right) }\left( \left(
X^{\prime }\right) ^{\prime },X_{m-1}\right) ...\hat{w}_{E}^{\left( 0\right)
}\left( X_{1},X^{\prime }\right) \right) ^{-\frac{1}{m}}
\end{equation*}%
where the coefficient $\hat{w}_{E}^{\left( 0\right) }\left( X^{\prime
},X\right) $\ was defined in section 4.

The weight of inverse uncertainty for investments in firms is given by:%
\begin{equation}
w\left( X,X\right) =1-\left\langle w\left( X^{\prime },X\right)
\right\rangle _{X^{\prime }}
\end{equation}%
Since the functions $\hat{w}$ and $w$\ both depend on the stakes, their
expression complete the minimization equations and yield $\hat{S}_{E}\left(
X^{\prime },X\right) $, $\hat{S}\left( X^{\prime },X\right) $, $S_{E}\left(
X,X\right) $, $S\left( X,X\right) $. \ 

Note incidentally that the above equations hold for a fixed value of the
uncertainty $\gamma $, but that since $\gamma $ can vary across the sector
space, the groups of investors can be relatively interconnected. However, in
first-order approximation, we will assume them to be independent, and solve
the model for each group separately\footnote{%
The third part of this work will examine the interactions between groups in
greater detail.}.

\section{Resolution of the model: no-default scenario}

The resolution of the model consists in solving the return equation for
investors using the relationships between stakes and returns established in
section 5.2.2\footnote{%
More precisely, the resolution of the model consists in solving the return
equation (\ref{QDM}) using the equations (\ref{Shn}), (\ref{Sht}), (\ref{Shv}%
), and (\ref{SNp}) and (\ref{STp}).}.\textbf{\ }The solutions of these
minimization equations will define the collective states of the system. In
the following, we will explicit the resolution process and implement it
under a no-default scenario.

\subsection{Methodology}

\subsubsection{Groups and sub-collective states}

To solve the full system, we must determine the levels of capital and
returns for each sector of the sector space $S$, along with the distribution
of stakes across all these sectors. Taken together, these quantities define
a collective state of the system.

Since capital and returns can be derived from the level of stakes between
investors, a collective state is described by a set of values of stakes: 
\begin{equation*}
\left\{ \hat{S}_{E}\left( X^{\prime },X\right) ,\hat{S}\left( X^{\prime
},X\right) ,S_{E}\left( X,X\right) ,S\left( X,X\right) \right\} _{\left(
X^{\prime },X\right) \in S}
\end{equation*}%
These quantities are governed by equation (\ref{QDM}).\ These are non-local
equations that involve the network of connections within each group.

We have seen that, under uncertainty, agents connect to only a finite number
of neighbors, so that agents are organized into several and
loosely-connected groups. Each of these groups $G_{\alpha }$ can be in one
among several possible phase, each defined by a distribution of stakes
between the agents of $G_{\alpha }$:%
\begin{equation*}
\left\{ \hat{S}_{E}\left( X^{\prime },X\right) ,\hat{S}\left( X^{\prime
},X\right) ,S_{E}\left( X,X\right) ,S\left( X,X\right) \right\} _{\left(
X^{\prime },X\right) \in G_{\alpha }}
\end{equation*}

We will call a \emph{sub-collective state} the combination of one group $%
G_{\alpha }$ and one of its possible phases. Thus, collective states
organize into several distinct and weakly-interacting sub-collective
states.\ In first approximation, a collective state can be seen as a set of
sub-collective states that writes:%
\begin{equation*}
\cup _{\alpha }\left\{ \hat{S}_{E}\left( X^{\prime },X\right) ,\hat{S}\left(
X^{\prime },X\right) ,S_{E}\left( X,X\right) ,S\left( X,X\right) \right\}
_{\left( X^{\prime },X\right) \in G_{\alpha }}
\end{equation*}%
where the groups $G_{\alpha }$ describe a certain organization of the sector
space $S$.

The organization of the sector space into groups $\left\{ G_{\alpha
}\right\} $ is not unique: there are an infinite number of such
organizations. Moreover, for each organization, multiple distinct phases are
possible for each group: the combined effect of uncertainty and
interconnections among sectors leads to an infinite number of possible
solutions, among which some are default states\footnote{%
See Gosselin and Lotz (2024).}. Thus, there are an infinite number of
possible collective states, each composed of multiple groups, each being in
one among several possible phase. Any given group associated with one
specific phase is called a \emph{sub-collective state}.\ For any group,
there may be several sub-collective states.

\subsubsection{Principle of resolution}

The return equation in sector $X$, equation (\ref{QDM}), connects all
investors in the sector space, and as such comprises three types of
variables: the inward aggregate stakes on investors in sector $X$; the
returns of investors $X$ and firms $X$, taken independently; and the average
returns of all investors of the sector space.

The resolution follows a four-step process.\ First, we average the return
equation over all sectors.\ This yields the averages of all the variables of
the equation: average stakes, levels of capital and returns. Second, we then
replace these results in the return equation, and further replace the
aggregate stakes (\ref{Grstn}), (\ref{Grstd}) and (\ref{Grstf}) in terms of
returns, so that the return equation depends only on returns. Third, solving
this equation yields the returns per sector and, in turn, the aggregate
stakes and the disposable capital per sector. Fourth, using equations (\ref%
{Shn}), (\ref{Sht}), (\ref{Shv}), we reconstruct the remaining quantities of
the model.

\subsection{Solving the model}

\subsubsection{Step 1: solving for average stakes, capital and returns}

\paragraph{General resolution}

The return equation (\ref{QDM}), averaged, writes:%
\begin{equation}
\left( 1-\left\langle \hat{S}\left( X^{\prime }\right) \right\rangle \right)
\left\langle \hat{R}_{exc}\left( X^{\prime }\right) \right\rangle \simeq
\left\langle S_{E}\left( X,X\right) \right\rangle \left\langle R_{exc}\left(
X\right) \right\rangle  \label{QDN}
\end{equation}%
The solutions of (\ref{QDN}) will ultimately depend on the assumptions on
the firms' returns per unit of capital $\left\langle f\left( X\right)
\right\rangle $.\ As a first approximation, and to simplify the
computations, we will suppose constant returns to scale, so that:%
\begin{equation*}
\left\langle f\left( X\right) \right\rangle =\left\langle f_{1}\left(
X\right) \right\rangle
\end{equation*}%
\textbf{\ }where $\left\langle f_{1}\left( X\right) \right\rangle $\ is a
net average productivity. Decreasing returns to scale will then be
introduced as corrections, especially when computing disposable capital, and
the firms' returns will be in this case\footnote{%
See Appendix 8.}:%
\begin{equation}
\left\langle f\left( X\right) \right\rangle =\frac{\left\langle f_{1}\left(
X\right) \right\rangle }{\left\langle K\right\rangle ^{r}}-\frac{C}{%
\left\langle K\right\rangle }-C_{0}  \label{DCr}
\end{equation}%
where $C$ stands for a fixed cost and $C_{0\text{ }}$ for a cost per unit of
capital\footnote{%
More precisely, considering the marginal productivity $\frac{1}{1-r}%
\left\langle f_{1}\left( X\right) \right\rangle \left\langle K\right\rangle
^{1-r}$, the net average return per unit of capital is:%
\begin{equation*}
\frac{1}{1-r}\left\langle f_{1}\left( X\right) \right\rangle \left\langle
K\right\rangle ^{1-r}-C-C_{0}\left\langle K\right\rangle
\end{equation*}%
Including the factor $\frac{1}{1-r}$\ in the definition of $\left\langle
f_{1}\left( X\right) \right\rangle $\ yields formula (\ref{DCr}).}.

The global average stakes in investors $\left\langle \hat{S}\left( X^{\prime
}\right) \right\rangle $, and in firms $\left\langle S_{E}\left( X,X\right)
\right\rangle $ and $\left\langle S\left( X,X\right) \right\rangle $, as
well as the ratio of average disposable capital of investors over that of
firms $\frac{\left\langle \hat{K}\right\rangle \left\Vert \hat{\Psi}%
\right\Vert ^{2}}{\left\langle K\right\rangle \left\Vert \Psi \right\Vert
^{2}}$, are expressed as functions of average shares in investors $%
\left\langle \hat{S}_{E}\left( X^{\prime }\right) \right\rangle $ and total
stakes in firms $\left\langle S\left( X\right) \right\rangle $. To solve the
average return equation for investors (\ref{QDN}), we use the equations for
the various stakes (\ref{Shn}), (\ref{Sht}), (\ref{Shv}), (\ref{SNp}) and (%
\ref{STp}), to reexpress all the variables, namely the global average stakes
in firms and the average returns in firms, $\left\langle S\left( X\right)
\right\rangle $ and $\left\langle f\left( X\right) \right\rangle $
respectively, as functions of average shares in investors $\left\langle \hat{%
S}_{E}\left( X^{\prime }\right) \right\rangle $, so that the full system is
expressed as a function of a single variable, the average shares in
investors.\ 

\paragraph{Average solutions}

Numerical analyses of the solutions reveal that investors' cross-investments 
$\left\langle \hat{S}_{E}\left( X^{\prime }\right) \right\rangle $ increase
with the excess average firm return $\left\langle f\left( X\right)
\right\rangle $ and decrease with uncertainty $\gamma $.

When average firm returns exceed the interest rate, $\left\langle f\left(
X\right) \right\rangle >r$: higher firm returns encourage sectoral firm
capital accumulation.\ Average capital held by firms $\left\langle
K\right\rangle \left\Vert \Psi \right\Vert ^{2}$ increase with returns.
Investors invest in their own sector and in firms specifically.\ Disposable
capital for investors diminishes, and the ratio of average disposable
capital of investors over that of firms, $\frac{\left\langle \hat{K}%
\right\rangle \left\Vert \hat{\Psi}\right\Vert ^{2}}{\left\langle
K\right\rangle \left\Vert \Psi \right\Vert ^{2}}$\ decreases with returns.

When average firm returns do not exceed the interest rate,\ $\left\langle
f\left( X\right) \right\rangle <r$: lower firm returns fail to compensate
for the higher investment risk, and capital invested in firms fall
accordingly. As global uncertainty $\gamma $\ increases, investments
diversify, thus increasing the capital ratio $\frac{\left\langle \hat{K}%
\right\rangle \left\Vert \hat{\Psi}\right\Vert ^{2}}{\left\langle
K\right\rangle \left\Vert \Psi \right\Vert ^{2}}$.

When firms' returns are high, uncertainty favors direct investment in firms,
whereas when uncertainty is low, it favors investment diversification among
investors.

\paragraph{Average variables under constant return to scale}

Solutions can be approximated when returns $\left\langle f\left( X\right)
\right\rangle $ are close to the interest rates $\left\langle \bar{r}\left(
X\right) \right\rangle $.

\subparagraph{Benchmark case}

In the benchmark case, average returns equal the average interest rate, $%
\left\langle f\left( X\right) \right\rangle -\left\langle \bar{r}\left(
X\right) \right\rangle =0$.\ Under this condition, the value of $%
\left\langle \hat{S}_{E}\left( X^{\prime },X\right) \right\rangle $, denoted 
$z_{0}$, directly depends on the level of uncertainty $\gamma $.\ 

When there is no uncertainty, $\gamma $ $=$ $0$, and we have:%
\begin{equation*}
\left\langle \hat{S}_{E}\left( X^{\prime },X\right) \right\rangle
=\left\langle \hat{S}_{L}\left( X^{\prime },X\right) \right\rangle
=\left\langle S_{E}\left( X^{\prime },X\right) \right\rangle =\left\langle
S_{L}\left( X^{\prime },X\right) \right\rangle =z_{0}=\frac{1}{4}
\end{equation*}%
When there are no excess returns and no uncertainty, the diversification in
portfolio is maximal.\ 

However, when there is uncertainty, $\gamma >0$: stakes remain equivalent in
terms of returns, but not in terms of risk, so that in this case, $z_{0}$
measures the impact of uncertainty on investment choices. When uncertainty
is high, the level of average shares in investors $\left\langle \hat{S}%
_{E}\left( X^{\prime },X\right) \right\rangle $ is correspondingly low: for
equivalent returns, investors favor direct over intermediated investment in
firms. When uncertainty is low, we are back to the benchmark case: investors
choose to diversify their investments.

\subparagraph{Impact of excess returns on cross-investors' stakes}

When average returns differ from the interest rate, average stakes depend
both on the average excess return $\left\langle R_{exc}\left( X\right)
\right\rangle $ and $z_{0}$, the measure of uncertainty.

The average shares taken in and loans granted by investors in other
investors are, respectively:%
\begin{equation}
\left\langle \hat{S}_{E}\left( X^{\prime },X\right) \right\rangle
=\left\langle \hat{S}_{E}\left( X^{\prime }\right) \right\rangle
=z_{0}\left( 1+\frac{1}{2}\frac{z_{0}^{2}}{d\left( z_{0}\right) }%
\left\langle R_{exc}\left( X\right) \right\rangle \right)  \label{VSt}
\end{equation}%
and:%
\begin{equation}
\left\langle \hat{S}_{L}\left( X^{\prime },X\right) \right\rangle
=z_{0}\left( 1-\frac{1}{2}\left( 1-\frac{z_{0}^{2}}{d\left( z_{0}\right) }%
\right) \left\langle R_{exc}\left( X\right) \right\rangle \right)
\label{VSk}
\end{equation}%
where:%
\begin{equation*}
\left\langle f\left( X\right) \right\rangle =\left\langle f_{1}\left(
X\right) \right\rangle
\end{equation*}%
and:%
\begin{equation*}
d\left( z_{0}\right) =1-5z_{0}+8z_{0}^{2}
\end{equation*}%
and the total average stake is:%
\begin{eqnarray}
&&\left\langle \hat{S}\left( X^{\prime },X\right) \right\rangle
=\left\langle \hat{S}_{E}\left( X^{\prime },X\right) \right\rangle
+\left\langle \hat{S}_{L}\left( X^{\prime },X\right) \right\rangle
\label{Vsn} \\
&=&2z_{0}\left( 1-\frac{1}{4d\left( z_{0}\right) }\left( 1-3z_{0}\right)
\left( 1-2z_{0}\right) \left\langle R_{exc}\left( X\right) \right\rangle
\right)  \notag
\end{eqnarray}%
For a given level of uncertainty, any increase in firms' excess returns will
induce investors to invest in firms.\ They will do so in their own sector,
but also in other sectors, via the intermediation of other investors. As a
result, average cross-sectoral investments in shares $\left\langle \hat{S}%
_{E}\left( X^{\prime },X\right) \right\rangle $ will increase. By
comparison, the attractivity of loans, along with their relative
profitability, will diminish and so will $\left\langle \hat{S}_{L}\left(
X^{\prime },X\right) \right\rangle $. The overall cross-investors' volume of
stakes $\left\langle \hat{S}\left( X^{\prime },X\right) \right\rangle $ will
nonetheless decline, since direct investment in firms remains more
profitable.

A relative increase in uncertainty will dampen cross-sectoral
intermediation. Portfolio management requires diversification through loans,
which limits investors' ability to fully benefit from higher returns.
Actually, investors' aggregate returns increase at a slower pace than that
of firms:

\begin{equation*}
\left\langle \hat{R}_{exc}\left( X\right) \right\rangle =\frac{1}{2}%
\left\langle R_{exc}\left( X\right) \right\rangle
\end{equation*}%
so that an increase in firms returns $\delta \left\langle f_{1}\left(
X\right) \right\rangle $ will only yield an increase of $\delta \left\langle
f_{1}\left( X\right) \right\rangle $ $/2$ of investors returns.

\subparagraph{Impact of excess returns on investors' stakes in firms}

The stakes in firms are given by: 
\begin{eqnarray}
\left\langle S_{E}\left( X,X\right) \right\rangle &=&\frac{1-2z_{0}}{2}%
\left( 1+\left( z_{0}\varepsilon \left( z_{0}\right) +\left( \frac{3}{4}%
-z_{0}\right) \right) \left\langle R_{exc}\left( X\right) \right\rangle
\right)  \label{Vsr} \\
\left\langle S_{L}\left( X,X\right) \right\rangle &=&\frac{1-2z_{0}}{2}%
\left( 1+\left( z_{0}-\varepsilon \left( z_{0}\right) \left( \frac{1}{2}%
-z_{0}\right) -\frac{3}{8}\right) \left\langle R_{exc}\left( X\right)
\right\rangle \right)  \notag
\end{eqnarray}%
and:%
\begin{eqnarray}
\left\langle S\left( X,X\right) \right\rangle &=&\left\langle S_{E}\left(
X,X\right) \right\rangle +\left\langle S_{L}\left( X,X\right) \right\rangle
\label{Vsw} \\
&=&\left( 1-2z_{0}\right) \left( 1+\left( \frac{1}{2}+\varepsilon \left(
z_{0}\right) \right) z_{0}\left\langle R_{exc}\left( X\right) \right\rangle
\right)  \notag
\end{eqnarray}%
where:%
\begin{equation*}
\varepsilon \left( z_{0}\right) =\frac{z_{0}^{2}\left( 1-4z_{0}\right) }{%
\left( 1-5z_{0}+8z_{0}^{2}\right) }
\end{equation*}%
Firms that experience excess returns and seek fresh capital will turn to
equity creation rather than debt. Shares in firms $\left\langle S_{E}\left(
X,X\right) \right\rangle $ increase, whereas loans $\left\langle S_{L}\left(
X,X\right) \right\rangle $ decline. Nonetheless, total investment in firms $%
\left\langle S\left( X,X\right) \right\rangle $ increase. Besides, a
relative increase in uncertainty will increase direct investment in firms.

\paragraph{Accounting for decreasing return to scale}

Decreasing returns to scale for firms amounts to replace, in the formulas
for stakes:%
\begin{equation*}
\left\langle f\left( X\right) \right\rangle =\left\langle f_{1}\left(
X\right) \right\rangle
\end{equation*}
by the adjusted firm returns defined in expression (\ref{DCr}):%
\begin{equation*}
\left\langle f\left( X\right) \right\rangle =\frac{\left\langle f_{1}\left(
X\right) \right\rangle }{\left\langle K\right\rangle ^{r}}-\frac{C}{%
\left\langle K\right\rangle }-C_{0}
\end{equation*}%
The previous interpretations still hold, but the impact of the productivity $%
\left\langle f_{1}\left( X\right) \right\rangle $ is now dampened by the
level of disposable capital for firms and for investors.

Average disposable capital for investors is:%
\begin{equation}
\left\langle \hat{K}\right\rangle \left\Vert \hat{\Psi}\right\Vert
^{2}\simeq \frac{\hat{\mu}V\sigma _{\hat{K}}^{2}}{2}\left( \frac{\frac{%
\left\Vert \hat{\Psi}_{0}\right\Vert ^{2}}{\hat{\mu}}\left( 1-\hat{S}\right)
+\tau \left\langle \hat{f}\right\rangle \hat{S}}{\left\langle \hat{f}%
\right\rangle \left( 1-\hat{S}\right) }\right) ^{2}  \label{VRK}
\end{equation}%
where $\left\langle \hat{f}\right\rangle $ is the average return for
investors:%
\begin{equation}
\left\langle \hat{f}\right\rangle \simeq \left\langle \hat{r}\left(
X^{\prime }\right) \right\rangle +\frac{f_{a}-f_{b}\left( \frac{%
C_{0}+\left\langle \bar{r}\left( X\right) \right\rangle }{f_{1}\left(
X\right) }\right) ^{\frac{2}{r}}}{2}  \label{Rpl}
\end{equation}%
and where $f_{a}$, $f_{b}$ and $\tau $ are some coefficients\footnote{%
These coefficients are defined in Apppendix 5.4.} $\ $Specifically, the
coefficient $f_{a}$\ is the average return, and depends on the interest rate
and variable costs.\ The coefficient $f_{b}$\ is an effective cost that
depends on fixed costs and the number of firms in the sector: the higher the
number of firms, the lower the cost. The coefficient $\tau $\ measures the
share of external investors in disposable capital, and $\frac{\left\Vert 
\hat{\Psi}_{0}\right\Vert ^{2}}{\hat{\mu}}$\ is the contribution to
disposable capital proportional to the average number of investors. The
parameter $\left\Vert \hat{\Psi}_{0}\right\Vert ^{2}$ represents the average
number of investors per sectors.

In the above,the parameter $\hat{\mu}$\ represents the fluctuations in the
number of agents within each sector. When $\hat{\mu}$\ is high, the number
of agents may fluctuate widely.\ In this case, the average is no longer a
defining term for the level of disposable capital.

Average disposable capital for firms is: 
\begin{equation}
\left\langle K\right\rangle \left\Vert \Psi \right\Vert ^{2}\simeq \left(
1-\left\langle S\left( X,X\right) \right\rangle \frac{\left\langle \hat{K}%
\right\rangle \left\Vert \hat{\Psi}\right\Vert ^{2}}{\left( \left\langle
K\right\rangle \left\Vert \Psi \right\Vert ^{2}\right) _{0}}\right) \left(
\left\langle K\right\rangle \left\Vert \Psi \right\Vert ^{2}\right) _{0}
\end{equation}%
where the factor:%
\begin{equation*}
\left( \left\langle K\right\rangle \left\Vert \Psi \right\Vert ^{2}\right)
_{0}=\left( \left( \frac{2\epsilon }{3\sigma _{\hat{K}}^{2}}\right) ^{\frac{r%
}{2}}\frac{\left\langle f_{1}\left( X\right) \right\rangle }{C_{0}+\bar{r}}%
\right) ^{\frac{2}{r}}
\end{equation*}%
is the firms' average disposable capital in the absence of investors'
stakes, that is when $\left\langle S\left( X,X\right) \right\rangle =0$.

The presence of investors reduces the share of firms' disposable capital.
There is an eviction effect. Indeed, due to decreasing returns, capital is
limited and caped at a certain level, a part of which will be taken by
investors. Supposing firms could reach this level of capital without
investors, they could only do so at a considerable cost and over a longer
period.

As a rule, disposable capital for firms $\left\langle K\right\rangle
\left\Vert \Psi \right\Vert ^{2}$\ increases with average productivity $%
\left\langle f_{1}\left( X\right) \right\rangle $: a global increase in
firms' productivity increases their capital levels and favors direct capital
accumulation in firms, but reduces intermediation. All in all, investors'
total average disposable capital $\left\langle \hat{K}\right\rangle
\left\Vert \hat{\Psi}\right\Vert ^{2}$\ decreases with $\left\langle
f_{1}\right\rangle $, but their private capital $\left( 1-\hat{S}\right)
\left\langle \hat{K}\right\rangle \left\Vert \hat{\Psi}\right\Vert ^{2}$
will increase\footnote{%
See Appendix 5.3.4.}.

Ultimately, introducing decreasing returns to scale does not modify formulas
(\ref{VSt}), (\ref{VSk}), (\ref{Vsn}), (\ref{Vsr}), (\ref{Vsw}) for stakes,
but \ amounts to replacing, in all these formula, the excess return by its
expression under decreasing returns to scale as\ obtained by equation (\ref%
{Rpl}):%
\begin{equation*}
\left\langle \hat{R}_{exc}\left( X\right) \right\rangle \simeq \frac{%
f_{a}-f_{b}\left( \frac{C_{0}+\left\langle \bar{r}\left( X\right)
\right\rangle }{f_{1}\left( X\right) }\right) ^{\frac{2}{r}}}{2}
\end{equation*}%
The previous interpretations still hold, but the results are dampenend by
decreasing returns.

\subsubsection{Step 2: Writing the return equations in terms of returns}

As a first approximation, integrals in equation (\ref{QDM}) can be replaced
by their averages.\ The return equation can thus be written:\textbf{\ }%
\begin{equation}
0=\widehat{DF}\left( X\right) \hat{R}_{exc}\left( X\right) -\left\langle 
\hat{S}_{E}\left( X^{\prime },X\right) \right\rangle _{X^{\prime
}}\left\langle \widehat{DF}\left( X^{\prime }\right) \right\rangle
\left\langle \hat{R}_{exc}\left( X^{\prime }\right) \right\rangle
-S_{E}\left( X,X\right) R_{exc}\left( X\right)  \label{DTS}
\end{equation}

This equation involves the averages $\left\langle \hat{S}\left( X^{\prime
}\right) \right\rangle $, $\left\langle \hat{S}\left( X^{\prime }\right)
\right\rangle $, $\left\langle \hat{f}\left( X^{\prime }\right)
\right\rangle $ computed in step 1, but also the returns $\hat{f}\left(
X\right) $ and $f\left( X\right) $, the aggregate stakes $\hat{S}\left(
X\right) $, $\hat{S}_{E}\left( X\right) $ and $\left\langle \hat{S}%
_{E}\left( X^{\prime },X\right) \right\rangle _{X^{\prime }}$,\ and the
shares in firms $S_{E}\left( X,X\right) $.

Using the equations for stakes $\hat{S}_{E}\left( X^{\prime },X\right) $, $%
\hat{S}_{L}\left( X^{\prime },X\right) $ and $\hat{S}\left( X^{\prime
},X\right) $ given by (\ref{Shn}), (\ref{Sht}), (\ref{Shv}), along with the
risk coefficients $\hat{w}\left( X^{\prime },X\right) $ and $w\left(
X\right) $ defined in (\ref{hb}) and (\ref{hc}),\ we can replace the average
stakes by functions of returns and average variables, and recast explicitly
the return equation (\ref{DTS}) as a return equation for $\hat{f}\left(
X\right) $\footnote{%
See Appendix 7.}:%
\begin{equation}
0=\frac{\left\langle \hat{f}\right\rangle ^{2}\left( \frac{\Xi }{1-\Xi }%
\right) ^{2}-\Xi \left( \hat{f}^{\prime }\left( X\right) \right) ^{2}}{%
\left\langle \hat{f}\right\rangle ^{2}\left( \frac{\Xi }{1-\Xi }\right) ^{2}-%
\frac{1}{2}A\left( 1+\hat{\Delta}^{E}\left( X\right) \right) \left( \hat{f}%
^{\prime }\left( X\right) \right) ^{2}}\hat{R}_{exc}\left( X\right) -h\left(
X\right)  \label{DTR}
\end{equation}%
where:%
\begin{equation*}
\hat{f}^{\prime }\left( X\right) =\hat{f}\left( X\right) \left(
1-\left\langle \hat{S}\right\rangle \right) +\left\langle \hat{S}\left(
X^{\prime },X\right) \right\rangle _{X^{\prime }}\left\langle \hat{f}%
\right\rangle
\end{equation*}%
\begin{equation*}
\Xi =A\left( 1+\hat{\Delta}\left( X\right) \right)
\end{equation*}%
and:%
\begin{eqnarray*}
h\left( X\right) &=&\hat{w}\left( X\right) \left( 1-w\left( X\right) \hat{%
\Delta}\left( X\right) +\frac{1}{2}\left\langle \hat{R}_{exc}\left(
X^{\prime }\right) \right\rangle \right) \\
&&\times \left\langle \widehat{DF}\left( X^{\prime }\right) \right\rangle
\left\langle \hat{R}_{exc}\left( X^{\prime }\right) \right\rangle
+S_{E}\left( X,X\right) R_{exc}\left( X\right)
\end{eqnarray*}%
where $\hat{\Delta}^{E}\left( X\right) $ and $\hat{\Delta}\left( X\right) $
are defined by\ (\ref{dtf}) and (\ref{dtR}) respectively\ and the parameter $%
A$\ is defined in Appendix 7.

\subsubsection{Step 3: solving for sectoral returns and aggregate stakes}

In general, equations (\ref{DTR}) admit in first approximation two solutions
per sector: one characterized by low, the other by high investors' returns.\
This multiplicity is consistent with the findings in Gosselin and Lotz
(2024).

For each sector, the first solution corresponds to high levels of
productivity and capital with low returns per unit of capital due to
decreasing returns to scale. The second solution corresponds to low levels
of productivity and capital, where investors' returns stem from investments
in other sectors.

\paragraph{Two types of returns per sector}

Defining:%
\begin{equation*}
z=\left\langle \hat{S}_{E}\left( X^{\prime },X\right) \right\rangle
\end{equation*}%
that measures the perception of uncertainty by investors\footnote{%
The link between $z$ and $z_{0}$, the value of $\left\langle \hat{S}%
_{E}\left( X^{\prime },X\right) \right\rangle $ under the constraint $%
\left\langle f\left( X\right) \right\rangle -\left\langle \bar{r}\left(
X\right) \right\rangle =0$, is given in Appendix 5.4.1. In first
approximation, we can consider $z=z_{0}$.}, the two solutions\ for
investors' excess returns depend on $z$, on firms' excess returns per sector
and average returns in the system. They write\footnote{%
See Appendix 7 for the derivation.}:%
\begin{equation}
\hat{R}_{exc}^{H}\left( X\right) =\frac{\hat{R}_{exc}^{H,0}\left( X\right) }{%
2}+\sqrt{\left( \frac{\hat{R}_{exc}^{H,0}\left( X\right) }{2}\right) ^{2}-%
\frac{1-2z}{1-z}\frac{\left\langle \bar{r}\left( X\right) \right\rangle }{%
a\left( z\right) }\hat{R}_{exc}^{L,0}\left( X\right) }  \label{SLn}
\end{equation}%
and:%
\begin{equation}
\hat{R}_{exc}^{L}\left( X\right) =\frac{\hat{R}_{exc}^{H,0}\left( X\right) }{%
2}-\sqrt{\left( \frac{\hat{R}_{exc}^{H,0}\left( X\right) }{2}\right) ^{2}-%
\frac{1-2z}{1-z}\frac{\left\langle \bar{r}\left( X\right) \right\rangle }{%
a\left( z\right) }\hat{R}_{exc}^{L,0}\left( X\right) }  \label{Slp}
\end{equation}%
with:%
\begin{equation}
\hat{R}_{exc}^{H,0}\left( X\right) =\left\langle \hat{R}_{exc}\left(
X^{\prime }\right) \right\rangle +\frac{z\left( 1-2z\right) }{\left(
1-z\right) ^{2}a\left( z\right) }\left\langle R_{exc}\left( X^{\prime
}\right) \right\rangle +\frac{\left( 1-2z\right) }{\left( 1-z\right) a\left(
z\right) }\left\langle \bar{r}\left( X\right) \right\rangle  \label{dernier1}
\end{equation}%
and:%
\begin{equation}
\hat{R}_{exc}^{L,0}\left( X\right) =z\left\langle \hat{R}_{exc}\left(
X^{\prime }\right) \right\rangle +\frac{\left( 1-z\right) }{2}R_{exc}\left(
X\right)  \label{dernier2}
\end{equation}%
where the function $a\left( z\right) $\ is defined in Appendix 7. Note that
the excess return of firms $R_{exc}\left( X\right) $\ is given by $\left(
f_{1}\left( X\right) -\left\langle \hat{r}\left( X^{\prime }\right)
\right\rangle \right) $\ under constant returns to scale, and by $f_{a}-%
\frac{1}{2}f_{b}\left( \frac{C_{0}+\bar{r}}{f_{1}\left( X\right) }\right) ^{%
\frac{1}{r}}$\ under decreasing returns to scale\footnote{%
See Appendix 7.2.}. The interpretations will be similar in both cases, but
with dampened amplitudes for decreasing returns.

Obviously, the two solutions (\ref{SLn}) and (\ref{Slp}) differ whether we
add or substract the second term:%
\begin{equation}
\sqrt{\left( \frac{\hat{R}_{exc}^{H,0}\left( X\right) }{2}\right) ^{2}-\frac{%
1-2z}{1-z}\frac{\left\langle \bar{r}\left( X\right) \right\rangle }{a\left(
z\right) }\hat{R}_{exc}^{L,0}\left( X\right) }  \label{dernier3}
\end{equation}%
When it is added, it is the high return solution (\ref{SLn}), in which
returns are relatively higher than otherwise. When it is substracted, it
yields the low return solution (\ref{Slp}).

When excess returns are relatively small compared to the interest rate,
these two solutions can be expressed as:%
\begin{equation}
\hat{R}_{exc}^{H}\left( X\right) =\hat{R}_{exc}^{H,0}\left( X\right) -\hat{R}%
_{exc}^{L,0}\left( X\right)  \label{HRet}
\end{equation}%
and:%
\begin{equation}
\hat{R}_{exc}^{L}\left( X\right) =\hat{R}_{exc}^{L,0}\left( X\right) \left(
1-\frac{2a\left( z\right) \left( 1-z\right) }{\left( 1-2z\right) }\left( 
\frac{\left\langle \hat{R}_{exc}\left( X\right) \right\rangle }{\left\langle 
\bar{r}\right\rangle }+\frac{R_{exc}\left( X\right) }{\left\langle \bar{r}%
\right\rangle }\right) \right)  \label{LRet}
\end{equation}%
Note that the high return solution only occurs under some conditions. For
instance, when uncertainty is high, returns $\hat{R}_{exc}^{H}\left(
X\right) $\ would have to be very high to compensate for high uncertainty,
which rules out this solution practically. The high return solution
therefore only occurs when the risk perception of investors is low. Besides,
it only arise for investors with very high level of disposable capital 
\footnote{%
See Appendix 7.} that can lure in other investors, and extract above-average
returns.

\paragraph{Interpretation of sectoral returns}

\subparagraph{Low return solution}

Under this solution, investors' excess returns are given, in last analysis,
by equation (\ref{LRet}). Using the definition (\ref{dernier2}) of $\hat{R}%
_{exc}^{L,0}\left( X\right) $ this solution can be analyzed as a sum of
returns over investors :%
\begin{equation}
z\left\langle \hat{R}_{exc}\left( X^{\prime }\right) \right\rangle
\label{CTn}
\end{equation}%
and over firms:%
\begin{equation}
\frac{\left( 1-z\right) }{2}R_{exc}\left( X\right)  \label{CT2}
\end{equation}%
both terms being dampened by the factor:%
\begin{equation}
1-\frac{2a\left( z\right) \left( 1-z\right) }{\left( 1-2z\right) }\left( 
\frac{\left\langle \hat{R}_{exc}\left( X\right) \right\rangle }{\left\langle 
\bar{r}\right\rangle }+\frac{R_{exc}\left( X\right) }{\left\langle \bar{r}%
\right\rangle }\right)  \label{Dfc}
\end{equation}%
The contributions (\ref{CTn}) and (\ref{CT2}) can be interpreted as follows:
when uncertainty is high, $z$ is low. Investors mainly invest on their
sector's firms, as expected. When uncertainty is low, investors diversify
their stakes between firms and investors. However, under the low return
solution, investors' returns are essentialy driven by their investments in
their sector's firms, and only marginally by other investors'
intermediation:\ they are less exposed to risk. Sectoral returns are thus
similar to the average returns and only differ from them in that they
replace average firm returns by sectoral ones.

As mentioned above, these contributions are dampened by the factor (\ref{Dfc}%
), which shows that the investors' average excess return under the low
return solution will be systematically lower than the overall average. The
low return investors experience a loss of a fraction of their average excess
returns, $\frac{\left\langle \hat{R}_{exc}\left( X\right) \right\rangle }{%
\left\langle \bar{r}\right\rangle }+\frac{R_{exc}\left( X\right) }{%
\left\langle \bar{r}\right\rangle }$. This fraction is directly related to
the existence of the high return solution, in which high return investors
get above-average returns at the expense of low return investors and firms.
Since the high return solution is favored by low uncertainty, this fraction
decreases with uncertainty.

\subparagraph{High return solution}

Under this solution, cross-sectoral investments - loans and participations
in investors - drive the returns, through the term:%
\begin{equation*}
\hat{R}_{exc}^{H,0}\left( X\right)
\end{equation*}%
However, to rightly assess these higher returns, one must deplete them from
the returns under the low return solution:%
\begin{equation*}
\hat{R}_{exc}^{L,0}\left( X\right)
\end{equation*}%
Overall, the solution $\hat{R}_{exc}^{H}\left( X\right) $ decomposes into
three terms:%
\begin{equation*}
\hat{R}_{exc}^{H}\left( X\right) =\left( 1-z\right) \left\langle \hat{R}%
_{exc}\left( X^{\prime }\right) \right\rangle +\left( \frac{z\left(
1-2z\right) }{\left( 1-z\right) ^{2}a\left( z\right) }-\frac{1-z}{2}\right)
\left\langle R_{exc}\left( X^{\prime }\right) \right\rangle +\frac{\left(
1-z\right) }{a\left( z\right) }\left\langle \bar{r}\left( X\right)
\right\rangle
\end{equation*}%
When the perception of risk is low, cross-investments (first term) exceeds
investments on firms (second term). \ Under this solution, because of firms'
low productivity and returns, investors invest more in other investors than
on the firms in their own sector. The third term represents the additional
profit investors with high disposable capital can negociate from loans to
other investors.

Structurally, the high return solution is systematically greater than the
low return solution, since investors with high disposable capital can ask
for higher interest rates and returns. As already noticed in the above,
these additional returns are made at the expense of low-return investors, so
that for any given average, it is this discrepency between high and low
returns that realizes the average solution.

\paragraph{Implications for the collective state}

The collective state thus consists of two types of solutions. The low return
solution is the standard solution. It is similar to the results found about
average returns, except for the presence of the dampening factor induced by
the presence of high return solutions. In comparison, the high return
solution corresponds to a deviation to these normal returns. They describe
investors that maintain a very high level of capital by attracting other
investors, and by extracting higher returns from firms and other investors.\
By comparison, the other investors in the system will receive
lower-than-average returns.\ As for loans, firms and investors will face
different levels of interest rates, whether they borrow from high or low
investors. \ 

This would also suggest that returns are, as a rule, driven by firms'
returns, but that superior returns are mainly driven by cross-sectoral
investments of investors with high level of disposable capital.

We will see later on that these two solutions, high and low, may impair the
stability of the collective states. Small changes in returns can induce
large flows of capital that may destabilize the collective state and
transition the system toward other - and possible default- states.

\paragraph{The level of stakes and capital per sector}

Now that we have stated the possible solutions of sectoral returns, we can
compute the associated levels of stakes and capital.

\subparagraph{Outward aggregate stakes in investors}

Outward aggregate stakes (\ref{Twn}) and (\ref{Tws}) compute the stakes of
investors $X$\ outside their sector.

The proportion of capital invested by investor $X$\ in other investors $%
\left\langle \hat{S}_{E}\left( X^{\prime },X\right) \right\rangle
_{X^{\prime }}$ is:\textbf{\ }%
\begin{equation}
\left\langle \hat{S}_{E}\left( X^{\prime },X\right) \right\rangle _{\hat{X}%
^{\prime }}\simeq \frac{1}{2}\left\langle \hat{w}\left( X^{\prime },X\right)
\right\rangle \left( 1+\left\langle \hat{\Delta}^{E}\left( X^{\prime
},X\right) \right\rangle _{X^{\prime }}\right)  \label{Shg}
\end{equation}%
and the proportion of loans granted by investor $X$\ to other investors $%
X^{\prime }$ $\left\langle \hat{S}_{E}\left( X^{\prime },X\right)
\right\rangle _{X^{\prime }}$ is:%
\begin{equation}
\left\langle \hat{S}_{L}\left( X^{\prime },X\right) \right\rangle _{\hat{X}%
^{\prime }}\simeq \frac{1}{2}\left\langle \hat{w}\left( X^{\prime },X\right)
\right\rangle \left( 1+\left\langle \hat{\Delta}^{L}\left( X^{\prime
},X\right) \right\rangle _{X^{\prime }}\right)  \notag
\end{equation}%
where the coefficient of proportionality of investors $X$ investments in
other investors and in firms are given by the average of formula (\ref{hb}),
and write respectively:%
\begin{equation}
\left\langle \hat{w}\left( X^{\prime },X\right) \right\rangle =\frac{\left(
1-\left( \gamma \left\langle \hat{S}_{E}\left( X\right) \right\rangle
\right) ^{2}\right) }{2-\left( \gamma \left\langle \hat{S}_{E}\left(
X\right) \right\rangle \right) ^{2}}  \label{CFt}
\end{equation}%
and:%
\begin{equation}
w\left( X\right) =1-\left\langle \hat{w}\left( X^{\prime },X\right)
\right\rangle  \label{CFtbis}
\end{equation}%
and:%
\begin{eqnarray*}
\left\langle \hat{\Delta}^{E}\left( X^{\prime },X\right) \right\rangle
_{X^{\prime }} &=&\left\langle \hat{f}\left( X^{\prime }\right)
\right\rangle -R^{w,\hat{w}}\left( X\right) \\
\left\langle \hat{\Delta}^{L}\left( X^{\prime },X\right) \right\rangle
_{X^{\prime }} &=&\left\langle \hat{r}\left( X^{\prime }\right)
\right\rangle -R^{w,\hat{w}}\left( X\right)
\end{eqnarray*}%
From formula (\ref{CFt}), we can infer that the higher the average perceived
risk $\left( \gamma \left\langle \hat{S}_{E}\left( X\right) \right\rangle
\right) ^{2}$, the lower the coefficient $\left\langle \hat{w}\left(
X^{\prime },X\right) \right\rangle $\ and the lower the participations
between investors.

Recall that the quantity $\Delta \left( X\right) $\ measures the relative
return of firms $X$\ across equity and loans, so that the higher the returns
of firms $X$ are, the less investors $X$ allocate their capital externally,
and this capital is further diversify among multiple investors $X^{\prime }$%
. This is the low return solution, where firms' productivity is sufficient
to attract investments and provide global levels of returns. Under the high
return solution, because of the high level of returns in investors $%
\left\langle \hat{f}\left( X^{\prime }\right) \right\rangle $, investors
will rather invest externally.\ 

\subparagraph{Capital level per sector}

Under decreasing returns to scale, firms' returns are:%
\begin{equation*}
f\left( X\right) =\frac{f_{1}\left( X\right) }{\left( K_{X}\left\vert \hat{%
\Psi}\left( X\right) \right\vert ^{2}\right) ^{r}}-\frac{C}{K_{X}}-C_{0}
\end{equation*}%
For each solution (\ref{SLn}), capital levels per sector are for investors:%
\begin{equation}
\hat{K}_{X}\left\vert \hat{\Psi}\left( X\right) \right\vert ^{2}\simeq
\left\langle \hat{K}\right\rangle \left\Vert \hat{\Psi}\right\Vert
^{2}\left( \frac{\left\langle \hat{f}\right\rangle I_{X/\left\langle
X^{\prime }\right\rangle }}{\hat{f}\left( X\right) \left( 1-\left\langle 
\hat{S}\right\rangle \right) +\left\langle \hat{S}\left( X^{\prime
},X\right) \right\rangle _{X^{\prime }}\left\langle \hat{f}\right\rangle }%
\right) ^{2}  \label{Dcn}
\end{equation}%
where the average investors return $\left\langle \hat{f}\right\rangle $\ is
given by (\ref{Rpl}), and investors $X$\ returns are given by:%
\begin{equation*}
\hat{f}\left( X\right) \simeq \hat{r}\left( X\right) +\frac{1}{2}f_{a}-\frac{%
1}{2}f_{b}\left( \frac{C_{0}+\hat{r}\left( X\right) }{f_{1}\left( X\right) }%
\right) ^{\frac{1}{r}}
\end{equation*}%
Disposable capital (\ref{Dcn}) is proportional to the ratio:%
\begin{equation}
I_{X/\left\langle X^{\prime }\right\rangle }=\frac{\frac{\left\langle \hat{S}%
\left( X,X^{\prime }\right) \right\rangle _{X^{\prime }}}{1-\left\langle 
\hat{S}\left( X,X^{\prime }\right) \right\rangle _{X^{\prime }}}}{\frac{%
\left\langle \hat{S}\left( X,X^{\prime }\right) \right\rangle }{%
1-\left\langle \hat{S}\left( X,X^{\prime }\right) \right\rangle }}
\label{Idef}
\end{equation}%
which measures the level of investment in sector $X$\ by other sectors, $%
\frac{\left\langle \hat{S}\left( X,X^{\prime }\right) \right\rangle
_{X^{\prime }}}{1-\left\langle \hat{S}\left( X,X^{\prime }\right)
\right\rangle _{X^{\prime }}}$, with respect to the level of investment in
the rest of the market, $\frac{\left\langle \hat{S}\left( X,X^{\prime
}\right) \right\rangle }{1-\left\langle \hat{S}\left( X,X^{\prime }\right)
\right\rangle }$.

It is also weighted by the ratio:%
\begin{equation*}
\frac{\left\langle \hat{f}\right\rangle }{\hat{f}\left( X\right) \left(
1-\left\langle \hat{S}\right\rangle \right) +\left\langle \hat{S}\left(
X^{\prime },X\right) \right\rangle _{X^{\prime }}\left\langle \hat{f}%
\right\rangle }
\end{equation*}%
that measures a saturation effect: a high level of returns limits the level
of capital that produces this level of return. It is not solely the return $%
\hat{f}\left( X\right) $ that matters, but its weighted combination with
average returns $\left\langle \hat{S}\left( X^{\prime },X\right)
\right\rangle _{X^{\prime }}\left\langle \hat{f}\right\rangle $\ which
translates the share of capital reinvested in the market by investors $X$.

The disposable capital for firms is given by:%
\begin{equation}
K_{X}\left\vert \Psi \left( X\right) \right\vert ^{2}\simeq \left( 1-S\left(
X,X\right) \frac{\hat{K}_{X}\left\vert \hat{\Psi}\left( X\right) \right\vert
^{2}}{\left( K_{X}\left\Vert \Psi \left( X\right) \right\Vert ^{2}\right)
_{0}}\right) \left( K_{X}\left\Vert \Psi \left( X\right) \right\Vert
^{2}\right) _{0}  \label{Dct}
\end{equation}%
where:%
\begin{equation*}
\left( K_{X}\left\Vert \Psi \left( X\right) \right\Vert ^{2}\right)
_{0}=\left( \left( \frac{2\epsilon }{3\sigma _{\hat{K}}^{2}}\right) ^{\frac{r%
}{2}}\frac{f_{1}\left( X\right) }{C_{0}+\bar{r}}\right) ^{\frac{2}{r}}
\end{equation*}%
is the firms $X$ disposable capital in the absence of investor
participation, i.e. when\textbf{\ }$S\left( X,X\right) =0$.

Under both solutions, firms' capital increases with firms' productivity $%
f_{1}\left( X\right) $, and investors' $X$ disposable capital increase with
firms' productivity $f_{1}\left( X\right) $. This refines our results for
the average capital found in step 1, since\ this increase in firms'
productivity in any given sector $X$ increases investments in investors $X$
from other sectors.\textbf{\ }However, investors' disposable capital is
higher under the high-return solution than under the low-return solution,
since investors attract more capital so\ that, under this solution the ratio 
$I_{X/\left\langle X^{\prime }\right\rangle }$\textbf{\ } is particularly
high.

\subparagraph{Inward aggregate stakes}

Now that we have computed the outward aggregate stakes in investors and the
level of disposable capital per sector, we can derive the inward aggregate
stakes, the stakes taken by all investors in one sector, to better grasp the
distribution of investments.\ Since firms $X$ can only be invested in by
investors $X$, the stakes in firms, that represent the proportion invested
in investors $X$, and the aggregate stakes in firms, that represent the part
of total capital invested in firms, only differ of a single factor, which is
the proportion of investors' disposable capital, both private and
intermediated, over the firm's disposable capital.

\subparagraph{Inward aggregate stakes in investors}

Recall that the general formula for the inward aggregate stakes in investors
are given by (\ref{Grstn}), (\ref{Grstw}) and (\ref{Grstd}):%
\begin{eqnarray*}
\hat{S}_{E}\left( X^{\prime }\right) &=&\frac{1}{2}\hat{w}\left( X^{\prime
}\right) \left( 1+\hat{\Delta}^{E}\left( X^{\prime }\right) \right) Q_{\hat{K%
}}\left( X^{\prime }\right) \\
\hat{S}_{L}\left( X^{\prime }\right) &=&\frac{1}{2}\hat{w}\left( X^{\prime
}\right) \left( 1+\hat{\Delta}^{L}\left( X^{\prime }\right) \right) Q_{\hat{K%
}}\left( X^{\prime }\right) \\
\hat{S}\left( X^{\prime }\right) &=&\hat{w}\left( X^{\prime }\right) \left(
1+\Delta \left( X^{\prime }\right) \right) Q_{\hat{K}}\left( X^{\prime
}\right)
\end{eqnarray*}%
They involve the quotient of aggregate disposable capital of all investors $%
X^{\prime }$ over that investors $X$ which is computed by (\ref{Dcn}):%
\begin{equation*}
Q_{\hat{K}}\left( X^{\prime }\right) =\frac{\left\langle \hat{K}%
\right\rangle \left\Vert \hat{\Psi}\right\Vert ^{2}}{\hat{K}_{X^{\prime
}}\left\vert \hat{\Psi}\left( X^{\prime }\right) \right\vert ^{2}}=\left( 
\frac{\hat{f}\left( X\right) \left( 1-\left\langle \hat{S}\right\rangle
\right) +\left\langle \hat{S}\left( X^{\prime },X\right) \right\rangle
_{X^{\prime }}\left\langle \hat{f}\right\rangle }{\left\langle \hat{f}%
\right\rangle I_{X/\left\langle X^{\prime }\right\rangle }}\right) ^{2}
\end{equation*}%
and the coefficient of proportionality:%
\begin{equation*}
\hat{w}\left( X^{\prime }\right) =\frac{\left( 1-\left( \gamma \left\langle 
\hat{S}_{E}\left( X\right) \right\rangle \right) ^{2}\right) }{2-\left(
\gamma \left\langle \hat{S}_{E}\left( X\right) \right\rangle \right)
^{2}-\gamma ^{2}\left\langle \hat{S}_{E}\left( X\right) \right\rangle
\left\langle \hat{w}\left( X^{\prime },X\right) \right\rangle \left\langle
w\left( X\right) \right\rangle \Delta \left( X^{\prime }\right) }
\end{equation*}%
which depends directly on uncertainty and returns.

The capital invested in the investors of a sector depends negatively on
uncertainty and positively on the relative returns of both investors and
firms in that sector. This translates the fact that investors $X^{\prime }$\
also invest in $X$\ to gain access to the sector firms' returns.

\subparagraph{Inward stakes in firms}

As expected, the proportion of shares invested in firms $X$, $S_{E}\left(
X,X\right) $, increases with their return $f_{1}\left( X\right) $:%
\begin{equation}
S_{E}\left( X,X\right) =\frac{1}{2}w\left( X\right) \left( 1+\left( \hat{w}%
\left( X\right) \left( f\left( X\right) -\hat{R}\right) +\frac{1}{2}w\left(
X\right) R_{exc}\left( X\right) \right) \right)  \label{Shr}
\end{equation}%
\begin{equation}
S_{L}\left( X,X\right) =\frac{1}{2}w\left( X\right) \left( 1+\left( \hat{w}%
\left( X\right) \left( r\left( X\right) -\hat{R}\right) \right) \right)
\label{LN}
\end{equation}%
The decision of investing in firms $X$\ depends negatively on the average
returns of cross-investments $\left\langle \hat{f}\left( X^{\prime }\right)
\right\rangle _{\hat{w}_{E}}$\ and interest rates between investors $%
\left\langle \hat{r}\left( X^{\prime }\right) \right\rangle _{\hat{w}_{L}}$.
The stakes in firms will thus be higher in the low return solution than in
the high one.

The inward total stake in firms, which is the sum of equations (\ref{Shr})
and (\ref{LN}), is given by:%
\begin{equation}
S\left( X,X\right) =w\left( X\right) \left( 1+\left( \hat{w}\left( X\right)
\left( R\left( X\right) -\hat{R}\right) \right) \right)  \label{Shl}
\end{equation}

\subparagraph{Inward aggregate stake in firms}

The aggregate stakes of investors in firms $X$ is obtained by multiplying (%
\ref{Shr}), (\ref{LN}) and (\ref{LN}) by the capital ratio, which yields:%
\begin{equation*}
S_{E}\left( X\right) \equiv S_{E}\left( X,X\right) \frac{\hat{K}%
_{X}\left\vert \hat{\Psi}\left( X\right) \right\vert ^{2}}{K_{X}\left\vert
\Psi \left( X\right) \right\vert ^{2}}
\end{equation*}%
\begin{equation*}
S_{L}\left( X\right) \equiv S_{L}\left( X,X\right) \frac{\hat{K}%
_{X}\left\vert \hat{\Psi}\left( X\right) \right\vert ^{2}}{K_{X}\left\vert
\Psi \left( X\right) \right\vert ^{2}}
\end{equation*}%
and:%
\begin{equation*}
S\left( X\right) =S\left( X,X\right) \frac{\hat{K}_{\hat{X}}\left\vert \hat{%
\Psi}\left( X\right) \right\vert ^{2}}{K_{\hat{X}}\left\vert \Psi \left(
X\right) \right\vert ^{2}}
\end{equation*}%
These formula combine two opposing effects: the increased investment, due to
higher returns $f_{1}\left( X\right) $, and the decrease in the capital
density ratio $\frac{\hat{K}_{X}\left\vert \hat{\Psi}\left( X\right)
\right\vert ^{2}}{K_{X}\left\vert \Psi \left( X\right) \right\vert ^{2}}$,
due to decreasing marginal returns. Thus the overall proportion of
investors' capital in firms depends on the model's structural parameters.

\subsubsection{Step 4: solving for stakes}

Now that the solution defined by the return function $\hat{f}\left( X\right) 
$ have been established, we close the resolution by computing the stakes
invested sector by sector\footnote{%
See equations (\ref{Shn}), (\ref{Sht}), (\ref{Shv}).}:%
\begin{eqnarray}
\hat{S}_{E}\left( X^{\prime },X\right) &=&\frac{1}{2}\hat{w}\left( X^{\prime
},X\right) \left( 1+\hat{\Delta}^{E}\left( X^{\prime },X\right) \right)
\label{SNXPX} \\
\hat{S}_{L}\left( X^{\prime },X\right) &=&\frac{1}{2}\hat{w}\left( X^{\prime
},X\right) \left( 1+\hat{\Delta}^{L}\left( X^{\prime },X\right) \right) 
\notag \\
&&  \notag \\
\hat{S}\left( X^{\prime },X\right) &=&\hat{w}\left( X^{\prime },X\right)
\left( 1+\hat{\Delta}\left( X^{\prime },X\right) \right)  \label{SNP}
\end{eqnarray}%
The differences $\hat{\Delta}^{E}\left( X^{\prime },X\right) $ and $\hat{%
\Delta}^{L}\left( X^{\prime },X\right) $ represent the deviations of \ $\hat{%
f}\left( X^{\prime }\right) $ and $\hat{R}\left( X^{\prime }\right) $ from
their respective averages, and the expressions for the stakes invested in
firms $S_{E}\left( X,X\right) $ and $S\left( X,X\right) $ are provided in (%
\ref{Shr}) and (\ref{Shl}), respectively.

Any given collective state will be characterized by the stakes $\hat{S}%
_{E}\left( X^{\prime },X\right) $, $\hat{S}_{L}\left( X^{\prime },X\right) $
and $\hat{S}\left( X^{\prime },X\right) $, given by (\ref{SNXPX}) and (\ref%
{SNP}). These solutions do not differ fundamentaly from the aggregate stakes
described in (\ref{Shg}). The behaviors of agents sector by sector follow a
two-solution pattern. The difference between stakes (\ref{SNXPX}) and (\ref%
{SNP}) and aggregate stakes result from local variations in the perception
of risks $\hat{w}\left( X^{\prime },X\right) $ and $w\left( X^{\prime
},X\right) $ defined by (\ref{hb}) and (\ref{hc}). The sector dependence of
these two parameters impact the solutions and induce local variations around
the aggregate stakes, returns, and capital levels.

\section{Accounting for the default states}

The model has been resolved under a no-default scenario. In this light,
defaults states can be defined as modifications of non-default states.
Assuming one or several firms defaulting in one sector initially, the
default state is built recursively from this initial impact.\ Once this
initial default has propagated, the returns in each sector will be
deviations from what would have been the returns in a non-default scenario,
equation (\ref{QDM}).

We will detail below this propagation mechanism and the conditions for the
default to materialize and spread. The loss incured by investors and the
fraction of defaulting investors will then be computed.

\subsection{Uncertainty and default}

Note that the role of uncertainty changes depending on the agent who bares
the risk of default. When uncertainty increases, the threshold for default
diminishes and the risk of default increases for the agent that is invested
in, since uncertainty deters other investors.

In our model, uncertainty is defined very broadly, and corresponds to a
global perception of risk. The parameter $\gamma $, when not otherwise
mentionned, refers to an average perception of risk of the whole system on
all the sectors of the system. It would nonetheless be easy to render this
parameter sector-dependent, and we will do so when needed in our
interpretations.

Besides, the fact that the uncertainty is high does not necessarily imply
that the risk has risen, and inversely.\ Uncertainty and risk are not
correlated. Risk can exist without necessarily being perceived. In this
case, Investors may act under a false presomption of full certainty, and
risk is all the more great.

In the study of defaults, the notion of risk perception is fundamental, and
its role depends on the perception agents have of it and where and from
where it is seen. The default of a firm will be less likely to propagate to
other investors in a climate of confidence, when $\gamma $ is small.
Alternately, as mentioned above, the system as a whole is in a greater zone
of default when it benefits from a false climate of confidence.\ The
intermediation being stronger, if a default occurs, intermediation will
favor its propagation. The propagation of a default is more likely when,
locally, one sector is being perceived as risky, and thus locally $\gamma $
increases, investors in the sector default and this default propagates
because the rest of the system is impaired by an intermediation
corresponding to a $\gamma $ small. Default is more likely when the
differential between the $\gamma $ of the system and the $\gamma $ of a
specific sector is wider.

\subsection{Principle of propagation}

Under the default-scenario, the investors' return equation takes the form%
\footnote{%
See Appendix 1.}: 
\begin{eqnarray}
0 &=&\int \left( \delta \left( X-X^{\prime }\right) -\hat{S}_{E}\left(
X^{\prime },X\right) \right) \widehat{DF}\left( X^{\prime }\right) \hat{R}%
_{exc}\left( X^{\prime }\right) -S_{E}\left( X,X\right) R_{exc}\left(
X\right)  \label{RTdf} \\
&&-\int \max \left( -1,\left( 1+\hat{f}\left( X^{\prime }\right) \right) 
\frac{1-\hat{S}\left( X^{\prime }\right) }{\hat{S}_{L}\left( X^{\prime
}\right) }\right) H\left( -\left( 1+f\left( X^{\prime }\right) \right)
\right) \hat{S}_{L}\left( X^{\prime },X\right) dX^{\prime }  \notag \\
&&-\int \max \left( -1,\left( 1+f\left( X^{\prime }\right) \right) \frac{%
\left( 1-S\left( X\right) \right) }{S_{L}\left( X\right) }\right) H\left(
-\left( 1+f\left( X\right) \right) \right) S_{L}\left( X\right)  \notag
\end{eqnarray}%
where $H$ is the Heaviside function.

The solution to equation (\ref{RTdf}) is:%
\begin{eqnarray}
\hat{R}_{exc}\left( X\right) &=&\left( \delta \left( X-X^{\prime }\right) -%
\hat{S}_{E}\left( X^{\prime },X\right) \widehat{DF}\left( X^{\prime }\right)
\right) ^{-1}  \label{Df} \\
&&\left\{ \int_{\hat{S}_{-}}\max \left( -1,\left( 1+\hat{f}\left( X^{\prime
}\right) \right) \frac{1-\hat{S}\left( X^{\prime }\right) }{\hat{S}%
_{L}\left( X^{\prime }\right) }\right) \hat{S}_{L}\left( X^{\prime
},X\right) dX^{\prime }\right.  \notag \\
&&+\left. \int_{S_{-}}\max \left( -1,\left( 1+f\left( X\right) \right) \frac{%
\left( 1-S\left( X\right) \right) }{S_{L}\left( X\right) }\right)
S_{L}\left( X\right) +S_{E}\left( X,X\right) \max \left( R_{exc}\left(
X\right) ,0\right) \right\}  \notag
\end{eqnarray}%
where $S$ and $\hat{S}$ are the possible sets of defaulted firms and
investors, respectively. These sets are found recursively: we first assume
some realized initial default states $S_{0}$ and $\hat{S}_{0}$, when a set
of firms and investors respectively\ experience a complete loss of their
private capital, leading to:

For investors in $\hat{S}_{0}$:%
\begin{equation}
\hat{f}\left( X^{\prime }\right) =-1  \label{Cdf}
\end{equation}%
and for firms in $S_{0}$:

\begin{equation*}
f\left( \hat{X}\right) =-1
\end{equation*}%
where the number $-1$ in the right-hand side of both equations stands for a
return of $-100$ percent, so that private capital for defaulting firms and
investors is $0$. The impact of these initial defaults on the returns of
other investors is then found by inserting these conditions in (\ref{Df}),
which in turn modifies the return equation of the remaining investors'
returns, leading to:%
\begin{eqnarray*}
\hat{f}\left( X\right) &=&\left( \delta \left( X-X^{\prime }\right) -\hat{S}%
_{E}\left( X^{\prime },X\right) \right) ^{-1} \\
&&\left\{ -\int_{\hat{S}_{0-}}\hat{S}_{L}\left( X^{\prime },X\right)
dX^{\prime }-\int_{S_{0-}}S_{L}\left( X\right) +S_{E}\left( X,X\right)
f\left( X\right) \right\}
\end{eqnarray*}%
Some of the solutions $\hat{f}\left( X\right) $ of this new return equation
may themselves satisfy the default condition (\ref{Cdf}), and reveal new
defaulting investors that will further increase the default set.

The possible sets of defaulting firms are therefore obtained by the limit of
the sequence of equations:%
\begin{eqnarray*}
\hat{f}_{n+1}\left( \hat{X}\right) &=&\left( \delta \left( X-X^{\prime
}\right) -\hat{S}_{E}\left( X^{\prime },X\right) \right) ^{-1} \\
&&\left\{ -\int_{\hat{S}_{n}}\hat{S}_{L}\left( X^{\prime },X\right)
dX^{\prime }-\int_{S_{n}}S_{L}\left( X\right) +S_{E}\left( X,X\right)
f\left( X\right) \right\}
\end{eqnarray*}%
where $\hat{S}_{n}$ and $S_{n}$ are the overall set of investors and firms
having defaulted after $n$ iterations, and the remaining sets of agents are
denoted $S/\hat{S}_{n}$ and $S/S_{n}$.

Once the iteration stabilizes, the remaining sets of non-defaulted agents $%
\left( \hat{S}_{\infty },S_{\infty }\right) $ is defined by the limit $%
\left( S/\hat{S}_{n},S/S_{n}\right) \underset{n\rightarrow \infty }{%
\rightarrow }$ $\left( \hat{S}_{\infty },S_{\infty }\right) $ with returns $%
\left\{ \hat{f}_{n}\left( X\right) \underset{n\rightarrow \infty }{%
\rightarrow }\hat{f}\left( X\right) \right\} $, for which the resulting
disposable capital is given by equations (\ref{Dcn}) and (\ref{Dct}).

\subsection{Conditions for propagation}

Firms default when they meet the conditions for initial defaults. But for
this default to propagate to the whole set of investors, some additional
conditions must be met. First, the defaut must propagate to the firm's
immediate investors. Then, it must propagate from these investors to other
investors\footnote{%
See Appendix 8.}.

\subsubsection{Propagation from firms to investors}

Some firms may lack the disposable capital to face their costs, leading them
to default. Whether these initial defaults may push their intra-sectoral
investors into default depends on the magnitude of loss incured by the
defaulting firms.

For the default to spread to the firm's investors, the level of loss that a
firm has to experience, by lack of capital or increase in costs, must be
below a negative threshold\footnote{%
See Appendix 8.}:%
\begin{equation*}
R\left( X\right) <D_{Tw}
\end{equation*}%
where the threshold $D_{Tw}$ is: 
\begin{eqnarray*}
&&D_{Tw}=-\frac{4\left( 1+\frac{1}{2}\hat{w}\left( X\right) w\left( X\right)
\left\langle \widehat{DF}\left( X^{\prime }\right) \right\rangle
\left\langle \hat{R}_{exc}\left( X^{\prime }\right) \right\rangle \right) }{%
w\left( X\right) -\hat{w}\left( X\right) \left\langle \hat{R}_{exc}\left(
X^{\prime }\right) \right\rangle }\frac{\hat{w}\left( X\right) }{w\left(
X\right) } \\
&&\times \left( \left( 1+\left\langle \bar{r}\left( X\right) \right\rangle
\right) +\frac{1+\frac{1}{2}\left\langle \hat{R}_{exc}\left( X^{\prime
}\right) \right\rangle -w\left( X\right) \Delta r\left( X\right) }{1+\frac{1%
}{2}\hat{w}\left( X\right) w\left( X\right) \left\langle \widehat{DF}\left(
X^{\prime }\right) \right\rangle \left\langle \hat{R}_{exc}\left( X^{\prime
}\right) \right\rangle }\left\langle \widehat{DF}\left( X^{\prime }\right)
\right\rangle \left\langle \hat{R}_{exc}\left( X^{\prime }\right)
\right\rangle \right)
\end{eqnarray*}%
Since the threshold is negative, the default zone encompasses all returns
falling below this level.

When average excess returns among the whole set of investors $\left\langle 
\hat{R}_{exc}\left( X^{\prime }\right) \right\rangle $ increase or
uncertainty $\gamma $ decreases, the default zone decreases, and so does the
risk of default in the sector: the sector is safer financially. On the
contrary, when the average excess returns of the whole set of investors
decrease or uncertainty increases, the sector as a whole becomes more risky,
and investors diversify away from their sector by investing with other
investors, which further increases the default risk.

For a given level of returns among investors, any increase in interest rates
or in the level of loans between investors increases the default zone and
the risk of default for investors.

\subsubsection{Propagation from investors to other investors}

Given an initial default by firms and investors in certain sectors, the
condition under which the default may spread to other investors is:%
\begin{equation}
\left( \left\langle \hat{S}_{L}\left( X^{\prime },X\right) \right\rangle
+\left\langle S_{L}\left( X,X\right) \right\rangle \right) \Pi \left(
X\right) >2\left\langle \hat{f}\right\rangle \left( 1-\left\langle \hat{S}%
\left( X^{\prime }\right) \right\rangle \right)  \label{Dflt}
\end{equation}%
where the expression $\Pi \left( X\right) $ is defined by:%
\begin{equation*}
\Pi \left( X\right) =\frac{1+N\left\langle S_{E}\left( X^{\prime },X^{\prime
}\right) \right\rangle }{\left( 1-\left\langle \hat{S}\left( X\right)
\right\rangle \right) \left( 1-N\left( 1-\frac{\left\langle S_{E}\left(
X,X\right) \right\rangle }{1-\left\langle \hat{S}\left( X\right)
\right\rangle }\right) \right) }
\end{equation*}%
describes the level of interconnections within the system, not only between
investors but also between investors and firms. It involves average loans,
average investment stakes $\left\langle S_{E}\left( X^{\prime },X^{\prime
}\right) \right\rangle $ and $\left\langle \hat{S}\left( X\right)
\right\rangle $, and a parameter $N$.\ This last parameter\footnote{%
Defined in Appendix 8.3.} measures the level of intermediation within any
given group.\ It depends positively on the relative levels of disposable
capital of investors and firms, $\frac{\left\langle \hat{K}_{X}\left\vert 
\hat{\Psi}\left( X\right) \right\vert ^{2}\right\rangle }{\left\langle
K_{X}\left\vert \Psi \left( X\right) \right\vert ^{2}\right\rangle }$, on
the inverse uncertainty coefficients.\ Recall that the values of the
parameters in the equation are those of a non-default scenario.

Equation (\ref{Dflt}) shows that the potential for systemic default is a
global property of a group of connected agents: a default may begin in a
single sector, but its full propagation will depend on the structural
configuration of the system. A high volume of loans, $\left\langle \hat{S}%
_{L}\left( X^{\prime },X\right) \right\rangle +\left\langle S_{L}\left(
X,X\right) \right\rangle $, combined with a high level of intermediation
between investors ---$\left\langle \hat{S}\left( X^{\prime }\right)
\right\rangle $ close to $1$ ---set the system in the default zone.\ In this
default zone, any further increase in average loans or investors' average
returns $\left\langle \hat{f}\right\rangle $, even simply expected, increase
the probability of defaults\textbf{.}

Note that the expression $\Pi \left( X\right) $ is increasing in $N$: the
level of intermediation within the group amplifies the probability of
systemic default. The expression increases when investor disposable capital
increases and uncertainty decreases: the lower the perception of risk, the
higher the coefficients $\left\langle w\left( X\right) \right\rangle $ and $%
\left\langle \hat{w}\left( X\right) \right\rangle $, thereby increasing the
likelihood of cascading defaults within the system.

When risk is perceived as low\textbf{, }investments between investors
increase, and so does the probability of default, under adverse conditions.
This tendency is further magnified by the relative level of financial
disposable capital $\frac{\left\langle \hat{K}_{X}\left\vert \hat{\Psi}%
\left( X\right) \right\vert ^{2}\right\rangle }{\left\langle K_{X}\left\vert
\Psi \left( X\right) \right\vert ^{2}\right\rangle }$.\ When it is high,
defaults are more probable, structurally.

These various causes of structural fragility tend to reinforce one another.
A high level of capital invested in the financial sector typically coincides
with a low perception of risk and higher inter-investor stakes $\left\langle 
\hat{S}\left( X\right) \right\rangle $. Taken together, these factors lower
the threshold of average loan exposure ---$\left( \left\langle \hat{S}%
_{L}\left( X^{\prime },X\right) \right\rangle +\left\langle S_{L}\left(
X,X\right) \right\rangle \right) $--- required to trigger a structural
default.

\subsection{Description of the default state}

Once the default has materialized, the system is in a default state. This
state can be analyzed as a deviation from a no-default state. To do so, we
will compute, for the default state, both the average loss and the fraction
of investors affected.

When defaults occur, investors' returns are shifted by:

\begin{equation}
\hat{f}\left( X\right) \rightarrow \hat{f}\left( X\right) -\mu d\hat{f}
\label{Sh}
\end{equation}%
where $d\hat{f}$ is the loss incured by the remaining investors for each
investor defaulting, and $\mu $ is the fraction of investors impacted by the
default.

We find that:%
\begin{equation*}
d\hat{f}\left( X\right) =\frac{\left( \left\langle \hat{S}_{L}\left(
X^{\prime },X\right) \right\rangle +\left\langle S_{L}\left( X,X\right)
\right\rangle \right) \mu +\left\langle S_{E}\left( X^{\prime },X^{\prime
}\right) \right\rangle \left\langle df\left( X^{\prime }\right)
\right\rangle }{1-\left\langle \hat{S}\left( X^{\prime }\right)
\right\rangle }
\end{equation*}%
and that the fraction of defaulting investors $\mu $ is:%
\begin{equation}
\mu =\frac{1-\left\langle \hat{S}\left( X^{\prime }\right) \right\rangle }{%
\left( \left\langle \hat{S}_{L}\left( X^{\prime },X\right) \right\rangle
+\left\langle S_{L}\left( X,X\right) \right\rangle \right) +\left\langle
S_{E}\left( X^{\prime },X^{\prime }\right) \right\rangle \left\langle
df\left( X^{\prime }\right) \right\rangle -2\left\langle \hat{f}%
\right\rangle \left( 1-\left\langle \hat{S}\left( X^{\prime }\right)
\right\rangle \right) }  \label{frc}
\end{equation}%
where all parameter values are those of a non-default scenario.

The average loss that results from the default for the entire group of firms
can also be derived:%
\begin{equation*}
\left\langle df\left( X^{\prime }\right) \right\rangle =-\frac{\left(
\left\langle \hat{S}_{L}\left( X^{\prime },X\right) \right\rangle
+\left\langle S_{L}\left( X,X\right) \right\rangle \right) \frac{N}{%
1-\left\langle \hat{S}\left( X\right) \right\rangle }dC}{1-N\left( 1-\frac{%
\left\langle S_{E}\left( X,X\right) \right\rangle }{1-\left\langle \hat{S}%
\left( X\right) \right\rangle }\right) dC}
\end{equation*}%
where the parameter $dC$ writes:%
\begin{equation}
dC=C-\frac{rf_{1}\left( \hat{X}\right) }{\left( K_{\hat{X}}\right) ^{r-1}}
\label{Dcr}
\end{equation}%
This parameter reflects the impact of defaults on the firms' cost
structure.\ It measures the loss in returns per unit of stake divested from
firm $X$. Thus, higher values of $dC$ are associated with larger capital
loss and an increased likelihood of default.

The above expressions reveal that, the higher the investment stake $S\left(
X\right) $ in sector $X$, the smaller the value of $dC$. Financially
vulnerable states---characterized by a high $dC$---are more prone to
initiate a cascade of defaults. The resulting fraction of defaulting
investors will be given by equation (\ref{frc}).

\section{Stability of sub-collective states}

We have defined a sub-collective state as a given group of sectors of the
system (and hence their agents) in a specific phase, i.e. the set of stakes,
disposable capital and returns per sector, etc. and all their relevant
variables.

So far, we have considered sub-collective states as static. But for any
group of investors, multiple sub-collective states exist, including multiple
default states.\ Besides, our study of the default states has shown that the
set of defaulting agent is ultimately the result of a series of cascading
default among firms and investors. This suggest that shocks could trigger
new phases in the collective states and that some underlying dynamics exist
between these phases. The unstability in the dynamics of a phase will then
reveal the possibility of tansitions between phases.

To study the dynamics and the existence of transitions between these phases,
we must transform the static return equation (\ref{QDM}) into a dynamic
one.\ We will do so by introducing some time parameter in the equation.\ We
will then use this dynamic version of the return equation to analyze the
deviations of any sub-group around a given static phase and, in turn, its
(in)stability.

\subsection{Internal dynamics}

To reveal the internal dynamics of the system, we must transform the static
return equation (\ref{QDM}) for each group into a dynamic system. To do so,
we perform a first-order perturbation of this return equation.

\subsubsection{Dynamic return equation}

To study the distribution of stakes within groups dynamically, we must
detail the temporal sequence of investments by replacing the static return
equation (\ref{QDM}) with a dynamical formulation:%
\begin{eqnarray}
&&\left( \delta \left( X-X^{\prime }\right) -\hat{S}_{E}\left( X^{\prime
},X,\theta -1\right) \right) \frac{\left( 1-\hat{S}\left( X^{\prime },\theta
-1\right) \right) \left( \hat{f}\left( X^{\prime },\theta \right) -\bar{r}%
\right) }{1-\hat{S}_{E}\left( X^{\prime },\theta \right) }  \label{CPR} \\
&=&S_{E}\left( X,X,\theta -1\right) \left( f\left( X,\theta -1\right) -\bar{r%
}\right)  \notag
\end{eqnarray}%
which captures the lag between the investments $\hat{S}_{E}\left( X^{\prime
},X,\theta -1\right) $ and $S_{E}\left( X,X,\theta -1\right) $, and the
returns they provide:%
\begin{equation*}
\hat{f}\left( X^{\prime },\theta \right) -\bar{r}
\end{equation*}%
The returns generated by firms are themselves driven by lagged productivity:%
\begin{equation*}
f\left( X,\theta -1\right) -\bar{r}
\end{equation*}%
so that capital reallocation delays the impact on firms' returns\footnote{%
In this study, we focus on the case of decreasing returns to scale.}%
\footnote{%
The formula for \ $f\left( \hat{X},\theta -1\right) -\bar{r}$ is provided in
Appendix 9.}.

\subsubsection{Dynamic fluctuations of stakes}

Let us consider first-order fluctuations in investors returns, $\delta \hat{f%
}\left( X\right) $, and stakes $\delta \hat{S}_{E}\left( X^{\prime
},X\right) $ and $\delta S_{E}\left( X,X\right) $. These fluctuations depend
on a discrete time parameter $\theta $. Expanding the dynamic return
equation (\ref{CPR}) around a static solution yields\footnote{%
Derived in Appendix 9.} a dynamic equation for $\delta \hat{f}\left(
X\right) $:\ 
\begin{eqnarray*}
0 &=&\left( \delta \left( X-X^{\prime }\right) -\hat{S}_{E}\left( X^{\prime
},X,\theta -1\right) \right) \frac{1-\left( \hat{S}\left( X^{\prime },\theta
-1\right) +\delta \hat{S}\left( X^{\prime },\theta -1\right) \right) }{%
1-\left( \hat{S}_{E}\left( X^{\prime },\theta -1\right) +\delta \hat{S}%
_{E}\left( X^{\prime },\theta -1\right) \right) }\left( \hat{R}_{exc}\left(
X\right) +\delta \hat{f}\left( X\right) \right) \\
&&-\delta \hat{S}_{E}\left( X^{\prime },X,\theta -1\right) \frac{\left( 1-%
\hat{S}\left( X^{\prime },\theta -1\right) \right) \hat{R}_{exc}\left(
X^{\prime }\right) }{1-\hat{S}_{E}\left( X^{\prime },\theta -1\right) } \\
&&-\delta S_{E}\left( X,X,\theta -1\right) \left( f\left( X,\theta -1\right)
-\bar{r}\right) +S_{E}\left( X,X,\theta -1\right) \frac{\partial f\left(
X,\theta -1\right) }{\partial S\left( X,\theta -1\right) }
\end{eqnarray*}%
where the fluctuation in aggregate stake allocation, $\delta S\left(
X\right) $ varies both with intra-sectoral investments $\delta S\left(
X,X\right) $ and total disposable capital for investment: 
\begin{equation}
\delta S\left( X\right) =\delta S\left( X,X\right) \frac{\hat{K}_{\hat{X}%
}\left\vert \hat{\Psi}\left( X\right) \right\vert ^{2}}{K_{\hat{X}%
}\left\vert \Psi \left( X\right) \right\vert ^{2}}+S\left( X,X\right) \delta 
\frac{\hat{K}_{X}\left\vert \hat{\Psi}\left( X\right) \right\vert ^{2}}{%
K_{X}\left\vert \Psi \left( X\right) \right\vert ^{2}}
\end{equation}%
The variations in stakes in each sector is described by a coupled system of
equations involving investor returns and aggregate sectoral stake allocations%
\footnote{%
See Appendix 9.}:%
\begin{equation}
\left( 
\begin{array}{c}
\delta \hat{f}\left( X,\theta \right) -\delta \hat{f}\left( X,\theta
-1\right) \\ 
\delta S\left( X,\theta -1\right) -\delta S\left( X,\theta -2\right)%
\end{array}%
\right) =\left( 
\begin{array}{cc}
\alpha & c \\ 
h & \beta%
\end{array}%
\right) \left( 
\begin{array}{c}
\delta \hat{f}\left( X,\theta -1\right) \\ 
\delta S\left( X,\theta -2\right)%
\end{array}%
\right)  \label{SV}
\end{equation}%
that can alternately be reformulated solely in terms of stakes in firms and
other investors\footnote{%
See Appendix 11.}:%
\begin{equation*}
\left( 
\begin{array}{c}
\delta \hat{S}\left( X^{\prime },X,\theta \right) -\delta \hat{S}\left(
X^{\prime },X,\theta -1\right) \\ 
\delta S\left( X,\theta -1\right) -\delta S\left( X,\theta -2\right)%
\end{array}%
\right) =\left( 
\begin{array}{cc}
\alpha & \frac{\left\langle \hat{w}\left( X^{\prime },X\right) \right\rangle 
}{2}c \\ 
\frac{w}{\frac{\left\langle \hat{w}\left( X^{\prime },X\right) \right\rangle 
}{2}} & \beta%
\end{array}%
\right) \left( 
\begin{array}{c}
\delta \hat{S}\left( X^{\prime },X,\theta -1\right) \\ 
\delta S\left( X,\theta -2\right)%
\end{array}%
\right)
\end{equation*}%
The coefficients $\alpha $, $\beta $, $c$ and $w$ are defined explicitly in
Appendix 11.

\subsection{Internal stability}

To study the stability of any group under fluctuations, we first consider
that, for each sector $X^{\prime }$, we can aggregate other investors'
stakes in $X^{\prime }$. This amounts to replace the stakes from each
alternate sector $X$ $\ $to $X^{\prime }$, $\hat{S}\left( X^{\prime
},X,\theta -1\right) $, by their average $\left\langle \hat{S}\left(
X^{\prime },X,\theta -1\right) \right\rangle _{X}$ over the whole sector
space. When we do so, the resulting return equations for $X^{\prime }$
solely depend on $X^{\prime }$ and the inward aggregated averages. This
allows the study of deviations in the return equation (\ref{SV}) sector by
sector, independently. We will then detail more acurately the interactions
between sectors to study the propagation of perturbations within the sector
space.

\subsubsection{Independent fluctuation of stakes}

\paragraph{Eigenvalues}

The eigenvalues of equation (\ref{SV}) determine the stability of each
sector\ under some perturbations. When stakes are approximated by inward
aggregate stakes, these eigenvalues write:%
\begin{eqnarray}
&&\frac{1}{2}\left( \alpha +\beta \right) -\frac{1}{2}\sqrt{\left( \alpha
-\beta \right) ^{2}+4ch}  \label{gnv} \\
&&\frac{1}{2}\left( \alpha +\beta \right) +\frac{1}{2}\sqrt{\left( \alpha
-\beta \right) ^{2}+4ch}  \notag
\end{eqnarray}%
They are typically negative\footnote{%
This stability analysis is conducted in Appendix 10.}, so that the system is
stable in general. But since the dominant eigenvalue generally increases
with $\left\langle \hat{f}\left( X\right) \right\rangle -\bar{r}$,
instability arises when returns are high, which corresponds to the high
return solution presented above\footnote{%
See section 6.2.3.} and is coherent with the higher level of capital
circulation in this case.

\paragraph{Stability}

Instability primarily emerges when investors and firms' returns exhibit a
significant discrepancy, particularly when the returns of investors $X$ are
significantly higher than the returns of firms $X$, that is:%
\begin{equation}
\hat{R}_{exc}\left( X\right) >>1\text{ and }R_{exc}\left( X\right) \simeq 1
\label{CD}
\end{equation}%
Under these conditions, investors diversify capital away from their own
sector, undermining the stability of its equilibrium. When investors'
returns in other sectors are significantly high $\left\langle \hat{f}\left(
X^{\prime }\right) \right\rangle >>1$, capital flows out toward more
profitable sectors, especially when the perceived risk is low ($\gamma <<1$%
)\ which further amplifies such reallocations. This corresponds to high
returns solutions described in the previous section.

Under these conditions (\ref{CD}), even minor changes in - whether
anticipated or not - returns, can cause capital to shift in or out of the
sector, potentially leading to either a decline in returns and subsequent
capital accumulation, or conversely, a concentration of capital and higher
returns.

Another case of instability arises when the returns of firms $X$ do not
differ significantly from the interest rates, and when investors $X$ are
heavily invested in these firms, that is: 
\begin{equation*}
R_{exc}\left( X\right) <<1\mathbf{\ }\text{and}\mathbf{\ }S\left( X,\theta
-1\right) \rightarrow 1
\end{equation*}%
for instance when uncertainty is very high.

Under these conditions, sector $X$ becomes unstable only if $\left\langle 
\hat{f}\left( X^{\prime }\right) \right\rangle $\ is sufficiently low: even
expected small shifts in returns can induce significant shifts in capital.
This corresponds to the low return solution studied earlier: the potential
returns across the whole sector space $\left\langle \hat{f}\left( X^{\prime
}\right) \right\rangle $ are too low to attract investors $X$.\ They will
rather invest in firms $X$.\ This induces a high level of capital in firms $%
X $ and in turn low returns per unit of capital. However, any fluctuation in
returns $\left\langle \hat{f}\left( X^{\prime }\right) \right\rangle $\ may
shift the system towards an other equilibrium.

\paragraph{Dynamics of perturbations}

For a small initial perturbation in investors' returns and stakes, the
dominant term in the dynamics is\footnote{%
See Appendix 10.}:%
\begin{equation}
\left( 
\begin{array}{c}
\delta \hat{f}\left( X,\theta \right) \\ 
\delta S\left( X,\theta -1\right)%
\end{array}%
\right) \simeq \left( 
\begin{array}{c}
\left( A\delta \hat{f}+C\delta S\right) \exp \left( \lambda _{+}\theta
\right) \\ 
B\left( A\delta \hat{f}+C\delta S\right) \exp \left( \lambda _{+}\theta
\right)%
\end{array}%
\right)  \label{st}
\end{equation}%
where $\lambda _{+}$ denotes the largest eigenvalue of the system\footnote{%
The coefficients are detailed in Appendix 9.}.\ 

Depending on the initial disturbances $\delta \hat{f}$ and $\delta S$, two
cases may arise: either investors $X$ returns and total stakes in firms $X$ $%
\delta \hat{f}\left( X\right) $ $\ $and $\delta S\left( X,\theta -1\right) $
respectively both decay to zero, leading the system back to its equilibrium;
or they are amplified, inducing a transition of the system towards another
equilibrium.

In this second case, a positive perturbation either in $\delta \hat{f}$ or $%
\delta S$ pushes the sector towards a new equilibrium of higher returns and
larger capital accumulation in investors $X$: higher investors' returns
attract capital and increase the firms' capitalization in the sector.
Conversely, a negative perturbation shifts the sector towards lower returns
and lower firm capitalization.

\subsubsection{Interacting fluctuations between sectors}

In general, fluctuations in cross-sectoral investments are correlated, which
implies that the dynamics of several or all sectors and must consequently be
considered in interaction.

This implies that the stability of the system depends on specific
interactions between sub-groups of agents must be considered through total
sector-dependent stakes (\ref{SNXPX}) and (\ref{Shl}) rather than their
averages:%
\begin{eqnarray*}
\hat{S}\left( X^{\prime },X\right) &=&\hat{w}\left( X^{\prime },X\right)
\left( 1+\hat{\Delta}\left( X^{\prime },X\right) \right) \\
S\left( X,X\right) &=&w\left( X\right) \left( 1+\hat{w}\left( X\right)
\Delta \left( X\right) \right)
\end{eqnarray*}%
The sector-dependent variations of the total stakes are:%
\begin{eqnarray*}
\delta \hat{S}\left( X^{\prime },X,\theta -1\right) &\rightarrow &\frac{1}{2}%
\hat{w}\left( X^{\prime },X\right) \left( \frac{1}{2}\delta \hat{f}\left(
X^{\prime }\right) -\left\langle w\left( X\right) \right\rangle \delta
R\left( X\right) \right) \\
\delta S\left( X,X,\theta -1\right) &\rightarrow &\frac{1}{2}w\left(
X\right) \hat{w}\left( X\right) \delta f\left( X\right)
\end{eqnarray*}%
These formula compute the impact of a modification $\delta \hat{f}\left(
X^{\prime }\right) $ in investor $X^{\prime }$\ returns on the stakes $\hat{S%
}\left( X^{\prime },X,\theta -1\right) $\ of investor $X$, and reveal the
interconnections between sectors $X$\ and $X^{\prime }$\ fluctuations.

This correction modifies\footnote{%
See Appendix 11.}, for each sector, the matrix of the system with an
additional term:

\begin{equation*}
\left( 
\begin{array}{cc}
\alpha & c \\ 
w & \beta%
\end{array}%
\right) \rightarrow \left( 
\begin{array}{cc}
\alpha & c \\ 
h & \beta%
\end{array}%
\right) +\left( 
\begin{array}{cc}
0 & W \\ 
0 & 0%
\end{array}%
\right)
\end{equation*}%
where:

\begin{equation*}
W=-\hat{S}_{E}\left( X^{\prime },X,\theta -1\right) \widehat{DF}\left(
X^{\prime }\right)
\end{equation*}%
and alters the dynamic coefficients at each point in the group, such that
equation (\ref{SV}) becomes:%
\begin{equation}
\left( 
\begin{array}{c}
\delta \hat{f}\left( X,\theta \right) -\delta \hat{f}\left( X,\theta
-1\right) \\ 
\delta S\left( X,\theta -1\right) -\delta S\left( X,\theta -2\right)%
\end{array}%
\right) =\left( 
\begin{array}{cc}
\alpha & c\left( X\right) \\ 
h & \beta%
\end{array}%
\right) \left( 
\begin{array}{c}
\delta \hat{f}\left( X,\theta -1\right) \\ 
\delta S\left( X,\theta -2\right)%
\end{array}%
\right)
\end{equation}%
The above modifications turn the eigenvalues (\ref{gnv}) into local
quantities, so that the stability of individual sectors now depends on the
value of $c\left( X\right) $.

\paragraph{Unilateral interactions: Modification of eigenvalues and
instability propagation}

To study how subgroups of intercating agents depart from non interacting
ones, we must assume stakes to be both sector-dependent and dynamically
interacting.\ We can then measure the impact of sector-specific variations $%
w $ on neighboring sectors and on the whole system.

A typical case of unilateral interactions between two sectors can be
expressed as a dynamical system of two distinct sectors, coupled such that :%
\begin{equation*}
\left( 
\begin{array}{c}
\delta \hat{f}\left( X,\theta \right) -\delta \hat{f}\left( X,\theta
-1\right) \\ 
\delta S\left( X,\theta -1\right) -\delta S\left( X,\theta -2\right) \\ 
\delta \hat{f}\left( X^{\prime },\theta \right) -\delta \hat{f}\left(
X^{\prime },\theta -1\right) \\ 
\delta S\left( X^{\prime },\theta -1\right) -\delta S\left( X^{\prime
},\theta -2\right)%
\end{array}%
\right) =\left( 
\begin{array}{cccc}
\alpha & c & 0 & W \\ 
h & \beta & 0 & 0 \\ 
0 & 0 & \alpha ^{\prime } & c^{\prime } \\ 
0 & 0 & h^{\prime } & \beta ^{\prime }%
\end{array}%
\right) \left( 
\begin{array}{c}
\delta \hat{f}\left( X,\theta -1\right) \\ 
\delta S\left( X,\theta -2\right) \\ 
\delta \hat{f}\left( X^{\prime },\theta -1\right) \\ 
\delta S\left( X^{\prime },\theta -2\right)%
\end{array}%
\right)
\end{equation*}%
This system exhibits non-reciprocal interactions: variations in one sector
unilaterally impact the other sector.\ Investment behavior in sector $1$ is
unchanged, but causes investment in sector $2$ to deviate from its average.
Such interactions do not, in themselves, alter the eigenvalues and the
stability of the overall system. As long as sector $1$ remains stable, the
dynamics of sector $2$ is impacted but still stable: fluctuations in both
blocks remain damped and contained. Instability emerges only when sector $1$
itself becomes unstable, leading to a reallocation of investment stakes and
disposable capital across sectors.\ This instability may propagate to
neighboring sectors and the broader system.

\paragraph{Multilateral interactions: Modification of eigenvalues and higher
instability}

To study the reciprocal interactions between two sectors, we will modify
both dynamical matrices symmetrically:

\begin{equation*}
\left( 
\begin{array}{cccc}
\alpha & c & 0 & W \\ 
h & \beta & 0 & 0 \\ 
0 & W^{\prime } & \alpha ^{\prime } & c^{\prime } \\ 
0 & 0 & h^{\prime } & \beta ^{\prime }%
\end{array}%
\right)
\end{equation*}%
and refer to this configuration as a multilateral deviation from average.
The modification of each sector eigenvalues and stability under this
simplifying hypothesis\footnote{%
See Appendix 11.} is $\left( \alpha ,\beta ,w,c\right) \simeq \left( \alpha
^{\prime },\beta ^{\prime },w^{\prime },c^{\prime }\right) $, which isolates
the effect of the reciprocal extra-shares $W$ and $W^{\prime }$.%
\begin{equation*}
\Delta \lambda _{+}\left( X\right) =\frac{\hat{S}_{E}\left( X^{\prime
},X,\theta -1\right) \widehat{DF}\left( X^{\prime }\right) \hat{S}_{E}\left(
X,X^{\prime },\theta -1\right) \widehat{DF}\left( X\right) }{\lambda
_{X}-\lambda _{X^{\prime }}}\frac{h^{2}}{\left( \alpha -\beta \right)
^{2}+4ch}
\end{equation*}%
Under these conditions, the largest eigenvalue may indirectly drag the
system to an unstable zone, where $\lambda _{X}>>1$.

More generally, circular investment deviations, i.e., loops of modified
investment patterns involving multiple sectors, modify the eigenvalues by
sums of contributions of the type:

\begin{equation*}
\Delta \lambda _{i\alpha }=\prod \frac{W_{j_{k+1}\alpha _{k+1},j_{k}\alpha
_{k}}}{\lambda _{i\alpha }-\lambda _{j_{k+1}\alpha _{k+1}}}
\end{equation*}%
where $\lambda _{i\alpha }${} are the eigenvalues associated with sector $i$%
, and $W_{j_{k+1}\alpha _{k+1},j_{k}\alpha _{k}}${}represents the
interaction between $j_{k}\alpha _{k}$ and $j_{k+1}\alpha _{k+1}$.

The global effect is the sum over all possible paths and computed by: 
\begin{equation*}
\Delta \lambda _{i\alpha }=\sum_{\left( j_{k}\alpha _{k}\right) }\prod \frac{%
W_{j_{k+1}\alpha _{k+1},j_{k}\alpha _{k}}}{\lambda _{i\alpha }-\lambda
_{j_{k+1}\alpha _{k+1}}}
\end{equation*}%
These loops directly affect the eigenvalues and the stability of the
collective state. When the largest eigenvalue turns positive, the system
becomes unstable and may transition to another equilibrium. In this regime,
any drop in returns toward zero or below may drag the system into default.
In general, these looped interactions increase the systemic fragility and
contribute to the propagation of instabilities along the investment circuit.

\subsection{Transitions}

Within groups of stably interconnected agents, external fluctuations may
change connectivity patterns which may, in turn, create circulation loops
among specific groups of agents and reinforce the internal investments
within these groups.

Such feedback loops may trigger instability when fluctuations in investors'
returns drive capital away from firms and push the system toward a default
state. This new equilibrium typically involves a reduced, more concentrated
group of investors.

\subsubsection{Group modification}

Under greater certainty, an anticipated increase in returns may induce new
investment towards new groups of agents. As a result, several groups may
merge to form a larger collective structure.

Let us consider a shock on both returns and stakes in one group $G_{a}$:%
\begin{equation*}
\left\{ \left( 
\begin{array}{c}
\delta \hat{f}_{a}\left( X,\theta \right) \\ 
\delta S_{a}\left( X,\theta -1\right)%
\end{array}%
\right) \right\} _{a}
\end{equation*}%
where $a$ is the group index. This may induce two groups $G_{a}${} and $%
G_{b} $ to merge, creating a joint vector of returns $\left( \hat{f}%
_{a}+\delta \hat{f}_{a}\right) \left( X,\theta \right) $ and stakes $\left(
S_{a}+\delta S_{a}\right) \left( X,\theta -1\right) $:%
\begin{equation*}
\left( 
\begin{array}{c}
\left( \mathbf{\hat{f}+\delta \hat{f}}\right) \left( X,\theta \right) \\ 
\left( \mathbf{S+\delta S}\right) \left( X,\theta -1\right)%
\end{array}%
\right)
\end{equation*}%
which represents the evolving state of the newly formed group.

Dynamically, the transition unfolds through the propagation mechanism:%
\begin{eqnarray*}
\delta \hat{f}_{a}\left( X_{a},\theta \right) &\rightarrow &\delta
S_{ab}\left( X_{a},\theta -1\right) \\
\delta S_{ab}\left( X_{a},\theta -1\right) &\rightarrow &\delta S_{b}\left(
X_{b},\theta -1\right) \rightarrow \delta \lambda \left( X_{b},\theta
-1\right)
\end{eqnarray*}

Even if all the elements of both original groups were incorporated into the
new structure $\left\{ X_{a}\right\} \cup \left\{ X_{b}\right\} $, the
resulting group could nonetheless contain unstable components, driving the
system toward new equilibria, including default configurations.

\subsubsection{Interpretation}

The evolution towards instability can be analyzed following the framework
described in (\ref{CD}). When new connections are introduced, the system
evolves such that $\hat{f}\left( X\right) \rightarrow \hat{f}\left( X\right)
+\delta \hat{f}\left( X\right) $, potentially triggering transitions toward
instability in specific sectors. While the impacted sector begins to exhibit
fluctuations, the rest of the system may remain stable in the short term.
Gradually, however, the real economic activity of the affected group
contracts, and its financial robustness deteriorates under the cumulative
effect of these fluctuations. In time, this instability can lead to default
in some sectors, thereby altering the structure and connectivity of the
broader system\footnote{%
The parameters characterizing the emergence of such unstable regimes, as
well as the range of fluctuations capable of triggering these transitions,
are detailed in Appendix 10.}.

\section{Results}

The present paper has showed that within collective states, there may exist
some sub-collective states.\ We have used this notion of sub-collective
states to study the present model, first under the hypothesis of no-default,
and then, by introducing the possibility of defaults.

\subsection{Sub-collective states}

We have found that when stakes and uncertainty are endogenized agents tend
to self-organize into more or less independent groups, for which the various
variables of each sector can be derived. Any of these given groups can
potentially exist in multiple phases, characterized by different connection
structures, capital levels, and so forth, among which one or several may
correspond to default states. We refer to the combination of a group in one
of its possible phases as a \emph{sub-collective state}. Each arbitrary
group is thus associated with a multiplicity of sub-collective states and,
to a first approximation, analyzing a collective state amounts to studying
agents' connections within each of these groups independently.

Collective states thus exhibit a twofold multiplicity: in the decomposition
of agents into groups, and in the infinite combinations of\emph{\ }%
sub-collective states associated with any given group partition. This
multiplicity of decompositions into sub-collective states suggests that the
stability of these states cannot be assumed \textit{a priori}. This paper
therefore develops a dynamic framework around each collective state, which
allows to analyze\textbf{\ }the transitions between different possible
collective states.

In the present paper, collective states describe a global configuration of
the whole set of agents that include their returns, capital levels, mutual
stakes, and other relevant quantities. Three factors determine the
equilibria of the model: investment-related uncertainty, interest rates, and
firm returns---i.e., essentially their productivity. It is the distribution
of these three factors within the sectors and among agents of a group that
will determine the possible distributions of all other variables in the
model, the stability of the collective state and the possibility of default.

The model yields two possible types of equilibria: a no-default equilibrium
and a default equilibrium. Since this paper focuses on static collective
states, they represent two clearly distinct configurations, each associated
with a different resolution mechanism. The no-default equilibrium is solved
under the assumption that all firms have access to all the capital they
require. The default scenario comes as a deviation from this equilibrium:
assuming some initial defaults, a recursive resolution of the model
highlights the mechanism through which they may propagate across the system.

\subsection{Equilibrium without default}

Since fields encompass both local and global features,\ and and that they
interact with one another, we first solve the model on average, then use
these relevant averages to solve the model sector by sector.

\subsubsection{Average equilibrium}

The model's average equilibrium is described by the average stakes between
investors, their average stakes in firms, and the level of average
disposable capital and excess returns for firms and investors, respectively.
These averages reveal the main features of the collective state
inter-sectoral equilibria.\ 

These averages are themselves determined by the firms' average excess return
over interest rates---the difference between the average productivity of
firms and prevailing interest rates---and the uncertainty of investors about
their investments in other investors.\ Taken jointly, they determine firms'
average returns, investors' returns, and total available capital, with
uncertainty either amplifying or reducing diversification and intermediation.

Higher productivity reduces firms' reliance on external investor capital.\
There is a crowding-out effect: with greater productivity, firms
simultaneously reduce their dependence on investor funding and enhance
firms' ability to attract investor interest.

The nature of the capital to which a firm gains access depends on the level
of interest rates. The higher the interest rate, the more investors will
favor debt instruments; conversely, when interest rates are low, equity
investment becomes more attractive. By contrast, when firms' productivity
falls below the interest rate, resorting to external capital becomes costly
and ultimately unsustainable in the long run. More broadly, any increase in
average productivity or in interest rates tends to undermine access to
capital for the least productive firms.

Under constant productivity and interest rates, uncertainty reduces
intermediation---that is, investment flows between investors. It encourages
intra-sectoral investment and tends to reduce the capital available to
investors and, indirectly, to firms as well. This decline in intermediation
stabilizes the collective state, as capital fluctuations become less
pronounced for both firms and investors, although this more stable
equilibrium may be associated with lower returns for investors.

\subsubsection{Equilibria per sector}

At the sectoral level, it is the distribution among sectors of firms'
productivity, interest rates and uncertainty that will determine the
possible distributions of all other variables in the model, as well as the
stability of the collective state and the possibility of default.

In each sector, regardless of the firms' actual returns, two types of
investors may exist: low or high return investors. In practice and on
average, each sector will be characterized by its main type of investor, so
that we can speak of a high- or low-return sector.\textbf{\ }This collective
strategy will determine both the capital available to firms and the firms'
resulting returns within that sector.

Under the low-return strategy, investors diversify between the firms of
their sector and oher sectors' investors, depending on the level of
uncertainty and the returns of the market and their respective sector
firms'returns. Their returns are close but below the level of average
returns.

Under the high-return strategy, investors with high level of disposable
capital divert capital away from their own sector toward other sectors,
seeking \textit{ex ante }higher returns. Alternatively, this behavior might
also be motivated by a desire to diversify away from a sector perceived as
too risky. Due to their high level of capital, these investors can ask
above-average returns for their participations or loans.

This high-return strategy can emerge in any sector, but will be less
relevant in sectors where firms returns are high, as investors must resist
the premia of other investors to reinvest locally, and face the additional
layer of uncertainty\textbf{\ }associated with a limited marginal gain. The
high-return strategy is therefore bound to be marginal in sectors with high
return firms. Conversely, the low-return strategy is bound to be marginal
among investors in sectors with low return firms. Incidentally, note that in
our model, uncertainty does not concern firm returns \textit{per se}, but
rather investor returns. However, this assumption can easily be relaxed by
lifting the constraints on investment shares.

Ultimately, it is the global environment of firm returns that shapes both
the low-return and high-return strategies adopted by investors. And
reciprocally, it is the investment strategies of investors that determine
the nature of firm returns within a given sector.

\subsubsection{Stability}

The multiplicity of possible equilibria suggests that these equilibria can
be modified, and are thus not necessarily stable. When there is indeed
instability, The conditions of emergence and propagation must be studied.

\paragraph{Stability per sector}

In general, the low-return strategy, which prioritizes one's own sector
while diversifying only moderately, tends to be the norm. By contrast, the
high-return strategy is more unstable, since it depends on fluctuations in
global returns. This instability is particularly strong when there is a
large gap between investor and firm returns within the same sector, which
indicates that the sector's investors have adopted a high-return strategy
and are heavily invested outside their own sector and will be more exposed
to global return fluctuations, let alone their own potential investment
misjudgments.

The high-return strategy is generally unstable when it produces very high
returns for investors in a context of standard returns for firms. Such
structural imbalance, in which investors in a given sector earn high returns
from investments made outside their own sector, cannot be sustained over
time. First, local investors lack the comparative advantage of a privileged
access to capital in other sectors. Second, their strategic choices deprive
their own firms of the capital they need, putting the whole sector at risk.

The low-return strategy is stable when firms returns in the sector are
strong, since even conservative investment behavior yields high or at least
sufficient returns. However, it becomes unstable when firm returns are poor.
In this case, a significant return gap between firms in the sector and those
in the broader economy makes diversification clearly profitable for local
investors. Even slight fluctuations in the sector's return environment can
destabilize the investors' low-return equilibrium, and---by
extension---undermine the capitalization of firms in that sector.

\paragraph{Propagation of instability across sectors}

In general, sectors being interconnected, fluctuations in returns and stakes
within one sector can propagate to others.\ When a sector experiences
significant fluctuations in returns or capital allocations, these
disturbances can spread to other - and potentially all - sectors within the
system.

Two types of sectors are particularly vulnerable to exogenous shocks:
sectors with low firm returns combined with a high-return strategy, which
are highly sensitive to shocks in global returns; and sectors with low firm
returns and a low-return strategy in an environment of high global returns,
which are vulnerable to increases in global returns or negative shocks
within their own sector.

Instability may also emerge endogenously within a group of sectors. Weakly
connected sectors, or those with diffuse rather than concentrated
connections to other sectors, tend to be stable. Inversely, strongly
connected sectors---especially those engaged in bilateral dependencies and
also extensively linked to the broader network---can act as channels of
systemic instability.

Bilateral interactions between sectors are, in themselves, a source of
instability. When investors are interdependent in stakes with investors in
other sectors, their respective sectors become more unstable than the
average. There is a feedback effect, whereby the dynamics of one sector
affect others and reverberate back. This instability will be the cause of
transitions between collective states.

\subsection{Equilibrium with defaults}

A default state is a static configuration in which the possibility of
default has been explicitly incorporated into the resolution process.
Default states are an integral part of the model's set of possible
solutions. Assuming that one or more firms are in default, the model
constructs the corresponding static solution under that hypothesis.

Defaults states are similar in their features to no-default states, and are
actually constructed on the no-default baseline: returns, stakes and capital
levels after default are obtained as a shift in the levels of returns,
stakes and capital of the no-default state. Specifically, returns are
shifted downward proportionally to the number of sectors in the group who
are in default.

As a result, the default state can be viewed as a configuration defined on a
smaller group with reduced capital, and so forth. The form of the default
states, and particularly the overall loss of the remaining agents, depends
directly on the non-default states they emerge from. The higher the level of
loans, i.e. the higher the loans between investors and banks or loans of
banks to investors, the higher the losses.

Since a default state includes the initial defaults, as well as all firms
and investors to whom the default has propagated, its structure depends both
on the initial defaults and on the conditions governing the propagation
dynamics. These are structural conditions on the group that depend on the
levels of lending, returns, intermediation, and consequently, uncertainty.

The first condition concerns the propagation of default from a firm to its
direct investors. The risk of propagation depends on the entire system. If
the average return of investors is high, and if the uncertainty regarding
the sector to which the defaulting firms belong is elevated, then the risk
of default among the investors in that sector increases: the system diverts
away from the defaulting sector, causing its investors to default.

The second condition determines the propagation of default to other
investors: a default will propagate more readily when the level of loans to
investors and firms is high, and when uncertainty is low. Low uncertainty
increases intermediation and thus the risk of propagation. Conversely, a
high average return among investors reduces the risk of default propagation,
as elevated returns help prevent a total loss of capital. These parameters
are not independent. For instance, when returns are low and uncertainty is
high, loans increase, and with it the risk of propagation.

In this context, uncertainty plays a dual role. Locally, high uncertainty
about investors in sectors with potentially defaulting firms weakens those
investors and facilitates the emergence of defaults. Conversely, too low
uncertainty among investors at the system-wide scale facilitates the
propagation of defaults throughout the system: confidence increases
intermediation and the possibility of contagion.

Therefore, defaults do not arise from an increase in overall uncertainty,
but rather from its decrease; this is a manifestation of the herd effect.
Uncertainty limits investment and reduces its concentration, while certainty
diverts investors from their initial sectors, concentrates capital, and
potentially increases systemic risk.

\subsection{Transitions}

The existence of multiple and possibly unstable equilibria implies, in a
dynamic context, the existence of transitions between collective states.
These transitions can occur from a non-default to a defaut state, or more
generally between non-default states.

\subsubsection{Possible transition towards a default}

Any sector whose firms generate structurally below-average returns will
attract below-average capital. This does not necessarily imply a default,
specifically when investors in this sector adopt low-return strategies.

However, when for various exogenous reasons, investors shift to a
high-return strategy, they introduce a potential in the sector.\ Indeed, a
high-return strategy depends heavily on fluctuations in the broader
environment.\ Crucially, these investors lack any comparative advantage to
lure in other investors.

Besides, pursuing a high-return strategy, they are now exposed to the
volatility of the return environment. Should a negative fluctuation occur,
its impact---combined with the investors' own instability---would lead to a
loss in their returns and capital. This loss, directly passed on to the
investor sector's firms, would lead to a further reduction in capital and
trigger a default, which in turn, may propagate to the sector's investors
and ultimately spread across part or all the system.

\subsubsection{Possible transition between non-default states}

The realization of a default state can be viewed as a transition from a
no-default state to a default state. A non-default equilibrium, once turned
unstable, may be driven towards default through the propagation mechanism
studied earlier: some sectors default, leading others to default.

This is in fact indicative of a more general mechanism. For each group,
multiple phases exist. By studying the stability of these phases, one can
determine whether fluctuations will alter them or not. When subject to
fluctuations, relatively unstable phases, i.e. phases whose stability domain
is relatively narrow, will be more likely to drift towards another phase, to
transition to a new phase. Thus, at the level of collective states,
transitions between group phases constitute a transition of the collective
state. Moreover, if groups are more or less interconnected, cascades of
group transitions may occur, inducing transitions of collective states.

The only difference with a default state is that, upon transitioning to a
default state, the group is ultimately reduced in size, whereas when moving
from one standard phase to another, the group remains unchanged.

\section{Discussion}

From a methodological standpoint, this paper has shown that additional
variables can be integrated into an existing field model by adding
supplementary fields to the model.\ We start from an investment field model
in which investment decisions are exogenous and endogenize these decisions
by transforming them into the arguments of a new field governed by an action
functional that reflects microeconomic behaviors. Any field model can thus
be extended and made more complex by incorporating additional fields.

The endogenization notably reveals the role of uncertainty in agents'
investment decisions, which, by constraining their equity stakes, leads to
specific sectoral allocation within each collective state.

Our results show that collective states, which until now seemed to be
defined over the entire sectoral space, are to a first approximation,
arbitrary collections of sectoral groups. Each sector is characterized both
by its connections with other sectors within its group and by the levels of
capital and returns associated with each agent type operating in that sector.

The specification of a group of sectors along with their specific
characteristics defines a sub-collective state. For given exogenous
parameters, each group is associated with a multiplicity of sub-collective
states, that have \textit{a priori} the same group-level averages for
inter-agent connections and capital and return levels by agent type across
the group. Each sub-collective state within a group is distinguished from
the others by the sector-level values taken by these variables. The set of
sub-collective states associated with a given group thus generally forms
variations around the group average values.

Nevertheless, because stakes are now determined endogenously by agents'
returns, uncertainty about these returns may lead to an unequal distribution
of capital, the undercapitalization of certain sectors, and, consequently,
the emergence of default.

Two types of equilibria emerge: those in which agents do not anticipate
default, and those in which agents have already internalized the possibility
of a default within the sectoral space. As long as default is not
anticipated, and in the absence of exogenous shocks, all sectors---even
those that are undercapitalized---may continue to access the capital they
need. However, once a default occurs, agents adjust their expectations
accordingly: a default state may then materialize, in which an initial wave
of realized defaults can propagate across the entire group. Some sectors may
default, impacting the group as a whole, since both the average
characteristics of the associated sub-collective state and the number of
remaining sectors within the group are reduced.

The transition to a default state unfolds as a gradual loss of
capitalization that may eventually push certain sectors into default.
However, this process alone does not necessarily imply contagion to the
entire group. The default state---characterized by the propagation of an
initial default---is by no means inevitable: it depends critically on the
specific features of the sub-collective state. Three characteristics are
fundamental to the emergence of a default state: a high degree of
interconnection among investors, a high level of lending between them, and
low firm returns in the sectors constituting the group. A default state can
thus be viewed as a transformation of a non-default state, which implicitly
reveals that the group itself, its internal configuration, and therefore the
collective state as a whole, have undergone a change.

Thus, collective states are not fixed structures: they are subject to
transitions driven either by internal fluctuations or by exogenous shocks.
These transitions may occur between similar states---for instance, from one
non-default collective state to another---but may also involve paradigm
shifts, such as the transition from a non-default collective state to a
default one. Moreover, such transitions may also unfold in cascades, at
varying speeds and intervals. A stable non-default state may therefore,
under the influence of internal or external fluctuations, evolve into
another stable state---albeit less robust---and gradually transition toward
a default state.

\section{Conclusion}

This paper has highlighted the critical role of available capital in the
emergence of defaults, and the structural instability that high investment
return strategies induce. This may be due to the fact that the model does
not incorporate monetary creation that could mitigate the lack of capital.
Indeed, banks, which were already present in Gosselin and Lotz (2024), were
deliberately excluded here.

Besides, in our model, investors have the exclusive right to invest in firms
within their own sector. When firms in their sector yield
higher-than-average returns, these investors benefit from the intermediation
demand of investors from other sectors, who depend on them to gain exposure
to those returns. In this sense, they hold a relative monopoly on the
returns generated by firms in their sector.

To address these issues, the second part of this series will focus on how
the inclusion of banks, and the monetary creation they enable, alters the
system's stability outcomes and influences the transition mechanisms between
collective states, particularly regarding the emergence of defaults.

\section*{References}


\begin{description}
\item Abergel, F., Chakraborti, A., Muni Toke, I., \& Patriarca, M. (2011a).
Econophysics review: I. Empirical facts. Quantitative Finance, 11(7),
991--1012.

\item Abergel, F., Chakraborti, A., Muni Toke, I., \& Patriarca, M. (2011b).
Econophysics review: II. Agent-based models. Quantitative Finance, 11(7),
1013--1041.

\item Acemoglu, D., Ozdaglar, A., \& Tahbaz-Salehi, A. (2015). Systemic risk
and stability in financial networks. American Economic Review, 105(2),
564--608.

\item Acharya, V. V., Pedersen, L. H., Philippon, T., \& Richardson, M.
(2017). Measuring systemic risk. The Review of Financial Studies, 30(1),
2--47.

\item Adrian, T., \& Brunnermeier, M. K. (2016). CoVaR. American Economic
Review, 106(7), 1705--1741.

\item Allen, F., \& Gale, D. (2000). Financial contagion. Journal of
Political Economy, 108(1), 1--33.

\item Bardoscia, M., Livan, G., \& Marsili, M. (2017). Statistical mechanics
of complex economies. Journal of Statistical Mechanics: Theory and
Experiment, 2017, 1--21.

\item Bardoscia, M., Battiston, S., Caccioli, F., \& Caldarelli, G. (2019).
Pathways towards instability in financial networks. Nature Communications,
10(1), 1--9.

\item Battiston, S., Puliga, M., Kaushik, R., Tasca, P., \& Caldarelli, G.
(2012). Debtrank: Too central to fail? Financial networks, the FED and
systemic risk. Scientific Reports, 2, 541.

\item Battiston, S., Caldarelli, G., May, R. M., Roukny, T., \& Stiglitz, J.
E. (2020). The price of complexity in financial networks. Proceedings of the
National Academy of Sciences, 117(52), 32779--32786.

\item Bernanke, B., Gertler, M., \& Gilchrist, S. (1999). The financial
accelerator in a quantitative business cycle framework. In J. B. Taylor \&
M. Woodford (Eds.), Handbook of Macroeconomics (Vol. 1, Part C, pp.
1341--1393). Elsevier.

\item Bensoussan, A., Frehse, J., \& Yam, P. (2018). Mean Field Games and
Mean Field Type Control Theory. Springer, New York.

\item B\"{o}hm, V., Kikuchi, T., \& Vachadze, G. (2008). Asset pricing and
productivity growth: The role of consumption scenarios. Computational
Economics, 32, 163--181.

\item Caggese, A., \& Orive, A. P. (2015). The interaction between household
and firm dynamics and the amplification of financial shocks. Barcelona GSE
Working Paper Series, Working Paper n%
${{}^o}$
866.

\item Campello, M., Graham, J., \& Harvey, C. R. (2010). The real effects of
financial constraints: Evidence from a financial crisis. Journal of
Financial Economics, 97(3), 470--487.

\item Cifuentes, R., Ferrucci, G., \& Shin, H. S. (2005). Liquidity risk and
contagion. Journal of the European Economic Association, 3(2--3), 556--566.

\item Cochrane, J. H. (Ed.). (2006). Financial markets and the real economy.
Edward Elgar, International Library of Critical Writings in Financial
Economics, Vol. 18.

\item Elliott, M., Golub, B., \& Jackson, M. O. (2014). Financial networks
and contagion. American Economic Review, 104(10), 3115--3153.

\item Gaffard, J.-L., \& Napoletano, M. (Eds.). (2012). Agent-based models
and economic policy. OFCE, Paris.

\item Gai, P., \& Kapadia, S. (2010). Contagion in financial networks.
Proceedings of the Royal Society A: Mathematical, Physical and Engineering
Sciences, 466(2120), 2401--2423.

\item Gennaioli, N., Shleifer, A., \& Vishny, R. W. (2012). Neglected risks,
financial innovation, and financial fragility. Journal of Financial
Economics, 104(3), 452--468.

\item Gomes, D. A., Nurbekyan, L., \& Pimentel, E. A. (2015). Economic
models and mean-field games theory. Publica\c{c}\~{o}es Matem\'{a}ticas do
IMPA, 30%
${{}^o}$
Col\'{o}quio Brasileiro de Matem\'{a}tica, Rio de Janeiro.

\item Gosselin P., \& Lotz, A. (2024) Financial Interactions and Capital
Accumulation. hal-04573829.

\item Gosselin, P., \& Lotz, A. (2023a). A statistical field theory and
neural structures dynamics I: Action functionals, background states and
external perturbations. hal-04301171.

\item Gosselin, P., \& Lotz, A. (2023b). Statistical field theory and neural
structures dynamics II: Signals propagation, interferences, bound states.
hal-04301181v1.

\item Gosselin, P., \& Lotz, A. (2023c). A statistical field theory and
neural structures dynamics III: Effective action for connectivities,
interactions and emerging collective states. hal-04301185.

\item Gosselin, P., \& Lotz, A. (2023d). A statistical field theory and
neural structures dynamics IV: Field-theoretic formalism for interacting
collective states. hal-04301191.

\item Gosselin, P., Lotz, A., \& Wambst, M. (2017). A path integral approach
to interacting economic systems with multiple heterogeneous agents.
IF\_PREPUB, hal-01549586v2.

\item Gosselin, P., Lotz, A., \& Wambst, M. (2020). A path integral approach
to business cycle models with large number of agents. Journal of Economic
Interaction and Coordination, 15, 899--942.

\item Gosselin, P., Lotz, A., \& Wambst, M. (2021). A statistical field
approach to capital accumulation. Journal of Economic Interaction and
Coordination, 16, 817--908.

\item Grassetti, F., Mammana, C., \& Michetti, E. (2022). A dynamical model
for real economy and finance. Mathematical Finance and Economics.
https://doi.org/10.1007/s11579-021-00311-3

\item Greenwood, R., \& Hanson, S. G. (2013). Issuer quality and corporate
bond returns. The Review of Financial Studies, 26(6), 1483--1525.

\item Grosshans, D., \& Zeisberger, S. (2018). All's well that ends well? On
the importance of how returns are achieved. Journal of Banking \& Finance,
87, 397--410.

\item Haldane, A. G., \& May, R. M. (2011). Systemic risk in banking
ecosystems. Nature, 469(7330), 351--355.

\item Holmstrom, B., \& Tirole, J. (1997). Financial intermediation,
loanable funds, and the real sector. Quarterly Journal of Economics, 112(3),
663--691.

\item Jackson, M. O. (2010). Social and economic networks. Princeton
University Press, Princeton.

\item Jermann, U. J., \& Quadrini, V. (2012). Macroeconomic effects of
financial shocks. American Economic Review, 102(1), 238--271.

\item Khan, A., \& Thomas, J. K. (2013). Credit shocks and aggregate
fluctuations in an economy with production heterogeneity. Journal of
Political Economy, 121(6), 1055--1107.

\item Kaplan, G., \& Violante, L. (2018). Microeconomic heterogeneity and
macroeconomic shocks. Journal of Economic Perspectives, 32(3), 167--194.

\item Kleinert, H. (1989). Gauge fields in condensed matter (Vols. I--II).
World Scientific, Singapore.

\item Kleinert, H. (2009). Path integrals in quantum mechanics, statistics,
polymer physics, and financial markets (5th ed.). World Scientific,
Singapore.

\item Krugman, P. (1991). Increasing returns and economic geography. Journal
of Political Economy, 99(3), 483--499.

\item Langfield, S., Liu, Z., \& Ota, T. (2020). Mapping the network of
financial linkages: An industry-level analysis of the UK. Journal of Banking
\& Finance, 118, 105946.

\item Lasry, J.-M., Lions, P.-L., \& Gu\'{e}ant, O. (2010a). Application of
mean field games to growth theory. hal-00348376.

\item Lasry, J.-M., Lions, P.-L., \& Gu\'{e}ant, O. (2010b). Mean field
games and applications. In Paris-Princeton lectures on Mathematical Finance.
Springer. hal-01393103.

\item Lux, T. (2008). Applications of statistical physics in finance and
economics. Kiel Institute for the World Economy (IfW), Kiel.

\item Lux, T. (2016). Applications of statistical physics methods in
economics: Current state and perspectives. European Physical Journal Special
Topics, 225, 3255. https://doi.org/10.1140/epjst/e2016-60101-x

\item Mandel, A., Jaeger, C., F\"{u}rst, S., Lass, W., Lincke, D., Meissner,
F., Pablo-Marti, F., \& Wolf, S. (2010). Agent-based dynamics in
disaggregated growth models. Documents de travail du Centre d'Economie de la
Sorbonne, Paris.

\item Mandel, A. (2012). Agent-based dynamics in the general equilibrium
model. Complexity Economics, 1, 105--121.

\item Monacelli, T., Quadrini, V., \& Trigari, A. (2011). Financial markets
and unemployment. NBER Working Papers, 17389. National Bureau of Economic
Research.

\item Reinhart, C. M., \& Rogoff, K. S. (2009). This time is different:
Eight centuries of financial folly. Princeton University Press.

\item Sims, C. A. (2006). Rational inattention: Beyond the linear quadratic
case. American Economic Review, 96(2), 158--163.

\item Yang, J. (2018). Information theoretic approaches to economics.
Journal of Economic Surveys, 32(3), 940--960.
\end{description}

\pagebreak

\section*{Appendices}

\section*{Appendix 1 Alternate formulation for return equations}

\subsection*{A1.1 Average field and capital per sector}

The average field $\left\Vert \hat{\Psi}\left( X_{1}\right) \right\Vert ^{2}$
and the amount of capital $\hat{K}_{X_{1}}\left\Vert \hat{\Psi}\left(
X_{1}\right) \right\Vert ^{2}$ and the average capital $\hat{K}_{X_{1}}$ in
sector $X$\ as functions of $\hat{K}_{0}^{2}$ and averages $\left\langle 
\hat{K}\right\rangle $ and $\left\langle \hat{K}_{0}\right\rangle $:

\begin{eqnarray}
\left\Vert \hat{\Psi}\left( X_{1}\right) \right\Vert ^{2} &\simeq &\hat{\mu}%
\frac{\hat{K}_{0}^{3}}{\sigma _{\hat{K}}^{2}}\left( \frac{\hat{g}^{2}\left(
X_{1}\right) }{3}-\frac{\left\langle \hat{K}\right\rangle \left\langle \hat{g%
}\right\rangle ^{2}}{2\left\langle \hat{K}_{0}\right\rangle }\underline{\hat{%
k}}\left( \left\langle X\right\rangle ,\left\langle X\right\rangle \right)
\right) \\
&=&\frac{\hat{\mu}}{\sigma _{\hat{K}}^{2}}\left( 2\frac{\sigma _{\hat{K}}^{2}%
}{\hat{g}^{2}\left( X_{1}\right) }\left( \frac{\left\Vert \hat{\Psi}%
_{0}\left( X_{1}\right) \right\Vert ^{2}}{\hat{\mu}}-D\left( X_{1}\right)
\right) \right) ^{\frac{3}{2}}\left( \frac{\hat{g}^{2}\left( X_{1}\right) }{3%
}-\frac{\left\langle \hat{K}\right\rangle \left\langle \hat{g}\right\rangle
^{2}}{2\left\langle \hat{K}_{0}\right\rangle }\hat{k}\right)  \notag
\end{eqnarray}%
\begin{eqnarray}
\hat{K}\left[ X_{1}\right] &=&\hat{K}_{X}\left\Vert \hat{\Psi}\left(
X_{1}\right) \right\Vert ^{2}\simeq \hat{\mu}\frac{\hat{K}_{0}^{4}}{2\sigma
_{\hat{K}}^{2}}\left( \frac{\hat{g}^{2}\left( X_{1}\right) }{4}-\frac{%
\left\langle \hat{K}\right\rangle \left\langle \hat{g}\right\rangle ^{2}}{%
3\left\langle \hat{K}_{0}\right\rangle }\hat{k}\left( \left\langle
X\right\rangle ,\left\langle X\right\rangle \right) \right) \\
&=&\frac{\hat{\mu}}{2\sigma _{\hat{K}}^{2}}\left( 2\frac{\sigma _{\hat{K}%
}^{2}}{\hat{g}^{2}\left( X_{1}\right) }\left( \frac{\left\Vert \hat{\Psi}%
_{0}\left( X_{1}\right) \right\Vert ^{2}}{\hat{\mu}}-D\left( X_{1}\right)
\right) \right) ^{2}\left( \frac{\hat{g}^{2}\left( X_{1}\right) }{4}-\frac{%
r\left\langle \hat{g}\right\rangle ^{2}}{3}\underline{\hat{k}}\right)  \notag
\end{eqnarray}%
and the average capital per investor in sector $X_{1}$ is: 
\begin{equation}
\hat{K}_{X_{1}}=\frac{\hat{K}\left[ X_{1}\right] }{\left\Vert \hat{\Psi}%
\left( X_{1}\right) \right\Vert ^{2}}=\frac{\sqrt{2\frac{\sigma _{\hat{K}%
}^{2}}{\hat{g}^{2}\left( X_{1}\right) }\left( \frac{\left\Vert \hat{\Psi}%
_{0}\left( X_{1}\right) \right\Vert ^{2}}{\hat{\mu}}-D\left( X_{1}\right)
\right) }\left( \frac{\hat{g}^{2}\left( X_{1}\right) }{4}-\frac{%
r\left\langle \hat{g}\right\rangle ^{2}}{3}\hat{k}\right) }{2\left( \frac{%
\hat{g}^{2}\left( X_{1}\right) }{3}-\frac{\left\langle \hat{K}\right\rangle
\left\langle \hat{g}\right\rangle ^{2}}{2\left\langle \hat{K}%
_{0}\right\rangle }\hat{k}\right) }
\end{equation}%
where the modified returns $\hat{g}$ are defined by: 
\begin{equation*}
\hat{g}\left( \hat{K},X\right) =\left( 1-M\left( \left( \hat{K},X\right)
,\left( \hat{K}^{\prime },X^{\prime }\right) \right) \left\vert \hat{\Psi}%
\left( \hat{K}^{\prime },X^{\prime }\right) \right\vert ^{2}\right) ^{-1}%
\hat{f}\left( \hat{K}^{\prime },X^{\prime }\right)
\end{equation*}%
with:%
\begin{equation*}
M\left( \hat{K},X,\hat{K}^{\prime },X^{\prime }\right) =\frac{\hat{k}\left(
X,X^{\prime }\right) \hat{K}}{1+\underline{\hat{k}}\left( X\right) }
\end{equation*}

\subsection*{A1.2 General form of the equation}

The return equation, including defaults writes:%
\begin{eqnarray}
&&\frac{f\left( X\right) -\bar{r}}{1+\underline{\hat{k}}_{L}\left( X\right) }%
-\int \frac{\hat{k}_{E}\left( X^{\prime },X\right) \hat{K}^{\prime
}\left\vert \hat{\Psi}\left( \hat{K}^{\prime },X^{\prime }\right)
\right\vert ^{2}}{1+\underline{\hat{k}}\left( X^{\prime }\right) }\frac{%
f\left( X^{\prime }\right) -\bar{r}}{1+\underline{\hat{k}}_{L}\left(
X^{\prime }\right) }d\hat{K}^{\prime }dX^{\prime } \\
&=&\int \left( \frac{1+f\left( X^{\prime }\right) }{\underline{\hat{k}}%
_{L}\left( X^{\prime }\right) }H\left( -\frac{1+f\left( X^{\prime }\right) }{%
\underline{\hat{k}}_{L}\left( X^{\prime }\right) }\right) \right) \frac{\hat{%
k}_{L}\left( X^{\prime },X\right) \hat{K}^{\prime }\left\vert \hat{\Psi}%
\left( \hat{K}^{\prime },X^{\prime }\right) \right\vert ^{2}}{1+\underline{%
\hat{k}}\left( X^{\prime }\right) }d\hat{K}^{\prime }dX^{\prime }  \notag \\
&&+\int \left( \frac{1+f_{1}^{\prime }\left( X^{\prime }\right) }{\underline{%
k}_{L}\left( X^{\prime }\right) }H\left( -\frac{1+f_{1}^{\prime }\left(
X^{\prime }\right) }{\underline{k}_{L}\left( X^{\prime }\right) }\right)
\right) \frac{k_{L}\left( X^{\prime },X\right) \left\vert \Psi \left(
K^{\prime },X^{\prime }\right) \right\vert ^{2}K^{\prime }}{1+\underline{k}%
\left( X^{\prime }\right) }dK^{\prime }dX^{\prime }  \notag \\
&&+\int \frac{\left\vert \Psi \left( K^{\prime },X^{\prime }\right)
\right\vert ^{2}k_{E}\left( X^{\prime },X\right) K^{\prime }}{1+\underline{k}%
\left( X^{\prime }\right) }\left( \frac{f_{1}^{\prime }\left( X^{\prime
}\right) -\bar{r}+\Delta F_{\tau }\left( \bar{R}\left( K^{\prime },X^{\prime
}\right) \right) }{1+\underline{k}_{L}\left( X^{\prime }\right) }\right)
dK^{\prime }dX^{\prime }  \notag
\end{eqnarray}%
The second term is the portion of return from other investors retrieved by
the investor in $X$. On the right side of the equation, the first term
calculates the loss due to the default of other investors, when it occurs.
The second term calculates the loss due to the default of firms in which the
investor had invested, and the last term is the return generated by the
firms in which the investor had invested.

The reformulation of return equation is performed straightforwardly by
rewriting all the coefficients as functions of the shares: 
\begin{equation*}
\hat{S}_{\eta }\left( X^{\prime },\hat{K}^{\prime },X\right) =\frac{\hat{k}%
_{\eta }\left( X^{\prime },X\right) \hat{K}^{\prime }\left\vert \hat{\Psi}%
\left( \hat{K}^{\prime },X^{\prime }\right) \right\vert ^{2}}{1+\underline{%
\hat{k}}\left( X^{\prime }\right) }
\end{equation*}%
and:%
\begin{equation*}
S_{\eta }\left( X^{\prime },K^{\prime },X\right) =\frac{k_{\eta }\left(
X^{\prime },X\right) \left\vert \Psi \left( K^{\prime },X^{\prime }\right)
\right\vert ^{2}K^{\prime }}{1+\underline{k}\left( X^{\prime }\right) }
\end{equation*}%
the equation for returns becomes:%
\begin{eqnarray}
0 &=&\int \left( \delta \left( X-X^{\prime }\right) -\hat{S}_{E}\left(
X^{\prime },\hat{K}^{\prime },X\right) \right) \frac{f\left( X^{\prime
}\right) -\bar{r}}{1+\underline{\hat{k}}_{L}\left( X^{\prime }\right) }%
dX^{\prime }d\hat{K}^{\prime }  \label{MPK} \\
&&-\int S_{E}\left( X^{\prime },K^{\prime },X\right) \left( \frac{%
f_{1}^{\prime }\left( X^{\prime }\right) -\bar{r}}{1+\underline{k}_{L}\left(
X^{\prime }\right) }+\Delta F_{\tau }\left( \bar{R}\left( K^{\prime
},X^{\prime }\right) \right) \right) dX^{\prime }dK^{\prime }  \notag \\
&&-\int \frac{1+f\left( X^{\prime }\right) }{\underline{\hat{k}}_{L}\left(
X^{\prime }\right) }H\left( -\frac{1+f\left( X^{\prime }\right) }{\underline{%
\hat{k}}_{L}\left( X^{\prime }\right) }\right) \hat{S}_{L}\left( X^{\prime },%
\hat{K}^{\prime },X\right) dX^{\prime }d\hat{K}^{\prime }  \notag \\
&&-\int \frac{1+f_{1}^{\prime }\left( X^{\prime }\right) }{\underline{k}%
_{L}\left( X^{\prime }\right) }H\left( -\frac{1+f_{1}^{\prime }\left(
X^{\prime }\right) }{\underline{k}_{L}\left( X^{\prime }\right) }\right)
S_{L}\left( X^{\prime },K^{\prime },X\right) dX^{\prime }d\hat{K}^{\prime } 
\notag
\end{eqnarray}%
with $\delta \left( X-X^{\prime }\right) $ the Dirac function.

Integrating this equation over $\hat{K}^{\prime }$ leads to define the
following quantities:%
\begin{eqnarray*}
\hat{S}_{\eta }\left( X^{\prime },\hat{K}^{\prime },X\right) &\rightarrow
&\int \frac{\hat{k}_{\eta }\left( X^{\prime },X\right) \hat{K}^{\prime
}\left\vert \hat{\Psi}\left( X^{\prime }\right) \right\vert ^{2}}{1+%
\underline{\hat{k}}\left( X^{\prime }\right) } \\
&=&\frac{\hat{k}_{\eta }\left( X^{\prime },X\right) \hat{K}_{X^{\prime
}}\left\vert \hat{\Psi}\left( X^{\prime }\right) \right\vert ^{2}}{1+%
\underline{\hat{k}}\left( X^{\prime }\right) }\equiv \hat{S}_{\eta }\left(
X^{\prime },X\right)
\end{eqnarray*}%
and:%
\begin{equation*}
S_{\eta }\left( X^{\prime },K^{\prime },X\right) \rightarrow \frac{k_{\eta
}\left( X^{\prime },X\right) K_{X^{\prime }}\left\vert \Psi \left( X^{\prime
}\right) \right\vert ^{2}}{1+\underline{k}\left( X^{\prime }\right) }\equiv
S_{\eta }\left( X^{\prime },X\right)
\end{equation*}%
and it yields to replace (\ref{MPK}) by:%
\begin{eqnarray}
0 &=&\int \left( \delta \left( X-X^{\prime }\right) -\hat{S}_{E}\left(
X^{\prime },X\right) \right) \frac{f\left( X^{\prime }\right) -\bar{r}}{1+%
\underline{\hat{k}}_{L}\left( X^{\prime }\right) }dX^{\prime }  \label{TRV}
\\
&&-\int S_{E}\left( X^{\prime },X\right) \left( \frac{f_{1}^{\prime }\left(
X^{\prime }\right) -\bar{r}+\Delta F_{\tau }\left( \bar{R}\left( K^{\prime
},X^{\prime }\right) \right) }{1+\underline{k}_{L}\left( X^{\prime }\right) }%
\right) dX^{\prime }  \notag \\
&&-\int \frac{1+f\left( X^{\prime }\right) }{\underline{\hat{k}}_{L}\left(
X^{\prime }\right) }H\left( -\frac{1+f\left( X^{\prime }\right) }{\underline{%
\hat{k}}_{L}\left( X^{\prime }\right) }\right) \hat{S}_{L}\left( X^{\prime
},X\right) dX^{\prime }-\int \frac{1+f_{1}^{\prime }\left( X^{\prime
}\right) }{\underline{k}_{L}\left( X^{\prime }\right) }H\left( -\frac{%
1+f_{1}^{\prime }\left( X^{\prime }\right) }{\underline{k}_{L}\left(
X^{\prime }\right) }\right) S_{L}\left( X^{\prime },X\right) dX^{\prime } 
\notag
\end{eqnarray}%
with the constraint:%
\begin{equation*}
\int \left( \hat{S}_{E}\left( X^{\prime },X\right) +\hat{S}_{L}\left(
X^{\prime },X\right) \right) dX^{\prime }+\int \left( S_{E}\left( X^{\prime
},X\right) +S_{L}\left( X^{\prime },X\right) \right) dX^{\prime }=1
\end{equation*}%
The replacement is completed by comptng the remaining coefficents in (\ref%
{TRV}) in terms of the $\hat{S}_{\eta }$ and $S_{\eta }$. Given that:%
\begin{equation*}
\hat{S}_{\eta }\left( X^{\prime },X\right) =\frac{\hat{k}_{\eta }\left(
X^{\prime },X\right) \hat{K}_{X^{\prime }}\left\vert \hat{\Psi}\left(
X^{\prime }\right) \right\vert ^{2}}{1+\underline{\hat{k}}\left( X^{\prime
}\right) }
\end{equation*}%
and:%
\begin{equation*}
\hat{S}\left( X^{\prime },X\right) =\hat{S}_{E}\left( X^{\prime },X\right) +%
\hat{S}_{L}\left( X^{\prime },X\right)
\end{equation*}%
we find:%
\begin{equation*}
\int \hat{S}\left( X^{\prime },X\right) \frac{\hat{K}_{X}\left\vert \hat{\Psi%
}\left( X\right) \right\vert ^{2}}{\hat{K}_{X^{\prime }}\left\vert \hat{\Psi}%
\left( X^{\prime }\right) \right\vert ^{2}}dX=\frac{\underline{\hat{k}}%
\left( X^{\prime }\right) }{1+\underline{\hat{k}}\left( X^{\prime }\right) }
\end{equation*}%
Moreover, defining averages:%
\begin{equation*}
\hat{S}_{\eta }\left( X^{\prime }\right) =\int \hat{S}_{\eta }\left(
X^{\prime },X\right) \frac{\hat{K}_{X}\left\vert \hat{\Psi}\left( X\right)
\right\vert ^{2}}{\hat{K}_{X^{\prime }}\left\vert \hat{\Psi}\left( X^{\prime
}\right) \right\vert ^{2}}dX,\hat{S}\left( X^{\prime }\right) =\int \hat{S}%
\left( X^{\prime },X\right) \frac{\hat{K}_{X}\left\vert \hat{\Psi}\left(
X\right) \right\vert ^{2}}{\hat{K}_{X^{\prime }}\left\vert \hat{\Psi}\left(
X^{\prime }\right) \right\vert ^{2}}dX
\end{equation*}%
\begin{equation*}
S_{\eta }\left( X^{\prime }\right) =\int S_{\eta }\left( X^{\prime
},X\right) \frac{\hat{K}_{X}\left\vert \hat{\Psi}\left( X\right) \right\vert
^{2}}{K_{X^{\prime }}\left\vert \Psi \left( X^{\prime }\right) \right\vert
^{2}}dX,S\left( X^{\prime }\right) =\int S\left( X^{\prime },X\right) \frac{%
\hat{K}_{X}\left\vert \hat{\Psi}\left( X\right) \right\vert ^{2}}{%
K_{X^{\prime }}\left\vert \Psi \left( X^{\prime }\right) \right\vert ^{2}}dX
\end{equation*}%
we find the expression of the initial set of parameters as functions of the
new parameters:%
\begin{equation*}
\frac{1}{1+\underline{\hat{k}}\left( X^{\prime }\right) }=1-\hat{S}\left(
X^{\prime }\right)
\end{equation*}%
\begin{equation*}
\underline{\hat{k}}\left( X^{\prime }\right) =\frac{\hat{S}\left( X^{\prime
}\right) }{1-\hat{S}\left( X^{\prime }\right) }
\end{equation*}%
\begin{equation*}
1+\underline{\hat{k}}_{L}\left( X^{\prime }\right) =\frac{1-\hat{S}%
_{E}\left( X^{\prime }\right) }{1-\hat{S}\left( X^{\prime }\right) }
\end{equation*}%
\begin{equation*}
\frac{\underline{k}\left( X^{\prime }\right) }{1+\underline{k}\left(
X^{\prime }\right) }=S\left( X^{\prime }\right)
\end{equation*}%
\begin{equation*}
\frac{1}{1+\underline{k}\left( X^{\prime }\right) }=1-S\left( X^{\prime
}\right)
\end{equation*}%
\begin{eqnarray*}
\underline{k}\left( X^{\prime }\right) &=&\frac{S\left( X^{\prime }\right) }{%
1-S\left( X^{\prime }\right) } \\
\underline{k}\left( X^{\prime }\right) &=&S\left( X^{\prime }\right)
\end{eqnarray*}%
\begin{equation*}
1+\underline{k}_{L}\left( X^{\prime }\right) =\frac{1-S_{E}\left( X^{\prime
}\right) }{1-S\left( X^{\prime }\right) }
\end{equation*}%
and equation (\ref{TRV}) writes:%
\begin{eqnarray*}
0 &=&\int \left( \delta \left( X-X^{\prime }\right) -\hat{S}_{E}\left(
X^{\prime },X\right) \right) \frac{1-\hat{S}\left( X^{\prime }\right) }{1-%
\hat{S}_{E}\left( X^{\prime }\right) }\left( f\left( X^{\prime }\right) -%
\bar{r}\right) dX^{\prime } \\
&&-\int S_{E}\left( X^{\prime },X\right) \left( \frac{1-S\left( X^{\prime
}\right) }{1-S_{E}\left( X^{\prime }\right) }\left( \left( f_{1}^{\prime
}\left( X^{\prime }\right) -\bar{r}\right) +\Delta F_{\tau }\left( \bar{R}%
\left( K^{\prime },X^{\prime }\right) \right) \right) \right) dX^{\prime } \\
&&-\int \left( 1+f\left( X^{\prime }\right) \right) \frac{1-\hat{S}\left(
X^{\prime }\right) }{\hat{S}_{L}\left( X^{\prime }\right) }H\left( -\frac{%
1+f\left( X^{\prime }\right) }{\underline{\hat{k}}_{L}\left( X^{\prime
}\right) }\right) \hat{S}_{L}\left( X^{\prime },X\right) dX^{\prime } \\
&&-\int \left( 1+f_{1}^{\prime }\left( X^{\prime }\right) \right) \frac{%
1-S\left( X^{\prime }\right) }{S_{L}\left( X^{\prime }\right) }H\left( -%
\frac{1+f_{1}^{\prime }\left( X^{\prime }\right) }{\underline{k}_{L}\left(
X^{\prime }\right) }\right) S_{L}\left( X^{\prime },X\right) dX^{\prime }
\end{eqnarray*}%
As in the text, we consider that investment take place to close location so
that:%
\begin{eqnarray*}
S_{E}\left( X^{\prime },X\right) &=&S_{E}\left( X,X\right) \delta \left(
X^{\prime }-X\right) \\
S_{L}\left( X^{\prime },X\right) &=&S_{L}\left( X,X\right) \delta \left(
X^{\prime }-X\right)
\end{eqnarray*}%
and the constraint simplifies as:%
\begin{equation*}
\int \left( \hat{S}_{E}\left( X^{\prime },X\right) +\hat{S}_{L}\left(
X^{\prime },X\right) \right) dX^{\prime }+S_{E}\left( X,X\right)
+S_{L}\left( X,X\right) =1
\end{equation*}%
whereas the return equation (\ref{TRV}) becomes:%
\begin{eqnarray}
0 &=&\int \left( \delta \left( X-X^{\prime }\right) -\hat{S}_{E}\left(
X^{\prime },X\right) \right) \frac{1-\hat{S}\left( X^{\prime }\right) }{1-%
\hat{S}_{E}\left( X^{\prime }\right) }\left( \hat{f}\left( X^{\prime
}\right) -\bar{r}\right) dX^{\prime }  \label{QDl} \\
&&-\int \left( 1+f\left( X^{\prime }\right) \right) \frac{1-\hat{S}\left(
X^{\prime }\right) }{\hat{S}_{L}\left( X^{\prime }\right) }H\left( -\left(
1+f\left( X^{\prime }\right) \right) \right) \hat{S}_{L}\left( X^{\prime
},X\right) dX^{\prime }  \notag \\
&&-\left( 1+f_{1}^{\prime }\left( X\right) \right) \left( 1-S\left( X\right)
\right) H\left( -\left( 1+f_{1}^{\prime }\left( X\right) \right) \right) 
\notag \\
&&-S_{E}\left( X,X\right) \left( \frac{1-S\left( X\right) }{1-S_{E}\left(
X\right) }\left( \left( f_{1}^{\prime }\left( X\right) -\bar{r}\right)
+\Delta F_{\tau }\left( \bar{R}\left( K,X\right) \right) \right) \right) 
\notag
\end{eqnarray}

\subsection*{A1.3 No default scenario}

Remark that without defaut, this reduces to:%
\begin{eqnarray}
0 &=&\int \left( \delta \left( X-X^{\prime }\right) -\hat{S}_{E}\left(
X^{\prime },X\right) \right) \frac{1-\hat{S}\left( X^{\prime }\right) }{1-%
\hat{S}_{E}\left( X^{\prime }\right) }\left( f\left( X^{\prime }\right) -%
\bar{r}\right) dX^{\prime }  \label{RT} \\
&&-S_{E}\left( X,X\right) \left( \frac{1-S\left( X\right) }{1-S_{E}\left(
X\right) }\left( \left( f_{1}^{\prime }\left( X\right) -\bar{r}\right)
+\Delta F_{\tau }\left( \bar{R}\left( K,X\right) \right) \right) \right) 
\notag
\end{eqnarray}%
with solution:%
\begin{eqnarray}
\hat{f}\left( X^{\prime }\right) &=&\bar{r}+\frac{1-\hat{S}_{E}\left(
X^{\prime }\right) }{1-\hat{S}\left( X^{\prime }\right) }\left( \delta
\left( X-X^{\prime }\right) -\hat{S}_{E}\left( X^{\prime },X\right) \right)
^{-1}  \label{RN} \\
&&\times \left( S_{E}\left( X,X\right) \left( \frac{1-S\left( X\right) }{%
1-S_{E}\left( X\right) }\left( \left( f_{1}^{\prime }\left( X\right) -\bar{r}%
\right) +\Delta F_{\tau }\left( \bar{R}\left( K,X\right) \right) \right)
\right) \right)  \notag
\end{eqnarray}%
\begin{equation*}
\frac{\frac{\hat{k}_{E}\left( X^{\prime },X\right) \hat{K}\left[ X^{\prime }%
\right] }{1+\underline{\hat{k}}\left( X^{\prime }\right) }}{\frac{%
\left\langle \hat{K}\left[ X^{\prime }\right] \right\rangle }{1+\left\langle 
\underline{\hat{k}}\left( X^{\prime }\right) \right\rangle }}
\end{equation*}%
\begin{equation*}
\frac{\frac{\int \hat{k}\left( X^{\prime },X\right) \left\langle \hat{K}%
\left[ X^{\prime }\right] \right\rangle }{1+\left\langle \underline{\hat{k}}%
\left( X^{\prime }\right) \right\rangle }}{\frac{\left\langle \hat{K}\left[
X^{\prime }\right] \right\rangle }{1+\left\langle \underline{\hat{k}}\left(
X^{\prime }\right) \right\rangle }}=1
\end{equation*}

\subsection*{A1.3 Default scenario}

Coming back to the equation with default, the loss realized when some
default arise is:%
\begin{equation*}
\max \left( -\left( 1+\bar{r}\right) ,\frac{1+f_{1}^{\prime }\left(
X^{\prime }\right) }{\underline{k}_{L}\left( X^{\prime }\right) }\right)
\end{equation*}%
which translates that an investor can not lose more than its amounts of
loans plus the potential return on them.

When the loss is maximal equation (\ref{QDl}) becomes:

\begin{eqnarray*}
0 &=&\int \left( \delta \left( X-X^{\prime }\right) -\hat{S}_{E}\left(
X^{\prime },X\right) \right) \frac{1-\hat{S}\left( X^{\prime }\right) }{1-%
\hat{S}_{E}\left( X^{\prime }\right) }\left( f\left( X^{\prime }\right) -%
\bar{r}\right) dX^{\prime } \\
&&+\int \left( 1+\bar{r}\right) \frac{1-\hat{S}\left( X^{\prime }\right) }{%
\hat{S}_{L}\left( X^{\prime }\right) }\hat{S}_{L}\left( X^{\prime },X\right)
dX^{\prime }+\left( 1+\bar{r}\right) \left( 1-S\left( X\right) \right) \\
&&-S_{E}\left( X,X\right) \left( \frac{1-S\left( X\right) }{1-S_{E}\left(
X\right) }\left( \left( f_{1}^{\prime }\left( X\right) -\bar{r}\right)
+\Delta F_{\tau }\left( \bar{R}\left( K,X\right) \right) \right) \right)
\end{eqnarray*}%
whereas, in the general case, the return equation (\ref{QDl}) is:%
\begin{eqnarray*}
0 &=&\int \left( \delta \left( X-X^{\prime }\right) -\hat{S}_{E}\left(
X^{\prime },X\right) \right) \frac{1-\hat{S}\left( X^{\prime }\right) }{1-%
\hat{S}_{E}\left( X^{\prime }\right) }f\left( X^{\prime }\right) dX^{\prime }
\\
&&-\int \max \left( -1,\left( 1+f\left( X^{\prime }\right) \right) \frac{1-%
\hat{S}\left( X^{\prime }\right) }{\hat{S}_{L}\left( X^{\prime }\right) }%
\right) H\left( -\left( 1+f\left( X^{\prime }\right) \right) \right) \hat{S}%
_{L}\left( X^{\prime },X\right) dX^{\prime } \\
&&-\int \max \left( -1,\left( 1+f\left( X^{\prime }\right) \right) \frac{%
\left( 1-S\left( X\right) \right) }{S_{L}\left( X\right) }\right) H\left(
-\left( 1+f_{1}^{\prime }\left( X\right) \right) \right) S_{L}\left(
X\right) -S_{E}\left( X,X\right) \frac{1-S\left( X\right) }{1-S_{E}\left(
X\right) }f_{1}^{\prime }\left( X\right)
\end{eqnarray*}%
Defining $\hat{S}_{-}$and $S_{-}$ the default sets for investors and firms
respectively, so that $1+f\left( X\right) <0$ nd $1+f_{1}^{\prime }\left(
X\right) <0$, the equation also writes: 
\begin{eqnarray}
0 &=&\int \left( \delta \left( X-X^{\prime }\right) -\hat{S}_{E}\left(
X^{\prime },X\right) \right) \frac{1-\hat{S}\left( X^{\prime }\right) }{1-%
\hat{S}_{E}\left( X^{\prime }\right) }f\left( X^{\prime }\right) dX^{\prime }
\label{dfn} \\
&&-\int_{\hat{S}_{-}}\max \left( -1,\left( 1+f\left( X^{\prime }\right)
\right) \frac{1-\hat{S}\left( X^{\prime }\right) }{\hat{S}_{L}\left(
X^{\prime }\right) }\right) \hat{S}_{L}\left( X^{\prime },X\right)
dX^{\prime }  \notag \\
&&-\int_{S_{-}}\max \left( -1,\left( 1+f_{1}^{\prime }\left( X\right)
\right) \frac{\left( 1-S\left( X\right) \right) }{S_{L}\left( X\right) }%
\right) S_{L}\left( X\right) -S_{E}\left( X,X\right) \frac{1-S\left(
X\right) }{1-S_{E}\left( X\right) }f_{1}^{\prime }\left( X\right)  \notag
\end{eqnarray}%
with solution:%
\begin{eqnarray*}
\hat{f}\left( X\right) &=&\left( \delta \left( X-X^{\prime }\right) -\hat{S}%
_{E}\left( X^{\prime },X\right) \right) ^{-1} \\
&&\left\{ \int_{\hat{S}_{-}}\max \left( -1,\left( 1+f\left( X^{\prime
}\right) \right) \frac{1-\hat{S}\left( X^{\prime }\right) }{\hat{S}%
_{L}\left( X^{\prime }\right) }\right) \hat{S}_{L}\left( X^{\prime
},X\right) dX^{\prime }\right. \\
&&+\left. \int_{S_{-}}\max \left( -1,\left( 1+f_{1}^{\prime }\left( X\right)
\right) \frac{\left( 1-S\left( X\right) \right) }{S_{L}\left( X\right) }%
\right) S_{L}\left( X\right) +S_{E}\left( X,X\right) \frac{1-S\left(
X\right) }{1-S_{E}\left( X\right) }\max \left( f_{1}^{\prime }\left(
X\right) ,0\right) \right\}
\end{eqnarray*}%
If 
\begin{equation*}
\max \left( f_{1}^{\prime }\left( X\right) ,0\right) =0
\end{equation*}%
we have:%
\begin{eqnarray}
f\left( X\right) &=&\left( \delta \left( X-X^{\prime }\right) -\hat{S}%
_{E}\left( X^{\prime },X\right) \right) ^{-1}  \label{DFt} \\
&&\left\{ \int_{\hat{S}_{-}}\max \left( -1,\left( 1+f\left( X^{\prime
}\right) \right) \frac{1-\hat{S}\left( X^{\prime }\right) }{\hat{S}%
_{L}\left( X^{\prime }\right) }\right) \hat{S}_{L}\left( X^{\prime
},X\right) dX^{\prime }\right.  \notag \\
&&+\left. \int_{S_{-}}\max \left( -1,\left( 1+f_{1}^{\prime }\left( X\right)
\right) \frac{\left( 1-S\left( X\right) \right) }{S_{L}\left( X\right) }%
\right) S_{L}\left( X\right) \right\}  \notag
\end{eqnarray}

\subsection*{A1.4 Finding default states}

Default states may arise if there is a non empty minimal domain $D$, such
that the return of the sector $\hat{f}\left( X_{1}\right) $, given by (\ref%
{DFt}), satisfies:%
\begin{equation*}
\hat{f}\left( X_{1}\right) -\bar{r}<-1
\end{equation*}%
This corresponds to the set of sectors such defaults occur. The return is so
low that the firm's or investor's equity alone is insufficient to repay the
loans

The return equation, in terms of average capital per sector, is modified by
this default set, denoted as $DF_{0}$ (see Gosselin and Lotz (2024)): 
\begin{eqnarray}
\hat{f}_{0}\left( X_{1}\right) -\bar{r}^{\prime } &\simeq &\int \left(
1-\left( 1+\underline{\hat{k}}_{L}\left( X_{1}\right) \right) \hat{S}%
_{1}^{E}\left( X^{\prime },X_{1}\right) -1_{DF_{0}}\frac{\hat{k}_{L}\left(
X^{\prime },X_{1}\right) }{1+\underline{\hat{k}}\left( X^{\prime }\right) }%
\right) ^{-1}\left( 1+\underline{\hat{k}}_{L}\left( X_{1}\right) \right)
\label{rtn} \\
&&\times \left( \left( \frac{A}{\left( f_{1}\left( X^{\prime },\hat{K}\left[
X^{\prime }\right] \right) \right) ^{2}}+\frac{B}{\left( f_{1}\left(
X^{\prime },\hat{K}\left[ X^{\prime }\right] \right) \right) ^{3}}\right)
\left( \left( f_{1}\left( X^{\prime }\right) -\bar{r}^{\prime }\right) +\tau
F\left( X\right) \left( f_{1}\left( X^{\prime }\right) -\left\langle
f_{1}\left( X^{\prime }\right) \right\rangle \right) \right) \right.  \notag
\\
&&\left. -\frac{1_{DF_{0}}\hat{k}_{L}\left( X^{\prime },X_{1}\right) \left(
1+\bar{r}^{\prime }\right) }{1+\underline{\hat{k}}\left( X^{\prime }\right) }%
\right)  \notag
\end{eqnarray}%
The diffusion of returns is constrained to the sectors with no default,
denoted as $V/DF_{0}$ and propagates to the whole sector space. Other
default states are defined recursively. Beginning with an initial default
set, we recursively define \ the $n$-th default set: 
\begin{equation*}
DF_{n}=\left\{ X_{1},\hat{f}_{n-1}\left( X_{1}\right) <-1\right\}
\end{equation*}%
where $\hat{f}_{n}\left( X_{1}\right) $ is defined by:%
\begin{eqnarray}
&&\hat{f}_{n+1}\left( X_{1}\right) -\bar{r}^{\prime }  \label{rtd} \\
&=&\int \left( 1-\left( 1+\underline{\hat{k}}_{L}\left( X_{1}\right) \right) 
\hat{S}_{1}^{E}\left( X^{\prime },X_{1}\right) -1_{DF_{n}}\frac{\hat{k}%
_{L}\left( X^{\prime },X_{1}\right) }{1+\underline{\hat{k}}\left( X^{\prime
}\right) }\right) ^{-1}  \notag \\
&&\times \left( \left( \frac{A}{\left( f_{1}\left( X^{\prime },\hat{K}\left[
X^{\prime }\right] \right) \right) ^{2}}+\frac{B}{\left( f_{1}\left(
X^{\prime },\hat{K}\left[ X^{\prime }\right] \right) \right) ^{3}}\right)
\left( \left( f_{1}\left( X^{\prime }\right) -\bar{r}^{\prime }\right) +\tau
F\left( X\right) \left( f_{1}\left( X^{\prime }\right) -\left\langle
f_{1}\left( X^{\prime }\right) \right\rangle \right) \right) \right.  \notag
\\
&&\left. -\frac{1_{DF_{n}}\hat{k}_{L}\left( X^{\prime },X_{1}\right) \left(
1+\bar{r}^{\prime }\right) }{1+\underline{\hat{k}}\left( X^{\prime }\right) }%
\right)  \notag
\end{eqnarray}%
Ultimately, the return for a collective state with default is given by the
limit:%
\begin{equation*}
\hat{f}_{n+1}\left( X_{1}\right) \rightarrow \hat{f}\left( X_{1}\right)
\end{equation*}%
The same group may thus be connected in several states, including some
sequence of default states.

To inspect the form of the solutions we start by considering groups with
averaged connectivities solutions, and then compute the deviations from
these averages. The inspection of approximate solution with default is
studied in appendix 8, as modification of non default solution.

\section*{Appendix 2. Action functional for stakes fields}

\subsection*{A2.1 Translation of the micro set-up and minimization equations}

We start with the sum over points of objective functions integrated over
time:%
\begin{equation*}
\sum_{j}\hat{S}_{Eij}\hat{f}_{j}+\sum_{j}\hat{S}_{Lij}\hat{r}_{j}-\frac{1}{2}%
\sum_{ij}\frac{\left( \hat{S}_{\eta ij}\right) ^{2}}{\hat{w}_{\eta
_{i}}\left( X_{j}\right) }+\sum_{k}S_{Eik}f_{k}+\sum_{k}S_{Lik}\bar{r}_{k}-%
\frac{1}{2}\sum_{ik}\frac{\left( S_{\eta ik}\right) ^{2}}{w_{\eta ik}\left(
X_{k}\right) }
\end{equation*}%
\begin{equation}
\int dt\left( \sum_{i,j,k}\hat{S}_{Eij}\hat{f}_{j}+\sum_{i,j,k}\hat{S}_{Lij}%
\hat{r}_{j}-\frac{1}{2}\sum_{i,j,k}\frac{\left( \hat{S}_{\eta ij}\right) ^{2}%
}{\hat{w}_{\eta _{ij}}\left( X_{j}\right) }+\sum_{i,k,j}S_{Eik}f_{k}+%
\sum_{i,k,j}S_{Lik}\bar{r}_{k}-\frac{1}{2}\sum_{ji,k}\frac{\left( S_{\eta
ik}\right) ^{2}}{w_{\eta ik}\left( X_{k}\right) }\right)  \label{Smt}
\end{equation}%
Then, we add an inertial term for all agents and a term corresponding to the
time scale in which collective states are considered:%
\begin{equation}
-\alpha \sum_{i,j}\left( \frac{d}{dt}\hat{S}_{\eta ij}\right) ^{2}-\alpha
\sum_{i,k}\left( \frac{d}{dt}S_{\eta ik}\right) ^{2}-\beta \int dt
\label{NTS}
\end{equation}%
Considering the field: 
\begin{equation*}
\Gamma \left( S_{E},\hat{S}_{E},S_{L},\hat{S}_{L},X^{\prime },X\right)
\end{equation*}%
written:%
\begin{equation*}
\Gamma \left( \hat{S}^{\left( T\right) },X^{\prime },X\right)
\end{equation*}%
with:%
\begin{equation*}
\hat{S}^{\left( T\right) }=\left( S_{E},\hat{S}_{E},S_{L},\hat{S}_{L}\right)
\end{equation*}%
the potential (\ref{Smt}) becomes in terms of field:%
\begin{equation*}
-\sum_{\eta }\int \Gamma ^{\dag }\left( \hat{S}^{\left( T\right) },X^{\prime
},X\right) \left( -\frac{\left( \hat{S}_{\eta }^{\left( T\right) }\right)
^{2}}{2\hat{w}_{\eta }^{T}\left( X^{\prime },X\right) }+\hat{V}_{\eta }\hat{S%
}_{\eta }^{\left( T\right) }\right) \Gamma \left( \hat{S}^{\left( T\right)
},X^{\prime },X\right) d\left( \hat{S}^{\left( T\right) },X^{\prime
},X\right)
\end{equation*}%
The coefficients $\hat{w}_{\eta }^{T}\left( X^{\prime },X\right) $
translates the uncertainty coefficients $\hat{w}_{\eta _{i}}\left(
X_{j}\right) $, $w_{\eta ik}\left( X_{k}\right) $ in inter, or intra,
sectoral functions, while $\hat{V}_{\eta }$ encompass the returns. We have: 
\begin{eqnarray*}
\hat{V}_{1} &=&\hat{f}\left( X^{\prime }\right) ,\hat{w}_{E}\left( X^{\prime
},X\right) =\hat{w}_{E}\left( X^{\prime },X\right) \\
\hat{V}_{2} &=&\hat{r}\left( X^{\prime }\right) ,\hat{w}_{L}\left( X^{\prime
},X\right) =\hat{w}_{L}\left( X^{\prime },X\right)
\end{eqnarray*}%
\begin{eqnarray*}
\hat{V}_{3} &=&f\left( X\right) ,\hat{w}_{3}\left( X^{\prime },X\right)
=w_{E}\left( X,X\right) \\
\hat{V}_{4} &=&r\left( X\right) ,\hat{w}_{4}\left( X^{\prime },X\right)
=w_{L}\left( X,X\right)
\end{eqnarray*}%
The contributions (\ref{NTS}) are translated by:%
\begin{equation*}
-\sigma _{\hat{K}}^{2}\sum_{\eta }\int \Gamma ^{\dag }\left( \hat{S}^{\left(
T\right) },X^{\prime },X\right) \nabla _{\hat{S}_{\eta }^{\left( T\right)
}}^{2}\Gamma \left( \hat{S}^{\left( T\right) },X^{\prime },X\right) d\left( 
\hat{S}^{\left( T\right) },X^{\prime },X\right) -\int \beta \left\vert
\Gamma \left( \hat{S}^{\left( T\right) },X^{\prime },X\right) \right\vert
^{2}d\left( \hat{S}^{\left( T\right) },X^{\prime },X\right)
\end{equation*}%
We also add the contraint:%
\begin{equation*}
\int \lambda \left( X\right) \left( \sum_{\eta }\int \hat{S}_{\eta }^{\left(
T\right) }\left\vert \Gamma \left( \hat{S}^{\left( T\right) },X^{\prime
},X\right) \right\vert ^{2}dX^{\prime }d\hat{S}^{\left( T\right) }-1\right)
\left\vert \Gamma \left( \hat{S}^{\left( T\right) },X^{\prime },X\right)
\right\vert ^{2}d\left( \hat{S}^{\left( T\right) },X^{\prime },X\right)
\end{equation*}%
The factor $\left\vert \Gamma \left( \hat{S}^{\left( T\right) },X^{\prime
},X\right) \right\vert ^{2}$ accounts for the number of agents in sector $%
X^{\prime }$ connected to$,X$.

The full action functional is the sum of these contributions:%
\begin{eqnarray*}
&&-\sigma _{\hat{K}}^{2}\sum_{\eta }\int \Gamma ^{\dag }\left( \hat{S}%
^{\left( T\right) },X^{\prime },X\right) \nabla _{\hat{S}_{\eta }^{\left(
T\right) }}^{2}\Gamma \left( \hat{S}^{\left( T\right) },X^{\prime },X\right)
d\left( \hat{S}^{\left( T\right) },X^{\prime },X\right) -\int \beta
\left\vert \Gamma \left( \hat{S}^{\left( T\right) },X^{\prime },X\right)
\right\vert ^{2}d\left( \hat{S}^{\left( T\right) },X^{\prime },X\right) \\
&&+\sum_{\eta }\int \left( \frac{\left( \hat{S}_{\eta }^{\left( T\right)
}\right) ^{2}}{2\hat{w}_{\eta }}-\hat{V}_{\eta }\hat{S}_{\eta }^{\left(
T\right) }\right) \left\vert \Gamma \left( \hat{S}^{\left( T\right)
},X^{\prime },X\right) \right\vert ^{2}d\left( \hat{S}^{\left( T\right)
},X^{\prime },X\right) \\
&&\int \lambda \left( X\right) \left( \sum_{\eta }\int \hat{S}_{\eta
}^{\left( T\right) }\left\vert \Gamma \left( \hat{S}^{\left( T\right)
},X^{\prime },X\right) \right\vert ^{2}dX^{\prime }d\hat{S}^{\left( T\right)
}-1\right) \left\vert \Gamma \left( \hat{S}^{\left( T\right) },X^{\prime
},X\right) \right\vert ^{2}d\left( \hat{S}^{\left( T\right) },X^{\prime
},X\right)
\end{eqnarray*}%
and the minimization equations is obtained by the derivative with respect to 
$\Gamma \left( \hat{S}^{\left( T\right) },X^{\prime },X\right) $: 
\begin{eqnarray}
0 &=&\left( \sum_{\eta }\left( -\sigma _{\hat{K}}^{2}\nabla _{\hat{S}_{\eta
}^{\left( T\right) }}^{2}+\frac{\left( \hat{S}_{\eta }^{\left( T\right)
}\right) ^{2}}{2\hat{w}_{\eta }}-\hat{V}_{\eta }\hat{S}_{\eta }^{\left(
T\right) }\right) -\beta \right) \Gamma \left( \hat{S}^{\left( T\right)
},X^{\prime },X\right)  \label{MNT} \\
&&+\lambda \left( X\right) \left( \sum_{\eta }\int \hat{S}_{\eta }^{\left(
T\right) }\left\vert \Gamma \left( \hat{S}^{\left( T\right) },X^{\prime
},X\right) \right\vert ^{2}dX^{\prime }d\hat{S}^{\left( T\right) }-1\right)
\Gamma \left( \hat{S}^{\left( T\right) },X^{\prime },X\right)  \notag \\
&&+\lambda \left( X\right) \Gamma \left( \hat{S}^{\left( T\right)
},X^{\prime },X\right) \hat{S}_{\eta }^{\left( T\right) }\left\Vert \Gamma
\left( \hat{S}^{\left( T\right) },X^{\prime },X\right) \right\Vert ^{2} 
\notag
\end{eqnarray}%
lng wth the derivtv wth respect to $\lambda \left( X\right) $ wch impls:%
\begin{equation}
\left( \sum_{\eta }\int \hat{S}_{\eta }^{\left( T\right) }\left\vert \Gamma
\left( \hat{S}^{\left( T\right) },X^{\prime },X\right) \right\vert
^{2}dX^{\prime }d\hat{S}^{\left( T\right) }-1\right) =0  \label{CNTs}
\end{equation}%
and (\ref{MNT}) simplifies as:%
\begin{eqnarray}
0 &=&\left( \sum_{\eta }\left( -\sigma _{\hat{K}}^{2}\nabla _{\hat{S}_{\eta
}^{\left( T\right) }}^{2}+\frac{\left( \hat{S}_{\eta }^{\left( T\right)
}\right) ^{2}}{2\hat{w}_{\eta }}-\hat{V}_{\eta }\hat{S}_{\eta }^{\left(
T\right) }+\lambda \left( X\right) \left\Vert \Gamma \left( \hat{S}^{\left(
T\right) },X^{\prime },X\right) \right\Vert _{X}^{2}\hat{S}_{\eta }^{\left(
T\right) }\right) -\beta \right) \\
&&\times \Gamma \left( \hat{S}^{\left( T\right) },X^{\prime },X\right) 
\notag
\end{eqnarray}%
where:%
\begin{equation*}
\left\Vert \Gamma \left( \hat{S}^{\left( T\right) },X^{\prime },X\right)
\right\Vert _{X}^{2}=\int \left\vert \Gamma \left( \hat{S}^{\left( T\right)
},X^{\prime },X\right) \right\vert ^{2}dX^{\prime }d\hat{S}^{\left( T\right)
}
\end{equation*}%
We can rescale $\lambda \left( X\right) $ and replace:%
\begin{equation*}
\lambda \left( X\right) \left\Vert \Gamma \left( \hat{S}^{\left( T\right)
},X^{\prime },X\right) \right\Vert _{X}^{2}\rightarrow \lambda \left(
X\right)
\end{equation*}%
Solutions of this equation have the form:%
\begin{equation*}
\Gamma _{0,X^{\prime },X}\left( \hat{S}^{\left( T\right) }\right) \Gamma
\left( X^{\prime },X\right)
\end{equation*}%
The function $\Gamma _{0,X}\left( \hat{S}^{\left( T\right) }\right) $ is a
parabolic cylinder function depending on $\beta $. To simplify, we can
adjust the parameter $\beta $ so that:%
\begin{equation*}
\Gamma _{0,X^{\prime },X}\left( \hat{S}^{\left( T\right) }\right) =N\exp
\left( -\sum_{\eta }\frac{\left( \hat{S}_{\eta }^{\left( T\right) }-%
\overline{\hat{S}_{\eta }^{\left( T\right) }}\left( X^{\prime },X\right)
\right) ^{2}}{2\sigma _{\hat{K}}^{2}}\right)
\end{equation*}%
with $N$ a normalizatn factor. The equation for $\overline{\hat{S}^{\left(
T\right) }}$ is the minimum of:%
\begin{equation*}
\frac{\left( \hat{S}_{\eta }^{\left( T\right) }\right) ^{2}}{2\hat{w}_{\eta
}\left( X^{\prime },X\right) }-\hat{V}_{\eta }\left( X^{\prime },X\right) 
\hat{S}_{\eta }^{\left( T\right) }-\beta +\lambda \left( X\right) \hat{S}%
_{\eta }^{\left( T\right) }
\end{equation*}%
so that:%
\begin{equation}
\overline{\hat{S}_{\eta }^{\left( T\right) }}\left( X^{\prime },X\right) =%
\hat{w}_{\eta }\left( X^{\prime },X\right) \left( \hat{V}_{\eta }\left(
X^{\prime },X\right) +\lambda \left( X\right) \right)  \label{MNS}
\end{equation}%
The average $\overline{\hat{S}^{\left( T\right) }}$ is also the average:%
\begin{equation}
\overline{\hat{S}_{\eta }^{\left( T\right) }}\left( X^{\prime },X\right) =%
\frac{\int \hat{S}_{\eta }^{\left( T\right) }\left\vert \Gamma _{0,X^{\prime
},X}\left( \hat{S}^{\left( T\right) }\right) \right\vert ^{2}d\hat{S}%
^{\left( T\right) }}{\int \left\vert \Gamma _{0,X^{\prime },X}\left( \hat{S}%
^{\left( T\right) }\right) \right\vert ^{2}d\hat{S}^{\left( T\right) }}=%
\frac{\int \hat{S}_{\eta }^{\left( T\right) }\left\vert \Gamma \left( \hat{S}%
^{\left( T\right) },X^{\prime },X\right) \right\vert ^{2}d\hat{S}^{\left(
T\right) }}{\int \left\vert \Gamma \left( \hat{S}^{\left( T\right)
},X^{\prime },X\right) \right\vert ^{2}d\hat{S}^{\left( T\right) }}
\label{VRD}
\end{equation}%
and (\ref{MNS}) writes:%
\begin{equation*}
\int \hat{S}_{\eta }^{\left( T\right) }\left\vert \Gamma _{0,X^{\prime
},X}\left( \hat{S}^{\left( T\right) }\right) \right\vert ^{2}d\hat{S}%
^{\left( T\right) }=\hat{w}_{\eta }\left( X^{\prime },X\right) \left( \hat{V}%
_{\eta }\left( X^{\prime },X\right) +\lambda \left( X\right) \right) \int
\left\vert \Gamma _{0,X^{\prime },X}\left( \hat{S}^{\left( T\right) }\right)
\right\vert ^{2}d\hat{S}^{\left( T\right) }
\end{equation*}%
Multiplying by $\Gamma \left( X^{\prime },X\right) $, we integrate with
respect to $X^{\prime }$ and sum over $\eta $ and we find:%
\begin{eqnarray*}
&&\sum_{\eta }\int \hat{S}_{\eta }^{\left( T\right) }\left\vert \Gamma
\left( \hat{S}^{\left( T\right) },X^{\prime },X\right) \right\vert
^{2}dX^{\prime }d\hat{S}^{\left( T\right) } \\
&=&\sum_{\eta }\int \hat{w}_{\eta }\left( X^{\prime },X\right) \hat{V}_{\eta
}\left( X^{\prime },X\right) \left\vert \Gamma \left( \hat{S}^{\left(
T\right) },X^{\prime },X\right) \right\vert ^{2}dX^{\prime }d\hat{S}^{\left(
T\right) }+\lambda \left( X\right) \int \hat{w}_{\eta }\left( X^{\prime
},X\right) \left\vert \Gamma \left( \hat{S}^{\left( T\right) },X^{\prime
},X\right) \right\vert ^{2}dX^{\prime }d\hat{S}^{\left( T\right) }
\end{eqnarray*}%
Then using the constraint (\ref{CNTs}) yields:%
\begin{equation*}
1=\sum_{\eta }\int \hat{w}_{\eta }\left( X^{\prime },X\right) \hat{V}_{\eta
}\left( X^{\prime },X\right) \left\vert \Gamma \left( \hat{S}^{\left(
T\right) },X^{\prime },X\right) \right\vert ^{2}dX^{\prime }d\hat{S}^{\left(
T\right) }+\lambda \left( X\right) \int \hat{w}_{\eta }\left( X^{\prime
},X\right) \left\vert \Gamma \left( \hat{S}^{\left( T\right) },X^{\prime
},X\right) \right\vert ^{2}dX^{\prime }d\hat{S}^{\left( T\right) }
\end{equation*}%
so that:%
\begin{equation*}
\lambda \left( X\right) =\frac{1-\sum_{\eta }\int \hat{w}_{\eta }\left(
X^{\prime },X\right) \hat{V}_{\eta }\left( X^{\prime },X\right) \left\vert
\Gamma \left( \hat{S}^{\left( T\right) },X^{\prime },X\right) \right\vert
^{2}dX^{\prime }d\hat{S}^{\left( T\right) }}{\sum_{\eta }\int \hat{w}_{\eta
}\left( X^{\prime },X\right) \left\vert \Gamma \left( \hat{S}^{\left(
T\right) },X^{\prime },X\right) \right\vert ^{2}dX^{\prime }d\hat{S}^{\left(
T\right) }}
\end{equation*}%
using the normalization over $\hat{S}^{\left( T\right) }$, this becomes:%
\begin{equation*}
\lambda \left( X\right) =\frac{1-\sum_{\eta }\int \hat{w}_{\eta }\left(
X^{\prime },X\right) \hat{V}_{\eta }\left( X^{\prime },X\right) \left\vert
\Gamma \left( X^{\prime },X\right) \right\vert ^{2}dX^{\prime }}{\sum_{\eta
}\int \hat{w}_{\eta }\left( X^{\prime },X\right) \left\vert \Gamma \left(
X^{\prime },X\right) \right\vert ^{2}dX^{\prime }}
\end{equation*}%
In the sequel we consider tht the distribution $\left\vert \Gamma \left(
X^{\prime },X\right) \right\vert ^{2}$ varies slowly, so that:%
\begin{equation*}
\lambda \left( X\right) =\frac{1-\sum_{\eta }\int \hat{w}_{\eta }\left(
X^{\prime },X\right) \hat{V}_{\eta }\left( X^{\prime },X\right) dX^{\prime }%
}{\sum_{\eta }\int \hat{w}_{\eta }\left( X^{\prime },X\right) dX^{\prime }}
\end{equation*}%
Using the explicit form for $\hat{V}_{\eta }\left( X^{\prime },X\right) $
leads to the expanded form of $\lambda \left( X\right) $: 
\begin{eqnarray}
&&\lambda \left( X\right)  \label{LMN} \\
&\rightarrow &\frac{1-\int \hat{w}_{E}\left( X^{\prime },X\right) \hat{f}%
\left( X^{\prime }\right) -\int \hat{w}_{L}\left( X^{\prime },X\right) \hat{r%
}\left( X^{\prime }\right) -w_{E}\left( X,X\right) f\left( X\right)
-w_{L}\left( X,X\right) \bar{r}}{\hat{w}_{E}\left( X\right) +\hat{w}%
_{L}\left( X\right) +w_{E}\left( X\right) +w_{L}\left( X\right) }  \notag
\end{eqnarray}%
where we set:%
\begin{equation*}
\hat{w}_{\eta }\left( X\right) =\int \hat{w}_{\eta }\left( X^{\prime
},X\right) dX^{\prime }
\end{equation*}%
In the sequel we will simplify the notation, and $\overline{\hat{S}_{\eta
}^{\left( T\right) }}\left( X^{\prime },X\right) \rightarrow \hat{S}_{\eta
}^{\left( T\right) }\left( X^{\prime },X\right) $, so that the average
shares equations write:%
\begin{equation}
\hat{S}_{\eta }^{\left( T\right) }\left( X^{\prime },X\right) =\hat{w}_{\eta
}\left( X^{\prime },X\right) \left( \hat{V}_{\eta }\left( X^{\prime
},X\right) +\lambda \left( X\right) \right)  \label{VRT}
\end{equation}%
These minimizations equations for the shares will be studied in appendix 3.

Ultimately, remark that (\ref{VRD}) also writes:%
\begin{equation*}
\hat{S}_{\eta }\left( X^{\prime },X\right) =\int \hat{S}_{\eta }\left\vert
\Gamma \left( S_{E},\hat{S}_{E},S_{L},\hat{S}_{L},X^{\prime },X\right)
\right\vert ^{2}d\left( S_{E},\hat{S}_{E},S_{L},\hat{S}_{L}\right)
\end{equation*}%
for the shares in investors, and:%
\begin{equation*}
S_{\eta }\left( X\right) =\int S_{\eta }\left\vert \Gamma \left( S_{E},\hat{S%
}_{E},S_{L},\hat{S}_{L},X^{\prime },X\right) \right\vert ^{2}d\left( S_{E},%
\hat{S}_{E},S_{L},\hat{S}_{L},X^{\prime }\right)
\end{equation*}%
for the shares in firms.

With these notations, the constraint on allocation writes:%
\begin{equation*}
\int \left( \hat{S}_{E}\left( X^{\prime },X\right) +\hat{S}_{L}\left(
X^{\prime },X\right) \right) dX^{\prime }+\int \left( S_{E}\left( X^{\prime
},X\right) +S_{L}\left( X^{\prime },X\right) \right) dX^{\prime }=1
\end{equation*}

\subsection*{A2.2 Equivalent formulation}

We could directly consider the following functional:%
\begin{eqnarray*}
S\left( \Gamma \right) &=&-\sum_{\eta }\int \Gamma ^{\dag }\left( \hat{S}%
^{\left( T\right) },X^{\prime },X\right) \left( \sigma _{\hat{K}}^{2}\nabla
_{\hat{S}_{\eta }^{\left( T\right) }}^{2}-V\left( \Gamma \right) \right)
\Gamma \left( \hat{S}^{\left( T\right) },X^{\prime },X\right) d\left( \hat{S}%
^{\left( T\right) },X^{\prime },X\right) \\
&&+\int \lambda \left( X\right) \left( \sum_{\eta }\int \hat{S}_{\eta
}^{\left( T\right) }\left( X^{\prime },X\right) dX^{\prime }-1\right)
\left\vert \Gamma \left( \hat{S}_{\eta }^{\left( T\right) },X^{\prime
},X\right) \right\vert ^{2}d\left( \hat{S}_{\eta }^{\left( T\right)
},X^{\prime },X\right)
\end{eqnarray*}%
where $V\left( \Gamma \right) $ implement the micr preferences:

\begin{eqnarray*}
&&V\left( \Gamma \right) \\
&\rightarrow &\int \left( \hat{S}_{E}\hat{f}\left( X^{\prime }\right) +\hat{S%
}_{L}\hat{r}\left( X^{\prime }\right) -\frac{1}{2}\frac{\left( \hat{S}_{E}+%
\hat{S}_{L}\right) ^{2}\left\vert \Gamma \left( \bar{S},X^{\prime },X\right)
\right\vert ^{2}}{\hat{w}_{\eta }\left( X^{\prime },X\right) }\right. \\
&&\left. +S_{E}f\left( X\right) +S_{L}\bar{r}\left( X\right) -\frac{1}{2}%
\frac{\left( S_{E}+S_{L}\right) ^{2}\left\vert \Gamma \left( \bar{S}%
,X^{\prime },X\right) \right\vert ^{2}}{w_{\eta }\left( X\right) }\right)
\left\vert \Gamma \left( \bar{S},X^{\prime },X\right) \right\vert ^{2}
\end{eqnarray*}%
This can be approximated by:%
\begin{eqnarray*}
V\left( \Gamma \right) &\simeq &\int \left( \hat{w}_{E}\left( X^{\prime
},X\right) \hat{S}_{E}\left( X^{\prime },X\right) \hat{f}\left( X^{\prime
}\right) +\hat{w}_{L}\left( X^{\prime },X\right) \hat{S}_{L}\left( X^{\prime
},X\right) \hat{r}\left( X^{\prime }\right) -\frac{\left( \left( \hat{S}_{E}+%
\hat{S}_{L}\right) \left( X^{\prime },X\right) \right) ^{2}}{2\alpha }\right.
\\
&&\left. +w_{E}\left( X\right) S_{E}\left( X,X\right) f\left( X\right)
+w_{L}\left( X\right) S_{L}\left( X,X\right) \bar{r}\left( X\right) -\frac{%
\left( \left( S_{E}+S_{L}\right) \left( X,X\right) \right) ^{2}}{2\alpha }%
\right) ^{2}\left\vert \Gamma \left( X^{\prime },X\right) \right\vert ^{2}
\end{eqnarray*}%
Including the constraint, the full potential to consider is:%
\begin{eqnarray*}
V\left( \Gamma \right) &\simeq &\int \left( \hat{w}_{E}\left( X^{\prime
},X\right) \hat{S}_{E}\left( X^{\prime },X\right) \hat{f}\left( X^{\prime
}\right) +\hat{w}_{L}\left( X^{\prime },X\right) \hat{S}_{L}\left( X^{\prime
},X\right) \hat{r}\left( X^{\prime }\right) -\frac{\left( \left( \hat{S}_{E}+%
\hat{S}_{L}\right) \left( X^{\prime },X\right) \right) ^{2}}{2\alpha }\right.
\\
&&\left. +w_{E}\left( X\right) S_{E}\left( X,X\right) f\left( X\right)
+w_{L}\left( X\right) S_{L}\left( X,X\right) \bar{r}\left( X\right) -\frac{%
\left( \left( S_{E}+S_{L}\right) \left( X,X\right) \right) ^{2}}{2\alpha }%
\right) ^{2}\left\vert \Gamma \left( X^{\prime },X\right) \right\vert ^{2} \\
&&+\int \lambda \left( X\right) \left( \sum_{\eta }\int \hat{S}_{\eta
}^{\left( T\right) }\left( X^{\prime },X\right) dX^{\prime }-1\right)
\left\vert \Gamma \left( \bar{S},X^{\prime },X\right) \right\vert ^{2}
\end{eqnarray*}

To simplify we can consider that $\left\vert \Gamma \left( \bar{S},X^{\prime
},X\right) \right\vert ^{2}$ varies slowly (being proportional to the number
of agents in $X$ times the number of agents in $X^{\prime }$, so that we
minimize:%
\begin{eqnarray*}
V\left( \Gamma \right) &\simeq &\int \left( \hat{w}_{E}\left( X^{\prime
},X\right) \hat{S}_{E}\left( X^{\prime },X\right) \hat{f}\left( X^{\prime
}\right) +\hat{w}_{L}\left( X^{\prime },X\right) \hat{S}_{L}\left( X^{\prime
},X\right) \hat{r}\left( X^{\prime }\right) -\frac{\left( \left( \hat{S}_{E}+%
\hat{S}_{L}\right) \left( X^{\prime },X\right) \right) ^{2}}{2\alpha }\right.
\\
&&\left. +w_{E}\left( X\right) S_{E}\left( X,X\right) f\left( X\right)
+w_{L}\left( X\right) S_{L}\left( X,X\right) \bar{r}\left( X\right) -\frac{%
\left( \left( S_{E}+S_{L}\right) \left( X,X\right) \right) ^{2}}{2}\right)
^{2} \\
&&+\int \lambda \left( X\right) \left( \sum_{\eta }\int \hat{S}_{\eta
}^{\left( T\right) }\left( X^{\prime },X\right) dX^{\prime }-1\right)
\end{eqnarray*}%
and this leads to the same saddle point as the action in the text.

Note that in absence of inertia measured by $\alpha $, the potential $%
V\left( \Gamma \right) $ describes allocation $\bar{S}_{\eta }$. As for the
micro description $\hat{w}_{\eta }$ and $w_{\eta }$ measure the
uncertainties associated to the returns $\hat{f}\left( X^{\prime }\right) $

\section*{Appendix 3. Saddle points equations. Collective states.}

\subsection*{A3.1 Saddle point equations for stakes and solutions for given
uncertainty}

Collective states satisfy the saddle point equations for $S\left( \Gamma
\right) $. The expanded form of equation (\ref{VRT}) is: 
\begin{eqnarray*}
\hat{S}_{E}\left( X^{\prime },X\right) &=&\hat{w}_{E}\left( X^{\prime
},X\right) \left( \hat{f}\left( X^{\prime }\right) +\lambda \left( X\right)
\right) \\
\hat{S}_{L}\left( X^{\prime },X\right) &=&\hat{w}_{L}\left( X^{\prime
},X\right) \left( \bar{r}\left( X^{\prime },X\right) +\lambda \left(
X\right) \right)
\end{eqnarray*}%
for investors-investors shares, and:%
\begin{eqnarray*}
S_{E}\left( X,X\right) &=&w_{E}\left( X\right) \left( f\left( X\right)
+\lambda \left( X\right) \right) \\
S_{L}\left( X,X\right) &=&w_{E}\left( X\right) \left( \bar{r}\left(
X,X\right) +\lambda \left( X\right) \right)
\end{eqnarray*}%
for investors-firms shares. The Lagrange multiplier $\lambda \left( X\right) 
$ was given in (\ref{LMN}),: 
\begin{eqnarray*}
&&\lambda \left( X\right) \\
&\rightarrow &\frac{1-\int \hat{w}_{E}\left( X^{\prime },X\right) \hat{f}%
\left( X^{\prime }\right) -\int \hat{w}_{L}\left( X^{\prime },X\right) \hat{r%
}\left( X^{\prime }\right) -w_{E}\left( X,X\right) f\left( X\right)
-w_{L}\left( X,X\right) \bar{r}}{\hat{w}_{E}\left( X\right) +\hat{w}%
_{L}\left( X\right) +w_{E}\left( X\right) +w_{L}\left( X\right) }
\end{eqnarray*}%
where we set:%
\begin{equation*}
\hat{w}_{\eta }\left( X\right) =\int \hat{w}_{\eta }\left( X^{\prime
},X\right)
\end{equation*}%
The solutions for shares are:%
\begin{eqnarray*}
&&\hat{S}_{E}\left( X^{\prime },X\right) \\
&=&\underline{\hat{S}}_{E}\left( X^{\prime },X\right) +\frac{\hat{w}%
_{E}\left( X^{\prime },X\right) }{\hat{w}_{E}\left( X\right) +\hat{w}%
_{L}\left( X\right) +w_{E}\left( X\right) +w_{L}\left( X\right) } \\
&&\times \left\{ \hat{w}_{E}\left( X\right) \left( \hat{f}\left( X^{\prime
}\right) -\left\langle \hat{f}\left( X^{\prime }\right) \right\rangle _{\hat{%
w}_{E}}\right) +\hat{w}_{L}\left( X\right) \left( \hat{f}\left( X^{\prime
}\right) -\left\langle \hat{r}\left( X^{\prime }\right) \right\rangle _{\hat{%
w}_{L}}\right) \right. \\
&&\left. +w_{E}\left( X\right) \left( \hat{f}\left( X^{\prime }\right)
-f\left( X\right) \right) +w_{L}\left( X\right) \left( \hat{f}\left(
X^{\prime }\right) -\bar{r}\right) \right\}
\end{eqnarray*}%
\begin{eqnarray*}
&&\hat{S}_{E}\left( X^{\prime },X\right) \\
&=&\underline{\hat{S}}_{E}\left( X^{\prime },X\right) +\frac{\hat{w}%
_{E}\left( X^{\prime },X\right) }{\hat{w}_{E}\left( X\right) +\hat{w}%
_{L}\left( X\right) +w_{E}\left( X\right) +w_{L}\left( X\right) } \\
&&\times \left\{ \hat{w}_{E}\left( X\right) \left( \hat{f}\left( X^{\prime
}\right) -\left\langle \hat{f}\left( X^{\prime }\right) \right\rangle _{\hat{%
w}_{E}}\right) +\hat{w}_{L}\left( X\right) \left( \hat{f}\left( X^{\prime
}\right) -\left\langle \hat{r}\left( X^{\prime }\right) \right\rangle _{\hat{%
w}_{L}}\right) \right. \\
&&\left. +w_{E}\left( X\right) \left( \hat{f}\left( X^{\prime }\right)
-f\left( X\right) \right) +w_{L}\left( X\right) \left( \hat{f}\left(
X^{\prime }\right) -\bar{r}\right) \right\}
\end{eqnarray*}%
\begin{eqnarray*}
\hat{S}_{L}\left( X^{\prime },X\right) &=&\underline{\hat{S}}_{L}\left(
X^{\prime },X\right) +\frac{\hat{w}_{L}\left( X^{\prime },X\right) }{\hat{w}%
_{E}\left( X\right) +\hat{w}_{L}\left( X\right) +w_{E}\left( X\right)
+w_{L}\left( X\right) } \\
&&\times \left\{ \hat{w}_{E}\left( X\right) \left( \hat{r}\left( X^{\prime
}\right) -\left\langle \hat{f}\left( X^{\prime }\right) \right\rangle _{\hat{%
w}_{E}}\right) +\hat{w}_{L}\left( X\right) \left( \hat{r}\left( X^{\prime
}\right) -\left\langle \hat{r}\left( X^{\prime }\right) \right\rangle _{\hat{%
w}_{L}}\right) \right. \\
&&\left. +w_{E}\left( X\right) \left( \hat{r}\left( X^{\prime }\right)
-f\left( X\right) \right) +w_{L}\left( X\right) \left( \hat{r}\left(
X^{\prime }\right) -\bar{r}\right) \right\}
\end{eqnarray*}%
with:%
\begin{equation*}
\left\langle F\left( X^{\prime }\right) \right\rangle _{\hat{w}_{\eta }}=%
\frac{\int F\left( X^{\prime }\right) \hat{w}_{\eta }\left( X^{\prime
},X\right) }{\hat{w}_{\eta }\left( X\right) }
\end{equation*}%
for any function $F\left( X^{\prime }\right) $ and:%
\begin{equation*}
\underline{\hat{S}}_{\eta }\left( X^{\prime },X\right) =\frac{\hat{w}_{\eta
}\left( X^{\prime },X\right) }{\hat{w}_{E}\left( X\right) +\hat{w}_{L}\left(
X\right) +w_{E}\left( X\right) +w_{L}\left( X\right) }
\end{equation*}%
The shares of investment in firms are given by:%
\begin{eqnarray*}
&&S_{E}\left( X,X\right) \\
&=&\underline{S}_{E}\left( X,X\right) +w_{E}\left( X\right) \\
&&\times \frac{\hat{w}_{E}\left( X\right) \left( f\left( X\right)
-\left\langle \hat{f}\left( X^{\prime }\right) \right\rangle _{\hat{w}%
_{E}}\right) +\hat{w}_{L}\left( X\right) \left( f\left( X\right)
-\left\langle \hat{r}\left( X^{\prime }\right) \right\rangle _{\hat{w}%
_{L}}\right) +w_{L}\left( X\right) \left( f\left( X\right) -\bar{r}\right) }{%
\hat{w}_{E}\left( X\right) +\hat{w}_{L}\left( X\right) +w_{E}\left( X\right)
+w_{L}\left( X\right) }
\end{eqnarray*}%
for participations, and:%
\begin{eqnarray*}
&&S_{L}\left( X,X\right) \\
&=&\underline{S}_{L}\left( X,X\right) +w_{L}\left( X\right) \\
&&\times \frac{\hat{w}_{E}\left( X\right) \left( \bar{r}\left( X\right)
-\left\langle \hat{f}\left( X^{\prime }\right) \right\rangle _{\hat{w}%
_{E}}\right) +\hat{w}_{L}\left( X\right) \left( \bar{r}\left( X\right)
-\left\langle \hat{r}\left( X^{\prime }\right) \right\rangle _{\hat{w}%
_{L}}\right) +w_{L}\left( X\right) \left( \bar{r}\left( X\right) -f\left(
X\right) \right) }{\hat{w}_{E}\left( X\right) +\hat{w}_{L}\left( X\right)
+w_{E}\left( X\right) +w_{L}\left( X\right) }
\end{eqnarray*}%
for loans, with:%
\begin{equation*}
\underline{S}_{\eta }\left( X,X\right) =\frac{w_{\eta }\left( X^{\prime
},X\right) }{\int \hat{w}_{E}\left( X^{\prime },X\right) +\int \hat{w}%
_{L}\left( X^{\prime },X\right) +w_{E}\left( X\right) +w_{L}\left( X\right) }
\end{equation*}%
The coefficients $\hat{w}_{\alpha }$, $w_{\alpha }$ are endogeneous since
they depend on the uncertainty on returns depending themselves on th $%
\underline{\hat{S}}_{\eta }$ $S_{\eta }$. Solving will be done by detailling
this uncertainties. Before doing so, we simplify slightly by imposing
several assmptions.

First, we normalize $\hat{w}_{E}\left( X\right) +\hat{w}_{L}\left( X\right)
+w_{E}\left( X\right) +w_{L}\left( X\right) =1$ and choose:%
\begin{equation*}
\hat{w}_{E}\left( X^{\prime },X\right) =\hat{w}_{L}\left( X^{\prime
},X\right) =\frac{1}{2}\hat{w}\left( X^{\prime },X\right)
\end{equation*}%
\begin{equation*}
w_{E}\left( X,X\right) =w_{L}\left( X,X\right) =\frac{w\left( X,X\right) }{2}
\end{equation*}%
\begin{equation*}
\hat{w}_{E}\left( X\right) =\hat{w}_{L}\left( X\right) =\frac{\hat{w}\left(
X\right) }{2}
\end{equation*}%
\begin{equation*}
w_{E}\left( X\right) =w_{L}\left( X\right) =\frac{w\left( X\right) }{2}
\end{equation*}%
The constraint on coefficients implies:%
\begin{equation*}
w\left( X\right) =1-\hat{w}\left( X\right)
\end{equation*}%
and the solutions writes ultimately:%
\begin{eqnarray*}
&&\hat{S}_{E}\left( X^{\prime },X\right) \\
&=&\frac{\underline{\hat{S}}\left( X^{\prime },X\right) }{2}+\frac{\hat{w}%
\left( X^{\prime },X\right) }{2}\left( \hat{w}\left( X\right) \left( \hat{f}%
\left( X^{\prime }\right) -\frac{\left\langle \hat{f}\left( X^{\prime
}\right) \right\rangle _{\hat{w}_{E}}+\left\langle \hat{r}\left( X^{\prime
}\right) \right\rangle _{\hat{w}_{L}}}{2}\right) +w\left( X\right) \left( 
\hat{f}\left( X^{\prime }\right) -\frac{f\left( X\right) +r\left( X\right) }{%
2}\right) \right) \\
&=&\frac{\underline{\hat{S}}\left( X^{\prime },X\right) }{2}+\frac{\hat{w}%
\left( X^{\prime },X\right) }{2}\left( \hat{f}\left( X^{\prime }\right) -%
\hat{w}\left( X\right) \frac{\left\langle \hat{f}\left( X^{\prime }\right)
\right\rangle _{\hat{w}_{E}}+\left\langle \hat{r}\left( X^{\prime }\right)
\right\rangle _{\hat{w}_{L}}}{2}-w\left( X\right) \frac{f\left( X\right)
+r\left( X\right) }{2}\right)
\end{eqnarray*}%
\begin{eqnarray*}
&&\hat{S}_{L}\left( X^{\prime },X\right) \\
&=&\frac{\underline{\hat{S}}\left( X^{\prime },X\right) }{2}+\frac{\hat{w}%
\left( X^{\prime },X\right) }{2}\left( \hat{w}\left( X\right) \left( \hat{r}%
\left( X^{\prime }\right) -\frac{\left\langle \hat{f}\left( X^{\prime
}\right) \right\rangle _{\hat{w}_{E}}+\left\langle \hat{r}\left( X^{\prime
}\right) \right\rangle _{\hat{w}_{L}}}{2}\right) +w\left( X\right) \left( 
\hat{r}\left( X^{\prime }\right) -\frac{f\left( X\right) +r\left( X\right) }{%
2}\right) \right) \\
&=&\frac{\underline{\hat{S}}\left( X^{\prime },X\right) }{2}+\frac{\hat{w}%
\left( X^{\prime },X\right) }{2}\left( \hat{r}\left( X^{\prime }\right) -%
\hat{w}\left( X\right) \frac{\left\langle \hat{f}\left( X^{\prime }\right)
\right\rangle _{\hat{w}_{E}}+\left\langle \hat{r}\left( X^{\prime }\right)
\right\rangle _{\hat{w}_{L}}}{2}-w\left( X\right) \frac{f\left( X\right)
+r\left( X\right) }{2}\right)
\end{eqnarray*}%
with:%
\begin{equation*}
\underline{\hat{S}}_{E}\left( X^{\prime },X\right) =\underline{\hat{S}}%
_{L}\left( X^{\prime },X\right) =\frac{\underline{\hat{S}}\left( X^{\prime
},X\right) }{2}=\frac{1}{2}\hat{w}\left( X^{\prime },X\right)
\end{equation*}%
\begin{eqnarray*}
&&\hat{S}\left( X^{\prime },X\right) \\
&=&\underline{\hat{S}}\left( X^{\prime },X\right) +\hat{w}\left( X^{\prime
},X\right) \left( \frac{\hat{f}\left( X^{\prime }\right) +\hat{r}\left(
X^{\prime }\right) }{2}-\hat{w}\left( X\right) \frac{\left\langle \hat{f}%
\left( X^{\prime }\right) \right\rangle _{\hat{w}_{E}}+\left\langle \hat{r}%
\left( X^{\prime }\right) \right\rangle _{\hat{w}_{L}}}{2}-w\left( X\right) 
\frac{f\left( X\right) +r\left( X\right) }{2}\right)
\end{eqnarray*}%
\begin{eqnarray}
&&S_{E}\left( X,X\right)  \label{SNpa} \\
&=&\frac{\underline{S}\left( X,X\right) }{2}+\frac{w\left( X\right) }{2}%
\left( \hat{w}\left( X\right) \left( f\left( X\right) -\frac{\left\langle 
\hat{f}\left( X^{\prime }\right) \right\rangle _{\hat{w}_{E}}+\left\langle 
\hat{r}\left( X^{\prime }\right) \right\rangle _{\hat{w}_{L}}}{2}\right) +%
\frac{w\left( X\right) }{2}\left( f\left( X\right) -\bar{r}\left( X\right)
\right) \right)  \notag \\
&=&\frac{\underline{S}\left( X,X\right) }{2}+\frac{w\left( X\right) }{2}%
\left( f\left( X\right) -\hat{w}\left( X\right) \frac{\left\langle \hat{f}%
\left( X^{\prime }\right) \right\rangle _{\hat{w}_{E}}+\left\langle \hat{r}%
\left( X^{\prime }\right) \right\rangle _{\hat{w}_{L}}}{2}-w\left( X\right) 
\frac{f\left( X\right) +\bar{r}\left( X\right) }{2}\right)
\end{eqnarray}%
\begin{eqnarray*}
&&S_{L}\left( X,X\right) \\
&=&\frac{\underline{S}\left( X,X\right) }{2}+\frac{w\left( X\right) }{2}%
\left( \hat{w}\left( X\right) \left( \bar{r}\left( X\right) -\frac{%
\left\langle \hat{f}\left( X^{\prime }\right) \right\rangle _{\hat{w}%
_{E}}+\left\langle \hat{r}\left( X^{\prime }\right) \right\rangle _{\hat{w}%
_{L}}}{2}\right) +\frac{w\left( X\right) }{2}\left( \bar{r}\left( X\right)
-f\left( X\right) \right) \right) \\
&=&\frac{\underline{S}\left( X,X\right) }{2}+\frac{w\left( X\right) }{2}%
\left( \bar{r}\left( X\right) -\hat{w}\left( X\right) \frac{\left\langle 
\hat{f}\left( X^{\prime }\right) \right\rangle _{\hat{w}_{E}}+\left\langle 
\hat{r}\left( X^{\prime }\right) \right\rangle _{\hat{w}_{L}}}{2}-w\left(
X\right) \frac{f\left( X\right) +\bar{r}\left( X\right) }{2}\right)
\end{eqnarray*}%
with:%
\begin{equation*}
\underline{S}_{E}\left( X,X\right) =\underline{S}_{L}\left( X,X\right) =%
\frac{\underline{S}\left( X,X\right) }{2}=\frac{w\left( X,X\right) }{2}
\end{equation*}%
\begin{equation}
S\left( X,X\right) =\underline{S}\left( X,X\right) +w\left( X\right) \left( 
\hat{w}\left( X\right) \left( \frac{f\left( X\right) +\bar{r}\left( X\right) 
}{2}-\frac{\left\langle \hat{f}\left( X^{\prime }\right) \right\rangle _{%
\hat{w}_{E}}+\left\langle \hat{r}\left( X^{\prime }\right) \right\rangle _{%
\hat{w}_{L}}}{2}\right) \right)  \label{STpa}
\end{equation}%
The solutions for the saddle points depend both on sectors average
disposable capital and on the coefficients $\hat{w}\left( X^{\prime
},X\right) $, $w\left( X\right) $ and their averages. The sectors'
disposable capital depend on the connectivities and were derived in Gosselin
and Lotz (2024). Their expression will be recalled below. The coefficients $%
\hat{w}\left( X^{\prime },X\right) $, $w\left( X\right) $ are themselves
partly endogeneous. They depend on subjective uncertainties between sectors,
but also on the various connections: each neighbour investing itself in
other neighbours, uncertainty spreads among the whole sytem. We will
estimate the coeficients $\hat{w}\left( X^{\prime },X\right) $, $w\left(
X\right) $ in the sequel.

\subsection{Total shares per sector}

We have found the invested shares as functions of returns and uncertainty.
In the sequel, we will also need to express the total invested shares per
sector, that is the functions $\hat{S}_{\eta }\left( X^{\prime }\right) $ as
functions of the same variable. Recall that:%
\begin{eqnarray*}
\hat{S}_{\eta }\left( X^{\prime }\right) &=&\int \hat{S}_{\eta }\left(
X^{\prime },X\right) \frac{\hat{K}_{X}\left\vert \hat{\Psi}\left( X\right)
\right\vert ^{2}}{\hat{K}_{X^{\prime }}\left\vert \hat{\Psi}\left( X^{\prime
}\right) \right\vert ^{2}}dX \\
&\simeq &\int \hat{S}_{\eta }\left( X^{\prime },X\right) dX\frac{%
\left\langle \hat{K}\right\rangle \left\Vert \hat{\Psi}\right\Vert ^{2}}{%
\hat{K}_{X^{\prime }}\left\vert \hat{\Psi}\left( X^{\prime }\right)
\right\vert ^{2}}
\end{eqnarray*}%
that are the total invested stakes (participation or loans) invested in
investors located $X^{\prime }$: 
\begin{eqnarray*}
&&\hat{S}_{E}\left( X^{\prime }\right) \\
&\simeq &\left( \frac{\underline{\hat{S}}\left( X^{\prime }\right) }{2}+%
\frac{\hat{w}\left( X^{\prime }\right) }{2}\left( \hat{f}\left( X^{\prime
}\right) -\hat{w}\frac{\left\langle \hat{f}\left( X^{\prime }\right)
\right\rangle _{\hat{w}_{E}}+\left\langle \hat{r}\left( X^{\prime }\right)
\right\rangle _{\hat{w}_{L}}}{2}-w\frac{\left\langle f\left( X\right)
+r\left( X\right) \right\rangle _{w}}{2}\right) \right) \frac{\left\langle 
\hat{K}\right\rangle \left\Vert \hat{\Psi}\right\Vert ^{2}}{\hat{K}%
_{X^{\prime }}\left\vert \hat{\Psi}\left( X^{\prime }\right) \right\vert ^{2}%
}
\end{eqnarray*}%
with:%
\begin{eqnarray*}
\hat{w}\left( X^{\prime }\right) &=&\int \hat{w}\left( X^{\prime },X\right)
dX \\
\hat{w} &=&\int \hat{w}\left( X\right) dX
\end{eqnarray*}%
and:%
\begin{equation*}
w\left( X,X\right) =1-\hat{w}\left( X\right)
\end{equation*}%
\begin{eqnarray*}
&&\hat{S}\left( X^{\prime }\right) \\
&\simeq &\left( \underline{\hat{S}}\left( X^{\prime }\right) +\hat{w}\left(
X^{\prime }\right) \left( \frac{\hat{f}\left( X^{\prime }\right) +\hat{r}%
\left( X^{\prime }\right) }{2}-\hat{w}\frac{\left\langle \hat{f}\left(
X^{\prime }\right) \right\rangle _{\hat{w}_{E}}+\left\langle \hat{r}\left(
X^{\prime }\right) \right\rangle _{\hat{w}_{L}}}{2}-w\frac{\left\langle
f\left( X\right) +r\left( X\right) \right\rangle _{w}}{2}\right) \right) 
\frac{\left\langle \hat{K}\right\rangle \left\Vert \hat{\Psi}\right\Vert ^{2}%
}{\hat{K}_{X^{\prime }}\left\vert \hat{\Psi}\left( X^{\prime }\right)
\right\vert ^{2}}
\end{eqnarray*}%
Similarly we consider the total share $S_{\eta }\left( X^{\prime }\right) $
invested in firms located at $X^{\prime }$: 
\begin{equation*}
S_{\eta }\left( X^{\prime }\right) =\int S_{\eta }\left( X^{\prime
},X\right) \frac{\hat{K}_{X}\left\vert \hat{\Psi}\left( X\right) \right\vert
^{2}}{K_{X^{\prime }}\left\vert \Psi \left( X^{\prime }\right) \right\vert
^{2}}dX
\end{equation*}%
Under our approximation of local investment, this reduces to:%
\begin{equation*}
S_{\eta }\left( X\right) =S_{\eta }\left( X,X\right) \frac{\hat{K}%
_{X}\left\vert \hat{\Psi}\left( X\right) \right\vert ^{2}}{K_{X}\left\vert
\Psi \left( X\right) \right\vert ^{2}}
\end{equation*}%
and similarly:%
\begin{equation*}
S\left( X\right) =S\left( X,X\right) \frac{\hat{K}_{X}\left\vert \hat{\Psi}%
\left( X\right) \right\vert ^{2}}{K_{X}\left\vert \Psi \left( X\right)
\right\vert ^{2}}
\end{equation*}

\subsection*{A3.2 Form of uncertainty and blocks of connected agents}

Once the equations for the shares have been found, we can complete the
system by deriving formulas for uncertainty (i.e. the risk). This is done by
using the return equations (\ref{RT})\footnote{%
Only participations are considered. Loans are included in the next paragraph.
\par
{}}:%
\begin{equation*}
\left( \delta \left( X-X^{\prime }\right) -\hat{S}_{E}\left( X^{\prime
},X\right) \right) \frac{\left( 1-\hat{S}\left( X^{\prime }\right) \right)
\left( \hat{f}\left( X^{\prime }\right) -\bar{r}\right) }{1-\hat{S}%
_{E}\left( X^{\prime }\right) }=S_{E}\left( X,X\right) \frac{\left(
1-S\left( X\right) \right) \left( f_{1}^{\prime }\left( X\right) -\bar{r}%
\right) }{1-S_{E}\left( X\right) }
\end{equation*}%
and expanding it in series:%
\begin{eqnarray*}
\hat{f}\left( X\right) -\bar{r} &=&\frac{1-\hat{S}_{E}\left( X\right) }{1-%
\hat{S}\left( X\right) }\left[ S_{E}\left( X,X\right) \frac{\left( 1-S\left(
X\right) \right) \left( f_{1}^{\prime }\left( X\right) -\bar{r}\right) }{%
1-S_{E}\left( X\right) }\right] \\
&&+\frac{1-\hat{S}_{E}\left( X\right) }{1-\hat{S}\left( X\right) }\sum \hat{S%
}_{E}^{m}\left( X^{\prime },X\right) \left[ S_{E}\left( X^{\prime
},X^{\prime }\right) \frac{\left( 1-S\left( X^{\prime }\right) \right)
\left( f_{1}^{\prime }\left( X^{\prime }\right) -\bar{r}\right) }{%
1-S_{E}\left( X^{\prime }\right) }\right]
\end{eqnarray*}%
This series shows that due to the diffusion, investing in one other
investors leads to a chain of intermediated investments, which increases the
uncertainty, from the point of view of the investor based in $X$:

\begin{equation*}
\rho \left( X,\hat{f}\left( X^{\prime }\right) \right) =\rho \left[ \frac{1-%
\hat{S}_{E}\left( X^{\prime }\right) }{1-\hat{S}\left( X^{\prime }\right) }%
\sum \hat{S}_{E}^{m}\left( \left( X^{\prime }\right) ^{\prime },X^{\prime
}\right) \left[ S_{E}\left( \left( X^{\prime }\right) ^{\prime }\right) 
\frac{\left( 1-S\left( \left( X^{\prime }\right) ^{\prime }\right) \right)
\left( f_{1}^{\prime }\left( \left( X^{\prime }\right) ^{\prime }\right) -%
\bar{r}\right) }{1-S_{E}\left( \left( X^{\prime }\right) ^{\prime }\right) }%
\right] \right]
\end{equation*}%
We consider that the uncertainty is multiplicative:%
\begin{eqnarray*}
&&\rho \left( \frac{1-\hat{S}_{E}\left( X^{\prime }\right) }{1-\hat{S}\left(
X^{\prime }\right) }\hat{S}_{E}^{m}\left( \left( X^{\prime }\right) ^{\prime
},X^{\prime }\right) \left[ S_{E}\left( \left( X^{\prime }\right) ^{\prime
}\right) \frac{\left( 1-S\left( \left( X^{\prime }\right) ^{\prime }\right)
\right) \left( f_{1}^{\prime }\left( \left( X^{\prime }\right) ^{\prime
}\right) -\bar{r}\right) }{1-S_{E}\left( \left( X^{\prime }\right) ^{\prime
}\right) }\right] \right) \\
&\rightarrow &\zeta ^{2}\left( \frac{1}{\hat{w}_{E}^{\left( 0\right) }\left(
\left( X^{\prime }\right) ^{\prime },X_{m-1}\right) ...\hat{w}_{E}^{\left(
0\right) }\left( X_{1},X^{\prime }\right) }\right) \hat{S}_{E}^{2m}\left(
X^{\prime },X\right)
\end{eqnarray*}%
$\zeta ^{2}$ being the variance of:%
\begin{equation*}
\frac{1-\hat{S}_{E}\left( X^{\prime }\right) }{1-\hat{S}\left( X^{\prime
}\right) }S_{E}\left( X^{\prime }\right) \frac{\left( 1-S\left( X^{\prime
}\right) \right) \left( f_{1}^{\prime }\left( X^{\prime }\right) -\bar{r}%
\right) }{1-S_{E}\left( X^{\prime }\right) }
\end{equation*}%
We also assume that for paths that are disconnected the uncertainty is
additive, so that a first approximtion for uncertainty is:%
\begin{eqnarray*}
&&\rho \left( X,\hat{f}\left( X^{\prime }\right) \right) \\
&\rightarrow &\sum \zeta ^{2}\left( \frac{1}{\hat{w}_{E}^{\left( 0\right)
}\left( X^{\prime },X_{m-1}\right) ...\hat{w}_{E}^{\left( 0\right) }\left(
X_{m-1},X\right) }\right) \hat{S}_{E}^{2m}\left( X^{\prime },X\right)
\end{eqnarray*}%
These assumptions have the following consequences for shares:%
\begin{eqnarray*}
&&\hat{S}_{E}\left( X^{\prime },X\right) \\
&=&\frac{\underline{\hat{S}}\left( X^{\prime },X\right) }{2}+\frac{\hat{w}%
\left( X^{\prime },X\right) }{2}\left( \hat{w}\left( X\right) \left( \hat{f}%
\left( X^{\prime }\right) -\frac{\left\langle \hat{f}\left( X^{\prime
}\right) \right\rangle _{\hat{w}_{E}}+\left\langle \hat{r}\left( X^{\prime
}\right) \right\rangle _{\hat{w}_{L}}}{2}\right) +w\left( X\right) \left( 
\hat{f}\left( X^{\prime }\right) -\frac{f\left( X\right) +r\left( X\right) }{%
2}\right) \right) \\
&=&\frac{\underline{\hat{S}}\left( X^{\prime },X\right) }{2}+\frac{\hat{w}%
\left( X^{\prime },X\right) }{2}\left( \hat{f}\left( X^{\prime }\right) -%
\hat{w}\left( X\right) \frac{\left\langle \hat{f}\left( X^{\prime }\right)
\right\rangle _{\hat{w}_{E}}+\left\langle \hat{r}\left( X^{\prime }\right)
\right\rangle _{\hat{w}_{L}}}{2}-w\left( X\right) \frac{f\left( X\right)
+r\left( X\right) }{2}\right)
\end{eqnarray*}%
The coefficient $\hat{w}\left( X^{\prime },X\right) $ being inversely
proportional to the uncertainty:%
\begin{eqnarray*}
&&\hat{w}\left( X^{\prime },X\right) \\
&\rightarrow &\left( \sum \zeta ^{2}\left( \frac{1}{\hat{w}_{E}^{\left(
0\right) }\left( X^{\prime },X_{m-1}\right) ...\hat{w}_{E}^{\left( 0\right)
}\left( X_{m-1},X\right) }\right) \hat{S}_{E}^{2m}\left( X^{\prime
},X\right) \right) ^{-1}
\end{eqnarray*}%
If the series is not bounded, then;%
\begin{equation*}
\underline{\hat{S}}_{E}\left( X^{\prime },X\right) =\frac{\underline{\hat{S}}%
\left( X^{\prime },X\right) }{2}=\frac{\hat{w}_{E}\left( X^{\prime
},X\right) }{\hat{w}_{E}\left( X\right) +\hat{w}_{L}\left( X\right)
+w_{E}\left( X\right) +w_{L}\left( X\right) }\rightarrow 0
\end{equation*}%
and $\hat{S}_{E}\left( X^{\prime },X\right) \rightarrow 0$.

This implies that connections are organized mainly in limited zone of
independent elements: if one element of connected group connects to an other
group, it increases the uncertainty of the group as a whole. If we assume a
minimal value of the shares, this implies that agents are organized in
finite groups of relatively close investors. As a consequence, there are
organizations in several regions, and connections between these groups are
limited. However, as explained in the text, consdering for example that the
uncertainty is sector dependent, we can consider groups interact to some
extent. These interactions will induce some transitions and reoganizations.

Due to inertia in modification, there are many possibilities of reorganizing
connections. This possibility will be the studied in the context of fields
of groups.

\subsection*{A3.3 Precise form of uncertainty and computations of $\hat{w}$}

Th uncertainty is evaluatd. We consider only participations at first and
include loans later that $\zeta ^{2}$ is the uncertainty for the$\
S_{E}\left( X^{\prime }\right) \frac{\left( 1-S\left( X^{\prime }\right)
\right) \left( f_{1}^{\prime }\left( X^{\prime }\right) -\bar{r}\right) }{%
1-S_{E}\left( X^{\prime }\right) }$, so that for choosing to take share in $%
\hat{f}\left( X^{\prime }\right) $ we sum over disconnected paths:%
\begin{eqnarray*}
\hat{f}\left( X^{\prime }\right) -\bar{r} &=&\frac{1-\hat{S}_{E}\left(
X^{\prime }\right) }{1-\hat{S}\left( X^{\prime }\right) }\left[ S_{E}\left(
X^{\prime }\right) \frac{\left( 1-S\left( X^{\prime }\right) \right) \left(
f_{1}^{\prime }\left( X\right) -\bar{r}\right) }{1-S_{E}\left( X^{\prime
}\right) }\right] \\
&&+\frac{1-\hat{S}_{E}\left( X^{\prime }\right) }{1-\hat{S}\left( X^{\prime
}\right) }\sum \hat{S}_{E}^{m}\left( \left( X^{\prime }\right) ^{\prime
},X^{\prime }\right) \left[ S_{E}\left( \left( X^{\prime }\right) ^{\prime
}\right) \frac{\left( 1-S\left( \left( X^{\prime }\right) ^{\prime }\right)
\right) \left( f_{1}^{\prime }\left( \left( X^{\prime }\right) ^{\prime
}\right) -\bar{r}\right) }{1-S_{E}\left( \left( X^{\prime }\right) ^{\prime
}\right) }\right]
\end{eqnarray*}%
and we assume as before that the uncertainty is multiplicative along paths
so that $\rho \left( X,\hat{f}\left( X^{\prime }\right) \right) $ becomes:

\begin{eqnarray}
&&\frac{\zeta ^{2}}{\hat{w}_{E}^{\left( 0\right) }\left( X^{\prime
},X\right) }+\sum \frac{\zeta ^{2}}{\hat{w}_{E}^{\left( 0\right) }\left(
X^{\prime },X\right) }\left( \frac{1}{\hat{w}_{E}^{\left( 0\right) }\left(
X^{\prime },X_{1}\right) ...\hat{w}_{E}^{\left( 0\right) }\left(
X_{m-1},X_{m}\right) }\right)  \label{PRV} \\
&&\times \hat{S}_{E}^{2}\left( X_{1},X^{\prime }\right) ...\hat{S}%
_{E}^{2}\left( X_{k+1},X_{k}\right)  \notag \\
&\simeq &\frac{1}{\hat{w}_{E}^{\left( 0\right) }\left( X^{\prime },X\right) }%
\left( \zeta ^{2}+\left\langle \hat{S}_{E}\left( X_{1},X^{\prime }\right)
\right\rangle _{X_{1}}^{2}\sum \zeta ^{2}\left( \gamma ^{2}\right)
^{m}\left\langle \hat{S}_{E}^{2}\left( X^{\prime },X\right) \right\rangle
^{m-1}\right)  \notag \\
&\simeq &\frac{1}{\hat{w}_{E}^{\left( 0\right) }\left( X^{\prime },X\right) }%
\left( \zeta ^{2}+\zeta ^{2}\frac{\left( \gamma \left\langle \hat{S}%
_{E}\left( X_{1},X^{\prime }\right) \right\rangle _{X_{1}}\right) ^{2}}{%
1-\left( \gamma \left\langle \hat{S}_{E}\left( X^{\prime },X\right)
\right\rangle \right) ^{2}}\right)  \notag
\end{eqnarray}%
where $\left\langle \hat{S}_{E}^{2}\left( X^{\prime },X_{1}\right)
\right\rangle _{X_{1}}$ is the average taken over $X_{1}$ and with $\gamma $
representing the averag distance dependent uncertainty $\gamma ^{2}\simeq
\left( \frac{1}{\hat{w}_{E}^{\left( 0\right) }\left( X^{\prime
},X_{1}\right) ...\hat{w}_{E}^{\left( 0\right) }\left( X_{m-1},X_{m}\right) }%
\right) ^{\frac{1}{m}}$. Moreover $\zeta ^{2}$ is the variance of:%
\begin{equation*}
\frac{1-\hat{S}_{E}\left( X^{\prime }\right) }{1-\hat{S}\left( X^{\prime
}\right) }S_{E}\left( \left( X^{\prime }\right) ^{\prime }\right) \frac{%
\left( 1-S\left( \left( X^{\prime }\right) ^{\prime }\right) \right) \left(
f_{1}^{\prime }\left( \left( X^{\prime }\right) ^{\prime }\right) -\bar{r}%
\right) }{1-S_{E}\left( \left( X^{\prime }\right) ^{\prime }\right) }
\end{equation*}%
and is assumed to be constant for all sectors to consider the effct of
distance so that it is approximated by:%
\begin{equation*}
\left\langle \frac{1-\hat{S}_{E}\left( X^{\prime }\right) }{1-\hat{S}\left(
X^{\prime }\right) }S_{E}\left( \left( X^{\prime }\right) ^{\prime }\right)
\right\rangle ^{2}\left\langle \frac{\left( 1-S\left( \left( X^{\prime
}\right) ^{\prime }\right) \right) }{1-S_{E}\left( \left( X^{\prime }\right)
^{\prime }\right) }\right\rangle ^{2}Var\left( f_{1}^{\prime }\left( \left(
X^{\prime }\right) ^{\prime }\right) -\bar{r}\right)
\end{equation*}%
To include loans, we assume the same uncertainty for $\hat{S}_{L}$ and $%
S_{L} $ for the sake of simplicity. This leads to consider that the
uncertainty of participation is propagated through the sum:

\begin{eqnarray}
&&\frac{\zeta ^{2}}{\hat{w}_{E}^{\left( 0\right) }\left( X^{\prime
},X\right) }+\sum \frac{\zeta ^{2}}{\hat{w}_{E}^{\left( 0\right) }\left(
X^{\prime },X\right) }\left( \frac{1}{\hat{w}_{E}^{\left( 0\right) }\left(
X^{\prime },X_{1}\right) ...\hat{w}_{E}^{\left( 0\right) }\left(
X_{m-1},X_{m}\right) }\right) \hat{S}^{2}\left( X_{1},X^{\prime }\right) ...%
\hat{S}^{2}\left( X_{k+1},X_{k}\right)  \label{RSC} \\
&\simeq &\frac{1}{\hat{w}_{E}^{\left( 0\right) }\left( X^{\prime },X\right) }%
\left( \zeta ^{2}+\left\langle \hat{S}\left( X_{1},X^{\prime }\right)
\right\rangle _{X_{1}}^{2}\sum \zeta ^{2}\left( \gamma ^{2}\right)
^{m}\left\langle \hat{S}^{2}\left( X^{\prime },X\right) \right\rangle
^{m-1}\right)  \notag \\
&\simeq &\frac{1}{\hat{w}_{E}^{\left( 0\right) }\left( X^{\prime },X\right) }%
\left( \zeta ^{2}+\zeta ^{2}\frac{\left( \gamma \left\langle \hat{S}\left(
X_{1},X^{\prime }\right) \right\rangle _{X_{1}}\right) ^{2}}{1-\left( \gamma
\left\langle \hat{S}\left( X^{\prime },X\right) \right\rangle \right) ^{2}}%
\right)  \notag
\end{eqnarray}%
In first approximation $\left\langle \hat{S}\left( X_{1},X^{\prime }\right)
\right\rangle _{X_{1}}\simeq 2\left\langle \hat{S}_{E}\left( X_{1},X^{\prime
}\right) \right\rangle _{X_{1}}$ so that, the sum reduces to the previous
formula (\ref{PRV}) if we rescale $\gamma \rightarrow \frac{\gamma }{2}$.

Moreover, given our assumption, the uncertainty due to the loans is also
given by (\ref{RSC}). Consequently $\hat{w}_{E}\left( X^{\prime },X\right) =%
\hat{w}_{L}\left( X^{\prime },X\right) =\frac{\hat{w}}{2}\left( X^{\prime
},X\right) $ 
\begin{eqnarray*}
\hat{w}\left( X^{\prime },X\right) &\rightarrow &\frac{2\zeta ^{2}}{2\left(
\zeta ^{2}+\frac{\zeta ^{2}}{\hat{w}_{E}^{\left( 0\right) }\left( X^{\prime
},X\right) }\left( 1+\frac{\left( \gamma \left\langle \hat{S}_{E}\left(
X_{1},X^{\prime }\right) \right\rangle _{X_{1}}\right) ^{2}}{1-\left( \gamma
\left\langle \hat{S}_{E}\left( X^{\prime },X\right) \right\rangle \right)
^{2}}\right) \right) } \\
&=&\frac{\left( 1-\left( \gamma \left\langle \hat{S}_{E}\left( X\right)
\right\rangle \right) ^{2}\right) \hat{w}_{E}^{\left( 0\right) }\left(
X^{\prime },X\right) }{1+\hat{w}_{E}^{\left( 0\right) }\left( X^{\prime
},X\right) \left( 1-\left( \gamma \left\langle \hat{S}_{E}\left( X\right)
\right\rangle \right) ^{2}\right) +\left( \gamma \left\langle \hat{S}%
_{E}\left( X_{1},X^{\prime }\right) \right\rangle _{X_{1}}\right)
^{2}-\left( \gamma \left\langle \hat{S}_{E}\left( X\right) \right\rangle
\right) ^{2}}
\end{eqnarray*}

\section*{Appendix 4. Averages computations}

The saddle point equations are first solved in averages. This is done by
first computing the average quantities of the model as functions of shares.
Then, we write the average return equations as an equation for average
shares. The solutions yield the averages of the model.

\subsection*{A4.1 Formula for averages $\left\langle \hat{S}_{E}\left(
X^{\prime }\right) \right\rangle =\left\langle \hat{S}_{E}\left( X^{\prime
},X\right) \right\rangle $ and $\hat{S}\left( X^{\prime }\right) =\hat{S}%
\left( X^{\prime },X\right) $}

We will use the approximation:%
\begin{eqnarray}
\hat{S}_{E}\left( X^{\prime }\right) &=&\int \hat{S}_{E}\left( X^{\prime
},X\right) \frac{\hat{K}_{X}\left\vert \hat{\Psi}\left( X\right) \right\vert
^{2}}{\hat{K}_{X^{\prime }}\left\vert \hat{\Psi}\left( X^{\prime }\right)
\right\vert ^{2}}dX  \label{PRT} \\
&\rightarrow &\int \hat{S}_{E}\left( X^{\prime },X\right) dX\frac{%
\left\langle \hat{K}\right\rangle \left\Vert \hat{\Psi}\right\Vert ^{2}}{%
\hat{K}_{X^{\prime }}\left\vert \hat{\Psi}\left( X^{\prime }\right)
\right\vert ^{2}}  \notag
\end{eqnarray}%
\begin{eqnarray}
\hat{S}\left( X^{\prime }\right) &=&\int \hat{S}\left( X^{\prime },X\right) 
\frac{\hat{K}_{X}\left\vert \hat{\Psi}\left( X\right) \right\vert ^{2}}{\hat{%
K}_{X^{\prime }}\left\vert \hat{\Psi}\left( X^{\prime }\right) \right\vert
^{2}}dX  \label{PRZ} \\
&\rightarrow &\int \hat{S}\left( X^{\prime },X\right) dX\frac{\left\langle 
\hat{K}\right\rangle \left\Vert \hat{\Psi}\right\Vert ^{2}}{\hat{K}%
_{X^{\prime }}\left\vert \hat{\Psi}\left( X^{\prime }\right) \right\vert ^{2}%
}  \notag
\end{eqnarray}%
To derive the equations for average shares, we first use:%
\begin{eqnarray}
&&\left\langle \hat{S}_{E}\left( X^{\prime }\right) \right\rangle
\label{Snf} \\
&\simeq &\frac{\left\langle \underline{\hat{S}}\left( X^{\prime }\right)
\right\rangle }{2}  \notag \\
&&+\frac{\left\langle \hat{w}\left( X^{\prime }\right) \right\rangle }{2}%
\left( \hat{w}\left( \left\langle \hat{f}\left( X^{\prime }\right)
\right\rangle -\frac{\left\langle \hat{f}\left( X^{\prime }\right)
\right\rangle _{\hat{w}_{E}}+\left\langle \hat{r}\left( X^{\prime }\right)
\right\rangle _{\hat{w}_{L}}}{2}\right) +w\left( \left\langle \hat{f}\left(
X^{\prime }\right) \right\rangle -\frac{\left\langle f\left( X\right)
+r\left( X\right) \right\rangle _{w}}{2}\right) \right)  \notag \\
&\rightarrow &\frac{\left\langle \underline{\hat{S}}\left( X^{\prime
}\right) \right\rangle }{2}  \notag \\
&&+\frac{\left\langle \hat{w}\left( X^{\prime }\right) \right\rangle }{2}%
\left( \hat{w}\left( \frac{\left\langle \hat{f}\left( X^{\prime }\right)
\right\rangle -\left\langle \hat{r}\left( X^{\prime }\right) \right\rangle _{%
\hat{w}_{L}}}{2}\right) +w\left( \left\langle \hat{f}\left( X^{\prime
}\right) \right\rangle -\frac{\left\langle f\left( X\right) +r\left(
X\right) \right\rangle _{w}}{2}\right) \right)  \notag
\end{eqnarray}%
and:%
\begin{eqnarray}
&&\left\langle \hat{S}\left( X^{\prime }\right) \right\rangle  \label{Stf} \\
&\simeq &\left( \left\langle \underline{\hat{S}}\left( X^{\prime }\right)
\right\rangle +\left\langle \hat{w}\left( X^{\prime }\right) \right\rangle
\left( \frac{\left\langle \hat{f}\left( X^{\prime }\right) \right\rangle
+\left\langle \hat{r}\left( X^{\prime }\right) \right\rangle }{2}-\hat{w}%
\frac{\left\langle \hat{f}\left( X^{\prime }\right) \right\rangle _{\hat{w}%
_{E}}+\left\langle \hat{r}\left( X^{\prime }\right) \right\rangle _{\hat{w}%
_{L}}}{2}-w\frac{\left\langle f\left( X\right) +r\left( X\right)
\right\rangle _{w}}{2}\right) \right)  \notag \\
&&\frac{\left\langle \hat{K}\right\rangle \left\Vert \hat{\Psi}\right\Vert
^{2}}{\left\langle \hat{K}_{X^{\prime }}\right\rangle \left\langle
\left\vert \hat{\Psi}\left( X^{\prime }\right) \right\vert ^{2}\right\rangle 
}  \notag \\
&\simeq &\underline{\hat{S}}\left( X^{\prime }\right) +\left\langle \hat{w}%
\left( X^{\prime }\right) \right\rangle w\left( \frac{\left\langle \hat{f}%
\left( X^{\prime }\right) +\hat{r}\left( X^{\prime }\right) \right\rangle }{2%
}-\frac{\left\langle f\left( X\right) +r\left( X\right) \right\rangle _{w}}{2%
}\right)  \notag
\end{eqnarray}%
and replace $\left\langle \hat{S}\left( X^{\prime }\right) \right\rangle $
with $\left\langle \hat{S}_{E}\left( X^{\prime }\right) \right\rangle $.
Writing: 
\begin{eqnarray*}
&&\left\langle \hat{S}\left( X^{\prime }\right) \right\rangle -2\left\langle 
\hat{S}_{E}\left( X^{\prime }\right) \right\rangle \\
&&\left\langle \hat{w}\left( X^{\prime }\right) \right\rangle \left( -\hat{w}%
\left( \frac{\left\langle \hat{f}\left( X^{\prime }\right) \right\rangle
-\left\langle \hat{r}\left( X^{\prime }\right) \right\rangle _{\hat{w}_{L}}}{%
2}\right) +w\frac{\left\langle \hat{r}\left( X^{\prime }\right)
\right\rangle -\left\langle \hat{f}\left( X^{\prime }\right) \right\rangle }{%
2}\right) \\
&&\left\langle \hat{w}\left( X^{\prime }\right) \right\rangle \left( -\hat{w}%
\left( \frac{\left\langle \hat{f}\left( X^{\prime }\right) \right\rangle
-\left\langle \hat{r}\left( X^{\prime }\right) \right\rangle _{\hat{w}_{L}}}{%
2}\right) +w\frac{\left\langle \hat{r}\left( X^{\prime }\right)
\right\rangle -\left\langle \hat{f}\left( X^{\prime }\right) \right\rangle }{%
2}\right)
\end{eqnarray*}%
we find the relation:%
\begin{equation*}
\left\langle \hat{S}\left( X^{\prime }\right) \right\rangle =2\left\langle 
\hat{S}_{E}\left( X^{\prime }\right) \right\rangle +\left\langle \hat{w}%
\left( X^{\prime }\right) \right\rangle \frac{\left\langle \hat{r}\left(
X^{\prime }\right) \right\rangle -\left\langle \hat{f}\left( X^{\prime
}\right) \right\rangle }{2}
\end{equation*}%
Th coefficient $\left\langle \hat{w}\right\rangle $ depends on the average $%
\left\langle S_{E}\left( X\right) \right\rangle $. We assume that $%
\left\langle \frac{1}{\hat{w}_{E}^{\left( 0\right) }\left( X,X^{\prime
}\right) }\right\rangle =1$ to simplify. Given that: 
\begin{equation*}
\gamma ^{2}\simeq \left( \frac{1}{\hat{w}_{E}^{\left( 0\right) }\left(
X^{\prime },X_{1}\right) ...\hat{w}_{E}^{\left( 0\right) }\left(
X_{m-1},X_{m}\right) }\right) ^{\frac{1}{m}}
\end{equation*}%
and since the functions $\frac{1}{\hat{w}_{E}^{\left( 0\right) }\left(
X_{m-1},X_{m}\right) }$ increase, we may replace in first approximation: 
\begin{equation*}
\left\langle \hat{w}\right\rangle \rightarrow \frac{\zeta ^{2}}{\zeta
^{2}+\zeta ^{2}\frac{1}{1-\left( \gamma \left\langle \hat{S}_{E}\left(
X\right) \right\rangle \right) ^{2}}}=\frac{1-\left( \gamma \left\langle 
\hat{S}_{E}\left( X\right) \right\rangle \right) ^{2}}{2-\left( \gamma
\left\langle \hat{S}_{E}\left( X\right) \right\rangle \right) ^{2}}
\end{equation*}%
with $\gamma $ representing the average distance-dependent uncertainty.

Given the $\left\langle \underline{S}\left( X,X\right) \right\rangle
=\left\langle \hat{w}\right\rangle $, we find the following equation:%
\begin{eqnarray}
\left\langle \hat{S}_{E}\left( X\right) \right\rangle &=&\frac{\frac{%
1-\left( \gamma \left\langle \hat{S}_{E}\left( X\right) \right\rangle
\right) ^{2}}{2-\left( \gamma \left\langle \hat{S}_{E}\left( X\right)
\right\rangle \right) ^{2}}}{2}+\frac{\frac{1-\left( \gamma \left\langle 
\hat{S}_{E}\left( X\right) \right\rangle \right) ^{2}}{2-\left( \gamma
\left\langle \hat{S}_{E}\left( X\right) \right\rangle \right) ^{2}}}{2}
\label{Snt} \\
&&\times \left( \frac{1-\left( \gamma \left\langle \hat{S}_{E}\left(
X\right) \right\rangle \right) ^{2}}{2-\left( \gamma \left\langle \hat{S}%
_{E}\left( X\right) \right\rangle \right) ^{2}}\left( \frac{\left\langle 
\hat{f}\left( X^{\prime }\right) \right\rangle -\left\langle \hat{r}\left(
X^{\prime }\right) \right\rangle _{\hat{w}_{L}}}{2}\right) \right.  \notag \\
&&\left. +\frac{1}{2-\left( \gamma \left\langle \hat{S}_{E}\left( X\right)
\right\rangle \right) ^{2}}\left( \left\langle \hat{f}\left( X^{\prime
}\right) \right\rangle -\frac{\left\langle f\left( X\right) +r\left(
X\right) \right\rangle _{w}}{2}\right) \right)  \notag
\end{eqnarray}

\subsection*{A4.2 Average Return $\left\langle \hat{f}\left( X^{\prime
}\right) \right\rangle $ in function of shares}

Using equation (\ref{Snt}) yields the following formula for average returns:

\begin{equation}
\left\langle \hat{f}\left( X^{\prime }\right) \right\rangle =\frac{\frac{%
2\left( 2-\left( \gamma \left\langle \hat{S}_{E}\left( X\right)
\right\rangle \right) ^{2}\right) }{1-\left( \gamma \left\langle \hat{S}%
_{E}\left( X\right) \right\rangle \right) ^{2}}\left\langle \hat{S}%
_{E}\left( X\right) \right\rangle -1+\frac{\left\langle \hat{r}\left(
X^{\prime }\right) \right\rangle _{\hat{w}_{L}}}{2}+\frac{\left\langle
f\left( X\right) \right\rangle }{2\left( 2-\left( \gamma \left\langle \hat{S}%
_{E}\left( X\right) \right\rangle \right) ^{2}\right) }}{\left( 1-\frac{1}{2}%
\frac{1-\left( \gamma \left\langle \hat{S}_{E}\left( X\right) \right\rangle
\right) ^{2}}{2-\left( \gamma \left\langle \hat{S}_{E}\left( X\right)
\right\rangle \right) ^{2}}\right) }  \label{Fsh}
\end{equation}%
Note for the sequel that:%
\begin{eqnarray*}
&&\left\langle \hat{f}\left( X^{\prime }\right) \right\rangle -\left\langle 
\hat{r}\left( X^{\prime }\right) \right\rangle _{\hat{w}_{L}} \\
&=&\frac{\frac{2\left( 2-\left( \gamma \left\langle \hat{S}_{E}\left(
X\right) \right\rangle \right) ^{2}\right) }{1-\left( \gamma \left\langle 
\hat{S}_{E}\left( X\right) \right\rangle \right) ^{2}}\left\langle \hat{S}%
_{E}\left( X\right) \right\rangle -1+\frac{\left\langle \hat{r}\left(
X^{\prime }\right) \right\rangle _{\hat{w}_{L}}}{2}+\frac{\left\langle
f\left( X\right) \right\rangle }{2\left( 2-\left( \gamma \left\langle \hat{S}%
_{E}\left( X\right) \right\rangle \right) ^{2}\right) }-\left( 1-\frac{1}{2}%
\frac{1-\left( \gamma \left\langle \hat{S}_{E}\left( X\right) \right\rangle
\right) ^{2}}{2-\left( \gamma \left\langle \hat{S}_{E}\left( X\right)
\right\rangle \right) ^{2}}\right) \left\langle \hat{r}\left( X^{\prime
}\right) \right\rangle _{\hat{w}_{L}}}{\left( 1-\frac{1}{2}\frac{1-\left(
\gamma \left\langle \hat{S}_{E}\left( X\right) \right\rangle \right) ^{2}}{%
2-\left( \gamma \left\langle \hat{S}_{E}\left( X\right) \right\rangle
\right) ^{2}}\right) }
\end{eqnarray*}%
\begin{equation}
\left\langle \hat{f}\left( X^{\prime }\right) \right\rangle -\left\langle 
\hat{r}\left( X^{\prime }\right) \right\rangle _{\hat{w}_{L}}=\frac{\frac{%
2\left( 2-\left( \gamma \left\langle \hat{S}_{E}\left( X\right)
\right\rangle \right) ^{2}\right) }{1-\left( \gamma \left\langle \hat{S}%
_{E}\left( X\right) \right\rangle \right) ^{2}}\left\langle \hat{S}%
_{E}\left( X\right) \right\rangle -1+\frac{\left\langle f\left( X\right)
\right\rangle -\left\langle \hat{r}\left( X^{\prime }\right) \right\rangle _{%
\hat{w}_{L}}}{2\left( 2-\left( \gamma \left\langle \hat{S}_{E}\left(
X\right) \right\rangle \right) ^{2}\right) }}{\left( 1-\frac{1}{2}\frac{%
1-\left( \gamma \left\langle \hat{S}_{E}\left( X\right) \right\rangle
\right) ^{2}}{2-\left( \gamma \left\langle \hat{S}_{E}\left( X\right)
\right\rangle \right) ^{2}}\right) }  \label{Fsk}
\end{equation}

\subsection*{A4.3 Formula for $\left\langle S_{E}\left( X,X\right)
\right\rangle $ and $\left\langle S\left( X,X\right) \right\rangle $}

\subsubsection*{A4.3.1 Constant return to scale}

The averages of equations (\ref{SNpa}) and (\ref{STpa}) are:%
\begin{eqnarray*}
&&\left\langle S_{E}\left( X,X\right) \right\rangle \\
&=&\frac{w\left( X\right) }{2}\left( 1+\left( \left\langle \hat{w}\left(
X\right) \right\rangle \left( \left\langle f\left( X\right) \right\rangle -%
\frac{\left\langle \hat{f}\left( X^{\prime }\right) \right\rangle _{\hat{w}%
_{E}}+\left\langle \hat{r}\left( X^{\prime }\right) \right\rangle _{\hat{w}%
_{L}}}{2}\right) +\frac{w\left( X\right) }{2}\left( f\left( X\right) -\bar{r}%
\left( X\right) \right) \right) \right)
\end{eqnarray*}%
\begin{equation}
\left\langle S\left( X,X\right) \right\rangle =\left\langle w\left( X\right)
\right\rangle \left( 1+\left( \hat{w}\left( X\right) \left( \frac{%
\left\langle f\left( X\right) \right\rangle +\left\langle \bar{r}\left(
X\right) \right\rangle }{2}-\frac{\left\langle \hat{f}\left( X^{\prime
}\right) \right\rangle _{\hat{w}_{E}}+\left\langle \hat{r}\left( X^{\prime
}\right) \right\rangle _{\hat{w}_{L}}}{2}\right) \right) \right)  \label{Qs}
\end{equation}%
with:%
\begin{equation*}
\left\langle \hat{w}\right\rangle \rightarrow \frac{\zeta ^{2}}{\zeta
^{2}+\zeta ^{2}\frac{1}{1-\left( \gamma \left\langle \hat{S}_{E}\left(
X\right) \right\rangle \right) ^{2}}}=\frac{1-\left( \gamma \left\langle 
\hat{S}_{E}\left( X\right) \right\rangle \right) ^{2}}{2-\left( \gamma
\left\langle \hat{S}_{E}\left( X\right) \right\rangle \right) ^{2}}
\end{equation*}%
and:%
\begin{equation*}
\left\langle w\right\rangle \rightarrow \frac{1}{2-\left( \gamma
\left\langle \hat{S}_{E}\left( X\right) \right\rangle \right) ^{2}}
\end{equation*}%
The averages of return are replaced by their constant return to scale values:%
\begin{equation}
\frac{\left\langle f\left( X\right) \right\rangle +\left\langle \bar{r}%
\left( X\right) \right\rangle }{2}\rightarrow \frac{\left\langle f_{1}\left(
X\right) \right\rangle -\left( 1-\left\langle S\left( X\right) \right\rangle
\right) C+\left\langle \bar{r}\left( X\right) \right\rangle }{2}  \label{vr}
\end{equation}%
leading to:%
\begin{eqnarray*}
\left\langle S\left( X,X\right) \right\rangle &=&\frac{\left\langle w\left(
X\right) \right\rangle }{\left( 1-\frac{\left\langle w\left( X\right)
\right\rangle \hat{w}\left( X\right) C}{2}\frac{\left\langle \hat{K}%
\right\rangle \left\Vert \hat{\Psi}\right\Vert ^{2}}{\left\langle
K\right\rangle \left\Vert \Psi \right\Vert ^{2}}\right) } \\
&&\times \left( 1+\left( \hat{w}\left( X\right) \left( \frac{\left\langle
f_{1}\left( X\right) \right\rangle -C+\left\langle \bar{r}\left( X\right)
\right\rangle }{2}-\frac{\left\langle \hat{f}\left( X^{\prime }\right)
\right\rangle _{\hat{w}_{E}}+\left\langle \hat{r}\left( X^{\prime }\right)
\right\rangle _{\hat{w}_{L}}}{2}\right) \right) \right)
\end{eqnarray*}%
or which is equivalent:%
\begin{equation*}
1-\left\langle S\left( X\right) \right\rangle =1-\frac{\left\langle w\left(
X\right) \right\rangle \left( 1+\left( \hat{w}\left( X\right) \left( \frac{%
\left\langle f_{1}\left( X\right) \right\rangle -C+\left\langle \bar{r}%
\left( X\right) \right\rangle }{2}-\frac{\left\langle \hat{f}\left(
X^{\prime }\right) \right\rangle _{\hat{w}_{E}}+\left\langle \hat{r}\left(
X^{\prime }\right) \right\rangle _{\hat{w}_{L}}}{2}\right) \right) \right) }{%
\left( 1-\frac{\left\langle w\left( X\right) \right\rangle \hat{w}\left(
X\right) C}{2}\frac{\left\langle \hat{K}\right\rangle \left\Vert \hat{\Psi}%
\right\Vert ^{2}}{\left\langle K\right\rangle \left\Vert \Psi \right\Vert
^{2}}\right) }\frac{\left\langle \hat{K}\right\rangle \left\Vert \hat{\Psi}%
\right\Vert ^{2}}{\left\langle K\right\rangle \left\Vert \Psi \right\Vert
^{2}}
\end{equation*}%
and:%
\begin{eqnarray*}
&&\left\langle S_{E}\left( X,X\right) \right\rangle \\
&=&\frac{w\left( X\right) }{2}\left( 1+\left( \left\langle \hat{w}\left(
X\right) \right\rangle \left( \left\langle f_{1}\left( X\right)
\right\rangle -\left( 1-\left\langle S\left( X\right) \right\rangle \right)
C-\frac{\left\langle \hat{f}\left( X^{\prime }\right) \right\rangle _{\hat{w}%
_{E}}+\left\langle \hat{r}\left( X^{\prime }\right) \right\rangle _{\hat{w}%
_{L}}}{2}\right) \right. \right. \\
&&\left. \left. +\frac{w\left( X\right) }{2}\left( \left\langle f_{1}\left(
X\right) \right\rangle -\left( 1-\left\langle S\left( X\right) \right\rangle
\right) C-\bar{r}\left( X\right) \right) \right) \right)
\end{eqnarray*}

\subsection*{A4.4 Capital ratio}

Average capital for investors is given by (see Gosselin and Lotz (2024)):%
\begin{equation}
\left\langle \hat{K}\right\rangle \left\Vert \hat{\Psi}\right\Vert ^{2}=%
\frac{\hat{\mu}V\sigma _{\hat{K}}^{2}}{2\left\langle \hat{g}\right\rangle
^{2}}\left( \frac{\left( \frac{\left\Vert \hat{\Psi}_{0}\right\Vert ^{2}}{%
\hat{\mu}}+\frac{\left\langle \hat{g}\right\rangle }{2}\left( \frac{1-r}{r}%
\right) \underline{\hat{k}}\right) }{\left( 1-2r\left( 1-r\right) \underline{%
\hat{k}}\right) }\right) ^{2}\left( 1-2\frac{2+\underline{\hat{k}}-\sqrt{%
\left( 2+\underline{\hat{k}}\right) ^{2}-\underline{\hat{k}}}}{9}\right)
\label{Vl}
\end{equation}%
with:%
\begin{equation*}
r=\frac{2+\hat{k}-\sqrt{\left( 2+\hat{k}\right) ^{2}-\hat{k}}}{6\underline{%
\hat{k}}}
\end{equation*}%
In terms of new variables:%
\begin{equation*}
\left\langle \hat{g}\right\rangle \rightarrow \left( 1-\hat{S}\left(
X^{\prime },X\right) \right) ^{-1}\left\langle \hat{f}\right\rangle
\end{equation*}%
and:%
\begin{equation*}
\hat{k}=\frac{\hat{S}}{1-\hat{S}}
\end{equation*}%
\begin{equation*}
\underline{\hat{k}}\left( X^{\prime }\right) =\frac{\hat{S}\left( X^{\prime
}\right) }{1-\hat{S}\left( X^{\prime }\right) },\hat{S}\left( X^{\prime
}\right) =\frac{\underline{\hat{k}}\left( X^{\prime }\right) }{1+\underline{%
\hat{k}}\left( X^{\prime }\right) },\frac{1}{1+\underline{\hat{k}}\left(
X^{\prime }\right) }=1-\hat{S}\left( X^{\prime }\right)
\end{equation*}

\begin{equation*}
\hat{S}_{L}\left( X^{\prime }\right) =\frac{\underline{\hat{k}}_{L}\left(
X^{\prime }\right) }{1+\underline{\hat{k}}\left( X^{\prime }\right) }
\end{equation*}%
\begin{equation*}
\underline{\hat{k}}_{L}\left( X^{\prime }\right) =\hat{S}_{L}\left(
X^{\prime }\right) \left( 1+\underline{\hat{k}}\left( X^{\prime }\right)
\right) =\frac{\hat{S}_{L}\left( X^{\prime }\right) }{1-\hat{S}\left(
X^{\prime }\right) }
\end{equation*}%
\begin{equation*}
1+\underline{\hat{k}}_{L}\left( X^{\prime }\right) =\frac{1-\hat{S}\left(
X^{\prime }\right) +\hat{S}_{L}\left( X^{\prime }\right) }{1-\hat{S}\left(
X^{\prime }\right) }=\frac{1-\hat{S}_{E}\left( X^{\prime }\right) }{1-\hat{S}%
\left( X^{\prime }\right) }
\end{equation*}%
\begin{equation*}
\frac{1+\underline{\hat{k}}_{L}\left( X^{\prime }\right) }{1+\underline{\hat{%
k}}\left( X^{\prime }\right) }=1-\hat{S}_{E}\left( X^{\prime }\right)
\end{equation*}%
average capital becomes:%
\begin{eqnarray}
&&\left\langle \hat{K}\right\rangle \left\Vert \hat{\Psi}\right\Vert ^{2} \\
&\rightarrow &\frac{\hat{\mu}V\sigma _{\hat{K}}^{2}\left( 1-\hat{S}\right)
^{2}}{2\left\langle \hat{f}\right\rangle ^{2}}\left( \frac{\left( \frac{%
\left\Vert \hat{\Psi}_{0}\right\Vert ^{2}}{\hat{\mu}}+\left( \frac{1-r}{r}%
\right) \frac{\left\langle \hat{f}\right\rangle }{2\left( 1-\hat{S}\right) }%
\hat{S}\right) }{\left( 1-2r\left( 1-r\right) \frac{\hat{S}}{\left( 1-\hat{S}%
\right) }\right) }\right) ^{2}\left( 1-2\frac{2+\hat{S}-\sqrt{\left( 2+\hat{S%
}\right) ^{2}-\hat{S}}}{9}\right)  \notag
\end{eqnarray}%
which writes also:%
\begin{eqnarray}
&&\left\langle \hat{K}\right\rangle \left\Vert \hat{\Psi}\right\Vert ^{2}
\label{K1} \\
&=&\frac{\hat{\mu}V\sigma _{\hat{K}}^{2}}{2\left\langle \hat{f}\right\rangle
^{2}}\left( \frac{\left( \frac{\left\Vert \hat{\Psi}_{0}\right\Vert ^{2}}{%
\hat{\mu}}\left( 1-\hat{S}\right) +\left( \frac{1-r}{r}\right) \frac{%
\left\langle \hat{f}\right\rangle \hat{S}}{2}\right) }{\left( 1-\hat{S}%
\right) -2r\left( 1-r\right) \frac{\hat{S}}{1-\hat{S}}}\right) ^{2}\left( 1-2%
\frac{2+\frac{\hat{S}}{1-\hat{S}}-\sqrt{\left( 2+\frac{\hat{S}}{1-\hat{S}}%
\right) ^{2}-\frac{\hat{S}}{1-\hat{S}}}}{9}\right)  \notag
\end{eqnarray}%
Here:%
\begin{equation*}
\frac{1-r}{r}=\frac{1-\frac{2+\hat{k}-\sqrt{\left( 2+\hat{k}\right) ^{2}-%
\hat{k}}}{6\underline{\hat{k}}}}{\frac{2+\hat{k}-\sqrt{\left( 2+\hat{k}%
\right) ^{2}-\hat{k}}}{6\underline{\hat{k}}}}=\frac{1-\frac{2+\frac{\hat{S}}{%
1-\hat{S}}-\sqrt{\left( 2+\frac{\hat{S}}{1-\hat{S}}\right) ^{2}-\frac{\hat{S}%
}{1-\hat{S}}}}{6\frac{\hat{S}}{1-\hat{S}}}}{\frac{2+\frac{\hat{S}}{1-\hat{S}}%
-\sqrt{\left( 2+\frac{\hat{S}}{1-\hat{S}}\right) ^{2}-\frac{\hat{S}}{1-\hat{S%
}}}}{6\frac{\hat{S}}{1-\hat{S}}}}
\end{equation*}%
$\allowbreak $

To find $\left\langle K\right\rangle \left\Vert \Psi \right\Vert ^{2}$, we
use that:%
\begin{equation*}
S\left( X^{\prime }\right) =\frac{\int \underline{k}\left( X^{\prime
},X\right) dX\frac{\left\langle \hat{K}\right\rangle \left\Vert \hat{\Psi}%
\right\Vert ^{2}}{\left\langle K\right\rangle \left\Vert \Psi \right\Vert
^{2}}}{1+\underline{k}\left( X^{\prime }\right) }
\end{equation*}

\begin{equation*}
\frac{1}{1+\underline{\hat{k}}\left( X^{\prime }\right) }=1-\hat{S}\left(
X^{\prime }\right) ,\underline{\hat{k}}\left( X^{\prime }\right) =\frac{\hat{%
S}\left( X^{\prime }\right) }{1-\hat{S}\left( X^{\prime }\right) },%
\underline{\hat{k}}=\frac{\hat{S}}{1-\hat{S}}
\end{equation*}

\begin{equation*}
\underline{k}\left( X^{\prime }\right) =\frac{S\left( X^{\prime }\right) }{%
1-S\left( X^{\prime }\right) }
\end{equation*}%
\begin{equation*}
\frac{1}{1+\underline{k}\left( X^{\prime }\right) }=\frac{1-S\left(
X^{\prime }\right) }{1+\left\langle S\right\rangle }
\end{equation*}%
\begin{eqnarray*}
\underline{k}_{L}\left( X^{\prime }\right) &=&S_{L}\left( X^{\prime }\right) 
\frac{\underline{k}\left( X^{\prime }\right) }{S\left( X^{\prime }\right) }
\\
&=&\frac{S_{L}\left( X^{\prime }\right) }{1-S\left( X^{\prime }\right) }
\end{eqnarray*}%
\begin{equation*}
\underline{k}_{1}\left( X^{\prime }\right) =\frac{S_{E}\left( X^{\prime
}\right) }{1-S\left( X^{\prime }\right) }
\end{equation*}%
\begin{equation*}
1+\underline{\hat{k}}_{L}\left( X^{\prime }\right) =\frac{1-\hat{S}%
_{E}\left( X^{\prime }\right) }{1-\hat{S}\left( X^{\prime }\right) }
\end{equation*}%
\begin{equation*}
\frac{1+\underline{k}_{L}\left( X^{\prime }\right) }{1+\underline{k}\left(
X^{\prime }\right) }\rightarrow 1-S_{E}\left( X^{\prime }\right)
\end{equation*}

\begin{equation*}
C\rightarrow \frac{C}{1+\underline{k}\left( X^{\prime }\right) }=\left(
1-S\left( X^{\prime }\right) \right) C
\end{equation*}%
\begin{equation*}
\frac{\left( 1-\left( S\left( X^{\prime }\right) \right) \right) \left(
f_{1}^{\prime }\left( X\right) -\bar{r}\right) }{1-\left( S_{E}\left(
X^{\prime }\right) \right) }=f_{1}\left( X\right) -\left( 1-\left( S\left(
X^{\prime }\right) \right) \right) C-\bar{r}
\end{equation*}%
so that $\left\langle K\right\rangle \left\Vert \Psi \right\Vert ^{2}$ is
given by%
\begin{eqnarray}
&&\left\langle K\right\rangle \left\Vert \Psi \right\Vert ^{2} \\
&=&\frac{\epsilon K_{0}}{\sigma _{\hat{K}}^{2}f_{1}^{\left( e\right) }\left(
X\right) }\left\{ \frac{1}{4}\left( 2X^{2}\left( \bar{C}+X\right) -\left( X-%
\bar{C}\right) \left( X^{2}+\left( \bar{C}\right) ^{2}\right) \right) -\frac{%
\bar{C}}{3}\left( X\left( X-\bar{C}\right) +C^{2}\right) \right\}  \notag \\
&=&\frac{\epsilon K_{0}}{\sigma _{\hat{K}}^{2}f_{1}^{\left( e\right) }\left(
X\right) }\frac{1}{12}\left( 3X-\bar{C}\right) \left( \bar{C}+X\right) ^{2} 
\notag \\
&\rightarrow &\frac{\epsilon K_{0}\frac{1}{12}\left( 3X-C\left(
1-S_{E}\left( X^{\prime }\right) \right) \right) \left( C\left( 1-\hat{S}%
_{E}\left( X^{\prime }\right) \right) +X\right) ^{2}}{\sigma _{\hat{K}%
}^{2}\left( \frac{1-S_{E}\left( X\right) }{1-S\left( X\right) }\left(
f_{1}\left( X\right) -r\left( X\right) \right) +r\left( X\right) \right) } 
\notag
\end{eqnarray}%
where the maximal capital is given by:%
\begin{equation*}
K_{0}=\frac{\bar{C}\left( X\right) +\sqrt{\sigma _{\hat{K}}^{2}\left( \frac{%
\left\vert \Psi _{0}\left( X\right) \right\vert ^{2}}{\epsilon }-\frac{%
f_{1}\left( X\right) }{2}\right) }}{f_{1}^{\left( e\right) }\left( X\right) }%
\simeq \frac{\sqrt{\sigma _{\hat{K}}^{2}\frac{\left\vert \Psi _{0}\left(
X\right) \right\vert ^{2}}{\epsilon }}}{f_{1}^{\left( e\right) }\left(
X\right) }
\end{equation*}%
Using that in first approximation:%
\begin{equation*}
\left( 3X-C\left( 1-S\right) \right) \left( C\left( 1-S\right) +X\right)
^{2}\simeq 3X^{3}
\end{equation*}%
we have:%
\begin{equation}
\left\langle K\right\rangle \left\Vert \Psi \right\Vert ^{2}=\frac{1}{12}%
\frac{\epsilon \sqrt{\sigma _{\hat{K}}^{2}\frac{\left\vert \Psi _{0}\left(
X\right) \right\vert ^{2}}{\epsilon }}}{\sigma _{\hat{K}}^{2}\left( \frac{%
1-\left\langle S_{E}\left( X\right) \right\rangle }{1-\left\langle S\left(
X\right) \right\rangle }\left( f_{1}\left( X\right) -r\left( X\right)
\right) +r\left( X\right) \right) ^{2}}\left( 3X-C\left( 1-S\right) \right)
\left( C\left( 1-S\right) +X\right) ^{2}  \label{Cp}
\end{equation}%
Given (\ref{K1}) and (\ref{Cp}) the average capital ratio writes:

\begin{eqnarray}
\frac{\left\langle \hat{K}\right\rangle \left\Vert \hat{\Psi}\right\Vert ^{2}%
}{\left\langle K\right\rangle \left\Vert \Psi \right\Vert ^{2}} &=&\frac{6%
\hat{\mu}V\sigma _{\hat{K}}^{4}\left( \frac{1-\left\langle S_{E}\left(
X\right) \right\rangle }{1-\left\langle S\left( X\right) \right\rangle }%
\left( \left\langle f_{1}\right\rangle -\left\langle r\left( X\right)
\right\rangle \right) +\left\langle r\left( X\right) \right\rangle \right)
^{2}}{\left\langle \hat{f}\right\rangle ^{2}\epsilon \sqrt{\sigma _{\hat{K}%
}^{2}\frac{\left\vert \Psi _{0}\left( X\right) \right\vert ^{2}}{\epsilon }}%
\left( 3X-C\left( 1-S\right) \right) \left( C\left( 1-S\right) +X\right) ^{2}%
}  \label{RtK} \\
&&\times \left( \frac{\left( \frac{\left\Vert \hat{\Psi}_{0}\right\Vert ^{2}%
}{\hat{\mu}}\left( 1-\hat{S}\right) +\left( \frac{1-r}{r}\right) \frac{%
\left\langle \hat{f}\right\rangle \hat{S}}{2}\right) }{\left( 1-\hat{S}%
\right) -2r\left( 1-r\right) \frac{\hat{S}}{1-\hat{S}}}\right) ^{2}\left( 1-2%
\frac{2+\frac{\hat{S}}{1-\hat{S}}-\sqrt{\left( 2+\frac{\hat{S}}{1-\hat{S}}%
\right) ^{2}-\frac{\hat{S}}{1-\hat{S}}}}{9}\right)  \notag
\end{eqnarray}%
Given (\ref{PRT}) and (\ref{PRZ}), this formula is in fact an equation for $%
\frac{\left\langle \hat{K}\right\rangle \left\Vert \hat{\Psi}\right\Vert ^{2}%
}{\left\langle K\right\rangle \left\Vert \Psi \right\Vert ^{2}}$:%
\begin{eqnarray}
0 &=&\frac{\left\langle \hat{K}\right\rangle \left\Vert \hat{\Psi}%
\right\Vert ^{2}}{\left\langle K\right\rangle \left\Vert \Psi \right\Vert
^{2}}\left( 1-\left\langle S\left( X,X\right) \right\rangle \frac{%
\left\langle \hat{K}\right\rangle \left\Vert \hat{\Psi}\right\Vert ^{2}}{%
\left\langle K\right\rangle \left\Vert \Psi \right\Vert ^{2}}\right) ^{2}-%
\frac{6\hat{\mu}V\sigma _{\hat{K}}^{4}\left( 1-\hat{S}\right) ^{2}}{%
\left\langle \hat{f}\right\rangle ^{2}\epsilon \sqrt{\sigma _{\hat{K}}^{2}%
\frac{\left\vert \Psi _{0}\left( X\right) \right\vert ^{2}}{\epsilon }}3X^{3}%
}  \label{Kq} \\
&&\times \left( \left( 1-\left\langle S_{E}\left( X,X\right) \right\rangle 
\frac{\left\langle \hat{K}\right\rangle \left\Vert \hat{\Psi}\right\Vert ^{2}%
}{\left\langle K\right\rangle \left\Vert \Psi \right\Vert ^{2}}\right)
\left( 1-\frac{\left\langle r\left( X\right) \right\rangle }{\left\langle
f_{1}\right\rangle }\right) +\left( 1-\left\langle S\left( X,X\right)
\right\rangle \frac{\left\langle \hat{K}\right\rangle \left\Vert \hat{\Psi}%
\right\Vert ^{2}}{\left\langle K\right\rangle \left\Vert \Psi \right\Vert
^{2}}\right) \frac{\left\langle r\left( X\right) \right\rangle }{%
\left\langle f_{1}\right\rangle }\right) ^{2}  \notag \\
&&\times \left\langle f_{1}\right\rangle ^{2}\left( \frac{\left( \frac{%
\left\Vert \hat{\Psi}_{0}\right\Vert ^{2}}{\hat{\mu}}+\left( \frac{1-r}{r}%
\right) \frac{\left\langle \hat{f}\right\rangle }{2\left( 1-\hat{S}\right) }%
\frac{\hat{S}}{1-\hat{S}}\right) }{\left( 1-2r\left( 1-r\right) \frac{\hat{S}%
}{1-\hat{S}}\right) }\right) ^{2}\left( 1-2\frac{2+\frac{\hat{S}}{1-\hat{S}}-%
\sqrt{\left( 2+\frac{\hat{S}}{1-\hat{S}}\right) ^{2}-\frac{\hat{S}}{1-\hat{S}%
}}}{9}\right)  \notag
\end{eqnarray}%
Defining:%
\begin{eqnarray*}
A &=&\frac{2\hat{\mu}V\sigma _{\hat{K}}^{4}\left( 1-\hat{S}\right)
^{2}\left\langle f_{1}\right\rangle ^{2}}{\left\langle \hat{f}\right\rangle
^{2}\epsilon \sigma _{\hat{K}}^{2}\left( \frac{\left\vert \Psi _{0}\left(
X\right) \right\vert ^{2}}{\epsilon }\right) ^{2}}\left( \frac{\left( \frac{%
\left\Vert \hat{\Psi}_{0}\right\Vert ^{2}}{\hat{\mu}}+\left( \frac{1-r}{r}%
\right) \frac{\left\langle \hat{f}\right\rangle }{2\left( 1-\hat{S}\right) }%
\frac{\hat{S}}{1-\hat{S}}\right) }{\left( 1-2r\left( 1-r\right) \frac{\hat{S}%
}{1-\hat{S}}\right) }\right) ^{2}\left( 1-2\frac{2+\frac{\hat{S}}{1-\hat{S}}-%
\sqrt{\left( 2+\frac{\hat{S}}{1-\hat{S}}\right) ^{2}-\frac{\hat{S}}{1-\hat{S}%
}}}{9}\right) \\
&\simeq &\frac{2\hat{\mu}V\sigma _{\hat{K}}^{4}\left( 1-\hat{S}\right)
^{2}\left\langle f_{1}\right\rangle ^{2}}{\left\langle \hat{f}\right\rangle
^{2}\epsilon \sigma _{\hat{K}}^{2}\left( \frac{\left\vert \Psi _{0}\left(
X\right) \right\vert ^{2}}{\epsilon }\right) ^{2}}\left( \frac{\left( \frac{%
\left\Vert \hat{\Psi}_{0}\right\Vert ^{2}}{\hat{\mu}}+\left( \frac{1-r}{r}%
\right) \frac{\left\langle \hat{f}\right\rangle }{2\left( 1-\hat{S}\right) }%
\frac{\hat{S}}{1-\hat{S}}\right) }{\left( 1-2r\left( 1-r\right) \frac{\hat{S}%
}{1-\hat{S}}\right) }\right) ^{2} \\
&\simeq &\frac{2\hat{\mu}\epsilon V\sigma _{\hat{K}}^{2}\left( 1-\hat{S}%
\right) ^{2}\left\langle f_{1}\right\rangle ^{2}}{\left\langle \hat{f}%
\right\rangle ^{2}}\left( \frac{\left\Vert \hat{\Psi}_{0}\right\Vert ^{2}}{%
\left\vert \Psi _{0}\left( X\right) \right\vert ^{2}}\right) ^{2}
\end{eqnarray*}%
and considering the case $r=\left\langle S_{E}\left( X,X\right)
\right\rangle <<1$ and $s=\left\langle S\left( X,X\right) \right\rangle <<1$
the capital equation writes: 
\begin{equation*}
x^{3}+\left( \frac{1}{s^{2}}\left( 2Ar+1\right) -\frac{1}{3s^{4}}\left(
Ar^{2}+2s\right) ^{2}\right) x+\left( \frac{1}{3s^{4}}\left(
Ar^{2}+2s\right) \left( 2Ar+1\right) -\frac{2}{27s^{6}}\left(
Ar^{2}+2s\right) ^{3}-\frac{A}{s^{2}}\right) =0
\end{equation*}%
This equation has one feasible solution:$\allowbreak $%
\begin{eqnarray}
&&\frac{\left\langle \hat{K}\right\rangle \left\Vert \hat{\Psi}\right\Vert
^{2}}{\left\langle K\right\rangle \left\Vert \Psi \right\Vert ^{2}}
\label{RTC} \\
&=&2\sqrt{\frac{\left( B\left( X\right) \right) ^{2}-3\left\langle S\left(
X,X\right) \right\rangle ^{2}\left( 2A\left\langle \tilde{S}_{E}\left(
X,X\right) \right\rangle +1\right) }{9\left\langle S\left( X,X\right)
\right\rangle ^{4}}}  \notag \\
&&\cosh \left( \frac{1}{3}\func{arccosh}\left( \frac{\left( \frac{B\left(
X\right) }{3}\left( \left( 2A\left\langle \tilde{S}_{E}\left( X,X\right)
\right\rangle +1\right) -\frac{2\left( B\left( X\right) \right) ^{2}}{%
9\left\langle \tilde{S}_{E}\left( X,X\right) \right\rangle ^{2}}\right)
-A\left\langle \tilde{S}_{E}\left( X,X\right) \right\rangle ^{2}\right) }{%
2\left\langle \tilde{S}_{E}\left( X,X\right) \right\rangle ^{4}\left( \frac{%
\left( B\left( X\right) \right) ^{2}-3\left\langle S\left( X,X\right)
\right\rangle ^{2}\left( 2A\left\langle \tilde{S}_{E}\left( X,X\right)
\right\rangle +1\right) }{9\left\langle S\left( X,X\right) \right\rangle ^{4}%
}\right) ^{\frac{3}{2}}}\right) \right)  \notag \\
&&+\frac{B\left( X\right) }{3\left\langle S\left( X,X\right) \right\rangle
^{2}}  \notag
\end{eqnarray}%
where:%
\begin{equation*}
\tilde{S}_{E}\left( X,X\right) =\left\langle \tilde{S}_{E}\left( X,X\right)
\right\rangle \frac{\left\langle \hat{K}\right\rangle \left\Vert \hat{\Psi}%
\right\Vert ^{2}}{\left\langle K\right\rangle \left\Vert \Psi \right\Vert
^{2}}\left( 1-\frac{\left\langle r\left( X\right) \right\rangle }{%
\left\langle f_{1}\right\rangle }\right) +\left\langle S\left( X,X\right)
\right\rangle \frac{\left\langle \hat{K}\right\rangle \left\Vert \hat{\Psi}%
\right\Vert ^{2}}{\left\langle K\right\rangle \left\Vert \Psi \right\Vert
^{2}}\frac{\left\langle r\left( X\right) \right\rangle }{\left\langle
f_{1}\right\rangle }
\end{equation*}%
\begin{equation*}
B\left( X\right) =\left( A\left\langle \tilde{S}_{E}\left( X,X\right)
\right\rangle ^{2}+2\left\langle S\left( X,X\right) \right\rangle \right)
\end{equation*}%
We can find an simpler expression for $\frac{\left\langle \hat{K}%
\right\rangle \left\Vert \hat{\Psi}\right\Vert ^{2}}{\left\langle
K\right\rangle \left\Vert \Psi \right\Vert ^{2}}$ if we consider in first
approximation $x\simeq A$, so that capital equation becomes:%
\begin{eqnarray*}
x &=&A\frac{\left( 1-rx\right) ^{2}}{\left( 1-sx\right) ^{2}} \\
&\simeq &A\frac{\left( 1-rA\right) ^{2}}{\left( 1-sA\right) ^{2}}=\frac{2%
\hat{\mu}\epsilon V\sigma _{\hat{K}}^{2}\left( 1-\left\langle \hat{S}%
\right\rangle \right) ^{2}\left\langle f_{1}\right\rangle ^{2}}{\left\langle 
\hat{f}\right\rangle ^{2}}\left( \frac{\left\Vert \hat{\Psi}_{0}\right\Vert
^{2}}{\left\Vert \Psi _{0}\left( X\right) \right\Vert ^{2}}\right) ^{2}\frac{%
\left( 1-r\frac{2\hat{\mu}\epsilon V\sigma _{\hat{K}}^{2}\left(
1-\left\langle \hat{S}\right\rangle \right) ^{2}\left\langle
f_{1}\right\rangle ^{2}}{\left\langle \hat{f}\right\rangle ^{2}}\left( \frac{%
\left\Vert \hat{\Psi}_{0}\right\Vert ^{2}}{\left\Vert \Psi _{0}\left(
X\right) \right\Vert ^{2}}\right) ^{2}\right) ^{2}}{\left( 1-s\frac{2\hat{\mu%
}\epsilon V\sigma _{\hat{K}}^{2}\left( 1-\left\langle \hat{S}\right\rangle
\right) ^{2}\left\langle f_{1}\right\rangle ^{2}}{\left\langle \hat{f}%
\right\rangle ^{2}}\left( \frac{\left\Vert \hat{\Psi}_{0}\right\Vert ^{2}}{%
\left\Vert \Psi _{0}\left( X\right) \right\Vert ^{2}}\right) ^{2}\right) ^{2}%
}
\end{eqnarray*}%
with solution:%
\begin{eqnarray}
&&\frac{\left\langle \hat{K}\right\rangle \left\Vert \hat{\Psi}\right\Vert
^{2}}{\left\langle K\right\rangle \left\Vert \Psi \right\Vert ^{2}}
\label{Qtk} \\
&\rightarrow &\frac{2\hat{\mu}\epsilon V\sigma _{\hat{K}}^{2}\left(
1-\left\langle \hat{S}\right\rangle \right) ^{2}\left\langle
f_{1}\right\rangle ^{2}}{\left\langle \hat{f}\right\rangle ^{2}}\left( \frac{%
\left\Vert \hat{\Psi}_{0}\right\Vert ^{2}}{\left\Vert \Psi _{0}\left(
X\right) \right\Vert ^{2}}\right) ^{2}  \notag \\
&&\times \left[ \frac{\left( 1-\left\langle \tilde{S}_{E}\left( X,X\right)
\right\rangle \frac{2\hat{\mu}\epsilon V\sigma _{\hat{K}}^{2}\left(
1-\left\langle \hat{S}\right\rangle \right) ^{2}\left\langle
f_{1}\right\rangle ^{2}}{\left\langle \hat{f}\right\rangle ^{2}}\left( \frac{%
\left\Vert \hat{\Psi}_{0}\right\Vert ^{2}}{\left\Vert \Psi _{0}\left(
X\right) \right\Vert ^{2}}\right) ^{2}\right) ^{2}}{\left( 1-\left\langle
S\left( X,X\right) \right\rangle \frac{2\hat{\mu}\epsilon V\sigma _{\hat{K}%
}^{2}\left( 1-\left\langle \hat{S}\right\rangle \right) ^{2}\left\langle
f_{1}\right\rangle ^{2}}{\left\langle \hat{f}\right\rangle ^{2}}\left( \frac{%
\left\Vert \hat{\Psi}_{0}\right\Vert ^{2}}{\left\Vert \Psi _{0}\left(
X\right) \right\Vert ^{2}}\right) ^{2}\right) ^{2}}+\frac{\left\langle 
\tilde{S}_{E}\left( X,X\right) \right\rangle ^{2}+2\left\langle S\left(
X,X\right) \right\rangle }{3\left\langle S\left( X,X\right) \right\rangle
^{2}}\right]  \notag
\end{eqnarray}%
where we used the notation:%
\begin{equation*}
\left\langle \hat{S}_{E}\right\rangle =\left\langle \hat{S}_{E}\left(
X^{\prime },X\right) \right\rangle
\end{equation*}

\subsubsection*{A4.4.1 Particular case}

A further simplifcation arises in the case of low returns $\frac{%
\left\langle r\left( X\right) \right\rangle }{\left\langle
f_{1}\right\rangle }\simeq 1$ and : $1-\frac{\left\langle r\left( X\right)
\right\rangle }{\left\langle f_{1}\right\rangle }<<\frac{\left\langle
r\left( X\right) \right\rangle }{\left\langle f_{1}\right\rangle }$ Equation
(\ref{Kq}) writes:

\begin{eqnarray*}
0 &=&\frac{\left\langle \hat{K}\right\rangle \left\Vert \hat{\Psi}%
\right\Vert ^{2}}{\left\langle K\right\rangle \left\Vert \Psi \right\Vert
^{2}}-\frac{6\hat{\mu}V\sigma _{\hat{K}}^{4}\left( 1-\hat{S}\right) ^{2}}{%
\left\langle \hat{f}\right\rangle ^{2}\epsilon \sqrt{\sigma _{\hat{K}}^{2}%
\frac{\left\vert \Psi _{0}\left( X\right) \right\vert ^{2}}{\epsilon }}3X^{3}%
} \\
&&\times \left\langle r\left( X\right) \right\rangle ^{2}\left( \frac{\left( 
\frac{\left\Vert \hat{\Psi}_{0}\right\Vert ^{2}}{\hat{\mu}}+\left( \frac{1-r%
}{r}\right) \frac{\left\langle \hat{f}\right\rangle }{2\left( 1-\hat{S}%
\right) }\frac{\hat{S}}{1-\hat{S}}\right) }{\left( 1-2r\left( 1-r\right) 
\frac{\hat{S}}{1-\hat{S}}\right) }\right) ^{2}\left( 1-2\frac{2+\frac{\hat{S}%
}{1-\hat{S}}-\sqrt{\left( 2+\frac{\hat{S}}{1-\hat{S}}\right) ^{2}-\frac{\hat{%
S}}{1-\hat{S}}}}{9}\right)
\end{eqnarray*}%
and the solution is:%
\begin{eqnarray*}
\frac{\left\langle \hat{K}\right\rangle \left\Vert \hat{\Psi}\right\Vert ^{2}%
}{\left\langle K\right\rangle \left\Vert \Psi \right\Vert ^{2}} &=&\frac{6%
\hat{\mu}V\sigma _{\hat{K}}^{4}\left( 1-\hat{S}\right) ^{2}\left\langle
r\left( X\right) \right\rangle ^{2}}{\left\langle \hat{f}\right\rangle
^{2}\epsilon \sqrt{\sigma _{\hat{K}}^{2}\frac{\left\vert \Psi _{0}\left(
X\right) \right\vert ^{2}}{\epsilon }}3X^{3}}\left( \frac{\left( \frac{%
\left\Vert \hat{\Psi}_{0}\right\Vert ^{2}}{\hat{\mu}}+\left( \frac{1-r}{r}%
\right) \frac{\left\langle \hat{f}\right\rangle }{2\left( 1-\hat{S}\right) }%
\frac{\hat{S}}{1-\hat{S}}\right) }{\left( 1-2r\left( 1-r\right) \frac{\hat{S}%
}{1-\hat{S}}\right) }\right) ^{2} \\
&&\times \left( 1-2\frac{2+\frac{\hat{S}}{1-\hat{S}}-\sqrt{\left( 2+\frac{%
\hat{S}}{1-\hat{S}}\right) ^{2}-\frac{\hat{S}}{1-\hat{S}}}}{9}\right) \\
&\simeq &\frac{6\hat{\mu}V\sigma _{\hat{K}}^{4}\left( 1-\hat{S}\right)
^{2}\left\langle r\left( X\right) \right\rangle ^{2}}{\left\langle \hat{f}%
\right\rangle ^{2}\epsilon \sqrt{\sigma _{\hat{K}}^{2}\frac{\left\vert \Psi
_{0}\left( X\right) \right\vert ^{2}}{\epsilon }}3X^{3}}\left( \left( \frac{%
\left\Vert \hat{\Psi}_{0}\right\Vert ^{2}}{\hat{\mu}}+\left( \frac{1-r}{r}%
\right) \frac{\left\langle \hat{f}\right\rangle }{2\left( 1-\hat{S}\right) }%
\frac{\hat{S}}{1-\hat{S}}\right) \right) ^{2}
\end{eqnarray*}%
with:%
\begin{equation*}
\left\langle \hat{S}_{E}\right\rangle =\left\langle \hat{S}_{E}\left(
X^{\prime },X\right) \right\rangle
\end{equation*}

\subsection*{A4.5 Formula for $\left\langle S_{E}\left( X\right)
\right\rangle $ and $\left\langle S\left( X\right) \right\rangle $}

Given the definition of $\left\langle S_{E}\left( X\right) \right\rangle $
and $\left\langle S\left( X\right) \right\rangle $, we have:%
\begin{eqnarray*}
\left\langle S_{E}\left( X\right) \right\rangle &=&\left\langle S_{E}\left(
X,X\right) \right\rangle \frac{\left\langle \hat{K}\right\rangle \left\Vert 
\hat{\Psi}\right\Vert ^{2}}{\left\langle K\right\rangle \left\Vert \Psi
\right\Vert ^{2}} \\
\left\langle S\left( X\right) \right\rangle &=&\left\langle S\left(
X,X\right) \right\rangle \frac{\left\langle \hat{K}\right\rangle \left\Vert 
\hat{\Psi}\right\Vert ^{2}}{\left\langle K\right\rangle \left\Vert \Psi
\right\Vert ^{2}}
\end{eqnarray*}

\subsection*{A4.6 Shares as functions of returns}

For the sequel, we derive the formula for the shares in function of return.
We set:

\begin{equation*}
x=\left\langle \hat{S}_{E}\left( X\right) \right\rangle
\end{equation*}%
\begin{eqnarray}
C &=&\frac{\left\langle \hat{f}\left( X^{\prime }\right) \right\rangle
-\left\langle \hat{r}\left( X^{\prime }\right) \right\rangle _{\hat{w}_{L}}}{%
2}  \label{Cf} \\
D &=&\left\langle \hat{f}\left( X^{\prime }\right) \right\rangle -\frac{%
\left\langle f\left( X\right) +r\left( X\right) \right\rangle _{w}}{2} 
\notag
\end{eqnarray}%
and the relation between shares and return writes:%
\begin{equation*}
x-\frac{\frac{1-\left( \gamma x\right) ^{2}}{2-\left( \gamma x\right) ^{2}}}{%
2}-\frac{\frac{1-\left( \gamma x\right) ^{2}}{2-\left( \gamma x\right) ^{2}}%
}{2}\left( \frac{1-\left( \gamma x\right) ^{2}}{2-\left( \gamma x\right) ^{2}%
}C+\frac{1}{2-\left( \gamma x\right) ^{2}}D\right)
\end{equation*}%
Considering that $\gamma x<1$, we can neglect the terms of order $4$ and $5$%
, and the equation reduces to:$\allowbreak $%
\begin{equation*}
x^{3}-\frac{1}{8}\left( 2C+D+3\right) x^{2}-\frac{1}{\gamma ^{2}}x+\frac{%
C+D+2}{8\gamma ^{2}}
\end{equation*}%
written as:%
\begin{equation*}
x^{3}-\frac{1}{8}Vx^{2}-\frac{1}{\gamma ^{2}}x+\frac{R}{8\gamma ^{2}}
\end{equation*}%
with $V=\left( 2C+D+3\right) $ and $R=C+D+2$. Changing variable $%
x\rightarrow x+\frac{1}{24}V$ leads to: 
\begin{equation*}
x^{3}-\left( \frac{1}{\gamma ^{2}}+\frac{1}{192}V^{2}\right) x+\left( \frac{1%
}{8}\frac{R}{\gamma ^{2}}-\frac{1}{24}\frac{V}{\gamma ^{2}}-\frac{1}{6912}%
V^{3}\right) =0
\end{equation*}%
with solution: 
\begin{eqnarray}
\left\langle \hat{S}_{E}\left( X\right) \right\rangle &=&2\sqrt{\frac{1}{3}%
\left( \frac{1}{\gamma ^{2}}+\frac{1}{192}V^{2}\right) }\cos \left( \frac{%
\arccos \left( -\frac{3\sqrt{3}\left( \frac{1}{8}\frac{R}{\gamma ^{2}}-\frac{%
1}{24}\frac{V}{\gamma ^{2}}-\frac{1}{6912}V^{3}\right) }{2\left( \frac{1}{%
\gamma ^{2}}+\frac{1}{192}V^{2}\right) ^{\frac{3}{2}}}\right) -2\pi }{3}%
\right) +\frac{3+2C+D}{24}  \label{Snh} \\
&\rightarrow &2\sqrt{\frac{1}{3}\left( \frac{1}{\gamma ^{2}}+\frac{\left(
\left( 3+2C+D\right) \right) ^{2}}{192}\right) }\cos \left( \frac{\arccos
\left( -\frac{\sqrt{3}\left( \frac{C+2D+3}{\gamma ^{2}}-\frac{\left(
3+2C+D\right) ^{3}}{288}\right) }{16\left( \frac{1}{\gamma ^{2}}+\frac{%
\left( 3+2C+D\right) ^{2}}{192}\right) ^{\frac{3}{2}}}\right) -2\pi }{3}%
\right) +\frac{3+2C+D}{24}  \notag
\end{eqnarray}

\section*{Appendix 5 Solving average return equation}

Once average shares equations have been found, we can consider the return
equation in the no default scenario:%
\begin{equation}
\left( \delta \left( X-X^{\prime }\right) -\hat{S}_{E}\left( X^{\prime
},X\right) \right) \frac{\left( 1-\hat{S}\left( X^{\prime }\right) \right)
\left( f\left( X^{\prime }\right) -\bar{r}\right) }{1-\hat{S}_{E}\left(
X^{\prime }\right) }=S_{E}\left( X,X\right) \frac{\left( 1-S\left( X\right)
\right) \left( f_{1}^{\prime }\left( X\right) -\bar{r}\right) }{%
1-S_{E}\left( X\right) }  \label{NDS}
\end{equation}

\subsection*{A5.1 Form of the equation}

Using that:%
\begin{equation*}
\frac{1}{1+\underline{k}\left( X^{\prime }\right) }=1-S\left( X^{\prime
}\right)
\end{equation*}%
\begin{equation*}
\underline{k}\left( X^{\prime }\right) =\frac{S\left( X^{\prime }\right) }{%
1-S\left( X^{\prime }\right) }
\end{equation*}%
\begin{equation*}
\underline{k}_{L}\left( X^{\prime }\right) =S_{L}\left( X^{\prime }\right) 
\frac{\underline{k}\left( X^{\prime }\right) }{S\left( X^{\prime }\right) }=%
\frac{S_{L}\left( X^{\prime }\right) }{1-S\left( X^{\prime }\right) },1+%
\underline{k}_{L}\left( X^{\prime }\right) =\frac{1-S_{E}\left( X^{\prime
}\right) }{1-S\left( X^{\prime }\right) }
\end{equation*}%
and replacing the fixed cost by its normalized value compared to disposable
capital:%
\begin{equation*}
C\rightarrow \frac{C}{1+\underline{k}\left( X^{\prime }\right) }=\left(
1-S\left( X^{\prime }\right) \right) C
\end{equation*}%
the firms return term in equation (\ref{NDS}) becomes:%
\begin{equation*}
\frac{\left( 1-S\left( X\right) \right) \left( f_{1}^{\prime }\left(
X\right) -\bar{r}\right) }{1-S_{E}\left( X\right) }=f_{1}\left( X\right)
-\left( 1-S\left( X\right) \right) \frac{C}{K_{X}}-\bar{r}
\end{equation*}%
As a consequence, the average of return equation (\ref{NDS}) writes:%
\begin{equation*}
\left( \delta \left( X-X^{\prime }\right) -\hat{S}_{E}\left( X^{\prime
},X\right) \right) \frac{\left( 1-\hat{S}\left( X^{\prime }\right) \right)
\left( f\left( X^{\prime }\right) -\bar{r}\right) }{1-\hat{S}_{E}\left(
X^{\prime }\right) }=S_{E}\left( X,X\right) \left[ \left( 1-S\left( X\right)
\right) ^{r}f_{1}\left( X\right) -\left( 1-S\left( X\right) \right) \frac{C}{%
K_{X}}-C_{0}-\bar{r}\right]
\end{equation*}%
In average, this equation simplifies as:%
\begin{equation}
\left( 1-\left\langle \hat{S}\left( X^{\prime }\right) \right\rangle \right)
\left( \left\langle \hat{f}\left( X^{\prime }\right) \right\rangle -\bar{r}%
\right) \simeq \left\langle S_{E}\left( X,X\right) \right\rangle \left[
\left( 1-S\left( X\right) \right) ^{r}f_{1}\left( X\right) -C_{0}-\bar{r}%
\right]  \label{Qrd}
\end{equation}%
where:%
\begin{equation*}
\left\langle \hat{S}\left( X^{\prime }\right) \right\rangle =2\left\langle 
\hat{S}_{E}\left( X^{\prime }\right) \right\rangle +\left\langle \hat{w}%
\left( X^{\prime }\right) \right\rangle \frac{\left\langle \hat{r}\left(
X^{\prime }\right) \right\rangle -\left\langle \hat{f}\left( X^{\prime
}\right) \right\rangle }{2}
\end{equation*}

\bigskip To solve this equation, we have to express $1-\left\langle \hat{S}%
\left( X^{\prime }\right) \right\rangle $ and $\left( 1-S\left( X\right)
\right) $ in terms of the model variables. Given (\ref{Fsk}):%
\begin{equation}
\left\langle \hat{f}\left( X^{\prime }\right) \right\rangle -\left\langle 
\hat{r}\left( X^{\prime }\right) \right\rangle _{\hat{w}_{L}}=\frac{\frac{%
2\left( 2-\left( \gamma \left\langle \hat{S}_{E}\left( X\right)
\right\rangle \right) ^{2}\right) }{1-\left( \gamma \left\langle \hat{S}%
_{E}\left( X\right) \right\rangle \right) ^{2}}\left\langle \hat{S}%
_{E}\left( X\right) \right\rangle -1+\frac{\left\langle f\left( X\right)
\right\rangle -\left\langle \hat{r}\left( X^{\prime }\right) \right\rangle _{%
\hat{w}_{L}}}{2\left( 2-\left( \gamma \left\langle \hat{S}_{E}\left(
X\right) \right\rangle \right) ^{2}\right) }}{\left( 1-\frac{1}{2}\frac{%
1-\left( \gamma \left\langle \hat{S}_{E}\left( X\right) \right\rangle
\right) ^{2}}{2-\left( \gamma \left\langle \hat{S}_{E}\left( X\right)
\right\rangle \right) ^{2}}\right) }  \label{Fsl}
\end{equation}%
we fnd:%
\begin{eqnarray}
\left\langle \hat{S}\left( X^{\prime }\right) \right\rangle &=&2\left\langle 
\hat{S}_{E}\left( X^{\prime }\right) \right\rangle -\frac{1-\left( \gamma
\left\langle \hat{S}_{E}\left( X\right) \right\rangle \right) ^{2}}{3-\left(
\gamma \left\langle \hat{S}_{E}\left( X\right) \right\rangle \right) ^{2}}
\label{SHV} \\
&&\times \left( \frac{2\left( 2-\left( \gamma \left\langle \hat{S}_{E}\left(
X\right) \right\rangle \right) ^{2}\right) }{1-\left( \gamma \left\langle 
\hat{S}_{E}\left( X\right) \right\rangle \right) ^{2}}\left\langle \hat{S}%
_{E}\left( X\right) \right\rangle -1+\frac{\left\langle f\left( X\right)
\right\rangle -\left\langle \hat{r}\left( X^{\prime }\right) \right\rangle _{%
\hat{w}_{L}}}{2\left( 2-\left( \gamma \left\langle \hat{S}_{E}\left(
X\right) \right\rangle \right) ^{2}\right) }\right)  \notag
\end{eqnarray}%
where:%
\begin{equation*}
1-\left\langle S\left( X\right) \right\rangle =1-\frac{\left\langle w\left(
X\right) \right\rangle \left( 1+\left( \hat{w}\left( X\right) \left( \frac{%
\left\langle f_{1}\left( X\right) \right\rangle +\left\langle \bar{r}\left(
X\right) \right\rangle }{2}-\frac{\left\langle \hat{f}\left( X^{\prime
}\right) \right\rangle _{\hat{w}_{E}}+\left\langle \hat{r}\left( X^{\prime
}\right) \right\rangle _{\hat{w}_{L}}}{2}\right) \right) \right) }{\left( 1-%
\frac{\left\langle w\left( X\right) \right\rangle \hat{w}\left( X\right) C}{2%
}\frac{\left\langle \hat{K}\right\rangle \left\Vert \hat{\Psi}\right\Vert
^{2}}{\left\langle K\right\rangle \left\Vert \Psi \right\Vert ^{2}}\right) }%
\frac{\left\langle \hat{K}\right\rangle \left\Vert \hat{\Psi}\right\Vert ^{2}%
}{\left\langle K\right\rangle \left\Vert \Psi \right\Vert ^{2}}
\end{equation*}%
\begin{equation*}
\frac{\left\langle \hat{K}\right\rangle \left\Vert \hat{\Psi}\right\Vert ^{2}%
}{\left\langle K\right\rangle \left\Vert \Psi \right\Vert ^{2}}\simeq \frac{6%
\hat{\mu}V\sigma _{\hat{K}}^{4}\left( 1-\hat{S}\right) ^{2}\left\langle
r\left( X\right) \right\rangle ^{2}}{\left\langle \hat{f}\right\rangle
^{2}\epsilon \sqrt{\sigma _{\hat{K}}^{2}\frac{\left\vert \Psi _{0}\left(
X\right) \right\vert ^{2}}{\epsilon }}3X^{3}}\left( \left( \frac{\left\Vert 
\hat{\Psi}_{0}\right\Vert ^{2}}{\hat{\mu}}+\left( \frac{1-\left\langle
r\left( X\right) \right\rangle }{\left\langle r\left( X\right) \right\rangle 
}\right) \frac{\left\langle \hat{f}\right\rangle }{2\left( 1-\hat{S}\right) }%
\frac{\hat{S}}{1-\hat{S}}\right) \right) ^{2}
\end{equation*}%
We first solve the simplest case of constant return to scale in which firms
returns are independent on the shares of investments. This simplifies the
resolution, since in this case, the average shares can be directly replaced
as functions of average returns.

\subsection*{A5.2 Neglecting $C$ and constant return to scale}

\subsubsection*{A5.2.1 Equation in terms of shares}

In this case, the return equation reduces to:

\begin{equation}
\left( 1-\left\langle \hat{S}\left( X^{\prime }\right) \right\rangle \right)
\left( \left\langle \hat{f}\left( X^{\prime }\right) \right\rangle -\bar{r}%
\right) \simeq \left\langle S_{E}\left( X,X\right) \right\rangle \left[
\left\langle f_{1}\left( X\right) \right\rangle -\left\langle \bar{r}\left(
X\right) \right\rangle \right]  \label{sq}
\end{equation}%
and:%
\begin{equation}
\frac{\left\langle f\left( X\right) \right\rangle +\left\langle \bar{r}%
\left( X\right) \right\rangle }{2}\rightarrow \frac{\left\langle f_{1}\left(
X\right) \right\rangle +\left\langle \bar{r}\left( X\right) \right\rangle }{2%
}
\end{equation}%
\begin{eqnarray*}
&&\left\langle S_{E}\left( X,X\right) \right\rangle \\
&=&\frac{w\left( X\right) }{2}\left( 1+\left( \left\langle \hat{w}\left(
X\right) \right\rangle \left( \left\langle f_{1}\left( X\right)
\right\rangle -\frac{\left\langle \hat{f}\left( X^{\prime }\right)
\right\rangle _{\hat{w}_{E}}+\left\langle \hat{r}\left( X^{\prime }\right)
\right\rangle _{\hat{w}_{L}}}{2}\right) +\frac{w\left( X\right) }{2}\left(
\left\langle f_{1}\left( X\right) \right\rangle -\bar{r}\left( X\right)
\right) \right) \right)
\end{eqnarray*}%
The left hand side of (\ref{sq}) is (given (\ref{SHV}) and (\ref{Fsl}):%
\begin{eqnarray}
\left\langle \hat{S}\left( X^{\prime }\right) \right\rangle &=&2\left\langle 
\hat{S}_{E}\left( X^{\prime }\right) \right\rangle -\frac{1-\left( \gamma
\left\langle \hat{S}_{E}\left( X\right) \right\rangle \right) ^{2}}{3-\left(
\gamma \left\langle \hat{S}_{E}\left( X\right) \right\rangle \right) ^{2}} \\
&&\times \left( \frac{2\left( 2-\left( \gamma \left\langle \hat{S}_{E}\left(
X\right) \right\rangle \right) ^{2}\right) }{1-\left( \gamma \left\langle 
\hat{S}_{E}\left( X\right) \right\rangle \right) ^{2}}\left\langle \hat{S}%
_{E}\left( X\right) \right\rangle -1+\frac{\left\langle f\left( X\right)
\right\rangle -\left\langle \hat{r}\left( X^{\prime }\right) \right\rangle _{%
\hat{w}_{L}}}{2\left( 2-\left( \gamma \left\langle \hat{S}_{E}\left(
X\right) \right\rangle \right) ^{2}\right) }\right)  \notag
\end{eqnarray}%
\begin{eqnarray}
&&\left( 1-\left\langle \hat{S}\left( X^{\prime }\right) \right\rangle
\right) \left( \left\langle \hat{f}\left( X^{\prime }\right) \right\rangle -%
\bar{r}\right)  \label{lh} \\
&=&\left( 1-2\left\langle \hat{S}_{E}\left( X^{\prime }\right) \right\rangle
+\frac{1-\left( \gamma \left\langle \hat{S}_{E}\left( X\right) \right\rangle
\right) ^{2}}{3-\left( \gamma \left\langle \hat{S}_{E}\left( X\right)
\right\rangle \right) ^{2}}\right.  \notag \\
&&\left. \times \left( \frac{2\left( 2-\left( \gamma \left\langle \hat{S}%
_{E}\left( X\right) \right\rangle \right) ^{2}\right) }{1-\left( \gamma
\left\langle \hat{S}_{E}\left( X\right) \right\rangle \right) ^{2}}%
\left\langle \hat{S}_{E}\left( X\right) \right\rangle -1+\frac{\left\langle
f\left( X\right) \right\rangle -\left\langle \hat{r}\left( X^{\prime
}\right) \right\rangle _{\hat{w}_{L}}}{2\left( 2-\left( \gamma \left\langle 
\hat{S}_{E}\left( X\right) \right\rangle \right) ^{2}\right) }\right) \right)
\notag \\
&&\times \frac{\frac{2\left( 2-\left( \gamma \left\langle \hat{S}_{E}\left(
X\right) \right\rangle \right) ^{2}\right) }{1-\left( \gamma \left\langle 
\hat{S}_{E}\left( X\right) \right\rangle \right) ^{2}}\left\langle \hat{S}%
_{E}\left( X\right) \right\rangle -1+\frac{\left\langle f\left( X\right)
\right\rangle -\left\langle \hat{r}\left( X^{\prime }\right) \right\rangle _{%
\hat{w}_{L}}}{2\left( 2-\left( \gamma \left\langle \hat{S}_{E}\left(
X\right) \right\rangle \right) ^{2}\right) }}{\left( 1-\frac{1}{2}\frac{%
1-\left( \gamma \left\langle \hat{S}_{E}\left( X\right) \right\rangle
\right) ^{2}}{2-\left( \gamma \left\langle \hat{S}_{E}\left( X\right)
\right\rangle \right) ^{2}}\right) }  \notag
\end{eqnarray}%
while the right hand side is:%
\begin{eqnarray}
&&\left\langle S_{E}\left( X,X\right) \right\rangle \left[ \left\langle
f_{1}\left( X\right) \right\rangle -\left\langle \bar{r}\left( X\right)
\right\rangle \right]  \label{rh} \\
&=&\frac{w\left( X\right) }{2}\left( 1+\left( \left\langle \hat{w}\left(
X\right) \right\rangle \left( \left\langle f_{1}\left( X\right)
\right\rangle -\frac{\left\langle \hat{f}\left( X^{\prime }\right)
\right\rangle _{\hat{w}_{E}}+\left\langle \hat{r}\left( X^{\prime }\right)
\right\rangle _{\hat{w}_{L}}}{2}\right) \right. \right.  \notag \\
&&\left. \left. +\frac{w\left( X\right) }{2}\left( \left\langle f_{1}\left(
X\right) \right\rangle -\bar{r}\left( X\right) \right) \right) \right) \left[
\left\langle f_{1}\left( X\right) \right\rangle -\left\langle \bar{r}\left(
X\right) \right\rangle \right]  \notag \\
&=&\frac{w\left( X\right) }{2}\left[ \left\langle f_{1}\left( X\right)
\right\rangle -\left\langle \bar{r}\left( X\right) \right\rangle \right] 
\notag \\
&&\times \left( 1+\left( \left\langle \hat{w}\left( X\right) \right\rangle +%
\frac{w\left( X\right) }{2}\right) \left( \left\langle f_{1}\left( X\right)
\right\rangle -\bar{r}\left( X\right) \right) -\left\langle \hat{w}\left(
X\right) \right\rangle \left( \frac{\left\langle \hat{f}\left( X^{\prime
}\right) \right\rangle _{\hat{w}_{E}}-\left\langle \hat{r}\left( X^{\prime
}\right) \right\rangle _{\hat{w}_{L}}}{2}\right) \right)  \notag \\
&=&\frac{1}{2}\frac{1}{2-\left( \gamma \left\langle \hat{S}_{E}\left(
X\right) \right\rangle \right) ^{2}}\left( 1+\left( \frac{3-2\left( \gamma
\left\langle \hat{S}_{E}\left( X\right) \right\rangle \right) ^{2}}{2\left(
2-\left( \gamma \left\langle \hat{S}_{E}\left( X\right) \right\rangle
\right) ^{2}\right) }\right) \left( \left\langle f_{1}\left( X\right)
\right\rangle -\left\langle \hat{r}\left( X^{\prime }\right) \right\rangle _{%
\hat{w}_{L}}\right) \right.  \notag \\
&&\left. -\frac{\frac{2\left( 2-\left( \gamma \left\langle \hat{S}_{E}\left(
X\right) \right\rangle \right) ^{2}\right) }{1-\left( \gamma \left\langle 
\hat{S}_{E}\left( X\right) \right\rangle \right) ^{2}}\left\langle \hat{S}%
_{E}\left( X\right) \right\rangle -1+\frac{\left\langle f\left( X\right)
\right\rangle -\left\langle \hat{r}\left( X^{\prime }\right) \right\rangle _{%
\hat{w}_{L}}}{2\left( 2-\left( \gamma \left\langle \hat{S}_{E}\left(
X\right) \right\rangle \right) ^{2}\right) }}{2\left( 1-\frac{1}{2}\frac{%
1-\left( \gamma \left\langle \hat{S}_{E}\left( X\right) \right\rangle
\right) ^{2}}{2-\left( \gamma \left\langle \hat{S}_{E}\left( X\right)
\right\rangle \right) ^{2}}\right) }\right) \left[ \left\langle f_{1}\left(
X\right) \right\rangle -\left\langle \bar{r}\left( X\right) \right\rangle %
\right]  \notag
\end{eqnarray}%
Defining $z=\left\langle \hat{S}_{E}\left( X\right) \right\rangle $, this
equation writes:%
\begin{eqnarray}
&&\left( 1-2z+\frac{1-\left( \gamma z\right) ^{2}}{3-\left( \gamma z\right)
^{2}}\left( \frac{2\left( 2-\left( \gamma z\right) ^{2}\right) }{1-\left(
\gamma z\right) ^{2}}z-1+\frac{\left\langle f\left( X\right) \right\rangle
-\left\langle \hat{r}\left( X^{\prime }\right) \right\rangle _{\hat{w}_{L}}}{%
2\left( 2-\left( \gamma z\right) ^{2}\right) }\right) \right)  \label{zq} \\
&&\times \frac{\frac{2\left( 2-\left( \gamma z\right) ^{2}\right) }{1-\left(
\gamma z\right) ^{2}}z-1+\frac{\left\langle f\left( X\right) \right\rangle
-\left\langle \hat{r}\left( X^{\prime }\right) \right\rangle _{\hat{w}_{L}}}{%
2\left( 2-\left( \gamma z\right) ^{2}\right) }}{\left( 1-\frac{1}{2}\frac{%
1-\left( \gamma z\right) ^{2}}{2-\left( \gamma z\right) ^{2}}\right) } 
\notag \\
&=&\frac{1}{2}\frac{1}{2-\left( \gamma z\right) ^{2}}\left( 1+\left( \frac{%
3-\left( \gamma z\right) ^{2}}{2\left( 2-\left( \gamma z\right) ^{2}\right) }%
\right) \left( \left\langle f_{1}\left( X\right) \right\rangle -\left\langle 
\hat{r}\left( X^{\prime }\right) \right\rangle _{\hat{w}_{L}}\right) \right.
\notag \\
&&\left. -\frac{\frac{2\left( 2-\left( \gamma z\right) ^{2}\right) }{%
1-\left( \gamma z\right) ^{2}}z-1+\frac{\left\langle f\left( X\right)
\right\rangle -\left\langle \hat{r}\left( X^{\prime }\right) \right\rangle _{%
\hat{w}_{L}}}{2\left( 2-\left( \gamma z\right) ^{2}\right) }}{2\left( 1-%
\frac{1}{2}\frac{1-\left( \gamma z\right) ^{2}}{2-\left( \gamma z\right) ^{2}%
}\right) }\right) \left[ \left\langle f_{1}\left( X\right) \right\rangle
-\left\langle \bar{r}\left( X\right) \right\rangle \right]  \notag
\end{eqnarray}%
\begin{equation*}
-\frac{1-\left( \gamma \left\langle \hat{S}_{E}\left( X\right) \right\rangle
\right) ^{2}}{2-\left( \gamma \left\langle \hat{S}_{E}\left( X\right)
\right\rangle \right) ^{2}}\frac{\frac{\left\langle f\left( X\right)
\right\rangle -\left\langle \hat{r}\left( X^{\prime }\right) \right\rangle _{%
\hat{w}_{L}}}{2\left( 2-\left( \gamma \left\langle \hat{S}_{E}\left(
X\right) \right\rangle \right) ^{2}\right) }}{2\left( 1-\frac{1}{2}\frac{%
1-\left( \gamma \left\langle \hat{S}_{E}\left( X\right) \right\rangle
\right) ^{2}}{2-\left( \gamma \left\langle \hat{S}_{E}\left( X\right)
\right\rangle \right) ^{2}}\right) }
\end{equation*}%
This equation is defined as long as $1-\left( \gamma z\right) ^{2}>0$ and $%
\left\langle K\right\rangle \left\Vert \Psi \right\Vert ^{2}>0$ that is:%
\begin{equation*}
\frac{1-\left\langle S_{E}\left( X\right) \right\rangle }{1-\left\langle
S\left( X\right) \right\rangle }\left( \left\langle f_{1}\left( X\right)
\right\rangle -\left\langle r\left( X\right) \right\rangle \right)
+\left\langle r\left( X\right) \right\rangle >0
\end{equation*}%
as implied by (\ref{Cp}). Numerical studies shows that $\left\langle \hat{S}%
_{E}\left( X\right) \right\rangle $ decreases with $\gamma $ and increases
with $\left\langle f_{1}\left( X\right) \right\rangle $.

\subsubsection*{A5.2.3 Dependency of disposable capital in parameters}

The dependency of:%
\begin{equation*}
\frac{1-\left\langle S_{E}\left( X\right) \right\rangle }{1-\left\langle
S\left( X\right) \right\rangle }\left( \left\langle f_{1}\left( X\right)
\right\rangle -\left\langle r\left( X\right) \right\rangle \right)
+\left\langle r\left( X\right) \right\rangle
\end{equation*}%
in parameters is needed to obtain the variations in capital. It is obtained
by writing:%
\begin{eqnarray*}
&&\left\langle S_{E}\left( X,X\right) \right\rangle \\
&=&\frac{w\left( X\right) }{2}\left( 1+\left( \left\langle \hat{w}\left(
X\right) \right\rangle \left( \left\langle f\left( X\right) \right\rangle -%
\frac{\left\langle \hat{f}\left( X^{\prime }\right) \right\rangle _{\hat{w}%
_{E}}+\left\langle \hat{r}\left( X^{\prime }\right) \right\rangle _{\hat{w}%
_{L}}}{2}\right) +\frac{w\left( X\right) }{2}\left( f\left( X\right) -\bar{r}%
\left( X\right) \right) \right) \right)
\end{eqnarray*}%
\begin{equation}
\left\langle S\left( X,X\right) \right\rangle =\left\langle w\left( X\right)
\right\rangle \left( 1+\left( \hat{w}\left( X\right) \left( \frac{%
\left\langle f\left( X\right) \right\rangle +\left\langle \bar{r}\left(
X\right) \right\rangle }{2}-\frac{\left\langle \hat{f}\left( X^{\prime
}\right) \right\rangle _{\hat{w}_{E}}+\left\langle \hat{r}\left( X^{\prime
}\right) \right\rangle _{\hat{w}_{L}}}{2}\right) \right) \right)
\end{equation}%
\begin{eqnarray*}
&&\left\langle S\left( X,X\right) \right\rangle -2\left\langle S_{E}\left(
X,X\right) \right\rangle \\
&&\left\langle w\left( X\right) \right\rangle \hat{w}\left( X\right) \left( 
\frac{\left\langle f\left( X\right) \right\rangle +\left\langle \bar{r}%
\left( X\right) \right\rangle }{2}-\frac{\left\langle \hat{f}\left(
X^{\prime }\right) \right\rangle _{\hat{w}_{E}}+\left\langle \hat{r}\left(
X^{\prime }\right) \right\rangle _{\hat{w}_{L}}}{2}\right) \\
&&-w\left( X\right) \left\langle \hat{w}\left( X\right) \right\rangle \left(
\left\langle f\left( X\right) \right\rangle -\frac{\left\langle \hat{f}%
\left( X^{\prime }\right) \right\rangle _{\hat{w}_{E}}+\left\langle \hat{r}%
\left( X^{\prime }\right) \right\rangle _{\hat{w}_{L}}}{2}\right) -w\left(
X\right) \frac{w\left( X\right) }{2}\left( f\left( X\right) -\bar{r}\left(
X\right) \right)
\end{eqnarray*}%
\begin{equation*}
\left\langle S\left( X,X\right) \right\rangle -2\left\langle S_{E}\left(
X,X\right) \right\rangle =\left\langle w\left( X\right) \right\rangle \hat{w}%
\left( X\right) \left( \left\langle \bar{r}\left( X\right) \right\rangle
-\left\langle f\left( X\right) \right\rangle \right)
\end{equation*}%
with:%
\begin{equation*}
\left\langle \hat{w}\right\rangle \rightarrow \frac{\zeta ^{2}}{\zeta
^{2}+\zeta ^{2}\frac{1}{1-\left( \gamma \left\langle \hat{S}_{E}\left(
X\right) \right\rangle \right) ^{2}}}=\frac{1-\left( \gamma \left\langle 
\hat{S}_{E}\left( X\right) \right\rangle \right) ^{2}}{2-\left( \gamma
\left\langle \hat{S}_{E}\left( X\right) \right\rangle \right) ^{2}}
\end{equation*}%
and:%
\begin{equation*}
\left\langle w\right\rangle \rightarrow \frac{1}{2-\left( \gamma
\left\langle \hat{S}_{E}\left( X\right) \right\rangle \right) ^{2}}
\end{equation*}%
As a consequence:%
\begin{equation*}
\left\langle S\left( X,X\right) \right\rangle -2\left\langle S_{E}\left(
X,X\right) \right\rangle =\frac{1}{2-\left( \gamma \left\langle \hat{S}%
_{E}\left( X\right) \right\rangle \right) ^{2}}\frac{1-\left( \gamma
\left\langle \hat{S}_{E}\left( X\right) \right\rangle \right) ^{2}}{2-\left(
\gamma \left\langle \hat{S}_{E}\left( X\right) \right\rangle \right) ^{2}}%
\left( \left\langle \bar{r}\left( X\right) \right\rangle -\left\langle
f\left( X\right) \right\rangle \right)
\end{equation*}%
and we find:%
\begin{eqnarray*}
\frac{1-\left\langle S_{E}\left( X\right) \right\rangle }{1-\left\langle
S\left( X\right) \right\rangle } &=&\frac{1-\left\langle S_{E}\left(
X\right) \right\rangle }{1-\left( 2\left\langle S_{E}\left( X\right)
\right\rangle -\frac{1-\left( \gamma \left\langle \hat{S}_{E}\left( X\right)
\right\rangle \right) ^{2}}{\left( 2-\left( \gamma \left\langle \hat{S}%
_{E}\left( X\right) \right\rangle \right) ^{2}\right) ^{2}}\left(
\left\langle f\left( X\right) \right\rangle -\left\langle \bar{r}\left(
X\right) \right\rangle \right) \frac{\left\langle \hat{K}\right\rangle
\left\Vert \hat{\Psi}\right\Vert ^{2}}{\left\langle K\right\rangle
\left\Vert \Psi \right\Vert ^{2}}\right) } \\
&=&\frac{1-\left\langle S_{E}\left( X,X\right) \right\rangle \frac{%
\left\langle \hat{K}\right\rangle \left\Vert \hat{\Psi}\right\Vert ^{2}}{%
\left\langle K\right\rangle \left\Vert \Psi \right\Vert ^{2}}}{1-\left(
2\left\langle S_{E}\left( X,X\right) \right\rangle -\frac{1-\left( \gamma
\left\langle \hat{S}_{E}\left( X\right) \right\rangle \right) ^{2}}{\left(
2-\left( \gamma \left\langle \hat{S}_{E}\left( X\right) \right\rangle
\right) ^{2}\right) ^{2}}\left( \left\langle f\left( X\right) \right\rangle
-\left\langle \bar{r}\left( X\right) \right\rangle \right) \right) \frac{%
\left\langle \hat{K}\right\rangle \left\Vert \hat{\Psi}\right\Vert ^{2}}{%
\left\langle K\right\rangle \left\Vert \Psi \right\Vert ^{2}}}
\end{eqnarray*}%
Now, writing:%
\begin{eqnarray*}
\frac{\left\langle \hat{K}\right\rangle \left\Vert \hat{\Psi}\right\Vert ^{2}%
}{\left\langle K\right\rangle \left\Vert \Psi \right\Vert ^{2}} &=&\frac{%
\left\langle \hat{K}\right\rangle \left\Vert \hat{\Psi}\right\Vert ^{2}}{%
\frac{1}{12}\frac{\epsilon \sqrt{\sigma _{\hat{K}}^{2}\frac{\left\vert \Psi
_{0}\left( X\right) \right\vert ^{2}}{\epsilon }}}{\sigma _{\hat{K}%
}^{2}\left( \frac{1-\left\langle S_{E}\left( X\right) \right\rangle }{%
1-\left\langle S\left( X\right) \right\rangle }\left( f_{1}\left( X\right)
-r\left( X\right) \right) +r\left( X\right) \right) ^{2}}\left( 3X-C\left(
1-S\right) \right) \left( C\left( 1-S\right) +X\right) ^{2}} \\
&\equiv &\left( \frac{1-\left\langle S_{E}\left( X\right) \right\rangle }{%
1-\left\langle S\left( X\right) \right\rangle }\left( \left\langle f\left(
X\right) \right\rangle -\left\langle \bar{r}\left( X\right) \right\rangle
\right) +\left\langle \bar{r}\left( X\right) \right\rangle \right) ^{2}R
\end{eqnarray*}%
the quantity $\frac{1-\left\langle S_{E}\left( X\right) \right\rangle }{%
1-\left\langle S\left( X\right) \right\rangle }$ satifies:%
\begin{eqnarray*}
&&\frac{1-\left\langle S_{E}\left( X\right) \right\rangle }{1-\left\langle
S\left( X\right) \right\rangle } \\
&=&\frac{1-\left\langle S_{E}\left( X,X\right) \right\rangle \left( \frac{%
1-\left\langle S_{E}\left( X\right) \right\rangle }{1-\left\langle S\left(
X\right) \right\rangle }\left( \left\langle f\left( X\right) \right\rangle
-\left\langle \bar{r}\left( X\right) \right\rangle \right) +\left\langle 
\bar{r}\left( X\right) \right\rangle \right) ^{2}R}{1-\left( 2\left\langle
S_{E}\left( X,X\right) \right\rangle -\frac{1-\left( \gamma \left\langle 
\hat{S}_{E}\left( X\right) \right\rangle \right) ^{2}}{\left( 2-\left(
\gamma \left\langle \hat{S}_{E}\left( X\right) \right\rangle \right)
^{2}\right) ^{2}}\left( \left\langle f\left( X\right) \right\rangle
-\left\langle \bar{r}\left( X\right) \right\rangle \right) \right) \left( 
\frac{1-\left\langle S_{E}\left( X\right) \right\rangle }{1-\left\langle
S\left( X\right) \right\rangle }\left( \left\langle f\left( X\right)
\right\rangle -\left\langle \bar{r}\left( X\right) \right\rangle \right)
+\left\langle \bar{r}\left( X\right) \right\rangle \right) ^{2}R}
\end{eqnarray*}%
which is given by:%
\begin{equation*}
\frac{1-\left\langle S_{E}\left( X\right) \right\rangle }{1-\left\langle
S\left( X\right) \right\rangle }=\frac{1-\left\langle S_{E}\left( X,X\right)
\right\rangle \left( \left\langle \bar{r}\left( X\right) \right\rangle
\right) ^{2}R}{1-2\left\langle S_{E}\left( X,X\right) \right\rangle
\left\langle \bar{r}\left( X\right) \right\rangle ^{2}R}
\end{equation*}%
at the lowest order, and:%
\begin{eqnarray*}
&&\frac{1-\left\langle S_{E}\left( X\right) \right\rangle }{1-\left\langle
S\left( X\right) \right\rangle } \\
&=&\frac{1-\left\langle S_{E}\left( X,X\right) \right\rangle VR}{1-\left(
2\left\langle S_{E}\left( X,X\right) \right\rangle -\frac{1-\left( \gamma
\left\langle \hat{S}_{E}\left( X\right) \right\rangle \right) ^{2}}{\left(
2-\left( \gamma \left\langle \hat{S}_{E}\left( X\right) \right\rangle
\right) ^{2}\right) ^{2}}\left( \left\langle f\left( X\right) \right\rangle
-\left\langle \bar{r}\left( X\right) \right\rangle \right) \right) VR}
\end{eqnarray*}%
with:%
\begin{equation*}
V=\left( \frac{1-\left\langle S_{E}\left( X,X\right) \right\rangle r^{2}R}{%
1-2\left\langle S_{E}\left( X,X\right) \right\rangle r^{2}R}\left(
\left\langle f\left( X\right) \right\rangle -\left\langle \bar{r}\left(
X\right) \right\rangle \right) +\left\langle \bar{r}\left( X\right)
\right\rangle \right) ^{2}
\end{equation*}%
at the first order and:%
\begin{eqnarray*}
&&\frac{1-\left\langle S_{E}\left( X\right) \right\rangle }{1-\left\langle
S\left( X\right) \right\rangle }\left( \left\langle f_{1}\left( X\right)
\right\rangle -\left\langle r\left( X\right) \right\rangle \right)
+\left\langle r\left( X\right) \right\rangle \\
&=&\frac{1-\left\langle S_{E}\left( X,X\right) \right\rangle VR\left(
\left\langle f_{1}\left( X\right) \right\rangle -\left\langle r\left(
X\right) \right\rangle \right) +\left\langle r\left( X\right) \right\rangle 
}{1-\left( 2\left\langle S_{E}\left( X,X\right) \right\rangle -\frac{\left(
1-\left( \gamma \left\langle \hat{S}_{E}\left( X\right) \right\rangle
\right) ^{2}\right) \left( \left\langle f\left( X\right) \right\rangle
-\left\langle \bar{r}\left( X\right) \right\rangle \right) }{\left( 2-\left(
\gamma \left\langle \hat{S}_{E}\left( X\right) \right\rangle \right)
^{2}\right) ^{2}}\right) VR}
\end{eqnarray*}%
Quantity $\left\langle S_{E}\left( X,X\right) \right\rangle $ is evaluated
by (\ref{rh}) at the solution of return equation. Numerical studies show
that:%
\begin{equation*}
\frac{1-\left\langle S_{E}\left( X\right) \right\rangle }{1-\left\langle
S\left( X\right) \right\rangle }\left( \left\langle f_{1}\left( X\right)
\right\rangle -\left\langle r\left( X\right) \right\rangle \right)
+\left\langle r\left( X\right) \right\rangle
\end{equation*}%
is decreasing with $\left\langle f_{1}\left( X\right) \right\rangle $ and
increasing with $\gamma $ if $\left\langle f_{1}\left( X\right)
\right\rangle >r$. As a consequence, given (\ref{Cp}), (\ref{RtK}): 
\begin{equation*}
\left\langle K\right\rangle \left\Vert \Psi \right\Vert ^{2}=\frac{1}{12}%
\frac{\epsilon \sqrt{\sigma _{\hat{K}}^{2}\frac{\left\vert \Psi _{0}\left(
X\right) \right\vert ^{2}}{\epsilon }}}{\sigma _{\hat{K}}^{2}\left( \frac{%
1-\left\langle S_{E}\left( X\right) \right\rangle }{1-\left\langle S\left(
X\right) \right\rangle }\left( f_{1}\left( X\right) -r\left( X\right)
\right) +r\left( X\right) \right) ^{2}}\left( 3X-C\left( 1-S\right) \right)
\left( C\left( 1-S\right) +X\right) ^{2}
\end{equation*}%
is increasing with $f_{1}\left( X\right) $ and decreasing with $\gamma $ for 
$\left\langle f_{1}\left( X\right) \right\rangle >r$, while:%
\begin{equation*}
\frac{\left\langle \hat{K}\right\rangle \left\Vert \hat{\Psi}\right\Vert ^{2}%
}{\left\langle K\right\rangle \left\Vert \Psi \right\Vert ^{2}}
\end{equation*}%
is decreasing with $\left\langle f_{1}\left( X\right) \right\rangle $ and
increasing with $\gamma $ for $\left\langle f_{1}\left( X\right)
\right\rangle >r$. Investors invest more directly in their sectors which
reduces the level of disposable capital. Higher returns for firms allow for
more capital. Below the threshold returns do not offset the higher risk
linked to the investment, and reduces the level of capital. For an
increasing global uncertainty, investors have the tendency to diversify, so
that $\frac{\left\langle \hat{K}\right\rangle \left\Vert \hat{\Psi}%
\right\Vert ^{2}}{\left\langle K\right\rangle \left\Vert \Psi \right\Vert
^{2}}$ increases. More precise results will be obtained below with an
approximate resolution.

\subsection*{A5.3 Including $C$-corrections and decreasing return}

The case of decreasing return to scales modifies the previous resolution
since in this case, the returns depend on the shares, so that the average
shares equations have to be solved before using them in the return equation.
This is done in several steps.

\subsubsection*{A5.3.1 Principle of resolution}

Decreasing returns modifies the returns produced by firms:%
\begin{eqnarray}
&&K_{pX}\frac{f_{1}\left( X\right) \left( \left( 1+\frac{\underline{k}\left(
X\right) }{\left\langle K_{p}\right\rangle }\hat{K}_{X}\right) K_{pX}\right)
^{1-r}-C_{0}\left( 1+\frac{\underline{k}\left( X\right) }{\left\langle
K\right\rangle }\hat{K}_{X}\right) K_{pX}-C}{\left( 1+\frac{\underline{k}%
\left( X\right) }{\left\langle K\right\rangle }\hat{K}_{X}\right) K_{pX}}
\label{Dc} \\
&\rightarrow &\frac{f_{1}\left( X\right) \left( \left( 1+\frac{\underline{k}%
\left( X\right) }{\left\langle K_{p}\right\rangle }\hat{K}_{X}\right)
K_{pX}\right) ^{1-r}-C_{0}\left( 1+\frac{\underline{k}\left( X\right) }{%
\left\langle K\right\rangle }\hat{K}_{X}\left\vert \hat{\Psi}\left( X\right)
\right\vert ^{2}\right) K_{pX}-C}{\left( 1+\frac{\underline{k}\left(
X\right) }{\left\langle K\right\rangle }\hat{K}_{X}\left\vert \hat{\Psi}%
\left( X\right) \right\vert ^{2}\right) }  \notag \\
&=&\left( \frac{f_{1}\left( X\right) }{\left( \left( 1+\frac{\underline{k}%
\left( X\right) }{\left\langle K\right\rangle }\hat{K}_{X}\left\vert \hat{%
\Psi}\left( X\right) \right\vert ^{2}\right) K_{pX}\right) ^{r}}%
-C_{0}\right) K_{p}-\frac{C}{1+\frac{\underline{k}\left( X\right) }{%
\left\langle K\right\rangle }\hat{K}_{X}\left\vert \hat{\Psi}\left( X\right)
\right\vert ^{2}}  \notag
\end{eqnarray}%
where $K_{pX}$ is the private capital for firms at $X$.

Compared to the constant return to scale case, this amounts to replace $%
f_{1}\left( X\right) $ by:%
\begin{eqnarray*}
&&\frac{f_{1}\left( X\right) }{\left( \left( 1+\frac{\underline{k}\left(
X\right) }{\left\langle K\right\rangle }\hat{K}_{X}\left\vert \hat{\Psi}%
\left( X\right) \right\vert ^{2}\right) K_{pX}\right) ^{r}}-\frac{C}{\left(
1+\frac{\underline{k}\left( X\right) }{\left\langle K\right\rangle }\hat{K}%
_{X}\left\vert \hat{\Psi}\left( X\right) \right\vert ^{2}\right) K_{pX}}%
-C_{0} \\
&\rightarrow &\frac{f_{1}\left( X\right) }{\left( K_{X}\right) ^{r}}-\frac{C%
}{K_{X}}-C_{0}
\end{eqnarray*}%
The first term produces a total return proportional to:%
\begin{equation*}
\left( K_{X}\right) ^{1-r}f_{1}\left( X\right)
\end{equation*}%
and per unit of disposable capital:%
\begin{equation*}
\frac{f_{1}\left( X\right) }{\left( \left( 1+\frac{\underline{k}\left(
X\right) }{\left\langle K\right\rangle }\hat{K}_{X}\left\vert \hat{\Psi}%
\left( X\right) \right\vert ^{2}\right) K_{X}\right) ^{r}}
\end{equation*}%
and this corresponding to replace:

\begin{equation}
f_{1}\left( X\right) \rightarrow \frac{f_{1}\left( X\right) }{\left[ K_{X}%
\right] ^{r}}-\frac{C}{K_{X}}-C_{0}  \label{FDT}
\end{equation}%
in the return equation.

As in Gosselin and Lotz (2024), to render dimensionless the decreasing
factor in (\ref{FDT}), we should include a refernce value and replace $K_{X}$
by $\frac{K_{X}}{K_{ref}}$. In first approximation, if $K_{ref}$ is close to
a normalized value of $1$, this amounts to set $\frac{K_{X}}{K_{ref}}%
\rightarrow K_{X}$ . When the dependency in $K_{X}$ is needed, we will
reintroduce explicitly $\frac{K_{X}}{K_{ref}}$.

Using formula (\ref{FDT}), the return equation (\ref{sq}) writes:%
\begin{equation}
\left( 1-\left\langle \hat{S}\left( X^{\prime }\right) \right\rangle \right)
\left( \left\langle f\left( X^{\prime }\right) \right\rangle -\bar{r}\right)
=\left\langle S_{E}\left( X\right) \right\rangle \left( \frac{f_{1}\left(
X\right) }{\left[ K_{X}\right] ^{r}}-\frac{C}{K_{X}}-C_{0}-\bar{r}\right)
\end{equation}%
with:%
\begin{equation*}
\left\langle \hat{S}\left( X^{\prime }\right) \right\rangle =2\left\langle 
\hat{S}_{E}\left( X^{\prime }\right) \right\rangle +\left\langle \hat{w}%
\left( X^{\prime }\right) \right\rangle \frac{\left\langle \hat{r}\left(
X^{\prime }\right) \right\rangle -\left\langle \hat{f}\left( X^{\prime
}\right) \right\rangle }{2}
\end{equation*}%
is a function of $\left\langle \hat{S}_{E}\left( X^{\prime }\right)
\right\rangle $.

Thus, in the case of decreasing return, $\left\langle f\left( X\right)
\right\rangle $, the resolution is similar by replacing:

\begin{equation*}
\left\langle f\left( X\right) \right\rangle \rightarrow \left[ \frac{%
\left\langle f_{1}\left( X\right) \right\rangle }{\left[ \left\langle
K_{X}\right\rangle \right] ^{r}}-\frac{C}{\left\langle K_{X}\right\rangle }%
-C_{0}\right]
\end{equation*}%
Functions $\left\langle \hat{S}\left( X^{\prime }\right) \right\rangle $, $%
\left\langle S_{E}\left( X,X\right) \right\rangle $, $\left\langle S\left(
X,X\right) \right\rangle $ and $\frac{\left\langle \hat{K}\right\rangle
\left\Vert \hat{\Psi}\right\Vert ^{2}}{\left\langle K\right\rangle
\left\Vert \Psi \right\Vert ^{2}}$ are functions of $\left\langle \hat{S}%
_{E}\left( X^{\prime }\right) \right\rangle $ and $\left\langle S\left(
X\right) \right\rangle $. We solve $\left\langle S\left( X\right)
\right\rangle $ as a function of $\left\langle \hat{S}_{E}\left( X^{\prime
}\right) \right\rangle $ by using (\ref{Qtk}), (\ref{Qs}) and (\ref{vr}).
Then $\left\langle f\left( X\right) \right\rangle $ becoms function of $%
\left\langle \hat{S}_{E}\left( X^{\prime }\right) \right\rangle $.

Then $\left\langle \hat{S}\left( X^{\prime }\right) \right\rangle $, $%
\left\langle S_{E}\left( X,X\right) \right\rangle $, $\left\langle S\left(
X,X\right) \right\rangle $ and $\frac{\left\langle \hat{K}\right\rangle
\left\Vert \hat{\Psi}\right\Vert ^{2}}{\left\langle K\right\rangle
\left\Vert \Psi \right\Vert ^{2}}$ are functions of $\left\langle \hat{S}%
_{E}\left( X^{\prime }\right) \right\rangle $

Equations (\ref{Cf}) and (\ref{Snh}) allow to write $\left\langle \hat{f}%
\left( X^{\prime }\right) \right\rangle $ as function of $\left\langle \hat{S%
}_{E}\left( X^{\prime },X\right) \right\rangle $ and external parameters $%
\left\langle f_{1}\left( X\right) \right\rangle $ and $\left\langle r\left(
X\right) \right\rangle $. Given (\ref{Fsh}):%
\begin{equation}
\left\langle \hat{f}\left( X^{\prime }\right) \right\rangle -\left\langle 
\hat{r}\left( X^{\prime }\right) \right\rangle _{\hat{w}_{L}}=\frac{\frac{%
2\left( 2-\left( \gamma \left\langle \hat{S}_{E}\left( X\right)
\right\rangle \right) ^{2}\right) }{1-\left( \gamma \left\langle \hat{S}%
_{E}\left( X\right) \right\rangle \right) ^{2}}\left\langle \hat{S}%
_{E}\left( X\right) \right\rangle -1+\frac{\left\langle f\left( X\right)
\right\rangle -\left\langle \hat{r}\left( X^{\prime }\right) \right\rangle _{%
\hat{w}_{L}}}{2\left( 2-\left( \gamma \left\langle \hat{S}_{E}\left(
X\right) \right\rangle \right) ^{2}\right) }}{\left( 1-\frac{1}{2}\frac{%
1-\left( \gamma \left\langle \hat{S}_{E}\left( X\right) \right\rangle
\right) ^{2}}{2-\left( \gamma \left\langle \hat{S}_{E}\left( X\right)
\right\rangle \right) ^{2}}\right) }
\end{equation}

\begin{equation}
\left\langle \hat{f}\left( X^{\prime }\right) \right\rangle =\frac{\frac{%
2\left( 2-\left( \gamma \left\langle \hat{S}_{E}\left( X\right)
\right\rangle \right) ^{2}\right) }{1-\left( \gamma \left\langle \hat{S}%
_{E}\left( X\right) \right\rangle \right) ^{2}}\left\langle \hat{S}%
_{E}\left( X\right) \right\rangle -1+\frac{\left\langle \hat{r}\left(
X^{\prime }\right) \right\rangle _{\hat{w}_{L}}}{2}+\frac{\left\langle
f\left( X\right) \right\rangle }{2\left( 2-\left( \gamma \left\langle \hat{S}%
_{E}\left( X\right) \right\rangle \right) ^{2}\right) }}{\left( 1-\frac{1}{2}%
\frac{1-\left( \gamma \left\langle \hat{S}_{E}\left( X\right) \right\rangle
\right) ^{2}}{2-\left( \gamma \left\langle \hat{S}_{E}\left( X\right)
\right\rangle \right) ^{2}}\right) }  \label{fsh}
\end{equation}%
so that (\ref{RTC}) yields $\frac{\left\langle \hat{K}\right\rangle
\left\Vert \hat{\Psi}\right\Vert ^{2}}{\left\langle K\right\rangle
\left\Vert \Psi \right\Vert ^{2}}$ as a function of $\hat{S}_{E}\left(
X\right) $.

\subsubsection*{A5.3.2 Capital ratio including decreasing returns}

We first consider the modifications to the capital ratio formula due to
decreasing returns. In this case (Gosselin and Lotz (2024)), the average
private capital is:%
\begin{eqnarray*}
K_{Xp} &\simeq &\frac{1}{\left( 1+\frac{\underline{k}\left( X\right) }{%
\left\langle K\right\rangle }\hat{K}_{X}\left\vert \hat{\Psi}\left( X\right)
\right\vert ^{2}\right) }\left( \frac{\left( 1+\underline{k}_{L}\left(
X\right) \right) f_{1}\left( X\right) }{\left( 1+\underline{k}_{L}\left(
X\right) \right) C_{0}+\underline{k}_{L}\left( X\right) \bar{r}}\right) ^{%
\frac{1}{r}}=\left( 1-S\left( X\right) \right) \left( \frac{f_{1}\left(
X\right) }{C_{0}+\frac{\underline{k}_{L}\left( X\right) }{\left( 1+%
\underline{k}_{L}\left( X\right) \right) }\bar{r}}\right) ^{\frac{1}{r}} \\
&=&\left( 1-S\left( X\right) \right) \left( \frac{f_{1}\left( X\right) }{%
C_{0}+\frac{S_{L}\left( X^{\prime }\right) }{1-S_{E}\left( X^{\prime
}\right) }\bar{r}}\right) ^{\frac{1}{r}}\simeq \left( 1-S\left( X\right)
\right) \left( \frac{f_{1}\left( X\right) }{C_{0}+\bar{r}}\right) ^{\frac{1}{%
r}}
\end{eqnarray*}%
so that the disposable capital is:%
\begin{equation*}
K_{X}\simeq \left( \frac{f_{1}\left( X\right) }{C_{0}+\bar{r}}\right) ^{%
\frac{1}{r}}
\end{equation*}%
while:%
\begin{equation*}
\left\vert \Psi \left( X\right) \right\vert ^{2}=\epsilon \frac{\bar{C}%
\left( X\right) +\sqrt{\sigma _{\hat{K}}^{2}\left( \frac{\left\vert \Psi
_{0}\left( X\right) \right\vert ^{2}}{\epsilon }-\frac{f_{1}\left( X\right) 
}{2}\right) }}{\sigma _{\hat{K}}^{2}f_{1}^{e}\left( X\right) }=\frac{%
\epsilon \left( C\left( 1-S\right) +X\right) }{\sigma _{\hat{K}%
}^{2}f_{1}^{e}\left( X\right) }
\end{equation*}%
with:%
\begin{equation*}
f_{1}^{e}\left( X\right) =\left( 1+\frac{\underline{k}\left( X\right) }{%
\left\langle K\right\rangle }\hat{K}_{X}\left\vert \hat{\Psi}\left( X\right)
\right\vert ^{2}\right) \left( \frac{C_{0}+\bar{r}}{f_{1}\left( X\right) }%
\right) ^{\frac{1}{r}}\frac{\left( 3X^{\left( e\right) }-C^{\left( e\right)
}\right) \left( C^{\left( e\right) }+X^{\left( e\right) }\right) }{%
2X^{\left( e\right) }-C^{\left( e\right) }}
\end{equation*}%
that is:%
\begin{eqnarray*}
\left\vert \Psi \left( X\right) \right\vert ^{2} &=&\frac{\epsilon \left(
C\left( 1-S\right) +X\right) }{\sigma _{\hat{K}}^{2}\left( 1+\frac{%
\underline{k}\left( X\right) }{\left\langle K\right\rangle }\hat{K}%
_{X}\left\vert \hat{\Psi}\left( X\right) \right\vert ^{2}\right) \left( 
\frac{C_{0}+\bar{r}}{f_{1}\left( X\right) }\right) ^{\frac{1}{r}}\frac{%
\left( 3X^{\left( e\right) }-C^{\left( e\right) }\right) \left( C^{\left(
e\right) }+X^{\left( e\right) }\right) }{2X^{\left( e\right) }-C^{\left(
e\right) }}} \\
&=&\left( 1-S\left( X\right) \right) \left( \frac{f_{1}\left( X\right) }{%
C_{0}+\bar{r}}\right) ^{\frac{1}{r}}\frac{\epsilon \left( 2X^{\left(
e\right) }-C^{\left( e\right) }\right) }{\sigma _{\hat{K}}^{2}\left( \frac{%
C_{0}+\bar{r}}{f_{1}\left( X\right) }\right) ^{\frac{1}{r}}\left( 3X^{\left(
e\right) }-C^{\left( e\right) }\right) }
\end{eqnarray*}%
and:%
\begin{eqnarray*}
K_{X}\left\vert \Psi \left( X\right) \right\vert ^{2} &=&\left( 1-S\left(
X\right) \right) \left( \left( \frac{f_{1}\left( X\right) }{C_{0}+C_{0}+%
\frac{S_{L}\left( X^{\prime }\right) }{1-S_{E}\left( X^{\prime }\right) }%
\bar{r}}\right) ^{\frac{1}{r}}\right) ^{2}\frac{\epsilon \left( 2X^{\left(
e\right) }-C^{\left( e\right) }\right) }{\sigma _{\hat{K}}^{2}\left(
3X^{\left( e\right) }-C^{\left( e\right) }\right) } \\
&\simeq &\left( \left( 1-S\left( X\right) \right) \left( \frac{f_{1}\left(
X\right) }{C_{0}+C_{0}+\frac{S_{L}\left( X^{\prime }\right) }{1-S_{E}\left(
X^{\prime }\right) }\bar{r}}\right) ^{\frac{2}{r}}\right) \frac{2\epsilon }{%
3\sigma _{\hat{K}}^{2}}
\end{eqnarray*}%
The average:%
\begin{equation*}
\left\langle K\right\rangle \left\Vert \Psi \right\Vert ^{2}=\frac{2\epsilon 
}{3\sigma _{\hat{K}}^{2}}\left( \left( 1-\left\langle S\left( X\right)
\right\rangle \right) \left( \frac{\left\langle f_{1}\right\rangle }{C_{0}+%
\bar{r}}\right) ^{\frac{2}{r}}\right)
\end{equation*}%
can be written as:%
\begin{equation*}
\left\langle K\right\rangle \left\Vert \Psi \right\Vert ^{2}=\left(
1-\left\langle S\left( X,X\right) \right\rangle \frac{\left\langle \hat{K}%
\right\rangle \left\Vert \hat{\Psi}\right\Vert ^{2}}{\left\langle
K\right\rangle \left\Vert \Psi \right\Vert ^{2}}\right) \left( \left( \frac{%
2\epsilon }{3\sigma _{\hat{K}}^{2}}\right) ^{\frac{r}{2}}\frac{\left\langle
f_{1}\right\rangle }{C_{0}+\bar{r}}\right) ^{\frac{2}{r}}
\end{equation*}%
or:%
\begin{equation*}
\left( \left\langle K\right\rangle \left\Vert \Psi \right\Vert ^{2}\right)
^{2}=\left( \left\langle K\right\rangle \left\Vert \Psi \right\Vert
^{2}-\left\langle S\left( X,X\right) \right\rangle \left\langle \hat{K}%
\right\rangle \left\Vert \hat{\Psi}\right\Vert ^{2}\right) \left( \left( 
\frac{2\epsilon }{3\sigma _{\hat{K}}^{2}}\right) ^{\frac{r}{2}}\frac{%
\left\langle f_{1}\right\rangle }{C_{0}+\bar{r}}\right) ^{\frac{2}{r}}
\end{equation*}%
with solution:$\allowbreak $%
\begin{equation*}
\left( \left( \frac{2\epsilon }{3\sigma _{\hat{K}}^{2}}\right) ^{\frac{r}{2}}%
\frac{\left\langle f_{1}\right\rangle }{C_{0}+\bar{r}}\right) ^{\frac{2}{r}%
}+\left( \left( \frac{2\epsilon }{3\sigma _{\hat{K}}^{2}}\right) ^{\frac{r}{2%
}}\frac{\left\langle f_{1}\right\rangle }{C_{0}+\bar{r}}\right) ^{\frac{1}{r}%
}\sqrt{\left( \left( \frac{2\epsilon }{3\sigma _{\hat{K}}^{2}}\right) ^{%
\frac{r}{2}}\frac{\left\langle f_{1}\right\rangle }{C_{0}+\bar{r}}\right) ^{%
\frac{1}{r}}-4\left\langle S\left( X,X\right) \right\rangle \left\langle 
\hat{K}\right\rangle \left\Vert \hat{\Psi}\right\Vert ^{2}}
\end{equation*}

Alternatively, the equation for $\left\langle K\right\rangle \left\Vert \Psi
\right\Vert ^{2}$ writes in first approximation: 
\begin{equation*}
\left\langle K\right\rangle \left\Vert \Psi \right\Vert ^{2}\rightarrow
\left( \left( \frac{2\epsilon }{3\sigma _{\hat{K}}^{2}}\right) ^{\frac{r}{2}}%
\frac{\left\langle f_{1}\right\rangle }{C_{0}+\bar{r}}\right) ^{\frac{2}{r}}
\end{equation*}%
with first order correction due to $\left\langle S\left( X,X\right)
\right\rangle $:%
\begin{equation}
\left\langle K\right\rangle \left\Vert \Psi \right\Vert ^{2}\simeq \left(
1-\left\langle S\left( X,X\right) \right\rangle \frac{\left\langle \hat{K}%
\right\rangle \left\Vert \hat{\Psi}\right\Vert ^{2}}{\left( \left( \frac{%
2\epsilon }{3\sigma _{\hat{K}}^{2}}\right) ^{\frac{r}{2}}\frac{\left\langle
f_{1}\right\rangle }{C_{0}+\bar{r}}\right) ^{\frac{2}{r}}}\right) \left(
\left( \frac{2\epsilon }{3\sigma _{\hat{K}}^{2}}\right) ^{\frac{r}{2}}\frac{%
\left\langle f_{1}\right\rangle }{C_{0}+\bar{r}}\right) ^{\frac{2}{r}}
\label{Mrf}
\end{equation}%
Using (\ref{Vl}) and $\hat{k}=\frac{\hat{S}}{1-\hat{S}}$ along wth:%
\begin{equation*}
r=\frac{2+\hat{k}-\sqrt{\left( 2+\hat{k}\right) ^{2}-\hat{k}}}{6\underline{%
\hat{k}}}=\frac{1}{6\left( 2+\hat{k}+\sqrt{\left( 2+\hat{k}\right) ^{2}-\hat{%
k}}\right) }\simeq \frac{1}{24}<<1
\end{equation*}%
the average disposable capital for investors becomes:%
\begin{equation}
\left\langle \hat{K}\right\rangle \left\Vert \hat{\Psi}\right\Vert
^{2}\rightarrow \frac{\hat{\mu}V\sigma _{\hat{K}}^{2}}{2\left\langle \hat{f}%
\right\rangle ^{2}}\left( \frac{\left( \frac{\left\Vert \hat{\Psi}%
_{0}\right\Vert ^{2}}{\hat{\mu}}\left( 1-\hat{S}\right) +\left( \frac{1-r}{r}%
\right) \frac{\left\langle \hat{f}\right\rangle \hat{S}}{2}\right) }{\left(
1-\hat{S}\right) -2r\left( 1-r\right) \frac{\hat{S}}{1-\hat{S}}}\right) ^{2}
\label{NVK}
\end{equation}%
and the capital ratio is:%
\begin{equation}
\frac{\left\langle \hat{K}\right\rangle \left\Vert \hat{\Psi}\right\Vert ^{2}%
}{\left\langle K\right\rangle \left\Vert \Psi \right\Vert ^{2}}\simeq \frac{%
\frac{\hat{\mu}V\sigma _{\hat{K}}^{2}}{2\left\langle \hat{f}\right\rangle
^{2}}\left( \frac{\left( \frac{\left\Vert \hat{\Psi}_{0}\right\Vert ^{2}}{%
\hat{\mu}}\left( 1-\hat{S}\right) +\left( \frac{1-r}{r}\right) \frac{%
\left\langle \hat{f}\right\rangle \hat{S}}{2}\right) }{\left( 1-\hat{S}%
\right) -2r\left( 1-r\right) \frac{\hat{S}}{1-\hat{S}}}\right) ^{2}}{\left(
1-\left\langle S\left( X,X\right) \right\rangle \frac{\frac{\hat{\mu}V\sigma
_{\hat{K}}^{2}}{2\left\langle \hat{f}\right\rangle ^{2}}\left( \frac{\left( 
\frac{\left\Vert \hat{\Psi}_{0}\right\Vert ^{2}}{\hat{\mu}}\left( 1-\hat{S}%
\right) +\left( \frac{1-r}{r}\right) \frac{\left\langle \hat{f}\right\rangle 
\hat{S}}{2}\right) }{\left( 1-\hat{S}\right) -2r\left( 1-r\right) \frac{\hat{%
S}}{1-\hat{S}}}\right) ^{2}}{\left( \left( \frac{2\epsilon }{3\sigma _{\hat{K%
}}^{2}}\right) ^{\frac{r}{2}}\frac{\left\langle f_{1}\right\rangle }{C_{0}+%
\bar{r}}\right) ^{\frac{2}{r}}}\right) \left( \left( \frac{2\epsilon }{%
3\sigma _{\hat{K}}^{2}}\right) ^{\frac{r}{2}}\frac{\left\langle
f_{1}\right\rangle }{C_{0}+\bar{r}}\right) ^{\frac{2}{r}}}  \label{frm}
\end{equation}

\subsubsection*{A5.3.3 Expression of $\left\langle S\left( X,X\right)
\right\rangle $ and $\left\langle S\left( X\right) \right\rangle $}

\paragraph{A5.3.3.1 Equation for $\left\langle S\left( X,X\right)
\right\rangle $ and $\left\langle S\left( X\right) \right\rangle $}

Decreasing return to scale modify the equation for $\left\langle S\left(
X,X\right) \right\rangle $ in the followng way: 
\begin{eqnarray}
&&\left\langle S\left( X,X\right) \right\rangle \\
&=&\left\langle w\left( X\right) \right\rangle \left( 1+\left( \hat{w}\left(
X\right) \left( \frac{\frac{\left\langle f_{1}\left( X\right) \right\rangle 
}{\left( \left\langle K\right\rangle \left\Vert \Psi \right\Vert ^{2}\right)
^{r}}-C_{0}-\frac{C}{\left\langle K\right\rangle \left\Vert \Psi \right\Vert
^{2}}}{2}-\frac{\left\langle \hat{f}\left( X^{\prime }\right) \right\rangle
_{\hat{w}_{E}}+\left\langle \hat{r}\left( X^{\prime }\right) \right\rangle _{%
\hat{w}_{L}}}{2}\right) \right) \right)  \notag
\end{eqnarray}%
with:%
\begin{equation*}
\left\langle S\left( X\right) \right\rangle =\left\langle S\left( X,X\right)
\right\rangle \frac{\left\langle \hat{K}\right\rangle \left\Vert \hat{\Psi}%
\right\Vert ^{2}}{\left\langle K\right\rangle \left\Vert \Psi \right\Vert
^{2}}
\end{equation*}%
and:%
\begin{equation}
\frac{\left\langle f\left( X\right) \right\rangle +\left\langle \bar{r}%
\left( X\right) \right\rangle }{2}\rightarrow \frac{\frac{\left\langle
f_{1}\left( X\right) \right\rangle }{\left( \left\langle K\right\rangle
\left\Vert \Psi \right\Vert ^{2}\right) ^{r}}-C_{0}-\frac{C}{\left\langle
K\right\rangle \left\Vert \Psi \right\Vert ^{2}}-\left( 1-\left\langle
S\left( X\right) \right\rangle \right) C+\left\langle \bar{r}\left( X\right)
\right\rangle }{2}
\end{equation}%
Where we can use formula (\ref{frm}) for $\frac{\left\langle \hat{K}%
\right\rangle \left\Vert \hat{\Psi}\right\Vert ^{2}}{\left\langle
K\right\rangle \left\Vert \Psi \right\Vert ^{2}}$: The equations for $%
\left\langle S\left( X,X\right) \right\rangle $ and $\left\langle S\left(
X\right) \right\rangle $ also write: 
\begin{eqnarray*}
\left\langle S\left( X,X\right) \right\rangle &=&\frac{\left\langle w\left(
X\right) \right\rangle }{\left( 1-\frac{\left\langle w\left( X\right)
\right\rangle \hat{w}\left( X\right) C}{2}\frac{\left\langle \hat{K}%
\right\rangle \left\Vert \hat{\Psi}\right\Vert ^{2}}{\left\langle
K\right\rangle \left\Vert \Psi \right\Vert ^{2}}\right) } \\
&&\times \left( 1+\left( \hat{w}\left( X\right) \left( \frac{\frac{%
\left\langle f_{1}\left( X\right) \right\rangle }{\left( \left\langle
K\right\rangle \left\Vert \Psi \right\Vert ^{2}\right) ^{r}}-C_{0}-\frac{C}{%
\left\langle K\right\rangle \left\Vert \Psi \right\Vert ^{2}}}{2}-\frac{%
\left\langle \hat{f}\left( X^{\prime }\right) \right\rangle _{\hat{w}%
_{E}}+\left\langle \hat{r}\left( X^{\prime }\right) \right\rangle _{\hat{w}%
_{L}}}{2}\right) \right) \right)
\end{eqnarray*}%
\begin{eqnarray*}
\left\langle S\left( X\right) \right\rangle &=&\frac{\left\langle w\left(
X\right) \right\rangle }{\left( 1-\frac{\left\langle w\left( X\right)
\right\rangle \hat{w}\left( X\right) C}{2}\frac{\left\langle \hat{K}%
\right\rangle \left\Vert \hat{\Psi}\right\Vert ^{2}}{\left\langle
K\right\rangle \left\Vert \Psi \right\Vert ^{2}}\right) } \\
&&\times \left( 1+\left( \hat{w}\left( X\right) \left( \frac{\frac{%
\left\langle f_{1}\left( X\right) \right\rangle }{\left( \left\langle
K\right\rangle \left\Vert \Psi \right\Vert ^{2}\right) ^{r}}-C_{0}-\frac{C}{%
\left\langle K\right\rangle \left\Vert \Psi \right\Vert ^{2}}}{2}-\frac{%
\left\langle \hat{f}\left( X^{\prime }\right) \right\rangle _{\hat{w}%
_{E}}+\left\langle \hat{r}\left( X^{\prime }\right) \right\rangle _{\hat{w}%
_{L}}}{2}\right) \right) \right) \frac{\left\langle \hat{K}\right\rangle
\left\Vert \hat{\Psi}\right\Vert ^{2}}{\left\langle K\right\rangle
\left\Vert \Psi \right\Vert ^{2}}
\end{eqnarray*}%
which allows to consider first order corrections due to $\left( \left\langle
K\right\rangle \left\Vert \Psi \right\Vert ^{2}\right) ^{r}$.

The return $\frac{\left\langle f_{1}\left( X\right) \right\rangle }{\left(
\left\langle K\right\rangle \left\Vert \Psi \right\Vert ^{2}\right) ^{r}}%
-C_{0}-\frac{C}{\left\langle K\right\rangle \left\Vert \Psi \right\Vert ^{2}}
$ has been estimated in Gosselin and Lotz (2024) by:%
\begin{equation*}
f_{1}^{e}\left( X\right) =\left( 1+\frac{\underline{k}\left( X\right) }{%
\left\langle K\right\rangle }\hat{K}_{X}\left\vert \hat{\Psi}\left( X\right)
\right\vert ^{2}\right) \left( \frac{C_{0}+\bar{r}}{f_{1}\left( X\right) }%
\right) ^{\frac{1}{r}}\frac{\left( 3X^{\left( e\right) }-C^{\left( e\right)
}\right) \left( C^{\left( e\right) }+X^{\left( e\right) }\right) }{%
2X^{\left( e\right) }-C^{\left( e\right) }}
\end{equation*}%
and this formula will be used while considering approximate solutions. Here
we rather consider the solutions with decreasing returns as deviations from
constant return to scale.

\paragraph*{A5.3.3.2 Lowest order and first order approximation}

For $r<<1$ and $C<<f_{1}$, at the lowest order we can consider that $%
\left\langle S\left( X,X\right) \right\rangle $ is given by its formula with
constant return $\left\langle S\left( X,X\right) \right\rangle _{cr}$ so
that: 
\begin{equation}
\left\langle S\left( X\right) \right\rangle =\left\langle S\left( X,X\right)
\right\rangle _{cr}\frac{\left\langle \hat{K}\right\rangle \left\Vert \hat{%
\Psi}\right\Vert ^{2}}{\left\langle K\right\rangle \left\Vert \Psi
\right\Vert ^{2}}  \label{md}
\end{equation}%
where $\frac{\left\langle \hat{K}\right\rangle \left\Vert \hat{\Psi}%
\right\Vert ^{2}}{\left\langle K\right\rangle \left\Vert \Psi \right\Vert
^{2}}$ is computed with equation (\ref{frm}):%
\begin{equation}
\frac{\left\langle \hat{K}\right\rangle \left\Vert \hat{\Psi}\right\Vert ^{2}%
}{\left\langle K\right\rangle \left\Vert \Psi \right\Vert ^{2}}\simeq \frac{%
\frac{\hat{\mu}V\sigma _{\hat{K}}^{2}\left( 1-\hat{S}\right) ^{2}}{%
2\left\langle \hat{f}\right\rangle ^{2}}\left( \frac{\left( \frac{\left\Vert 
\hat{\Psi}_{0}\right\Vert ^{2}}{\hat{\mu}}+\left( \frac{1-r}{r}\right) \frac{%
\left\langle \hat{f}\right\rangle }{2\left( 1-\hat{S}\right) }\frac{\hat{S}}{%
1-\hat{S}}\right) }{\left( 1-2r\left( 1-r\right) \frac{\hat{S}}{1-\hat{S}}%
\right) }\right) }{\left( 1-\left\langle S\left( X,X\right) \right\rangle 
\frac{\frac{\hat{\mu}V\sigma _{\hat{K}}^{2}\left( 1-\hat{S}\right) ^{2}}{%
2\left\langle \hat{f}\right\rangle ^{2}}\left( \frac{\left( \frac{\left\Vert 
\hat{\Psi}_{0}\right\Vert ^{2}}{\hat{\mu}}+\left( \frac{1-r}{r}\right) \frac{%
\left\langle \hat{f}\right\rangle }{2\left( 1-\hat{S}\right) }\frac{\hat{S}}{%
1-\hat{S}}\right) }{\left( 1-2r\left( 1-r\right) \frac{\hat{S}}{1-\hat{S}}%
\right) }\right) }{\left( \left( \frac{2\epsilon }{3\sigma _{\hat{K}}^{2}}%
\right) ^{\frac{r}{2}}\frac{\left\langle f_{1}\right\rangle }{C_{0}+\bar{r}}%
\right) ^{\frac{2}{r}}}\right) \left( \left( \frac{2\epsilon }{3\sigma _{%
\hat{K}}^{2}}\right) ^{\frac{r}{2}}\frac{\left\langle f_{1}\right\rangle }{%
C_{0}+\bar{r}}\right) ^{\frac{2}{r}}}  \label{KRC}
\end{equation}%
\begin{equation}
\left\langle S\left( X,X\right) \right\rangle _{cr}=\left\langle w\left(
X\right) \right\rangle \left( 1+\left( \hat{w}\left( X\right) \left( \frac{%
\left\langle f_{1}\left( X\right) \right\rangle }{2}-\frac{\left\langle \hat{%
f}\left( X^{\prime }\right) \right\rangle _{\hat{w}_{E}}+\left\langle \hat{r}%
\left( X^{\prime }\right) \right\rangle _{\hat{w}_{L}}}{2}\right) \right)
\right)  \label{Scr}
\end{equation}%
and to the first order:%
\begin{equation}
\left\langle S\left( X\right) \right\rangle =\left\langle S\left( X,X\right)
\right\rangle \frac{\left\langle \hat{K}\right\rangle \left\Vert \hat{\Psi}%
\right\Vert ^{2}}{\left\langle K\right\rangle \left\Vert \Psi \right\Vert
^{2}}
\end{equation}%
\begin{eqnarray}
\left\langle S\left( X,X\right) \right\rangle &=&\frac{\left\langle w\left(
X\right) \right\rangle }{\left( 1-\frac{\left\langle w\left( X\right)
\right\rangle \hat{w}\left( X\right) C}{2}\frac{\left\langle \hat{K}%
\right\rangle \left\Vert \hat{\Psi}\right\Vert ^{2}}{\left\langle
K\right\rangle \left\Vert \Psi \right\Vert ^{2}}\right) }  \label{SDT} \\
&&\times \left( 1+\left( \hat{w}\left( X\right) \left( \frac{\left(
1-\left\langle S\left( X\right) \right\rangle \right) ^{r}\left\langle
f_{1}\left( X\right) \right\rangle -C}{2}-\frac{\left\langle \hat{f}\left(
X^{\prime }\right) \right\rangle _{\hat{w}_{E}}+\left\langle \hat{r}\left(
X^{\prime }\right) \right\rangle _{\hat{w}_{L}}}{2}\right) \right) \right) 
\frac{\left\langle \hat{K}\right\rangle \left\Vert \hat{\Psi}\right\Vert ^{2}%
}{\left\langle K\right\rangle \left\Vert \Psi \right\Vert ^{2}}  \notag
\end{eqnarray}%
where we replace $\left\langle S\left( X\right) \right\rangle \rightarrow
\left\langle S\left( X,X\right) \right\rangle _{cr}\frac{\left\langle \hat{K}%
\right\rangle \left\Vert \hat{\Psi}\right\Vert ^{2}}{\left\langle
K\right\rangle \left\Vert \Psi \right\Vert ^{2}}$.

\subsubsection*{A5.3.4 Expression of $\left\langle \hat{S}\left( X^{\prime
},X\right) \right\rangle $ and $\left\langle \hat{S}_{E}\left( X^{\prime
},X\right) \right\rangle $}

The formula (\ref{Stf}), (\ref{Snt}) and (\ref{Fsh}) for $\left\langle \hat{S%
}\left( X^{\prime },X\right) \right\rangle $ and $\left\langle \hat{S}%
_{E}\left( X^{\prime },X\right) \right\rangle $ and $\left\langle \hat{f}%
\left( X^{\prime }\right) \right\rangle $\ are the same as before, with $%
\left\langle f_{1}\left( X\right) \right\rangle $ replaced by:%
\begin{equation*}
\frac{f_{1}\left( X\right) +\tau \left( \left\langle f_{1}\left( X\right)
\right\rangle -\left\langle f_{1}\left( X^{\prime }\right) \right\rangle
\right) }{\left( \left\langle K\right\rangle \left\Vert \Psi \right\Vert
^{2}\right) ^{r}}-\frac{C}{\left\langle K\right\rangle \left\Vert \Psi
\right\Vert ^{2}}-\bar{r}
\end{equation*}

\subsubsection*{A5.3.5 Solution of average return equation}

The average return equation is similar to that under\ constant return:%
\begin{eqnarray}
&&\left( 1-\left\langle \hat{S}\left( X^{\prime }\right) \right\rangle
\right) \left( \left\langle \hat{f}\left( X^{\prime }\right) \right\rangle -%
\bar{r}\right) \\
&\simeq &\left\langle S_{E}\left( X,X\right) \right\rangle \left[
\left\langle f_{1}\left( X\right) \right\rangle ^{\left( dr\right)
}-\left\langle \bar{r}\left( X\right) \right\rangle \right]  \notag
\end{eqnarray}%
with the replacement of $\left\langle f_{1}\left( X\right) \right\rangle $
in (\ref{lh}) and (\ref{rh}) by:%
\begin{equation*}
\left\langle f_{1}\left( X\right) \right\rangle ^{\left( dr\right) }=\left( 
\frac{f_{1}\left( X\right) }{\left( \left\langle K\right\rangle \left\Vert
\Psi \right\Vert ^{2}\right) ^{r}}-\frac{C}{\left\langle K\right\rangle
\left\Vert \Psi \right\Vert ^{2}}-C_{0}\right)
\end{equation*}%
at the lowest order, and:%
\begin{equation}
\left\langle f_{1}\left( X\right) \right\rangle ^{\left( dr\right) }=\left( 
\frac{f_{1}\left( X\right) }{\left( \left\langle K\right\rangle \left\Vert
\Psi \right\Vert ^{2}\right) ^{r}}-\frac{C}{\left\langle K\right\rangle
\left\Vert \Psi \right\Vert ^{2}}-C_{0}\right)
\end{equation}%
At first order, with $\left\langle S\left( X\right) \right\rangle $ given by
(\ref{md}) at lowest order and (\ref{SDT}) at the first order.

Equation (\ref{zq}) is still valid with this replacement:%
\begin{eqnarray*}
&&\left( 1-2z+\frac{1-\left( \gamma z\right) ^{2}}{3-\left( \gamma z\right)
^{2}}\left( \frac{2\left( 2-\left( \gamma z\right) ^{2}\right) }{1-\left(
\gamma z\right) ^{2}}z-1+\frac{\left\langle f_{1}\left( X\right)
\right\rangle ^{\left( dr\right) }-\left\langle \hat{r}\left( X^{\prime
}\right) \right\rangle _{\hat{w}_{L}}}{2\left( 2-\left( \gamma z\right)
^{2}\right) }\right) \right) \\
&&\times \frac{\frac{2\left( 2-\left( \gamma z\right) ^{2}\right) }{1-\left(
\gamma z\right) ^{2}}z-1+\frac{\left\langle f_{1}\left( X\right)
\right\rangle ^{\left( dr\right) }-\left\langle \hat{r}\left( X^{\prime
}\right) \right\rangle _{\hat{w}_{L}}}{2\left( 2-\left( \gamma z\right)
^{2}\right) }}{\left( 1-\frac{1}{2}\frac{1-\left( \gamma z\right) ^{2}}{%
2-\left( \gamma z\right) ^{2}}\right) } \\
&=&\frac{1}{2}\frac{1}{2-\left( \gamma z\right) ^{2}}\left( 1+\left( \frac{%
3-\left( \gamma z\right) ^{2}}{2\left( 2-\left( \gamma z\right) ^{2}\right) }%
\right) \left( \left\langle f_{1}\left( X\right) \right\rangle ^{\left(
dr\right) }-\left\langle \hat{r}\left( X^{\prime }\right) \right\rangle _{%
\hat{w}_{L}}\right) \right. \\
&&\left. -\frac{\frac{2\left( 2-\left( \gamma z\right) ^{2}\right) }{%
1-\left( \gamma z\right) ^{2}}z-1+\frac{\left\langle f\left( X\right)
\right\rangle -\left\langle \hat{r}\left( X^{\prime }\right) \right\rangle _{%
\hat{w}_{L}}}{2\left( 2-\left( \gamma z\right) ^{2}\right) }}{2\left( 1-%
\frac{1}{2}\frac{1-\left( \gamma z\right) ^{2}}{2-\left( \gamma z\right) ^{2}%
}\right) }\right) \left[ \left\langle f_{1}\left( X\right) \right\rangle
^{\left( dr\right) }-\left\langle \hat{r}\left( X^{\prime }\right)
\right\rangle _{\hat{w}_{L}}\right]
\end{eqnarray*}%
The replacement amounts to modify $\left\langle f_{1}\left( X\right)
\right\rangle $ by a factor $\frac{1}{\left( \left\langle K\right\rangle
\left\Vert \Psi \right\Vert ^{2}\right) ^{r}}$ and shifting $\bar{r}$ by $%
\frac{C}{\left\langle K\right\rangle \left\Vert \Psi \right\Vert ^{2}}+C_{0}$
where $\left( \left\langle K\right\rangle \left\Vert \Psi \right\Vert
^{2}\right) $ is computed using the values of shares obtained under constant
return to scale. This does not modify the results, but diminishes the
dependency of shares in $\left\langle f_{1}\left( X\right) \right\rangle $
by $\frac{1}{\left( \left\langle K\right\rangle \left\Vert \Psi \right\Vert
^{2}\right) ^{r}}$.

\subsection*{A5.4 Approximate equation and solutions}

\subsubsection*{A5.4.1 Average stakes}

At the lowest order in $\left\langle f_{1}\left( X\right) \right\rangle
-\left\langle \bar{r}\left( X\right) \right\rangle $, equation (\ref{zq}) is:%
\begin{equation}
\left( 1-2z+\frac{1-\left( \gamma z\right) ^{2}}{3-\left( \gamma z\right)
^{2}}\left( \frac{2\left( 2-\left( \gamma z\right) ^{2}\right) }{1-\left(
\gamma z\right) ^{2}}z-1\right) \right) \frac{\frac{2\left( 2-\left( \gamma
z\right) ^{2}\right) }{1-\left( \gamma z\right) ^{2}}z-1}{\left( 1-\frac{1}{2%
}\frac{1-\left( \gamma z\right) ^{2}}{2-\left( \gamma z\right) ^{2}}\right) }%
=0
\end{equation}%
and the share $z$ satisfies:%
\begin{equation}
\frac{2\left( 2-\left( \gamma z\right) ^{2}\right) }{1-\left( \gamma
z\right) ^{2}}z-1=0  \label{NCT}
\end{equation}%
that is:%
\begin{equation*}
1-4z-z^{2}\gamma ^{2}+2z^{3}\gamma ^{2}=0
\end{equation*}%
There is one solution $z_{0}\left( \gamma \right) $ satisfying $z\gamma $
and $z<1$. For $\gamma \rightarrow 0$, $z\rightarrow \frac{1}{4}$. In the
sequel, we use that for this solution:%
\begin{equation*}
\gamma ^{2}z_{0}^{2}=\frac{1-4z_{0}}{1-2z_{0}},1-\left( \gamma z_{0}\right)
^{2}=1-\frac{1-4z_{0}}{1-2z_{0}}=2\frac{z_{0}}{1-2z_{0}}
\end{equation*}%
\begin{equation*}
2-\left( \gamma z_{0}\right) ^{2}=\frac{1}{1-2z_{0}},3-\frac{1-4z_{0}}{%
1-2z_{0}}=2\frac{1-z_{0}}{1-2z_{0}}
\end{equation*}%
We can find a solution to the first order in $\left\langle f\left( X\right)
\right\rangle -\left\langle \hat{r}\left( X^{\prime }\right) \right\rangle _{%
\hat{w}_{L}}$ by setting:%
\begin{equation*}
z=z_{0}\left( \gamma \right) +v
\end{equation*}%
Developping the LHS of (\ref{NCT}) around:%
\begin{eqnarray*}
&&\frac{2\left( 2-\left( \gamma \left( z_{0}+v\right) \right) ^{2}\right) }{%
1-\left( \gamma \left( z_{0}+v\right) \right) ^{2}}\left( z_{0}+v\right) -1
\\
&=&2\frac{\left( \gamma ^{2}z_{0}^{2}\right) ^{2}-\gamma ^{2}z_{0}^{2}+2}{%
\left( 1-\gamma ^{2}z_{0}^{2}\right) ^{2}}v
\end{eqnarray*}%
equation (\ref{zq}) writes$:$

\begin{equation}
\left( 1-2z_{0}\right) \frac{2\frac{\left( \gamma ^{2}z_{0}^{2}\right)
^{2}-\gamma ^{2}z_{0}^{2}+2}{\left( 1-\gamma ^{2}z_{0}^{2}\right) ^{2}}v+%
\frac{\left\langle f\left( X\right) \right\rangle -\left\langle \hat{r}%
\left( X^{\prime }\right) \right\rangle _{\hat{w}_{L}}}{2\left( 2-\left(
\gamma z_{0}\right) ^{2}\right) }}{\left( 1-z_{0}\right) }=\frac{z_{0}}{%
1-\left( \gamma z_{0}\right) ^{2}}\left( \left\langle f_{1}\left( X\right)
\right\rangle -\left\langle \bar{r}\left( X\right) \right\rangle \right)
\end{equation}%
with:%
\begin{eqnarray}
v &=&\left( \frac{\frac{\left( 1-z_{0}\right) z_{0}\left( 1-\left( \gamma
z_{0}\right) ^{2}\right) }{\left( 1-2z_{0}\right) }-\frac{\left( 1-\gamma
^{2}z_{0}^{2}\right) ^{2}}{2\left( 2-\left( \gamma z_{0}\right) ^{2}\right) }%
}{2\left( \left( \gamma ^{2}z_{0}^{2}\right) ^{2}-\gamma
^{2}z_{0}^{2}+2\right) }\right) \left( \left\langle f_{1}\left( X\right)
\right\rangle -\left\langle \bar{r}\left( X\right) \right\rangle \right)
\label{NF} \\
&=&\left( \frac{z_{0}^{2}\left( 1-\gamma ^{2}z_{0}^{2}\right) }{2\left(
\left( \gamma ^{2}z_{0}^{2}\right) ^{2}-\gamma ^{2}z_{0}^{2}+2\right) \left(
1-2z_{0}\right) }\right) \left( \left\langle f_{1}\left( X\right)
\right\rangle -\left\langle \bar{r}\left( X\right) \right\rangle \right) 
\notag \\
&=&\left( \frac{z_{0}^{2}\left( 2\frac{z_{0}}{1-2z_{0}}\right) }{2\left(
\left( \frac{1-4z_{0}}{1-2z_{0}}\right) ^{2}-\frac{1-4z_{0}}{1-2z_{0}}%
+2\right) \left( 1-2z_{0}\right) }\right) \left( \left\langle f_{1}\left(
X\right) \right\rangle -\left\langle \bar{r}\left( X\right) \right\rangle
\right)  \notag \\
&=&\frac{z_{0}^{3}\left( \left\langle f_{1}\left( X\right) \right\rangle
-\left\langle \bar{r}\left( X\right) \right\rangle \right) }{2\left(
1-5z_{0}+8z_{0}^{2}\right) }  \notag
\end{eqnarray}%
As a consequence:%
\begin{eqnarray}
&&\left\langle \hat{S}_{E}\left( X^{\prime },X\right) \right\rangle
\label{SHN} \\
&=&\left\langle \hat{S}_{E}\left( X^{\prime }\right) \right\rangle
=z_{0}\left( \gamma \right) \left( 1+\left( \frac{z_{0}\left( 1-\gamma
^{2}z_{0}^{2}\right) }{2\left( \left( \gamma ^{2}z_{0}^{2}\right)
^{2}-\gamma ^{2}z_{0}^{2}+2\right) \left( 1-2z_{0}\right) }\right) \left(
\left\langle f_{1}\left( X\right) \right\rangle -\left\langle \bar{r}\left(
X\right) \right\rangle \right) \right)  \notag \\
&=&z_{0}\left( \gamma \right) \left( 1+\frac{1}{2}\frac{z_{0}^{2}}{%
1-5z_{0}+8z_{0}^{2}}\left( \left\langle f_{1}\left( X\right) \right\rangle
-\left\langle \bar{r}\left( X\right) \right\rangle \right) \right)  \notag
\end{eqnarray}%
$\allowbreak $

We start by compting:%
\begin{eqnarray*}
\left\langle \hat{f}\left( X^{\prime }\right) \right\rangle -\left\langle 
\hat{r}\left( X^{\prime }\right) \right\rangle _{\hat{w}_{L}} &=&\frac{\frac{%
2\left( 2-\left( \gamma z\right) ^{2}\right) }{1-\left( \gamma z\right) ^{2}}%
z-1+\frac{\left\langle f_{1}\left( X\right) \right\rangle ^{\left( dr\right)
}-\left\langle \hat{r}\left( X^{\prime }\right) \right\rangle _{\hat{w}_{L}}%
}{2\left( 2-\left( \gamma z\right) ^{2}\right) }}{\left( 1-\frac{1}{2}\frac{%
1-\left( \gamma z\right) ^{2}}{2-\left( \gamma z\right) ^{2}}\right) } \\
&=&\frac{2\frac{\left( \gamma ^{2}z_{0}^{2}\right) ^{2}-\gamma
^{2}z_{0}^{2}+2}{\left( 1-\gamma ^{2}z_{0}^{2}\right) ^{2}}v+\frac{%
\left\langle f_{1}\left( X\right) \right\rangle ^{\left( dr\right)
}-\left\langle \hat{r}\left( X^{\prime }\right) \right\rangle _{\hat{w}_{L}}%
}{2\left( 2-\left( \gamma \left\langle \hat{S}_{E}\left( X\right)
\right\rangle \right) ^{2}\right) }}{\left( 1-\frac{1}{2}\frac{1-\left(
\gamma \left\langle \hat{S}_{E}\left( X\right) \right\rangle \right) ^{2}}{%
2-\left( \gamma \left\langle \hat{S}_{E}\left( X\right) \right\rangle
\right) ^{2}}\right) }
\end{eqnarray*}%
Given (\ref{NF}), we find:%
\begin{eqnarray}
&&\left\langle \hat{f}\left( X^{\prime }\right) \right\rangle -\left\langle 
\hat{r}\left( X^{\prime }\right) \right\rangle _{\hat{w}_{L}}  \label{RHt} \\
&=&\frac{\left( 2\frac{\left( \gamma ^{2}z_{0}^{2}\right) ^{2}-\gamma
^{2}z_{0}^{2}+2}{\left( 1-\gamma ^{2}z_{0}^{2}\right) ^{2}}\left( \frac{%
z_{0}^{2}\left( 1-\gamma ^{2}z_{0}^{2}\right) }{2\left( \left( \gamma
^{2}z_{0}^{2}\right) ^{2}-\gamma ^{2}z_{0}^{2}+2\right) \left(
1-2z_{0}\right) }\right) +\frac{1}{2\left( 2-\left( \gamma \left\langle \hat{%
S}_{E}\left( X\right) \right\rangle \right) ^{2}\right) }\right) \left(
\left\langle f_{1}\left( X\right) \right\rangle ^{\left( dr\right)
}-\left\langle \hat{r}\left( X^{\prime }\right) \right\rangle _{\hat{w}%
_{L}}\right) }{\left( 1-\frac{1}{2}\frac{1-\left( \gamma \left\langle \hat{S}%
_{E}\left( X\right) \right\rangle \right) ^{2}}{2-\left( \gamma \left\langle 
\hat{S}_{E}\left( X\right) \right\rangle \right) ^{2}}\right) }  \notag \\
&=&\frac{z_{0}\left( \left\langle f_{1}\left( X\right) \right\rangle
^{\left( dr\right) }-\left\langle \hat{r}\left( X^{\prime }\right)
\right\rangle _{\hat{w}_{L}}\right) }{\left( 1-\gamma ^{2}z_{0}^{2}\right)
\left( 1-2z_{0}\right) }=\frac{z_{0}\left( \left\langle f_{1}\left( X\right)
\right\rangle ^{\left( dr\right) }-\left\langle \hat{r}\left( X^{\prime
}\right) \right\rangle _{\hat{w}_{L}}\right) }{\left( 1-\frac{\left(
1-4z_{0}\right) }{\left( 1-2z_{0}\right) }\right) \left( 1-2z_{0}\right) } 
\notag \\
&=&\frac{\left( \left\langle f_{1}\left( X\right) \right\rangle ^{\left(
dr\right) }-\left\langle \hat{r}\left( X^{\prime }\right) \right\rangle _{%
\hat{w}_{L}}\right) }{2}  \notag
\end{eqnarray}

We can derive $\left\langle \hat{S}\left( X^{\prime },X\right) \right\rangle 
$, by using:%
\begin{eqnarray}
\left\langle \hat{S}\left( X^{\prime },X\right) \right\rangle
&=&2\left\langle \hat{S}_{E}\left( X^{\prime }\right) \right\rangle -\frac{%
1-\left( \gamma \left\langle \hat{S}_{E}\left( X\right) \right\rangle
\right) ^{2}}{3-\left( \gamma \left\langle \hat{S}_{E}\left( X\right)
\right\rangle \right) ^{2}} \\
&&\times \left( \frac{2\left( 2-\left( \gamma \left\langle \hat{S}_{E}\left(
X\right) \right\rangle \right) ^{2}\right) }{1-\left( \gamma \left\langle 
\hat{S}_{E}\left( X\right) \right\rangle \right) ^{2}}\left\langle \hat{S}%
_{E}\left( X\right) \right\rangle -1+\frac{\left\langle f\left( X\right)
\right\rangle -\left\langle \hat{r}\left( X^{\prime }\right) \right\rangle _{%
\hat{w}_{L}}}{2\left( 2-\left( \gamma \left\langle \hat{S}_{E}\left(
X\right) \right\rangle \right) ^{2}\right) }\right)  \notag
\end{eqnarray}%
The second term writes:%
\begin{eqnarray*}
&&\frac{1-\left( \gamma \left\langle \hat{S}_{E}\left( X\right)
\right\rangle \right) ^{2}}{3-\left( \gamma \left\langle \hat{S}_{E}\left(
X\right) \right\rangle \right) ^{2}}\left( \frac{2\left( 2-\left( \gamma
\left\langle \hat{S}_{E}\left( X\right) \right\rangle \right) ^{2}\right) }{%
1-\left( \gamma \left\langle \hat{S}_{E}\left( X\right) \right\rangle
\right) ^{2}}\left\langle \hat{S}_{E}\left( X\right) \right\rangle -1+\frac{%
\left\langle f\left( X\right) \right\rangle -\left\langle \hat{r}\left(
X^{\prime }\right) \right\rangle _{\hat{w}_{L}}}{2\left( 2-\left( \gamma
\left\langle \hat{S}_{E}\left( X\right) \right\rangle \right) ^{2}\right) }%
\right) \\
&=&\frac{1-\left( \gamma z_{0}\right) ^{2}}{3-\left( \gamma z_{0}\right) ^{2}%
}\left( 1-\frac{1}{2}\frac{1-\left( \gamma z_{0}\right) ^{2}}{2-\left(
\gamma z_{0}\right) ^{2}}\right) \left( \left\langle \hat{f}\left( X^{\prime
}\right) \right\rangle -\left\langle \hat{r}\left( X^{\prime }\right)
\right\rangle _{\hat{w}_{L}}\right) \\
&=&\frac{1}{2}\frac{1-\left( \gamma z_{0}\right) ^{2}}{2-\left( \gamma
z_{0}\right) ^{2}}\left( \left\langle \hat{f}\left( X^{\prime }\right)
\right\rangle -\left\langle \hat{r}\left( X^{\prime }\right) \right\rangle _{%
\hat{w}_{L}}\right) =z_{0}\frac{z_{0}\left( \left\langle f_{1}\left(
X\right) \right\rangle ^{\left( dr\right) }-\left\langle \hat{r}\left(
X^{\prime }\right) \right\rangle _{\hat{w}_{L}}\right) }{2}
\end{eqnarray*}%
and $\left\langle \hat{S}\left( X^{\prime },X\right) \right\rangle $
simplifies:%
\begin{eqnarray}
&&\left\langle \hat{S}\left( X^{\prime },X\right) \right\rangle  \label{SHt}
\\
&=&2z_{0}\left( \gamma \right) \left( 1+\left( \frac{z_{0}\left( 1-\gamma
^{2}z_{0}^{2}\right) }{2\left( \left( \gamma ^{2}z_{0}^{2}\right)
^{2}-\gamma ^{2}z_{0}^{2}+2\right) \left( 1-2z_{0}\right) }\right) \left(
\left\langle f_{1}\left( X\right) \right\rangle -\left\langle \bar{r}\left(
X\right) \right\rangle \right) \right)  \notag \\
&&-z_{0}\frac{z_{0}\left( \left\langle f_{1}\left( X\right) \right\rangle
^{\left( dr\right) }-\left\langle \hat{r}\left( X^{\prime }\right)
\right\rangle _{\hat{w}_{L}}\right) }{\left( 1-\gamma ^{2}z_{0}^{2}\right)
\left( 1-2z_{0}\right) }  \notag \\
&=&2z_{0}\left( \gamma \right) \left( 1+\frac{z_{0}}{2\left( 1-2z_{0}\right) 
}\left( \frac{\left( 1-\gamma ^{2}z_{0}^{2}\right) }{\left( \left( \gamma
^{2}z_{0}^{2}\right) ^{2}-\gamma ^{2}z_{0}^{2}+2\right) }-\frac{1}{\left(
1-\gamma ^{2}z_{0}^{2}\right) }\right) \left( \left\langle f_{1}\left(
X\right) \right\rangle ^{\left( dr\right) }-\left\langle \hat{r}\left(
X^{\prime }\right) \right\rangle _{\hat{w}_{L}}\right) \right)  \notag
\end{eqnarray}%
so that, using $\gamma ^{2}z_{0}^{2}=\frac{1-4z_{0}}{1-2z_{0}}$:%
\begin{eqnarray}
&&\left\langle \hat{S}\left( X^{\prime },X\right) \right\rangle  \label{SHT}
\\
&=&2z_{0}\left( \gamma \right) \left( 1-\frac{1}{4}\frac{\left(
1-3z_{0}\right) \left( 1-2z_{0}\right) }{-5z_{0}+8z_{0}^{2}+1}\left(
\left\langle f_{1}\left( X\right) \right\rangle ^{\left( dr\right)
}-\left\langle \hat{r}\left( X^{\prime }\right) \right\rangle _{\hat{w}%
_{L}}\right) \right)  \notag
\end{eqnarray}%
The shares invested in real sector are:%
\begin{eqnarray*}
&&\left\langle S_{E}\left( X,X\right) \right\rangle \\
&=&\frac{1}{2}\frac{1}{2-\left( \gamma z\right) ^{2}}\left( 1+\frac{\left(
3-\left( \gamma z\right) ^{2}\right) \left( \left\langle f_{1}\left(
X\right) \right\rangle ^{\left( dr\right) }-\left\langle \hat{r}\left(
X^{\prime }\right) \right\rangle _{\hat{w}_{L}}\right) }{2\left( 2-\left(
\gamma z\right) ^{2}\right) }-\frac{\frac{2\left( 2-\left( \gamma z\right)
^{2}\right) }{1-\left( \gamma z\right) ^{2}}z-1+\frac{\left\langle f\left(
X\right) \right\rangle -\left\langle \hat{r}\left( X^{\prime }\right)
\right\rangle _{\hat{w}_{L}}}{2\left( 2-\left( \gamma z\right) ^{2}\right) }%
}{2\left( 1-\frac{1}{2}\frac{1-\left( \gamma z\right) ^{2}}{2-\left( \gamma
z\right) ^{2}}\right) }\right) \\
&=&\left( \frac{z_{0}}{\left( 1-\left( \gamma z_{0}\right) ^{2}\right) }+%
\frac{v\gamma ^{2}z_{0}^{2}}{\left( 2-\gamma ^{2}z_{0}^{2}\right) ^{2}}%
\right) \left( 1+\frac{\left( 2\frac{1-z_{0}}{1-2z_{0}}\right) \left(
\left\langle f_{1}\left( X\right) \right\rangle ^{\left( dr\right)
}-\left\langle \hat{r}\left( X^{\prime }\right) \right\rangle _{\hat{w}%
_{L}}\right) }{2\left( \frac{1}{1-2z_{0}}\right) }-\frac{\left( \left\langle
f_{1}\left( X\right) \right\rangle ^{\left( dr\right) }-\left\langle \hat{r}%
\left( X^{\prime }\right) \right\rangle _{\hat{w}_{L}}\right) }{4}\right) \\
&=&\left( \frac{z_{0}}{\left( 1-\left( \gamma z_{0}\right) ^{2}\right) }%
\right) \left( 1+2\left( 1-4z_{0}\right) v+\left( 1-z_{0}\right) \left(
\left\langle f_{1}\left( X\right) \right\rangle ^{\left( dr\right)
}-\left\langle \hat{r}\left( X^{\prime }\right) \right\rangle _{\hat{w}%
_{L}}\right) -\frac{\left( \left\langle f_{1}\left( X\right) \right\rangle
^{\left( dr\right) }-\left\langle \hat{r}\left( X^{\prime }\right)
\right\rangle _{\hat{w}_{L}}\right) }{4}\right)
\end{eqnarray*}%
and this reduces to:%
\begin{eqnarray}
&&\left\langle S_{E}\left( X,X\right) \right\rangle  \label{SNF} \\
&=&\frac{1-2z_{0}}{2}\left( 1+\left( \frac{z_{0}^{3}\left( 1-4z_{0}\right) }{%
\left( 1-5z_{0}+8z_{0}^{2}\right) }+\left( \frac{3}{4}-z_{0}\right) \right)
\left( \left\langle f_{1}\left( X\right) \right\rangle ^{\left( dr\right)
}-\left\langle \hat{r}\left( X^{\prime }\right) \right\rangle _{\hat{w}%
_{L}}\right) \right)  \notag
\end{eqnarray}%
Ultimately, given that the total share $\left\langle S\left( X,X\right)
\right\rangle $ satisfies: 
\begin{eqnarray*}
\left\langle S\left( X,X\right) \right\rangle &=&\left\langle w\left(
X\right) \right\rangle \left( 1+\left( \hat{w}\left( X\right) \left( \frac{%
\left\langle f\left( X\right) \right\rangle +\left\langle \bar{r}\left(
X\right) \right\rangle }{2}-\frac{\left\langle \hat{f}\left( X^{\prime
}\right) \right\rangle _{\hat{w}_{E}}+\left\langle \hat{r}\left( X^{\prime
}\right) \right\rangle _{\hat{w}_{L}}}{2}\right) \right) \right) \\
&=&\frac{1}{2-\left( \gamma z\right) ^{2}}\left( 1+\left( \frac{1-\left(
\gamma z\right) ^{2}}{2-\left( \gamma z\right) ^{2}}\frac{\left\langle
f\left( X\right) \right\rangle -\left\langle \bar{r}\left( X\right)
\right\rangle }{2}\right) \right)
\end{eqnarray*}%
we obtain:%
\begin{eqnarray*}
&&\left\langle S\left( X,X\right) \right\rangle \\
&=&\frac{1}{2-\left( \gamma z\right) ^{2}}\left( 1+\left( \frac{1-\left(
\gamma z\right) ^{2}}{2-\left( \gamma z\right) ^{2}}\left( \frac{%
\left\langle f\left( X\right) \right\rangle -\left\langle \bar{r}\left(
X\right) \right\rangle }{2}-\frac{\frac{2\left( 2-\left( \gamma z\right)
^{2}\right) }{1-\left( \gamma z\right) ^{2}}z-1+\frac{\left\langle
f_{1}\left( X\right) \right\rangle ^{\left( dr\right) }-\left\langle \hat{r}%
\left( X^{\prime }\right) \right\rangle _{\hat{w}_{L}}}{2\left( 2-\left(
\gamma z\right) ^{2}\right) }}{2\left( 1-\frac{1}{2}\frac{1-\left( \gamma
z\right) ^{2}}{2-\left( \gamma z\right) ^{2}}\right) }\right) \right) \right)
\\
&=&2\left( \frac{z_{0}}{\left( 1-\left( \gamma z_{0}\right) ^{2}\right) }+%
\frac{v\gamma ^{2}z_{0}^{2}}{\left( 2-\gamma ^{2}z_{0}^{2}\right) ^{2}}%
\right) \left( 1+\left( z_{0}\left( \left( \left\langle f\left( X\right)
\right\rangle -\left\langle \bar{r}\left( X\right) \right\rangle \right) -%
\frac{\left( \left\langle f_{1}\left( X\right) \right\rangle ^{\left(
dr\right) }-\left\langle \hat{r}\left( X^{\prime }\right) \right\rangle _{%
\hat{w}_{L}}\right) }{2}\right) \right) \right) \\
&=&2\left( \frac{1-2z_{0}}{2}+\frac{z_{0}^{3}\left( 1-4z_{0}\right) \left(
1-2z_{0}\right) }{2\left( 1-5z_{0}+8z_{0}^{2}\right) }\left( \left\langle
f_{1}\left( X\right) \right\rangle -\left\langle \bar{r}\left( X\right)
\right\rangle \right) \right) \left( 1+z_{0}\frac{\left( \left\langle
f_{1}\left( X\right) \right\rangle ^{\left( dr\right) }-\left\langle \hat{r}%
\left( X^{\prime }\right) \right\rangle _{\hat{w}_{L}}\right) }{2}\right)
\end{eqnarray*}%
and this writes to the first order:

\begin{equation}
\left\langle S\left( X,X\right) \right\rangle =\left( 1-2z_{0}\right) \left(
1+\left( \frac{1}{2}+\frac{z_{0}^{2}\left( 1-4z_{0}\right) }{\left(
1-5z_{0}+8z_{0}^{2}\right) }\right) z_{0}\left( \left\langle f_{1}\left(
X\right) \right\rangle ^{\left( dr\right) }-\left\langle \hat{r}\left(
X^{\prime }\right) \right\rangle _{\hat{w}_{L}}\right) \right)  \label{STF}
\end{equation}%
To evaluate\ $\left\langle f_{1}\left( X\right) \right\rangle ^{\left(
dr\right) }$ in the case of decreasing returns, we first consider levels of
capital and capital ratios.%
\begin{equation*}
\left\langle S\left( X\right) \right\rangle =\left\langle S\left( X,X\right)
\right\rangle \frac{\left\langle \hat{K}\right\rangle \left\Vert \hat{\Psi}%
\right\Vert ^{2}}{\left\langle K\right\rangle \left\Vert \Psi \right\Vert
^{2}}
\end{equation*}%
Remark that we have:%
\begin{eqnarray}
\left\langle S_{L}\left( X,X\right) \right\rangle &=&\left\langle S\left(
X,X\right) \right\rangle -\left\langle S_{E}\left( X,X\right) \right\rangle
\label{STH} \\
&=&\frac{1-2z_{0}}{2}\left( 1-\left( \frac{3}{8}-z_{0}-\frac{1}{2}\frac{%
z_{0}^{3}\left( 1-4z_{0}\right) }{\left( 1-5z_{0}+8z_{0}^{2}\right) }\right)
\left( \left\langle f_{1}\left( X\right) \right\rangle ^{\left( dr\right)
}-\left\langle \hat{r}\left( X^{\prime }\right) \right\rangle _{\hat{w}%
_{L}}\right) \right)  \notag
\end{eqnarray}

\subsubsection*{A5.4.2 Corrections due to decreasing returns; principle}

As explained in the previous paragraph, including decreasing return to
scales amounts to replace:%
\begin{eqnarray*}
&&\left\langle f_{1}\left( X\right) \right\rangle ^{\left( dr\right)
}-\left\langle \hat{r}\left( X^{\prime }\right) \right\rangle _{\hat{w}_{L}}
\\
&\rightarrow &\left( \frac{\frac{\left\langle f_{1}\left( X\right)
\right\rangle }{1-r}}{\left( \left\langle K\right\rangle \right) ^{r}}-\frac{%
C}{\left\langle K\right\rangle }-\left\langle \hat{r}\left( X^{\prime
}\right) \right\rangle _{\hat{w}_{L}}-C_{0}\right)
\end{eqnarray*}%
To study the impact of a modification in returns $\left\langle f_{1}\left(
X\right) \right\rangle $, we reintroduce the dimensionless factor $\frac{%
\left\langle K\right\rangle }{K_{ref}}$ (see discussion after (\ref{FDT})).
Given that averages are considered, we will replace:%
\begin{equation*}
\frac{\frac{\left\langle f_{1}\left( X\right) \right\rangle }{1-r}}{\left(
\left( 1+\frac{\underline{k}\left( X\right) }{\left\langle K\right\rangle }%
\hat{K}_{X}\left\vert \hat{\Psi}\left( \hat{X}\right) \right\vert
^{2}\right) K_{X}\right) ^{r}}-C_{0}-\frac{C}{\left\langle K\right\rangle }-%
\bar{r}=\frac{f_{1}^{\left( e\right) }\left( X\right) -\bar{r}}{\left( 1+%
\underline{k}_{L}\left( X\right) \right) }
\end{equation*}%
The factor $\left( 1+\frac{\underline{k}\left( X\right) }{\left\langle
K\right\rangle }\hat{K}_{X}\left\vert \hat{\Psi}\left( \hat{X}\right)
\right\vert ^{2}\right) $ is given by $1-$ $\left\langle S\left( X,X\right)
\right\rangle $. The factor $\left\langle S\left( X,X\right) \right\rangle $
has been obtained in (\ref{STF}) and $\left\langle \hat{S}\left( X^{\prime
},X\right) \right\rangle $ in (\ref{SHT}). The formula for $\left\langle 
\hat{f}\right\rangle $ is given in (\ref{RHt}). To study the variation of $%
\left\langle f_{1}\left( X\right) \right\rangle ^{\left( dr\right)
}-\left\langle \hat{r}\left( X^{\prime }\right) \right\rangle _{\hat{w}_{L}}$
with respect to $\left\langle f_{1}\left( X\right) \right\rangle $, we can
neglect $\frac{C}{\left\langle K\right\rangle \left\Vert \Psi \right\Vert
^{2}}$ and thus consider in first approximation that the variation of $%
\left\langle f_{1}\left( X\right) \right\rangle ^{\left( dr\right)
}-\left\langle \hat{r}\left( X^{\prime }\right) \right\rangle _{\hat{w}_{L}}$
reduces to the variation of $\frac{\left\langle f_{1}\left( X\right)
\right\rangle }{\left( \left\langle K\right\rangle \left\Vert \Psi
\right\Vert ^{2}\right) ^{r}}$.

Since $\left( \left\langle K\right\rangle \left\Vert \Psi \right\Vert
^{2}\right) ^{r}$ increases with $\left\langle f_{1}\left( X\right)
\right\rangle $, the rate of increase of $\left\langle f_{1}\left( X\right)
\right\rangle ^{\left( dr\right) }-\left\langle \hat{r}\left( X^{\prime
}\right) \right\rangle _{\hat{w}_{L}}$ with $\left\langle f_{1}\left(
X\right) \right\rangle $ is at rate lower than $1$. The results are
preserved, with lower amplitude. We give more details in the subsequent
paragraphs.

\subsubsection*{A5.4.3 Corrections to firms returns}

A more precise analysis and formulas for disposable capital can be obtained
using the formulas given in (Gosselin and Lotz (2024)). Disposable capital
for firms satisfy: 
\begin{equation}
K_{X}\rightarrow \frac{1}{4f_{1}^{\left( e\right) }\left( X\right) }\frac{%
\left( 3X^{\left( e\right) }-C^{\left( e\right) }\right) \left( C^{\left(
e\right) }+X^{\left( e\right) }\right) }{2X^{\left( e\right) }-C^{\left(
e\right) }}  \label{KQ}
\end{equation}%
with:%
\begin{equation*}
f_{1}^{\left( e\right) }\left( X\right) \rightarrow \frac{\left( 1+%
\underline{k}_{L}\left( X\right) \right) \frac{f_{1}\left( X\right) }{1-r}}{%
\left( \left( 1+\frac{\underline{k}\left( X\right) }{\left\langle
K\right\rangle }\hat{K}_{X}\left\vert \hat{\Psi}\left( \hat{X}\right)
\right\vert ^{2}\right) K_{X}\right) ^{r}}-C_{0}^{\left( e\right) }
\end{equation*}%
We have found in Gosselin and Lotz (2024) the return by firm:%
\begin{eqnarray*}
\frac{\frac{f_{1}^{\left( e\right) }\left( X\right) }{1-r}-r}{\left( 1+%
\underline{k}_{L}\left( X\right) \right) } &\rightarrow &rC_{0}+\frac{\left(
1+\frac{\underline{k}\left( X\right) }{\left\langle K\right\rangle }\hat{K}%
_{X}\left\vert \hat{\Psi}\left( \hat{X}\right) \right\vert ^{2}\right)
\left( 3X^{\left( e\right) }-C^{\left( e\right) }\right) \left( C^{\left(
e\right) }+X^{\left( e\right) }\right) }{\left( 1+\frac{\underline{k}\left(
X\right) }{\left\langle K\right\rangle }\hat{K}_{X}\left\vert \hat{\Psi}%
\left( \hat{X}\right) \right\vert ^{2}\right) K_{X}\left( 1+\underline{k}%
_{L}\left( X\right) \right) \left( 2X^{\left( e\right) }-C^{\left( e\right)
}\right) } \\
&&-\frac{C}{\left( 1+\underline{k}_{L}\left( X\right) \right) \left( 1+\frac{%
\underline{k}\left( X\right) }{\left\langle K\right\rangle }\hat{K}%
_{X}\left\vert \hat{\Psi}\left( \hat{X}\right) \right\vert ^{2}\right) K_{X}}
\end{eqnarray*}%
leading to:%
\begin{equation*}
\frac{f_{1}^{\left( e\right) }\left( X\right) -r}{\left( 1+\underline{k}%
_{L}\left( X\right) \right) }\rightarrow rC_{0}+\left( \frac{C_{0}+\frac{%
S_{L}\left( X^{\prime }\right) }{1-S_{E}\left( X^{\prime }\right) }\bar{r}}{%
f_{1}\left( X\right) }\right) ^{\frac{1}{r}}\left( 1-S_{L}\left( X^{\prime
}\right) \right) \left( \frac{\left( 3X^{\left( e\right) }-C^{\left(
e\right) }\right) \left( C^{\left( e\right) }+X^{\left( e\right) }\right) }{%
1-S\left( X^{\prime }\right) }-C\right)
\end{equation*}%
Using (\ref{STF}) and (\ref{STH}):%
\begin{equation*}
1-S\left( X^{\prime }\right) \simeq 1-\left( 1-2z_{0}\right) \frac{%
\left\langle \hat{K}\right\rangle \left\Vert \hat{\Psi}\right\Vert ^{2}}{%
\left\langle K\right\rangle \left\Vert \Psi \right\Vert ^{2}}-d\frac{%
f_{1}^{\left( e\right) }\left( X\right) -r}{\left( 1+\underline{k}_{L}\left(
X\right) \right) }
\end{equation*}%
\begin{equation*}
1-S_{L}\left( X^{\prime }\right) =1-\frac{\left( 1-2z_{0}\right) }{2}\frac{%
\left\langle \hat{K}\right\rangle \left\Vert \hat{\Psi}\right\Vert ^{2}}{%
\left\langle K\right\rangle \left\Vert \Psi \right\Vert ^{2}}+e\frac{%
f_{1}^{\left( e\right) }\left( X\right) -r}{\left( 1+\underline{k}_{L}\left(
X\right) \right) }
\end{equation*}%
with:%
\begin{equation*}
d=\left( 1-2z_{0}\right) \left( \frac{1}{2}+\frac{z_{0}^{2}\left(
1-4z_{0}\right) }{\left( 1-5z_{0}+8z_{0}^{2}\right) }\right) z_{0}\frac{%
\left\langle \hat{K}\right\rangle \left\Vert \hat{\Psi}\right\Vert ^{2}}{%
\left\langle K\right\rangle \left\Vert \Psi \right\Vert ^{2}}
\end{equation*}%
\begin{equation*}
e=\frac{1-2z_{0}}{2}\left( \frac{3}{8}-z_{0}-\frac{1}{2}\frac{%
z_{0}^{3}\left( 1-4z_{0}\right) }{\left( 1-5z_{0}+8z_{0}^{2}\right) }\right) 
\frac{\left\langle \hat{K}\right\rangle \left\Vert \hat{\Psi}\right\Vert ^{2}%
}{\left\langle K\right\rangle \left\Vert \Psi \right\Vert ^{2}}
\end{equation*}%
We define $V=\frac{f_{1}^{\left( e\right) }\left( X\right) -r}{\left( 1+%
\underline{k}_{L}\left( X\right) \right) }$ so that $V$ satisfies:%
\begin{equation*}
V=rC_{0}+A\left( 1+EV\right) \left( \frac{G}{1-DV}-CB\right)
\end{equation*}%
with:%
\begin{equation*}
A=\left( \frac{C_{0}+\frac{S_{L}\left( X^{\prime }\right) }{1-S_{E}\left(
X^{\prime }\right) }\bar{r}}{f_{1}\left( X\right) }\right) ^{\frac{1}{r}}%
\frac{1-\frac{\left( 1-2z_{0}\right) }{2}\frac{\left\langle \hat{K}%
\right\rangle \left\Vert \hat{\Psi}\right\Vert ^{2}}{\left\langle
K\right\rangle \left\Vert \Psi \right\Vert ^{2}}}{1-\left( 1-2z_{0}\right) 
\frac{\left\langle \hat{K}\right\rangle \left\Vert \hat{\Psi}\right\Vert ^{2}%
}{\left\langle K\right\rangle \left\Vert \Psi \right\Vert ^{2}}}
\end{equation*}%
\begin{equation*}
B\rightarrow 1-\left( 1-2z_{0}\right) \frac{\left\langle \hat{K}%
\right\rangle \left\Vert \hat{\Psi}\right\Vert ^{2}}{\left\langle
K\right\rangle \left\Vert \Psi \right\Vert ^{2}}
\end{equation*}%
\begin{equation*}
D=\frac{d}{1-\left( 1-2z_{0}\right) \frac{\left\langle \hat{K}\right\rangle
\left\Vert \hat{\Psi}\right\Vert ^{2}}{\left\langle K\right\rangle
\left\Vert \Psi \right\Vert ^{2}}}
\end{equation*}%
\begin{equation*}
G=\left( 3X^{\left( e\right) }-C^{\left( e\right) }\right) \left( C^{\left(
e\right) }+X^{\left( e\right) }\right)
\end{equation*}%
\begin{equation*}
E=\frac{e}{1-\frac{\left( 1-2z_{0}\right) }{2}\frac{\left\langle \hat{K}%
\right\rangle \left\Vert \hat{\Psi}\right\Vert ^{2}}{\left\langle
K\right\rangle \left\Vert \Psi \right\Vert ^{2}}}
\end{equation*}%
The following approximation holds:%
\begin{eqnarray*}
&&\frac{\left( 1-2z_{0}\right) \left( \frac{1}{2}+\frac{z_{0}^{2}\left(
1-4z_{0}\right) }{\left( 1-5z_{0}+8z_{0}^{2}\right) }\right) z_{0}\frac{%
\left\langle \hat{K}\right\rangle \left\Vert \hat{\Psi}\right\Vert ^{2}}{%
\left\langle K\right\rangle \left\Vert \Psi \right\Vert ^{2}}}{1-\left(
1-2z_{0}\right) \frac{\left\langle \hat{K}\right\rangle \left\Vert \hat{\Psi}%
\right\Vert ^{2}}{\left\langle K\right\rangle \left\Vert \Psi \right\Vert
^{2}}} \\
&\rightarrow &\frac{\frac{1}{2}\left( 1-2z_{0}\right) z_{0}\frac{%
\left\langle \hat{K}\right\rangle \left\Vert \hat{\Psi}\right\Vert ^{2}}{%
\left\langle K\right\rangle \left\Vert \Psi \right\Vert ^{2}}}{1-\left(
1-2z_{0}\right) \frac{\left\langle \hat{K}\right\rangle \left\Vert \hat{\Psi}%
\right\Vert ^{2}}{\left\langle K\right\rangle \left\Vert \Psi \right\Vert
^{2}}}
\end{eqnarray*}%
and the solution for $V$:

$\allowbreak $%
\begin{eqnarray}
V &=&\frac{f_{1}^{\left( e\right) }\left( X\right) -r}{\left( 1+\underline{k}%
_{L}\left( X\right) \right) }  \label{RTD} \\
&=&\frac{1+ABC\left( E-D\right) -AGE+rC_{0}D}{2D\left( 1+ABCE\right) } 
\notag \\
&&+\sqrt{\left( \frac{1+ABC\left( E-D\right) -AGE+rC_{0}D}{2D\left(
1+ABCE\right) }\right) ^{2}-\frac{A\left( G-BC\right) +rC_{0}}{\allowbreak
D\left( 1+ABCE\right) }}  \notag
\end{eqnarray}%
is at the lowest order:%
\begin{eqnarray*}
\frac{f_{1}^{\left( e\right) }\left( X\right) -r}{\left( 1+\underline{k}%
_{L}\left( X\right) \right) } &\simeq &rC_{0}+A\left( G-BC\right) \\
&=&rC_{0}-\left( \frac{C_{0}+\frac{S_{L}\left( X^{\prime }\right) }{%
1-S_{E}\left( X^{\prime }\right) }\bar{r}}{f_{1}\left( X\right) }\right) ^{%
\frac{1}{r}}\frac{1-\frac{\left( 1-2z_{0}\right) }{2}\frac{\left\langle \hat{%
K}\right\rangle \left\Vert \hat{\Psi}\right\Vert ^{2}}{\left\langle
K\right\rangle \left\Vert \Psi \right\Vert ^{2}}}{1-\left( 1-2z_{0}\right) 
\frac{\left\langle \hat{K}\right\rangle \left\Vert \hat{\Psi}\right\Vert ^{2}%
}{\left\langle K\right\rangle \left\Vert \Psi \right\Vert ^{2}}}\left(
C\left( 1-\left( 1-2z_{0}\right) \frac{\left\langle \hat{K}\right\rangle
\left\Vert \hat{\Psi}\right\Vert ^{2}}{\left\langle K\right\rangle
\left\Vert \Psi \right\Vert ^{2}}\right) -3X^{2}\right)
\end{eqnarray*}

\subsubsection*{A5.3.4 Disposable capital and capital ratios under
decreasing returns}

Investors disposable capital $\left\langle \hat{K}\right\rangle \left\Vert 
\hat{\Psi}\right\Vert ^{2}$ is defined in (\ref{NVK}): 
\begin{equation}
\left\langle \hat{K}\right\rangle \left\Vert \hat{\Psi}\right\Vert
^{2}\rightarrow \frac{\hat{\mu}V\sigma _{\hat{K}}^{2}}{2\left\langle \hat{f}%
\right\rangle ^{2}}\left( \frac{\left( \frac{\left\Vert \hat{\Psi}%
_{0}\right\Vert ^{2}}{\hat{\mu}}\left( 1-\hat{S}\right) +\left( \frac{1-r}{r}%
\right) \frac{\left\langle \hat{f}\right\rangle \hat{S}}{2}\right) }{\left(
1-\hat{S}\right) -2r\left( 1-r\right) \frac{\hat{S}}{1-\hat{S}}}\right) ^{2}
\end{equation}%
with $\hat{S}=\left\langle \hat{S}\left( X^{\prime },X\right) \right\rangle $
given in (\ref{SHt}). To compute $\left\langle \hat{f}\right\rangle $, we
use (\ref{RHt}), and the average return per unit of capital is given by
Thus, we can replace:%
\begin{equation}
\left\langle \hat{f}\right\rangle =\left\langle \hat{r}\left( X^{\prime
}\right) \right\rangle +\frac{\left\langle f\right\rangle -\left\langle \hat{%
r}\right\rangle }{2}  \label{FHt}
\end{equation}%
$\allowbreak $

As a consequence:%
\begin{eqnarray}
&&\left\langle \hat{K}\right\rangle \left\Vert \hat{\Psi}\right\Vert ^{2}
\label{VRk} \\
&\rightarrow &\frac{\hat{\mu}V\sigma _{\hat{K}}^{2}}{2\left\langle \hat{f}%
\right\rangle ^{2}}\left( \frac{\left( \frac{\left\Vert \hat{\Psi}%
_{0}\right\Vert ^{2}}{\hat{\mu}}\left( 1-\hat{S}\right) +\left( \frac{1-r}{r}%
\right) \frac{\left\langle \hat{f}\right\rangle \hat{S}}{2}\right) }{\left(
1-\hat{S}\right) -2r\left( 1-r\right) \frac{\hat{S}}{1-\hat{S}}}\right) ^{2}
\notag \\
&\simeq &\frac{\hat{\mu}V\sigma _{\hat{K}}^{2}}{2}\left( \frac{\frac{%
\left\Vert \hat{\Psi}_{0}\right\Vert ^{2}}{\hat{\mu}}\left( 1-\hat{S}\right)
+\left( \frac{1-r}{2r}\right) \left( \left\langle \hat{r}\left( X^{\prime
}\right) \right\rangle +\frac{\left\langle f\right\rangle -\left\langle \hat{%
r}\right\rangle }{2}\right) \hat{S}}{\left( \left\langle \hat{r}\left(
X^{\prime }\right) \right\rangle +\frac{\left\langle f\right\rangle
-\left\langle \hat{r}\right\rangle }{2}\right) \left( 1-\hat{S}\right) }%
\right) ^{2}  \notag
\end{eqnarray}%
To solve for $\left\langle \hat{K}\right\rangle \left\Vert \hat{\Psi}%
\right\Vert ^{2}$, we consider in first approximation:%
\begin{equation}
\left\langle \hat{K}\right\rangle \left\Vert \hat{\Psi}\right\Vert ^{2}=%
\frac{V\sigma _{\hat{K}}^{2}}{2\hat{\mu}}\left( \frac{\left\Vert \hat{\Psi}%
_{0}\right\Vert ^{2}}{\left\langle \hat{r}\left( X^{\prime }\right)
\right\rangle +\frac{\left\langle f\right\rangle -\left\langle \hat{r}%
\right\rangle }{2}}\right) ^{2}  \label{FRp}
\end{equation}%
Disposable capital for firms is given by $\left\langle K\right\rangle
\left\Vert \Psi \right\Vert ^{2}$ so that using (\ref{Mrf}), we find:%
\begin{equation}
\left\langle K\right\rangle \left\Vert \Psi \right\Vert ^{2}\simeq \left(
1-\left\langle S\left( X,X\right) \right\rangle \frac{\left\langle \hat{K}%
\right\rangle \left\Vert \hat{\Psi}\right\Vert ^{2}}{\left( \left( \frac{%
2\epsilon }{3\sigma _{\hat{K}}^{2}}\right) ^{\frac{r}{2}}\frac{\left\langle
f_{1}\right\rangle }{C_{0}+\bar{r}}\right) ^{\frac{2}{r}}}\right) \left(
\left( \frac{2\epsilon }{3\sigma _{\hat{K}}^{2}}\right) ^{\frac{r}{2}}\frac{%
\left\langle f_{1}\right\rangle }{C_{0}+\bar{r}}\right) ^{\frac{2}{r}}
\end{equation}%
We write:%
\begin{eqnarray*}
\Xi &=&\left( \frac{C_{0}+\frac{S_{L}\left( X^{\prime }\right) }{%
1-S_{E}\left( X^{\prime }\right) }\bar{r}}{f_{1}\left( X\right) }\right) ^{%
\frac{1}{r}} \\
v &=&\left( \frac{2\epsilon }{3\sigma _{\hat{K}}^{2}}\right)
\end{eqnarray*}%
and the ratio of disposable capital is:%
\begin{eqnarray*}
&&\frac{\left\langle \hat{K}\right\rangle \left\Vert \hat{\Psi}\right\Vert
^{2}}{\left\langle K\right\rangle \left\Vert \Psi \right\Vert ^{2}} \\
&\simeq &\frac{\left\langle \hat{K}\right\rangle \left\Vert \hat{\Psi}%
\right\Vert ^{2}\Xi ^{2}}{\left( 1-\left\langle S\left( X,X\right)
\right\rangle \frac{\left\langle \hat{K}\right\rangle \left\Vert \hat{\Psi}%
\right\Vert ^{2}\Xi ^{2}}{v}\right) v}
\end{eqnarray*}%
Using that:%
\begin{eqnarray*}
1-\left\langle S\left( X,X\right) \right\rangle \frac{\left\langle \hat{K}%
\right\rangle \left\Vert \hat{\Psi}\right\Vert ^{2}}{\left\langle
K\right\rangle \left\Vert \Psi \right\Vert ^{2}} &\simeq &1-\left(
1-2z_{0}\right) \frac{\left\langle \hat{K}\right\rangle \left\Vert \hat{\Psi}%
\right\Vert ^{2}}{\left\langle K\right\rangle \left\Vert \Psi \right\Vert
^{2}} \\
&\simeq &1-\left( 1-2z_{0}\right) \frac{\left\langle \hat{K}\right\rangle
\left\Vert \hat{\Psi}\right\Vert ^{2}\Xi ^{2}}{v}
\end{eqnarray*}%
we have:%
\begin{equation}
\frac{\left\langle \hat{K}\right\rangle \left\Vert \hat{\Psi}\right\Vert ^{2}%
}{\left\langle K\right\rangle \left\Vert \Psi \right\Vert ^{2}}\simeq \frac{%
\left\langle \hat{K}\right\rangle \left\Vert \hat{\Psi}\right\Vert ^{2}\Xi
^{2}}{\left( 1-\left( 1-2z_{0}\right) \frac{\left\langle \hat{K}%
\right\rangle \left\Vert \hat{\Psi}\right\Vert ^{2}\Xi ^{2}}{v}\right) v}
\label{Crt}
\end{equation}%
and the formula for rturns becomes:%
\begin{eqnarray*}
\left\langle f\right\rangle -\left\langle \hat{r}\right\rangle &\simeq
&rC_{0}-\left( 1-\left( 1-2z_{0}\right) \frac{\left\langle \hat{K}%
\right\rangle \left\Vert \hat{\Psi}\right\Vert ^{2}\Xi ^{2}}{\left( 1-\left(
1-2z_{0}\right) \frac{\left\langle \hat{K}\right\rangle \left\Vert \hat{\Psi}%
\right\Vert ^{2}\Xi ^{2}}{v}\right) v}\right) \Xi \\
&&\times \frac{1-\frac{\left( 1-2z_{0}\right) }{2}\frac{\left\langle \hat{K}%
\right\rangle \left\Vert \hat{\Psi}\right\Vert ^{2}\Xi ^{2}}{\left( 1-\left(
1-2z_{0}\right) \frac{\left\langle \hat{K}\right\rangle \left\Vert \hat{\Psi}%
\right\Vert ^{2}\Xi ^{2}}{v}\right) v}}{1-\left( 1-2z_{0}\right) \frac{%
\left\langle \hat{K}\right\rangle \left\Vert \hat{\Psi}\right\Vert ^{2}\Xi
^{2}}{\left( 1-\left( 1-2z_{0}\right) \frac{\left\langle \hat{K}%
\right\rangle \left\Vert \hat{\Psi}\right\Vert ^{2}\Xi ^{2}}{v}\right) v}}%
\left( C\left( 1-\left( 1-2z_{0}\right) \frac{\left\langle \hat{K}%
\right\rangle \left\Vert \hat{\Psi}\right\Vert ^{2}\Xi ^{2}}{\left( 1-\left(
1-2z_{0}\right) \frac{\left\langle \hat{K}\right\rangle \left\Vert \hat{\Psi}%
\right\Vert ^{2}\Xi ^{2}}{v}\right) v}\right) -3X^{2}\right)
\end{eqnarray*}%
At the second order in $\Xi $, we find that:%
\begin{equation*}
\left\langle f\right\rangle -\left\langle \hat{r}\right\rangle \simeq
rC_{0}-\Xi \left( C-3X^{2}\right) +o\left( \Xi ^{2}\right)
\end{equation*}

Using (\ref{FRp})\bigskip :%
\begin{equation}
\left\langle \hat{K}\right\rangle \left\Vert \hat{\Psi}\right\Vert ^{2}=%
\frac{V\sigma _{\hat{K}}^{2}}{2\hat{\mu}}\left( \frac{\left\Vert \hat{\Psi}%
_{0}\right\Vert ^{2}}{\left\langle \hat{r}\left( X^{\prime }\right)
\right\rangle +\frac{\left\langle f\right\rangle -\left\langle \hat{r}%
\right\rangle }{2}}\right) ^{2}
\end{equation}%
we are led to the equation for $\left\langle \hat{K}\right\rangle \left\Vert 
\hat{\Psi}\right\Vert ^{2}$:%
\begin{equation}
\left\langle \hat{K}\right\rangle \left\Vert \hat{\Psi}\right\Vert ^{2}=%
\frac{V\sigma _{\hat{K}}^{2}}{2\hat{\mu}}\left( \frac{\left\Vert \hat{\Psi}%
_{0}\right\Vert ^{2}}{\left\langle \hat{r}\left( X^{\prime }\right)
\right\rangle +\frac{1}{2}\left( rC_{0}-\Xi \left( C-3X^{2}\right) \right) }%
\right) ^{2}
\end{equation}%
Replacing: 
\begin{equation*}
\Xi ^{2}\rightarrow \left( \frac{C_{0}+\frac{S_{L}\left( X^{\prime }\right) 
}{1-S_{E}\left( X^{\prime }\right) }\bar{r}}{f_{1}\left( X\right) }\right) ^{%
\frac{2}{r}}
\end{equation*}%
this yields in first approximation:%
\begin{equation}
\left\langle \hat{K}\right\rangle \left\Vert \hat{\Psi}\right\Vert ^{2}=%
\frac{V\sigma _{\hat{K}}^{2}\left\Vert \hat{\Psi}_{0}\right\Vert ^{4}}{2\hat{%
\mu}\left( \left\langle \hat{r}\left( X^{\prime }\right) \right\rangle +%
\frac{rC_{0}-\left( \frac{C_{0}+\frac{S_{L}\left( X^{\prime }\right) }{%
1-S_{E}\left( X^{\prime }\right) }\bar{r}}{f_{1}\left( X\right) }\right) ^{%
\frac{1}{r}}\left( C-3X^{2}\right) }{2}\right) ^{2}}
\end{equation}%
which is decreasing with $\left\langle f_{1}\right\rangle $. The return
writes:%
\begin{equation}
\left\langle f\right\rangle -\left\langle \hat{r}\right\rangle \simeq
rC_{0}-\left( C-3X^{2}\right) \left( \frac{C_{0}+\frac{S_{L}\left( X^{\prime
}\right) }{1-S_{E}\left( X^{\prime }\right) }\bar{r}}{f_{1}\left( X\right) }%
\right) ^{\frac{1}{r}}  \label{Zrth}
\end{equation}%
which is increasing with $\left\langle f_{1}\right\rangle $ for $C>3X^{2}$.

Moreover, the investors return writes:%
\begin{eqnarray*}
\left\langle \hat{f}\right\rangle &=&\left\langle \hat{r}\left( X^{\prime
}\right) \right\rangle +\frac{\left\langle f\right\rangle -\left\langle \hat{%
r}\right\rangle }{2} \\
&=&f_{a}-f_{b}\left( \frac{C_{0}+\frac{S_{L}\left( X^{\prime }\right) }{%
1-S_{E}\left( X^{\prime }\right) }\bar{r}}{f_{1}\left( X\right) }\right) ^{%
\frac{1}{r}}
\end{eqnarray*}%
with:%
\begin{eqnarray}
f_{a} &=&\left\langle \hat{r}\left( X^{\prime }\right) \right\rangle +\frac{1%
}{2}rC_{0}  \label{LSf} \\
f_{b} &=&\frac{1}{2}\left( C-3X^{2}\right)  \notag
\end{eqnarray}%
Then at the first order, we can write%
\begin{equation*}
\left\langle \hat{K}\right\rangle \left\Vert \hat{\Psi}\right\Vert
^{2}\simeq \frac{\hat{\mu}V\sigma _{\hat{K}}^{2}}{2}\left( \frac{\frac{%
\left\Vert \hat{\Psi}_{0}\right\Vert ^{2}}{\hat{\mu}}\left( 1-\hat{S}\right)
+\left( \frac{1-r}{2r}\right) \left( \left\langle \hat{r}\left( X^{\prime
}\right) \right\rangle +\frac{\left\langle f\right\rangle -\left\langle \hat{%
r}\right\rangle }{2}\right) \hat{S}}{\left( \left\langle \hat{r}\left(
X^{\prime }\right) \right\rangle +\frac{\left\langle f\right\rangle
-\left\langle \hat{r}\right\rangle }{2}\right) \left( 1-\hat{S}\right) }%
\right) ^{2}
\end{equation*}%
with $\left\langle f\right\rangle -\left\langle \hat{r}\right\rangle $ given
by the zeroth order formula (\ref{Zrth}).

Private capital for investors is given by:%
\begin{eqnarray}
&&\left( 1-\hat{S}\right) \left\langle \hat{K}\right\rangle \left\Vert \hat{%
\Psi}\right\Vert ^{2}  \label{IPK} \\
&\simeq &\left( 1-\hat{S}\right) \frac{\hat{\mu}V\sigma _{\hat{K}}^{2}}{2}%
\left( \frac{\frac{\left\Vert \hat{\Psi}_{0}\right\Vert ^{2}}{\hat{\mu}}%
\left( 1-\hat{S}\right) +\left( \frac{1-r}{2r}\right) \left( \left\langle 
\hat{r}\left( X^{\prime }\right) \right\rangle +\frac{\left\langle
f\right\rangle }{2}\right) \hat{S}}{\left( \left\langle \hat{r}\left(
X^{\prime }\right) \right\rangle +\frac{\left\langle f\right\rangle }{2}%
\right) \left( 1-\hat{S}\right) }\right) ^{2}  \notag
\end{eqnarray}%
and in general, for $\frac{\left\Vert \hat{\Psi}_{0}\right\Vert ^{2}}{\hat{%
\mu}}>1$, this is increasing with $\left\langle f_{1}\right\rangle $.

The capital ratio (\ref{Crt}) becomes:%
\begin{eqnarray}
\frac{\left\langle \hat{K}\right\rangle \left\Vert \hat{\Psi}\right\Vert ^{2}%
}{\left\langle K\right\rangle \left\Vert \Psi \right\Vert ^{2}} &\simeq &%
\frac{\left\langle \hat{K}\right\rangle \left\Vert \hat{\Psi}\right\Vert
^{2}\Xi ^{2}}{\left( 1-\left( 1-2z_{0}\right) \frac{\left\langle \hat{K}%
\right\rangle \left\Vert \hat{\Psi}\right\Vert ^{2}\Xi ^{2}}{v}\right) v} \\
&=&\frac{\frac{\hat{\mu}V\sigma _{\hat{K}}^{2}}{2}\left( \frac{\frac{%
\left\Vert \hat{\Psi}_{0}\right\Vert ^{2}}{\hat{\mu}}\left( 1-\hat{S}\right)
+\left( \frac{1-r}{2r}\right) \left( \left\langle \hat{r}\left( X^{\prime
}\right) \right\rangle +\frac{\left\langle f\right\rangle }{2}\right) \hat{S}%
}{\left( \left\langle \hat{r}\left( X^{\prime }\right) \right\rangle +\frac{%
\left\langle f\right\rangle }{2}\right) \left( 1-\hat{S}\right) }\right) ^{2}%
}{\left( 1-\left( 1-2z_{0}\right) \frac{\frac{\hat{\mu}V\sigma _{\hat{K}}^{2}%
}{2}\left( \frac{\frac{\left\Vert \hat{\Psi}_{0}\right\Vert ^{2}}{\hat{\mu}}%
\left( 1-\hat{S}\right) +\left( \frac{1-r}{2r}\right) \left( \left\langle 
\hat{r}\left( X^{\prime }\right) \right\rangle +\frac{\left\langle
f\right\rangle }{2}\right) \hat{S}}{\left( \left\langle \hat{r}\left(
X^{\prime }\right) \right\rangle +\frac{\left\langle f\right\rangle }{2}%
\right) \left( 1-\hat{S}\right) }\right) ^{2}}{\left( \left( \frac{2\epsilon 
}{3\sigma _{\hat{K}}^{2}}\right) ^{\frac{r}{2}}\frac{\left\langle
f_{1}\right\rangle }{C_{0}+\bar{r}}\right) ^{\frac{2}{r}}}\right) \left(
\left( \frac{2\epsilon }{3\sigma _{\hat{K}}^{2}}\right) ^{\frac{r}{2}}\frac{%
\left\langle f_{1}\right\rangle }{C_{0}+\bar{r}}\right) ^{\frac{2}{r}}} 
\notag
\end{eqnarray}

\section*{Appendix 6. Return equations for half averages of shares}

\subsection*{A6.1 Principle}

Having found the averages shares, capital ratio and returns, we can use the
return equation to find $S\left( X\right) $ 
\begin{equation}
\left( \delta \left( X-X^{\prime }\right) -\hat{S}_{E}\left( X^{\prime
},X\right) \right) \frac{\left( 1-\hat{S}\left( X^{\prime }\right) \right)
\left( \hat{f}\left( X^{\prime }\right) -\bar{r}\right) }{1-\hat{S}%
_{E}\left( X^{\prime }\right) }=S_{E}\left( X,X\right) \frac{\left(
1-S\left( X\right) \right) \left( f_{1}^{\prime }\left( X\right) -\bar{r}%
\right) }{1-S_{E}\left( X\right) }  \label{RTns}
\end{equation}%
with:%
\begin{equation*}
C\rightarrow \frac{C}{\left( 1+\underline{k}\left( X^{\prime }\right)
\right) K_{pX}}=\frac{\left( 1-S\left( X^{\prime }\right) \right) }{K_{pX}}C=%
\frac{C}{K_{X}}
\end{equation*}%
and:%
\begin{equation*}
\Delta F_{\tau }\left( \bar{R}\left( K,X\right) \right) =\tau \left(
\left\langle f_{1}\left( X\right) \right\rangle -\left\langle f_{1}\left(
X^{\prime }\right) \right\rangle \right)
\end{equation*}%
so that, for constant return to scale:%
\begin{eqnarray*}
&&\frac{\left( 1-S\left( X\right) \right) \left( f_{1}^{\prime }\left(
X\right) -\bar{r}\right) }{1-S_{E}\left( X\right) } \\
&\rightarrow &f_{1}\left( X\right) -\frac{C}{K_{X}}-\bar{r}+\tau \left(
\left\langle f_{1}\left( X\right) \right\rangle -\left\langle f_{1}\left(
X^{\prime }\right) \right\rangle \right)
\end{eqnarray*}%
In the case of decreasing return,%
\begin{eqnarray*}
&&\frac{\left( 1-S\left( X\right) \right) \left( f_{1}^{\prime }\left(
X\right) -\bar{r}\right) }{1-S_{E}\left( X\right) } \\
&\rightarrow &\frac{f_{1}\left( X\right) -\frac{C}{K_{X}}+\tau \left(
\left\langle f_{1}\left( X\right) \right\rangle -\left\langle f_{1}\left(
X^{\prime }\right) \right\rangle \right) }{\left( K_{X}\left\vert \hat{\Psi}%
\left( X\right) \right\vert ^{2}\right) ^{r}}-C_{0}-\bar{r}
\end{eqnarray*}%
for an equation:%
\begin{eqnarray*}
&&\left( 1-\left\langle \hat{S}\left( X^{\prime }\right) \right\rangle
\right) \left( \left\langle \hat{f}\left( X^{\prime }\right) \right\rangle -%
\bar{r}\right) \\
&=&S_{E}\left( X,X\right) \left[ \frac{f_{1}\left( X\right) -\frac{C}{K_{X}}%
+\tau \left( \left\langle f_{1}\left( X\right) \right\rangle -\left\langle
f_{1}\left( X^{\prime }\right) \right\rangle \right) }{\left(
K_{X}\left\vert \hat{\Psi}\left( X\right) \right\vert ^{2}\right) ^{r}}%
-C_{0}-\bar{r}\right]
\end{eqnarray*}%
Given (\ref{Scr}) and (\ref{Fsh}) $S\left( X\right) $ depend on $%
\left\langle \hat{S}_{E}\left( X^{\prime },X\right) \right\rangle $ We can
consider that:%
\begin{eqnarray*}
\frac{\partial }{\partial \left\langle \hat{S}_{E}\left( X^{\prime
},X\right) \right\rangle }\left( \left\langle K\right\rangle \left\Vert \Psi
\right\Vert ^{2}\right) ^{r} &<&<1 \\
\frac{\partial }{\partial \left\langle \hat{S}_{E}\left( X^{\prime
},X\right) \right\rangle }\frac{C}{\left\langle K\right\rangle \left\Vert
\Psi \right\Vert ^{2}} &<&<1
\end{eqnarray*}%
and the solution with decreasing return to scale corresponds in first
proximation to shift: 
\begin{eqnarray*}
&&f_{1}\left( X\right) -C \\
&\rightarrow &\frac{f_{1}\left( X\right) -\frac{C}{K_{X}}+\tau \left(
\left\langle f_{1}\left( X\right) \right\rangle -\left\langle f_{1}\left(
X^{\prime }\right) \right\rangle \right) }{\left( K_{X}\left\vert \hat{\Psi}%
\left( X\right) \right\vert ^{2}\right) ^{r}}
\end{eqnarray*}%
in the constant return to scale equation.

\subsection*{A6.2 Computation of half-averages}

for $r<<1$ and $C<1$, we can consider in first approximation that:%
\begin{equation*}
f\left( X\right) \simeq \frac{f_{1}\left( X\right) -\frac{C}{K_{X}\left\vert 
\hat{\Psi}\left( X\right) \right\vert ^{2}}}{\left( K_{X}\left\vert \hat{\Psi%
}\left( X\right) \right\vert ^{2}\right) ^{r}}-C_{0}
\end{equation*}%
We compute \ $\left\langle \hat{S}_{E}\left( X^{\prime },X\right)
\right\rangle _{X^{\prime }}$, the capital ratio $\frac{\left\langle \hat{K}%
\right\rangle \left\Vert \hat{\Psi}\right\Vert ^{2}}{\hat{K}_{X^{\prime
}}\left\vert \hat{\Psi}\left( X^{\prime }\right) \right\vert ^{2}}$, $\hat{S}%
_{E}\left( X^{\prime }\right) $ and $\hat{S}\left( X^{\prime }\right) $, $%
S_{E}\left( X,X\right) $, $S\left( X,X\right) $, $S_{E}\left( X\right) $ and 
$S\left( X\right) $.

\subsubsection*{A6.2.1 Calculus of $\left\langle \hat{S}_{E}\left( X^{\prime
},X\right) \right\rangle _{X^{\prime }}$}

Given the formula for $\hat{S}_{E}\left( X^{\prime },X\right) $: 
\begin{eqnarray*}
&&\hat{S}_{E}\left( X^{\prime },X\right) \\
&=&\frac{\underline{\hat{S}}\left( X^{\prime },X\right) }{2} \\
&&+\frac{\hat{w}\left( X^{\prime },X\right) }{2}\left( \hat{w}\left(
X\right) \left( \hat{f}\left( X^{\prime }\right) -\frac{\left\langle \hat{f}%
\left( X^{\prime }\right) \right\rangle _{\hat{w}_{E}}+\left\langle \hat{r}%
\left( X^{\prime }\right) \right\rangle _{\hat{w}_{L}}}{2}\right) +w\left(
X\right) \left( \hat{f}\left( X^{\prime }\right) -\frac{f\left( X\right)
+r\left( X\right) }{2}\right) \right) \\
&=&\frac{\hat{w}\left( X^{\prime },X\right) }{2}\left( 1+\left( \hat{f}%
\left( X^{\prime }\right) -\hat{w}\left( X\right) \frac{\left\langle \hat{f}%
\left( X^{\prime }\right) \right\rangle _{\hat{w}_{E}}+\left\langle \hat{r}%
\left( X^{\prime }\right) \right\rangle _{\hat{w}_{L}}}{2}-w\left( X\right) 
\frac{f\left( X\right) +r\left( X\right) }{2}\right) \right)
\end{eqnarray*}%
and the coefficient of uncertainty $\hat{w}\left( X^{\prime },X\right) $,
given by:%
\begin{eqnarray*}
\frac{\hat{w}}{2}\left( X^{\prime },X\right) &\rightarrow &\frac{\zeta ^{2}}{%
2\left( \zeta ^{2}+\frac{\zeta ^{2}}{\hat{w}_{E}^{\left( 0\right) }\left(
X,X^{\prime }\right) }\left( 1+\frac{\left( \gamma \left\langle \hat{S}%
_{E}\left( X_{1},X^{\prime }\right) \right\rangle _{X_{1}}\right) ^{2}}{%
1-\left( \gamma \left\langle \hat{S}_{E}\left( X^{\prime },X\right)
\right\rangle \right) ^{2}}\right) \right) } \\
&=&\frac{1}{2}\frac{\left( 1-\left( \gamma \left\langle \hat{S}_{E}\left(
X\right) \right\rangle \right) ^{2}\right) \hat{w}_{E}^{\left( 0\right)
}\left( X^{\prime },X\right) }{1+\hat{w}_{E}^{\left( 0\right) }\left(
X^{\prime },X\right) \left( 1-\left( \gamma \left\langle \hat{S}_{E}\left(
X\right) \right\rangle \right) ^{2}\right) +\left( \gamma \left\langle \hat{S%
}_{E}\left( X_{1},X^{\prime }\right) \right\rangle _{X_{1}}\right)
^{2}-\left( \gamma \left\langle \hat{S}_{E}\left( X\right) \right\rangle
\right) ^{2}}
\end{eqnarray*}%
We write the formula for the half average $\left\langle \hat{S}_{E}\left(
X^{\prime },X\right) \right\rangle _{X^{\prime }}$:%
\begin{eqnarray*}
&&\left\langle \hat{S}_{E}\left( X^{\prime },X\right) \right\rangle
_{X^{\prime }} \\
&\approx &\frac{\left\langle \hat{w}\left( X^{\prime },X\right)
\right\rangle _{X^{\prime }}}{2}\left( 1+\left( \hat{w}\left( X\right)
\left( \frac{\left\langle \hat{f}\left( X^{\prime }\right) \right\rangle
-\left\langle \hat{r}\left( X^{\prime }\right) \right\rangle _{\hat{w}_{L}}}{%
2}\right) +w\left( X\right) \left( \left\langle \hat{f}\left( X^{\prime
}\right) \right\rangle -\frac{f\left( X\right) +r\left( X\right) }{2}\right)
\right) \right)
\end{eqnarray*}%
The average $\left\langle \frac{\hat{w}\left( X^{\prime },X\right) }{2}%
\right\rangle _{X^{\prime }}$ is obtained straightforwardly: 
\begin{equation*}
\left\langle \frac{\hat{w}\left( X^{\prime },X\right) }{2}\right\rangle
_{X^{\prime }}\simeq \frac{1}{2}\frac{\left( 1-\left( \gamma \left\langle 
\hat{S}_{E}\left( X\right) \right\rangle \right) ^{2}\right) \left\langle 
\hat{w}_{E}^{\left( 0\right) }\left( X^{\prime },X\right) \right\rangle
_{X^{\prime }}}{1+\left\langle \hat{w}_{E}^{\left( 0\right) }\left(
X^{\prime },X\right) \right\rangle _{X^{\prime }}\left( 1-\left( \gamma
\left\langle \hat{S}_{E}\left( X\right) \right\rangle \right) ^{2}\right) }
\end{equation*}%
Assuming that the distance factor is normalized:%
\begin{equation*}
\left\langle \hat{w}_{E}^{\left( 0\right) }\left( X,X^{\prime }\right)
\right\rangle _{X^{\prime }}\simeq 1
\end{equation*}%
we have:%
\begin{equation*}
\hat{w}\left( X\right) \simeq \left\langle \hat{w}\left( X^{\prime
},X\right) \right\rangle _{X^{\prime }}
\end{equation*}%
and:%
\begin{equation*}
\left\langle \hat{w}\left( X^{\prime },X\right) \right\rangle _{X^{\prime
}}\simeq \left\langle \hat{w}\left( X^{\prime },X\right) \right\rangle
\end{equation*}%
\begin{equation*}
\left\langle \underline{\hat{S}}\left( X^{\prime },X\right) \right\rangle
_{X^{\prime }}\simeq \left\langle \underline{\hat{S}}\left( X^{\prime
},X\right) \right\rangle
\end{equation*}%
As a consequence, the expanded form for $\left\langle \hat{S}_{E}\left(
X^{\prime },X\right) \right\rangle _{X^{\prime }}$ is:%
\begin{eqnarray}
&&\left\langle \hat{S}_{E}\left( X^{\prime },X\right) \right\rangle
_{X^{\prime }}  \label{hv} \\
&\simeq &\frac{\left\langle \underline{\hat{S}}\left( X^{\prime },X\right)
\right\rangle }{2}  \notag \\
&&+\frac{\left\langle \hat{w}\left( X^{\prime },X\right) \right\rangle }{2}%
\left( \hat{w}\left( X\right) \left( \frac{\left\langle \hat{f}\left(
X^{\prime }\right) \right\rangle -\left\langle \hat{r}\left( X^{\prime
}\right) \right\rangle _{\hat{w}_{L}}}{2}\right) +w\left( X\right) \left(
\left\langle \hat{f}\left( X^{\prime }\right) \right\rangle -\frac{f\left(
X\right) +r\left( X\right) }{2}\right) \right)  \notag \\
&\simeq &\frac{\left\langle \hat{w}\left( X^{\prime },X\right) \right\rangle 
}{2}-\frac{\left\langle \hat{w}\left( X^{\prime },X\right) \right\rangle }{2}%
\left\langle w\left( X\right) \right\rangle \Delta \left( \frac{f\left(
X\right) +r\left( X\right) }{2}\right)  \notag \\
&=&\frac{\left\langle \hat{w}\left( X^{\prime },X\right) \right\rangle }{2}%
\left( 1-\left\langle w\left( X\right) \right\rangle \Delta \left( \frac{%
f\left( X\right) +r\left( X\right) }{2}\right) +\frac{\left\langle \hat{f}%
\left( X^{\prime }\right) \right\rangle -\left\langle \hat{r}\left(
X^{\prime }\right) \right\rangle _{\hat{w}_{L}}}{2}\right)  \notag
\end{eqnarray}%
with:%
\begin{equation*}
\Delta \left( \frac{f\left( X\right) +r\left( X\right) }{2}\right) =\left( 
\frac{f\left( X\right) +r\left( X\right) }{2}-\frac{\left\langle \hat{f}%
\left( X^{\prime }\right) \right\rangle +\left\langle \hat{r}\left(
X^{\prime }\right) \right\rangle _{\hat{w}_{L}}}{2}\right)
\end{equation*}%
\begin{equation*}
\left\langle \hat{w}\left( X^{\prime },X\right) \right\rangle =\frac{\left(
1-\left( \gamma \left\langle \hat{S}_{E}\left( X\right) \right\rangle
\right) ^{2}\right) }{1+\left( 1-\left( \gamma \left\langle \hat{S}%
_{E}\left( X\right) \right\rangle \right) ^{2}\right) }
\end{equation*}%
\begin{equation*}
\left\langle w\left( X\right) \right\rangle =\frac{1}{1+\left( 1-\left(
\gamma \left\langle \hat{S}_{E}\left( X\right) \right\rangle \right)
^{2}\right) }
\end{equation*}

\subsubsection*{A6.2.2 Capital ratio with constant return to scale}

The average $\left\langle \hat{K}\right\rangle \left\Vert \hat{\Psi}%
\right\Vert ^{2}$ is given by (\ref{VRk}).

To derive $\hat{K}_{X^{\prime }}\left\vert \hat{\Psi}\left( X^{\prime
}\right) \right\vert ^{2}$,\ we can expand formula (\ref{RTns}) by computing
the action of the non local operator $\frac{1}{1-\hat{S}}$ on $\hat{f}$:%
\begin{eqnarray*}
&&\left( \frac{1}{1-\hat{S}}\hat{f}\right) \left( X^{\prime }\right) \\
&=&\hat{f}\left( X^{\prime }\right) +\frac{\left\langle \hat{S}\left(
X,X^{\prime }\right) \right\rangle _{X}\left\langle \hat{f}\right\rangle }{%
1-\left\langle \hat{S}\right\rangle }=\frac{\hat{f}\left( X^{\prime }\right)
\left( 1-\left\langle \hat{S}\right\rangle \right) +\left\langle \hat{S}%
\left( X,X^{\prime }\right) \right\rangle _{X}\left\langle \hat{f}%
\right\rangle }{1-\left\langle \hat{S}\right\rangle }
\end{eqnarray*}%
In Gosselin and Lotz (2024), we obtained:%
\begin{equation*}
\hat{K}\left[ \hat{X}^{\prime }\right] \simeq \left\langle \hat{K}%
\right\rangle \left\Vert \hat{\Psi}\right\Vert ^{2}\left( \frac{\left\langle 
\hat{g}\right\rangle }{\hat{g}\left( \hat{X}^{\prime }\right) }\frac{\hat{k}%
\left( X^{\prime },\left\langle X\right\rangle \right) }{\hat{k}}\right) ^{2}
\end{equation*}%
We use that:%
\begin{equation*}
\hat{k}=\frac{\left\langle \hat{S}\left( X^{\prime }\right) \right\rangle }{%
1-\left\langle \hat{S}\left( X^{\prime }\right) \right\rangle }=\frac{%
\left\langle \hat{S}\left( X^{\prime },X\right) \right\rangle }{%
1-\left\langle \hat{S}\left( X^{\prime },X\right) \right\rangle }
\end{equation*}%
and:%
\begin{equation*}
\hat{S}\left( X^{\prime }\right) =\frac{\hat{k}\left( X^{\prime
},\left\langle X\right\rangle \right) \frac{\left\langle \hat{K}%
\right\rangle \left\Vert \hat{\Psi}\right\Vert ^{2}}{\hat{K}\left[ \hat{X}%
^{\prime }\right] }}{1+\hat{k}\left( X^{\prime },\left\langle X\right\rangle
\right) \frac{\left\langle \hat{K}\right\rangle \left\Vert \hat{\Psi}%
\right\Vert ^{2}}{\hat{K}\left[ \hat{X}^{\prime }\right] }}
\end{equation*}%
so that:%
\begin{equation*}
\hat{k}\left( X^{\prime },\left\langle X\right\rangle \right) =\frac{\hat{S}%
\left( X^{\prime }\right) }{1-\hat{S}\left( X^{\prime }\right) }\frac{\hat{K}%
\left[ \hat{X}^{\prime }\right] }{\left\langle \hat{K}\right\rangle
\left\Vert \hat{\Psi}\right\Vert ^{2}}=\frac{\left\langle \hat{S}\left(
X^{\prime },X\right) \right\rangle _{X}}{1-\hat{S}\left( X^{\prime }\right) }
\end{equation*}%
and the ratio $\frac{\hat{k}\left( X^{\prime },\left\langle X\right\rangle
\right) }{\hat{k}}$ becomes: 
\begin{equation*}
\frac{\hat{k}\left( X^{\prime },\left\langle X\right\rangle \right) }{\hat{k}%
}=\frac{\frac{\left\langle \hat{S}\left( X^{\prime },X\right) \right\rangle
_{X}}{1-\hat{S}\left( X^{\prime }\right) }}{\frac{\left\langle \hat{S}\left(
X^{\prime },X\right) \right\rangle }{1-\left\langle \hat{S}\left( X^{\prime
},X\right) \right\rangle }}
\end{equation*}%
The disposable capital that follows the equation:%
\begin{equation}
\hat{K}\left[ \hat{X}^{\prime }\right] =\left\langle \hat{K}\right\rangle
\left\Vert \hat{\Psi}\right\Vert ^{2}\left( \frac{\left\langle \hat{g}%
\right\rangle }{\hat{g}\left( \hat{X}^{\prime }\right) }\frac{\frac{%
\left\langle \hat{S}\left( X^{\prime },X\right) \right\rangle _{X}}{%
1-\left\langle \hat{S}\left( X^{\prime },X\right) \right\rangle _{X}\frac{%
\left\langle \hat{K}\right\rangle \left\Vert \hat{\Psi}\right\Vert ^{2}}{%
\hat{K}\left[ \hat{X}^{\prime }\right] }}}{\frac{\left\langle \hat{S}\left(
X^{\prime },X\right) \right\rangle }{1-\left\langle \hat{S}\left( X^{\prime
},X\right) \right\rangle }}\right) ^{2}  \label{Gnr}
\end{equation}%
with solution depending on average disposable capital:%
\begin{equation*}
\hat{K}\left[ \hat{X}^{\prime }\right] =\frac{1}{2}\frac{\left( \frac{%
\left\langle \hat{g}\right\rangle }{\hat{g}\left( \hat{X}^{\prime }\right) }%
\frac{\left\langle \hat{S}\left( X^{\prime },X\right) \right\rangle
_{X}\left( 1-\left\langle \hat{S}\left( X^{\prime },X\right) \right\rangle
\right) }{\left\langle \hat{S}\left( X^{\prime },X\right) \right\rangle }%
\right) ^{2}\left( 1-\sqrt{1+\frac{4\left\langle \hat{S}\left( X^{\prime
},X\right) \right\rangle _{X}}{\left( \frac{\left\langle \hat{g}%
\right\rangle }{\hat{g}\left( \hat{X}^{\prime }\right) }\frac{\left\langle 
\hat{S}\left( X^{\prime },X\right) \right\rangle _{X}\left( 1-\left\langle 
\hat{S}\left( X^{\prime },X\right) \right\rangle \right) }{\left\langle \hat{%
S}\left( X^{\prime },X\right) \right\rangle }\right) ^{2}}}\right)
+2\left\langle \hat{S}\left( X^{\prime },X\right) \right\rangle _{X}}{\left(
\left\langle \hat{S}\left( X^{\prime },X\right) \right\rangle _{X}\right)
^{2}}
\end{equation*}%
$\allowbreak $The capital ratio at the lowest order is:%
\begin{equation*}
\frac{\left\langle \hat{K}\right\rangle \left\Vert \hat{\Psi}\right\Vert ^{2}%
}{\hat{K}\left[ \hat{X}^{\prime }\right] }\simeq \left( \frac{\hat{g}\left( 
\hat{X}^{\prime }\right) }{\left\langle \hat{g}\right\rangle }\frac{\frac{%
\left\langle \hat{S}\left( X^{\prime },X\right) \right\rangle }{%
1-\left\langle \hat{S}\left( X^{\prime },X\right) \right\rangle }}{\frac{%
\left\langle \hat{S}\left( X^{\prime },X\right) \right\rangle _{X}}{1-\hat{S}%
\left( X^{\prime }\right) }}\right) ^{2}
\end{equation*}%
which writes:%
\begin{eqnarray}
\frac{\left\langle \hat{K}\right\rangle \left\Vert \hat{\Psi}\right\Vert ^{2}%
}{\hat{K}_{X^{\prime }}\left\vert \hat{\Psi}\left( X^{\prime }\right)
\right\vert ^{2}} &\simeq &\frac{\left( \left( \hat{f}\left( X^{\prime
}\right) +\frac{\left\langle \hat{S}\left( X,X^{\prime }\right)
\right\rangle _{X}\left\langle \hat{f}\right\rangle }{1-\left\langle \hat{S}%
\right\rangle }\right) \frac{\left\langle \hat{S}\left( X^{\prime },X\right)
\right\rangle }{1-\left\langle \hat{S}\left( X^{\prime },X\right)
\right\rangle }\right) ^{2}}{\left( \frac{\left\langle \hat{f}\right\rangle 
}{1-\left\langle \hat{S}\right\rangle }\frac{\left\langle \hat{S}\left(
X^{\prime },X\right) \right\rangle _{X}}{1-\left\langle \hat{S}\left(
X^{\prime },X\right) \right\rangle _{X}}\right) ^{2}}  \label{RV} \\
&=&\left( \frac{\left( \hat{f}\left( X^{\prime }\right) \left(
1-\left\langle \hat{S}\right\rangle \right) +\left\langle \hat{S}\left(
X,X^{\prime }\right) \right\rangle _{X}\left\langle \hat{f}\right\rangle
\right) \frac{\left\langle \hat{S}\left( X^{\prime },X\right) \right\rangle 
}{1-\left\langle \hat{S}\left( X^{\prime },X\right) \right\rangle }}{%
\left\langle \hat{f}\right\rangle \frac{\left\langle \hat{S}\left( X^{\prime
},X\right) \right\rangle _{X}}{1-\left\langle \hat{S}\left( X^{\prime
},X\right) \right\rangle _{X}}}\right) ^{2}  \notag
\end{eqnarray}%
and at the first order the capital ratio is:%
\begin{equation*}
\frac{\left\langle \hat{K}\right\rangle \left\Vert \hat{\Psi}\right\Vert ^{2}%
}{\hat{K}_{X^{\prime }}\left\vert \hat{\Psi}\left( X^{\prime }\right)
\right\vert ^{2}}\simeq \left( \frac{\hat{g}\left( \hat{X}^{\prime }\right) 
}{\left\langle \hat{g}\right\rangle }\frac{\frac{\left\langle \hat{S}\left(
X^{\prime },X\right) \right\rangle }{1-\left\langle \hat{S}\left( X^{\prime
},X\right) \right\rangle }}{\frac{\left\langle \hat{S}\left( X^{\prime
},X\right) \right\rangle _{X}}{1-\left\langle \hat{S}\left( X^{\prime
},X\right) \right\rangle _{X}\frac{\hat{g}\left( \hat{X}^{\prime }\right) 
\frac{\left\langle \hat{S}\left( X^{\prime },X\right) \right\rangle }{%
1-\left\langle \hat{S}\left( X^{\prime },X\right) \right\rangle }}{%
\left\langle \hat{g}\right\rangle \frac{\left\langle \hat{S}\left( X^{\prime
},X\right) \right\rangle _{X}}{1-\left\langle \hat{S}\left( X^{\prime
},X\right) \right\rangle _{X}}}}}\right) ^{2}
\end{equation*}%
The capital ratio between investors and firms to the first order first order
writes:%
\begin{eqnarray*}
&&\frac{\hat{K}_{X}\left\vert \hat{\Psi}\left( X\right) \right\vert ^{2}}{%
K_{X}\left\vert \Psi \left( X\right) \right\vert ^{2}} \\
&\rightarrow &\frac{2\hat{\mu}\epsilon V\sigma _{\hat{K}}^{2}\left(
1-\left\langle \hat{S}\right\rangle \right) ^{2}f_{1}^{2}\left( X\right) }{%
\left( \hat{f}\left( X^{\prime }\right) \left( 1-\left\langle \hat{S}%
\right\rangle \right) +\left\langle \hat{S}\left( X,X^{\prime }\right)
\right\rangle _{X}\left\langle \hat{f}\right\rangle \right) ^{2}}\left( 
\frac{\left\vert \hat{\Psi}_{0}\left( X\right) \right\vert ^{2}}{\left\vert
\Psi _{0}\left( X\right) \right\vert ^{2}}\right) ^{2} \\
&&\times \left[ \frac{\left( 1-S_{E}\left( X,X\right) \frac{2\hat{\mu}%
\epsilon V\sigma _{\hat{K}}^{2}\left( 1-\left\langle \hat{S}\right\rangle
\right) ^{2}f_{1}^{2}\left( X\right) }{\left( \hat{f}\left( X^{\prime
}\right) \left( 1-\left\langle \hat{S}\right\rangle \right) +\left\langle 
\hat{S}\left( X,X^{\prime }\right) \right\rangle _{X}\left\langle \hat{f}%
\right\rangle \right) ^{2}}\left( \frac{\left\vert \hat{\Psi}_{0}\left(
X\right) \right\vert ^{2}}{\left\vert \Psi _{0}\left( X\right) \right\vert
^{2}}\right) ^{2}\right) ^{2}}{\left( 1-S_{E}\left( X,X\right) \frac{2\hat{%
\mu}\epsilon V\sigma _{\hat{K}}^{2}\left( 1-\left\langle \hat{S}%
\right\rangle \right) ^{2}f_{1}^{2}\left( X\right) }{\left( \hat{f}\left(
X^{\prime }\right) \left( 1-\left\langle \hat{S}\right\rangle \right)
+\left\langle \hat{S}\left( X,X^{\prime }\right) \right\rangle
_{X}\left\langle \hat{f}\right\rangle \right) ^{2}}\left( \frac{\left\vert 
\hat{\Psi}_{0}\left( X\right) \right\vert ^{2}}{\left\vert \Psi _{0}\left(
X\right) \right\vert ^{2}}\right) ^{2}\right) ^{2}}+\frac{S_{E}^{2}\left(
X,X\right) +2S\left( X,X\right) }{3S^{2}\left( X,X\right) }\right]
\end{eqnarray*}

\subsubsection*{A6.2.3 Capital ratio for decreasing return to scale}

For decreasing return to scale we obtain:%
\begin{eqnarray*}
K_{X}\left\vert \Psi \left( X\right) \right\vert ^{2} &=&\left( 1-S\left(
X\right) \right) \left( \left( \frac{2\epsilon }{3\sigma _{\hat{K}}^{2}}%
\right) ^{\frac{r}{2}}\left( \frac{f_{1}\left( X\right) }{C_{0}+\bar{r}}%
\right) \right) ^{\frac{2}{r}} \\
&\simeq &\left( 1-S\left( X,X\right) \frac{\hat{K}_{X}\left\vert \hat{\Psi}%
\left( X\right) \right\vert ^{2}}{\left( \left( \frac{2\epsilon }{3\sigma _{%
\hat{K}}^{2}}\right) ^{\frac{r}{2}}\left( \frac{f_{1}\left( X\right) }{C_{0}+%
\bar{r}}\right) \right) ^{\frac{2}{r}}}\right) \left( \left( \frac{2\epsilon 
}{3\sigma _{\hat{K}}^{2}}\right) ^{\frac{r}{2}}\left( \frac{f_{1}\left(
X\right) }{C_{0}+\bar{r}}\right) \right) ^{\frac{2}{r}}
\end{eqnarray*}%
We use (\ref{Gnr}) in first approximation:%
\begin{eqnarray}
\hat{K}_{X}\left\vert \hat{\Psi}\left( X\right) \right\vert ^{2} &\simeq
&\left\langle \hat{K}\right\rangle \left\Vert \hat{\Psi}\right\Vert
^{2}\left( \frac{\left\langle \hat{g}\right\rangle }{\hat{g}\left( \hat{X}%
\right) }\frac{\frac{\left\langle \hat{S}\left( X,X^{\prime }\right)
\right\rangle _{X^{\prime }}}{1-\left\langle \hat{S}\left( X,X^{\prime
}\right) \right\rangle _{X^{\prime }}}}{\frac{\left\langle \hat{S}\left(
X,X^{\prime }\right) \right\rangle }{1-\left\langle \hat{S}\left(
X,X^{\prime }\right) \right\rangle }}\right) ^{2}  \label{PRkh} \\
&\simeq &\left\langle \hat{K}\right\rangle \left\Vert \hat{\Psi}\right\Vert
^{2}\left( \frac{\left\langle \hat{f}\right\rangle }{\hat{f}\left( X\right)
\left( 1-\left\langle \hat{S}\right\rangle \right) +\left\langle \hat{S}%
\left( X^{\prime },X\right) \right\rangle _{X^{\prime }}\left\langle \hat{f}%
\right\rangle }\frac{\frac{\left\langle \hat{S}\left( X,X^{\prime }\right)
\right\rangle _{X^{\prime }}}{1-\left\langle \hat{S}\left( X,X^{\prime
}\right) \right\rangle _{X^{\prime }}}}{\frac{\left\langle \hat{S}\left(
X,X^{\prime }\right) \right\rangle }{1-\left\langle \hat{S}\left(
X,X^{\prime }\right) \right\rangle }}\right) ^{2}  \notag
\end{eqnarray}%
and the ratio becomes:%
\begin{equation}
\frac{\hat{K}_{X}\left\vert \hat{\Psi}\left( X\right) \right\vert ^{2}}{%
K_{X}\left\vert \Psi \left( X\right) \right\vert ^{2}}\simeq \frac{\hat{K}%
_{X}\left\vert \hat{\Psi}\left( X\right) \right\vert ^{2}}{\left( 1-\frac{%
S\left( X,X\right) \hat{K}_{X}\left\vert \hat{\Psi}\left( X\right)
\right\vert ^{2}}{\left( \left( \frac{2\epsilon }{3\sigma _{\hat{K}}^{2}}%
\right) ^{\frac{r}{2}}\left( \frac{f_{1}\left( X\right) }{C_{0}+\bar{r}}%
\right) \right) ^{\frac{2}{r}}}\right) \left( \left( \frac{2\epsilon }{%
3\sigma _{\hat{K}}^{2}}\right) ^{\frac{r}{2}}\left( \frac{f_{1}\left(
X\right) }{C_{0}+\bar{r}}\right) \right) ^{\frac{2}{r}}}  \label{crd}
\end{equation}

\subsubsection*{A6.2.4 Formula for $\hat{S}_{E}\left( X^{\prime }\right) $
nd $\hat{S}\left( X^{\prime }\right) $}

Using:%
\begin{equation*}
\hat{S}_{E}\left( X\right) =\left\langle \hat{S}_{E}\left( X,X^{\prime
}\right) \right\rangle _{X^{\prime }}\frac{\left\langle \hat{K}\right\rangle
\left\Vert \hat{\Psi}\right\Vert ^{2}}{K_{X}\left\vert \Psi \left( X\right)
\right\vert ^{2}}
\end{equation*}%
\begin{equation*}
\hat{S}\left( X\right) =\left\langle \hat{S}\left( X,X^{\prime }\right)
\right\rangle _{X^{\prime }}\frac{\left\langle \hat{K}\right\rangle
\left\Vert \hat{\Psi}\right\Vert ^{2}}{K_{X}\left\vert \Psi \left( X\right)
\right\vert ^{2}}
\end{equation*}%
and:%
\begin{eqnarray*}
&&\left\langle \hat{w}\left( X^{\prime },X\right) \right\rangle _{X} \\
&=&\frac{\left( 1-\left( \gamma \left\langle \hat{S}_{E}\left( X\right)
\right\rangle \right) ^{2}\right) }{2-\left( \gamma \left\langle \hat{S}%
_{E}\left( X\right) \right\rangle \right) ^{2}+\left( \gamma \left\langle 
\hat{S}_{E}\left( X_{1},X^{\prime }\right) \right\rangle _{X_{1}}\right)
^{2}-\left( \gamma \left\langle \hat{S}_{E}\left( X\right) \right\rangle
\right) ^{2}} \\
&\simeq &\frac{\left( 1-\left( \gamma \left\langle \hat{S}_{E}\left(
X\right) \right\rangle \right) ^{2}\right) }{2-\left( \gamma \left\langle 
\hat{S}_{E}\left( X\right) \right\rangle \right) ^{2}-\gamma \left\langle 
\hat{S}_{E}\left( X\right) \right\rangle \gamma \left\langle \hat{w}\left(
X^{\prime },X\right) \right\rangle \left\langle w\left( X\right)
\right\rangle \Delta \left( \frac{f\left( X^{\prime }\right) +r\left(
X^{\prime }\right) }{2}\right) }
\end{eqnarray*}%
\begin{eqnarray}
&&\hat{S}_{E}\left( X^{\prime }\right)  \label{SNx} \\
&\simeq &\frac{\left\langle \hat{w}\left( X^{\prime },X\right) \right\rangle
_{X}}{2}  \notag \\
&&\times \left( 1+\hat{f}\left( X^{\prime }\right) -\left( \left\langle \hat{%
w}\left( X\right) \right\rangle \frac{\left\langle \hat{f}\left( X^{\prime
}\right) \right\rangle _{\hat{w}_{E}}+\left\langle \hat{r}\left( X^{\prime
}\right) \right\rangle _{\hat{w}_{L}}}{2}+\left\langle w\left( X\right)
\right\rangle \frac{\left\langle f\left( X\right) \right\rangle
+\left\langle r\left( X\right) \right\rangle }{2}\right) \right) \frac{%
\left\langle \hat{K}\right\rangle \left\Vert \hat{\Psi}\right\Vert ^{2}}{%
\hat{K}_{X^{\prime }}\left\vert \hat{\Psi}\left( X^{\prime }\right)
\right\vert ^{2}}  \notag \\
&\rightarrow &\frac{1}{2}\frac{\left( 1-\left( \gamma \left\langle \hat{S}%
_{E}\left( X\right) \right\rangle \right) ^{2}\right) \left( 1+\Delta \hat{f}%
\left( X^{\prime }\right) \right) }{2-\left( \gamma \left\langle \hat{S}%
_{E}\left( X\right) \right\rangle \right) ^{2}+\left( \gamma \left\langle 
\hat{S}_{E}\left( X_{1},X^{\prime }\right) \right\rangle _{X_{1}}\right)
^{2}-\left( \gamma \left\langle \hat{S}_{E}\left( X\right) \right\rangle
\right) ^{2}}\frac{\left\langle \hat{K}\right\rangle \left\Vert \hat{\Psi}%
\right\Vert ^{2}}{\hat{K}_{X^{\prime }}\left\vert \hat{\Psi}\left( X^{\prime
}\right) \right\vert ^{2}}  \notag \\
&\rightarrow &\frac{1}{2}\frac{\left( 1-\left( \gamma \left\langle \hat{S}%
_{E}\left( X\right) \right\rangle \right) ^{2}\right) \left( 1+\Delta \hat{f}%
\left( X^{\prime }\right) \right) }{2-\left( \gamma \left\langle \hat{S}%
_{E}\left( X\right) \right\rangle \right) ^{2}-\gamma ^{2}\left\langle \hat{S%
}_{E}\left( X\right) \right\rangle \left\langle \hat{w}\left( X^{\prime
},X\right) \right\rangle \left\langle w\left( X\right) \right\rangle \Delta
\left( \frac{f\left( X^{\prime }\right) +r\left( X^{\prime }\right) }{2}%
\right) }\frac{\left\langle \hat{K}\right\rangle \left\Vert \hat{\Psi}%
\right\Vert ^{2}}{\hat{K}_{X^{\prime }}\left\vert \hat{\Psi}\left( X^{\prime
}\right) \right\vert ^{2}}  \notag
\end{eqnarray}%
where:%
\begin{equation}
\Delta \hat{f}\left( X^{\prime }\right) =\hat{f}\left( X^{\prime }\right)
-\left( \left\langle \hat{w}\left( X\right) \right\rangle \frac{\left\langle 
\hat{f}\left( X^{\prime }\right) \right\rangle _{\hat{w}_{E}}+\left\langle 
\hat{r}\left( X^{\prime }\right) \right\rangle _{\hat{w}_{L}}}{2}%
+\left\langle w\left( X\right) \right\rangle \frac{\left\langle f\left(
X\right) \right\rangle +\left\langle r\left( X\right) \right\rangle }{2}%
\right)  \label{DF}
\end{equation}

\begin{eqnarray}
&&\hat{S}\left( X^{\prime }\right)  \label{St} \\
&\simeq &\left\langle \hat{w}\left( X^{\prime },X\right) \right\rangle
_{X}\left( 1+\left( \frac{\hat{f}\left( X^{\prime }\right) +\hat{r}\left(
X^{\prime }\right) }{2}-\left\langle \hat{w}\left( X\right) \right\rangle 
\frac{\left\langle \hat{f}\left( X^{\prime }\right) \right\rangle _{\hat{w}%
_{E}}+\left\langle \hat{r}\left( X^{\prime }\right) \right\rangle _{\hat{w}%
_{L}}}{2}\right. \right.  \notag \\
&&\left. \left. -\left\langle w\left( X\right) \right\rangle \frac{%
\left\langle f\left( X\right) +r\left( X\right) \right\rangle _{w}}{2}%
\right) \right) \frac{\left\langle \hat{K}\right\rangle \left\Vert \hat{\Psi}%
\right\Vert ^{2}}{\hat{K}_{X^{\prime }}\left\vert \hat{\Psi}\left( X^{\prime
}\right) \right\vert ^{2}} \\
&\rightarrow &\frac{\left( 1-\left( \gamma \left\langle \hat{S}_{E}\left(
X\right) \right\rangle \right) ^{2}\right) \left( 1+\frac{\Delta \hat{f}%
\left( X^{\prime }\right) +\Delta \hat{r}\left( X^{\prime }\right) }{2}%
\right) }{2-\left( \gamma \left\langle \hat{S}_{E}\left( X\right)
\right\rangle \right) ^{2}-\gamma ^{2}\left\langle \hat{S}_{E}\left(
X\right) \right\rangle \left\langle \hat{w}\left( X^{\prime },X\right)
\right\rangle \left\langle w\left( X\right) \right\rangle \Delta \left( 
\frac{f\left( X^{\prime }\right) +r\left( X^{\prime }\right) }{2}\right) }%
\frac{\left\langle \hat{K}\right\rangle \left\Vert \hat{\Psi}\right\Vert ^{2}%
}{\hat{K}_{X^{\prime }}\left\vert \hat{\Psi}\left( X^{\prime }\right)
\right\vert ^{2}}  \notag
\end{eqnarray}%
with:%
\begin{equation}
\frac{\Delta \hat{f}\left( X^{\prime }\right) +\Delta \hat{r}\left(
X^{\prime }\right) }{2}=\frac{\hat{f}\left( X^{\prime }\right) +\hat{r}%
\left( X^{\prime }\right) }{2}-\left\langle \hat{w}\left( X\right)
\right\rangle \frac{\left\langle \hat{f}\left( X^{\prime }\right)
\right\rangle _{\hat{w}_{E}}+\left\langle \hat{r}\left( X^{\prime }\right)
\right\rangle _{\hat{w}_{L}}}{2}-\left\langle w\left( X\right) \right\rangle 
\frac{\left\langle f\left( X\right) +r\left( X\right) \right\rangle _{w}}{2}
\label{Dfr}
\end{equation}%
We als have:%
\begin{equation*}
\hat{S}_{L}\left( X^{\prime }\right) \simeq \frac{1}{2}\frac{\left( 1-\left(
\gamma \left\langle \hat{S}_{E}\left( X\right) \right\rangle \right)
^{2}\right) \left( 1+\Delta \hat{r}\left( X^{\prime }\right) \right) }{%
2-\left( \gamma \left\langle \hat{S}_{E}\left( X\right) \right\rangle
\right) ^{2}-\gamma ^{2}\left\langle \hat{S}_{E}\left( X\right)
\right\rangle \left\langle \hat{w}\left( X^{\prime },X\right) \right\rangle
\left\langle w\left( X\right) \right\rangle \Delta \left( \frac{f\left(
X^{\prime }\right) +r\left( X^{\prime }\right) }{2}\right) }\frac{%
\left\langle \hat{K}\right\rangle \left\Vert \hat{\Psi}\right\Vert ^{2}}{%
\hat{K}_{X^{\prime }}\left\vert \hat{\Psi}\left( X^{\prime }\right)
\right\vert ^{2}}
\end{equation*}

\subsubsection*{A6.2.5 Formula for $S_{E}\left( X,X\right) $, $S\left(
X,X\right) $, $S_{E}\left( X\right) $ and $S\left( X\right) $}

\begin{eqnarray}
&&S_{E}\left( X\right)  \label{Snx} \\
&=&\frac{w\left( X\right) }{2}\left( 1+\left( \hat{w}\left( X\right) \left(
f\left( X\right) -\frac{\left\langle \hat{f}\left( X^{\prime }\right)
\right\rangle _{\hat{w}_{E}}+\left\langle \hat{r}\left( X^{\prime }\right)
\right\rangle _{\hat{w}_{L}}}{2}\right) +\frac{w\left( X\right) }{2}\left(
f\left( X\right) -\bar{r}\left( X\right) \right) \right) \right) \frac{\hat{K%
}_{X}\left\vert \hat{\Psi}\left( X\right) \right\vert ^{2}}{K_{X}\left\vert
\Psi \left( X\right) \right\vert ^{2}}  \notag \\
&\equiv &S_{E}\left( X,X\right) \frac{\hat{K}_{X}\left\vert \hat{\Psi}\left(
X\right) \right\vert ^{2}}{K_{X}\left\vert \Psi \left( X\right) \right\vert
^{2}}  \notag
\end{eqnarray}%
\begin{eqnarray}
S\left( X\right) &=&w\left( X\right) \left( 1+\left( \hat{w}\left( X\right)
\left( \frac{f\left( X\right) +\bar{r}\left( X\right) }{2}-\frac{%
\left\langle \hat{f}\left( X^{\prime }\right) \right\rangle _{\hat{w}%
_{E}}+\left\langle \hat{r}\left( X^{\prime }\right) \right\rangle _{\hat{w}%
_{L}}}{2}\right) \right) \right) \frac{\hat{K}_{X}\left\vert \hat{\Psi}%
\left( X\right) \right\vert ^{2}}{K_{X}\left\vert \Psi \left( X\right)
\right\vert ^{2}}  \label{Sx} \\
&=&S\left( X,X\right) \frac{\hat{K}_{X}\left\vert \hat{\Psi}\left( X\right)
\right\vert ^{2}}{K_{X}\left\vert \Psi \left( X\right) \right\vert ^{2}} 
\notag
\end{eqnarray}%
\begin{equation*}
\left\langle \frac{\hat{w}\left( X^{\prime },X\right) }{2}\right\rangle
_{X^{\prime }}\simeq \frac{\left( 1-\left( \gamma \left\langle \hat{S}%
_{E}\left( X\right) \right\rangle \right) ^{2}\right) \left\langle \hat{w}%
_{E}^{\left( 0\right) }\left( X^{\prime },X\right) \right\rangle _{X^{\prime
}}}{1+\left\langle \hat{w}_{E}^{\left( 0\right) }\left( X^{\prime },X\right)
\right\rangle _{X^{\prime }}\left( 1-\left( \gamma \left\langle \hat{S}%
_{E}\left( X\right) \right\rangle \right) ^{2}\right) }
\end{equation*}%
\begin{equation*}
\left\langle \hat{w}_{E}^{\left( 0\right) }\left( X,X^{\prime }\right)
\right\rangle _{X^{\prime }}\simeq 1
\end{equation*}%
\begin{eqnarray*}
\hat{w}\left( X\right) &\simeq &\left\langle \hat{w}\left( X^{\prime
},X\right) \right\rangle _{X^{\prime }} \\
&\rightarrow &\left\langle \hat{w}\left( X^{\prime },X\right) \right\rangle
\end{eqnarray*}%
\begin{eqnarray*}
&&w\left( X\right) \\
&\rightarrow &\left\langle w\left( X\right) \right\rangle =1-\left\langle 
\hat{w}\left( X\right) \right\rangle
\end{eqnarray*}

\subsubsection{A6.2.6 Formula for $\hat{S}_{E}\left( X\right) $ in terms of $%
\hat{f}\left( X^{\prime }\right) $}

Using (\ref{SNx}):%
\begin{equation}
\left\langle \hat{S}_{E}\left( X^{\prime },X\right) \right\rangle _{X}\simeq 
\frac{1}{2}\frac{\left( 1-\left( \gamma \left\langle \hat{S}_{E}\left(
X\right) \right\rangle \right) ^{2}\right) \left( 1+\Delta \hat{f}\left(
X^{\prime }\right) \right) }{2-\left( \gamma \left\langle \hat{S}_{E}\left(
X\right) \right\rangle \right) ^{2}+\left( \gamma \left\langle \hat{S}%
_{E}\left( X_{1},X^{\prime }\right) \right\rangle _{X_{1}}\right)
^{2}-\left( \gamma \left\langle \hat{S}_{E}\left( X\right) \right\rangle
\right) ^{2}}
\end{equation}%
The expression for $\left\langle \hat{S}_{E}\left( X,X^{\prime }\right)
\right\rangle _{X}$ is given by (\ref{hv}):%
\begin{equation}
\left\langle \hat{S}_{E}\left( X^{\prime },X\right) \right\rangle
_{X^{\prime }}\simeq \frac{\left\langle \hat{w}\left( X^{\prime },X\right)
\right\rangle }{2}\left( 1-\left\langle w\left( X\right) \right\rangle
\Delta \left( \frac{f\left( X\right) +r\left( X\right) }{2}\right) +\frac{%
\left\langle \hat{f}\left( X^{\prime }\right) \right\rangle -\left\langle 
\hat{r}\left( X^{\prime }\right) \right\rangle _{\hat{w}_{L}}}{2}\right)
\end{equation}%
where:%
\begin{equation*}
\Delta \hat{f}\left( X^{\prime }\right) =\hat{f}\left( X^{\prime }\right)
-\left( \left\langle \hat{w}\left( X\right) \right\rangle \frac{\left\langle 
\hat{f}\left( X^{\prime }\right) \right\rangle _{\hat{w}_{E}}+\left\langle 
\hat{r}\left( X^{\prime }\right) \right\rangle _{\hat{w}_{L}}}{2}%
+\left\langle w\left( X\right) \right\rangle \frac{\left\langle f\left(
X\right) \right\rangle +\left\langle r\left( X\right) \right\rangle }{2}%
\right)
\end{equation*}%
Using (\ref{RV}):%
\begin{eqnarray}
\hat{S}_{E}\left( X^{\prime }\right) &\simeq &\frac{1}{2}\frac{\left(
1-\left( \gamma \left\langle \hat{S}_{E}\left( X\right) \right\rangle
\right) ^{2}\right) \left( 1+\Delta \hat{f}\left( X^{\prime }\right) \right) 
}{2-\left( \gamma \left\langle \hat{S}_{E}\left( X\right) \right\rangle
\right) ^{2}+\left( \gamma \left\langle \hat{S}_{E}\left( X_{1},X^{\prime
}\right) \right\rangle _{X_{1}}\right) ^{2}-\left( \gamma \left\langle \hat{S%
}_{E}\left( X\right) \right\rangle \right) ^{2}}\frac{\left\langle \hat{K}%
\right\rangle \left\Vert \hat{\Psi}\right\Vert ^{2}}{\hat{K}_{X^{\prime
}}\left\vert \hat{\Psi}\left( X^{\prime }\right) \right\vert ^{2}}
\label{SNht} \\
&\simeq &\frac{1}{2}\frac{\left( 1-\left( \gamma \left\langle \hat{S}%
_{E}\left( X\right) \right\rangle \right) ^{2}\right) \left( 1+\Delta \hat{f}%
\left( X^{\prime }\right) \right) }{2-\left( \gamma \left\langle \hat{S}%
_{E}\left( X\right) \right\rangle \right) ^{2}+\left( \gamma \left\langle 
\hat{S}_{E}\left( X_{1},X^{\prime }\right) \right\rangle _{X_{1}}\right)
^{2}-\left( \gamma \left\langle \hat{S}_{E}\left( X\right) \right\rangle
\right) ^{2}}  \notag \\
&&\times \frac{\left( \frac{\left\Vert \hat{\Psi}_{0}\right\Vert ^{2}}{\hat{%
\mu}}+\left( \frac{1-r}{r}\right) \frac{\left\langle \hat{f}\right\rangle }{%
2\left( 1-\hat{S}\right) }\frac{\hat{S}}{1-\hat{S}}\right) ^{2}}{\left( 
\frac{\left\Vert \hat{\Psi}_{0}\left( X^{\prime }\right) \right\Vert ^{2}}{%
\hat{\mu}}+\left( \frac{1-r}{r}\right) \frac{\hat{f}\left( X^{\prime
}\right) \left( 1-\left\langle \hat{S}\right\rangle \right) +\left\langle 
\hat{S}\left( X,X^{\prime }\right) \right\rangle _{X}\left\langle \hat{f}%
\right\rangle }{1-\left\langle \hat{S}\right\rangle }\frac{\hat{S}}{1-\hat{S}%
}\right) ^{2}\left\langle \hat{f}\right\rangle ^{2}}  \notag
\end{eqnarray}

\section*{Appendix 7 Equation for returns in terms of $\hat{f}\left(
X\right) $}

\subsection*{A7.1 Constant returns}

\subsubsection*{A7.1.1 Solutions in terms of returns}

Once the set of average shares in one connected group is found, the
resolution is similar to the first part of this work. The equation is
written as a function for $\hat{f}\left( X\right) $, and using (\ref{SNht})
allows to find $\hat{S}_{E}\left( X\right) $ Return equation at the first
order is:%
\begin{eqnarray}
&&\left( \delta \left( X-X^{\prime }\right) -\hat{S}_{E}\left( X^{\prime
},X\right) \right) \frac{\left( 1-\hat{S}\left( X^{\prime }\right) \right)
\left( \hat{f}\left( X^{\prime }\right) -\bar{r}\right) }{1-\hat{S}%
_{E}\left( X^{\prime }\right) }  \label{rtq} \\
&=&S_{E}\left( X,X\right) \left( \frac{f_{1}\left( X\right) -\frac{C}{%
K_{X}\left\Vert \Psi \left( X\right) \right\Vert ^{2}}+\tau \left(
\left\langle f_{1}\left( X\right) \right\rangle -\left\langle f_{1}\left(
X^{\prime }\right) \right\rangle \right) }{\left( K_{X}\left\Vert \Psi
\left( X\right) \right\Vert ^{2}\right) ^{r}}-\bar{r}-C_{0}\right)  \notag
\end{eqnarray}%
Averaged vr $X^{\prime }$, this equation is:%
\begin{eqnarray*}
0 &=&\frac{1-\hat{S}\left( X\right) }{1-\hat{S}_{E}\left( X\right) }\left( 
\hat{f}\left( X\right) -\bar{r}\right) -\left\langle \hat{S}_{E}\left(
X^{\prime },X\right) \right\rangle _{X^{\prime }}\frac{1-\left\langle \hat{S}%
\left( X^{\prime }\right) \right\rangle }{1-\left\langle \hat{S}_{E}\left(
X^{\prime }\right) \right\rangle }\left( \left\langle \hat{f}\left(
X^{\prime }\right) \right\rangle -\left\langle \bar{r}\right\rangle \right)
\\
&&-S_{E}\left( X,X\right) \left( \frac{f_{1}\left( X\right) -\frac{C}{%
K_{X}\left\Vert \Psi \left( X\right) \right\Vert ^{2}}+\Delta F_{\tau
}\left( \bar{R}\left( K,X\right) \right) }{\left( K_{X}\left\Vert \Psi
\left( X\right) \right\Vert ^{2}\right) ^{r}}-\bar{r}\right)
\end{eqnarray*}%
whr:%
\begin{equation*}
\Delta F_{\tau }\left( \bar{R}\left( K,X\right) \right) =\tau \left(
\left\langle f_{1}\left( X\right) \right\rangle -\left\langle f_{1}\left(
X^{\prime }\right) \right\rangle \right)
\end{equation*}%
We can evaluate the various ratios using (\ref{St}) and (\ref{SNx}):%
\begin{equation*}
\frac{1-\hat{S}\left( X\right) }{1-\hat{S}_{E}\left( X\right) }=\frac{1-%
\frac{\left( 1-\left( \gamma \left\langle \hat{S}_{E}\left( X\right)
\right\rangle \right) ^{2}\right) \left( 1+\frac{\Delta \hat{f}\left(
X\right) +\Delta \hat{r}\left( X\right) }{2}\right) }{2-\left( \gamma
\left\langle \hat{S}_{E}\left( X\right) \right\rangle \right) ^{2}+\left(
\gamma \left\langle \hat{S}_{E}\left( X_{1},X\right) \right\rangle
_{X_{1}}\right) ^{2}-\left( \gamma \left\langle \hat{S}_{E}\left( X\right)
\right\rangle \right) ^{2}}\frac{\left\langle \hat{K}\right\rangle
\left\Vert \hat{\Psi}\right\Vert ^{2}}{\hat{K}_{X}\left\vert \hat{\Psi}%
\left( X\right) \right\vert ^{2}}}{1-\frac{1}{2}\frac{\left( 1-\left( \gamma
\left\langle \hat{S}_{E}\left( X\right) \right\rangle \right) ^{2}\right)
\left( 1+\Delta \hat{f}\left( X\right) \right) }{2-\left( \gamma
\left\langle \hat{S}_{E}\left( X\right) \right\rangle \right) ^{2}+\left(
\gamma \left\langle \hat{S}_{E}\left( X_{1},X\right) \right\rangle
_{X_{1}}\right) ^{2}-\left( \gamma \left\langle \hat{S}_{E}\left( X\right)
\right\rangle \right) ^{2}}\frac{\left\langle \hat{K}\right\rangle
\left\Vert \hat{\Psi}\right\Vert ^{2}}{\hat{K}_{X}\left\vert \hat{\Psi}%
\left( X\right) \right\vert ^{2}}}
\end{equation*}%
to rewrite the return equation:%
\begin{eqnarray}
&&0=\frac{1-\frac{\left( 1-\left( \gamma \left\langle \hat{S}_{E}\left(
X\right) \right\rangle \right) ^{2}\right) \left( 1+\frac{\Delta \hat{f}%
\left( X\right) +\Delta \hat{r}\left( X\right) }{2}\right) }{2-\left( \gamma
\left\langle \hat{S}_{E}\left( X\right) \right\rangle \right) ^{2}+\left(
\gamma \left\langle \hat{S}_{E}\left( X_{1},X\right) \right\rangle
_{X_{1}}\right) ^{2}-\left( \gamma \left\langle \hat{S}_{E}\left( X\right)
\right\rangle \right) ^{2}}\frac{\left\langle \hat{K}\right\rangle
\left\Vert \hat{\Psi}\right\Vert ^{2}}{\hat{K}_{X}\left\vert \hat{\Psi}%
\left( X\right) \right\vert ^{2}}}{1-\frac{1}{2}\frac{\left( 1-\left( \gamma
\left\langle \hat{S}_{E}\left( X\right) \right\rangle \right) ^{2}\right)
\left( 1+\Delta \hat{f}\left( X\right) \right) }{2-\left( \gamma
\left\langle \hat{S}_{E}\left( X\right) \right\rangle \right) ^{2}+\left(
\gamma \left\langle \hat{S}_{E}\left( X_{1},X\right) \right\rangle
_{X_{1}}\right) ^{2}-\left( \gamma \left\langle \hat{S}_{E}\left( X\right)
\right\rangle \right) ^{2}}\frac{\left\langle \hat{K}\right\rangle
\left\Vert \hat{\Psi}\right\Vert ^{2}}{\hat{K}_{X}\left\vert \hat{\Psi}%
\left( X\right) \right\vert ^{2}}}\left( \hat{f}\left( X\right) -\bar{r}%
\right)  \label{Rq} \\
&&-\frac{1}{2}\frac{\left( 1-\left( \gamma \left\langle \hat{S}_{E}\left(
X\right) \right\rangle \right) ^{2}\right) }{2-\left( \gamma \left\langle 
\hat{S}_{E}\left( X\right) \right\rangle \right) ^{2}}\left( 1-\frac{\Delta
\left( \frac{f\left( X\right) +r\left( X\right) }{2}\right) }{2-\left(
\gamma \left\langle \hat{S}_{E}\left( X\right) \right\rangle \right) ^{2}}+%
\frac{\left\langle \hat{f}\left( X^{\prime }\right) \right\rangle
-\left\langle \hat{r}\left( X^{\prime }\right) \right\rangle _{\hat{w}_{L}}}{%
2}\right)  \notag \\
&&\times \frac{1-\left\langle \hat{S}\left( X^{\prime }\right) \right\rangle 
}{1-\left\langle \hat{S}_{E}\left( X^{\prime }\right) \right\rangle }\left(
\left\langle \hat{f}\left( X^{\prime }\right) \right\rangle -\left\langle 
\bar{r}\right\rangle \right) -S_{E}\left( X,X\right) \left( f\left( X\right)
-r\right)  \notag
\end{eqnarray}%
with (\ref{DF}) and (\ref{Dfr}):%
\begin{equation}
\Delta \hat{f}\left( X^{\prime }\right) =\hat{f}\left( X^{\prime }\right)
-\left\langle \tilde{f}+\tilde{r}\right\rangle
\end{equation}%
\begin{equation*}
\left\langle \tilde{f}\right\rangle +\left\langle \tilde{r}\right\rangle
=\left\langle \hat{w}\left( X\right) \right\rangle \frac{\left\langle \hat{f}%
\left( X^{\prime }\right) \right\rangle _{\hat{w}_{E}}+\left\langle \hat{r}%
\left( X^{\prime }\right) \right\rangle _{\hat{w}_{L}}}{2}+\left\langle
w\left( X\right) \right\rangle \frac{\left\langle f\left( X\right)
\right\rangle +\left\langle r\left( X\right) \right\rangle }{2}
\end{equation*}%
Using that:%
\begin{eqnarray*}
&&\frac{\left\langle \hat{K}\right\rangle \left\Vert \hat{\Psi}\right\Vert
^{2}}{\hat{K}_{X^{\prime }}\left\vert \hat{\Psi}\left( X^{\prime }\right)
\right\vert ^{2}} \\
&\simeq &\left( \frac{\hat{g}\left( \hat{X}^{\prime }\right) }{\left\langle 
\hat{g}\right\rangle }\frac{\frac{\left\langle \hat{S}\left( X^{\prime
},X\right) \right\rangle }{1-\left\langle \hat{S}\left( X^{\prime },X\right)
\right\rangle }}{\frac{\left\langle \hat{S}\left( X^{\prime },X\right)
\right\rangle _{X}}{1-\left\langle \hat{S}\left( X^{\prime },X\right)
\right\rangle _{X}\frac{\left\langle \hat{K}\right\rangle \left\Vert \hat{%
\Psi}\right\Vert ^{2}}{\hat{K}_{X^{\prime }}\left\vert \hat{\Psi}\left(
X^{\prime }\right) \right\vert ^{2}}}}\right) ^{2} \\
&\simeq &\left( \frac{\left( \hat{f}\left( X^{\prime }\right) \left(
1-\left\langle \hat{S}\right\rangle \right) +\left\langle \hat{S}\left(
X,X^{\prime }\right) \right\rangle _{X}\left\langle \hat{f}\right\rangle
\right) \frac{\left\langle \hat{S}\left( X^{\prime },X\right) \right\rangle 
}{1-\left\langle \hat{S}\left( X^{\prime },X\right) \right\rangle }}{%
\left\langle \hat{f}\right\rangle \left\langle \hat{S}\left( X^{\prime
},X\right) \right\rangle _{X}}\right) ^{2} \\
&&\times \left( 1-\left\langle \hat{S}\left( X^{\prime },X\right)
\right\rangle _{X}\left( \frac{\left( \hat{f}\left( X^{\prime }\right)
\left( 1-\left\langle \hat{S}\right\rangle \right) +\left\langle \hat{S}%
\left( X,X^{\prime }\right) \right\rangle _{X}\left\langle \hat{f}%
\right\rangle \right) \frac{\left\langle \hat{S}\left( X^{\prime },X\right)
\right\rangle }{1-\left\langle \hat{S}\left( X^{\prime },X\right)
\right\rangle }}{\left\langle \hat{f}\right\rangle \frac{\left\langle \hat{S}%
\left( X^{\prime },X\right) \right\rangle _{X}}{1-\left\langle \hat{S}\left(
X^{\prime },X\right) \right\rangle _{X}}}\right) ^{2}\right) ^{2}
\end{eqnarray*}%
Return equation takes the following form:%
\begin{equation*}
0=\frac{1-\frac{\left( 1-\left( \gamma \left\langle \hat{S}_{E}\left(
X\right) \right\rangle \right) ^{2}\right) \left( 1+\frac{\Delta \hat{f}%
\left( X\right) +\Delta \hat{r}\left( X\right) }{2}\right) }{2-\left( \gamma
\left\langle \hat{S}_{E}\left( X\right) \right\rangle \right) ^{2}+\left(
\gamma \left\langle \hat{S}_{E}\left( X_{1},X\right) \right\rangle
_{X_{1}}\right) ^{2}-\left( \gamma \left\langle \hat{S}_{E}\left( X\right)
\right\rangle \right) ^{2}}W_{1}}{1-\frac{1}{2}\frac{\left( 1-\left( \gamma
\left\langle \hat{S}_{E}\left( X\right) \right\rangle \right) ^{2}\right)
\left( 1+\Delta \hat{f}\left( X\right) \right) }{2-\left( \gamma
\left\langle \hat{S}_{E}\left( X\right) \right\rangle \right) ^{2}+\left(
\gamma \left\langle \hat{S}_{E}\left( X_{1},X\right) \right\rangle
_{X_{1}}\right) ^{2}-\left( \gamma \left\langle \hat{S}_{E}\left( X\right)
\right\rangle \right) ^{2}}W_{1}}\left( \hat{f}\left( X\right) -\bar{r}%
\right) -h\left( X\right)
\end{equation*}%
with:%
\begin{equation*}
W_{1}=\left( \frac{\left( \hat{f}\left( X^{\prime }\right) \left(
1-\left\langle \hat{S}\right\rangle \right) +\left\langle \hat{S}\left(
X,X^{\prime }\right) \right\rangle _{X}\left\langle \hat{f}\right\rangle
\right) \frac{\left\langle \hat{S}\left( X^{\prime },X\right) \right\rangle 
}{1-\left\langle \hat{S}\left( X^{\prime },X\right) \right\rangle }}{%
\left\langle \hat{f}\right\rangle \frac{\left\langle \hat{S}\left( X^{\prime
},X\right) \right\rangle _{X}}{1-\left\langle \hat{S}\left( X^{\prime
},X\right) \right\rangle _{X}\frac{\left\langle \hat{K}\right\rangle
\left\Vert \hat{\Psi}\right\Vert ^{2}}{\hat{K}_{X^{\prime }}\left\vert \hat{%
\Psi}\left( X^{\prime }\right) \right\vert ^{2}}}}\right) ^{2}
\end{equation*}%
$_{{}}$and:%
\begin{eqnarray}
h\left( X\right) &=&\frac{1}{2}\frac{\left( 1-\left( \gamma \left\langle 
\hat{S}_{E}\left( X\right) \right\rangle \right) ^{2}\right) }{2-\left(
\gamma \left\langle \hat{S}_{E}\left( X\right) \right\rangle \right) ^{2}}%
\left( 1-\frac{\Delta \left( \frac{\hat{f}\left( X\right) +r\left( X\right) 
}{2}\right) }{2-\left( \gamma \left\langle \hat{S}_{E}\left( X\right)
\right\rangle \right) ^{2}}+\frac{\left\langle \hat{f}\left( X^{\prime
}\right) \right\rangle -\left\langle \hat{r}\left( X^{\prime }\right)
\right\rangle _{\hat{w}_{L}}}{2}\right)  \label{HDf} \\
&&\times \frac{1-\left\langle \hat{S}\left( X^{\prime }\right) \right\rangle 
}{1-\left\langle \hat{S}_{E}\left( X^{\prime }\right) \right\rangle }\left(
\left\langle \hat{f}\left( X^{\prime }\right) \right\rangle -\left\langle 
\bar{r}\right\rangle \right) +S_{E}\left( X,X\right) \left( f\left( X\right)
-r\right)  \notag
\end{eqnarray}%
Equivalently, the equation writes:%
\begin{eqnarray}
&&h\left( X\right) =\left( \hat{f}\left( X\right) -\bar{r}\right)  \label{Qh}
\\
&&\times \frac{\left\langle \hat{f}\right\rangle ^{2}\left( \frac{\frac{%
\left\langle \hat{S}\left( X^{\prime },X\right) \right\rangle _{X}}{%
1-\left\langle \hat{S}\left( X^{\prime },X\right) \right\rangle _{X}\frac{%
\left\langle \hat{K}\right\rangle \left\Vert \hat{\Psi}\right\Vert ^{2}}{%
\hat{K}_{X^{\prime }}\left\vert \hat{\Psi}\left( X^{\prime }\right)
\right\vert ^{2}}}}{\frac{\left\langle \hat{S}\left( X^{\prime },X\right)
\right\rangle }{1-\left\langle \hat{S}\left( X^{\prime },X\right)
\right\rangle }}\right) ^{2}-A\left( 1+\frac{\Delta \hat{f}\left( X\right)
+\Delta \hat{r}\left( X\right) }{2}\right) \left( \hat{f}\left( X\right)
\left( 1-\left\langle \hat{S}\right\rangle \right) +\left\langle \hat{S}%
\left( X^{\prime },X\right) \right\rangle _{X^{\prime }}\left\langle \hat{f}%
\right\rangle \right) ^{2}}{\left\langle \hat{f}\right\rangle ^{2}\left( 
\frac{\frac{\left\langle \hat{S}\left( X^{\prime },X\right) \right\rangle
_{X}}{1-\left\langle \hat{S}\left( X^{\prime },X\right) \right\rangle _{X}%
\frac{\left\langle \hat{K}\right\rangle \left\Vert \hat{\Psi}\right\Vert ^{2}%
}{\hat{K}_{X^{\prime }}\left\vert \hat{\Psi}\left( X^{\prime }\right)
\right\vert ^{2}}}}{\frac{\left\langle \hat{S}\left( X^{\prime },X\right)
\right\rangle }{1-\left\langle \hat{S}\left( X^{\prime },X\right)
\right\rangle }}\right) ^{2}-\frac{1}{2}A\left( 1+\Delta \hat{f}\left(
X\right) \right) \left( \hat{f}\left( X\right) \left( 1-\left\langle \hat{S}%
\right\rangle \right) +\left\langle \hat{S}\left( X^{\prime },X\right)
\right\rangle _{X^{\prime }}\left\langle \hat{f}\right\rangle \right) ^{2}} 
\notag
\end{eqnarray}%
with:%
\begin{eqnarray*}
\left\langle \hat{w}\left( X\right) \right\rangle &=&\frac{\left( 1-\left(
\gamma \left\langle \hat{S}_{E}\left( X\right) \right\rangle \right)
^{2}\right) }{2-\left( \gamma \left\langle \hat{S}_{E}\left( X\right)
\right\rangle \right) ^{2}} \\
\left\langle w\left( X\right) \right\rangle &=&\frac{1}{2-\left( \gamma
\left\langle \hat{S}_{E}\left( X\right) \right\rangle \right) ^{2}}
\end{eqnarray*}%
and:%
\begin{equation}
A=\frac{\left( 1-\left( \gamma \left\langle \hat{S}_{E}\left( X\right)
\right\rangle \right) ^{2}\right) }{2-\left( \gamma \left\langle \hat{S}%
_{E}\left( X\right) \right\rangle \right) ^{2}+\left( \gamma \left\langle 
\hat{S}_{E}\left( X_{1},X\right) \right\rangle _{X_{1}}\right) ^{2}-\left(
\gamma \left\langle \hat{S}_{E}\left( X\right) \right\rangle \right) ^{2}}
\label{cfr}
\end{equation}%
Using that:%
\begin{equation*}
\left\langle \hat{S}\left( X^{\prime },X\right) \right\rangle _{X}=\frac{%
\left( 1-\left( \gamma \left\langle \hat{S}_{E}\left( X\right) \right\rangle
\right) ^{2}\right) \left( 1+\frac{\Delta \hat{f}\left( X^{\prime }\right)
+\Delta \hat{r}\left( X^{\prime }\right) }{2}\right) }{2-\left( \gamma
\left\langle \hat{S}_{E}\left( X\right) \right\rangle \right) ^{2}+\frac{%
\left\langle \hat{w}\left( X^{\prime },X\right) \right\rangle }{2}%
\left\langle w\left( X\right) \right\rangle \Delta \left( \frac{f\left(
X^{\prime }\right) +r\left( X^{\prime }\right) }{2}\right) }
\end{equation*}%
where:%
\begin{equation*}
\Delta \left( \frac{f\left( X\right) +r\left( X\right) }{2}\right) =\left( 
\frac{f\left( X\right) +r\left( X\right) }{2}-\frac{\left\langle \hat{f}%
\left( X^{\prime }\right) \right\rangle +\left\langle \hat{r}\left(
X^{\prime }\right) \right\rangle _{\hat{w}_{L}}}{2}\right)
\end{equation*}%
\begin{equation*}
\frac{\Delta \hat{f}\left( X^{\prime }\right) +\Delta \hat{r}\left(
X^{\prime }\right) }{2}=\frac{\hat{f}\left( X^{\prime }\right) +\hat{r}%
\left( X^{\prime }\right) }{2}-\left\langle \hat{w}\left( X\right)
\right\rangle \frac{\left\langle \hat{f}\left( X^{\prime }\right)
\right\rangle _{\hat{w}_{E}}+\left\langle \hat{r}\left( X^{\prime }\right)
\right\rangle _{\hat{w}_{L}}}{2}-\left\langle w\left( X\right) \right\rangle 
\frac{\left\langle f\left( X\right) +r\left( X\right) \right\rangle _{w}}{2}
\end{equation*}%
we can write in first approximation:%
\begin{equation*}
\left\langle \hat{S}\left( X^{\prime },X\right) \right\rangle _{X}\simeq
A\left( 1+\frac{\Delta \hat{f}\left( X^{\prime }\right) +\Delta \hat{r}%
\left( X^{\prime }\right) }{2}\right)
\end{equation*}%
so that the equation for returns becomes:%
\begin{eqnarray}
&&h\left( X\right) =\left( \hat{f}\left( X\right) -\bar{r}\right) \\
&&\times \frac{\left\langle \hat{f}\right\rangle ^{2}W_{2}-A\left( 1+\frac{%
\Delta \hat{f}\left( X\right) +\Delta \hat{r}\left( X\right) }{2}\right)
\left( \hat{f}\left( X\right) \left( 1-\left\langle \hat{S}\right\rangle
\right) +\left\langle \hat{S}\left( X^{\prime },X\right) \right\rangle
_{X^{\prime }}\left\langle \hat{f}\right\rangle \right) ^{2}}{\left\langle 
\hat{f}\right\rangle ^{2}W_{2}-\frac{1}{2}A\left( 1+\Delta \hat{f}\left(
X\right) \right) \left( \hat{f}\left( X\right) \left( 1-\left\langle \hat{S}%
\right\rangle \right) +\left\langle \hat{S}\left( X^{\prime },X\right)
\right\rangle _{X^{\prime }}\left\langle \hat{f}\right\rangle \right) ^{2}} 
\notag
\end{eqnarray}

with:%
\begin{equation*}
W_{2}=\left( \frac{A\left( 1+\frac{\Delta \hat{f}\left( X^{\prime }\right)
+\Delta \hat{r}\left( X^{\prime }\right) }{2}\right) }{1-A\left( 1+\frac{%
\Delta \hat{f}\left( X^{\prime }\right) +\Delta \hat{r}\left( X^{\prime
}\right) }{2}\right) \frac{\left\langle \hat{K}\right\rangle \left\Vert \hat{%
\Psi}\right\Vert ^{2}}{\hat{K}_{X^{\prime }}\left\vert \hat{\Psi}\left(
X^{\prime }\right) \right\vert ^{2}}}\frac{1-A\left( 1+\frac{\left\langle
\Delta \hat{f}\left( X^{\prime }\right) +\Delta \hat{r}\left( X^{\prime
}\right) \right\rangle }{2}\right) }{A\left( 1+\frac{\left\langle \Delta 
\hat{f}\left( X^{\prime }\right) +\Delta \hat{r}\left( X^{\prime }\right)
\right\rangle }{2}\right) }\right) ^{2}
\end{equation*}%
This is an equation of the form:

\begin{equation*}
0=\frac{X_{0}^{2}W_{2}-A\left( 1+\frac{X+\Delta \hat{r}\left( X\right) }{2}%
\right) \left( VX+\left( 1-V\right) X_{0}\right) ^{2}}{X_{0}^{2}W_{2}-\frac{1%
}{2}A\left( 1+X\right) \left( VX+\left( 1-V\right) X_{0}\right) ^{2}}\left(
X+k-r\right) -h\left( X\right)
\end{equation*}%
For:%
\begin{equation*}
\frac{\left\langle \hat{K}\right\rangle \left\Vert \hat{\Psi}\right\Vert ^{2}%
}{\hat{K}_{X^{\prime }}\left\vert \hat{\Psi}\left( X^{\prime }\right)
\right\vert ^{2}}\simeq 1
\end{equation*}%
\begin{equation*}
X=\hat{f}\left( X\right)
\end{equation*}%
\begin{equation*}
V=\left( 1-\left\langle \hat{S}_{E}\right\rangle \right)
\end{equation*}%
\begin{equation*}
X_{0}\rightarrow \left\langle \hat{f}\right\rangle
\end{equation*}%
This writes:%
\begin{eqnarray*}
0 &=&\left( X_{0}^{2}W_{2}^{\prime }-A\left( 1+\frac{X+\Delta \hat{r}\left(
X\right) }{2}\right) \left( VX+\left( 1-V\right) X_{0}\right) ^{2}\right)
\left( X+k-r\right) \\
&&-h\left( X\right) \left( X_{0}^{2}\frac{\left\Vert \hat{\Psi}_{0}\left(
X\right) \right\Vert ^{4}}{\left\Vert \hat{\Psi}_{0}\right\Vert ^{4}}-\frac{%
A\left( 1+X\right) }{2}\left( VX+\left( 1-V\right) X_{0}\right) ^{2}\right)
\end{eqnarray*}%
where:%
\begin{equation*}
W_{2}^{\prime }=\left( \frac{A\left( 1+\frac{X+\Delta \hat{r}\left( X\right) 
}{2}\right) }{1-A\left( 1+\frac{X+\Delta \hat{r}\left( X\right) }{2}\right) }%
\frac{1-A\left( 1+\frac{\left\langle \Delta \hat{f}\left( X^{\prime }\right)
+\Delta \hat{r}\left( X^{\prime }\right) \right\rangle }{2}\right) }{A\left(
1+\frac{\left\langle \Delta \hat{f}\left( X^{\prime }\right) +\Delta \hat{r}%
\left( X^{\prime }\right) \right\rangle }{2}\right) }\right) ^{2}
\end{equation*}%
Numerical studies for some range of parameters show that in general two
possible solutions arise per sector. Considering decreasing return to scale,
one solution has relatively low returns, corresponding to a high level of
shares and disposable capital and one with relatively large return with low
level of shares and disposable capital.

\subsubsection*{A7.1.2 solutions in terms of shares}

We can change the representation and consider the returns in functions of
saddle point solutions for shares. Given that:%
\begin{equation}
\left\langle \hat{S}_{E}\left( X^{\prime },X\right) \right\rangle _{X}=\frac{%
\left\langle \hat{w}\left( X^{\prime },X\right) \right\rangle _{X}}{2}\left(
1+\Delta \hat{f}\left( X^{\prime }\right) \right)
\end{equation}%
with:%
\begin{equation*}
\Delta \hat{f}\left( X^{\prime }\right) =\hat{f}\left( X^{\prime }\right)
-\left( \left\langle \hat{w}\left( X\right) \right\rangle \frac{\left\langle 
\hat{f}\left( X^{\prime }\right) \right\rangle _{\hat{w}_{E}}+\left\langle 
\hat{r}\left( X^{\prime }\right) \right\rangle _{\hat{w}_{L}}}{2}%
+\left\langle w\left( X\right) \right\rangle \frac{\left\langle f\left(
X\right) \right\rangle +\left\langle r\left( X\right) \right\rangle }{2}%
\right)
\end{equation*}%
we can write:%
\begin{equation*}
\Delta \hat{f}\left( X^{\prime }\right) =\frac{\left\langle \hat{S}%
_{E}\left( X^{\prime },X\right) \right\rangle _{X}}{\frac{\left\langle \hat{w%
}\left( X^{\prime },X\right) \right\rangle _{X}}{2}}-1
\end{equation*}%
with the coefficient:%
\begin{equation*}
\left\langle \frac{\hat{w}\left( X^{\prime },X\right) }{2}\right\rangle
_{X}\simeq \frac{\left( 1-\left( \gamma \left\langle \hat{S}_{E}\left(
X\right) \right\rangle \right) ^{2}\right) }{2-\left( \gamma \left\langle 
\hat{S}_{E}\left( X\right) \right\rangle \right) ^{2}+\left( \gamma
\left\langle \hat{S}_{E}\left( X_{1},X^{\prime }\right) \right\rangle
_{X_{1}}\right) ^{2}-\left( \gamma \left\langle \hat{S}_{E}\left( X\right)
\right\rangle \right) ^{2}}
\end{equation*}

\begin{equation*}
\left\langle \hat{S}_{E}\left( X^{\prime },X\right) \right\rangle
_{X^{\prime }}\simeq \left\langle \hat{w}\left( X^{\prime },X\right)
\right\rangle \left( 1-\left\langle w\left( X\right) \right\rangle \Delta
\left( \frac{f\left( X\right) +r\left( X\right) }{2}\right) +\frac{%
\left\langle \hat{f}\left( X^{\prime }\right) \right\rangle -\left\langle 
\hat{r}\left( X^{\prime }\right) \right\rangle _{\hat{w}_{L}}}{2}\right)
\end{equation*}%
\begin{equation*}
\Delta \left( \frac{f\left( X\right) +r\left( X\right) }{2}\right) =\left( 
\frac{f\left( X\right) +r\left( X\right) }{2}-\frac{\left\langle f\left(
X\right) \right\rangle +\left\langle r\left( X\right) \right\rangle }{2}%
\right)
\end{equation*}

\subsection*{A7.2 Modification for decreasing return to scale.}

Including decreasing return to scale corresponds to modify:%
\begin{equation*}
f_{1}\left( X\right) \rightarrow \frac{f_{1}\left( X\right) +\tau \left(
\left\langle f_{1}\left( X\right) \right\rangle -\left\langle f_{1}\left(
X^{\prime }\right) \right\rangle \right) }{\left( K_{X}\left\vert \hat{\Psi}%
\left( X\right) \right\vert ^{2}\right) ^{r}}-\frac{C}{K_{X}}-C_{0}
\end{equation*}%
in (\ref{Rq}), the modification is both direct, and indirect by shifts in
coefficient.

\subsubsection*{A7.2.1 Expression of $S\left( X,X\right) $ and $S\left(
X\right) $}

Decreasing return to scale modify the equation for $S\left( X,X\right) $ in
the following way:%
\begin{eqnarray}
&&S\left( X,X\right)  \label{V} \\
&=&w\left( X\right) \left( 1+\left( \hat{w}\left( X\right) \left( \frac{%
f\left( X\right) +\bar{r}\left( X\right) }{2}-\frac{\left\langle \hat{f}%
\left( X^{\prime }\right) \right\rangle _{\hat{w}_{E}}+\left\langle \hat{r}%
\left( X^{\prime }\right) \right\rangle _{\hat{w}_{L}}}{2}\right) \right)
\right)  \notag \\
&=&w\left( X\right) \left( 1+\left( \hat{w}\left( X\right) \left( \frac{%
\frac{f_{1}\left( X\right) +\tau \left( \left\langle f_{1}\left( X\right)
\right\rangle -\left\langle f_{1}\left( X^{\prime }\right) \right\rangle
\right) }{\left( K_{X}\left\vert \hat{\Psi}\left( X\right) \right\vert
^{2}\right) ^{r}}-\frac{C}{K_{X}}-C_{0}+\bar{r}\left( X\right) }{2}-\frac{%
\left\langle \hat{f}\left( X^{\prime }\right) \right\rangle _{\hat{w}%
_{E}}+\left\langle \hat{r}\left( X^{\prime }\right) \right\rangle _{\hat{w}%
_{L}}}{2}\right) \right) \right)  \notag
\end{eqnarray}%
and:%
\begin{equation}
S\left( X\right) =S\left( X,X\right) \frac{\hat{K}_{X}\left\vert \hat{\Psi}%
\left( X\right) \right\vert ^{2}}{K_{X}\left\vert \Psi \left( X\right)
\right\vert ^{2}}  \label{Vb}
\end{equation}%
In first approximation we can replace:%
\begin{equation*}
S\left( X\right) _{dr}=S_{cr}\left( X,X\right) \frac{\hat{K}_{X}\left\vert 
\hat{\Psi}\left( X\right) \right\vert ^{2}}{K_{X}\left\vert \Psi \left(
X\right) \right\vert ^{2}}
\end{equation*}%
in (\ref{V}), with $\frac{\hat{K}_{X}\left\vert \hat{\Psi}\left( X\right)
\right\vert ^{2}}{K_{X}\left\vert \Psi \left( X\right) \right\vert ^{2}}$
given by (\ref{crd}) and $S_{cr}\left( X,X\right) $ was computed under
constant return.%
\begin{eqnarray}
&&S\left( X,X\right) \\
&=&w\left( X\right) \left( 1+\hat{w}\left( X\right) \frac{\frac{f_{1}\left(
X\right) +\tau \left( \left\langle f_{1}\left( X\right) \right\rangle
-\left\langle f_{1}\left( X^{\prime }\right) \right\rangle \right) }{\left(
K_{X}\left\vert \hat{\Psi}\left( X\right) \right\vert ^{2}\right) ^{r}}-%
\frac{C}{K_{X}}-C_{0}+\bar{r}\left( X\right) -\left( \left\langle \hat{f}%
\left( X^{\prime }\right) \right\rangle _{\hat{w}_{E}}+\left\langle \hat{r}%
\left( X^{\prime }\right) \right\rangle _{\hat{w}_{L}}\right) }{2}\right) 
\notag
\end{eqnarray}

\subsubsection*{A7.2.2 Equation for $S_{E}\left( X,X\right) $, $S_{E}\left(
X\right) $ and $\left\langle \hat{S}_{E}\left( X^{\prime },X\right)
\right\rangle _{X^{\prime }}$}

\begin{eqnarray}
&&S_{E}\left( X,X\right)  \label{Vc} \\
&=&\frac{w\left( X\right) }{2}\left( 1+\left( \hat{w}\left( X\right) \left(
f\left( X\right) -\frac{\left\langle \hat{f}\left( X^{\prime }\right)
\right\rangle _{\hat{w}_{E}}+\left\langle \hat{r}\left( X^{\prime }\right)
\right\rangle _{\hat{w}_{L}}}{2}\right) +\frac{w\left( X\right) }{2}\left(
f\left( X\right) -\bar{r}\left( X\right) \right) \right) \right)  \notag \\
&=&\frac{w\left( X\right) }{2}\left( 1+\left( \left( \hat{w}\left( X\right) +%
\frac{w\left( X\right) }{2}\right) f\left( X\right) -\left( \hat{w}\left(
X\right) \frac{\left\langle \hat{f}\left( X^{\prime }\right) \right\rangle _{%
\hat{w}_{E}}+\left\langle \hat{r}\left( X^{\prime }\right) \right\rangle _{%
\hat{w}_{L}}}{2}+\frac{w\left( X\right) }{2}\bar{r}\left( X\right) \right)
\right) \right)  \notag \\
&=&\frac{w\left( X\right) }{2}\left( 1+\left( \hat{w}\left( X\right) +\frac{%
w\left( X\right) }{2}\right) \left( \frac{f_{1}\left( X\right) +\tau \left(
\left\langle f_{1}\left( X\right) \right\rangle -\left\langle f_{1}\left(
X^{\prime }\right) \right\rangle \right) }{\left( K_{X}\left\vert \hat{\Psi}%
\left( X\right) \right\vert ^{2}\right) ^{r}}-\frac{C}{K_{X}}-C_{0}\right)
\right.  \notag \\
&&\left. -\left( \hat{w}\left( X\right) \frac{\left\langle \hat{f}\left(
X^{\prime }\right) \right\rangle _{\hat{w}_{E}}+\left\langle \hat{r}\left(
X^{\prime }\right) \right\rangle _{\hat{w}_{L}}}{2}+\frac{w\left( X\right) }{%
2}\bar{r}\left( X\right) \right) \right)  \notag
\end{eqnarray}

\begin{equation}
S_{E}\left( X\right) =S_{E}\left( X,X\right) \frac{\hat{K}_{X}\left\vert 
\hat{\Psi}\left( X\right) \right\vert ^{2}}{K_{X}\left\vert \Psi \left(
X\right) \right\vert ^{2}}  \label{Vcc}
\end{equation}%
The half average (\ref{hv}) is still valid:%
\begin{eqnarray}
&&\left\langle \hat{S}_{E}\left( X^{\prime },X\right) \right\rangle
_{X^{\prime }}  \label{hl} \\
&\simeq &\frac{\left\langle \hat{w}\left( X^{\prime },X\right) \right\rangle 
}{2}  \notag \\
&&\times \left( 1-\left\langle w\left( X\right) \right\rangle \Delta \left( 
\frac{\frac{f_{1}\left( X\right) +\tau \left( \left\langle f_{1}\left(
X\right) \right\rangle -\left\langle f_{1}\left( X^{\prime }\right)
\right\rangle \right) }{\left( K_{X}\left\vert \hat{\Psi}\left( X\right)
\right\vert ^{2}\right) ^{r}}-\frac{C}{K_{X}}-C_{0}+r\left( X\right) }{2}%
\right) +\frac{\left\langle \hat{f}\left( X^{\prime }\right) \right\rangle
-\left\langle \hat{r}\left( X^{\prime }\right) \right\rangle _{\hat{w}_{L}}}{%
2}\right)  \notag \\
&\simeq &\frac{\left\langle \hat{w}\left( X^{\prime },X\right) \right\rangle 
}{2}  \notag \\
&&\times \left( 1-\left\langle w\left( X\right) \right\rangle \Delta \left( 
\frac{\frac{f_{1}\left( X\right) +\tau \left( \left\langle f_{1}\left(
X\right) \right\rangle -\left\langle f_{1}\left( X^{\prime }\right)
\right\rangle \right) }{\left( K_{X}\left\vert \hat{\Psi}\left( X\right)
\right\vert ^{2}\right) ^{r}}-\frac{C}{K_{X}}+r\left( X\right) }{2}\right) +%
\frac{\left\langle \hat{f}\left( X^{\prime }\right) \right\rangle
-\left\langle \hat{r}\left( X^{\prime }\right) \right\rangle _{\hat{w}_{L}}}{%
2}\right)  \notag
\end{eqnarray}%
These formula can then be used in the return equations.

\subsubsection*{A7.2.3 Return equation}

The return equation for $\hat{f}\left( X\right) $ is given by:%
\begin{eqnarray}
&&0=\frac{\left\langle \hat{f}\right\rangle ^{2}W_{2}^{\prime \prime
}-A\left( 1+\frac{\Delta \hat{f}\left( X\right) +\Delta \hat{r}\left(
X\right) }{2}\right) \left( \hat{f}\left( X\right) \left( 1-\left\langle 
\hat{S}\right\rangle \right) +\left\langle \hat{S}\left( X^{\prime
},X\right) \right\rangle _{X^{\prime }}\left\langle \hat{f}\right\rangle
\right) ^{2}}{\left\langle \hat{f}\right\rangle ^{2}W_{2}^{\prime \prime }-%
\frac{1}{2}A\left( 1+\Delta \hat{f}\left( X\right) \right) \left( \hat{f}%
\left( X\right) \left( 1-\left\langle \hat{S}\right\rangle \right)
+\left\langle \hat{S}\left( X^{\prime },X\right) \right\rangle _{X^{\prime
}}\left\langle \hat{f}\right\rangle \right) ^{2}}  \label{Rtns} \\
&&\times \left( \hat{f}\left( X\right) -\bar{r}\right) -h\left( X\right) 
\notag
\end{eqnarray}%
with:%
\begin{equation*}
W_{2}^{\prime \prime }=\left( \frac{A\left( 1+\frac{X+\Delta \hat{r}\left(
X\right) }{2}\right) }{1-A\left( 1+\frac{X+\Delta \hat{r}\left( X\right) }{2}%
\right) \frac{\left\langle \hat{K}\right\rangle \left\Vert \hat{\Psi}%
\right\Vert ^{2}}{\hat{K}_{X^{\prime }}\left\vert \hat{\Psi}\left( X^{\prime
}\right) \right\vert ^{2}}}\frac{1-A\left( 1+\frac{\left\langle \Delta \hat{f%
}\left( X^{\prime }\right) +\Delta \hat{r}\left( X^{\prime }\right)
\right\rangle }{2}\right) }{A\left( 1+\frac{\left\langle \Delta \hat{f}%
\left( X^{\prime }\right) +\Delta \hat{r}\left( X^{\prime }\right)
\right\rangle }{2}\right) }\right) ^{2}
\end{equation*}%
\begin{eqnarray*}
\left\langle \hat{w}\left( X\right) \right\rangle &=&\frac{\left( 1-\left(
\gamma \left\langle \hat{S}_{E}\left( X\right) \right\rangle \right)
^{2}\right) }{2-\left( \gamma \left\langle \hat{S}_{E}\left( X\right)
\right\rangle \right) ^{2}} \\
\left\langle w\left( X\right) \right\rangle &=&\frac{1}{2-\left( \gamma
\left\langle \hat{S}_{E}\left( X\right) \right\rangle \right) ^{2}}
\end{eqnarray*}%
\begin{eqnarray*}
h\left( X\right) &=&\frac{1}{2}\frac{\left( 1-\left( \gamma \left\langle 
\hat{S}_{E}\left( X\right) \right\rangle \right) ^{2}\right) }{2-\left(
\gamma \left\langle \hat{S}_{E}\left( X\right) \right\rangle \right) ^{2}}%
\left( 1-\frac{\Delta \left( \frac{\hat{f}\left( X\right) +r\left( X\right) 
}{2}\right) }{2-\left( \gamma \left\langle \hat{S}_{E}\left( X\right)
\right\rangle \right) ^{2}}+\frac{\left\langle \hat{f}\left( X^{\prime
}\right) \right\rangle -\left\langle \hat{r}\left( X^{\prime }\right)
\right\rangle _{\hat{w}_{L}}}{2}\right) \\
&&\times \frac{1-\left\langle \hat{S}\left( X^{\prime }\right) \right\rangle 
}{1-\left\langle \hat{S}_{E}\left( X^{\prime }\right) \right\rangle }\left(
\left\langle \hat{f}\left( X^{\prime }\right) \right\rangle -\left\langle 
\bar{r}\right\rangle \right) +S_{E}\left( X,X\right) \left( \frac{%
f_{1}\left( X\right) -\frac{C}{K_{X}}+\tau \left( \left\langle f_{1}\left(
X\right) \right\rangle -\left\langle f_{1}\left( X^{\prime }\right)
\right\rangle \right) }{\left( K_{X}\left\vert \hat{\Psi}\left( X\right)
\right\vert ^{2}\right) ^{r}}-C_{0}-r\right)
\end{eqnarray*}%
and:%
\begin{equation*}
A=\frac{\left( 1-\left( \gamma \left\langle \hat{S}_{E}\left( X\right)
\right\rangle \right) ^{2}\right) }{2-\left( \gamma \left\langle \hat{S}%
_{E}\left( X\right) \right\rangle \right) ^{2}+\left( \gamma \left\langle 
\hat{S}_{E}\left( X_{1},X\right) \right\rangle _{X_{1}}\right) ^{2}-\left(
\gamma \left\langle \hat{S}_{E}\left( X\right) \right\rangle \right) ^{2}}
\end{equation*}%
This expression is computed with the averages $\left\langle \hat{S}%
_{E}\left( X\right) \right\rangle $ computed previously with decreasing
return to scale and $\left\langle \hat{S}_{E}\left( X^{\prime },X\right)
\right\rangle _{X^{\prime }}$ evaluated with (\ref{hl}).

\subsection*{A7.3 Approximate solution}

\subsubsection*{A7.3.1 Constant return to scale}

We rewrite equation (\ref{Rtns}) by replacing zeroth order averages:%
\begin{equation*}
\left\langle \hat{S}_{E}\left( X_{1},X\right) \right\rangle =z
\end{equation*}%
\begin{equation*}
\left\langle \hat{S}\left( X,X^{\prime }\right) \right\rangle =2z
\end{equation*}%
We also have:%
\begin{eqnarray*}
1-\left( \gamma \left\langle \hat{S}_{E}\left( X\right) \right\rangle
\right) ^{2} &=&\frac{2z}{1-2z} \\
\frac{1-\left( \gamma \left\langle \hat{S}_{E}\left( X\right) \right\rangle
\right) ^{2}}{2-\left( \gamma \left\langle \hat{S}_{E}\left( X\right)
\right\rangle \right) ^{2}} &=&\frac{1}{1-2z}
\end{eqnarray*}%
and we will approximate:%
\begin{equation*}
\left( \gamma \left\langle \hat{S}_{E}\left( X_{1},X\right) \right\rangle
_{X_{1}}\right) ^{2}\simeq \frac{1-4z}{1-2z}
\end{equation*}%
\begin{equation*}
\left\langle \hat{S}\left( X,X^{\prime }\right) \right\rangle _{X}\simeq 2z
\end{equation*}%
so that the equation reduces to:%
\begin{eqnarray*}
&&0=\frac{\left\langle \hat{f}\right\rangle ^{2}W_{3}-2z\left( 1+\frac{%
\Delta \hat{f}\left( X\right) +\Delta \hat{r}\left( X\right) }{2}\right)
\left( \hat{f}\left( X^{\prime }\right) \left( 1-2z\right) +2z\left\langle 
\hat{f}\right\rangle \right) ^{2}}{\left\langle \hat{f}\right\rangle
^{2}W_{3}-z\left( 1+\Delta \hat{f}\left( X\right) \right) \left( \hat{f}%
\left( X^{\prime }\right) \left( 1-2z\right) +2z\left\langle \hat{f}%
\right\rangle \right) ^{2}}\left( \hat{f}\left( X\right) -\bar{r}\right) \\
&&-z\left( 1-\left( 1-2z\right) \Delta \left( \frac{f\left( X\right)
+r\left( X\right) }{2}\right) +\frac{\left\langle \hat{f}\left( X^{\prime
}\right) \right\rangle -\left\langle \hat{r}\left( X^{\prime }\right)
\right\rangle _{\hat{w}_{L}}}{2}\right) \\
&&\times \frac{1-2z}{1-z}\left( \left\langle \hat{f}\left( X^{\prime
}\right) \right\rangle -\left\langle \bar{r}\right\rangle \right) -\frac{1-2z%
}{2}\left( f\left( X^{\prime }\right) -r\right)
\end{eqnarray*}%
where:%
\begin{equation*}
W_{3}=\left( \frac{2z\left( 1+\frac{\Delta \hat{f}\left( X\right) +\Delta 
\hat{r}\left( X\right) }{2}\right) }{1-2z\left( 1+\frac{\Delta \hat{f}\left(
X\right) +\Delta \hat{r}\left( X\right) }{2}\right) \frac{\left\langle \hat{K%
}\right\rangle \left\Vert \hat{\Psi}\right\Vert ^{2}}{\hat{K}_{X^{\prime
}}\left\vert \hat{\Psi}\left( X^{\prime }\right) \right\vert ^{2}}}\frac{%
1-2z\left( 1+\frac{\left\langle \Delta \hat{f}\left( X^{\prime }\right)
+\Delta \hat{r}\left( X^{\prime }\right) \right\rangle }{2}\right) }{%
2z\left( 1+\frac{\left\langle \Delta \hat{f}\left( X^{\prime }\right)
+\Delta \hat{r}\left( X^{\prime }\right) \right\rangle }{2}\right) }\right)
^{2}
\end{equation*}%
We also use that:%
\begin{equation}
\Delta \hat{f}\left( X^{\prime }\right) =\hat{f}\left( X^{\prime }\right)
-\left\langle \tilde{f}+\tilde{r}\right\rangle
\end{equation}%
with:%
\begin{equation*}
\left\langle \tilde{f}\right\rangle +\left\langle \tilde{r}\right\rangle
=\left\langle \hat{w}\left( X\right) \right\rangle \frac{\left\langle \hat{f}%
\left( X^{\prime }\right) \right\rangle _{\hat{w}_{E}}+\left\langle \hat{r}%
\left( X^{\prime }\right) \right\rangle _{\hat{w}_{L}}}{2}+\left\langle
w\left( X\right) \right\rangle \frac{\left\langle f\left( X\right)
\right\rangle +\left\langle r\left( X\right) \right\rangle }{2}
\end{equation*}%
along with:%
\begin{equation*}
\Delta \hat{f}\left( X^{\prime }\right) =\hat{f}\left( X^{\prime }\right)
-\left( \left\langle \hat{w}\left( X\right) \right\rangle \frac{\left\langle 
\hat{f}\left( X^{\prime }\right) \right\rangle _{\hat{w}_{E}}+\left\langle 
\hat{r}\left( X^{\prime }\right) \right\rangle _{\hat{w}_{L}}}{2}%
+\left\langle w\left( X\right) \right\rangle \frac{\left\langle f\left(
X\right) \right\rangle +\left\langle r\left( X\right) \right\rangle }{2}%
\right)
\end{equation*}%
\begin{equation}
\Delta \hat{f}\left( X^{\prime }\right) =\hat{f}\left( X^{\prime }\right)
-r-z\left( \left\langle \hat{f}\left( X^{\prime }\right) \right\rangle
-r\right) -\frac{1-2z}{2}\left( \left\langle f\left( X\right) \right\rangle
-r\right)
\end{equation}%
\begin{equation*}
\Delta \hat{r}\left( X\right) =-z\left( \left\langle \hat{f}\left( X^{\prime
}\right) \right\rangle -r\right) -\frac{\left( 1-2z\right) }{2}\left(
\left\langle f\left( X\right) \right\rangle -r\right)
\end{equation*}%
\begin{eqnarray*}
&&\frac{\Delta \hat{f}\left( X\right) +\Delta \hat{r}\left( X\right) }{2} \\
&\rightarrow &\frac{1}{2}\left( \hat{f}\left( X^{\prime }\right) -r\right)
-z\left( \left\langle \hat{f}\left( X^{\prime }\right) \right\rangle
-r\right) -\frac{1-2z}{2}\left( \left\langle f\left( X\right) \right\rangle
-r\right)
\end{eqnarray*}%
alternatively:%
\begin{equation*}
\frac{\Delta \hat{f}\left( X\right) +\Delta \hat{r}\left( X\right) }{2}%
\rightarrow z\left( \hat{f}\left( X^{\prime }\right) -\left\langle \hat{f}%
\left( X^{\prime }\right) \right\rangle \right) +\frac{\left( 1-2z\right) }{2%
}\left( \hat{f}\left( X^{\prime }\right) -\left\langle f\left( X\right)
\right\rangle \right)
\end{equation*}%
and:%
\begin{equation*}
\Delta \left( \frac{f\left( X\right) +r\left( X\right) }{2}\right)
\rightarrow -\left( f\left( X\right) -\left\langle f\left( X\right)
\right\rangle \right)
\end{equation*}%
and this leads to:%
\begin{eqnarray*}
&&0=\frac{\left\langle \hat{f}\right\rangle ^{2}D-2z\left( 1+z\left( \hat{f}%
\left( X^{\prime }\right) -\left\langle \hat{f}\left( X^{\prime }\right)
\right\rangle \right) +\frac{1-2z}{2}\left( \hat{f}\left( X^{\prime }\right)
-\left\langle f\left( X\right) \right\rangle \right) \right) \left( \hat{f}%
\left( X^{\prime }\right) \left( 1-2z\right) +2z\left\langle \hat{f}%
\right\rangle \right) ^{2}}{\left\langle \hat{f}\right\rangle ^{2}D-z\left(
1+\hat{f}\left( X^{\prime }\right) -r-z\left( \left\langle \hat{f}\left(
X^{\prime }\right) \right\rangle -r\right) -\frac{1-2z}{2}\left(
\left\langle f\left( X\right) \right\rangle -r\right) \right) \left( \hat{f}%
\left( X^{\prime }\right) \left( 1-2z\right) +2z\left\langle \hat{f}%
\right\rangle \right) ^{2}}\left( \hat{f}\left( X\right) -\bar{r}\right) \\
&&-z\left( 1+\left( 1-2z\right) \left( f\left( X\right) -\left\langle
f\left( X\right) \right\rangle \right) +\frac{\left\langle \hat{f}\left(
X^{\prime }\right) \right\rangle -\left\langle \hat{r}\left( X^{\prime
}\right) \right\rangle _{\hat{w}_{L}}}{2}\right) \\
&&\times \frac{1-2z}{1-z}\left( \left\langle \hat{f}\left( X^{\prime
}\right) \right\rangle -\left\langle \bar{r}\right\rangle \right) -\frac{1-2z%
}{2}\left( f\left( X^{\prime }\right) -r\right)
\end{eqnarray*}%
with:%
\begin{eqnarray*}
D &=&\left( \frac{2z\left( 1+z\left( \hat{f}\left( X^{\prime }\right)
-\left\langle \hat{f}\left( X^{\prime }\right) \right\rangle \right) +\frac{%
1-2z}{2}\left( \hat{f}\left( X^{\prime }\right) -\left\langle f\left(
X\right) \right\rangle \right) \right) }{1-2z\left( 1+z\left( \hat{f}\left(
X^{\prime }\right) -\left\langle \hat{f}\left( X^{\prime }\right)
\right\rangle \right) +\frac{1-2z}{2}\left( \hat{f}\left( X^{\prime }\right)
-\left\langle f\left( X\right) \right\rangle \right) \right) \frac{%
\left\langle \hat{K}\right\rangle \left\Vert \hat{\Psi}\right\Vert ^{2}}{%
\hat{K}_{X^{\prime }}\left\vert \hat{\Psi}\left( X^{\prime }\right)
\right\vert ^{2}}}\right. \\
&&\left. \frac{1-2z\left( 1+\frac{1-2z}{2}\left( \left\langle \hat{f}\left(
X^{\prime }\right) \right\rangle -\left\langle f\left( X\right)
\right\rangle \right) \right) }{2z\left( 1+\frac{1-2z}{2}t\right) }\right)
^{2}
\end{eqnarray*}%
We look for solutions around the state with:%
\begin{equation*}
\hat{f}\left( X^{\prime }\right) \simeq \left\langle \hat{f}\left( X^{\prime
}\right) \right\rangle \simeq \left\langle f\left( X\right) \right\rangle
\simeq \bar{r}
\end{equation*}%
so that we can consider second order expansion in $\frac{\hat{f}\left(
X^{\prime }\right) -r}{r}$ and $\frac{\left\langle \hat{f}\left( X^{\prime
}\right) \right\rangle -r}{r}$, by replacing:%
\begin{eqnarray*}
\frac{\left\langle \hat{f}\left( X^{\prime }\right) \right\rangle -r}{r}
&\rightarrow &\hat{f}\left( X^{\prime }\right) -r \\
\frac{\left\langle \hat{f}\left( X^{\prime }\right) \right\rangle -r}{r}
&\rightarrow &\left\langle \hat{f}\left( X^{\prime }\right) \right\rangle -r
\end{eqnarray*}%
and assuming:%
\begin{equation*}
f\left( X\right) -\left\langle f\left( X\right) \right\rangle <<1
\end{equation*}%
we are led to:%
\begin{eqnarray}
0 &=&\frac{\left( 1+\left\langle \hat{f}\left( X^{\prime }\right)
\right\rangle -r\right) ^{2}D-2z\left( 1+\frac{\hat{f}\left( X^{\prime
}\right) -\bar{r}}{2}-z\left( \left\langle \hat{f}\left( X^{\prime }\right)
\right\rangle -\left\langle \bar{r}\right\rangle \right) -\frac{1-2z}{2}%
\left( \left\langle f\left( X\right) \right\rangle -r\right) \right) V}{%
\left( 1+\left\langle \hat{f}\left( X^{\prime }\right) \right\rangle
-r\right) ^{2}D-z\left( 1+\left( \hat{f}\left( X^{\prime }\right) -r\right)
-z\left( \left\langle \hat{f}\left( X^{\prime }\right) \right\rangle
-r\right) -\frac{1-2z}{2}\left( \left\langle f\left( X\right) \right\rangle
-r\right) \right) V}\left( \hat{f}\left( X\right) -\bar{r}\right)
\label{PRqt} \\
&&-z\frac{1-2z}{1-z}\left( \left\langle \hat{f}\left( X^{\prime }\right)
\right\rangle -\left\langle \bar{r}\right\rangle \right) -z\frac{\left(
1-2z\right) ^{2}}{1-z}\left( \left\langle \hat{f}\left( X^{\prime }\right)
\right\rangle -\left\langle \bar{r}\right\rangle \right) \left( f\left(
X\right) -\left\langle f\left( X\right) \right\rangle \right) -\frac{1-2z}{2}%
\left( f\left( X^{\prime }\right) -r\right)  \notag
\end{eqnarray}%
with:%
\begin{equation*}
V=\left( 1+\left( \hat{f}\left( X^{\prime }\right) -\bar{r}\right) \left(
1-2z\right) +2z\left( \left\langle \hat{f}\left( X^{\prime }\right)
\right\rangle -\left\langle \bar{r}\right\rangle \right) \right) ^{2}
\end{equation*}%
or equivalently:%
\begin{eqnarray*}
0 &=&\frac{\left( 1+\left\langle \hat{f}\left( X^{\prime }\right)
\right\rangle -r\right) ^{2}D-2z\left( 1+\frac{1}{2}\left( \hat{f}\left(
X^{\prime }\right) -\bar{r}\right) -z\left( \left\langle \hat{f}\left(
X^{\prime }\right) \right\rangle -\left\langle \bar{r}\right\rangle \right) -%
\frac{1-2z}{2}\left( \left\langle f\left( X\right) \right\rangle -r\right)
\right) V^{\prime }}{\left( 1+\left\langle \hat{f}\left( X^{\prime }\right)
\right\rangle -r\right) ^{2}D-z\left( 1+\left( \hat{f}\left( X^{\prime
}\right) -\bar{r}\right) -z\left( \left\langle \hat{f}\left( X^{\prime
}\right) \right\rangle -r\right) -\frac{1-2z}{2}\left( \left\langle f\left(
X\right) \right\rangle -r\right) \right) V(}\left( \hat{f}\left( X\right) -%
\bar{r}\right) \\
&&-z\frac{1-2z}{1-z}\left( \left\langle \hat{f}\left( X^{\prime }\right)
\right\rangle -\left\langle \bar{r}\right\rangle \right) +z\frac{\left(
1-2z\right) ^{2}}{1-z}\left( \left\langle \hat{f}\left( X^{\prime }\right)
\right\rangle -\left\langle \bar{r}\right\rangle \right) \left( f\left(
X\right) -\bar{r}\right) -\frac{1-2z}{2}\left( f\left( X^{\prime }\right)
-r\right)
\end{eqnarray*}%
where:%
\begin{equation*}
V^{\prime }=\left( 1+\frac{\left( \hat{f}\left( X^{\prime }\right) -\bar{r}%
\right) }{\bar{r}}\left( 1-2z\right) +2z\left( \left\langle \hat{f}\left(
X^{\prime }\right) \right\rangle -\left\langle \bar{r}\right\rangle \right)
\right) ^{2}
\end{equation*}%
We set:%
\begin{equation*}
x=\frac{\hat{f}\left( X^{\prime }\right) -\bar{r}}{\bar{r}}
\end{equation*}%
\begin{equation*}
t=\frac{\left\langle \hat{f}\left( X^{\prime }\right) \right\rangle -\bar{r}%
}{\bar{r}}
\end{equation*}%
\begin{equation*}
w=\frac{\left\langle f\left( X\right) \right\rangle -\bar{r}}{\bar{r}}
\end{equation*}%
We also use that in first approximation:%
\begin{equation*}
\frac{\left\langle \hat{K}\right\rangle \left\Vert \hat{\Psi}\right\Vert ^{2}%
}{\hat{K}_{X^{\prime }}\left\vert \hat{\Psi}\left( X^{\prime }\right)
\right\vert ^{2}}\rightarrow \left( \frac{1+x\left( 1-2z\right) +2zt}{1+t}%
\frac{\frac{2z\left( 1+\frac{1-2z}{2}t-\frac{1-2z}{2}w\right) }{1-2z\left( 1+%
\frac{1-2z}{2}t-\frac{1-2z}{2}w\right) }}{\frac{2z\left( 1+\frac{1}{2}x-zt-%
\frac{1-2z}{2}w\right) }{1-2z\left( 1+\frac{1}{2}x-zt-\frac{1-2z}{2}w\right) 
}}\right) ^{2}
\end{equation*}%
and the equation becomes:%
\begin{eqnarray}
0 &=&\frac{\left( 1+t\right) ^{2}D-2z\left( 1+\frac{1}{2}x-zt-\frac{1-2z}{2}%
w\right) \left( 1+x\left( 1-2z\right) +2zt\right) ^{2}}{\left( 1+t\right)
^{2}D-z\left( 1+x-zt-\frac{1-2z}{2}w\right) \left( 1+x\left( 1-2z\right)
+2zt\right) ^{2}}  \label{qtb} \\
&&-z\frac{1-2z}{1-z}t-\frac{1-2z}{2}w  \notag
\end{eqnarray}%
with the expanded form for $D$:%
\begin{equation*}
D=\left( \frac{2z\left( 1+\frac{1}{2}x-zt-\frac{1-2z}{2}w\right) }{%
1-2z\left( 1+\frac{1}{2}x-zt-\frac{1-2z}{2}w\right) \left( \frac{1+x\left(
1-2z\right) +2zt}{1+t}\frac{\frac{2z\left( 1+\frac{1-2z}{2}t-\frac{1-2z}{2}%
w\right) }{1-2z\left( 1+\frac{1-2z}{2}t-\frac{1-2z}{2}w\right) }}{\frac{%
2z\left( 1+\frac{1}{2}x-zt-\frac{1-2z}{2}w\right) }{1-2z\left( 1+\frac{1}{2}%
x-zt-\frac{1-2z}{2}w\right) }}\right) ^{2}}\frac{1-2z\left( 1+\frac{1-2z}{2}%
t-\frac{1-2z}{2}w\right) }{2z\left( 1+\frac{1-2z}{2}t-\frac{1-2z}{2}w\right) 
}\right) ^{2}
\end{equation*}%
\bigskip

A second order expansion in $x$ yields:%
\begin{equation*}
0=a\left( z\right) x^{2}-\left( \frac{2z-1}{z-1}+a\left( z\right) t+\frac{1}{%
2}wz\frac{1-2z}{\left( z-1\right) ^{2}}\right) x+z\frac{1-2z}{1-z}t+\frac{%
1-2z}{2}w
\end{equation*}%
with:%
\begin{equation*}
a\left( z\right) =z\frac{1-13z+52z^{2}-44z^{3}}{\left( 1-2z\right)
^{2}\left( 1-z\right) ^{2}}
\end{equation*}

$\allowbreak $Coming back to the initial variables yields the solutions:%
\begin{eqnarray}
&&\hat{f}\left( X^{\prime }\right) -\bar{r} \\
&=&A_{1}\pm \sqrt{A_{1}^{2}-A_{2}\left\langle \bar{r}\right\rangle }  \notag
\end{eqnarray}%
with:%
\begin{eqnarray*}
A_{1} &=&\frac{\left\langle \hat{f}\left( X^{\prime }\right) \right\rangle
-\left\langle \bar{r}\right\rangle }{2}+\frac{z\left( 1-2z\right) }{\left(
1-z\right) ^{2}}\frac{\left( \left\langle f\left( X^{\prime }\right)
\right\rangle -\left\langle \bar{r}\right\rangle \right) }{2a\left( z\right) 
}+\frac{\left( 1-2z\right) }{\left( 1-z\right) }\frac{\left\langle \bar{r}%
\right\rangle }{2a\left( z\right) } \\
A_{2} &=&\frac{z\frac{1-2z}{1-z}\left( \left\langle \hat{f}\left( X^{\prime
}\right) \right\rangle -\left\langle \bar{r}\right\rangle \right) }{a\left(
z\right) }+\frac{\left( 1-2z\right) \left( f\left( X^{\prime }\right)
-r\right) }{2a\left( z\right) }
\end{eqnarray*}%
The condition of feasibility of the high return solution (with $+$), can be
understood by studying (\ref{qtb}) for:%
\begin{equation*}
t=w=0
\end{equation*}%
that is:%
\begin{equation}
0=\frac{D-2z\left( 1+\frac{1}{2}x\right) \left( 1+x\left( 1-2z\right)
\right) ^{2}}{D-z\left( 1+x\right) \left( 1+x\left( 1-2z\right) \right) ^{2}}
\end{equation}%
where:%
\begin{equation*}
D=\left( \frac{\left( 1+\frac{1}{2}x\right) \left( 1-2z\right) }{1-2z\left(
1+\frac{1}{2}x\right) \left( \left( 1+x\left( 1-2z\right) \right) \frac{%
1-2z\left( 1+\frac{1}{2}x\right) }{\left( 1-2z\right) \left( 1+\frac{1}{2}%
x\right) }\right) ^{2}}\right) ^{2}
\end{equation*}%
The numerical study of this equation shows that there is one feasible
solution $x$ (i.e. $x$ of magnitude $1$ or $2$) for $z$ relatively large,
that is $z>0.2$.

A series expansion yields the "$-$" solution if we consider it as negligible
compared to the high return solution.%
\begin{equation*}
\hat{f}_{-}\left( X^{\prime }\right) -\bar{r}\simeq \frac{\left( \frac{z%
\frac{1-2z}{1-z}\left( \left\langle \hat{f}\left( X^{\prime }\right)
\right\rangle -\left\langle \bar{r}\right\rangle \right) }{a\left( z\right) }%
+\frac{\left( 1-2z\right) \left( f\left( X^{\prime }\right) -r\right) }{%
2a\left( z\right) }\right) \left\langle \bar{r}\right\rangle }{2\left( \frac{%
\left\langle \hat{f}\left( X^{\prime }\right) \right\rangle -\left\langle 
\bar{r}\right\rangle }{2}+\frac{z\left( 1-2z\right) }{\left( 1-z\right) ^{2}}%
\frac{\left\langle f\left( X^{\prime }\right) \right\rangle -\left\langle 
\bar{r}\right\rangle }{2a\left( z\right) }+\frac{\left( 1-2z\right) }{\left(
1-z\right) }\frac{\left\langle \bar{r}\right\rangle }{2a\left( z\right) }%
\right) }-\frac{\left( \frac{z\frac{1-2z}{1-z}\left( \left\langle \hat{f}%
\left( X^{\prime }\right) \right\rangle -\left\langle \bar{r}\right\rangle
\right) }{a\left( z\right) }+\frac{\left( 1-2z\right) \left( f\left(
X^{\prime }\right) -r\right) }{2a\left( z\right) }\right) ^{2}\left\langle 
\bar{r}\right\rangle ^{2}}{8\left( \frac{\left\langle \hat{f}\left(
X^{\prime }\right) \right\rangle -\left\langle \bar{r}\right\rangle }{2}+%
\frac{z\left( 1-2z\right) }{\left( 1-z\right) ^{2}}\frac{\left\langle
f\left( X^{\prime }\right) \right\rangle -\left\langle \bar{r}\right\rangle 
}{2a\left( z\right) }+\frac{\left( 1-2z\right) }{\left( 1-z\right) }\frac{%
\left\langle \bar{r}\right\rangle }{2a\left( z\right) }\right) ^{3}}
\end{equation*}%
Assuming:%
\begin{equation*}
\frac{\left\langle \hat{f}\left( X^{\prime }\right) \right\rangle
-\left\langle \bar{r}\right\rangle }{2}+\frac{z\left( 1-2z\right) }{\left(
1-z\right) ^{2}}\frac{\left\langle f\left( X^{\prime }\right) \right\rangle
-\left\langle \bar{r}\right\rangle }{2a\left( z\right) }<<\frac{\left(
1-2z\right) }{\left( 1-z\right) }\frac{\left\langle \bar{r}\right\rangle }{%
2a\left( z\right) }
\end{equation*}%
the first term expands as:%
\begin{eqnarray*}
&&\frac{\left( \frac{z\frac{1-2z}{1-z}\left( \left\langle \hat{f}\left(
X^{\prime }\right) \right\rangle -\left\langle \bar{r}\right\rangle \right) 
}{a\left( z\right) }+\frac{\left( 1-2z\right) \left( f\left( X^{\prime
}\right) -r\right) }{2a\left( z\right) }\right) \left\langle \bar{r}%
\right\rangle }{2\left( \frac{\left\langle \hat{f}\left( X^{\prime }\right)
\right\rangle -\left\langle \bar{r}\right\rangle }{2}+\frac{z\left(
1-2z\right) }{\left( 1-z\right) ^{2}}\frac{\left\langle f\left( X^{\prime
}\right) \right\rangle -\left\langle \bar{r}\right\rangle }{2a\left(
z\right) }+\frac{\left( 1-2z\right) }{\left( 1-z\right) }\frac{\left\langle 
\bar{r}\right\rangle }{2a\left( z\right) }\right) } \\
&\simeq &\left( z\left( \left\langle \hat{f}\left( X^{\prime }\right)
\right\rangle -\left\langle \bar{r}\right\rangle \right) +\frac{1-z}{2}%
\left( f\left( X^{\prime }\right) -r\right) \right) \left( 1-\left( \frac{%
a\left( z\right) \left( 1-z\right) \left( \left\langle \hat{f}\left(
X^{\prime }\right) \right\rangle -\left\langle \bar{r}\right\rangle \right) 
}{\left\langle \bar{r}\right\rangle \left( 1-2z\right) }+\frac{z\left(
\left\langle f\left( X^{\prime }\right) \right\rangle -\left\langle \bar{r}%
\right\rangle \right) }{\left\langle \bar{r}\right\rangle \left( 1-z\right) }%
\right) \right)
\end{eqnarray*}%
and the second term writes:%
\begin{eqnarray*}
&&-\frac{\left( \frac{z\frac{1-2z}{1-z}\left( \left\langle \hat{f}\left(
X^{\prime }\right) \right\rangle -\left\langle \bar{r}\right\rangle \right) 
}{a\left( z\right) }+\frac{\left( 1-2z\right) \left( f\left( X^{\prime
}\right) -r\right) }{2a\left( z\right) }\right) ^{2}\left\langle \bar{r}%
\right\rangle ^{2}}{8\left( \frac{\left\langle \hat{f}\left( X^{\prime
}\right) \right\rangle -\left\langle \bar{r}\right\rangle }{2}+\frac{z\left(
1-2z\right) }{\left( 1-z\right) ^{2}}\frac{\left\langle f\left( X^{\prime
}\right) \right\rangle -\left\langle \bar{r}\right\rangle }{2a\left(
z\right) }+\frac{\left( 1-2z\right) }{\left( 1-z\right) }\frac{\left\langle 
\bar{r}\right\rangle }{2a\left( z\right) }\right) ^{3}} \\
&\simeq &-\frac{\left( \frac{z\frac{1-2z}{1-z}\left( \left\langle \hat{f}%
\left( X^{\prime }\right) \right\rangle -\left\langle \bar{r}\right\rangle
\right) }{a\left( z\right) }+\frac{\left( 1-2z\right) \left( f\left(
X^{\prime }\right) -r\right) }{2a\left( z\right) }\right) ^{2}\left\langle 
\bar{r}\right\rangle ^{2}}{8\left( \frac{\left( 1-2z\right) }{\left(
1-z\right) }\frac{\left\langle \bar{r}\right\rangle }{2a\left( z\right) }%
\right) ^{3}} \\
&\simeq &-\frac{\left( z\left( \left\langle \hat{f}\left( X^{\prime }\right)
\right\rangle -\left\langle \bar{r}\right\rangle \right) +\frac{1-z}{2}%
\left( f\left( X^{\prime }\right) -r\right) \right) ^{2}}{2\frac{\left(
1-2z\right) }{\left( 1-z\right) }\frac{\left\langle \bar{r}\right\rangle }{%
2a\left( z\right) }}
\end{eqnarray*}%
leading to:%
\begin{eqnarray*}
\hat{f}_{-}\left( X^{\prime }\right) -\bar{r} &\simeq &\left( z\left(
\left\langle \hat{f}\left( X^{\prime }\right) \right\rangle -\left\langle 
\bar{r}\right\rangle \right) +\frac{1-z}{2}\left( f\left( X^{\prime }\right)
-r\right) \right) \\
&&\left( 1-\left( \frac{2a\left( z\right) \left( 1-z\right) }{\left(
1-2z\right) }\frac{\left\langle \hat{f}\left( X^{\prime }\right)
\right\rangle -\left\langle \bar{r}\right\rangle }{\left\langle \bar{r}%
\right\rangle }+\left( \frac{z}{\left( 1-z\right) }+\frac{\left( 1-z\right)
a\left( z\right) }{\left( 1-2z\right) }\right) \frac{f\left( X^{\prime
}\right) -\bar{r}}{\left\langle \bar{r}\right\rangle }\right) \right) \\
&\rightarrow &\left( z\left( \left\langle \hat{f}\left( X^{\prime }\right)
\right\rangle -\left\langle \bar{r}\right\rangle \right) +\frac{1-z}{2}%
\left( f\left( X^{\prime }\right) -r\right) \right) \left( 1-\frac{2a\left(
z\right) \left( 1-z\right) }{\left( 1-2z\right) }\left( \frac{\left\langle 
\hat{f}\left( X^{\prime }\right) \right\rangle -\left\langle \bar{r}%
\right\rangle }{\left\langle \bar{r}\right\rangle }+\frac{f\left( X^{\prime
}\right) -\bar{r}}{\left\langle \bar{r}\right\rangle }\right) \right)
\end{eqnarray*}

\subsubsection*{A7.3.2 Decreasing return to scale}

As explained before this corresponds to replace:%
\begin{equation*}
f_{1}\left( X\right) \rightarrow \frac{f_{1}\left( X\right) +\tau \left(
f_{1}\left( X\right) -\left\langle f_{1}\left( X^{\prime }\right)
\right\rangle \right) }{\left( K_{X}\left\vert \hat{\Psi}\left( X\right)
\right\vert ^{2}\right) ^{r}}-\frac{C}{K_{X}}-C_{0}\simeq \frac{f_{1}\left(
X\right) }{\left( K_{X}\left\vert \hat{\Psi}\left( X\right) \right\vert
^{2}\right) ^{r}}-\frac{C}{K_{X}}-C_{0}
\end{equation*}%
and to keep the constant return values of shares.

Equation (\ref{PRqt}) is thus replaced by:%
\begin{eqnarray}
0 &=&\frac{\left( 1+\left\langle \hat{f}\left( X^{\prime }\right)
\right\rangle -r\right) ^{2}-2z\left( 1+\frac{1}{2}\left( \hat{f}\left(
X^{\prime }\right) -\bar{r}\right) -z\left( \left\langle \hat{f}\left(
X^{\prime }\right) \right\rangle -\left\langle \bar{r}\right\rangle \right)
\right) V^{\prime }}{\left( 1+\left\langle \hat{f}\left( X^{\prime }\right)
\right\rangle -r\right) ^{2}-z\left( 1+\left( \hat{f}\left( X^{\prime
}\right) -r\right) -z\left( \left\langle \hat{f}\left( X^{\prime }\right)
\right\rangle -r\right) \right) V^{\prime }}\left( \hat{f}\left( X\right) -%
\bar{r}\right) \\
&&-2z\frac{1-2z}{1-z}\left( \left\langle \hat{f}\left( X^{\prime }\right)
\right\rangle -\left\langle \bar{r}\right\rangle \right) -2z\frac{\left(
1-2z\right) ^{2}}{1-z}\left( \left\langle \hat{f}\left( X^{\prime }\right)
\right\rangle -\left\langle \bar{r}\right\rangle \right) \left( f\left(
X\right) -\left\langle f\left( X\right) \right\rangle \right)  \notag \\
&&-\frac{1-2z}{2}\left( \frac{f_{1}\left( X\right) }{\left( K_{X}\left\vert 
\hat{\Psi}\left( X\right) \right\vert ^{2}\right) ^{r}}-\frac{C}{K_{X}}%
-C_{0}-r\right)  \notag
\end{eqnarray}%
with:%
\begin{equation*}
V^{\prime }=\left( 1+\left( \hat{f}\left( X^{\prime }\right) -\bar{r}\right)
\left( 1-2z\right) +2z\left( \left\langle \hat{f}\left( X^{\prime }\right)
\right\rangle -\left\langle \bar{r}\right\rangle \right) \right) ^{2}\left( 
\frac{\frac{\left\langle \hat{S}\left( X^{\prime },X\right) \right\rangle }{%
1-\left\langle \hat{S}\left( X^{\prime },X\right) \right\rangle }}{%
\left\langle \hat{f}\right\rangle \frac{\left\langle \hat{S}\left( X^{\prime
},X\right) \right\rangle _{X}}{1-\left\langle \hat{S}\left( X^{\prime
},X\right) \right\rangle _{X}\frac{\left\langle \hat{K}\right\rangle
\left\Vert \hat{\Psi}\right\Vert ^{2}}{\hat{K}_{X^{\prime }}\left\vert \hat{%
\Psi}\left( X^{\prime }\right) \right\vert ^{2}}}}\right) ^{2}
\end{equation*}%
Given (\ref{PRkh}), we have:%
\begin{eqnarray*}
K_{X}\left\vert \Psi \left( X\right) \right\vert ^{2} &=&\left( \left(
1-S\left( X\right) \right) \right) ^{2}\left( \left( \frac{2\epsilon }{%
3\sigma _{\hat{K}}^{2}}\right) ^{\frac{r}{2}}\left( \frac{f_{1}\left(
X\right) }{C_{0}+\bar{r}}\right) \right) ^{\frac{2}{r}} \\
&\simeq &\left( 1-S\left( X,X\right) \frac{\hat{K}_{X}\left\vert \hat{\Psi}%
\left( X\right) \right\vert ^{2}}{\left( \left( \frac{2\epsilon }{3\sigma _{%
\hat{K}}^{2}}\right) ^{\frac{r}{2}}\left( \frac{f_{1}\left( X\right) }{C_{0}+%
\bar{r}}\right) \right) ^{\frac{2}{r}}}\right) ^{2}\left( \left( \frac{%
2\epsilon }{3\sigma _{\hat{K}}^{2}}\right) ^{\frac{r}{2}}\left( \frac{%
f_{1}\left( X\right) }{C_{0}+\bar{r}}\right) \right) ^{\frac{2}{r}}
\end{eqnarray*}%
and we replace:%
\begin{eqnarray*}
\hat{K}_{X^{\prime }}\left\vert \hat{\Psi}\left( X^{\prime }\right)
\right\vert ^{2} &\simeq &\left\langle \hat{K}\right\rangle \left\Vert \hat{%
\Psi}\right\Vert ^{2}\left( \frac{\left\langle \hat{g}\right\rangle }{\hat{g}%
\left( \hat{X}^{\prime }\right) }\frac{\frac{\left\langle \hat{S}\left(
X^{\prime },X\right) \right\rangle _{X}}{1-\left\langle \hat{S}\left(
X^{\prime },X\right) \right\rangle _{X}}}{\frac{\left\langle \hat{S}\left(
X^{\prime },X\right) \right\rangle }{1-\left\langle \hat{S}\left( X^{\prime
},X\right) \right\rangle }}\right) ^{2} \\
&\simeq &\left\langle \hat{K}\right\rangle \left\Vert \hat{\Psi}\right\Vert
^{2}\left( \frac{\left\langle \hat{f}\right\rangle \frac{\left\langle \hat{S}%
\left( X^{\prime },X\right) \right\rangle _{X}}{1-\left\langle \hat{S}\left(
X^{\prime },X\right) \right\rangle _{X}}}{\left( \hat{f}\left( X^{\prime
}\right) \left( 1-\left\langle \hat{S}\right\rangle \right) +\left\langle 
\hat{S}\left( X,X^{\prime }\right) \right\rangle _{X}\left\langle \hat{f}%
\right\rangle \right) \frac{\left\langle \hat{S}\left( X^{\prime },X\right)
\right\rangle }{1-\left\langle \hat{S}\left( X^{\prime },X\right)
\right\rangle }}\right) ^{2}
\end{eqnarray*}%
so that:%
\begin{eqnarray*}
K_{X}\left\vert \Psi \left( X\right) \right\vert ^{2} &\simeq &\left( 1-%
\frac{S\left( X,X\right) \left\langle \hat{K}\right\rangle \left\Vert \hat{%
\Psi}\right\Vert ^{2}\left( \frac{\left\langle \hat{f}\right\rangle \frac{%
\left\langle \hat{S}\left( X^{\prime },X\right) \right\rangle _{X}}{%
1-\left\langle \hat{S}\left( X^{\prime },X\right) \right\rangle _{X}}}{%
\left( \hat{f}\left( X^{\prime }\right) \left( 1-\left\langle \hat{S}%
\right\rangle \right) +\left\langle \hat{S}\left( X,X^{\prime }\right)
\right\rangle _{X}\left\langle \hat{f}\right\rangle \right) \frac{%
\left\langle \hat{S}\left( X^{\prime },X\right) \right\rangle }{%
1-\left\langle \hat{S}\left( X^{\prime },X\right) \right\rangle }}\right)
^{2}}{\left( \left( \frac{2\epsilon }{3\sigma _{\hat{K}}^{2}}\right) ^{\frac{%
r}{2}}\left( \frac{f_{1}\left( X\right) }{C_{0}+\bar{r}}\right) \right) ^{%
\frac{2}{r}}}\right) \left( \left( \frac{2\epsilon }{3\sigma _{\hat{K}}^{2}}%
\right) ^{\frac{r}{2}}\left( \frac{f_{1}\left( X\right) }{C_{0}+\bar{r}}%
\right) \right) ^{\frac{2}{r}} \\
&\simeq &\left( 1-\frac{S\left( X,X\right) \left\langle \hat{K}\right\rangle
\left\Vert \hat{\Psi}\right\Vert ^{2}}{\left( \left( \frac{2\epsilon }{%
3\sigma _{\hat{K}}^{2}}\right) ^{\frac{r}{2}}\left( \frac{f_{1}\left(
X\right) }{C_{0}+\bar{r}}\right) \right) ^{\frac{2}{r}}}\right) \left(
\left( \frac{2\epsilon }{3\sigma _{\hat{K}}^{2}}\right) ^{\frac{r}{2}}\left( 
\frac{f_{1}\left( X\right) }{C_{0}+\bar{r}}\right) \right) ^{\frac{2}{r}}
\end{eqnarray*}%
Similarly:%
\begin{equation*}
K_{X}\simeq \left( \left( \frac{2\epsilon }{3\sigma _{\hat{K}}^{2}}\right) ^{%
\frac{r}{2}}\left( \frac{f_{1}\left( X\right) }{C_{0}+\bar{r}}\right)
\right) ^{\frac{1}{r}}
\end{equation*}%
With $f_{1}\left( X\right) $ replaced by:%
\begin{eqnarray*}
&&\frac{1-2z}{2}\left( \frac{f_{1}\left( X\right) }{\left( K_{X}\left\vert 
\hat{\Psi}\left( X\right) \right\vert ^{2}\right) ^{r}}-C_{0}-r\right) \\
&\simeq &\frac{1-2z}{2}\left( f_{a}-\frac{1}{2}f_{b}\left( \frac{C_{0}+\frac{%
S_{L}\left( X^{\prime }\right) }{1-S_{E}\left( X^{\prime }\right) }\bar{r}}{%
f_{1}\left( X\right) }\right) ^{\frac{1}{r}}\right)
\end{eqnarray*}%
where the formula for $f_{a}$ and $f_{b}$ given by (\ref{LSf}) are still
valid.

\section*{Appendix 8 Evaluation of shares and firms' returns}

\subsection*{A8.1 Shares}

We can ultimately come back to the initial formula expressing the
uncertainty coefficients and the shares. For constant return to scale, the
results are: 
\begin{equation*}
\frac{\hat{w}\left( X^{\prime },X\right) }{2}=\frac{\left( 1-\left( \gamma
\left\langle \hat{S}_{E}\left( X\right) \right\rangle \right) ^{2}\right) 
\hat{w}_{E}^{\left( 0\right) }\left( X^{\prime },X\right) }{1+\hat{w}%
_{E}^{\left( 0\right) }\left( X^{\prime },X\right) \left( 1-\left( \gamma
\left\langle \hat{S}_{E}\left( X\right) \right\rangle \right) ^{2}\right)
+\left( \gamma \left\langle \hat{S}_{E}\left( X_{1},X^{\prime }\right)
\right\rangle _{X_{1}}\right) ^{2}-\left( \gamma \left\langle \hat{S}%
_{E}\left( X\right) \right\rangle \right) ^{2}}
\end{equation*}%
\begin{eqnarray}
&&\hat{S}_{E}\left( X^{\prime },X\right)  \label{SNXPXa} \\
&=&\frac{\underline{\hat{S}}\left( X^{\prime },X\right) }{2}+\frac{\hat{w}%
\left( X^{\prime },X\right) }{2}\left( \hat{w}\left( X\right) \left( \hat{f}%
\left( X^{\prime }\right) -\frac{\left\langle \hat{f}\left( X^{\prime
}\right) \right\rangle _{\hat{w}_{E}}+\left\langle \hat{r}\left( X^{\prime
}\right) \right\rangle _{\hat{w}_{L}}}{2}\right) +w\left( X\right) \left( 
\hat{f}\left( X^{\prime }\right) -\frac{f\left( X\right) +r\left( X\right) }{%
2}\right) \right)  \notag \\
&\simeq &\frac{\hat{w}\left( X^{\prime },X\right) }{2}\left( 1+\left( \hat{f}%
\left( X^{\prime }\right) -\left\langle \hat{w}\left( X\right) \right\rangle 
\frac{\left\langle \hat{f}\left( X^{\prime }\right) \right\rangle _{\hat{w}%
_{E}}+\left\langle \hat{r}\left( X^{\prime }\right) \right\rangle _{\hat{w}%
_{L}}}{2}-\left\langle w\left( X\right) \right\rangle \frac{f\left( X\right)
+r\left( X\right) }{2}\right) \right)  \notag \\
&=&\frac{\hat{w}\left( X^{\prime },X\right) }{2}\left( 1+\Delta \hat{f}%
\left( X^{\prime }\right) \right)  \notag \\
&\rightarrow &\frac{\left( 1-\left( \gamma \left\langle \hat{S}_{E}\left(
X\right) \right\rangle \right) ^{2}\right) \hat{w}_{E}^{\left( 0\right)
}\left( X^{\prime },X\right) \left( 1+\Delta \hat{f}\left( X^{\prime
}\right) \right) }{1+\hat{w}_{E}^{\left( 0\right) }\left( X^{\prime
},X\right) \left( 1-\left( \gamma \left\langle \hat{S}_{E}\left( X\right)
\right\rangle \right) ^{2}\right) +\left( \gamma \left\langle \hat{S}%
_{E}\left( X_{1},X^{\prime }\right) \right\rangle _{X_{1}}\right)
^{2}-\left( \gamma \left\langle \hat{S}_{E}\left( X\right) \right\rangle
\right) ^{2}}  \notag
\end{eqnarray}%
\begin{eqnarray*}
&&\hat{S}\left( X^{\prime },X\right) \\
&=&\underline{\hat{S}}\left( X^{\prime },X\right) +\hat{w}\left( X^{\prime
},X\right) \left( \frac{\hat{f}\left( X^{\prime }\right) +\hat{r}\left(
X^{\prime }\right) }{2}-\hat{w}\left( X\right) \frac{\left\langle \hat{f}%
\left( X^{\prime }\right) \right\rangle _{\hat{w}_{E}}+\left\langle \hat{r}%
\left( X^{\prime }\right) \right\rangle _{\hat{w}_{L}}}{2}-w\left( X\right) 
\frac{f\left( X\right) +r\left( X\right) }{2}\right) \\
&\simeq &\hat{w}\left( X^{\prime },X\right) \left( 1+\Delta \left( \frac{%
\hat{f}\left( X^{\prime }\right) +\hat{r}\left( X^{\prime }\right) }{2}%
\right) \right)
\end{eqnarray*}

\begin{eqnarray*}
&&\hat{S}\left( X^{\prime }\right) \\
&\simeq &\hat{w}\left( X^{\prime },X\right) \left( 1+\left( \frac{\hat{f}%
\left( X^{\prime }\right) +\hat{r}\left( X^{\prime }\right) }{2}%
-\left\langle \hat{w}\left( X\right) \right\rangle \frac{\left\langle \hat{f}%
\left( X^{\prime }\right) \right\rangle _{\hat{w}_{E}}+\left\langle \hat{r}%
\left( X^{\prime }\right) \right\rangle _{\hat{w}_{L}}}{2}-\left\langle
w\left( X\right) \right\rangle \frac{f\left( X\right) +r\left( X\right) }{2}%
\right) \right) \\
&\rightarrow &\frac{2\left( 1-\left( \gamma \left\langle \hat{S}_{E}\left(
X\right) \right\rangle \right) ^{2}\right) \hat{w}_{E}^{\left( 0\right)
}\left( X^{\prime },X\right) \left( 1+\frac{\Delta \hat{f}\left( X^{\prime
}\right) +\Delta \hat{r}\left( X^{\prime }\right) }{2}\right) }{1+\hat{w}%
_{E}^{\left( 0\right) }\left( X^{\prime },X\right) \left( 1-\left( \gamma
\left\langle \hat{S}_{E}\left( X\right) \right\rangle \right) ^{2}\right)
+\left( \gamma \left\langle \hat{S}_{E}\left( X_{1},X^{\prime }\right)
\right\rangle _{X_{1}}\right) ^{2}-\left( \gamma \left\langle \hat{S}%
_{E}\left( X\right) \right\rangle \right) ^{2}}
\end{eqnarray*}%
To include decreasing return, we replace in first approximation:

\begin{equation}
f\left( X\right) \rightarrow \frac{f_{1}\left( X\right) +\tau \left(
\left\langle f_{1}\left( X\right) \right\rangle -\left\langle f_{1}\left(
X^{\prime }\right) \right\rangle \right) }{\left( K_{X}\left\vert \hat{\Psi}%
\left( X\right) \right\vert ^{2}\right) ^{r}}-\frac{C}{K_{X}}-C_{0}
\label{rtrn}
\end{equation}

\subsection*{A8.2 Firms' returns}

Once the shares are found, we can obtain the returns for frms. They are
given by frst replacing in (\ref{rtrn}) the average shares by their sectors
values (\ref{V}):%
\begin{equation}
f\left( X\right) \rightarrow \frac{f_{1}\left( X\right) +\tau \left(
\left\langle f_{1}\left( X\right) \right\rangle -\left\langle f_{1}\left(
X^{\prime }\right) \right\rangle \right) }{\left( K_{X}\left\vert \hat{\Psi}%
\left( X\right) \right\vert ^{2}\right) ^{r}}-\frac{C}{K_{X}}-C_{0}
\label{fr}
\end{equation}%
The returns for firms are given using (\ref{Dc}) and:%
\begin{equation*}
\left( 1+\underline{k}_{L}\left( X\right) \right) \left( \frac{f_{1}\left(
X\right) }{\left( \left( 1+\frac{\underline{k}\left( X\right) }{\left\langle
K\right\rangle }\hat{K}_{X}\left\vert \hat{\Psi}\left( X\right) \right\vert
^{2}\right) K_{X}\right) ^{r}}-C_{0}-\frac{C}{\left( 1+\frac{\underline{k}%
\left( X\right) }{\left\langle K\right\rangle }\hat{K}_{X}\left\vert \hat{%
\Psi}\left( X\right) \right\vert ^{2}\right) K_{X}}\right) -\underline{k}%
_{L}\left( X\right) \bar{r}
\end{equation*}%
which corresponds to the net return of production plus variation in stocks,
from which the paiements of loans is substracted.

Using:%
\begin{equation*}
1+\underline{k}_{L}\left( X\right) =\frac{1-S_{E}\left( X\right) }{1-S\left(
X\right) }
\end{equation*}%
\begin{equation*}
\underline{k}_{L}\left( X\right) =\frac{S\left( X\right) -S_{E}\left(
X\right) }{1-S\left( X\right) }
\end{equation*}%
The return for firms is ultimately given by:%
\begin{equation}
f_{1}^{\prime }\left( X\right) =\frac{1-S_{E}\left( X\right) }{1-S\left(
X\right) }\left( \frac{f_{1}\left( X\right) +\tau \left( \left\langle
f_{1}\left( X\right) \right\rangle -\left\langle f_{1}\left( X^{\prime
}\right) \right\rangle \right) }{\left( K_{X}\right) ^{r}}-\frac{C}{K_{X}}%
-C_{0}\right) -\frac{S\left( X\right) -S_{E}\left( X\right) }{1-S\left(
X\right) }\bar{r}  \label{Fr}
\end{equation}

\subsection*{A8.3 Corrections with default}

\subsubsection*{A8.3.1 Initial default}

As explained in appendix 1, the equations for return have to be modified
when default arises. It may start from firms such that:%
\begin{equation*}
1+f_{1}^{\prime }\left( X\right) <0
\end{equation*}%
Using (\ref{Fr}), this arises if the inequality is satisfied:%
\begin{equation}
\frac{1-S_{E}\left( X\right) }{1-S\left( X\right) }\left( \frac{f_{1}\left(
X\right) +\tau \Delta F_{\tau }\left( \bar{R}\left( K,X\right) \right) }{%
\left( K_{X}\right) ^{r}}-\frac{C}{K_{X}}-C_{0}\right) -\frac{S\left(
X\right) -S_{E}\left( X\right) }{1-S\left( X\right) }\bar{r}<-1  \label{Ct}
\end{equation}%
where $S\left( X\right) $ is given by (\ref{V}) and $S_{E}\left( X\right) $
by (\ref{Vcc}). In this case, investors returns are modified by:%
\begin{equation*}
\hat{f}\left( X\right) \rightarrow \hat{f}\left( X\right) -\left( 1+\bar{r}%
\right) S_{L}\left( X\right)
\end{equation*}%
If this condition (\ref{Ct}) is satisfied, the result for return equation
has to be modified to include\ defaults. To do so, we write first (\ref{Ct})
as:%
\begin{equation*}
\frac{f_{1}\left( X\right) +\tau \Delta F_{\tau }\left( \bar{R}\left(
K,X\right) \right) }{\left( K_{X}\right) ^{r}}-\frac{C}{K_{X}}-C_{0}<\frac{%
\left( S\left( X\right) -S_{E}\left( X\right) \right) \bar{r}-\left(
1-S\left( X\right) \right) }{1-S_{E}\left( X\right) }
\end{equation*}%
For $\bar{r}<<1$, $\tau <<1$, we have an order of magnitude:%
\begin{equation*}
\frac{f_{1}\left( X\right) +\tau \Delta F_{\tau }\left( \bar{R}\left(
K,X\right) \right) }{\left( K_{X}\right) ^{r}}-\frac{C}{K_{X}}-C_{0}\simeq -1
\end{equation*}%
Given this condition, we can derive the values of shares that will be used
to find return equations:%
\begin{eqnarray}
&&S\left( X,X\right)  \label{Snr} \\
&=&w\left( X\right) \left( 1+\hat{w}\left( X\right) \frac{\left( 1-S\left(
X\right) _{dr}\right) ^{r}f_{1}\left( X\right) -C_{0}-\left( 1-S\left(
X\right) _{dr}\right) C+\bar{r}\left( X\right) -\left( \left\langle \hat{f}%
\left( X^{\prime }\right) \right\rangle _{\hat{w}_{E}}+\left\langle \hat{r}%
\left( X^{\prime }\right) \right\rangle _{\hat{w}_{L}}\right) }{2}\right) 
\notag \\
&<&w\left( X\right) \left( 1+\hat{w}\left( X\right) \frac{-1+\bar{r}\left(
X\right) -\left( \left\langle \hat{f}\left( X^{\prime }\right) \right\rangle
_{\hat{w}_{E}}+\left\langle \hat{r}\left( X^{\prime }\right) \right\rangle _{%
\hat{w}_{L}}\right) }{2}\right)  \notag \\
&\simeq &w\left( X\right) \left( 1-\hat{w}\left( X\right) \frac{%
1+\left\langle \hat{f}\left( X^{\prime }\right) \right\rangle _{\hat{w}_{E}}%
}{2}\right)  \notag
\end{eqnarray}%
Moreover:%
\begin{eqnarray}
&&S_{E}\left( X,X\right) \\
&=&\frac{w\left( X\right) }{2}\left( 1+\left( \hat{w}\left( X\right) +\frac{%
w\left( X\right) }{2}\right) \left( \left( 1-S\left( X\right) _{dr}\right)
^{r}f_{1}\left( X\right) -C_{0}-\left( 1-S\left( X\right) _{dr}\right)
C\right) \right.  \notag \\
&&\left. -\left( \hat{w}\left( X\right) \frac{\left\langle \hat{f}\left(
X^{\prime }\right) \right\rangle _{\hat{w}_{E}}+\left\langle \hat{r}\left(
X^{\prime }\right) \right\rangle _{\hat{w}_{L}}}{2}+\frac{w\left( X\right) }{%
2}\bar{r}\left( X\right) \right) \right)  \notag
\end{eqnarray}%
Using:%
\begin{equation*}
\hat{w}\left( X\right) =1-w\left( X\right)
\end{equation*}%
\begin{eqnarray}
&&S_{E}\left( X,X\right)  \label{Str} \\
&=&\frac{w\left( X\right) }{2}\left( 1-\left( \hat{w}\left( X\right) +\frac{%
w\left( X\right) }{2}\right) -\left( \hat{w}\left( X\right) \frac{%
\left\langle \hat{f}\left( X^{\prime }\right) \right\rangle _{\hat{w}%
_{E}}+\left\langle \hat{r}\left( X^{\prime }\right) \right\rangle _{\hat{w}%
_{L}}}{2}+\frac{w\left( X\right) }{2}\bar{r}\left( X\right) \right) \right) 
\notag \\
&=&\frac{w\left( X\right) }{2}\left( \frac{w\left( X\right) }{2}-\frac{%
\left( 1-w\left( X\right) \right) \left\langle \hat{f}\left( X^{\prime
}\right) \right\rangle _{\hat{w}_{E}}+\bar{r}\left( X\right) }{2}\right) 
\notag
\end{eqnarray}%
\begin{equation*}
\hat{w}\left( X\right) \rightarrow \frac{\left( 1-\left( \gamma \left\langle 
\hat{S}_{E}\left( X\right) \right\rangle \right) ^{2}\right) }{\left(
2-\left( \gamma \left\langle \hat{S}_{E}\left( X\right) \right\rangle
\right) ^{2}\right) ^{2}}
\end{equation*}%
\begin{equation*}
w\left( X\right) \simeq \frac{1}{2-\left( \gamma \left\langle \hat{S}%
_{E}\left( X\right) \right\rangle \right) ^{2}}
\end{equation*}%
These estimations of shares can be used to approximate equation (\ref{Qh})
for returns, which is:%
\begin{eqnarray}
&&w=\left( \hat{f}\left( X\right) -\bar{r}\right) \\
&&\times \frac{\left\langle \hat{f}\right\rangle ^{2}\left( \frac{%
\left\langle \hat{S}\left( X^{\prime },X\right) \right\rangle _{X}}{%
1-\left\langle \hat{S}\left( X^{\prime },X\right) \right\rangle _{X}}\right)
^{2}-A\left( 1+\frac{\Delta \hat{f}\left( X\right) +\Delta \hat{r}\left(
X\right) }{2}\right) \left( \hat{f}\left( X\right) \left( 1-\left\langle 
\hat{S}\right\rangle \right) +\left\langle \hat{S}\left( X^{\prime
},X\right) \right\rangle _{X^{\prime }}\left\langle \hat{f}\right\rangle
\right) ^{2}}{\left\langle \hat{f}\right\rangle ^{2}\left( \frac{%
\left\langle \hat{S}\left( X^{\prime },X\right) \right\rangle _{X}}{%
1-\left\langle \hat{S}\left( X^{\prime },X\right) \right\rangle _{X}}\right)
^{2}-\frac{1}{2}A\left( 1+\Delta \hat{f}\left( X\right) \right) \left( \hat{f%
}\left( X\right) \left( 1-\left\langle \hat{S}\right\rangle \right)
+\left\langle \hat{S}\left( X^{\prime },X\right) \right\rangle _{X^{\prime
}}\left\langle \hat{f}\right\rangle \right) ^{2}}  \notag
\end{eqnarray}%
with $w$ given by (\ref{HDf}):%
\begin{eqnarray}
w &=&\frac{\left( 1-\left( \gamma \left\langle \hat{S}_{E}\left( X\right)
\right\rangle \right) ^{2}\right) }{2-\left( \gamma \left\langle \hat{S}%
_{E}\left( X\right) \right\rangle \right) ^{2}}\left( 1-\frac{\Delta \left( 
\frac{\hat{f}\left( X\right) +r\left( X\right) }{2}\right) }{2-\left( \gamma
\left\langle \hat{S}_{E}\left( X\right) \right\rangle \right) ^{2}}+\frac{%
\left\langle \hat{f}\left( X^{\prime }\right) \right\rangle -\left\langle 
\hat{r}\left( X^{\prime }\right) \right\rangle _{\hat{w}_{L}}}{2}\right)
\label{Rs} \\
&&\times \frac{1-\left\langle \hat{S}\left( X^{\prime }\right) \right\rangle 
}{1-\left\langle \hat{S}_{E}\left( X^{\prime }\right) \right\rangle }\left(
\left\langle \hat{f}\left( X^{\prime }\right) \right\rangle -\left\langle 
\bar{r}\right\rangle \right) +S_{E}\left( X,X\right) \left( f\left( X\right)
-r\right)  \notag
\end{eqnarray}%
and $A$ defined in (\ref{cfr}).%
\begin{equation*}
\Delta \hat{f}\left( X\right) +\Delta \hat{r}\left( X\right) <0
\end{equation*}%
To obtain an order of magnitude of returns in the default scenario, we can
assume that:%
\begin{equation*}
\left( \hat{f}\left( X\right) -\bar{r}\right) \simeq w
\end{equation*}%
so that equation (\ref{Rs}) writes:%
\begin{eqnarray*}
&&\left( \hat{f}\left( X\right) -\bar{r}\right) \left( 1+\frac{\left(
1-\left( \gamma \left\langle \hat{S}_{E}\left( X\right) \right\rangle
\right) ^{2}\right) }{2\left( 2-\left( \gamma \left\langle \hat{S}_{E}\left(
X\right) \right\rangle \right) ^{2}\right) ^{2}}\frac{1-\left\langle \hat{S}%
\left( X^{\prime }\right) \right\rangle }{1-\left\langle \hat{S}_{E}\left(
X^{\prime }\right) \right\rangle }\left( \left\langle \hat{f}\left(
X^{\prime }\right) \right\rangle -\left\langle \bar{r}\right\rangle \right)
\right) \\
&=&\frac{\left( 1-\left( \gamma \left\langle \hat{S}_{E}\left( X\right)
\right\rangle \right) ^{2}\right) }{2-\left( \gamma \left\langle \hat{S}%
_{E}\left( X\right) \right\rangle \right) ^{2}}\left( 1-\frac{\Delta \left(
r\left( X\right) \right) }{2-\left( \gamma \left\langle \hat{S}_{E}\left(
X\right) \right\rangle \right) ^{2}}+\frac{\left\langle \hat{f}\left(
X^{\prime }\right) \right\rangle -\left\langle \hat{r}\left( X^{\prime
}\right) \right\rangle _{\hat{w}_{L}}}{2}\right) \frac{1-\left\langle \hat{S}%
\left( X^{\prime }\right) \right\rangle }{1-\left\langle \hat{S}_{E}\left(
X^{\prime }\right) \right\rangle }\left( \left\langle \hat{f}\left(
X^{\prime }\right) \right\rangle -\left\langle \bar{r}\right\rangle \right)
\\
&&+S_{E}\left( X,X\right) \left( f\left( X\right) -r\right)
\end{eqnarray*}%
which leads to:%
\begin{eqnarray}
&&\left( \hat{f}\left( X\right) -\bar{r}\right) \left( 1+\frac{\left(
1-\left( \gamma \left\langle \hat{S}_{E}\left( X\right) \right\rangle
\right) ^{2}\right) }{2\left( 2-\left( \gamma \left\langle \hat{S}_{E}\left(
X\right) \right\rangle \right) ^{2}\right) ^{2}}\frac{1-\left\langle \hat{S}%
\left( X^{\prime }\right) \right\rangle }{1-\left\langle \hat{S}_{E}\left(
X^{\prime }\right) \right\rangle }\left( \left\langle \hat{f}\left(
X^{\prime }\right) \right\rangle -\left\langle \bar{r}\right\rangle \right)
\right)  \label{Dvr} \\
&\simeq &\frac{\left( 1-\left( \gamma \left\langle \hat{S}_{E}\left(
X\right) \right\rangle \right) ^{2}\right) }{2-\left( \gamma \left\langle 
\hat{S}_{E}\left( X\right) \right\rangle \right) ^{2}}\left( 1-\frac{\Delta
\left( r\left( X\right) \right) }{2-\left( \gamma \left\langle \hat{S}%
_{E}\left( X\right) \right\rangle \right) ^{2}}+\frac{\left\langle \hat{f}%
\left( X^{\prime }\right) \right\rangle -\left\langle \hat{r}\left(
X^{\prime }\right) \right\rangle _{\hat{w}_{L}}}{2}\right) \frac{%
1-\left\langle \hat{S}\left( X^{\prime }\right) \right\rangle }{%
1-\left\langle \hat{S}_{E}\left( X^{\prime }\right) \right\rangle }\left(
\left\langle \hat{f}\left( X^{\prime }\right) \right\rangle -\left\langle 
\bar{r}\right\rangle \right)  \notag \\
&&+\frac{1}{4}\frac{1}{2-\left( \gamma \left\langle \hat{S}_{E}\left(
X\right) \right\rangle \right) ^{2}}\left( \frac{1-\left( 1-\left( \gamma
\left\langle \hat{S}_{E}\left( X\right) \right\rangle \right) ^{2}\right)
\left\langle \hat{f}\left( X^{\prime }\right) \right\rangle _{\hat{w}_{E}}}{%
2-\left( \gamma \left\langle \hat{S}_{E}\left( X\right) \right\rangle
\right) ^{2}}-\bar{r}\left( X\right) \right) \left( f\left( X\right)
-r\right)  \notag
\end{eqnarray}%
We can study the asymptotic behavior of (\ref{Dvr}).

For relatively large uncertainty:%
\begin{equation*}
S_{L}\left( X\right) \rightarrow \frac{3}{4}
\end{equation*}%
\begin{equation}
\hat{f}\left( X\right) -\bar{r}\rightarrow \frac{1}{4}\left( 1-\bar{r}\left(
X\right) \right) \left( f\left( X\right) -r\right)  \label{DFw}
\end{equation}%
so that the return for investors at sectors where firms default arise have
the following return:%
\begin{eqnarray*}
\hat{f}\left( X\right) &\rightarrow &\hat{f}\left( X\right) -\left( 1+\bar{r}%
\right) S_{L}\left( X\right) \\
&\rightarrow &\hat{f}\left( X\right) -\left( 1+\bar{r}\right) \frac{3}{4}
\end{eqnarray*}%
and replacing this value in (\ref{DFw}) yields $\hat{f}\left( X\right)
-\left( 1+\bar{r}\right) \frac{3}{4}$: 
\begin{eqnarray*}
&&\frac{1}{4}\left( 1-\bar{r}\left( X\right) \right) \left( f\left( X\right)
-r\right) -\left( 1+\bar{r}\right) \frac{3+\bar{r}}{4}+\bar{r} \\
&\rightarrow &\frac{1}{4}\left( 1-\bar{r}\left( X\right) \right) \left(
f\left( X\right) +1\right) -1
\end{eqnarray*}%
this is of order $-1$. If firms in a sector are led to default, so do the
investors in the same sector.

For relatively small uncertainty:%
\begin{equation*}
S_{E}\left( X,X\right) \left( f\left( X\right) -r\right) \rightarrow 0
\end{equation*}%
and there is no default.

More generally, the threshold for initial default:%
\begin{equation*}
\hat{f}\left( X\right) -\left( 1+\bar{r}\right) S_{L}\left( X\right) +1=0
\end{equation*}%
writes as a condition for investors' returns $\hat{f}\left( X\right) $: 
\begin{eqnarray*}
&&\hat{f}\left( X\right) \\
&\simeq &\hat{w}\left( X\right) \frac{1+\frac{\left\langle \hat{f}\left(
X^{\prime }\right) \right\rangle -\left\langle \hat{r}\left( X^{\prime
}\right) \right\rangle _{\hat{w}_{L}}}{2}-w\left( X\right) \Delta r\left(
X\right) }{1+\frac{\hat{w}\left( X\right) }{2^{2}}\frac{1-\left\langle \hat{S%
}\left( X^{\prime }\right) \right\rangle }{1-\left\langle \hat{S}_{E}\left(
X^{\prime }\right) \right\rangle }\left( \left\langle \hat{f}\left(
X^{\prime }\right) \right\rangle -\left\langle \bar{r}\right\rangle \right) }%
\frac{1-\left\langle \hat{S}\left( X^{\prime }\right) \right\rangle }{%
1-\left\langle \hat{S}_{E}\left( X^{\prime }\right) \right\rangle }\left(
\left\langle \hat{f}\left( X^{\prime }\right) \right\rangle -\left\langle 
\bar{r}\right\rangle \right) \\
&&+\frac{1}{4}w\left( X\right) \frac{w\left( X\right) -\hat{w}\left(
X\right) \left\langle \hat{f}\left( X^{\prime }\right) \right\rangle _{\hat{w%
}_{E}}-\bar{r}\left( X\right) }{1+\frac{\hat{w}\left( X\right) w\left(
X\right) }{2}\frac{1-\left\langle \hat{S}\left( X^{\prime }\right)
\right\rangle }{1-\left\langle \hat{S}_{E}\left( X^{\prime }\right)
\right\rangle }\left( \left\langle \hat{f}\left( X^{\prime }\right)
\right\rangle -\left\langle \bar{r}\right\rangle \right) }\left( f\left(
X\right) -r\right) -\left( 1+\bar{r}\right) w\left( X\right) +\bar{r}
\end{eqnarray*}%
and this is lower or equal to $-1$ depending on:%
\begin{eqnarray*}
&&f\left( X\right) -r \\
&<&-\frac{4\left( 1+\frac{\hat{w}\left( X\right) w\left( X\right) }{2}\frac{%
1-\left\langle \hat{S}\left( X^{\prime }\right) \right\rangle }{%
1-\left\langle \hat{S}_{E}\left( X^{\prime }\right) \right\rangle }\left(
\left\langle \hat{f}\left( X^{\prime }\right) \right\rangle -\left\langle 
\bar{r}\right\rangle \right) \right) }{w\left( X\right) -\hat{w}\left(
X\right) \left( \left\langle \hat{f}\left( X^{\prime }\right) \right\rangle
_{\hat{w}_{E}}-\bar{r}\left( X\right) \right) }\frac{\hat{w}\left( X\right) 
}{w\left( X\right) } \\
&&\times \left( \left( 1+\bar{r}\right) +\frac{1+\frac{\left\langle \hat{f}%
\left( X^{\prime }\right) \right\rangle -\left\langle \hat{r}\left(
X^{\prime }\right) \right\rangle _{\hat{w}_{L}}}{2}-w\left( X\right) \Delta
r\left( X\right) }{1+\frac{\hat{w}\left( X\right) w\left( X\right) }{2}\frac{%
1-\left\langle \hat{S}\left( X^{\prime }\right) \right\rangle }{%
1-\left\langle \hat{S}_{E}\left( X^{\prime }\right) \right\rangle }\left(
\left\langle \hat{f}\left( X^{\prime }\right) \right\rangle -\left\langle 
\bar{r}\right\rangle \right) }\frac{1-\left\langle \hat{S}\left( X^{\prime
}\right) \right\rangle }{1-\left\langle \hat{S}_{E}\left( X^{\prime }\right)
\right\rangle }\left( \left\langle \hat{f}\left( X^{\prime }\right)
\right\rangle -\left\langle \bar{r}\right\rangle \right) \right)
\end{eqnarray*}%
where $f\left( X\right) $ given by (\ref{fr}).

The threshold decreases with $\left\langle \hat{f}\left( X^{\prime }\right)
\right\rangle $ and increases with $\gamma $. We can replace: $\Delta
r\left( X\right) \rightarrow \frac{r\left( X\right) -\left\langle \hat{f}%
\left( X^{\prime }\right) \right\rangle }{2}$, $\left\langle \hat{S}%
_{E}\left( X^{\prime }\right) \right\rangle \rightarrow z$. We have in
average:%
\begin{eqnarray*}
&&f\left( X\right) -r \\
&<&-\frac{4\left( 1+\frac{\left( 1-2z\right) 2z}{2}\frac{1-2z}{1-z}\frac{%
\left\langle f\left( X\right) \right\rangle -r}{2}\right) }{1-2z-2z\left( 
\frac{\left\langle f\left( X\right) \right\rangle -r}{2}+r\right) -\bar{r}}%
\frac{2z}{1-2z} \\
&&\times \left( \left( 1+\bar{r}\right) +\frac{1+\frac{1}{2}\frac{%
\left\langle f\left( X\right) \right\rangle -r}{2}-\frac{1-2z}{2}\frac{%
\left\langle f\left( X\right) \right\rangle -r}{2}}{1-2z-2z\left( \frac{%
\left\langle f\left( X\right) \right\rangle -r}{2}+r\right) -\bar{r}}\frac{%
1-2z}{1-z}\frac{\left\langle f\left( X\right) \right\rangle -r}{2}\right)
\end{eqnarray*}%
If the uncertainty increases as well as return increases on average, the
threshold decreases, that is the loss has to be very low to induce default.

\subsubsection*{A8.3.2 Default Propagation}

\paragraph*{A8.3.2.1 Zeroth order approximation}

Given (\ref{dfn}) and considering default initially defined by: 
\begin{equation*}
\max \left( -1,\left( 1+f\left( X^{\prime }\right) \right) \frac{1-\hat{S}%
\left( X^{\prime }\right) }{\hat{S}_{L}\left( X^{\prime }\right) }\right) =-1
\end{equation*}

\begin{eqnarray}
0 &=&\int \left( \delta \left( X-X^{\prime }\right) -\hat{S}_{E}\left(
X^{\prime },X\right) \right) \frac{1-\hat{S}\left( X^{\prime }\right) }{1-%
\hat{S}_{E}\left( X^{\prime }\right) }f\left( X^{\prime }\right) dX^{\prime }
\\
&&+\int_{\hat{S}_{-}}\hat{S}_{L}\left( X^{\prime },X\right) dX^{\prime
}+1_{D}\left( X\right) S_{L}\left( X,X\right) -S_{E}\left( X,X\right)
f\left( X^{\prime }\right)  \notag
\end{eqnarray}%
where $1_{D}\left( X\right) =1$ if default occurs, and $0$ otherwise.

For investors that invest in neighbour firms:

\begin{eqnarray}
&&\int \left( \delta \left( X-X^{\prime }\right) -\hat{S}_{E}\left(
X^{\prime },X\right) \right) \frac{1-\hat{S}\left( X^{\prime }\right) }{1-%
\hat{S}_{E}\left( X^{\prime }\right) }f\left( X^{\prime }\right) dX^{\prime }
\\
&=&S_{E}\left( X,X\right) f\left( X^{\prime }\right) -\int_{\hat{S}_{-}}\hat{%
S}_{L}\left( X^{\prime },X\right) dX^{\prime }-1_{D}\left( X\right)
S_{L}\left( X,X\right)  \notag
\end{eqnarray}%
and then the resulting returns will be obtained by shifting $f\left(
X^{\prime }\right) $ with:%
\begin{equation*}
f\left( X^{\prime }\right) -df\left( X^{\prime }\right)
\end{equation*}%
and in first approximation:%
\begin{equation*}
\hat{f}\left( X\right) <-1\rightarrow \hat{f}\left( X\right) =-1
\end{equation*}%
in stable configurations. This simplification allows to get some results
about the stable default rates.

Consider shares set to their stable values: 
\begin{equation*}
\left( \hat{f}_{n}\left( X\right) -\bar{r}\right) \rightarrow \left( \hat{f}%
\left( X\right) -\bar{r}\right)
\end{equation*}%
and replace:

\begin{equation}
f\left( X\right) \rightarrow f\left( X\right) -df\left( X^{\prime }\right)
\label{RP}
\end{equation}%
in the return equation. We obtain in first approximation:%
\begin{eqnarray}
\hat{f}\left( X\right) &\rightarrow &\hat{f}\left( X\right) -\frac{%
1-\left\langle \hat{S}_{E}\left( X^{\prime }\right) \right\rangle }{%
1-\left\langle \hat{S}\left( X^{\prime }\right) \right\rangle }\frac{\left(
\int_{\hat{S}_{-}}\left( \hat{S}_{L}\left( X^{\prime },X\right) \right)
dX^{\prime }+\int_{S_{-}}S_{L}\left( X^{\prime },X^{\prime }\right)
+S_{E}\left( X^{\prime },X^{\prime }\right) df\left( X^{\prime }\right)
\right) }{1-\left\langle \hat{S}_{E}\left( X\right) \right\rangle }
\label{Rp} \\
&&-S_{E}\left( X,X\right) \delta f\left( X\right)  \notag \\
&\rightarrow &\hat{f}\left( X\right) -\frac{1}{1-\left\langle \hat{S}\left(
X^{\prime }\right) \right\rangle }\left( \int_{\hat{S}_{-}}\left( \hat{S}%
_{L}\left( X^{\prime },X\right) \right) dX^{\prime }+\int_{S_{-}}S_{L}\left(
X^{\prime },X^{\prime }\right) +S_{E}\left( X^{\prime },X^{\prime }\right)
df\left( X^{\prime }\right) \right)  \notag
\end{eqnarray}%
where we will consider that $\hat{S}=S$:

The factor $\frac{1}{1-\left\langle \hat{S}_{E}\left( X\right) \right\rangle 
}$ accounts for diffusion and where $f\left( X\right) $ are computed with
new values of 
\begin{equation*}
S_{E}\left( X,X\right) \text{, }S\left( X,X\right) \text{,}S_{E}\left(
X,X\right) \text{,}S_{E}\left( X,X\right) \text{...}
\end{equation*}%
that depends on the defaults.

These coefficients are computed with coefficients $w\left( X\right) $,$\hat{w%
}\left( X\right) $. These coefficients depend on relative averages and
uncertainties. We assume that they are constant in first approximation and
we compute $S_{E}\left( X,X\right) $, $S\left( X,X\right) $ with
modifications $f\left( X\right) $, $\hat{f}\left( X\right) $.

The system including default thus combines (\ref{RP}) and (\ref{Rp}) in
first approximation:%
\begin{equation}
\hat{f}\left( X\right) -\frac{1}{1-\left\langle \hat{S}\left( X^{\prime
}\right) \right\rangle }\left( \left( \left\langle \hat{S}_{L}\left(
X^{\prime },X\right) \right\rangle +\left\langle S_{L}\left( X,X\right)
\right\rangle \right) \mu +\left\langle S_{E}\left( X^{\prime },X^{\prime
}\right) \right\rangle \left\langle df\left( X^{\prime }\right)
\right\rangle \right)  \label{Rvt}
\end{equation}%
where $\mu $ represents the fraction of default in equilibrium and where $%
f\left( X\right) $ and $\hat{f}\left( X\right) $ are computed before default.

We also need the averages of this formula:%
\begin{equation*}
\left\langle f\left( X\right) \right\rangle \rightarrow \left\langle f\left(
X\right) \right\rangle -\left\langle df\left( X\right) \right\rangle
\end{equation*}%
\begin{equation}
\left\langle \hat{f}\left( X\right) \right\rangle -\frac{1}{1-\left\langle 
\hat{S}\left( X\right) \right\rangle }\left( \left( \left\langle \hat{S}%
_{L}\left( X^{\prime },X\right) \right\rangle +\left\langle S_{L}\left(
X,X\right) \right\rangle \right) \mu +\left\langle S_{E}\left( X^{\prime
},X^{\prime }\right) \right\rangle \left\langle df\left( X^{\prime }\right)
\right\rangle \right)  \label{Rv}
\end{equation}%
In first approximation we can neglect $d\hat{f}\left( X\right) $ and $%
\left\langle d\hat{f}\left( X\right) \right\rangle $ in (\ref{Rp}) and (\ref%
{Rv}) since the effect in default is directly measured by the effective $%
f\left( X\right) $.

We compute:%
\begin{equation*}
\frac{\left( \left\langle \hat{S}_{L}\left( X^{\prime },X\right)
\right\rangle +\left\langle S_{L}\left( X,X\right) \right\rangle \right) \mu 
}{\left\langle S_{E}\left( X,X\right) \right\rangle }
\end{equation*}%
in the default state as an expansion around the non default solution. This
is obtained by using that: 
\begin{eqnarray}
&&S\left( X\right) \\
&=&w\left( X\right) \left( 1+\left( \hat{w}\left( X\right) \left( \frac{%
f\left( X\right) +\bar{r}\left( X\right) }{2}-\frac{\left\langle \hat{f}%
\left( X^{\prime }\right) \right\rangle _{\hat{w}_{E}}+\left\langle \hat{r}%
\left( X^{\prime }\right) \right\rangle _{\hat{w}_{L}}}{2}\right) \right)
\right) \frac{\hat{K}_{X}\left\vert \hat{\Psi}\left( X\right) \right\vert
^{2}}{K_{X}\left\vert \Psi \left( X\right) \right\vert ^{2}}  \notag \\
&=&S\left( X,X\right) \frac{\hat{K}_{X}\left\vert \hat{\Psi}\left( X\right)
\right\vert ^{2}}{K_{X}\left\vert \Psi \left( X\right) \right\vert ^{2}} 
\notag
\end{eqnarray}%
that is, in first approximation:%
\begin{eqnarray*}
&&S\left( X\right) \\
&\rightarrow &w\left( X\right) \frac{\hat{K}_{X}\left\vert \hat{\Psi}\left(
X\right) \right\vert ^{2}}{K_{X}\left\vert \Psi \left( X\right) \right\vert
^{2}}\left( 1+\hat{w}\left( X\right) \left( \frac{f\left( X\right) -df\left(
X\right) +\bar{r}\left( X\right) }{2}\right. \right. \\
&&\left. \left. -\frac{\left\langle \hat{f}\left( X^{\prime }\right)
\right\rangle _{\hat{w}_{E}}-\frac{1}{1-\left\langle \hat{S}\left( X\right)
\right\rangle }\left( \left\langle \hat{S}_{L}\left( X^{\prime },X\right)
\right\rangle \mu +S_{L}\left( X,X\right) +\left\langle S_{E}\left(
X,X\right) \right\rangle \left\langle df\left( X\right) \right\rangle
\right) +\left\langle \hat{r}\left( X^{\prime }\right) \right\rangle _{\hat{w%
}_{L}}}{2}\right) \right)
\end{eqnarray*}%
and this writes:%
\begin{eqnarray}
S\left( X\right) &\rightarrow &S_{0}\left( X\right) -\frac{w\left( X\right) 
\hat{w}\left( X\right) }{2}\frac{\hat{K}_{X}\left\vert \hat{\Psi}\left(
X\right) \right\vert ^{2}}{K_{X}\left\vert \Psi \left( X\right) \right\vert
^{2}}  \label{Dtf} \\
&&\times \left( df\left( X\right) \left( 1-\frac{\left\langle S_{E}\left(
X,X\right) \right\rangle }{1-\left\langle \hat{S}\left( X\right)
\right\rangle }\right) -\frac{\left\langle \hat{S}_{L}\left( X^{\prime
},X\right) \right\rangle \mu +S_{L}\left( X,X\right) }{1-\left\langle \hat{S}%
\left( X\right) \right\rangle }\right) +S\left( X\right) \delta \frac{\hat{K}%
_{X}\left\vert \hat{\Psi}\left( X\right) \right\vert ^{2}}{K_{X}\left\vert
\Psi \left( X\right) \right\vert ^{2}}  \notag \\
&=&S_{0}\left( X\right) -\frac{\hat{w}\left( X\right) }{2\left( 1+\left( 
\hat{w}\left( X\right) \left( \frac{f\left( X\right) +\bar{r}\left( X\right) 
}{2}-\frac{\left\langle \hat{f}\left( X^{\prime }\right) \right\rangle _{%
\hat{w}_{E}}+\left\langle \hat{r}\left( X^{\prime }\right) \right\rangle _{%
\hat{w}_{L}}}{2}\right) \right) \right) }  \notag \\
&&\times S_{0}\left( X\right) \left( df\left( X\right) \left( 1-\frac{%
\left\langle S_{E}\left( X,X\right) \right\rangle }{1-\left\langle \hat{S}%
\left( X\right) \right\rangle }\right) -\frac{\left\langle \hat{S}_{L}\left(
X^{\prime },X\right) \right\rangle \mu +S_{L}\left( X,X\right) }{%
1-\left\langle \hat{S}\left( X\right) \right\rangle }\right) +S\left(
X\right) \delta \frac{\hat{K}_{X}\left\vert \hat{\Psi}\left( X\right)
\right\vert ^{2}}{K_{X}\left\vert \Psi \left( X\right) \right\vert ^{2}} 
\notag
\end{eqnarray}%
with $S_{0}\left( X\right) $ denoting the value of $S\left( X\right) $
without default and where the coefficient $\left\langle S_{E}\left(
X,X\right) \right\rangle $, $\left\langle \hat{S}\left( X\right)
\right\rangle $ are computed at zeroth order, that is without default. We
find the variation $\delta S\left( X\right) $ by computing $\delta \frac{%
\hat{K}_{X}\left\vert \hat{\Psi}\left( X\right) \right\vert ^{2}}{%
K_{X}\left\vert \Psi \left( X\right) \right\vert ^{2}}$ in the right hand
side of the previous equation. This will yield an equation for $\delta
S\left( X\right) $ by identification.

To find $\delta \frac{\hat{K}_{X}\left\vert \hat{\Psi}\left( X\right)
\right\vert ^{2}}{K_{X}\left\vert \Psi \left( X\right) \right\vert ^{2}}$,
we use that:%
\begin{equation}
\frac{\hat{K}_{X}\left\vert \hat{\Psi}\left( X\right) \right\vert ^{2}}{%
K_{X}\left\vert \Psi \left( X\right) \right\vert ^{2}}\simeq \frac{\hat{K}%
_{X}\left\vert \hat{\Psi}\left( X\right) \right\vert ^{2}}{\left( 1-S\left(
X,X\right) \frac{\hat{K}_{X}\left\vert \hat{\Psi}\left( X\right) \right\vert
^{2}}{\left( \left( \frac{2\epsilon }{3\sigma _{\hat{K}}^{2}}\right) ^{\frac{%
r}{2}}\left( \frac{f_{1}\left( X\right) }{C_{0}+\bar{r}}\right) \right) ^{%
\frac{2}{r}}}\right) ^{2}\left( \left( \frac{2\epsilon }{3\sigma _{\hat{K}%
}^{2}}\right) ^{\frac{r}{2}}\left( \frac{f_{1}\left( X\right) }{C_{0}+\bar{r}%
}\right) \right) ^{\frac{2}{r}}}  \label{Rw}
\end{equation}%
with:%
\begin{eqnarray*}
\hat{K}_{X}\left\vert \hat{\Psi}\left( X\right) \right\vert ^{2}
&=&\left\langle \hat{K}\right\rangle \left\Vert \hat{\Psi}\right\Vert
^{2}\left( \frac{\left\langle \hat{f}\right\rangle \frac{\left\langle \hat{S}%
\left( X^{\prime },X\right) \right\rangle _{X}}{1-\left\langle \hat{S}\left(
X^{\prime },X\right) \right\rangle _{X}}}{\left( \hat{f}\left( X^{\prime
}\right) \left( 1-\left\langle \hat{S}\right\rangle \right) +\left\langle 
\hat{S}\left( X,X^{\prime }\right) \right\rangle _{X}\left\langle \hat{f}%
\right\rangle \right) \frac{\left\langle \hat{S}\left( X^{\prime },X\right)
\right\rangle }{1-\left\langle \hat{S}\left( X^{\prime },X\right)
\right\rangle }}\right) ^{2} \\
&\simeq &\left\langle \hat{K}\right\rangle \left\Vert \hat{\Psi}\right\Vert
^{2}\simeq \frac{\hat{\mu}V\sigma _{\hat{K}}^{2}}{2\left\langle \hat{f}%
\right\rangle ^{2}}\left( \frac{\left( \frac{\left\Vert \hat{\Psi}%
_{0}\right\Vert ^{2}}{\hat{\mu}}\left( 1-\hat{S}\right) +\left( \frac{1-r}{r}%
\right) \frac{\left\langle \hat{f}\right\rangle \hat{S}}{2}\right) }{\left(
1-\hat{S}\right) -2r\left( 1-r\right) \frac{\hat{S}}{1-\hat{S}}}\right) ^{2}
\end{eqnarray*}%
In first approximation, for $r<<1$:%
\begin{equation*}
\hat{K}_{X}\left\vert \hat{\Psi}\left( X\right) \right\vert ^{2}\simeq \frac{%
\hat{\mu}V\sigma _{\hat{K}}^{2}\left( \left( \frac{1-r}{r}\right) \frac{\hat{%
S}}{1-\hat{S}}\right) ^{2}}{2}
\end{equation*}%
and given that:%
\begin{equation*}
\left\langle \hat{S}\left( X,X\right) \right\rangle =\hat{w}\left( X^{\prime
},X\right) \left( 1+\left( \frac{\hat{f}\left( X^{\prime }\right) +\hat{r}%
\left( X^{\prime }\right) }{2}-\left\langle \hat{w}\left( X\right)
\right\rangle \frac{\left\langle \hat{f}\left( X^{\prime }\right)
\right\rangle _{\hat{w}_{E}}+\left\langle \hat{r}\left( X^{\prime }\right)
\right\rangle _{\hat{w}_{L}}}{2}-\left\langle w\left( X\right) \right\rangle 
\frac{f\left( X\right) +r\left( X\right) }{2}\right) \right)
\end{equation*}%
We can compute the following quantities:%
\begin{equation*}
\left\langle \hat{S}\left( X,X\right) \right\rangle \rightarrow \left\langle
S\left( X,X\right) \right\rangle -\left\langle \hat{w}\left( X\right)
\right\rangle \left\langle w\left( X\right) \right\rangle \frac{1}{2}\left( 
\frac{\left\langle \hat{S}_{L}\left( X^{\prime },X\right) \right\rangle \mu
+S_{L}\left( X,X\right) }{1-\left\langle \hat{S}\left( X\right)
\right\rangle }+df\left( X\right) \left( \frac{\left\langle S_{E}\left(
X,X\right) \right\rangle }{1-\left\langle \hat{S}\left( X\right)
\right\rangle }-1\right) \right)
\end{equation*}%
\begin{equation*}
\frac{\hat{S}}{1-\hat{S}}\rightarrow \frac{\hat{S}}{1-\hat{S}}-\frac{%
\left\langle \hat{w}\left( X\right) \right\rangle \left\langle w\left(
X\right) \right\rangle }{2\left( 1-\hat{S}\right) ^{2}}\left( \frac{%
\left\langle \hat{S}_{L}\left( X^{\prime },X\right) \right\rangle \mu
+S_{L}\left( X,X\right) }{1-\left\langle \hat{S}\left( X\right)
\right\rangle }+df\left( X\right) \left( \frac{\left\langle S_{E}\left(
X,X\right) \right\rangle }{1-\left\langle \hat{S}\left( X\right)
\right\rangle }-1\right) \right)
\end{equation*}%
\begin{equation*}
\left( \frac{\hat{S}}{1-\hat{S}}\right) ^{2}\rightarrow \left( \frac{\hat{S}%
}{1-\hat{S}}\right) ^{2}-\frac{\hat{S}}{1-\hat{S}}\frac{\left\langle \hat{w}%
\left( X\right) \right\rangle \left\langle w\left( X\right) \right\rangle }{%
\left( 1-\hat{S}\right) ^{2}}\left( \frac{\left\langle \hat{S}_{L}\left(
X^{\prime },X\right) \right\rangle \mu +S_{L}\left( X,X\right) }{%
1-\left\langle \hat{S}\left( X\right) \right\rangle }+df\left( X\right)
\left( \frac{\left\langle S_{E}\left( X,X\right) \right\rangle }{%
1-\left\langle \hat{S}\left( X\right) \right\rangle }-1\right) \right)
\end{equation*}%
These quantities allow to derive the deviation of $\hat{K}_{X}\left\vert 
\hat{\Psi}\left( X\right) \right\vert ^{2}$ with respect to the no default
scenario: 
\begin{eqnarray*}
&&\hat{K}_{X}\left\vert \hat{\Psi}\left( X\right) \right\vert ^{2} \\
&\rightarrow &\hat{K}_{X}\left\vert \hat{\Psi}\left( X\right) \right\vert
^{2}-\frac{\left( 1-\hat{S}\right) }{\hat{S}}\left\langle \hat{w}\left(
X\right) \right\rangle \left\langle w\left( X\right) \right\rangle \\
&&\times \left( \frac{\left\langle \hat{S}_{L}\left( X^{\prime },X\right)
\right\rangle \mu +S_{L}\left( X,X\right) }{1-\left\langle \hat{S}\left(
X\right) \right\rangle }+df\left( X\right) \left( \frac{\left\langle
S_{E}\left( X,X\right) \right\rangle }{1-\left\langle \hat{S}\left( X\right)
\right\rangle }-1\right) \right) \hat{K}_{X}\left\vert \hat{\Psi}\left(
X\right) \right\vert ^{2}
\end{eqnarray*}%
and consequently:%
\begin{eqnarray*}
&&\frac{\delta \hat{K}_{X}\left\vert \hat{\Psi}\left( X\right) \right\vert
^{2}}{\hat{K}_{X}\left\vert \hat{\Psi}\left( X\right) \right\vert ^{2}} \\
&=&-\frac{\left( 1-\hat{S}\right) }{\hat{S}}\left\langle \hat{w}\left(
X\right) \right\rangle \left\langle w\left( X\right) \right\rangle \left( 
\frac{\left\langle \hat{S}_{L}\left( X^{\prime },X\right) \right\rangle \mu
+S_{L}\left( X,X\right) }{1-\left\langle \hat{S}\left( X\right)
\right\rangle }+df\left( X\right) \left( \frac{\left\langle S_{E}\left(
X,X\right) \right\rangle }{1-\left\langle \hat{S}\left( X\right)
\right\rangle }-1\right) \right)
\end{eqnarray*}%
Note that $\frac{\hat{S}}{1-\hat{S}}$ decreases without default, so that $%
\hat{K}_{X}\left\vert \hat{\Psi}\left( X\right) \right\vert ^{2}$ decreases.

The variation of (\ref{Rw}) is thus given by: 
\begin{equation*}
\delta \frac{\hat{K}_{X}\left\vert \hat{\Psi}\left( X\right) \right\vert ^{2}%
}{K_{X}\left\vert \Psi \left( X\right) \right\vert ^{2}}=\left( \frac{\delta 
\hat{K}_{X}\left\vert \hat{\Psi}\left( X\right) \right\vert ^{2}}{\hat{K}%
_{X}\left\vert \hat{\Psi}\left( X\right) \right\vert ^{2}}+\frac{2}{%
1-S\left( X\right) }\delta S\left( X\right) \right) \frac{\hat{K}%
_{X}\left\vert \hat{\Psi}\left( X\right) \right\vert ^{2}}{K_{X}\left\vert
\Psi \left( X\right) \right\vert ^{2}}
\end{equation*}%
and we obtain ultimately the variation of $\delta S\left( X\right) $ by
identifying the right hand side of (\ref{Dtf}) with $\delta S\left( X\right) 
$. This leads to: 
\begin{eqnarray*}
\delta S\left( X\right) &=&-\frac{w\left( X\right) \hat{w}\left( X\right) }{2%
}\frac{\hat{K}_{X}\left\vert \hat{\Psi}\left( X\right) \right\vert ^{2}}{%
K_{X}\left\vert \Psi \left( X\right) \right\vert ^{2}}\left( df\left(
X\right) \left( 1-\frac{\left\langle S_{E}\left( X,X\right) \right\rangle }{%
1-\left\langle \hat{S}\left( X\right) \right\rangle }\right) -\frac{%
\left\langle \hat{S}_{L}\left( X^{\prime },X\right) \right\rangle \mu
+S_{L}\left( X,X\right) }{1-\left\langle \hat{S}\left( X\right)
\right\rangle }\right) \\
&&+S\left( X,X\right) \left( \frac{\delta \hat{K}_{X}\left\vert \hat{\Psi}%
\left( X\right) \right\vert ^{2}}{\hat{K}_{X}\left\vert \hat{\Psi}\left(
X\right) \right\vert ^{2}}+\frac{2}{1-S\left( X\right) }\delta S\left(
X\right) \right) \frac{\hat{K}_{X}\left\vert \hat{\Psi}\left( X\right)
\right\vert ^{2}}{K_{X}\left\vert \Psi \left( X\right) \right\vert ^{2}}
\end{eqnarray*}%
with solution:%
\begin{equation*}
\delta S\left( X\right) =-\frac{\frac{w\left( X\right) \hat{w}\left(
X\right) }{2}\frac{\hat{K}_{X}\left\vert \hat{\Psi}\left( X\right)
\right\vert ^{2}}{K_{X}\left\vert \Psi \left( X\right) \right\vert ^{2}}%
\left( df\left( X\right) \left( 1-\frac{\left\langle S_{E}\left( X,X\right)
\right\rangle }{1-\left\langle \hat{S}\left( X\right) \right\rangle }\right)
-\frac{\left\langle \hat{S}_{L}\left( X^{\prime },X\right) \right\rangle \mu
+S_{L}\left( X,X\right) }{1-\left\langle \hat{S}\left( X\right)
\right\rangle }\right) -S\left( X\right) \frac{\delta \hat{K}_{X}\left\vert 
\hat{\Psi}\left( X\right) \right\vert ^{2}}{\hat{K}_{X}\left\vert \hat{\Psi}%
\left( X\right) \right\vert ^{2}}}{1-\frac{2S\left( X\right) }{1-S\left(
X\right) }}
\end{equation*}%
or in expanded form:%
\begin{eqnarray*}
&&\delta S\left( X\right) =-\frac{w\left( X\right) \hat{w}\left( X\right) 
\frac{\hat{K}_{X}\left\vert \hat{\Psi}\left( X\right) \right\vert ^{2}}{%
K_{X}\left\vert \Psi \left( X\right) \right\vert ^{2}}}{1-\frac{2S\left(
X\right) }{1-S\left( X\right) }}\left( \frac{1}{2}\left( df\left( X\right)
\left( 1-\frac{\left\langle S_{E}\left( X,X\right) \right\rangle }{%
1-\left\langle \hat{S}\left( X\right) \right\rangle }\right) -\frac{%
\left\langle \hat{S}_{L}\left( X^{\prime },X\right) \right\rangle \mu
+S_{L}\left( X,X\right) }{1-\left\langle \hat{S}\left( X\right)
\right\rangle }\right) \right. \\
&&\left. +\frac{S\left( X\right) \left( 1-\hat{S}\right) }{\hat{S}}\left( 
\frac{\left\langle \hat{S}_{L}\left( X^{\prime },X\right) \right\rangle \mu
+S_{L}\left( X,X\right) }{1-\left\langle \hat{S}\left( X\right)
\right\rangle }+df\left( X\right) \left( \frac{\left\langle S_{E}\left(
X,X\right) \right\rangle }{1-\left\langle \hat{S}\left( X\right)
\right\rangle }-1\right) \right) \right)
\end{eqnarray*}%
leading to:%
\begin{equation}
\delta S\left( X\right) \rightarrow -\frac{1}{2}w\left( X\right) \hat{w}%
\left( X\right) \frac{\hat{K}_{X}\left\vert \hat{\Psi}\left( X\right)
\right\vert ^{2}}{K_{X}\left\vert \Psi \left( X\right) \right\vert ^{2}}%
\frac{\left( 1-\frac{2S\left( X\right) \left( 1-\hat{S}\right) }{\hat{S}}%
\right) \left( df\left( X\right) \left( 1-\frac{\left\langle S_{E}\left(
X,X\right) \right\rangle }{1-\left\langle \hat{S}\left( X\right)
\right\rangle }\right) -\frac{\left\langle \hat{S}_{L}\left( X^{\prime
},X\right) \right\rangle \mu +S_{L}\left( X,X\right) }{1-\left\langle \hat{S}%
\left( X\right) \right\rangle }\right) }{1-\frac{2S\left( X\right) }{%
1-S\left( X\right) }}  \label{Ds}
\end{equation}%
In general $\frac{\hat{K}_{X}\left\vert \hat{\Psi}\left( X\right)
\right\vert ^{2}}{K_{X}\left\vert \Psi \left( X\right) \right\vert ^{2}}>>1$%
, $1-S\left( X\right) <<1$, and: 
\begin{equation*}
1-\frac{2S\left( X\right) }{1-S\left( X\right) }<0
\end{equation*}%
For $1-\hat{S}<<1$:%
\begin{equation*}
1-\frac{2S\left( X\right) \left( 1-\hat{S}\right) }{\hat{S}}>0
\end{equation*}%
Once $\delta S\left( X\right) $ is obtained, we can obtained the loss in
firms' returns in default scenario.

To find $f\left( X\right) $, recall that:

\begin{equation*}
f\left( X\right) =\left( 1-S\left( X\right) \right) ^{r}\left( f_{1}\left(
X\right) +\tau \Delta F_{\tau }\left( \bar{R}\left( K,X\right) \right)
\right) -C_{0}-\left( 1-S\left( X\right) \right) C
\end{equation*}%
Since:%
\begin{equation*}
f\left( X\right) \rightarrow f\left( X\right) -df\left( X\right)
\end{equation*}%
the following equation is satisfied:%
\begin{equation}
df\left( X\right) =-\left( C-\frac{r\left( f_{1}\left( X\right) +\tau \Delta
F_{\tau }\left( \bar{R}\left( K,X\right) \right) \right) }{\left( 1-S\left(
X\right) \right) ^{1-r}}\right) \delta S\left( X\right)  \label{dS}
\end{equation}%
Given (\ref{Ds}), equation (\ref{dS}) writes:%
\begin{eqnarray}
df\left( X\right) &=&\left( C-\frac{r\left( f_{1}\left( X\right) +\tau
\Delta F_{\tau }\left( \bar{R}\left( K,X\right) \right) \right) }{\left(
1-S\left( X\right) \right) ^{1-r}}\right)  \label{Dfw} \\
&&\times \frac{1}{2}w\left( X\right) \hat{w}\left( X\right) \frac{\hat{K}%
_{X}\left\vert \hat{\Psi}\left( X\right) \right\vert ^{2}}{K_{X}\left\vert
\Psi \left( X\right) \right\vert ^{2}}\frac{\left( 1-\frac{2S\left( X\right)
\left( 1-\hat{S}\right) }{\hat{S}}\right) \left( df\left( X\right) \left( 1-%
\frac{\left\langle S_{E}\left( X,X\right) \right\rangle }{1-\left\langle 
\hat{S}\left( X\right) \right\rangle }\right) -\frac{\left( \left\langle 
\hat{S}_{L}\left( X^{\prime },X\right) \right\rangle +\left\langle
S_{L}\left( X,X\right) \right\rangle \right) \mu }{1-\left\langle \hat{S}%
\left( X\right) \right\rangle }\right) }{1-\frac{2S\left( X\right) }{%
1-S\left( X\right) }}  \notag
\end{eqnarray}%
Note that, as expected, compared to the no default scenario $\frac{\hat{K}%
_{X}\left\vert \hat{\Psi}\left( X\right) \right\vert ^{2}}{K_{X}\left\vert
\Psi \left( X\right) \right\vert ^{2}}$ decreases, that is disposable
capital for firms reduces.

The solution of (\ref{Dfw}) is:%
\begin{equation*}
df\left( X\right) =-\frac{\frac{1}{2}dCw\left( X\right) \hat{w}\left(
X\right) \frac{\hat{K}_{X}\left\vert \hat{\Psi}\left( X\right) \right\vert
^{2}}{K_{X}\left\vert \Psi \left( X\right) \right\vert ^{2}}\frac{1-\frac{%
2S\left( X\right) \left( 1-\hat{S}\right) }{\hat{S}}}{1-\frac{2S\left(
X\right) }{1-S\left( X\right) }}\frac{\left( \left\langle \hat{S}_{L}\left(
X^{\prime },X\right) \right\rangle +\left\langle S_{L}\left( X,X\right)
\right\rangle \right) \mu }{1-\left\langle \hat{S}\left( X\right)
\right\rangle }}{1-\frac{1}{2}dCw\left( X\right) \hat{w}\left( X\right) 
\frac{\hat{K}_{X}\left\vert \hat{\Psi}\left( X\right) \right\vert ^{2}}{%
K_{X}\left\vert \Psi \left( X\right) \right\vert ^{2}}\frac{1-\frac{2S\left(
X\right) \left( 1-\hat{S}\right) }{\hat{S}}}{1-\frac{2S\left( X\right) }{%
1-S\left( X\right) }}\left( 1-\frac{\left\langle S_{E}\left( X,X\right)
\right\rangle }{1-\left\langle \hat{S}\left( X\right) \right\rangle }\right) 
}
\end{equation*}%
with:%
\begin{equation*}
dC=C-\frac{r\left( f_{1}\left( X\right) +\tau \Delta F_{\tau }\left( \bar{R}%
\left( K,X\right) \right) \right) }{\left( 1-S\left( X\right) \right) ^{1-r}}
\end{equation*}%
We can then derive the value of $d\hat{f}\left( X\right) $: 
\begin{equation*}
d\hat{f}\left( X\right) =\frac{\left( \left\langle \hat{S}_{L}\left(
X^{\prime },X\right) \right\rangle +\left\langle S_{L}\left( X,X\right)
\right\rangle \right) \mu +\left\langle S_{E}\left( X^{\prime },X^{\prime
}\right) \right\rangle \left\langle df\left( X^{\prime }\right)
\right\rangle }{1-\left\langle \hat{S}\left( X^{\prime }\right)
\right\rangle }
\end{equation*}%
with:%
\begin{eqnarray}
\left\langle df\left( X^{\prime }\right) \right\rangle &=&-\frac{\frac{1}{2}%
dC\left\langle w\left( X\right) \right\rangle \left\langle \hat{w}\left(
X\right) \right\rangle \frac{\left\langle \hat{K}_{X}\left\vert \hat{\Psi}%
\left( X\right) \right\vert ^{2}\right\rangle }{\left\langle K_{X}\left\vert
\Psi \left( X\right) \right\vert ^{2}\right\rangle }\frac{1-\frac{2S\left(
X\right) \left( 1-\hat{S}\right) }{\hat{S}}}{1-\frac{2S\left( X\right) }{%
1-S\left( X\right) }}\frac{\left( \left\langle \hat{S}_{L}\left( X^{\prime
},X\right) \right\rangle +\left\langle S_{L}\left( X,X\right) \right\rangle
\right) }{1-\left\langle \hat{S}\left( X\right) \right\rangle }}{1-\frac{1}{2%
}dC\left\langle w\left( X\right) \right\rangle \left\langle \hat{w}\left(
X\right) \right\rangle \frac{\left\langle \hat{K}_{X}\left\vert \hat{\Psi}%
\left( X\right) \right\vert ^{2}\right\rangle }{\left\langle K_{X}\left\vert
\Psi \left( X\right) \right\vert ^{2}\right\rangle }\frac{1-\frac{2S\left(
X\right) \left( 1-\hat{S}\right) }{\hat{S}}}{1-\frac{2S\left( X\right) }{%
1-S\left( X\right) }}\left( 1-\frac{\left\langle S_{E}\left( X,X\right)
\right\rangle }{1-\left\langle \hat{S}\left( X\right) \right\rangle }\right) 
}  \label{Dfv} \\
&=&-\frac{N\left( \left\langle \hat{S}_{L}\left( X^{\prime },X\right)
\right\rangle +\left\langle S_{L}\left( X,X\right) \right\rangle \right) }{%
\left( 1-\left\langle \hat{S}\left( X\right) \right\rangle \right) \left(
1-N\left( 1-\frac{\left\langle S_{E}\left( X,X\right) \right\rangle }{%
1-\left\langle \hat{S}\left( X\right) \right\rangle }\right) \right) } 
\notag
\end{eqnarray}%
where:%
\begin{equation*}
N=\frac{1}{2}\left\langle w\left( X\right) \right\rangle \left\langle \hat{w}%
\left( X\right) \right\rangle \frac{\left\langle \hat{K}_{X}\left\vert \hat{%
\Psi}\left( X\right) \right\vert ^{2}\right\rangle }{\left\langle
K_{X}\left\vert \Psi \left( X\right) \right\vert ^{2}\right\rangle }\frac{1-%
\frac{2S\left( X\right) \left( 1-\hat{S}\right) }{\hat{S}}}{1-\frac{2S\left(
X\right) }{1-S\left( X\right) }}dC
\end{equation*}%
leading to the investors returns after default:%
\begin{equation*}
\hat{f}\left( X\right) \rightarrow \hat{f}\left( X\right) -d\hat{f}
\end{equation*}%
so that the default of investors arise when:%
\begin{equation*}
\hat{f}\left( X\right) <-1+d\hat{f}
\end{equation*}%
We can then deduce the fraction of impacted sectors:%
\begin{equation*}
\frac{1}{2}+\frac{-1+\left\langle d\hat{f}\right\rangle -\left\langle \hat{f}%
\right\rangle }{2\left\langle \hat{f}\right\rangle }
\end{equation*}%
\begin{equation*}
\mu =\frac{1}{2}-\frac{1-\left\langle d\hat{f}\right\rangle +\left\langle 
\hat{f}\right\rangle }{2\left\langle \hat{f}\right\rangle }=\frac{%
\left\langle d\hat{f}\right\rangle -1}{2\left\langle \hat{f}\right\rangle }
\end{equation*}%
and we find:%
\begin{equation*}
\mu =\frac{1-\left\langle \hat{S}\left( X^{\prime }\right) \right\rangle }{%
\left( \left\langle \hat{S}_{L}\left( X^{\prime },X\right) \right\rangle
+\left\langle S_{L}\left( X,X\right) \right\rangle \right) +\left\langle
S_{E}\left( X^{\prime },X^{\prime }\right) \right\rangle \left\langle
df\left( X^{\prime }\right) \right\rangle -2\left\langle \hat{f}%
\right\rangle \left( 1-\left\langle \hat{S}\left( X^{\prime }\right)
\right\rangle \right) }
\end{equation*}%
so that the condition for default writes:%
\begin{equation*}
\left( \left\langle \hat{S}_{L}\left( X^{\prime },X\right) \right\rangle
+\left\langle S_{L}\left( X,X\right) \right\rangle \right) +\left\langle
S_{E}\left( X^{\prime },X^{\prime }\right) \right\rangle \left\langle
df\left( X^{\prime }\right) \right\rangle >2\left\langle \hat{f}%
\right\rangle \left( 1-\left\langle \hat{S}\left( X^{\prime }\right)
\right\rangle \right)
\end{equation*}%
Using (\ref{Dfv}), this also rewrites as:

\begin{equation*}
\left( \left\langle \hat{S}_{L}\left( X^{\prime },X\right) \right\rangle
+\left\langle S_{L}\left( X,X\right) \right\rangle \right) \left( \frac{%
1+N\left\langle S_{E}\left( X^{\prime },X^{\prime }\right) \right\rangle }{%
\left( 1-\left\langle \hat{S}\left( X\right) \right\rangle \right) \left(
1-N\left( 1-\frac{\left\langle S_{E}\left( X,X\right) \right\rangle }{%
1-\left\langle \hat{S}\left( X\right) \right\rangle }\right) \right) }%
\right) >2\left\langle \hat{f}\right\rangle \left( 1-\left\langle \hat{S}%
\left( X^{\prime }\right) \right\rangle \right)
\end{equation*}

\paragraph{A8.3.2.2 First order corrections}

We compute first order corrections to the shares arising in the returns
formula:%
\begin{eqnarray}
&&S_{L}\left( X,X\right) \\
&=&\frac{w\left( X\right) }{2}\left( 1+\frac{1}{2}\left( \bar{r}\left(
X\right) -\frac{\hat{w}\left( X\right) \left\langle \hat{f}\left( X^{\prime
}\right) \right\rangle _{\hat{w}_{E}}+w\left( X\right) f\left( X\right) }{2}%
\right) \right)  \notag
\end{eqnarray}%
\begin{eqnarray}
&&\left\langle \hat{S}_{L}\left( X^{\prime },X\right) \right\rangle
_{X^{\prime }} \\
&\simeq &\frac{\left\langle \hat{w}\left( X^{\prime },X\right) \right\rangle 
}{2}  \notag \\
&&+\frac{\left\langle \hat{w}\left( X^{\prime },X\right) \right\rangle }{2}%
\left( \left\langle \hat{r}\left( X^{\prime }\right) \right\rangle _{\hat{w}%
_{L}}-\frac{\hat{w}\left( X\right) \left\langle \hat{f}\left( X^{\prime
}\right) \right\rangle +w\left( X\right) f\left( X\right) }{2}\right)  \notag
\\
&\simeq &\frac{\left\langle \hat{w}\left( X^{\prime },X\right) \right\rangle 
}{2}\left( 1+\left\langle \hat{r}\left( X^{\prime }\right) \right\rangle _{%
\hat{w}_{L}}-\frac{\hat{w}\left( X\right) \left\langle \hat{f}\left(
X^{\prime }\right) \right\rangle +w\left( X\right) f\left( X\right) }{2}%
\right)  \notag
\end{eqnarray}%
\begin{eqnarray*}
&&\left\langle \hat{S}_{L}\left( X^{\prime },X\right) \right\rangle
_{X^{\prime }}\mu +S_{L}\left( X,X\right) \\
&\rightarrow &\left\langle \hat{S}_{L}\left( X^{\prime },X\right)
\right\rangle _{X^{\prime }}\mu +S_{L}\left( X,X\right) +\frac{1}{2}\left( 
\hat{w}\left( X\right) \left( \left\langle \hat{S}_{L}\left( X^{\prime
},X\right) \right\rangle _{X^{\prime }}\mu +S_{L}\left( X,X\right) \right)
+w\left( X\right) df\left( X\right) \right)
\end{eqnarray*}%
We also need to consider $S_{E}\left( X,X\right) $:%
\begin{eqnarray}
&&S_{E}\left( X,X\right) \\
&=&\frac{w\left( X\right) }{2}\left( 1+\left( \hat{w}\left( X\right) \left(
f\left( X\right) -\frac{\left\langle \hat{f}\left( X^{\prime }\right)
\right\rangle _{\hat{w}_{E}}+\left\langle \hat{r}\left( X^{\prime }\right)
\right\rangle _{\hat{w}_{L}}}{2}\right) +\frac{w\left( X\right) }{2}\left(
f\left( X\right) -\bar{r}\left( X\right) \right) \right) \right)  \notag
\end{eqnarray}%
\begin{eqnarray*}
&&S_{E}\left( X,X\right) \\
&\rightarrow &S_{E}\left( X,X\right) -\frac{w\left( X\right) }{2}\left(
\left( \hat{w}\left( X\right) +\frac{w\left( X\right) }{2}\right) df\left(
X\right) \right. \\
&&\left. -\frac{\hat{w}\left( X\right) }{2}\frac{\left( \left\langle \hat{S}%
_{L}\left( X^{\prime },X\right) \right\rangle +\left\langle S_{L}\left(
X,X\right) \right\rangle \right) \mu +\left\langle S_{E}\left( X^{\prime
},X^{\prime }\right) \right\rangle \left\langle df\left( X^{\prime }\right)
\right\rangle }{1-\left\langle \hat{S}\left( X^{\prime }\right)
\right\rangle }\right) \\
&\rightarrow &S_{E}\left( X,X\right) -\frac{w\left( X\right) }{2}\left(
\left( \hat{w}\left( X\right) \left( 1-\frac{\left\langle S_{E}\left(
X^{\prime },X^{\prime }\right) \right\rangle }{2\left( 1-\left\langle \hat{S}%
\left( X^{\prime }\right) \right\rangle \right) }\right) +\frac{w\left(
X\right) }{2}\right) df\left( X\right) \right. \\
&&\left. -\frac{\hat{w}\left( X\right) }{2}\frac{\left( \left\langle \hat{S}%
_{L}\left( X^{\prime },X\right) \right\rangle +\left\langle S_{L}\left(
X,X\right) \right\rangle \right) \mu }{1-\left\langle \hat{S}\left(
X^{\prime }\right) \right\rangle }\right)
\end{eqnarray*}%
and the combination involved in the default is given by:%
\begin{eqnarray*}
&&\left( \left\langle \hat{S}_{L}\left( X^{\prime },X\right) \right\rangle
+\left\langle S_{L}\left( X,X\right) \right\rangle \right) \mu +\left\langle
S_{E}\left( X^{\prime },X^{\prime }\right) \right\rangle \left\langle
df\left( X^{\prime }\right) \right\rangle \\
&&+\frac{1}{2}\left( \hat{w}\left( X\right) \left( \left\langle \hat{S}%
_{L}\left( X^{\prime },X\right) \right\rangle _{X^{\prime }}\mu +S_{L}\left(
X,X\right) \right) +w\left( X\right) df\left( X\right) \right) \\
&&-\frac{w\left( X\right) }{2}\left( \left( \hat{w}\left( X\right) \left( 1-%
\frac{\left\langle S_{E}\left( X^{\prime },X^{\prime }\right) \right\rangle 
}{2\left( 1-\left\langle \hat{S}\left( X^{\prime }\right) \right\rangle
\right) }\right) +\frac{w\left( X\right) }{2}\right) df\left( X\right)
\right. \\
&&\left. -\frac{\hat{w}\left( X\right) }{2}\frac{\left( \left\langle \hat{S}%
_{L}\left( X^{\prime },X\right) \right\rangle +\left\langle S_{L}\left(
X,X\right) \right\rangle \right) \mu }{1-\left\langle \hat{S}\left(
X^{\prime }\right) \right\rangle }df\left( X\right) \right)
\end{eqnarray*}%
So that the corrections of the following terms with respect to the first
order writes:%
\begin{equation*}
\frac{1}{2}\left( \hat{w}\left( X\right) \left( \left\langle \hat{S}%
_{L}\left( X^{\prime },X\right) \right\rangle _{X^{\prime }}\mu +S_{L}\left(
X,X\right) \right) \right) +\frac{w\left( X\right) }{2}\left( 1+\frac{\hat{w}%
\left( X\right) }{2}\frac{\left( \left\langle \hat{S}_{L}\left( X^{\prime
},X\right) \right\rangle +\left\langle S_{L}\left( X,X\right) \right\rangle
\right) \mu }{1-\left\langle \hat{S}\left( X^{\prime }\right) \right\rangle }%
\right) df\left( X\right)
\end{equation*}

\begin{eqnarray*}
&&\left\langle \hat{S}\left( X,X\right) \right\rangle \\
&=&\hat{w}\left( X^{\prime },X\right) \left( 1+\left( \frac{\hat{f}\left(
X^{\prime }\right) +\hat{r}\left( X^{\prime }\right) }{2}-\left\langle \hat{w%
}\left( X\right) \right\rangle \frac{\left\langle \hat{f}\left( X^{\prime
}\right) \right\rangle _{\hat{w}_{E}}+\left\langle \hat{r}\left( X^{\prime
}\right) \right\rangle _{\hat{w}_{L}}}{2}-\left\langle w\left( X\right)
\right\rangle \frac{f\left( X\right) +r\left( X\right) }{2}\right) \right) \\
&\rightarrow &\left\langle \hat{S}\left( X,X\right) \right\rangle
-\left\langle \hat{w}\left( X\right) \right\rangle \left\langle w\left(
X\right) \right\rangle \frac{1}{2}\left( \frac{\left( \left\langle \hat{S}%
_{L}\left( X^{\prime },X\right) \right\rangle +\left\langle S_{L}\left(
X,X\right) \right\rangle \right) \mu }{1-\left\langle \hat{S}\left( X\right)
\right\rangle }+df\left( X\right) \left( \frac{\left\langle S_{E}\left(
X,X\right) \right\rangle }{1-\left\langle \hat{S}\left( X\right)
\right\rangle }-1\right) \right)
\end{eqnarray*}%
and:%
\begin{equation*}
\frac{1}{1-\hat{S}}\rightarrow \frac{1}{1-\hat{S}}-\frac{\left\langle \hat{w}%
\left( X\right) \right\rangle \left\langle w\left( X\right) \right\rangle }{%
2\left( 1-\hat{S}\right) ^{2}}\left( \frac{\left( \left\langle \hat{S}%
_{L}\left( X^{\prime },X\right) \right\rangle +\left\langle S_{L}\left(
X,X\right) \right\rangle \right) \mu }{1-\left\langle \hat{S}\left( X\right)
\right\rangle }+df\left( X\right) \left( \frac{\left\langle S_{E}\left(
X,X\right) \right\rangle }{1-\left\langle \hat{S}\left( X\right)
\right\rangle }-1\right) \right)
\end{equation*}%
Ultimately, this modifies $d\hat{f}\left( X\right) $ and we find:

\begin{eqnarray*}
d\hat{f}\left( X\right) &=&-\frac{\left( \left( \left\langle \hat{S}%
_{L}\left( X^{\prime },X\right) \right\rangle +\left\langle S_{L}\left(
X,X\right) \right\rangle \right) \mu \right) \left( 1-\frac{\hat{w}\left(
X\right) }{2}\right) }{1-\left\langle \hat{S}\left( X^{\prime }\right)
\right\rangle } \\
&&-\frac{\left( \left\langle S_{E}\left( X^{\prime },X^{\prime }\right)
\right\rangle -\frac{w\left( X\right) }{2}\left( 1+\frac{\hat{w}\left(
X\right) }{2}\frac{\left( \left\langle \hat{S}_{L}\left( X^{\prime
},X\right) \right\rangle +\left\langle S_{L}\left( X,X\right) \right\rangle
\right) \mu }{1-\left\langle \hat{S}\left( X^{\prime }\right) \right\rangle }%
\right) \right) \left\langle df\left( X^{\prime }\right) \right\rangle }{%
1-\left\langle \hat{S}\left( X^{\prime }\right) \right\rangle } \\
&&-\frac{\left\langle \hat{w}\left( X\right) \right\rangle \left\langle
w\left( X\right) \right\rangle }{2\left( 1-\hat{S}\right) ^{2}}\left( \frac{%
\left( \left\langle \hat{S}_{L}\left( X^{\prime },X\right) \right\rangle
+\left\langle S_{L}\left( X,X\right) \right\rangle \right) \mu }{%
1-\left\langle \hat{S}\left( X\right) \right\rangle }+df\left( X\right)
\left( \frac{\left\langle S_{E}\left( X,X\right) \right\rangle }{%
1-\left\langle \hat{S}\left( X\right) \right\rangle }-1\right) \right) \\
&&\times \left( \left( \left\langle \hat{S}_{L}\left( X^{\prime },X\right)
\right\rangle +\left\langle S_{L}\left( X,X\right) \right\rangle \right) \mu
+\left\langle S_{E}\left( X^{\prime },X^{\prime }\right) \right\rangle
\left\langle df\left( X^{\prime }\right) \right\rangle \right)
\end{eqnarray*}%
\begin{eqnarray*}
d\hat{f}\left( X\right) &=&-\frac{\left( \left( \left\langle \hat{S}%
_{L}\left( X^{\prime },X\right) \right\rangle +\left\langle S_{L}\left(
X,X\right) \right\rangle \right) \mu \right) \left( 1-\frac{\hat{w}\left(
X\right) }{2}\right) }{1-\left\langle \hat{S}\left( X^{\prime }\right)
\right\rangle } \\
&&-\frac{\left( \left\langle S_{E}\left( X^{\prime },X^{\prime }\right)
\right\rangle -\frac{w\left( X\right) }{2}\left( 1+\frac{\hat{w}\left(
X\right) }{2}\frac{\left( \left\langle \hat{S}_{L}\left( X^{\prime
},X\right) \right\rangle +\left\langle S_{L}\left( X,X\right) \right\rangle
\right) \mu }{1-\left\langle \hat{S}\left( X^{\prime }\right) \right\rangle }%
\right) \right) \left\langle df\left( X^{\prime }\right) \right\rangle }{%
1-\left\langle \hat{S}\left( X^{\prime }\right) \right\rangle } \\
&&-\frac{\left( \frac{\left\langle \hat{w}\left( X\right) \right\rangle
\left\langle w\left( X\right) \right\rangle \left( \left\langle \hat{S}%
_{L}\left( X^{\prime },X\right) \right\rangle +\left\langle S_{L}\left(
X,X\right) \right\rangle \right) \mu }{2\left( 1-\hat{S}\right) }\left( 
\frac{\left\langle S_{E}\left( X,X\right) \right\rangle }{1-\left\langle 
\hat{S}\left( X\right) \right\rangle }-1\right) \right) \left\langle
df\left( X^{\prime }\right) \right\rangle }{1-\left\langle \hat{S}\left(
X^{\prime }\right) \right\rangle } \\
&&-\frac{\left\langle \hat{w}\left( X\right) \right\rangle \left\langle
w\left( X\right) \right\rangle }{2\left( 1-\hat{S}\right) ^{2}}\left( \frac{%
\left( \left\langle \hat{S}_{L}\left( X^{\prime },X\right) \right\rangle
+\left\langle S_{L}\left( X,X\right) \right\rangle \right) \mu }{%
1-\left\langle \hat{S}\left( X\right) \right\rangle }\right) \\
&&\times \left( \left( \left\langle \hat{S}_{L}\left( X^{\prime },X\right)
\right\rangle +\left\langle S_{L}\left( X,X\right) \right\rangle \right) \mu
+\left\langle S_{E}\left( X^{\prime },X^{\prime }\right) \right\rangle
\left\langle df\left( X^{\prime }\right) \right\rangle \right)
\end{eqnarray*}%
The share $\left\langle S_{E}\left( X^{\prime },X^{\prime }\right)
\right\rangle $ is modified by:%
\begin{eqnarray*}
&&\left\langle S_{E}\left( X^{\prime },X^{\prime }\right) \right\rangle -%
\frac{w\left( X\right) }{2}\left( 1+\frac{\hat{w}\left( X\right) }{2}\frac{%
\left( \left\langle \hat{S}_{L}\left( X^{\prime },X\right) \right\rangle
+\left\langle S_{L}\left( X,X\right) \right\rangle \right) \mu }{%
1-\left\langle \hat{S}\left( X^{\prime }\right) \right\rangle }\right) \\
&&+\frac{\left\langle \hat{w}\left( X\right) \right\rangle \left\langle
w\left( X\right) \right\rangle \left( \left\langle \hat{S}_{L}\left(
X^{\prime },X\right) \right\rangle +\left\langle S_{L}\left( X,X\right)
\right\rangle \right) \mu }{2\left( 1-\hat{S}\right) }\left( \frac{%
\left\langle S_{E}\left( X,X\right) \right\rangle }{1-\left\langle \hat{S}%
\left( X\right) \right\rangle }-1\right)
\end{eqnarray*}

which writes:%
\begin{eqnarray*}
&&\left\langle S_{E}\left( X^{\prime },X^{\prime }\right) \right\rangle +%
\frac{\left\langle \hat{w}\left( X\right) \right\rangle \left\langle w\left(
X\right) \right\rangle \left( \left\langle \hat{S}_{L}\left( X^{\prime
},X\right) \right\rangle +\left\langle S_{L}\left( X,X\right) \right\rangle
\right) \mu }{2\left( 1-\hat{S}\right) }\left( \frac{\left\langle
S_{E}\left( X,X\right) \right\rangle }{1-\left\langle \hat{S}\left( X\right)
\right\rangle }-\frac{3}{2}\right) \\
&&+\frac{\left\langle \hat{w}\left( X\right) \right\rangle \left\langle
w\left( X\right) \right\rangle }{2\left( 1-\hat{S}\right) }\left( \frac{%
\left( \left\langle \hat{S}_{L}\left( X^{\prime },X\right) \right\rangle
+\left\langle S_{L}\left( X,X\right) \right\rangle \right) \mu }{%
1-\left\langle \hat{S}\left( X\right) \right\rangle }\right) \left\langle
S_{E}\left( X^{\prime },X^{\prime }\right) \right\rangle \\
&=&\left\langle S_{E}\left( X^{\prime },X^{\prime }\right) \right\rangle +%
\frac{\left\langle \hat{w}\left( X\right) \right\rangle \left\langle w\left(
X\right) \right\rangle \left( \left\langle \hat{S}_{L}\left( X^{\prime
},X\right) \right\rangle +\left\langle S_{L}\left( X,X\right) \right\rangle
\right) \mu }{2\left( 1-\hat{S}\right) }\left( 2\frac{\left\langle
S_{E}\left( X,X\right) \right\rangle }{1-\left\langle \hat{S}\left( X\right)
\right\rangle }-\frac{3}{2}\right)
\end{eqnarray*}%
Inserting this formula in the expression for $d\hat{f}\left( X\right) $
leads to: 
\begin{eqnarray*}
d\hat{f}\left( X\right) &=&-\frac{\left( \left\langle \hat{S}_{L}\left(
X^{\prime },X\right) \right\rangle +\left\langle S_{L}\left( X,X\right)
\right\rangle \right) \mu \left( 1-\frac{\hat{w}\left( X\right) }{2}\right) 
}{1-\left\langle \hat{S}\left( X^{\prime }\right) \right\rangle } \\
&&-\frac{\frac{\left\langle \hat{w}\left( X\right) \right\rangle
\left\langle w\left( X\right) \right\rangle \left( \left\langle \hat{S}%
_{L}\left( X^{\prime },X\right) \right\rangle +\left\langle S_{L}\left(
X,X\right) \right\rangle \right) \mu }{2\left( 1-\hat{S}\right) }\left( 2%
\frac{\left\langle S_{E}\left( X,X\right) \right\rangle }{1-\left\langle 
\hat{S}\left( X\right) \right\rangle }-\frac{3}{2}\right) \left\langle
df\left( X^{\prime }\right) \right\rangle }{1-\left\langle \hat{S}\left(
X^{\prime }\right) \right\rangle } \\
&&-\frac{\left\langle \hat{w}\left( X\right) \right\rangle \left\langle
w\left( X\right) \right\rangle }{2\left( 1-\hat{S}\right) ^{2}}\left( \frac{%
\left( \left\langle \hat{S}_{L}\left( X^{\prime },X\right) \right\rangle
+\left\langle S_{L}\left( X,X\right) \right\rangle \right) \mu }{%
1-\left\langle \hat{S}\left( X\right) \right\rangle }\right) \left(
\left\langle \hat{S}_{L}\left( X^{\prime },X\right) \right\rangle
+\left\langle S_{L}\left( X,X\right) \right\rangle \right) \mu
\end{eqnarray*}

\section*{Appendix 9 \ Capital circulation}

We start with equation (\ref{CPR}) and consider perturbations $\delta \hat{f}%
\left( \hat{X},\theta -1\right) $, $\delta \hat{S}_{1}\left( \hat{X}^{\prime
},\hat{X},\theta -2\right) $ to deduce the expressions of $\delta \hat{f}%
\left( \hat{X},\theta \right) $ and $\delta \hat{S}_{1}\left( \hat{X}%
^{\prime },\hat{X},\theta -1\right) $. We consider the returns with
decreasing return to scale:%
\begin{eqnarray*}
&&f\left( X,\theta -1\right) -\bar{r} \\
&=&\left( 1-S\left( X,\theta -1\right) \right) ^{r}f_{1}\left( X\right)
-C_{0}-\left( 1-S\left( X,\theta -1\right) \right) C-\bar{r}+\left(
1-S\left( X,\theta -1\right) \right) ^{r}\tau \left( \left\langle
f_{1}\left( X\right) \right\rangle -\left\langle f_{1}\left( X^{\prime
}\right) \right\rangle \right)
\end{eqnarray*}

\subsection*{A9.1 Derivation of perturbation equations for averages}

The variation of (\ref{CPR}) around equilibrium values writes:%
\begin{eqnarray*}
0 &=&\left( \delta \left( X-X^{\prime }\right) -\hat{S}_{E}\left( X^{\prime
},X,\theta -1\right) \right) \frac{1-\left( \hat{S}\left( X^{\prime },\theta
-1\right) +\delta \hat{S}\left( X^{\prime },\theta -1\right) \right) }{%
1-\left( \hat{S}_{E}\left( X^{\prime },\theta -1\right) +\delta \hat{S}%
_{E}\left( X^{\prime },\theta -1\right) \right) }\left( \hat{f}\left(
X\right) +\delta \hat{f}\left( X\right) -\bar{r}\right) \\
&&-\delta \hat{S}_{E}\left( X^{\prime },X,\theta -1\right) \frac{\left( 1-%
\hat{S}\left( X^{\prime },\theta -1\right) \right) \left( \hat{f}\left(
X^{\prime }\right) -\bar{r}\right) }{1-\hat{S}_{E}\left( X^{\prime },\theta
-1\right) } \\
&&-\delta S_{E}\left( X,X,\theta -1\right) \\
&&\times \left( \left( 1-S\left( X,\theta -1\right) \right) ^{r}f_{1}\left(
X\right) -C_{0}-\left( 1-S\left( X,\theta -1\right) \right) C-\bar{r}+\left(
1-\delta S\left( X,\theta -1\right) \right) ^{r}\Delta F_{\tau }\left( \bar{R%
}\left( K,X\right) \right) \right) \\
&&+S_{E}\left( X,X,\theta -1\right) \left( r\left( 1-S\left( X,\theta
-1\right) \right) ^{r-1}\left( f_{1}\left( X\right) +\Delta F_{\tau }\left( 
\bar{R}\left( K,X\right) \right) \right) -C\right) \delta S\left( X,\theta
-1\right)
\end{eqnarray*}

\begin{eqnarray}
0 &=&\frac{1-\left( \hat{S}\left( X^{\prime },\theta -1\right) +\delta \hat{S%
}\left( X^{\prime },\theta -1\right) \right) }{1-\left( \hat{S}_{E}\left(
X^{\prime },\theta -1\right) +\delta \hat{S}_{E}\left( X^{\prime },\theta
-1\right) \right) }\left( \hat{f}\left( X\right) -\bar{r}\right) +\frac{%
1-\left( \hat{S}\left( X,\theta -1\right) \right) }{1-\left( \hat{S}\left(
X,\theta -1\right) \right) }\delta \hat{f}\left( X\right)  \label{Vr} \\
&&-\delta \left\langle \hat{S}_{E}\left( X^{\prime },X,\theta -1\right)
\right\rangle _{X^{\prime }}\frac{\left( 1-\left\langle \hat{S}\left(
X^{\prime },\theta -1\right) \right\rangle \right) \left( \left\langle \hat{f%
}\left( X^{\prime },\theta \right) \right\rangle -\bar{r}\right) }{%
1-\left\langle \hat{S}_{E}\left( X^{\prime },\theta -1\right) \right\rangle }
\notag \\
&&-\delta S_{E}\left( X,X,\theta -1\right)  \notag \\
&&\times \left( \left( 1-S\left( X,\theta -1\right) \right) ^{r}f_{1}\left(
X\right) -C_{0}-\left( 1-S\left( X,\theta -1\right) \right) C-\bar{r}+\left(
1-\delta S\left( X,\theta -1\right) \right) ^{r}\Delta F_{\tau }\left( \bar{R%
}\left( K,X\right) \right) \right)  \notag \\
&&+S_{E}\left( X,X,\theta -1\right) \left( r\left( 1-S\left( X,\theta
-1\right) \right) ^{r-1}\left( f_{1}\left( X\right) +\Delta F_{\tau }\left( 
\bar{R}\left( K,X\right) \right) \right) -C\right) \delta S\left( X,\theta
-1\right)  \notag
\end{eqnarray}%
We can compute the variations at first order of the various quantities
arising in (\ref{Vr}). First, an expansion ylds: 
\begin{equation*}
\frac{1-\left( \hat{S}\left( X,\theta \right) +\delta \hat{S}\left( X,\theta
\right) \right) }{1-\left( \hat{S}_{E}\left( X^{\prime },\theta \right)
+\delta \hat{S}_{E}\left( X,\theta \right) \right) }=\bigskip -\frac{\delta 
\hat{S}\left( X,\theta \right) }{1-\left( \hat{S}_{E}\left( X^{\prime
},\theta \right) \right) }+\frac{\left( 1-\left( \hat{S}\left( X,\theta
\right) \right) \right) }{\left( 1-\left( \hat{S}_{E}\left( X^{\prime
},\theta \right) \right) \right) ^{2}}\delta \hat{S}_{E}\left( X,\theta
\right)
\end{equation*}

Using that:

\begin{eqnarray}
&&\hat{S}_{E}\left( X^{\prime }\right) \\
&\rightarrow &\frac{1}{2}\frac{\left( 1-\left( \gamma \left\langle \hat{S}%
_{E}\left( X\right) \right\rangle \right) ^{2}\right) \left( 1+\Delta \hat{f}%
\left( X^{\prime }\right) \right) }{2-\left( \gamma \left\langle \hat{S}%
_{E}\left( X\right) \right\rangle \right) ^{2}-\gamma \left\langle \hat{S}%
_{E}\left( X\right) \right\rangle \gamma \left\langle \hat{w}\left(
X^{\prime },X\right) \right\rangle \left\langle w\left( X\right)
\right\rangle \Delta \left( \frac{f\left( X^{\prime }\right) +r\left(
X^{\prime }\right) }{2}\right) }\frac{\left\langle \hat{K}\right\rangle
\left\Vert \hat{\Psi}\right\Vert ^{2}}{\hat{K}_{X^{\prime }}\left\vert \hat{%
\Psi}\left( X^{\prime }\right) \right\vert ^{2}}  \notag
\end{eqnarray}%
and:

\begin{eqnarray}
&&\hat{S}\left( X^{\prime },\theta -1\right) \\
&\rightarrow &\frac{\left( 1-\left( \gamma \left\langle \hat{S}_{E}\left(
X\right) \right\rangle \right) ^{2}\right) \left( 1+\frac{\Delta \hat{f}%
\left( X^{\prime },\theta -1\right) +\Delta \hat{r}\left( X^{\prime },\theta
-1\right) }{2}\right) }{2-\left( \gamma \left\langle \hat{S}_{E}\left(
X\right) \right\rangle \right) ^{2}+\frac{\left\langle \hat{w}\left(
X^{\prime },X\right) \right\rangle }{2}\left\langle w\left( X\right)
\right\rangle \Delta \left( \frac{f\left( X^{\prime },\theta -1\right)
+r\left( X^{\prime },\theta -1\right) }{2}\right) }\frac{\left\langle \hat{K}%
\right\rangle \left\Vert \hat{\Psi}\right\Vert ^{2}}{\hat{K}_{X^{\prime
}}\left\vert \hat{\Psi}\left( X^{\prime }\right) \right\vert ^{2}}  \notag
\end{eqnarray}%
where:%
\begin{equation}
\Delta \hat{f}\left( X^{\prime }\right) =\hat{f}\left( X^{\prime }\right)
-\left( \left\langle \hat{w}\left( X\right) \right\rangle \frac{\left\langle 
\hat{f}\left( X^{\prime }\right) \right\rangle _{\hat{w}_{E}}+\left\langle 
\hat{r}\left( X^{\prime }\right) \right\rangle _{\hat{w}_{L}}}{2}%
+\left\langle w\left( X\right) \right\rangle \frac{\left\langle f\left(
X\right) \right\rangle +\left\langle r\left( X\right) \right\rangle }{2}%
\right)
\end{equation}%
with:%
\begin{equation}
\frac{\Delta \hat{f}\left( X^{\prime }\right) +\Delta \hat{r}\left(
X^{\prime }\right) }{2}=\frac{\hat{f}\left( X^{\prime }\right) +\hat{r}%
\left( X^{\prime }\right) }{2}-\left\langle \hat{w}\left( X\right)
\right\rangle \frac{\left\langle \hat{f}\left( X^{\prime }\right)
\right\rangle _{\hat{w}_{E}}+\left\langle \hat{r}\left( X^{\prime }\right)
\right\rangle _{\hat{w}_{L}}}{2}-\left\langle w\left( X\right) \right\rangle 
\frac{\left\langle f\left( X\right) +r\left( X\right) \right\rangle _{w}}{2}
\end{equation}%
we obtain:%
\begin{equation*}
\delta \hat{S}_{E}\left( X,\theta -1\right) \simeq \frac{\delta \Delta \hat{f%
}\left( X,\theta -1\right) }{1+\Delta \hat{f}\left( X,\theta -1\right) }\hat{%
S}_{E}\left( X\right) -\frac{\frac{\partial \hat{K}_{X^{\prime }}\left\vert 
\hat{\Psi}\left( X^{\prime }\right) \right\vert ^{2}}{\partial \hat{f}\left(
X,\theta -1\right) }}{\hat{K}_{X^{\prime }}\left\vert \hat{\Psi}\left(
X^{\prime }\right) \right\vert ^{2}}\hat{S}_{E}\left( X\right) \delta \hat{f}%
\left( X,\theta -1\right)
\end{equation*}%
\begin{equation*}
\delta \hat{S}\left( X,\theta -1\right) \simeq \frac{1}{2}\frac{\delta
\left( \Delta \hat{f}\left( X^{\prime },\theta -1\right) +\Delta \hat{r}%
\left( X^{\prime }\right) \right) }{1+\frac{\Delta \hat{f}\left( X^{\prime
},\theta -1\right) +\Delta \hat{r}\left( X^{\prime }\right) }{2}}\hat{S}%
\left( X\right) -\frac{\frac{\partial \hat{K}_{X^{\prime }}\left\vert \hat{%
\Psi}\left( X^{\prime }\right) \right\vert ^{2}}{\partial \hat{f}\left(
X,\theta -1\right) }}{\hat{K}_{X^{\prime }}\left\vert \hat{\Psi}\left(
X^{\prime }\right) \right\vert ^{2}}\hat{S}\left( X\right) \delta \hat{f}%
\left( X,\theta -1\right)
\end{equation*}

For the shares taken in firms, given that:%
\begin{eqnarray}
&&S_{E}\left( X\right) \\
&=&\frac{w\left( X\right) }{2}\left( 1+\left( \hat{w}\left( X\right) \left(
f\left( X\right) -\frac{\left\langle \hat{f}\left( X^{\prime }\right)
\right\rangle _{\hat{w}_{E}}+\left\langle \hat{r}\left( X^{\prime }\right)
\right\rangle _{\hat{w}_{L}}}{2}\right) +\frac{w\left( X\right) }{2}\left(
f\left( X\right) -\bar{r}\left( X\right) \right) \right) \right) \frac{\hat{K%
}_{X}\left\vert \hat{\Psi}\left( X\right) \right\vert ^{2}}{K_{X}\left\vert
\Psi \left( X\right) \right\vert ^{2}}  \notag
\end{eqnarray}%
and:%
\begin{equation}
S\left( X\right) =w\left( X\right) \left( 1+\left( \hat{w}\left( X\right)
\left( \frac{f\left( X\right) +\bar{r}\left( X\right) }{2}-\frac{%
\left\langle \hat{f}\left( X^{\prime }\right) \right\rangle _{\hat{w}%
_{E}}+\left\langle \hat{r}\left( X^{\prime }\right) \right\rangle _{\hat{w}%
_{L}}}{2}\right) \right) \right) \frac{\hat{K}_{X}\left\vert \hat{\Psi}%
\left( X\right) \right\vert ^{2}}{K_{X}\left\vert \Psi \left( X\right)
\right\vert ^{2}}
\end{equation}%
with:%
\begin{equation*}
\hat{w}\left( X\right) \rightarrow \left\langle \hat{w}\left( X^{\prime
},X\right) \right\rangle
\end{equation*}%
\begin{equation*}
w\left( X\right) \rightarrow 1-\left\langle \hat{w}\left( X\right)
\right\rangle
\end{equation*}%
We find the following vatiations:%
\begin{eqnarray*}
&&\delta S_{E}\left( X,X,\theta -1\right) \\
&=&\frac{\frac{w\left( X\right) }{2}\left( \hat{w}\left( X\right) \delta
f\left( X\right) +\frac{w\left( X\right) }{2}\delta \left( f\left( X\right) -%
\bar{r}\left( X\right) \right) \right) }{1+\left( \hat{w}\left( X\right)
\left( f\left( X\right) -\frac{\left\langle \hat{f}\left( X^{\prime }\right)
\right\rangle _{\hat{w}_{E}}+\left\langle \hat{r}\left( X^{\prime }\right)
\right\rangle _{\hat{w}_{L}}}{2}\right) +\frac{w\left( X\right) }{2}\left(
f\left( X\right) -\bar{r}\left( X\right) \right) \right) }S_{E}\left(
X,X\right) \\
&\rightarrow &\frac{\frac{w\left( X\right) }{2}\left( \hat{w}\left( X\right)
\delta f\left( X,\theta -1\right) +\frac{w\left( X\right) }{2}\delta \left(
f\left( X,\theta -1\right) -\bar{r}\left( X\right) \right) \right) }{%
1+\left( \hat{w}\left( X\right) \left( f\left( X\right) -\frac{\left\langle 
\hat{f}\left( X^{\prime }\right) \right\rangle _{\hat{w}_{E}}+\left\langle 
\hat{r}\left( X^{\prime }\right) \right\rangle _{\hat{w}_{L}}}{2}\right) +%
\frac{w\left( X\right) }{2}\left( f\left( X\right) -\bar{r}\left( X\right)
\right) \right) }S_{E}\left( X,X\right) +\frac{\delta \frac{\hat{K}%
_{X}\left\vert \hat{\Psi}\left( X\right) \right\vert ^{2}}{K_{X}\left\vert
\Psi \left( X\right) \right\vert ^{2}}}{\frac{\hat{K}_{X}\left\vert \hat{\Psi%
}\left( X\right) \right\vert ^{2}}{K_{X}\left\vert \Psi \left( X\right)
\right\vert ^{2}}}S_{E}\left( X,X\right) \\
&\rightarrow &\frac{\frac{w\left( X\right) }{2}\left( \hat{w}\left( X\right)
\delta f\left( X,\theta -1\right) +\frac{w\left( X\right) }{2}\delta \left(
f\left( X,\theta -1\right) -\bar{r}\left( X\right) \right) \right) }{%
1+\left( \hat{w}\left( X\right) \left( f\left( X\right) -\frac{\left\langle 
\hat{f}\left( X^{\prime }\right) \right\rangle _{\hat{w}_{E}}+\left\langle 
\hat{r}\left( X^{\prime }\right) \right\rangle _{\hat{w}_{L}}}{2}\right) +%
\frac{w\left( X\right) }{2}\left( f\left( X\right) -\bar{r}\left( X\right)
\right) \right) }S_{E}\left( X,X\right) \\
&&+\frac{\partial _{\hat{f}\left( X\right) }\left( \frac{\hat{K}%
_{X}\left\vert \hat{\Psi}\left( X\right) \right\vert ^{2}}{K_{X}\left\vert
\Psi \left( X\right) \right\vert ^{2}}\right) S_{E}\left( X\right) }{\frac{%
\hat{K}_{X}\left\vert \hat{\Psi}\left( X\right) \right\vert ^{2}}{%
K_{X}\left\vert \Psi \left( X\right) \right\vert ^{2}}}\delta \hat{f}\left(
X,\theta -1\right) +\frac{\partial _{f\left( X\right) }\left( \frac{\hat{K}%
_{X}\left\vert \hat{\Psi}\left( X\right) \right\vert ^{2}}{K_{X}\left\vert
\Psi \left( X\right) \right\vert ^{2}}\right) S_{E}\left( X\right) }{\frac{%
\hat{K}_{X}\left\vert \hat{\Psi}\left( X\right) \right\vert ^{2}}{%
K_{X}\left\vert \Psi \left( X\right) \right\vert ^{2}}}\frac{\partial
f\left( X\right) }{\partial S\left( X,\theta -2\right) }\delta S\left(
X,\theta -2\right)
\end{eqnarray*}%
\begin{equation*}
\delta S\left( X,\theta -1\right) =\frac{\hat{w}\left( X\right) \delta \frac{%
f\left( X,\theta -1\right) +\bar{r}\left( X\right) }{2}}{1+\left( \hat{w}%
\left( X\right) \left( \frac{f\left( X\right) +\bar{r}\left( X\right) }{2}-%
\frac{\left\langle \hat{f}\left( X^{\prime }\right) \right\rangle _{\hat{w}%
_{E}}+\left\langle \hat{r}\left( X^{\prime }\right) \right\rangle _{\hat{w}%
_{L}}}{2}\right) \right) }S\left( X\right) +\frac{\delta \frac{\hat{K}%
_{X}\left\vert \hat{\Psi}\left( X\right) \right\vert ^{2}}{K_{X}\left\vert
\Psi \left( X\right) \right\vert ^{2}}}{\frac{\hat{K}_{X}\left\vert \hat{\Psi%
}\left( X\right) \right\vert ^{2}}{K_{X}\left\vert \Psi \left( X\right)
\right\vert ^{2}}}S\left( X\right)
\end{equation*}%
The derivative $\frac{\partial f\left( X\right) }{\partial S\left( X,\theta
-2\right) }$ is obtaind straightforwardly:%
\begin{equation*}
\frac{\partial f\left( X\right) }{\partial S\left( X,\theta -2\right) }%
=-\left( r\left( 1-S\left( X,\theta -2\right) \right) ^{r-1}\left(
f_{1}\left( X\right) +\Delta F_{\tau }\left( \bar{R}\left( K,X\right)
\right) \right) -C\right)
\end{equation*}%
so that $\delta f\left( X,\theta -1\right) $ writes: 
\begin{equation*}
\delta f\left( X,\theta -1\right) =-\left( r\left( 1-S\left( X,\theta
-2\right) \right) ^{r-1}\left( f_{1}\left( X\right) +\Delta F_{\tau }\left( 
\bar{R}\left( K,X\right) \right) \right) -C\right) \delta S\left( X,\theta
-2\right)
\end{equation*}%
This expression allow to rewrite the variatns of shares as depending on $%
\delta S\left( X,\theta -2\right) $: 
\begin{eqnarray*}
\delta S\left( X,\theta -1\right) &=&\frac{\frac{\hat{w}\left( X\right) }{2}%
\frac{\partial f\left( X\right) }{\partial S\left( X,\theta -2\right) }%
\delta S\left( X,\theta -2\right) }{1+\left( \hat{w}\left( X\right) \left( 
\frac{f\left( X\right) +\bar{r}\left( X\right) }{2}-\frac{\left\langle \hat{f%
}\left( X^{\prime }\right) \right\rangle _{\hat{w}_{E}}+\left\langle \hat{r}%
\left( X^{\prime }\right) \right\rangle _{\hat{w}_{L}}}{2}\right) \right) }%
S\left( X\right) +\frac{\delta \frac{\hat{K}_{X}\left\vert \hat{\Psi}\left(
X\right) \right\vert ^{2}}{K_{X}\left\vert \Psi \left( X\right) \right\vert
^{2}}}{\frac{\hat{K}_{X}\left\vert \hat{\Psi}\left( X\right) \right\vert ^{2}%
}{K_{X}\left\vert \Psi \left( X\right) \right\vert ^{2}}}S\left( X\right) \\
&=&-\frac{\frac{\hat{w}\left( X\right) }{2}\left( r\left( 1-S\left( X,\theta
-2\right) \right) ^{r-1}\left( f_{1}\left( X\right) +\Delta F_{\tau }\left( 
\bar{R}\left( K,X\right) \right) \right) -C\right) S\left( X\right) }{%
1+\left( \hat{w}\left( X\right) \left( \frac{f\left( X\right) +\bar{r}\left(
X\right) }{2}-\frac{\left\langle \hat{f}\left( X^{\prime }\right)
\right\rangle _{\hat{w}_{E}}+\left\langle \hat{r}\left( X^{\prime }\right)
\right\rangle _{\hat{w}_{L}}}{2}\right) \right) }\delta S\left( X,\theta
-2\right) \\
&&+\frac{\partial _{\hat{f}\left( X\right) }\left( \frac{\hat{K}%
_{X}\left\vert \hat{\Psi}\left( X\right) \right\vert ^{2}}{K_{X}\left\vert
\Psi \left( X\right) \right\vert ^{2}}\right) S\left( X\right) }{\frac{\hat{K%
}_{X}\left\vert \hat{\Psi}\left( X\right) \right\vert ^{2}}{K_{X}\left\vert
\Psi \left( X\right) \right\vert ^{2}}}\delta \hat{f}\left( X,\theta
-1\right) +\frac{\partial _{f\left( X\right) }\left( \frac{\hat{K}%
_{X}\left\vert \hat{\Psi}\left( X\right) \right\vert ^{2}}{K_{X}\left\vert
\Psi \left( X\right) \right\vert ^{2}}\right) S\left( X\right) }{\frac{\hat{K%
}_{X}\left\vert \hat{\Psi}\left( X\right) \right\vert ^{2}}{K_{X}\left\vert
\Psi \left( X\right) \right\vert ^{2}}}\frac{\partial f\left( X\right) }{%
\partial S\left( X,\theta -2\right) }\delta S\left( X,\theta -2\right) \\
&=&-\left( \frac{\frac{\hat{w}\left( X\right) }{2}S\left( X\right) }{%
1+\left( \hat{w}\left( X\right) \left( \frac{f\left( X\right) +\bar{r}\left(
X\right) }{2}-\frac{\left\langle \hat{f}\left( X^{\prime }\right)
\right\rangle _{\hat{w}_{E}}+\left\langle \hat{r}\left( X^{\prime }\right)
\right\rangle _{\hat{w}_{L}}}{2}\right) \right) }+\frac{\partial _{f\left(
X\right) }\left( \frac{\hat{K}_{X}\left\vert \hat{\Psi}\left( X\right)
\right\vert ^{2}}{K_{X}\left\vert \Psi \left( X\right) \right\vert ^{2}}%
\right) S\left( X\right) }{\frac{\hat{K}_{X}\left\vert \hat{\Psi}\left(
X\right) \right\vert ^{2}}{K_{X}\left\vert \Psi \left( X\right) \right\vert
^{2}}}\right) \\
&&\times \left( r\left( 1-S\left( X,\theta -2\right) \right) ^{r-1}\left(
f_{1}\left( X\right) +\Delta F_{\tau }\left( \bar{R}\left( K,X\right)
\right) \right) -C\right) \delta S\left( X,\theta -2\right) \\
&&+\frac{\partial _{\hat{f}\left( X\right) }\left( \frac{\hat{K}%
_{X}\left\vert \hat{\Psi}\left( X\right) \right\vert ^{2}}{K_{X}\left\vert
\Psi \left( X\right) \right\vert ^{2}}\right) S\left( X\right) }{\frac{\hat{K%
}_{X}\left\vert \hat{\Psi}\left( X\right) \right\vert ^{2}}{K_{X}\left\vert
\Psi \left( X\right) \right\vert ^{2}}}\delta \hat{f}\left( X,\theta
-1\right)
\end{eqnarray*}%
\begin{eqnarray*}
&&\delta S_{E}\left( X,X,\theta -1\right) \\
&\rightarrow &\frac{\frac{w\left( X\right) }{4}\left( 1+\hat{w}\left(
X\right) \right) \delta f\left( X,\theta -1\right) }{1+\left( \hat{w}\left(
X\right) \left( f\left( X\right) -\frac{\left\langle \hat{f}\left( X^{\prime
}\right) \right\rangle _{\hat{w}_{E}}+\left\langle \hat{r}\left( X^{\prime
}\right) \right\rangle _{\hat{w}_{L}}}{2}\right) +\frac{w\left( X\right) }{2}%
\left( f\left( X\right) -\bar{r}\left( X\right) \right) \right) }S_{E}\left(
X\right) \\
&=&-\frac{\frac{w\left( X\right) }{4}\left( 1+\hat{w}\left( X\right) \right)
S_{E}\left( X\right) \left( r\left( 1-S\left( X,\theta -2\right) \right)
^{r-1}\left( f_{1}\left( X\right) +\Delta F_{\tau }\left( \bar{R}\left(
K,X\right) \right) \right) -C\right) \delta S\left( X,\theta -2\right) }{%
1+\left( \hat{w}\left( X\right) \left( f\left( X\right) -\frac{\left\langle 
\hat{f}\left( X^{\prime }\right) \right\rangle _{\hat{w}_{E}}+\left\langle 
\hat{r}\left( X^{\prime }\right) \right\rangle _{\hat{w}_{L}}}{2}\right) +%
\frac{w\left( X\right) }{2}\left( f\left( X\right) -\bar{r}\left( X\right)
\right) \right) }
\end{eqnarray*}

We also have to consider the variation of partial average:%
\begin{equation*}
\left\langle \hat{S}_{E}\left( X^{\prime },X\right) \right\rangle
_{X^{\prime }}\simeq \frac{\left\langle \hat{w}\left( X^{\prime },X\right)
\right\rangle }{2}\left( 1-\left\langle w\left( X\right) \right\rangle
\left( \frac{f\left( X\right) +r\left( X\right) }{2}-\frac{\left\langle \hat{%
f}\left( X^{\prime }\right) \right\rangle +\left\langle \hat{r}\left(
X^{\prime }\right) \right\rangle _{\hat{w}_{L}}}{2}\right) \right)
\end{equation*}%
which is given by:%
\begin{eqnarray*}
&&\delta \left\langle \hat{S}_{E}\left( X^{\prime },X,\theta -1\right)
\right\rangle _{X^{\prime }} \\
&\simeq &-\frac{\left\langle \hat{w}\left( X^{\prime },X\right)
\right\rangle \left\langle w\left( X\right) \right\rangle }{2}\delta \frac{%
f\left( X,\theta -1\right) +r\left( X,\theta -1\right) }{2} \\
&\rightarrow &\frac{\left\langle \hat{w}\left( X^{\prime },X\right)
\right\rangle \left\langle w\left( X\right) \right\rangle }{4}\left( r\left(
1-S\left( X,\theta -2\right) \right) ^{r-1}\left( f_{1}\left( X\right)
+\Delta F_{\tau }\left( \bar{R}\left( K,X\right) \right) \right) -C\right)
\delta S\left( X,\theta -2\right)
\end{eqnarray*}%
along with:%
\begin{eqnarray*}
&&\frac{1-\left( \hat{S}\left( X,\theta \right) +\delta \hat{S}\left(
X,\theta \right) \right) }{1-\left( \hat{S}_{E}\left( X^{\prime },\theta
\right) +\delta \hat{S}_{E}\left( X,\theta \right) \right) } \\
&=&-\frac{\hat{S}\left( X\right) }{2\left( 1-\left( \hat{S}_{E}\left(
X^{\prime },\theta \right) \right) \right) }\frac{\delta \left( \Delta \hat{f%
}\left( X^{\prime }\right) +\Delta \hat{r}\left( X^{\prime }\right) \right) 
}{1+\frac{\Delta \hat{f}\left( X^{\prime }\right) +\Delta \hat{r}\left(
X^{\prime }\right) }{2}}+\frac{\left( 1-\left( \hat{S}\left( X,\theta
\right) \right) \right) \hat{S}_{E}\left( X\right) }{\left( 1-\left( \hat{S}%
_{E}\left( X^{\prime },\theta \right) \right) \right) ^{2}}\frac{\delta
\Delta \hat{f}\left( X\right) }{1+\Delta \hat{f}\left( X\right) }
\end{eqnarray*}%
Gathering contributions, the variation $\delta \hat{f}\left( X,\theta
\right) $ satisfies:

\begin{eqnarray}
&&\frac{1-\left( \hat{S}\left( X\right) \right) }{1-\left( \hat{S}_{E}\left(
X\right) \right) }\delta \hat{f}\left( X,\theta \right)  \label{Tr} \\
&=&\left( \frac{\hat{S}\left( X\right) }{\left( 1-\left( \hat{S}_{E}\left(
X^{\prime }\right) \right) \right) }\left( \frac{1}{2\left( 1+\frac{\Delta 
\hat{f}\left( X^{\prime }\right) +\Delta \hat{r}\left( X^{\prime }\right) }{2%
}\right) }-\frac{\frac{\partial \hat{K}_{X^{\prime }}\left\vert \hat{\Psi}%
\left( X^{\prime }\right) \right\vert ^{2}}{\partial \hat{f}\left( X,\theta
-1\right) }}{\hat{K}_{X^{\prime }}\left\vert \hat{\Psi}\left( X^{\prime
}\right) \right\vert ^{2}}\right) \right.  \notag \\
&&\left. -\frac{\left( 1-\left( \hat{S}\left( X\right) \right) \right) \hat{S%
}_{E}\left( X\right) }{\left( 1-\left( \hat{S}_{E}\left( X^{\prime }\right)
\right) \right) ^{2}}\left( \frac{1}{1+\Delta \hat{f}\left( X\right) }-\frac{%
\frac{\partial \hat{K}_{X^{\prime }}\left\vert \hat{\Psi}\left( X^{\prime
}\right) \right\vert ^{2}}{\partial \hat{f}\left( X,\theta -1\right) }}{\hat{%
K}_{X^{\prime }}\left\vert \hat{\Psi}\left( X^{\prime }\right) \right\vert
^{2}}\right) \right) \left( \hat{f}\left( X\right) -\bar{r}\right) \delta 
\hat{f}\left( X,\theta -1\right)  \notag \\
&&-\frac{\left( 1-\left\langle \hat{S}\left( X^{\prime }\right)
\right\rangle \right) \left( \left\langle \hat{f}\left( X^{\prime }\right)
\right\rangle -\bar{r}\right) }{1-\left\langle \hat{S}_{E}\left( X^{\prime
}\right) \right\rangle }\frac{\left\langle \hat{w}\left( X^{\prime
},X\right) \right\rangle \left\langle w\left( X\right) \right\rangle }{4}%
\frac{\partial f\left( X\right) }{\partial S\left( X,\theta -1\right) }%
\delta S\left( X,\theta -2\right)  \notag \\
&&+\frac{\frac{w\left( X\right) }{4}\left( 1+\hat{w}\left( X\right) \right)
S_{E}\left( X\right) \frac{\partial f\left( X\right) }{\partial S\left(
X,\theta -1\right) }\left( f\left( X\right) -\bar{r}\right) }{1+\left( \hat{w%
}\left( X\right) \left( f\left( X\right) -\frac{\left\langle \hat{f}\left(
X^{\prime }\right) \right\rangle _{\hat{w}_{E}}+\left\langle \hat{r}\left(
X^{\prime }\right) \right\rangle _{\hat{w}_{L}}}{2}\right) +\frac{w\left(
X\right) }{2}\left( f\left( X\right) -\bar{r}\left( X\right) \right) \right) 
}\delta S\left( X,\theta -2\right)  \notag \\
&&+S_{E}\left( X,X,\theta -1\right) \frac{\partial f\left( X\right) }{%
\partial S\left( X,\theta -1\right) }\delta S\left( X,\theta -1\right) 
\notag
\end{eqnarray}%
with:%
\begin{equation*}
\frac{\partial f\left( X\right) }{\partial S\left( X,\theta -1\right) }%
=-\left( r\left( 1-S\left( X,\theta -1\right) \right) ^{r-1}\left(
f_{1}\left( X\right) +\Delta F_{\tau }\left( \bar{R}\left( K,X\right)
\right) \right) -C\right)
\end{equation*}%
The variation $\delta S\left( X,\theta -1\right) $ is:%
\begin{eqnarray}
&&\delta S\left( X,\theta -1\right)  \label{Vs} \\
&=&\left( \frac{\frac{\hat{w}\left( X\right) }{2}S\left( X\right) }{1+\left( 
\hat{w}\left( X\right) \left( \frac{f\left( X\right) +\bar{r}\left( X\right) 
}{2}-\frac{\left\langle \hat{f}\left( X^{\prime }\right) \right\rangle _{%
\hat{w}_{E}}+\left\langle \hat{r}\left( X^{\prime }\right) \right\rangle _{%
\hat{w}_{L}}}{2}\right) \right) }+\frac{\partial _{f\left( X\right) }\left( 
\frac{\hat{K}_{X}\left\vert \hat{\Psi}\left( X\right) \right\vert ^{2}}{%
K_{X}\left\vert \Psi \left( X\right) \right\vert ^{2}}\right) S\left(
X\right) }{\frac{\hat{K}_{X}\left\vert \hat{\Psi}\left( X\right) \right\vert
^{2}}{K_{X}\left\vert \Psi \left( X\right) \right\vert ^{2}}}\right) \frac{%
\partial f\left( X\right) }{\partial S\left( X,\theta -2\right) }\delta
S\left( X,\theta -2\right)  \notag \\
&&+\frac{\partial _{\hat{f}\left( X\right) }\left( \frac{\hat{K}%
_{X}\left\vert \hat{\Psi}\left( X\right) \right\vert ^{2}}{K_{X}\left\vert
\Psi \left( X\right) \right\vert ^{2}}\right) S\left( X\right) }{\frac{\hat{K%
}_{X}\left\vert \hat{\Psi}\left( X\right) \right\vert ^{2}}{K_{X}\left\vert
\Psi \left( X\right) \right\vert ^{2}}}\delta \hat{f}\left( X,\theta
-1\right)  \notag
\end{eqnarray}%
Inserting this result in (\ref{Tr}) leads to:%
\begin{eqnarray*}
&&\delta \hat{f}\left( X,\theta \right) -\delta \hat{f}\left( X,\theta
-1\right) \\
&=&\left\{ \left( \frac{\hat{S}\left( X\right) }{\left( 1-\left( \hat{S}%
\left( X\right) \right) \right) }\left( \frac{1}{2\left( 1+\frac{\Delta \hat{%
f}\left( X^{\prime }\right) +\Delta \hat{r}\left( X^{\prime }\right) }{2}%
\right) }-\frac{\frac{\partial \hat{K}_{X^{\prime }}\left\vert \hat{\Psi}%
\left( X^{\prime }\right) \right\vert ^{2}}{\partial \hat{f}\left( X,\theta
-1\right) }}{\hat{K}_{X^{\prime }}\left\vert \hat{\Psi}\left( X^{\prime
}\right) \right\vert ^{2}}\right) \right. \right. \\
&&\left. -\frac{\hat{S}_{E}\left( X\right) }{\left( 1-\left( \hat{S}%
_{E}\left( X^{\prime }\right) \right) \right) }\left( \frac{1}{1+\Delta \hat{%
f}\left( X\right) }-\frac{\frac{\partial \hat{K}_{X^{\prime }}\left\vert 
\hat{\Psi}\left( X^{\prime }\right) \right\vert ^{2}}{\partial \hat{f}\left(
X,\theta -1\right) }}{\hat{K}_{X^{\prime }}\left\vert \hat{\Psi}\left(
X^{\prime }\right) \right\vert ^{2}}\right) \right) \left( \hat{f}\left(
X\right) -\bar{r}\right) \\
&&\left. +\frac{1-\left( \hat{S}_{E}\left( X\right) \right) }{1-\left( \hat{S%
}\left( X\right) \right) }S_{E}\left( X,X,\theta -1\right) \frac{\partial
f\left( X\right) }{\partial S\left( X,\theta -1\right) }\frac{\partial _{%
\hat{f}\left( X\right) }\left( \frac{\hat{K}_{X}\left\vert \hat{\Psi}\left(
X\right) \right\vert ^{2}}{K_{X}\left\vert \Psi \left( X\right) \right\vert
^{2}}\right) }{\frac{\hat{K}_{X}\left\vert \hat{\Psi}\left( X\right)
\right\vert ^{2}}{K_{X}\left\vert \Psi \left( X\right) \right\vert ^{2}}}%
-1\right\} \delta \hat{f}\left( X,\theta -1\right) \\
&&+\left\{ -\frac{\left( 1-\left\langle \hat{S}\left( X^{\prime }\right)
\right\rangle \right) \left( \left\langle \hat{f}\left( X^{\prime }\right)
\right\rangle -\bar{r}\right) }{1-\left\langle \hat{S}_{E}\left( X^{\prime
}\right) \right\rangle }\frac{\left\langle \hat{w}\left( X^{\prime
},X\right) \right\rangle \left\langle w\left( X\right) \right\rangle }{4}%
\right. \\
&&+\frac{\frac{w\left( X\right) }{4}\left( 1+\hat{w}\left( X\right) \right)
S_{E}\left( X\right) \left( f\left( X\right) -\bar{r}\right) }{1+\left( \hat{%
w}\left( X\right) \left( f\left( X\right) -\frac{\left\langle \hat{f}\left(
X^{\prime }\right) \right\rangle _{\hat{w}_{E}}+\left\langle \hat{r}\left(
X^{\prime }\right) \right\rangle _{\hat{w}_{L}}}{2}\right) +\frac{w\left(
X\right) }{2}\left( f\left( X\right) -\bar{r}\left( X\right) \right) \right) 
}+S_{E}\left( X,X,\theta -1\right) \\
&&\left. \times \left( \frac{\frac{\hat{w}\left( X\right) }{2}S\left(
X\right) }{1+\left( \hat{w}\left( X\right) \left( \frac{f\left( X\right) +%
\bar{r}\left( X\right) }{2}-\frac{\left\langle \hat{f}\left( X^{\prime
}\right) \right\rangle _{\hat{w}_{E}}+\left\langle \hat{r}\left( X^{\prime
}\right) \right\rangle _{\hat{w}_{L}}}{2}\right) \right) }+\frac{\partial
_{f\left( X\right) }\left( \frac{\hat{K}_{X}\left\vert \hat{\Psi}\left(
X\right) \right\vert ^{2}}{K_{X}\left\vert \Psi \left( X\right) \right\vert
^{2}}\right) S\left( X\right) }{\frac{\hat{K}_{X}\left\vert \hat{\Psi}\left(
X\right) \right\vert ^{2}}{K_{X}\left\vert \Psi \left( X\right) \right\vert
^{2}}}\right) \frac{\partial f\left( X\right) }{\partial S\left( X,\theta
-2\right) }\right\} \\
&&\times \frac{1-\left( \hat{S}_{E}\left( X\right) \right) }{1-\left( \hat{S}%
\left( X\right) \right) }\frac{\partial f\left( X\right) }{\partial S\left(
X,\theta -1\right) }\delta S\left( X,\theta -2\right)
\end{eqnarray*}%
and (\ref{Vs}) becomes:%
\begin{eqnarray*}
&&\delta S\left( X,\theta -1\right) -\delta S\left( X,\theta -2\right) \\
&=&\left( \frac{\frac{\hat{w}\left( X\right) }{2}\frac{\partial f\left(
X\right) }{\partial S\left( X,\theta -1\right) }S\left( X\right) }{1+\left( 
\hat{w}\left( X\right) \left( \frac{f\left( X\right) +\bar{r}\left( X\right) 
}{2}-\frac{\left\langle \hat{f}\left( X^{\prime }\right) \right\rangle _{%
\hat{w}_{E}}+\left\langle \hat{r}\left( X^{\prime }\right) \right\rangle _{%
\hat{w}_{L}}}{2}\right) \right) }-1\right) \delta S\left( X,\theta -2\right)
\\
&&+\frac{\left[ \partial _{\hat{f}\left( X\right) }\left( \frac{\hat{K}%
_{X}\left\vert \hat{\Psi}\left( X\right) \right\vert ^{2}}{K_{X}\left\vert
\Psi \left( X\right) \right\vert ^{2}}\right) \right] S\left( X\right)
\delta \hat{f}\left( X,\theta -1\right) }{\frac{\hat{K}_{X}\left\vert \hat{%
\Psi}\left( X\right) \right\vert ^{2}}{K_{X}\left\vert \Psi \left( X\right)
\right\vert ^{2}}} \\
&=&\left( \left( \frac{\frac{\hat{w}\left( X\right) }{2}S\left( X\right) }{%
1+\left( \hat{w}\left( X\right) \left( \frac{f\left( X\right) +\bar{r}\left(
X\right) }{2}-\frac{\left\langle \hat{f}\left( X^{\prime }\right)
\right\rangle _{\hat{w}_{E}}+\left\langle \hat{r}\left( X^{\prime }\right)
\right\rangle _{\hat{w}_{L}}}{2}\right) \right) }\right. \right. \\
&&\left. \left. +\frac{\partial _{f\left( X\right) }\left( \frac{\hat{K}%
_{X}\left\vert \hat{\Psi}\left( X\right) \right\vert ^{2}}{K_{X}\left\vert
\Psi \left( X\right) \right\vert ^{2}}\right) S\left( X\right) }{\frac{\hat{K%
}_{X}\left\vert \hat{\Psi}\left( X\right) \right\vert ^{2}}{K_{X}\left\vert
\Psi \left( X\right) \right\vert ^{2}}}\right) \right) \frac{\partial
f\left( X\right) }{\partial S\left( X,\theta -2\right) }-1\delta S\left(
X,\theta -2\right) +\frac{\partial _{\hat{f}\left( X\right) }\left( \frac{%
\hat{K}_{X}\left\vert \hat{\Psi}\left( X\right) \right\vert ^{2}}{%
K_{X}\left\vert \Psi \left( X\right) \right\vert ^{2}}\right) S\left(
X\right) }{\frac{\hat{K}_{X}\left\vert \hat{\Psi}\left( X\right) \right\vert
^{2}}{K_{X}\left\vert \Psi \left( X\right) \right\vert ^{2}}}\delta \hat{f}%
\left( X,\theta -1\right)
\end{eqnarray*}%
correspndng to the system described in the text where the coefficients are:%
\begin{eqnarray}
\alpha &=&\left( \frac{\hat{S}\left( X\right) }{\left( 1-\left( \hat{S}%
\left( X\right) \right) \right) }\left( \frac{1}{2\left( 1+\frac{\Delta \hat{%
f}\left( X^{\prime }\right) +\Delta \hat{r}\left( X^{\prime }\right) }{2}%
\right) }-\frac{\frac{\partial \hat{K}_{X^{\prime }}\left\vert \hat{\Psi}%
\left( X^{\prime }\right) \right\vert ^{2}}{\partial \hat{f}\left( X,\theta
-1\right) }}{\hat{K}_{X^{\prime }}\left\vert \hat{\Psi}\left( X^{\prime
}\right) \right\vert ^{2}}\right) \right.  \label{LP} \\
&&\left. -\frac{\hat{S}_{E}\left( X\right) }{\left( 1-\left( \hat{S}%
_{E}\left( X^{\prime }\right) \right) \right) }\left( \frac{1}{1+\Delta \hat{%
f}\left( X\right) }-\frac{\frac{\partial \hat{K}_{X^{\prime }}\left\vert 
\hat{\Psi}\left( X^{\prime }\right) \right\vert ^{2}}{\partial \hat{f}\left(
X,\theta -1\right) }}{\hat{K}_{X^{\prime }}\left\vert \hat{\Psi}\left(
X^{\prime }\right) \right\vert ^{2}}\right) \right) \left( \hat{f}\left(
X\right) -\bar{r}\right)  \notag \\
&&+\frac{1-\left( \hat{S}_{E}\left( X\right) \right) }{1-\left( \hat{S}%
\left( X\right) \right) }S_{E}\left( X,X,\theta -1\right) \frac{\partial
f\left( X\right) }{\partial S\left( X,\theta -1\right) }\frac{\partial _{%
\hat{f}\left( X\right) }\left( \frac{\hat{K}_{X}\left\vert \hat{\Psi}\left(
X\right) \right\vert ^{2}}{K_{X}\left\vert \Psi \left( X\right) \right\vert
^{2}}\right) S\left( X\right) }{\frac{\hat{K}_{X}\left\vert \hat{\Psi}\left(
X\right) \right\vert ^{2}}{K_{X}\left\vert \Psi \left( X\right) \right\vert
^{2}}}-1  \notag
\end{eqnarray}%
\begin{equation*}
\beta =\left( \frac{\frac{\hat{w}\left( X\right) }{2}S\left( X\right) }{%
1+\left( \hat{w}\left( X\right) \left( \frac{f\left( X\right) +\bar{r}\left(
X\right) }{2}-\frac{\left\langle \hat{f}\left( X^{\prime }\right)
\right\rangle _{\hat{w}_{E}}+\left\langle \hat{r}\left( X^{\prime }\right)
\right\rangle _{\hat{w}_{L}}}{2}\right) \right) }+\frac{\partial _{f\left(
X\right) }\left( \frac{\hat{K}_{X}\left\vert \hat{\Psi}\left( X\right)
\right\vert ^{2}}{K_{X}\left\vert \Psi \left( X\right) \right\vert ^{2}}%
\right) S\left( X\right) }{\frac{\hat{K}_{X}\left\vert \hat{\Psi}\left(
X\right) \right\vert ^{2}}{K_{X}\left\vert \Psi \left( X\right) \right\vert
^{2}}}\right) \frac{\partial f\left( X\right) }{\partial S\left( X,\theta
-2\right) }-1
\end{equation*}%
\begin{eqnarray*}
&&c=\left\{ -\frac{\left( 1-\left\langle \hat{S}\left( X^{\prime }\right)
\right\rangle \right) \left( \left\langle \hat{f}\left( X^{\prime }\right)
\right\rangle -\bar{r}\right) }{1-\left\langle \hat{S}_{E}\left( X^{\prime
}\right) \right\rangle }\frac{\left\langle \hat{w}\left( X^{\prime
},X\right) \right\rangle \left\langle w\left( X\right) \right\rangle }{4}%
\right. \\
&&+\frac{\frac{w\left( X\right) }{4}\left( 1+\hat{w}\left( X\right) \right)
S_{E}\left( X\right) \left( f\left( X\right) -\bar{r}\right) }{1+\left( \hat{%
w}\left( X\right) \left( f\left( X\right) -\frac{\left\langle \hat{f}\left(
X^{\prime }\right) \right\rangle _{\hat{w}_{E}}+\left\langle \hat{r}\left(
X^{\prime }\right) \right\rangle _{\hat{w}_{L}}}{2}\right) +\frac{w\left(
X\right) }{2}\left( f\left( X\right) -\bar{r}\left( X\right) \right) \right) 
} \\
&&\left. +S_{E}\left( X\right) \left( \frac{\frac{\hat{w}\left( X\right) }{2}%
S\left( X\right) }{1+\left( \hat{w}\left( X\right) \left( \frac{f\left(
X\right) +\bar{r}\left( X\right) }{2}-\frac{\left\langle \hat{f}\left(
X^{\prime }\right) \right\rangle _{\hat{w}_{E}}+\left\langle \hat{r}\left(
X^{\prime }\right) \right\rangle _{\hat{w}_{L}}}{2}\right) \right) }+\frac{%
\partial _{f\left( X\right) }\left( \frac{\hat{K}_{X}\left\vert \hat{\Psi}%
\left( X\right) \right\vert ^{2}}{K_{X}\left\vert \Psi \left( X\right)
\right\vert ^{2}}\right) S\left( X\right) }{\frac{\hat{K}_{X}\left\vert \hat{%
\Psi}\left( X\right) \right\vert ^{2}}{K_{X}\left\vert \Psi \left( X\right)
\right\vert ^{2}}}\right) \frac{\partial f\left( X\right) }{\partial S\left(
X,\theta -2\right) }\right\} \\
&&\times \frac{\partial f\left( X\right) }{\partial S\left( X,\theta
-1\right) }
\end{eqnarray*}%
\begin{equation*}
h=\frac{\partial _{\hat{f}\left( X\right) }\left( \frac{\hat{K}%
_{X}\left\vert \hat{\Psi}\left( X\right) \right\vert ^{2}}{K_{X}\left\vert
\Psi \left( X\right) \right\vert ^{2}}\right) S\left( X\right) }{\frac{\hat{K%
}_{X}\left\vert \hat{\Psi}\left( X\right) \right\vert ^{2}}{K_{X}\left\vert
\Psi \left( X\right) \right\vert ^{2}}}
\end{equation*}

\section*{Appendix 10 Eigenvalues and stability of the system}

The eigenvalues of the system determine the stability of the group with
respect to perturbations. The eigenvalues and eigenvectors of the system are
given by:%
\begin{eqnarray*}
\frac{1}{2}\left( \alpha +\beta \right) -\frac{1}{2}\sqrt{\left( \alpha
-\beta \right) ^{2}+4ch} &\rightarrow &\left( 
\begin{array}{c}
-\frac{2c}{\sqrt{4c^{2}+\left( \alpha -\beta +\sqrt{\left( \alpha -\beta
\right) ^{2}+4ch}\right) ^{2}}} \\ 
\frac{\alpha -\beta +\sqrt{\left( \alpha -\beta \right) ^{2}+4ch}}{\sqrt{%
4c^{2}+\left( \alpha -\beta +\sqrt{\left( \alpha -\beta \right) ^{2}+4ch}%
\right) ^{2}}}%
\end{array}%
\right) \\
\frac{1}{2}\left( \alpha +\beta \right) +\frac{1}{2}\sqrt{\left( \alpha
-\beta \right) ^{2}+4ch} &\rightarrow &\left( 
\begin{array}{c}
-\frac{2c}{\sqrt{4c^{2}+\left( \alpha -\beta -\sqrt{\left( \alpha -\beta
\right) ^{2}+4ch}\right) ^{2}}} \\ 
\frac{\alpha -\beta -\sqrt{\left( \alpha -\beta \right) ^{2}+4ch}}{\sqrt{%
4c^{2}+\left( \alpha -\beta -\sqrt{\left( \alpha -\beta \right) ^{2}+4ch}%
\right) ^{2}}}%
\end{array}%
\right)
\end{eqnarray*}%
In general the largest eigenvalue is:$\frac{1}{2}\alpha +\frac{1}{2}\beta +%
\frac{1}{2}\sqrt{\alpha ^{2}-2\alpha \beta +\beta ^{2}+4ch}$. The sign of
this value will depend on the signs of the matrix elements. This section
studies the conditions for this eigenvalue to be positive or negative. To do
so we note first that:%
\begin{equation*}
h=\frac{\partial _{\hat{f}\left( X\right) }\left( \frac{\hat{K}%
_{X}\left\vert \hat{\Psi}\left( X\right) \right\vert ^{2}}{K_{X}\left\vert
\Psi \left( X\right) \right\vert ^{2}}\right) S\left( X\right) }{\frac{\hat{K%
}_{X}\left\vert \hat{\Psi}\left( X\right) \right\vert ^{2}}{K_{X}\left\vert
\Psi \left( X\right) \right\vert ^{2}}}<0
\end{equation*}%
and:%
\begin{equation*}
\frac{\partial f\left( X\right) }{\partial S\left( X,\theta -2\right) }<0
\end{equation*}%
so that $h<0$ and $\beta <0$.

The largest eigenvalue:%
\begin{equation*}
\frac{1}{2}\alpha +\frac{1}{2}\beta +\frac{1}{2}\sqrt{\alpha ^{2}-2\alpha
\beta +\beta ^{2}+4ch}>0
\end{equation*}%
is positive in two cases. The first case arises when:%
\begin{equation*}
\alpha +\beta >0
\end{equation*}%
and the second case when:%
\begin{equation*}
\alpha +\beta <0
\end{equation*}%
and;%
\begin{equation*}
\alpha \beta <ch
\end{equation*}%
We consider two cases.

\subsection*{A10.1 First case: $\hat{f}\left( X\right) -\bar{r}>>0$, $%
f\left( X\right) -\bar{r}>>0$}

For $f\left( X\right) -\bar{r}>>0$, $\hat{f}\left( X\right) -\bar{r}>>0$

\begin{eqnarray*}
\frac{\partial f\left( X\right) }{\partial S\left( X,\theta -1\right) }
&=&-\left( r\left( 1-S\left( X,\theta -1\right) \right) ^{r-1}\left(
f_{1}\left( X\right) +\Delta F_{\tau }\left( \bar{R}\left( K,X\right)
\right) \right) -C\right) \\
&=&-r\left( 1-S\left( X,\theta -1\right) \right) ^{-1}f\left( X\right)
\end{eqnarray*}%
for $S\left( X,\theta -1\right) \rightarrow 1$%
\begin{equation*}
\left\vert \frac{\partial f\left( X\right) }{\partial S\left( X,\theta
-1\right) }\right\vert >>f\left( X\right)
\end{equation*}

\subsubsection*{Estimation of the coefficients}

In (\ref{LP}) the term proportional to $\left( \hat{f}\left( X\right) -\bar{r%
}\right) $ is estimated by using $\hat{f}\left( X\right) -\bar{r}>>0$, $\hat{%
S}_{E}\left( X\right) \rightarrow \hat{S}\left( X\right) $ writing: 
\begin{eqnarray}
\alpha &=&\left( \frac{\hat{S}\left( X\right) }{\left( 1-\left( \hat{S}%
\left( X\right) \right) \right) }\left( \frac{1}{2\left( 1+\frac{\Delta \hat{%
f}\left( X^{\prime }\right) +\Delta \hat{r}\left( X^{\prime }\right) }{2}%
\right) }-\frac{\frac{\partial \hat{K}_{X^{\prime }}\left\vert \hat{\Psi}%
\left( X^{\prime }\right) \right\vert ^{2}}{\partial \hat{f}\left( X,\theta
-1\right) }}{\hat{K}_{X^{\prime }}\left\vert \hat{\Psi}\left( X^{\prime
}\right) \right\vert ^{2}}\right) \right. \\
&&\left. -\frac{\hat{S}_{E}\left( X\right) }{\left( 1-\left( \hat{S}%
_{E}\left( X^{\prime }\right) \right) \right) }\left( \frac{1}{1+\Delta \hat{%
f}\left( X\right) }-\frac{\frac{\partial \hat{K}_{X^{\prime }}\left\vert 
\hat{\Psi}\left( X^{\prime }\right) \right\vert ^{2}}{\partial \hat{f}\left(
X,\theta -1\right) }}{\hat{K}_{X^{\prime }}\left\vert \hat{\Psi}\left(
X^{\prime }\right) \right\vert ^{2}}\right) \right) \left( \hat{f}\left(
X\right) -\bar{r}\right)  \notag \\
&&+\frac{1-\left( \hat{S}_{E}\left( X\right) \right) }{1-\left( \hat{S}%
\left( X\right) \right) }S_{E}\left( X,X,\theta -1\right) \frac{\partial
f\left( X\right) }{\partial S\left( X,\theta -1\right) }\frac{\partial _{%
\hat{f}\left( X\right) }\left( \frac{\hat{K}_{X}\left\vert \hat{\Psi}\left(
X\right) \right\vert ^{2}}{K_{X}\left\vert \Psi \left( X\right) \right\vert
^{2}}\right) S\left( X\right) }{\frac{\hat{K}_{X}\left\vert \hat{\Psi}\left(
X\right) \right\vert ^{2}}{K_{X}\left\vert \Psi \left( X\right) \right\vert
^{2}}}-1  \notag
\end{eqnarray}%
\begin{eqnarray*}
&&\frac{\hat{S}\left( X\right) }{\left( 1-\left( \hat{S}\left( X\right)
\right) \right) }-\frac{\hat{S}_{E}\left( X\right) }{\left( 1-\left( \hat{S}%
_{E}\left( X^{\prime }\right) \right) \right) } \\
&&\frac{1}{\left( 1-\left( \hat{S}\left( X\right) \right) \right) }-\frac{1}{%
\left( 1-\left( \hat{S}_{E}\left( X^{\prime }\right) \right) \right) } \\
&=&\frac{\hat{S}_{L}\left( X\right) }{\left( 1-\left( \hat{S}\left( X\right)
\right) \right) \left( 1-\left( \hat{S}_{E}\left( X^{\prime }\right) \right)
\right) }
\end{eqnarray*}%
Given (\ref{SNx}) and (\ref{St}):%
\begin{equation*}
\frac{\hat{S}\left( X\right) }{\left( 1-\left( \hat{S}\left( X\right)
\right) \right) }-\frac{\hat{S}_{E}\left( X\right) }{\left( 1-\left( \hat{S}%
_{E}\left( X^{\prime }\right) \right) \right) }=\frac{\hat{S}_{E}\left(
X\right) }{\left( 1-\left( \hat{S}\left( X\right) \right) \right) \left(
1-\left( \hat{S}_{E}\left( X^{\prime }\right) \right) \right) }\frac{%
1+\Delta \hat{r}\left( X^{\prime }\right) }{1+\Delta \hat{f}\left( X^{\prime
}\right) }
\end{equation*}%
since:%
\begin{eqnarray*}
&&\frac{\frac{\partial \hat{K}_{X^{\prime }}\left\vert \hat{\Psi}\left(
X^{\prime }\right) \right\vert ^{2}}{\partial \hat{f}\left( X,\theta
-1\right) }}{\hat{K}_{X^{\prime }}\left\vert \hat{\Psi}\left( X^{\prime
}\right) \right\vert ^{2}} \\
&\rightarrow &-\frac{1}{\hat{f}\left( X\right) }
\end{eqnarray*}%
Moreover:%
\begin{eqnarray*}
&&\frac{\hat{S}_{L}\left( X\right) }{\left( 1-\left( \hat{S}\left( X\right)
\right) \right) \left( 1-\left( \hat{S}_{E}\left( X^{\prime }\right) \right)
\right) } \\
&\rightarrow &\frac{\frac{1+\Delta \hat{r}\left( X^{\prime }\right) }{\left(
1+\Delta \hat{f}\left( X^{\prime }\right) \right) }}{\left( 1+\Delta \hat{f}%
\left( X^{\prime }\right) \right) ^{-1}\left( 1+\Delta \hat{f}\left(
X^{\prime }\right) +\Delta \hat{r}\left( X^{\prime }\right) \right) ^{-1}}%
=\left( 1+\Delta \hat{r}\left( X^{\prime }\right) \right) \left( 1+\Delta 
\hat{f}\left( X^{\prime }\right) \right)
\end{eqnarray*}%
\begin{eqnarray*}
&&\left( \frac{\hat{S}\left( X\right) }{\left( 1-\left( \hat{S}\left(
X\right) \right) \right) }-\frac{\hat{S}_{E}\left( X\right) }{\left(
1-\left( \hat{S}_{E}\left( X^{\prime }\right) \right) \right) }\right)
\left( \frac{1}{1+\Delta \hat{f}\left( X\right) }-\frac{\frac{\partial \hat{K%
}_{X^{\prime }}\left\vert \hat{\Psi}\left( X^{\prime }\right) \right\vert
^{2}}{\partial \hat{f}\left( X,\theta -1\right) }}{\hat{K}_{X^{\prime
}}\left\vert \hat{\Psi}\left( X^{\prime }\right) \right\vert ^{2}}\right) \\
&=&\frac{\hat{S}_{L}\left( X\right) }{\left( 1-\left( \hat{S}\left( X\right)
\right) \right) \left( 1-\left( \hat{S}_{E}\left( X^{\prime }\right) \right)
\right) }\left( \frac{1}{1+\Delta \hat{f}\left( X\right) }-\frac{\frac{%
\partial \hat{K}_{X^{\prime }}\left\vert \hat{\Psi}\left( X^{\prime }\right)
\right\vert ^{2}}{\partial \hat{f}\left( X,\theta -1\right) }}{\hat{K}%
_{X^{\prime }}\left\vert \hat{\Psi}\left( X^{\prime }\right) \right\vert ^{2}%
}\right) \\
&\rightarrow &\left( 1+\Delta \hat{r}\left( X^{\prime }\right) \right)
\left( 1+\Delta \hat{f}\left( X^{\prime }\right) +\Delta \hat{r}\left(
X^{\prime }\right) \right) \left( \frac{1}{1+\Delta \hat{f}\left( X\right) }+%
\frac{1}{\hat{f}\left( X\right) }\right)
\end{eqnarray*}%
\begin{eqnarray*}
&&\frac{\hat{S}\left( X\right) }{\left( 1-\left( \hat{S}\left( X\right)
\right) \right) }\left( \frac{1}{2\left( 1+\frac{\Delta \hat{f}\left(
X^{\prime }\right) +\Delta \hat{r}\left( X^{\prime }\right) }{2}\right) }-%
\frac{1}{1+\Delta \hat{f}\left( X\right) }\right) \\
&&\frac{\hat{S}\left( X\right) }{\left( 1-\left( \hat{S}\left( X\right)
\right) \right) }\left( \frac{1}{1+\Delta \hat{f}\left( X^{\prime }\right)
+1+\Delta \hat{r}\left( X^{\prime }\right) }-\frac{1}{1+\Delta \hat{f}\left(
X\right) }\right) \\
&=&-\frac{\hat{S}\left( X\right) }{\left( 1-\left( \hat{S}\left( X\right)
\right) \right) }\frac{1+\Delta \hat{r}\left( X^{\prime }\right) }{\left(
1+\Delta \hat{f}\left( X^{\prime }\right) \right) ^{2}} \\
&\rightarrow &-\frac{1+\Delta \hat{r}\left( X^{\prime }\right) }{1+\Delta 
\hat{f}\left( X^{\prime }\right) }
\end{eqnarray*}%
\begin{eqnarray*}
&&\left( \frac{\hat{S}\left( X\right) }{\left( 1-\left( \hat{S}\left(
X\right) \right) \right) }\left( \frac{1}{2\left( 1+\frac{\Delta \hat{f}%
\left( X^{\prime }\right) +\Delta \hat{r}\left( X^{\prime }\right) }{2}%
\right) }-\frac{\frac{\partial \hat{K}_{X^{\prime }}\left\vert \hat{\Psi}%
\left( X^{\prime }\right) \right\vert ^{2}}{\partial \hat{f}\left( X,\theta
-1\right) }}{\hat{K}_{X^{\prime }}\left\vert \hat{\Psi}\left( X^{\prime
}\right) \right\vert ^{2}}\right) \right. \\
&&\left. -\frac{\hat{S}_{E}\left( X\right) }{\left( 1-\left( \hat{S}%
_{E}\left( X^{\prime }\right) \right) \right) }\left( \frac{1}{1+\Delta \hat{%
f}\left( X\right) }-\frac{\frac{\partial \hat{K}_{X^{\prime }}\left\vert 
\hat{\Psi}\left( X^{\prime }\right) \right\vert ^{2}}{\partial \hat{f}\left(
X,\theta -1\right) }}{\hat{K}_{X^{\prime }}\left\vert \hat{\Psi}\left(
X^{\prime }\right) \right\vert ^{2}}\right) \right) \left( \hat{f}\left(
X\right) -\bar{r}\right) \\
&\rightarrow &\left( \left( 1+\Delta \hat{r}\left( X^{\prime }\right)
\right) \left( 1+\Delta \hat{f}\left( X^{\prime }\right) +\Delta \hat{r}%
\left( X^{\prime }\right) \right) \left( \frac{1}{1+\Delta \hat{f}\left(
X\right) }+\frac{1}{\hat{f}\left( X\right) }\right) -\frac{1+\Delta \hat{r}%
\left( X^{\prime }\right) }{1+\Delta \hat{f}\left( X^{\prime }\right) }%
\right) \left( \hat{f}\left( X\right) -\bar{r}\right) \\
&&\left( 1+\Delta \hat{r}\left( X^{\prime }\right) \right) \left( 1+\Delta 
\hat{f}\left( X^{\prime }\right) \right) \left( \frac{1}{1+\Delta \hat{f}%
\left( X\right) }+\frac{1}{\hat{f}\left( X\right) }\right) \left( \hat{f}%
\left( X\right) -\bar{r}\right)
\end{eqnarray*}%
The second contribution to $\alpha $:%
\begin{eqnarray*}
&&\frac{1-\left( \hat{S}_{E}\left( X\right) \right) }{1-\left( \hat{S}\left(
X\right) \right) }S_{E}\left( X,X,\theta -1\right) \frac{\partial f\left(
X\right) }{\partial S\left( X,\theta -1\right) }\frac{\partial _{\hat{f}%
\left( X\right) }\left( \frac{\hat{K}_{X}\left\vert \hat{\Psi}\left(
X\right) \right\vert ^{2}}{K_{X}\left\vert \Psi \left( X\right) \right\vert
^{2}}\right) S\left( X\right) }{\frac{\hat{K}_{X}\left\vert \hat{\Psi}\left(
X\right) \right\vert ^{2}}{K_{X}\left\vert \Psi \left( X\right) \right\vert
^{2}}}-1 \\
&\simeq &\frac{\left( 1+\Delta \hat{f}\left( X^{\prime }\right) +\Delta \hat{%
r}\left( X^{\prime }\right) \right) }{1+\Delta \hat{f}\left( X^{\prime
}\right) }r\left( 1-S\left( X,\theta -1\right) \right) ^{-1}f\left( X\right) 
\frac{1}{\hat{f}\left( X\right) }-1 \\
&\simeq &r\left( 1-S\left( X,\theta -1\right) \right) ^{-1}f\left( X\right) 
\frac{1}{\hat{f}\left( X\right) }
\end{eqnarray*}%
\begin{eqnarray*}
\alpha &\simeq &\left( 1+\Delta \hat{r}\left( X^{\prime }\right) \right)
\left( 1+\Delta \hat{f}\left( X^{\prime }\right) \right) \left( \frac{1}{%
1+\Delta \hat{f}\left( X\right) }+\frac{1}{\hat{f}\left( X\right) }\right)
\left( \hat{f}\left( X\right) -\bar{r}\right) \\
&\simeq &\left( 1+\frac{1+\Delta \hat{f}\left( X^{\prime }\right) }{\hat{f}%
\left( X\right) }\right) \left( \hat{f}\left( X\right) -\bar{r}\right)
\end{eqnarray*}%
\begin{eqnarray*}
\beta &=&\left( \frac{\frac{\hat{w}\left( X\right) }{2}S\left( X\right) }{%
1+\left( \hat{w}\left( X\right) \left( \frac{f\left( X\right) +\bar{r}\left(
X\right) }{2}-\frac{\left\langle \hat{f}\left( X^{\prime }\right)
\right\rangle _{\hat{w}_{E}}+\left\langle \hat{r}\left( X^{\prime }\right)
\right\rangle _{\hat{w}_{L}}}{2}\right) \right) }+\frac{\partial _{f\left(
X\right) }\left( \frac{\hat{K}_{X}\left\vert \hat{\Psi}\left( X\right)
\right\vert ^{2}}{K_{X}\left\vert \Psi \left( X\right) \right\vert ^{2}}%
\right) S\left( X\right) }{\frac{\hat{K}_{X}\left\vert \hat{\Psi}\left(
X\right) \right\vert ^{2}}{K_{X}\left\vert \Psi \left( X\right) \right\vert
^{2}}}\right) \frac{\partial f\left( X\right) }{\partial S\left( X,\theta
-2\right) }-1 \\
&\simeq &-\frac{\frac{\hat{w}\left( X\right) }{2}S\left( X\right) }{1+\left( 
\hat{w}\left( X\right) \left( \frac{f\left( X\right) +\bar{r}\left( X\right) 
}{2}-\frac{\left\langle \hat{f}\left( X^{\prime }\right) \right\rangle _{%
\hat{w}_{E}}+\left\langle \hat{r}\left( X^{\prime }\right) \right\rangle _{%
\hat{w}_{L}}}{2}\right) \right) }r\left( 1-S\left( X,\theta -1\right)
\right) ^{-1}f\left( X\right) \\
&\simeq &-\frac{S\left( X\right) }{f\left( X\right) -\left\langle \hat{f}%
\left( X^{\prime }\right) \right\rangle }r\left( 1-S\left( X,\theta
-1\right) \right) ^{-1}f\left( X\right)
\end{eqnarray*}

\begin{eqnarray*}
&&c\rightarrow \left\{ -\frac{\left( 1-\left\langle \hat{S}\left( X^{\prime
}\right) \right\rangle \right) \left( \left\langle \hat{f}\left( X^{\prime
}\right) \right\rangle -\bar{r}\right) }{1-\left\langle \hat{S}_{E}\left(
X^{\prime }\right) \right\rangle }\frac{\left\langle \hat{w}\left( X^{\prime
},X\right) \right\rangle \left\langle w\left( X\right) \right\rangle }{4}%
\right. \\
&&\left. +\frac{\frac{1}{4}\left( 1+\hat{w}\left( X\right) \right)
S_{E}\left( X\right) }{\frac{3}{4}}-\frac{S_{E}\left( X\right) S\left(
X\right) }{f\left( X\right) -\left\langle \hat{f}\left( X^{\prime }\right)
\right\rangle }r\left( 1-S\left( X,\theta -2\right) \right) ^{-1}f\left(
X\right) \right\} \\
&&\times \frac{\partial f\left( X\right) }{\partial S\left( X,\theta
-1\right) } \\
&\simeq &\left( \frac{\left( 1-\left\langle \hat{S}\left( X^{\prime }\right)
\right\rangle \right) \left( \left\langle \hat{f}\left( X^{\prime }\right)
\right\rangle -\bar{r}\right) }{1-\left\langle \hat{S}_{E}\left( X^{\prime
}\right) \right\rangle }\frac{\left\langle \hat{w}\left( X^{\prime
},X\right) \right\rangle \left\langle w\left( X\right) \right\rangle }{4}%
\right. \\
&&\left. +\frac{S_{E}\left( X\right) S\left( X\right) }{f\left( X\right)
-\left\langle \hat{f}\left( X^{\prime }\right) \right\rangle }r\left(
1-S\left( X,\theta -2\right) \right) ^{-1}f\left( X\right) \right) r\left(
1-S\left( X,\theta -2\right) \right) ^{-1}f\left( X\right)
\end{eqnarray*}%
\begin{equation*}
w=-\frac{S\left( X\right) }{\hat{f}\left( X\right) }<0
\end{equation*}%
and:

\begin{equation*}
\alpha -\beta \simeq \left( 1+\frac{1+\Delta \hat{f}\left( X^{\prime
}\right) }{\hat{f}\left( X\right) }\right) \left( \hat{f}\left( X\right) -%
\bar{r}\right) +\frac{S\left( X\right) }{f\left( X\right) -\left\langle \hat{%
f}\left( X^{\prime }\right) \right\rangle }r\left( 1-S\left( X,\theta
-1\right) \right) ^{-1}f\left( X\right) >0
\end{equation*}%
\begin{equation*}
\alpha +\beta \simeq \left( 1+\frac{1+\Delta \hat{f}\left( X^{\prime
}\right) }{\hat{f}\left( X\right) }\right) \left( \hat{f}\left( X\right) -%
\bar{r}\right) -\frac{S\left( X\right) }{f\left( X\right) -\left\langle \hat{%
f}\left( X^{\prime }\right) \right\rangle }r\left( 1-S\left( X,\theta
-1\right) \right) ^{-1}f\left( X\right)
\end{equation*}%
\begin{equation*}
\alpha +\beta \simeq \left( 1+\frac{1+\Delta \hat{f}\left( X^{\prime
}\right) }{\hat{f}\left( X\right) }\right) \left( \hat{f}\left( X\right) -%
\bar{r}\right) -rS\left( X\right) f\left( X\right)
\end{equation*}%
The coefficients $\alpha $ and $\beta $ satisfy:%
\begin{eqnarray*}
\alpha &>&0 \\
\beta &<&0
\end{eqnarray*}

\subsubsection*{Eigenvectors and eigenvalues}

The signs of the coefficients of eigenvectors satisfy:

\ $\alpha -\beta >0$, $c>0$, $h<0$%
\begin{eqnarray*}
\frac{\alpha -\beta +\sqrt{\alpha ^{2}-2\alpha \beta +\beta ^{2}+4ch}}{2%
\sqrt{\alpha ^{2}-2\alpha \beta +\beta ^{2}+4ch}} &>&0,\frac{c}{\sqrt{\alpha
^{2}-2\alpha \beta +\beta ^{2}+4ch}}<0 \\
\frac{2h}{2\sqrt{\alpha ^{2}-2\alpha \beta +\beta ^{2}+4ch}} &<&0,\frac{%
\beta -\alpha +\sqrt{\alpha ^{2}-2\alpha \beta +\beta ^{2}+4ch}}{2\sqrt{%
\alpha ^{2}-2\alpha \beta +\beta ^{2}+4ch}}><0
\end{eqnarray*}%
The largest eigenvalue:%
\begin{equation*}
\frac{1}{2}\alpha +\frac{1}{2}\beta +\frac{1}{2}\sqrt{\alpha ^{2}-2\alpha
\beta +\beta ^{2}+4ch}>0
\end{equation*}%
is positive in two cases. The first case arises when:%
\begin{equation*}
\alpha +\beta >0
\end{equation*}%
which implies the following condition:%
\begin{equation*}
\left( 1+\frac{1+\Delta \hat{f}\left( X^{\prime }\right) }{\hat{f}\left(
X\right) }\right) \left( \hat{f}\left( X\right) -\bar{r}\right) >rS\left(
X\right) f\left( X\right)
\end{equation*}%
This is the most common case when $\hat{f}\left( X\right) \simeq f\left(
X\right) $.

The second case arises when:%
\begin{equation*}
\alpha +\beta <0
\end{equation*}%
that is:%
\begin{equation*}
\left( 1+\frac{1+\Delta \hat{f}\left( X^{\prime }\right) }{\hat{f}\left(
X\right) }\right) \left( \hat{f}\left( X\right) -\bar{r}\right) <rS\left(
X\right) f\left( X\right)
\end{equation*}%
whch arises when $f\left( X\right) =s\hat{f}\left( X\right) $ with $s>1$.

In this case, given that:

\begin{equation*}
-\alpha \beta \simeq \left( 1+\frac{1+\Delta \hat{f}\left( X^{\prime
}\right) }{\hat{f}\left( X\right) }\right) \left( \hat{f}\left( X\right) -%
\bar{r}\right) \frac{S\left( X\right) }{f\left( X\right) -\left\langle \hat{f%
}\left( X^{\prime }\right) \right\rangle }r\left( 1-S\left( X,\theta
-1\right) \right) ^{-1}f\left( X\right)
\end{equation*}%
\begin{eqnarray*}
-ch &\simeq &\frac{S\left( X\right) }{\hat{f}\left( X\right) }\left( \frac{%
\left( 1-\left\langle \hat{S}\left( X^{\prime }\right) \right\rangle \right)
\left( \left\langle \hat{f}\left( X^{\prime }\right) \right\rangle -\bar{r}%
\right) }{1-\left\langle \hat{S}_{E}\left( X^{\prime }\right) \right\rangle }%
\frac{\left\langle \hat{w}\left( X^{\prime },X\right) \right\rangle
\left\langle w\left( X\right) \right\rangle }{4}\right. \\
&&\left. +\frac{S_{E}\left( X\right) S\left( X\right) }{f\left( X\right)
-\left\langle \hat{f}\left( X^{\prime }\right) \right\rangle }r\left(
1-S\left( X,\theta -2\right) \right) ^{-1}f\left( X\right) \right) r\left(
1-S\left( X,\theta -2\right) \right) ^{-1}f\left( X\right)
\end{eqnarray*}%
and as a consequence:%
\begin{equation*}
\left( 1+\frac{1+\Delta \hat{f}\left( X^{\prime }\right) }{\hat{f}\left(
X\right) }\right) \left( \hat{f}\left( X\right) -\bar{r}\right) <\frac{%
S\left( X\right) }{\hat{f}\left( X\right) }S_{E}\left( X\right) r\left(
1-S\left( X,\theta -2\right) \right) ^{-1}f\left( X\right)
\end{equation*}%
which writes also:%
\begin{eqnarray*}
&&\left( 1+\frac{1+\Delta \hat{f}\left( X^{\prime }\right) }{\hat{f}\left(
X\right) }\right) \left( \hat{f}\left( X\right) -\bar{r}\right) <\frac{%
S\left( X\right) }{\hat{f}\left( X\right) }S_{E}\left( X\right) r\left(
f\left( X\right) -\hat{f}\left( X\right) \right) f\left( X\right) \\
&\simeq &\frac{S\left( X\right) }{\hat{f}\left( X\right) }S_{E}\left(
X\right) r\left( 1-S\left( X,\theta -2\right) \right) ^{-1}f\left( X\right)
\end{eqnarray*}%
Ultimately, this implies that: 
\begin{equation*}
-\alpha \beta <-ch
\end{equation*}%
which describes a stable state.

\subsection*{A10.1 First case bis: $\hat{f}\left( X\right) -\bar{r}>>0$, $%
f\left( X\right) -\bar{r}$ finite}

In this case, the coefficients rewrite:%
\begin{equation*}
\alpha \simeq \left( 1+\frac{1+\Delta \hat{f}\left( X^{\prime }\right) }{%
\hat{f}\left( X\right) }\right) \left( \hat{f}\left( X\right) -\bar{r}\right)
\end{equation*}

\begin{equation*}
\beta \simeq \left( \frac{\frac{\hat{w}\left( X\right) }{2}S\left( X\right) 
}{1+\left( \hat{w}\left( X\right) \left( \frac{f\left( X\right) +\bar{r}%
\left( X\right) }{2}-\frac{\left\langle \hat{f}\left( X^{\prime }\right)
\right\rangle _{\hat{w}_{E}}+\left\langle \hat{r}\left( X^{\prime }\right)
\right\rangle _{\hat{w}_{L}}}{2}\right) \right) }\right) \frac{\partial
f\left( X\right) }{\partial S\left( X,\theta -2\right) }-1>0
\end{equation*}%
and, as a consequence:%
\begin{equation*}
\alpha +\beta >0
\end{equation*}%
which describes an unstable state.

\subsection*{A10.2 Second case: $f\left( X\right) -\bar{r}<<1$}

For $f\left( X\right) -\bar{r}<<1$, $h<0$ and $\hat{S}\left( X\right) =2\hat{%
S}_{1}\left( X\right) $. Moreover: $S\left( X,\theta -1\right) =2S_{1}\left(
X,X,\theta -1\right) $

\subsubsection*{Estimation of the coefficients}

The coefficients can be estimated by:

\begin{eqnarray}
&&\alpha \\
&\rightarrow &\left( \frac{\hat{S}\left( X\right) }{\left( 1-\left( \hat{S}%
\left( X\right) \right) \right) }\left( \frac{1}{2}+\frac{1}{\hat{f}\left(
X\right) }\right) -\frac{\hat{S}_{E}\left( X\right) }{\left( 1-\left( \hat{S}%
_{E}\left( X^{\prime }\right) \right) \right) }\left( 1+\frac{1}{\hat{f}%
\left( X\right) }\right) \right) \left( \hat{f}\left( X\right) -\bar{r}%
\right)  \notag \\
&&+\frac{1-\left( \hat{S}_{E}\left( X\right) \right) }{1-\left( \hat{S}%
\left( X\right) \right) }S_{E}\left( X,X,\theta -1\right) r\left( 1-S\left(
X,\theta -1\right) \right) ^{-1}f\left( X\right) \frac{1}{\hat{f}\left(
X\right) }-1  \notag \\
&=&\frac{\hat{S}_{E}\left( X\right) \left( \hat{S}_{E}\left( X\right) +\frac{%
1}{\hat{f}\left( X\right) }\right) \left( \hat{f}\left( X\right) -\bar{r}%
\right) }{\left( 1-\left( \hat{S}_{E}\left( X^{\prime }\right) \right)
\right) \left( 1-\left( \hat{S}\left( X\right) \right) \right) }  \notag \\
&&+\frac{1-\left( \hat{S}_{E}\left( X\right) \right) }{1-\left( \hat{S}%
\left( X\right) \right) }S_{E}\left( X,X,\theta -1\right) r\left( 1-S\left(
X,\theta -1\right) \right) ^{-1}f\left( X\right) \frac{1}{\hat{f}\left(
X\right) }-1  \notag \\
&\rightarrow &\frac{1-\left( \hat{S}_{E}\left( X\right) \right) }{1-\left( 
\hat{S}\left( X\right) \right) }S_{E}\left( X,X,\theta -1\right) r\left(
1-S\left( X,\theta -1\right) \right) ^{-1}\frac{f\left( X\right) }{\hat{f}%
\left( X\right) }-1  \notag
\end{eqnarray}

\begin{eqnarray*}
\beta &\simeq &-\frac{\frac{\hat{w}\left( X\right) }{2}S\left( X\right) }{%
1+\left( \hat{w}\left( X\right) \left( \frac{f\left( X\right) +\bar{r}\left(
X\right) }{2}-\frac{\left\langle \hat{f}\left( X^{\prime }\right)
\right\rangle _{\hat{w}_{E}}+\left\langle \hat{r}\left( X^{\prime }\right)
\right\rangle _{\hat{w}_{L}}}{2}\right) \right) }r\left( 1-S\left( X,\theta
-1\right) \right) ^{-1}f\left( X\right) -1 \\
&\simeq &-\frac{\hat{w}\left( X\right) }{2}S\left( X\right) r\left(
1-S\left( X,\theta -1\right) \right) ^{-1}f\left( X\right) -1<0
\end{eqnarray*}%
\begin{eqnarray*}
&&c\simeq \left\{ \frac{\left( 1-\left\langle \hat{S}\left( X^{\prime
}\right) \right\rangle \right) \left( \left\langle \hat{f}\left( X^{\prime
}\right) \right\rangle -\bar{r}\right) }{1-\left\langle \hat{S}_{E}\left(
X^{\prime }\right) \right\rangle }\frac{\left\langle \hat{w}\left( X^{\prime
},X\right) \right\rangle \left\langle w\left( X\right) \right\rangle }{4}%
\right. \\
&&\left. +S_{E}\left( X\right) \left( \frac{\hat{w}\left( X\right) }{2}%
S\left( X\right) \right) r\left( 1-S\left( X,\theta -1\right) \right)
^{-1}f\left( X\right) \right\} \\
&&\times r\left( 1-S\left( X,\theta -1\right) \right) ^{-1}f\left( X\right)
\\
&>&0
\end{eqnarray*}%
and:%
\begin{eqnarray*}
\alpha -\beta &\rightarrow &\frac{1-\left( \hat{S}_{E}\left( X\right)
\right) }{1-\left( \hat{S}\left( X\right) \right) }S_{E}\left( X,X,\theta
-1\right) r\left( 1-S\left( X,\theta -1\right) \right) ^{-1} \\
&&+\frac{\hat{w}\left( X\right) }{2}S\left( X\right) r\left( 1-S\left(
X,\theta -1\right) \right) ^{-1}f\left( X\right)
\end{eqnarray*}%
As a consequence:%
\begin{equation*}
\alpha -\beta >0
\end{equation*}%
and:%
\begin{eqnarray*}
\alpha +\beta &\simeq &\frac{1-\left( \hat{S}_{E}\left( X\right) \right) }{%
1-\left( \hat{S}\left( X\right) \right) }S_{E}\left( X,X,\theta -1\right)
r\left( 1-S\left( X,\theta -1\right) \right) ^{-1}\frac{f\left( X\right) }{%
\hat{f}\left( X\right) }-1 \\
&&-\frac{\hat{w}\left( X\right) }{2}S\left( X\right) r\left( 1-S\left(
X,\theta -1\right) \right) ^{-1}f\left( X\right) -1
\end{eqnarray*}%
Thus, in general:%
\begin{equation*}
\alpha +\beta <0
\end{equation*}%
with the exception when:%
\begin{equation*}
\left( \frac{1-\left( \hat{S}_{E}\left( X\right) \right) }{1-\left( \hat{S}%
\left( X\right) \right) }S_{E}\left( X,X,\theta -1\right) -\frac{\hat{w}%
\left( X\right) }{2}S\left( X\right) \right) r\left( 1-S\left( X,\theta
-1\right) \right) ^{-1}\frac{f\left( X\right) }{\hat{f}\left( X\right) }>2
\end{equation*}%
Given that, in the case considered:%
\begin{eqnarray*}
&&\frac{1-\left( \hat{S}_{E}\left( X\right) \right) }{1-\left( \hat{S}\left(
X\right) \right) }S_{E}\left( X,X,\theta -1\right) -\frac{\hat{w}\left(
X\right) }{2}S\left( X\right) \\
&\simeq &\left( \frac{1-\left( \hat{S}_{E}\left( X\right) \right) }{1-\left( 
\hat{S}\left( X\right) \right) }-\hat{w}\left( X\right) \right) \frac{%
S\left( X\right) }{2}
\end{eqnarray*}%
we have:%
\begin{equation*}
\alpha +\beta >0
\end{equation*}%
under the condition that the quantity:%
\begin{equation*}
\left( 1-S\left( X,\theta -1\right) \right) ^{-1}\frac{f\left( X\right) }{%
\hat{f}\left( X\right) }
\end{equation*}%
is large enough, that is, when $S\left( X,\theta -1\right) $ is close to $1$.

\subsubsection*{Eigenvectors and eigenvalues}

The first case to consider is:%
\begin{equation*}
\left( \frac{1-\left( \hat{S}_{E}\left( X\right) \right) }{1-\left( \hat{S}%
\left( X\right) \right) }-\hat{w}\left( X\right) \right) \frac{S\left(
X\right) }{2}\left( 1-S\left( X,\theta -1\right) \right) ^{-1}\frac{f\left(
X\right) }{\hat{f}\left( X\right) }>2
\end{equation*}%
This condition implies relatively large $\frac{f\left( X\right) }{\hat{f}%
\left( X\right) }$ and $S\left( X\right) $ close to $1$. In this case $%
\alpha +\beta $ is unstable.

The second possibility is for $r<1$. Snc $S_{E}\left( X,X,\theta -1\right)
<1 $, $\hat{S}_{E}\left( X\right) <1$, $\hat{S}\left( X\right) <1$, $\alpha
<0$. This is the most general case.

In this case $\alpha <0$, and given that%
\begin{equation*}
\left\vert \alpha +\beta \right\vert >\left\vert \alpha -\beta \right\vert
\end{equation*}%
and:%
\begin{equation*}
ch<0
\end{equation*}%
As a consequence:%
\begin{equation*}
\left\vert \alpha +\beta \right\vert >\sqrt{\alpha ^{2}-2\alpha \beta +\beta
^{2}+4ch}
\end{equation*}%
and thus the largest eigenvalue is negative:%
\begin{equation*}
\frac{1}{2}\alpha +\frac{1}{2}\beta +\frac{1}{2}\sqrt{\alpha ^{2}-2\alpha
\beta +\beta ^{2}+4ch}<0
\end{equation*}%
As a consequence, the perturbation is stable.

The coefficients for eigenvectors satisfy the following inequalities:%
\begin{eqnarray*}
\frac{\alpha -\beta +\sqrt{\alpha ^{2}-2\alpha \beta +\beta ^{2}+4ch}}{2%
\sqrt{\alpha ^{2}-2\alpha \beta +\beta ^{2}+4ch}} &>&0,\frac{c}{\sqrt{\alpha
^{2}-2\alpha \beta +\beta ^{2}+4ch}}>0 \\
\frac{2h}{2\sqrt{\alpha ^{2}-2\alpha \beta +\beta ^{2}+4ch}} &<&0,\frac{%
\beta -\alpha +\sqrt{\alpha ^{2}-2\alpha \beta +\beta ^{2}+4ch}}{2\sqrt{%
\alpha ^{2}-2\alpha \beta +\beta ^{2}+4ch}}><0
\end{eqnarray*}%
The third possibility is $\alpha >0$, $\alpha +\beta <0$, and this means $%
\frac{f\left( X\right) }{\hat{f}\left( X\right) }$ relatively large. In this
case where $S\left( X,\theta -1\right) $ close to $1$, $\alpha >0$.

The perturbation can be unstable when:%
\begin{equation}
\alpha +\beta <0  \label{Cb}
\end{equation}%
and: 
\begin{equation}
-\alpha \beta >-ch  \label{Cd}
\end{equation}%
The condition (\ref{Cb}) translates as:%
\begin{equation*}
\left( 1-S\left( X,\theta -1\right) \right) ^{-1}\frac{f\left( X\right) }{%
\hat{f}\left( X\right) }
\end{equation*}%
is large enough, which fits with the conditn $\frac{f\left( X\right) }{\hat{f%
}\left( X\right) }$ relativly large in the case where $S\left( X,\theta
-1\right) $ is close to $1$.

To rewrite (\ref{Cd}), we use that in this case:%
\begin{eqnarray*}
&&-\alpha \beta \\
&>&\left( \frac{1-\left( \hat{S}_{E}\left( X\right) \right) }{1-\left( \hat{S%
}\left( X\right) \right) }S_{E}\left( X,X,\theta -1\right) r\left( 1-S\left(
X,\theta -1\right) \right) ^{-1}\frac{f\left( X\right) }{\hat{f}\left(
X\right) }-1\right) \\
&&\times \left( 1+\frac{\frac{\hat{w}\left( X\right) }{2}S\left( X\right) }{%
1+\left( \hat{w}\left( X\right) \left( \frac{f\left( X\right) +\bar{r}\left(
X\right) }{2}-\frac{\left\langle \hat{f}\left( X^{\prime }\right)
\right\rangle _{\hat{w}_{E}}+\left\langle \hat{r}\left( X^{\prime }\right)
\right\rangle _{\hat{w}_{L}}}{2}\right) \right) }r\left( 1-S\left( X,\theta
-1\right) \right) ^{-1}f\left( X\right) \right) \\
&>&\left( \frac{1-\left( \hat{S}_{E}\left( X\right) \right) }{1-\left( \hat{S%
}\left( X\right) \right) }S_{E}\left( X,X,\theta -1\right) r\left( 1-S\left(
X,\theta -1\right) \right) ^{-1}\frac{f\left( X\right) }{\hat{f}\left(
X\right) }-1\right) \\
&&\times \frac{\frac{\hat{w}\left( X\right) }{2}S\left( X\right) }{1+\left( 
\hat{w}\left( X\right) \left( \frac{f\left( X\right) +\bar{r}\left( X\right) 
}{2}-\frac{\left\langle \hat{f}\left( X^{\prime }\right) \right\rangle _{%
\hat{w}_{E}}+\left\langle \hat{r}\left( X^{\prime }\right) \right\rangle _{%
\hat{w}_{L}}}{2}\right) \right) }r\left( 1-S\left( X,\theta -1\right)
\right) ^{-1}f\left( X\right)
\end{eqnarray*}%
And thus, (\ref{Cd}) leads to the condition for instability:%
\begin{eqnarray}
&&\left( \frac{1-\left( \hat{S}_{E}\left( X\right) \right) }{1-\left( \hat{S}%
\left( X\right) \right) }S_{E}\left( X,X,\theta -1\right) r\left( 1-S\left(
X,\theta -1\right) \right) ^{-1}\frac{f\left( X\right) }{\hat{f}\left(
X\right) }-1\right)  \label{Cg} \\
&>&\frac{S_{E}\left( X\right) \left( \frac{\hat{w}\left( X\right) }{2}%
S\left( X\right) \right) \left( r\left( 1-S\left( X,\theta -1\right) \right)
^{-1}f\left( X\right) \right) ^{2}}{\hat{f}\left( X\right) }  \notag \\
&&+\frac{\left( 1-\left\langle \hat{S}\left( X^{\prime }\right)
\right\rangle \right) \left( \left\langle \hat{f}\left( X^{\prime }\right)
\right\rangle -\bar{r}\right) }{1-\left\langle \hat{S}_{E}\left( X^{\prime
}\right) \right\rangle }\frac{\left\langle \hat{w}\left( X^{\prime
},X\right) \right\rangle \left\langle h\left( X\right) \right\rangle }{4\hat{%
f}\left( X\right) }r\left( 1-S\left( X,\theta -1\right) \right) ^{-1}f\left(
X\right)  \notag
\end{eqnarray}%
or in first approximation:%
\begin{eqnarray}
&&\frac{1-\left( \hat{S}_{E}\left( X\right) \right) }{1-\left( \hat{S}\left(
X\right) \right) }S_{E}\left( X,X,\theta -1\right) \\
&>&S_{E}\left( X\right) \left( \frac{\hat{w}\left( X\right) }{2}S\left(
X\right) \right) r\left( 1-S\left( X,\theta -1\right) \right) ^{-1}f\left(
X\right)  \notag \\
&&+\frac{\left( 1-\left\langle \hat{S}\left( X^{\prime }\right)
\right\rangle \right) \left( \left\langle \hat{f}\left( X^{\prime }\right)
\right\rangle -\bar{r}\right) }{1-\left\langle \hat{S}_{E}\left( X^{\prime
}\right) \right\rangle }\frac{\left\langle \hat{w}\left( X^{\prime
},X\right) \right\rangle \left\langle h\left( X\right) \right\rangle }{4} 
\notag
\end{eqnarray}%
The first condition is not possible under our hypothesis except when $%
\left\langle \hat{f}\left( X^{\prime }\right) \right\rangle -\bar{r}<0$,
that is when the investor returns do not compensate the interst rate.

\subsection*{A10.3 Dependency of largest eigenvalue in parameters}

Using the decomposition:%
\begin{equation*}
\alpha +\beta =2\beta +\alpha -\beta
\end{equation*}%
and definining $X=\alpha -\beta $, the variation of the largest eigenvalue
is given by: 
\begin{equation*}
\frac{d}{dX}\left( X+2\beta +\sqrt{X^{2}+4ch}\right) =\frac{X+\sqrt{X^{2}+4ch%
}}{\sqrt{X^{2}+4ch}}
\end{equation*}%
gngrl $\alpha -\beta >0$, so that:

\begin{equation*}
\frac{d}{dX}\left( X+2\beta +\sqrt{X^{2}+4ch}\right) >0
\end{equation*}%
The eigenvalues increase with $\alpha -\beta $ that is decreases with $\hat{f%
}\left( X\right) -\bar{r}$. In general the largest eigenvalue decreases with 
$\left( \hat{f}\left( X\right) -\bar{r}\right) $.

\subsection*{A10.4 Formula for perturbations}

The initial variations $\delta \hat{f}$, $\delta S$, write in terms of
eigenvectrs:

\begin{equation*}
\left( 
\begin{array}{c}
\delta \hat{f} \\ 
\delta S%
\end{array}%
\right) =a\left( 
\begin{array}{c}
1 \\ 
\frac{1}{2c}\left( \beta -\alpha +\sqrt{\alpha ^{2}-2\alpha \beta +\beta
^{2}+4ch}\right)%
\end{array}%
\right) +b\left( 
\begin{array}{c}
1 \\ 
\frac{1}{2c}\left( \beta -\alpha -\sqrt{\alpha ^{2}-2\alpha \beta +\beta
^{2}+4ch}\right)%
\end{array}%
\right)
\end{equation*}%
where the coefficients $a$ nd $b$ depend on the initial variations.

We find the following relations: 
\begin{eqnarray*}
\frac{1}{2c}\left( \alpha -\beta +\sqrt{\alpha ^{2}-2\alpha \beta +\beta
^{2}+4ch}\right) \delta \hat{f}+\delta S &=&a\frac{1}{c}\left( \sqrt{\alpha
^{2}-2\alpha \beta +\beta ^{2}+4ch}\right) \\
\frac{1}{2c}\left( \alpha -\beta -\sqrt{\alpha ^{2}-2\alpha \beta +\beta
^{2}+4ch}\right) \delta \hat{f}+\delta S &=&-b\frac{1}{c}\left( \sqrt{\alpha
^{2}-2\alpha \beta +\beta ^{2}+4ch}\right)
\end{eqnarray*}%
which leads $a$:%
\begin{equation*}
a=\frac{\alpha -\beta +\sqrt{\alpha ^{2}-2\alpha \beta +\beta ^{2}+4ch}}{2%
\sqrt{\alpha ^{2}-2\alpha \beta +\beta ^{2}+4ch}}\delta \hat{f}+\frac{c}{%
\sqrt{\alpha ^{2}-2\alpha \beta +\beta ^{2}+4ch}}\delta S
\end{equation*}%
and:%
\begin{equation*}
b=\frac{\beta -\alpha +\sqrt{\alpha ^{2}-2\alpha \beta +\beta ^{2}+4ch}}{2%
\sqrt{\alpha ^{2}-2\alpha \beta +\beta ^{2}+4ch}}\delta \hat{f}-\frac{c}{%
\sqrt{\alpha ^{2}-2\alpha \beta +\beta ^{2}+4ch}}\delta S
\end{equation*}%
and the formula for perturbation in the continuous approximation writes: 
\begin{eqnarray*}
\left( 
\begin{array}{c}
\delta \hat{f}\left( X,\theta \right) \\ 
\delta S\left( X,\theta -1\right)%
\end{array}%
\right) &\rightarrow &a\exp \left( \lambda _{+}\theta \right) \left( 
\begin{array}{c}
1 \\ 
\frac{1}{2c}\left( \beta -\alpha +\sqrt{\alpha ^{2}-2\alpha \beta +\beta
^{2}+4ch}\right)%
\end{array}%
\right) \\
&&+b\exp \left( \lambda _{-}\theta \right) \left( 
\begin{array}{c}
1 \\ 
\frac{1}{2c}\left( \beta -\alpha -\sqrt{\alpha ^{2}-2\alpha \beta +\beta
^{2}+4ch}\right)%
\end{array}%
\right)
\end{eqnarray*}%
\begin{eqnarray*}
\delta \hat{f}\left( X,\theta \right) &=&\left( \frac{\alpha -\beta +\sqrt{%
\alpha ^{2}-2\alpha \beta +\beta ^{2}+4ch}}{2\sqrt{\alpha ^{2}-2\alpha \beta
+\beta ^{2}+4ch}}\delta \hat{f}+\frac{c}{\sqrt{\alpha ^{2}-2\alpha \beta
+\beta ^{2}+4ch}}\delta S\right) \exp \left( \lambda _{+}\theta \right) \\
&&+\left( \frac{\beta -\alpha +\sqrt{\alpha ^{2}-2\alpha \beta +\beta
^{2}+4ch}}{2\sqrt{\alpha ^{2}-2\alpha \beta +\beta ^{2}+4ch}}\delta \hat{f}-%
\frac{c}{\sqrt{\alpha ^{2}-2\alpha \beta +\beta ^{2}+4ch}}\delta S\right)
\exp \left( \lambda _{-}\theta \right)
\end{eqnarray*}%
\begin{eqnarray*}
\delta S\left( X,\theta -1\right) &=&\frac{2c\left( \frac{\alpha -\beta +%
\sqrt{\alpha ^{2}-2\alpha \beta +\beta ^{2}+4ch}}{2\sqrt{\alpha ^{2}-2\alpha
\beta +\beta ^{2}+4ch}}\delta \hat{f}+\frac{c}{\sqrt{\alpha ^{2}-2\alpha
\beta +\beta ^{2}+4ch}}\delta S\right) }{\left( \beta -\alpha +\sqrt{\alpha
^{2}-2\alpha \beta +\beta ^{2}+4ch}\right) }\exp \left( \lambda _{+}\theta
\right) \\
&&+\frac{2c\left( \frac{\beta -\alpha +\sqrt{\alpha ^{2}-2\alpha \beta
+\beta ^{2}+4ch}}{2\sqrt{\alpha ^{2}-2\alpha \beta +\beta ^{2}+4ch}}\delta 
\hat{f}-\frac{c}{\sqrt{\alpha ^{2}-2\alpha \beta +\beta ^{2}+4ch}}\delta
S\right) }{\left( \beta -\alpha -\sqrt{\alpha ^{2}-2\alpha \beta +\beta
^{2}+4ch}\right) }\exp \left( \lambda _{-}\theta \right)
\end{eqnarray*}%
which leads to the dominant contribution in the text.

\subsection*{A 10.5 Transitions induced by group connections}

As in the text, we can assume some reduction in uncertainty, expected
increase in return... some new connections arise, connecting several groups.

In the frst case $f\left( X\right) -\bar{r}>>0$, $\hat{f}\left( X\right) -%
\bar{r}>>0$ the transition from stability:%
\begin{equation*}
-\alpha \beta >-ch
\end{equation*}%
towards unstability:%
\begin{equation*}
-\alpha \beta <-ch
\end{equation*}%
arises from a change starting from:%
\begin{equation*}
f\left( X\right) \simeq s\hat{f}\left( X\right) ,s>1
\end{equation*}%
towards: 
\begin{equation*}
f\left( X\right) -\hat{f}\left( X\right) =O\left( 1\right)
\end{equation*}%
or:%
\begin{eqnarray*}
\hat{f}\left( X\right) &>&>1 \\
f\left( X\right) &=&O\left( 1\right)
\end{eqnarray*}%
This corresponds to an increase of $\hat{f}\left( X^{\prime }\right) $
through new investments.

In the second case, the transition arises from:%
\begin{eqnarray*}
f\left( X\right) &=&O\left( 1\right) \\
\hat{f}\left( X\right) &=&O\left( 1\right)
\end{eqnarray*}%
towards:%
\begin{eqnarray*}
f\left( X\right) &=&O\left( 1\right) \\
\hat{f}\left( X\right) &>&>1
\end{eqnarray*}%
and this reflects, for example, the connection to a group such that $%
\left\langle \hat{f}\left( X\right) \right\rangle >>1$.

\section*{Appendix 11. Including the investors local interactions.}

We can include deviations around average connexions in the group. This
amounts to include corrections to the perturbation equations.

\subsection*{A11.1 Modification of equations\ }

Specific interactions between some subgroups of agnts may included be
considering the inclusion of $\hat{S}_{E}\left( X^{\prime },X\right) $
rather its average in the return equation. Using that:

\begin{eqnarray*}
&&\hat{S}_{E}\left( X^{\prime },X\right) \\
&\rightarrow &\frac{\hat{w}\left( X^{\prime },X\right) }{2}\left( 1+\left( 
\hat{f}\left( X^{\prime }\right) -\left\langle \hat{w}\left( X\right)
\right\rangle \frac{\left\langle \hat{f}\left( X^{\prime }\right)
\right\rangle _{\hat{w}_{E}}+\left\langle \hat{r}\left( X^{\prime }\right)
\right\rangle _{\hat{w}_{L}}}{2}-\left\langle w\left( X\right) \right\rangle 
\frac{f\left( X\right) +r\left( X\right) }{2}\right) \right)
\end{eqnarray*}%
the variation of $\hat{S}_{E}\left( X^{\prime },X\right) $ is:%
\begin{eqnarray*}
&&\delta \hat{S}_{E}\left( X^{\prime },X,\theta -1\right) \\
&\rightarrow &\frac{\hat{w}\left( X^{\prime },X\right) }{2}\left( 1+\left(
\delta \hat{f}\left( X^{\prime }\right) -\left\langle w\left( X\right)
\right\rangle \delta \frac{f\left( X\right) +r\left( X\right) }{2}\right)
\right)
\end{eqnarray*}%
This modifies the variation of the return equation by:%
\begin{eqnarray*}
&&\frac{1-\left( \hat{S}\left( X\right) \right) }{1-\left( \hat{S}_{E}\left(
X\right) \right) }\delta \hat{f}\left( X,\theta \right) -\hat{S}_{E}\left(
X^{\prime },X,\theta -1\right) \frac{1-\hat{S}\left( X^{\prime }\right) }{1-%
\hat{S}_{E}\left( X^{\prime }\right) }\delta \hat{f}\left( X^{\prime
},\theta \right) \\
&=&\left( \Delta \left( X^{\prime },X\right) -\hat{S}_{E}\left( X^{\prime
},X,\theta -1\right) \right) \\
&&\times \left( \frac{\hat{S}\left( X^{\prime }\right) }{\left( 1-\left( 
\hat{S}_{E}\left( X^{\prime }\right) \right) \right) }\left( \frac{\left( 
\hat{f}\left( X^{\prime }\right) -\bar{r}\right) }{2\left( 1+\frac{\Delta 
\hat{f}\left( X^{^{\prime }\prime }\right) +\Delta \hat{r}\left( X^{\prime
^{\prime }}\right) }{2}\right) }-\frac{\frac{\partial \hat{K}_{X^{\prime
}}\left\vert \hat{\Psi}\left( X^{\prime }\right) \right\vert ^{2}}{\partial 
\hat{f}\left( X,\theta -1\right) }}{\hat{K}_{X^{\prime }}\left\vert \hat{\Psi%
}\left( X^{\prime }\right) \right\vert ^{2}}\right) \right. \\
&&\left. -\frac{\left( 1-\left( \hat{S}\left( X^{\prime }\right) \right)
\right) \hat{S}_{E}\left( X\right) }{\left( 1-\left( \hat{S}_{E}\left(
X^{\prime }\right) \right) \right) ^{2}}\left( \frac{1}{1+\Delta \hat{f}%
\left( X^{\prime }\right) }-\frac{\frac{\partial \hat{K}_{X^{\prime
}}\left\vert \hat{\Psi}\left( X^{\prime }\right) \right\vert ^{2}}{\partial 
\hat{f}\left( X,\theta -1\right) }}{\hat{K}_{X^{\prime }}\left\vert \hat{\Psi%
}\left( X^{\prime }\right) \right\vert ^{2}}\right) \right) \left( \hat{f}%
\left( X^{\prime }\right) -\bar{r}\right) \delta \hat{f}\left( X^{\prime
},\theta -1\right) \\
&&-\frac{\left( 1-\hat{S}\left( X^{\prime }\right) \right) \left( \hat{f}%
\left( X^{\prime }\right) -\bar{r}\right) }{1-\hat{S}_{E}\left( X^{\prime
}\right) }\frac{\hat{w}\left( X^{\prime },X\right) }{2}\delta \hat{f}\left(
X^{\prime },\theta -1\right) \\
&&-\frac{\left( 1-\hat{S}\left( X^{\prime }\right) \right) \left( \hat{f}%
\left( X^{\prime }\right) -\bar{r}\right) }{1-\hat{S}_{E}\left( X^{\prime
}\right) }\frac{\hat{w}\left( X^{\prime },X\right) h\left( X\right) }{4} \\
&&\times \left( r\left( 1-S\left( X,\theta -2\right) \right) ^{r-1}\left(
f_{1}\left( X\right) +\Delta F_{\tau }\left( \bar{R}\left( K,X\right)
\right) \right) -C\right) \delta S\left( X,\theta -2\right) \\
&&+\frac{\frac{w\left( X\right) }{4}\left( 1+\hat{w}\left( X\right) \right)
S_{E}\left( X\right) \left( r\left( 1-S\left( X,\theta -2\right) \right)
^{r-1}\left( f_{1}\left( X\right) +\Delta F_{\tau }\left( \bar{R}\left(
K,X\right) \right) \right) -C\right) }{1+\left( \hat{w}\left( X\right)
\left( f\left( X\right) -\frac{\left\langle \hat{f}\left( X^{\prime }\right)
\right\rangle _{\hat{w}_{E}}+\left\langle \hat{r}\left( X^{\prime }\right)
\right\rangle _{\hat{w}_{L}}}{2}\right) +\frac{w\left( X\right) }{2}\left(
f\left( X\right) -\bar{r}\left( X\right) \right) \right) } \\
&&\times \left( f\left( X\right) -\bar{r}\right) \delta S\left( X,\theta
-2\right) \\
&&-S_{E}\left( X,X,\theta -1\right) \left( r\left( 1-S\left( X,\theta
-1\right) \right) ^{r-1}\left( f_{1}\left( X\right) +\Delta F_{\tau }\left( 
\bar{R}\left( K,X\right) \right) \right) -C\right) \delta S\left( X,\theta
-1\right)
\end{eqnarray*}

Regrouping terms and, as before, inserting the equation for $\delta S\left(
X,\theta -1\right) $:%
\begin{eqnarray*}
\delta S\left( X,\theta -1\right) &=&\frac{\partial _{\hat{f}\left( X\right)
}\left( \frac{\hat{K}_{X}\left\vert \hat{\Psi}\left( X\right) \right\vert
^{2}}{K_{X}\left\vert \Psi \left( X\right) \right\vert ^{2}}\right) }{\frac{%
\hat{K}_{X}\left\vert \hat{\Psi}\left( X\right) \right\vert ^{2}}{%
K_{X}\left\vert \Psi \left( X\right) \right\vert ^{2}}}\delta \hat{f}\left(
X,\theta -1\right) \\
&&+\left( \frac{\frac{\hat{w}\left( X\right) }{2}S\left( X\right) }{1+\left( 
\hat{w}\left( X\right) \left( \frac{f\left( X\right) +\bar{r}\left( X\right) 
}{2}-\frac{\left\langle \hat{f}\left( X^{\prime }\right) \right\rangle _{%
\hat{w}_{E}}+\left\langle \hat{r}\left( X^{\prime }\right) \right\rangle _{%
\hat{w}_{L}}}{2}\right) \right) }+\frac{\partial _{f\left( X\right) }\left( 
\frac{\hat{K}_{X}\left\vert \hat{\Psi}\left( X\right) \right\vert ^{2}}{%
K_{X}\left\vert \Psi \left( X\right) \right\vert ^{2}}\right) S\left(
X\right) }{\frac{\hat{K}_{X}\left\vert \hat{\Psi}\left( X\right) \right\vert
^{2}}{K_{X}\left\vert \Psi \left( X\right) \right\vert ^{2}}}\right) \\
&&\times \left( r\left( 1-S\left( X,\theta -2\right) \right) ^{r-1}\left(
f_{1}\left( X\right) +\Delta F_{\tau }\left( \bar{R}\left( K,X\right)
\right) \right) -C\right) \delta S\left( X,\theta -2\right)
\end{eqnarray*}%
leads to:%
\begin{eqnarray*}
&&\frac{1-\left( \hat{S}\left( X\right) \right) }{1-\left( \hat{S}_{E}\left(
X\right) \right) }\delta \hat{f}\left( X,\theta \right) -\hat{S}_{E}\left(
X^{\prime },X,\theta -1\right) \frac{1-\hat{S}\left( X^{\prime }\right) }{1-%
\hat{S}_{E}\left( X^{\prime }\right) }\delta \hat{f}\left( X^{\prime
},\theta \right) \\
&=&\left\{ \left( \Delta \left( X^{\prime },X\right) -\hat{S}_{E}\left(
X^{\prime },X,\theta -1\right) \right) \left( \frac{\hat{S}\left( X^{\prime
}\right) }{\left( 1-\left( \hat{S}_{E}\left( X^{\prime }\right) \right)
\right) }\left( \frac{\left( \hat{f}\left( X^{\prime }\right) -\bar{r}%
\right) }{2\left( 1+\frac{\Delta \hat{f}\left( X^{^{\prime }\prime }\right)
+\Delta \hat{r}\left( X^{\prime ^{\prime }}\right) }{2}\right) }-\frac{\frac{%
\partial \hat{K}_{X^{\prime }}\left\vert \hat{\Psi}\left( X^{\prime }\right)
\right\vert ^{2}}{\partial \hat{f}\left( X,\theta -1\right) }}{\hat{K}%
_{X^{\prime }}\left\vert \hat{\Psi}\left( X^{\prime }\right) \right\vert ^{2}%
}\right) \right) \right. \\
&&\left. -\frac{\left( 1-\left( \hat{S}\left( X^{\prime }\right) \right)
\right) \hat{S}_{E}\left( X\right) }{\left( 1-\left( \hat{S}_{E}\left(
X^{\prime }\right) \right) \right) ^{2}}\left( \frac{1}{1+\Delta \hat{f}%
\left( X^{\prime }\right) }-\frac{\frac{\partial \hat{K}_{X^{\prime
}}\left\vert \hat{\Psi}\left( X^{\prime }\right) \right\vert ^{2}}{\partial 
\hat{f}\left( X,\theta -1\right) }}{\hat{K}_{X^{\prime }}\left\vert \hat{\Psi%
}\left( X^{\prime }\right) \right\vert ^{2}}\right) \right) \left( \hat{f}%
\left( X^{\prime }\right) -\bar{r}\right) \\
&&\left. -\frac{\left( 1-\hat{S}\left( X^{\prime }\right) \right) \left( 
\hat{f}\left( X^{\prime }\right) -\bar{r}\right) }{1-\hat{S}_{E}\left(
X^{\prime }\right) }\frac{\hat{w}\left( X^{\prime },X\right) }{2}%
+S_{E}\left( X,X,\theta -1\right) \frac{\partial f\left( X\right) }{\partial
S\left( X,\theta -1\right) }\frac{\partial _{\hat{f}\left( X\right) }\left( 
\frac{\hat{K}_{X}\left\vert \hat{\Psi}\left( X\right) \right\vert ^{2}}{%
K_{X}\left\vert \Psi \left( X\right) \right\vert ^{2}}\right) }{\frac{\hat{K}%
_{X}\left\vert \hat{\Psi}\left( X\right) \right\vert ^{2}}{K_{X}\left\vert
\Psi \left( X\right) \right\vert ^{2}}}\right\} \delta \hat{f}\left(
X^{\prime },\theta -1\right) \\
&&-\left\{ -\frac{\left( 1-\hat{S}\left( X^{\prime }\right) \right) \left( 
\hat{f}\left( X^{\prime }\right) -\bar{r}\right) }{1-\hat{S}_{E}\left(
X^{\prime }\right) }\frac{\hat{w}\left( X^{\prime },X\right) h\left(
X\right) }{4}\right. \\
&&+\frac{\frac{h\left( X\right) }{4}\left( 1+\hat{w}\left( X\right) \right)
S_{E}\left( X\right) \left( f\left( X\right) -\bar{r}\right) }{1+\left( \hat{%
w}\left( X\right) \left( f\left( X\right) -\frac{\left\langle \hat{f}\left(
X^{\prime }\right) \right\rangle _{\hat{w}_{E}}+\left\langle \hat{r}\left(
X^{\prime }\right) \right\rangle _{\hat{w}_{L}}}{2}\right) +\frac{h\left(
X\right) }{2}\left( f\left( X\right) -\bar{r}\left( X\right) \right) \right) 
} \\
&&\left. +\left( \frac{\frac{\hat{w}\left( X\right) }{2}S\left( X\right) }{%
1+\left( \hat{w}\left( X\right) \left( \frac{f\left( X\right) +\bar{r}\left(
X\right) }{2}-\frac{\left\langle \hat{f}\left( X^{\prime }\right)
\right\rangle _{\hat{w}_{E}}+\left\langle \hat{r}\left( X^{\prime }\right)
\right\rangle _{\hat{w}_{L}}}{2}\right) \right) }+\frac{\partial _{f\left(
X\right) }\left( \frac{\hat{K}_{X}\left\vert \hat{\Psi}\left( X\right)
\right\vert ^{2}}{K_{X}\left\vert \Psi \left( X\right) \right\vert ^{2}}%
\right) S\left( X\right) }{\frac{\hat{K}_{X}\left\vert \hat{\Psi}\left(
X\right) \right\vert ^{2}}{K_{X}\left\vert \Psi \left( X\right) \right\vert
^{2}}}\right) \frac{S_{E}\left( X,X,\theta -1\right) \partial f\left(
X\right) }{\partial S\left( X,\theta -2\right) }\right\} \\
&&\times \frac{\partial f\left( X\right) }{\partial S\left( X,\theta
-2\right) }\delta S\left( X,\theta -2\right)
\end{eqnarray*}%
where:%
\begin{equation*}
\frac{\partial f\left( X\right) }{\partial S\left( X,\theta -1\right) }%
=-\left( r\left( 1-S\left( X,\theta -1\right) \right) ^{r-1}\left(
f_{1}\left( X\right) +\Delta F_{\tau }\left( \bar{R}\left( K,X\right)
\right) \right) -C\right)
\end{equation*}%
In first approximatn we can neglect:%
\begin{eqnarray*}
&&\left( \Delta \left( X^{\prime },X\right) -\hat{S}_{E}\left( X^{\prime
},X,\theta -1\right) \right) \\
&&\times \left( \frac{\hat{S}\left( X^{\prime }\right) }{\left( 1-\left( 
\hat{S}_{E}\left( X^{\prime }\right) \right) \right) }\left( \frac{\left( 
\hat{f}\left( X^{\prime }\right) -\bar{r}\right) }{2\left( 1+\frac{\Delta 
\hat{f}\left( X^{^{\prime }\prime }\right) +\Delta \hat{r}\left( X^{\prime
^{\prime }}\right) }{2}\right) }-\frac{\frac{\partial \hat{K}_{X^{\prime
}}\left\vert \hat{\Psi}\left( X^{\prime }\right) \right\vert ^{2}}{\partial 
\hat{f}\left( X,\theta -1\right) }}{\hat{K}_{X^{\prime }}\left\vert \hat{\Psi%
}\left( X^{\prime }\right) \right\vert ^{2}}\right) \right. \\
&&\left. -\frac{\left( 1-\left( \hat{S}\left( X^{\prime }\right) \right)
\right) \hat{S}_{E}\left( X\right) }{\left( 1-\left( \hat{S}_{E}\left(
X^{\prime }\right) \right) \right) ^{2}}\left( \frac{1}{1+\Delta \hat{f}%
\left( X^{\prime }\right) }-\frac{\frac{\partial \hat{K}_{X^{\prime
}}\left\vert \hat{\Psi}\left( X^{\prime }\right) \right\vert ^{2}}{\partial 
\hat{f}\left( X,\theta -1\right) }}{\hat{K}_{X^{\prime }}\left\vert \hat{\Psi%
}\left( X^{\prime }\right) \right\vert ^{2}}\right) \right) \left( \hat{f}%
\left( X^{\prime }\right) -\bar{r}\right)
\end{eqnarray*}%
and the average perturbation equation is modified by the matrial term:%
\begin{equation*}
\left( 
\begin{array}{cc}
0 & h \\ 
0 & 0%
\end{array}%
\right)
\end{equation*}%
with:%
\begin{equation*}
-\hat{S}_{E}\left( X^{\prime },X,\theta -1\right) \frac{1-\hat{S}\left(
X^{\prime }\right) }{1-\hat{S}_{E}\left( X^{\prime }\right) }=h
\end{equation*}

\subsection*{A11.2 Modification of eigenvalues from loops}

\subsubsection*{A11.2.1 2 sectors reciprocal interaction}

Consider reciprocal interaction:%
\begin{equation*}
\left( 
\begin{array}{cccc}
\alpha & c & 0 & W \\ 
h & \beta & 0 & 0 \\ 
0 & W^{\prime } & \alpha ^{\prime } & c^{\prime } \\ 
0 & 0 & h^{\prime } & \beta ^{\prime }%
\end{array}%
\right)
\end{equation*}%
In the diagonalization process, the additional contribution $\left( 
\begin{array}{cc}
0 & W \\ 
0 & 0%
\end{array}%
\right) $ is modified in first approximation by the average matrix change of
basis:%
\begin{equation*}
\left( 
\begin{array}{cc}
\alpha & \eta \\ 
\beta & \theta%
\end{array}%
\right) =\left( 
\begin{array}{cc}
-\frac{2c}{\sqrt{4c^{2}+\left( \alpha -\beta +\sqrt{\left( \alpha -\beta
\right) ^{2}+4ch}\right) ^{2}}} & -\frac{2c}{\sqrt{4c^{2}+\left( \alpha
-\beta -\sqrt{\left( \alpha -\beta \right) ^{2}+4ch}\right) ^{2}}} \\ 
\frac{\alpha -\beta +\sqrt{\left( \alpha -\beta \right) ^{2}+4ch}}{\sqrt{%
4c^{2}+\left( \alpha -\beta +\sqrt{\left( \alpha -\beta \right) ^{2}+4ch}%
\right) ^{2}}} & \frac{\alpha -\beta -\sqrt{\left( \alpha -\beta \right)
^{2}+4ch}}{\sqrt{4c^{2}+\left( \alpha -\beta -\sqrt{\left( \alpha -\beta
\right) ^{2}+4ch}\right) ^{2}}}%
\end{array}%
\right)
\end{equation*}%
with formula:%
\begin{equation*}
\left( 
\begin{array}{cc}
\alpha & \eta \\ 
\beta & \theta%
\end{array}%
\right) ^{-1}\left( 
\begin{array}{cc}
0 & W \\ 
0 & 0%
\end{array}%
\right) \left( 
\begin{array}{cc}
\alpha & \eta \\ 
\beta & \theta%
\end{array}%
\right) =:\left( 
\begin{array}{cc}
W\theta \frac{\beta }{\theta \alpha -\beta \eta } & W\frac{\theta ^{2}}{%
\theta \alpha -\beta \eta } \\ 
-W\frac{\beta ^{2}}{\theta \alpha -\beta \eta } & -W\theta \frac{\beta }{%
\theta \alpha -\beta \eta }%
\end{array}%
\right)
\end{equation*}%
The correction to the eigenvalues for this reciprocal interactions between
two agents is obtained by considering the matrix, already diagonalized by
blocks:

\begin{equation*}
\begin{array}{cccc}
a-x & 0 & v & W \\ 
0 & b-x & j & y \\ 
v^{\prime } & W^{\prime } & a^{\prime }-x & 0 \\ 
j^{\prime } & y^{\prime } & 0 & b^{\prime }-x%
\end{array}%
\end{equation*}%
where $a$, $b$, are the averaged eigenvalues for each sectors. The modified
eigenvalues $x$ are first order deviations from one of the averaged value.

The deviation from eigenvalue $a$ and \ $b$\ is given by:%
\begin{equation*}
a+\frac{1}{\left( a-b^{\prime }\right) \left( a-a^{\prime }\right) }\left(
\left( a-a^{\prime }\right) Wj^{\prime }+\left( a-b^{\prime }\right)
vv^{\prime }\right)
\end{equation*}%
\begin{equation*}
b+\frac{1}{\left( b-b^{\prime }\right) \left( b-a^{\prime }\right) }\left(
\left( b-a^{\prime }\right) jW^{\prime }+\left( b-b^{\prime }\right)
yy^{\prime }\right)
\end{equation*}%
and more general, the correction to the eigenvalues $\lambda _{-}$ $\lambda
_{+}$ are:

\begin{eqnarray*}
&&\left( 
\begin{array}{cc}
W\theta \frac{\beta }{\theta \alpha -\beta \eta } & W\frac{\theta ^{2}}{%
\theta \alpha -\beta \eta }%
\end{array}%
\right) 
\begin{array}{cc}
\left( \lambda _{-}-\lambda _{-}^{\prime }\right) ^{-1} & 0 \\ 
0 & \left( \lambda _{-}-\lambda _{+}^{\prime }\right) ^{-1}%
\end{array}%
\left( 
\begin{array}{c}
W^{\prime }\theta \frac{\beta }{\theta \alpha -\beta \eta } \\ 
-W^{\prime }\frac{\beta ^{2}}{\theta \alpha -\beta \eta }%
\end{array}%
\right) \\
&&\left( 
\begin{array}{cc}
-W\frac{\beta ^{2}}{\theta \alpha -\beta \eta } & -W\theta \frac{\beta }{%
\theta \alpha -\beta \eta }%
\end{array}%
\right) 
\begin{array}{cc}
\left( \lambda _{+}-\lambda _{-}^{\prime }\right) ^{-1} & 0 \\ 
0 & \left( \lambda _{+}-\lambda _{+}^{\prime }\right) ^{-1}%
\end{array}%
\left( 
\begin{array}{c}
W^{\prime }\frac{\theta ^{2}}{\theta \alpha -\beta \eta } \\ 
-W^{\prime }\theta \frac{\beta }{\theta \alpha -\beta \eta }%
\end{array}%
\right)
\end{eqnarray*}%
leading to the modification:%
\begin{equation*}
\rightarrow \lambda _{-}-WW^{\prime }\theta ^{2}\beta ^{2}\frac{\lambda
_{+}^{\prime }-\lambda _{-}^{\prime }}{\left( \theta \alpha -\beta \eta
\right) ^{2}\left( \lambda _{-}-\lambda _{+}^{\prime }\right) \left( \lambda
-\lambda _{-}^{\prime }\right) }=\lambda _{-}+\Delta \lambda _{-}\left(
X\right)
\end{equation*}%
\begin{equation*}
\lambda _{+}+WW^{\prime }\theta ^{2}\beta ^{2}\frac{\lambda _{+}^{\prime
}-\lambda _{-}^{\prime }}{\left( \theta \alpha -\beta \eta \right)
^{2}\left( \lambda _{+}-\lambda _{+}^{\prime }\right) \left( \lambda
_{+}-\lambda _{-}^{\prime }\right) }=\lambda _{+}+\Delta \lambda _{+}\left(
X\right)
\end{equation*}%
In terms of the parameter of the system, the shifts write:%
\begin{eqnarray*}
&&\Delta \lambda _{-}\left( X\right) \\
&\rightarrow &\hat{S}_{E}\left( X^{\prime },X,\theta -1\right) \frac{1-\hat{S%
}\left( X^{\prime }\right) }{1-\hat{S}_{E}\left( X^{\prime }\right) }\hat{S}%
_{E}\left( X,X^{\prime },\theta -1\right) \frac{1-\hat{S}\left( X\right) }{1-%
\hat{S}_{E}\left( X\right) }\theta ^{2}\beta ^{2}\frac{\lambda _{+}^{\prime
}-\lambda _{-}^{\prime }}{\left( \theta \alpha -\beta \eta \right)
^{2}\left( \lambda _{-}-\lambda _{+}^{\prime }\right) \left( \lambda
-\lambda _{-}^{\prime }\right) }
\end{eqnarray*}

\begin{eqnarray*}
&&\Delta \lambda _{+}\left( X\right) \\
&=&\hat{S}_{E}\left( X^{\prime },X,\theta -1\right) \frac{1-\hat{S}\left(
X^{\prime }\right) }{1-\hat{S}_{E}\left( X^{\prime }\right) }\hat{S}%
_{E}\left( X,X^{\prime },\theta -1\right) \frac{1-\hat{S}\left( X\right) }{1-%
\hat{S}_{E}\left( X\right) }\theta ^{2}\beta ^{2}\frac{\lambda _{+}^{\prime
}-\lambda _{-}^{\prime }}{\left( \theta \alpha -\beta \eta \right)
^{2}\left( \lambda _{+}-\lambda _{+}^{\prime }\right) \left( \lambda
_{+}-\lambda _{-}^{\prime }\right) } \\
&=&\frac{\hat{S}_{E}\left( X^{\prime },X,\theta -1\right) \frac{1-\hat{S}%
\left( X^{\prime }\right) }{1-\hat{S}_{E}\left( X^{\prime }\right) }\hat{S}%
_{E}\left( X,X^{\prime },\theta -1\right) \frac{1-\hat{S}\left( X\right) }{1-%
\hat{S}_{E}\left( X\right) }}{\lambda _{X}-\lambda _{X^{\prime }}}\frac{h^{2}%
}{\left( \alpha -\beta \right) ^{2}+4ch}
\end{eqnarray*}%
The largest eigenvalue may switch towards $\lambda _{X}>>1$ which leads the
sector to the unstable case.

\subsubsection*{A11.2.2 Loops of several interacting sectors}

More genrally, we can consider deviations of shares with respect to average
participations that draw a loop among the system of investors. For such
loops of investment, the modification in eigenvalues are composed of
contributions:

\begin{equation*}
\Delta \lambda _{i\alpha }=\prod \frac{W_{j_{k+1}\alpha _{k+1},j_{k}\alpha
_{k}}}{\lambda _{i\alpha }-\lambda _{j_{k+1}\alpha _{k+1}}}
\end{equation*}%
where $\lambda _{i\alpha }$ r eigenvalues for sector $i$ $\alpha =\pm $ and $%
W_{j_{k+1}\alpha _{k+1},j_{k}\alpha _{k}}$ the interaction between $%
j_{k+1}\alpha _{k+1}$ nd $j_{k}\alpha _{k}$. Summing over all possible path
yields the modifications: 
\begin{equation*}
\Delta \lambda _{i\alpha }=\sum_{\left( j_{k}\alpha _{k}\right) }\prod \frac{%
W_{j_{k+1}\alpha _{k+1},j_{k}\alpha _{k}}}{\lambda _{i\alpha }-\lambda
_{j_{k+1}\alpha _{k+1}}}
\end{equation*}%
Such loops modify the eigenvalues and the stability of the state. If the
largest eigenvalue is shifted to a positive value, the resulting intability
shifts the state towards an other one. If returns are driven towards $0$ or
less, a default state is activated.

\subsubsection*{A11.2.3 Variations in terms of share}

The variations of the system and transitions can directly be written in
terms of shares between investors. Actually:%
\begin{eqnarray*}
\delta \hat{S}_{E}\left( X^{\prime },X,\theta \right) &=&\frac{\hat{w}\left(
X^{\prime },X\right) }{2}\left( 1+\left( \delta \hat{f}\left( X^{\prime
}\right) -\left\langle h\left( X\right) \right\rangle \delta \frac{f\left(
X\right) +r\left( X\right) }{2}\right) \right) \\
&\rightarrow &\frac{\hat{w}\left( X^{\prime },X\right) }{2}\delta \hat{f}%
\left( X^{\prime },\theta \right)
\end{eqnarray*}%
\begin{eqnarray*}
\delta \hat{S}\left( X^{\prime },X,\theta \right) &=&\hat{w}\left( X^{\prime
},X\right) \left( 1+\delta \Delta \left( \frac{\hat{f}\left( X^{\prime
}\right) +\hat{r}\left( X^{\prime }\right) }{2}\right) \right) \\
&\rightarrow &\hat{w}\left( X^{\prime },X\right) \frac{\delta \hat{f}\left(
X^{\prime },\theta \right) +\hat{r}\left( X^{\prime },,\theta \right) }{2}
\end{eqnarray*}%
where:%
\begin{equation*}
\hat{w}\left( X^{\prime },X\right) \rightarrow \frac{\left( 1-\left( \gamma
\left\langle \hat{S}_{E}\left( X\right) \right\rangle \right) ^{2}\right) 
\hat{w}_{E}^{\left( 0\right) }\left( X^{\prime },X\right) }{1+\hat{w}%
_{E}^{\left( 0\right) }\left( X^{\prime },X\right) \left( 1-\left( \gamma
\left\langle \hat{S}_{E}\left( X\right) \right\rangle \right) ^{2}\right)
+\left( \gamma \left\langle \hat{S}_{E}\left( X_{1},X^{\prime }\right)
\right\rangle _{X_{1}}\right) ^{2}-\left( \gamma \left\langle \hat{S}%
_{E}\left( X\right) \right\rangle \right) ^{2}}
\end{equation*}%
The formula for $\hat{w}\left( X^{\prime },X\right) $ shows that this
parameter is mainly exogeneous.

The dynamic equations for $\delta \hat{f}\left( X,\theta \right) $, $\delta
S\left( X,\theta -1\right) $ can be replaced by a dynamic equation: 
\begin{equation*}
\left( 
\begin{array}{c}
\delta \hat{S}\left( X^{\prime },X,\theta \right) -\delta \hat{S}\left(
X^{\prime },X,\theta -1\right) \\ 
\delta S\left( X,\theta -1\right) -\delta S\left( X,\theta -2\right)%
\end{array}%
\right) =\left( 
\begin{array}{cc}
\frac{\hat{w}\left( X^{\prime },X\right) }{2}\alpha & \frac{\hat{w}\left(
X^{\prime },X\right) }{2}c \\ 
h & \beta%
\end{array}%
\right) \left( 
\begin{array}{c}
\frac{\delta \hat{S}\left( X^{\prime },X,\theta -1\right) }{\frac{\hat{w}%
\left( X^{\prime },X\right) }{2}} \\ 
\delta S\left( X,\theta -2\right)%
\end{array}%
\right)
\end{equation*}%
\begin{equation*}
\left( 
\begin{array}{c}
\delta \hat{S}\left( X^{\prime },X,\theta \right) -\delta \hat{S}\left(
X^{\prime },X,\theta -1\right) \\ 
\delta S\left( X,\theta -1\right) -\delta S\left( X,\theta -2\right)%
\end{array}%
\right) =\left( 
\begin{array}{cc}
\alpha & \frac{\hat{w}\left( X^{\prime },X\right) }{2}c \\ 
\frac{h}{\frac{\hat{w}\left( X^{\prime },X\right) }{2}} & \beta%
\end{array}%
\right) \left( 
\begin{array}{c}
\delta \hat{S}\left( X^{\prime },X,\theta -1\right) \\ 
\delta S\left( X,\theta -2\right)%
\end{array}%
\right)
\end{equation*}%
leading to the same eigenvalues.

\end{document}